\documentclass[twoside,a4paper]{amsart}
\PassOptionsToPackage{pdfauthor={Vladimir V. Kisil},%
  pdftitle={Schwerdtfeger--Fillmore--Springer--Cnops Construction Implemented in GiNaC},%
  pdfsubject={Library description},%
  backref=page,%
  pagebackref=true, %
  pdfkeywords={GiNaC, cycle, CAS},%
  breaklinks=true,%
  bookmarks=true,%
}{hyperref}
\usepackage{xr-hyper}
\usepackage[printedin,extnum]{myamsart}
\input{mydef}
\ppnum{LEEDS--MATH--PURE--2005--29\\
}{arXiv: \href{http://arXiv.org/abs/cs.MS/0512073}{cs.MS/0512073}}
{2005}
\usepackage{noweb}
\usepackage{graphicx}
\usepackage{marvosym}
\nwfilename{parab-ortho1.nw}
\makeatletter
\let\tab=&
\def\idxexample#1{\nwix@id@uses#1}
\makeatother
\def\nwlbrace{\textbf{\texttt{\char123}}}
\def\nwrbrace{\textbf{\texttt{\char125}}}

\newcommand{\NoWEB}{\texttt{noweb}}
\providecommand{\MetaPost}{\texttt{Meta}\-\texttt{Post}}
\providecommand{\GiNaC}{\textsf{GiNaC}}

\providecommand{\cycle}[3][]{{#1 C^{#2}_{#3}}}
\providecommand{\realline}[3][]{#1 R^{#2}_{#3}}
\providecommand{\Asymptote}{\texttt{Asymptote}}
\providecommand{\bs}{\breve{\sigma}}
\providecommand{\rs}{\mathring{\sigma}}
\newif\iftth

\providecommand{\tr}{\mathop{tr}}
\providecommand{\vecbf}[1]{\mathbf{#1}}
\providecommand{\clifford}[2][]{\ifcase #1 #2\or \tilde{#2} \or \breve{#2} \fi}
\hypersetup{colorlinks=true,bookmarks=true}
\begin{document}
\def\LA{\begingroup\maybehbox\bgroup\setupmodname\Rm$\langle$}\def\RA{$\rangle$\egroup\endgroup}\providecommand{\MM}{\kern.5pt\raisebox{.4ex}{\begin{math}\scriptscriptstyle-\kern-1pt-\end{math}}\kern.5pt}\providecommand{\PP}{\kern.5pt\raisebox{.4ex}{\begin{math}\scriptscriptstyle+\kern-1pt+\end{math}}\kern.5pt}\def\commopen{/\begin{math}\ast\,\end{math}}\def\commclose{\,\begin{math}\ast\end{math}\kern-.5pt/}\def\begcomm{\begingroup\maybehbox\bgroup\setupmodname}\def\endcomm{\egroup\endgroup}\nwfilename{cycle.nw}\nwbegindocs{1}%
\nwenddocs{}\nwbegindocs{2}%
\nwenddocs{}\nwbegindocs{3}%
\nwenddocs{}\nwbegindocs{4}%
\nwenddocs{}\nwbegindocs{5}%
\nwenddocs{}\nwbegindocs{6}%
\nwenddocs{}\nwbegindocs{7}%
\nwenddocs{}\nwbegindocs{8}%
\nwenddocs{}\nwbegindocs{9}%
\nwenddocs{}\nwbegindocs{10}%
\nwenddocs{}\nwbegindocs{11}%
\nwenddocs{}\nwbegindocs{12}%
\nwenddocs{}\nwbegindocs{13}%
\nwenddocs{}\nwbegindocs{14}%
\nwenddocs{}\nwbegindocs{15}%
\nwenddocs{}\nwbegindocs{16}%
\nwdocspar
\title[Schwerdtfeger--Fillmore--Springer--Cnops Construction Implemented in
\GiNaC]{Schwerdtfeger--Fillmore--Springer--Cnops Construction\\ Implemented in \GiNaC}

\author[Vladimir V. Kisil]%
{\href{http://www.maths.leeds.ac.uk/~kisilv/}{Vladimir V. Kisil}}
\thanks{On  leave from Odessa University.}

\address{%
School of Mathematics\\
University of Leeds\\
Leeds LS2\,9JT\\
UK
}

\email{\href{mailto:kisilv@maths.leeds.ac.uk}{kisilv@maths.leeds.ac.uk}}

\urladdr{\href{http://www.maths.leeds.ac.uk/~kisilv/}%
{http://www.maths.leeds.ac.uk/\~{}kisilv/}}

\begin{abstract}
  This is an implementation of the Schwerdtfeger--Fillmore--Springer--Cnops
  construction (SFSCc) based on the Clifford algebra
  capacities~\cite{Kisil04c} of the \GiNaC\ computer algebra system.
  SFSCc linearises the linear-fraction action of the M\"obius group.
  This turns to be very useful in several theoretical and applied fields
  including engineering.  The package is realised as a C++ library and
  there are several Python wrapper of it, which can be used in
  interactive mode. 

  The core of this realisation of SFSCc is done for an arbitrary
  dimension, while a subclass for two dimensional cycles add some
  2D-specific routines including a visualisation to PostScript files
  through the \MetaPost\ or \Asymptote\ software.

  This library is a backbone of many results published
  in~\cite{Kisil05a}, which serve as illustrations of its usage. It can be
  ported (with various level of required changes) to other CAS with
  Clifford algebras capabilities.

  There is an ISO image of a Live Debian DVD attached to this paper
  at \texttt{arXiv} and the \href{https://docs.google.com/file/d/0BzfWNH9hAT3VMFl6Z3Z4aVJmcW8}{Google drive} (an updated version).

  The software is distributed under GNU GPLv3, see Appendix~\ref{sec:license}.
\end{abstract}
\dedicatory{Dedicated to the memory of Dennis Ritchie}
\date{\today\ (v2.3)}
\maketitle

\tableofcontents

\section{Introduction}
\label{sec:introduction}
The usage of computer algebra system (CAS) in Clifford Algebra
research has an established history with the famous ``Green
book''~\cite{DelSomSou92} already accompanied by a floppy disk with a
\texttt{REDUCE} package. This tradition is very much alive, see for
example the recent
books~\cites{GuerlebeckHabetaSproessig06,Kisil12a,Kisil11d}
accompanied by a software CD/DVD.  Numerous new packages are developed
by various research teams across the world to work with Clifford
algebras generally or address specific tasks, see on-line proceedings
of the recent IKM-2006 conference~\cite{IKM2006}.

Along this lines the present paper presents an implementation of the
Schwerdtfeger--Fill\-more--Springer--Cnops construction\footnote{In
  the case of circles this technique was already spectacularly
  developed by H.~Schwerdtfeger in 1960-ies, see
  \cite{Schwerdtfeger79a}. Unfortunately, that beautiful book was not
  known to the present author until he accomplished his own
  works~\cites{Kisil05a,Kisil05b,Kisil12a}.} (SFSCc) along with
illustrations of its usage.
SFSCc~\citelist{\cite{Schwerdtfeger79a}*{\S~1.1}
  \cite{Cnops02a}*{\S~4.1} \cite{Porteous95}*{\S~18}
  \cite{FillmoreSpringer90a} \cite{Kirillov06}*{\S~4.2}
  \cite{Kisil05a} \cite{Kisil12a}*{\S~4.2}} linearises the
linear-fraction action of the M\"obius group in \(\Space{R}{n}\).
This has clear advantages in several theoretical and applied fields
including engineering.  Our implementation is based on the Clifford
algebra capacities of the \GiNaC\ computer algebra
system~\cite{GiNaC}, which were described in~\cite{Kisil04c}. The code
is written using \NoWEB\
\href{http://en.wikipedia.org/wiki/Literate_programming}{literate
  programming} tool~\cite{NoWEB}

The core of this realisation of SFSCc is done for an arbitrary
dimension of \(\Space{R}{n}\) with a metric given by an arbitrary
bilinear form.  We also present a subclass for two dimensional cycles
(i.e. circles, parabolas and hyperbolas), which add some 2D specific
routines including a visualisation to PostScript files through the
\MetaPost~\cite{MetaPost} or \Asymptote~\cite{Asymptote} packages.
This software is the backbone of many results published
in~\cites{Kisil05a,Kisil12a,Kisil11d} and we use its application to~\cite{Kisil05a}
for the demonstration purpose.

There is a Python wrapper~\cite{pycycle} for this library. It
is based on \texttt{BoostPython} and \texttt{pyGiNaC} packages. The
wrapper allows to use all functions and methods from the library in
Python scripts or Python interactive shell. The drawing of object from
{\Tt{}\Rm{}{\bf{}cycle2D}\nwendquote} may be instantly seen in the interactive mode through the
\Asymptote. The live DVD supplied with book~\cite{Kisil12a} is based
on the library presented in this paper and its Python wrapper.

This library is now a part of MoebInv project
(\url{http://moebinv.sourceforge.net/})~\cite{Kisil14b}. Please look
there for latest updates, source and binary distributions. ISO images of
live DVD may be referred there as well. We do not
plan to use arXiv for these purposes anymore. 

The present package can be ported (with various level of required
changes) to other CAS with Clifford algebras capabilities similar to
\GiNaC.

The software is distributed under GNU GPLv3, see
Appendix~\ref{sec:license} and~\cite{GNUGPL}.

\nwenddocs{}\nwbegindocs{17}\nwdocspar

\section{User interface to classes cycle and cycle2D}
\label{sec:user-interf-class}
The {\Tt{}\Rm{}{\bf{}cycle} {\bf{}class}\nwendquote} describes loci of points \(\vecbf{x}\in\Space{R}{n}\) defined by a
quadratic equation
\begin{equation}
  \label{eq:quadric-def}
  k\vecbf{x}^2-2\scalar{\vecbf{l}}{\vecbf{x}}+m=0, \quad \textrm{
    where } k,m\in\Space{R}{}, \ \vecbf{l}\in\Space{R}{n}.
\end{equation}
The class {\Tt{}\Rm{}{\bf{}cycle}\nwendquote} correspondingly has member variables {\Tt{}\Rm{}{\it{}k}\nwendquote}, {\Tt{}\Rm{}{\it{}l}\nwendquote},
{\Tt{}\Rm{}{\it{}m}\nwendquote} to describe the equation~\eqref{eq:quadric-def} and the Clifford
algebra {\Tt{}\Rm{}{\it{}unit}\nwendquote} to describe the metric of surrounding space. The
plenty of methods are supplied for various tasks within SFSCc.

We also define a subclass {\Tt{}\Rm{}{\bf{}cycle2D}\nwendquote} which has more methods specific
to two dimensional environment.

\nwenddocs{}\nwbegindocs{18}\nwdocspar
\subsection[Constructors of cycle]{Constructors of {\Tt{}\Rm{}{\bf{}cycle}\nwendquote}}
\label{sec:constructors-cycle}

Here is various constructors for the {\Tt{}\Rm{}{\bf{}cycle}\nwendquote}s. The first one takes
values of \(k\), \(\vecbf{l}\), \(m\) as well as {\Tt{}\Rm{}{\it{}metric}\nwendquote} supplied
directly. Note that \(\vecbf{l}\) is admitted either in form of a {\Tt{}\Rm{}{\bf{}lst}\nwendquote}, {\Tt{}\Rm{}{\bf{}matrix}\nwendquote}
or {\Tt{}\Rm{}{\bf{}indexed}\nwendquote} objects from \GiNaC. Similarly {\Tt{}\Rm{}{\it{}metric}\nwendquote} can be given
by an object from either {\Tt{}\Rm{}{\bf{}tensor}\nwendquote}, {\Tt{}\Rm{}{\bf{}indexed}\nwendquote}, {\Tt{}\Rm{}{\bf{}matrix}\nwendquote} or {\Tt{}\Rm{}{\bf{}clifford}\nwendquote} classes
exactly in the same way as metric is provided for a
{\Tt{}\Rm{}{\it{}clifford\_unit}()\nwendquote} constructors~\cite{Kisil04c}.
\nwenddocs{}\nwbegincode{19}\sublabel{NW3gGP3e-2EuhSt-1}\nwmargintag{{\nwtagstyle{}\subpageref{NW3gGP3e-2EuhSt-1}}}\moddef{cycle class constructors~{\nwtagstyle{}\subpageref{NW3gGP3e-2EuhSt-1}}}\endmoddef\Rm{}\nwstartdeflinemarkup\nwusesondefline{\\{NW3gGP3e-2wwyff-1}}\nwprevnextdefs{\relax}{NW3gGP3e-2EuhSt-2}\nwenddeflinemarkup
{\bf{}public}:
{\bf{}cycle}({\bf{}const} {\bf{}ex} & {\it{}k}, {\bf{}const} {\bf{}ex} & {\it{}l}, {\bf{}const} {\bf{}ex} & {\it{}m},\nwindexdefn{\nwixident{cycle}}{cycle}{NW3gGP3e-2EuhSt-1}
   {\bf{}const} {\bf{}ex} & {\it{}metr} = -({\bf{}new} {\it{}tensdelta})\begin{math}\rightarrow\end{math}{\it{}setflag}({\it{}status\_flags}::{\it{}dynallocated}));
\nwindexdefn{\nwixident{cycle}}{cycle}{NW3gGP3e-2EuhSt-1}\nwindexdefn{\nwixident{k}}{k}{NW3gGP3e-2EuhSt-1}\nwindexdefn{\nwixident{l}}{l}{NW3gGP3e-2EuhSt-1}\nwindexdefn{\nwixident{m}}{m}{NW3gGP3e-2EuhSt-1}\nwindexdefn{\nwixident{metr}}{metr}{NW3gGP3e-2EuhSt-1}\eatline
\nwalsodefined{\\{NW3gGP3e-2EuhSt-2}\\{NW3gGP3e-2EuhSt-3}\\{NW3gGP3e-2EuhSt-4}}\nwused{\\{NW3gGP3e-2wwyff-1}}\nwidentdefs{\\{{\nwixident{cycle}}{cycle}}\\{{\nwixident{k}}{k}}\\{{\nwixident{l}}{l}}\\{{\nwixident{m}}{m}}\\{{\nwixident{metr}}{metr}}}\nwidentuses{\\{{\nwixident{ex}}{ex}}}\nwindexuse{\nwixident{ex}}{ex}{NW3gGP3e-2EuhSt-1}\nwendcode{}\nwbegindocs{20}\nwdocspar
\nwenddocs{}\nwbegindocs{21} Constructor for a {\Tt{}\Rm{}{\bf{}cycle}\nwendquote}~\eqref{eq:quadric-def} with \(k=1\) and
given {\Tt{}\Rm{}{\it{}l}\nwendquote} defined by the condition that square of its ``radius''
(which is \(\det C\), see~\cite[Defn.~\ref{E-de:radius}]{Kisil05a}) is
{\Tt{}\Rm{}{\it{}r\_squared}\nwendquote}. If a non-zero {\Tt{}\Rm{}{\it{}e}\nwendquote} is provided, then it is used to
calculate {\Tt{}\Rm{}{\it{}C}.{\it{}det}({\it{}e})\nwendquote}, otherwise the default value is {\Tt{}\Rm{}{\it{}C}.{\it{}det}({\it{}metr})\nwendquote}.
Note that for the default value of the {\Tt{}\Rm{}{\it{}metr}\nwendquote} the
value of {\Tt{}\Rm{}{\it{}l}\nwendquote} coincides with the centre of this {\Tt{}\Rm{}{\bf{}cycle}\nwendquote}.

\nwenddocs{}\nwbegincode{22}\sublabel{NW3gGP3e-2EuhSt-2}\nwmargintag{{\nwtagstyle{}\subpageref{NW3gGP3e-2EuhSt-2}}}\moddef{cycle class constructors~{\nwtagstyle{}\subpageref{NW3gGP3e-2EuhSt-1}}}\plusendmoddef\Rm{}\nwstartdeflinemarkup\nwusesondefline{\\{NW3gGP3e-2wwyff-1}}\nwprevnextdefs{NW3gGP3e-2EuhSt-1}{NW3gGP3e-2EuhSt-3}\nwenddeflinemarkup
{\bf{}cycle}({\bf{}const} {\bf{}lst} & {\it{}l},\nwindexdefn{\nwixident{cycle}}{cycle}{NW3gGP3e-2EuhSt-2}
   {\bf{}const} {\bf{}ex} & {\it{}metr} = -({\bf{}new} {\it{}tensdelta})\begin{math}\rightarrow\end{math}{\it{}setflag}({\it{}status\_flags}::{\it{}dynallocated}),
   {\bf{}const} {\bf{}ex} & {\it{}r\_squared} = 0, {\bf{}const} {\bf{}ex} & {\it{}e} = 0,
   {\bf{}const} {\bf{}ex} & {\it{}sign} = ({\bf{}new} {\it{}tensdelta})\begin{math}\rightarrow\end{math}{\it{}setflag}({\it{}status\_flags}::{\it{}dynallocated}));
 
\nwused{\\{NW3gGP3e-2wwyff-1}}\nwidentdefs{\\{{\nwixident{cycle}}{cycle}}}\nwidentuses{\\{{\nwixident{ex}}{ex}}\\{{\nwixident{l}}{l}}\\{{\nwixident{lst}}{lst}}\\{{\nwixident{metr}}{metr}}}\nwindexuse{\nwixident{ex}}{ex}{NW3gGP3e-2EuhSt-2}\nwindexuse{\nwixident{l}}{l}{NW3gGP3e-2EuhSt-2}\nwindexuse{\nwixident{lst}}{lst}{NW3gGP3e-2EuhSt-2}\nwindexuse{\nwixident{metr}}{metr}{NW3gGP3e-2EuhSt-2}\nwendcode{}\nwbegindocs{23}If we want to have a cycle identical to to a given one {\Tt{}\Rm{}{\it{}C}\nwendquote}
up to a space metric which should be replaced by a new one {\Tt{}\Rm{}{\it{}metr}\nwendquote},
we can use the next constructor.
\nwenddocs{}\nwbegincode{24}\sublabel{NW3gGP3e-2EuhSt-3}\nwmargintag{{\nwtagstyle{}\subpageref{NW3gGP3e-2EuhSt-3}}}\moddef{cycle class constructors~{\nwtagstyle{}\subpageref{NW3gGP3e-2EuhSt-1}}}\plusendmoddef\Rm{}\nwstartdeflinemarkup\nwusesondefline{\\{NW3gGP3e-2wwyff-1}}\nwprevnextdefs{NW3gGP3e-2EuhSt-2}{NW3gGP3e-2EuhSt-4}\nwenddeflinemarkup
{\bf{}cycle}({\bf{}const} {\bf{}cycle} & {\it{}C}, {\bf{}const} {\bf{}ex} & {\it{}metr});

\nwused{\\{NW3gGP3e-2wwyff-1}}\nwidentuses{\\{{\nwixident{cycle}}{cycle}}\\{{\nwixident{ex}}{ex}}\\{{\nwixident{metr}}{metr}}}\nwindexuse{\nwixident{cycle}}{cycle}{NW3gGP3e-2EuhSt-3}\nwindexuse{\nwixident{ex}}{ex}{NW3gGP3e-2EuhSt-3}\nwindexuse{\nwixident{metr}}{metr}{NW3gGP3e-2EuhSt-3}\nwendcode{}\nwbegindocs{25}To any cycle SFSCc associates a matrix, which is of the
form~\eqref{eq:matrix-from-cycle}~\cite[~\eqref{E-eq:matrix-for-cycle}]{Kisil05a}.
The following constructor make a {\Tt{}\Rm{}{\bf{}cycle}\nwendquote} from its matrix
representation, i.e. it is the realisation of the inverse of the map
\(Q\)~\cite[\eqref{E-eq:matrix-for-cycle}]{Kisil05a}.

\nwenddocs{}\nwbegincode{26}\sublabel{NW3gGP3e-2EuhSt-4}\nwmargintag{{\nwtagstyle{}\subpageref{NW3gGP3e-2EuhSt-4}}}\moddef{cycle class constructors~{\nwtagstyle{}\subpageref{NW3gGP3e-2EuhSt-1}}}\plusendmoddef\Rm{}\nwstartdeflinemarkup\nwusesondefline{\\{NW3gGP3e-2wwyff-1}}\nwprevnextdefs{NW3gGP3e-2EuhSt-3}{\relax}\nwenddeflinemarkup
{\bf{}cycle}({\bf{}const} {\bf{}matrix} & {\it{}M}, {\bf{}const} {\bf{}ex} & {\it{}metr}, {\bf{}const} {\bf{}ex} & {\it{}e} = 0, {\bf{}const} {\bf{}ex} & {\it{}sign} = 0);

\nwused{\\{NW3gGP3e-2wwyff-1}}\nwidentuses{\\{{\nwixident{cycle}}{cycle}}\\{{\nwixident{ex}}{ex}}\\{{\nwixident{matrix}}{matrix}}\\{{\nwixident{metr}}{metr}}}\nwindexuse{\nwixident{cycle}}{cycle}{NW3gGP3e-2EuhSt-4}\nwindexuse{\nwixident{ex}}{ex}{NW3gGP3e-2EuhSt-4}\nwindexuse{\nwixident{matrix}}{matrix}{NW3gGP3e-2EuhSt-4}\nwindexuse{\nwixident{metr}}{metr}{NW3gGP3e-2EuhSt-4}\nwendcode{}\nwbegindocs{27}\nwdocspar
\subsection[Accessing parameters of a cycle]{Accessing parameters of a {\Tt{}\Rm{}{\bf{}cycle}\nwendquote}}
\label{sec:access-param-cycle}

The following set of methods {\Tt{}\Rm{}{\it{}get\_}\begin{math}\ast\end{math}()\nwendquote} provide a reading access to
the various data in the class.
\nwenddocs{}\nwbegincode{28}\sublabel{NW3gGP3e-rnSJR-1}\nwmargintag{{\nwtagstyle{}\subpageref{NW3gGP3e-rnSJR-1}}}\moddef{accessing the data of a cycle~{\nwtagstyle{}\subpageref{NW3gGP3e-rnSJR-1}}}\endmoddef\Rm{}\nwstartdeflinemarkup\nwusesondefline{\\{NW3gGP3e-2wwyff-1}}\nwprevnextdefs{\relax}{NW3gGP3e-rnSJR-2}\nwenddeflinemarkup
{\bf{}public}:
 {\bf{}virtual} {\bf{}inline} {\bf{}ex} {\it{}get\_dim}() {\bf{}const} {\nwlbrace} {\bf{}return} {\it{}ex\_to}\begin{math}<\end{math}{\bf{}varidx}\begin{math}>\end{math}({\it{}unit}.{\it{}op}(1)).{\it{}get\_dim}(); {\nwrbrace}
 {\bf{}virtual} {\bf{}inline} {\bf{}ex} {\it{}get\_metric}() {\bf{}const} {\nwlbrace} {\bf{}return} {\it{}ex\_to}\begin{math}<\end{math}{\bf{}clifford}\begin{math}>\end{math}({\it{}unit}).{\it{}get\_metric}(); {\nwrbrace}
 {\bf{}virtual} {\bf{}inline} {\bf{}ex} {\it{}get\_metric}({\bf{}const} {\bf{}ex} &{\it{}i0}, {\bf{}const} {\bf{}ex} &{\it{}i1}) {\bf{}const}
      {\nwlbrace} {\bf{}return} {\it{}ex\_to}\begin{math}<\end{math}{\bf{}clifford}\begin{math}>\end{math}({\it{}unit}).{\it{}get\_metric}({\it{}i0}, {\it{}i1}); {\nwrbrace}
 {\bf{}virtual} {\bf{}inline} {\bf{}ex} {\it{}get\_k}() {\bf{}const} {\nwlbrace} {\bf{}return} {\it{}k}; {\nwrbrace}
\nwindexdefn{\nwixident{get{\_}dim}}{get:undim}{NW3gGP3e-rnSJR-1}\nwindexdefn{\nwixident{get{\_}metric}}{get:unmetric}{NW3gGP3e-rnSJR-1}\nwindexdefn{\nwixident{get{\_}k}}{get:unk}{NW3gGP3e-rnSJR-1}\eatline
\nwalsodefined{\\{NW3gGP3e-rnSJR-2}\\{NW3gGP3e-rnSJR-3}\\{NW3gGP3e-rnSJR-4}}\nwused{\\{NW3gGP3e-2wwyff-1}}\nwidentdefs{\\{{\nwixident{get{\_}dim}}{get:undim}}\\{{\nwixident{get{\_}k}}{get:unk}}\\{{\nwixident{get{\_}metric}}{get:unmetric}}}\nwidentuses{\\{{\nwixident{ex}}{ex}}\\{{\nwixident{k}}{k}}\\{{\nwixident{op}}{op}}\\{{\nwixident{varidx}}{varidx}}}\nwindexuse{\nwixident{ex}}{ex}{NW3gGP3e-rnSJR-1}\nwindexuse{\nwixident{k}}{k}{NW3gGP3e-rnSJR-1}\nwindexuse{\nwixident{op}}{op}{NW3gGP3e-rnSJR-1}\nwindexuse{\nwixident{varidx}}{varidx}{NW3gGP3e-rnSJR-1}\nwendcode{}\nwbegindocs{29}\nwdocspar
\nwenddocs{}\nwbegindocs{30}The member {\Tt{}\Rm{}{\it{}l}\nwendquote} can be obtained as the whole by the call
{\Tt{}\Rm{}{\it{}get\_l}()\nwendquote}, or its individual component is read, for example, by {\Tt{}\Rm{}{\it{}get\_l}(1)\nwendquote}.
\nwenddocs{}\nwbegincode{31}\sublabel{NW3gGP3e-rnSJR-2}\nwmargintag{{\nwtagstyle{}\subpageref{NW3gGP3e-rnSJR-2}}}\moddef{accessing the data of a cycle~{\nwtagstyle{}\subpageref{NW3gGP3e-rnSJR-1}}}\plusendmoddef\Rm{}\nwstartdeflinemarkup\nwusesondefline{\\{NW3gGP3e-2wwyff-1}}\nwprevnextdefs{NW3gGP3e-rnSJR-1}{NW3gGP3e-rnSJR-3}\nwenddeflinemarkup
{\bf{}inline} {\bf{}ex} {\it{}get\_l}() {\bf{}const} {\nwlbrace} {\bf{}return} {\it{}l}; {\nwrbrace}
{\bf{}inline} {\bf{}ex} {\it{}get\_l}({\bf{}const} {\bf{}ex} & {\it{}i}) {\bf{}const}
{\nwlbrace} {\bf{}return} ({\it{}l}.{\it{}is\_zero}()?0:{\it{}l}.{\it{}subs}({\it{}l}.{\it{}op}(1) \begin{math}\equiv\end{math} {\it{}i}, {\it{}subs\_options}::{\it{}no\_pattern})); {\nwrbrace}
{\bf{}inline} {\bf{}ex} {\it{}get\_m}() {\bf{}const} {\nwlbrace}{\bf{}return} {\it{}m};{\nwrbrace}
{\bf{}inline} {\bf{}ex} {\it{}get\_unit}() {\bf{}const} {\nwlbrace}{\bf{}return} {\it{}unit};{\nwrbrace}
\nwindexdefn{\nwixident{get{\_}l}}{get:unl}{NW3gGP3e-rnSJR-2}\nwindexdefn{\nwixident{get{\_}m}}{get:unm}{NW3gGP3e-rnSJR-2}\nwindexdefn{\nwixident{get{\_}unit}}{get:ununit}{NW3gGP3e-rnSJR-2}\eatline
\nwused{\\{NW3gGP3e-2wwyff-1}}\nwidentdefs{\\{{\nwixident{get{\_}l}}{get:unl}}\\{{\nwixident{get{\_}m}}{get:unm}}\\{{\nwixident{get{\_}unit}}{get:ununit}}}\nwidentuses{\\{{\nwixident{ex}}{ex}}\\{{\nwixident{is{\_}zero}}{is:unzero}}\\{{\nwixident{l}}{l}}\\{{\nwixident{m}}{m}}\\{{\nwixident{op}}{op}}\\{{\nwixident{subs}}{subs}}}\nwindexuse{\nwixident{ex}}{ex}{NW3gGP3e-rnSJR-2}\nwindexuse{\nwixident{is{\_}zero}}{is:unzero}{NW3gGP3e-rnSJR-2}\nwindexuse{\nwixident{l}}{l}{NW3gGP3e-rnSJR-2}\nwindexuse{\nwixident{m}}{m}{NW3gGP3e-rnSJR-2}\nwindexuse{\nwixident{op}}{op}{NW3gGP3e-rnSJR-2}\nwindexuse{\nwixident{subs}}{subs}{NW3gGP3e-rnSJR-2}\nwendcode{}\nwbegindocs{32} 
\nwenddocs{}\nwbegindocs{33}Methods {\Tt{}\Rm{}{\it{}nops}()\nwendquote}, {\Tt{}\Rm{}{\it{}op}()\nwendquote}, {\Tt{}\Rm{}{\it{}let\_op}()\nwendquote}, {\Tt{}\Rm{}{\it{}is\_equal}()\nwendquote},
{\Tt{}\Rm{}{\it{}subs}()\nwendquote} are standard for expression in \GiNaC\ and described in the
\GiNaC\ tutorial.
The first three methods are rarely called by a user.
In many cases the method {\Tt{}\Rm{}{\it{}subs}()\nwendquote}
may replaced by more suitable {\Tt{}\Rm{}{\it{}subject\_to}()\nwendquote}~\ref{sec:geom-meth-cycle}.

\nwenddocs{}\nwbegincode{34}\sublabel{NW3gGP3e-rnSJR-3}\nwmargintag{{\nwtagstyle{}\subpageref{NW3gGP3e-rnSJR-3}}}\moddef{accessing the data of a cycle~{\nwtagstyle{}\subpageref{NW3gGP3e-rnSJR-1}}}\plusendmoddef\Rm{}\nwstartdeflinemarkup\nwusesondefline{\\{NW3gGP3e-2wwyff-1}}\nwprevnextdefs{NW3gGP3e-rnSJR-2}{NW3gGP3e-rnSJR-4}\nwenddeflinemarkup
{\it{}size\_t} {\it{}nops}() {\bf{}const} {\nwlbrace}{\bf{}return} 4;{\nwrbrace}
{\bf{}ex} {\it{}op}({\it{}size\_t} {\it{}i}) {\bf{}const};
{\bf{}ex} & {\it{}let\_op}({\it{}size\_t} {\it{}i});
{\bf{}bool} {\it{}is\_equal}({\bf{}const} {\bf{}basic} & {\it{}other}, {\bf{}bool} {\it{}projectively} = {\bf{}true}) {\bf{}const};
{\bf{}bool} {\it{}is\_zero}() {\bf{}const};
{\bf{}cycle} {\it{}subs}({\bf{}const} {\bf{}ex} & {\it{}e}, {\bf{}unsigned} {\it{}options} = 0) {\bf{}const};
{\bf{}inline} {\bf{}cycle} {\it{}normal}() {\bf{}const}
 {\nwlbrace} {\bf{}return} {\bf{}cycle}({\it{}k}.{\it{}normal}(), {\it{}l}.{\it{}normal}(), {\it{}m}.{\it{}normal}(), {\it{}unit}.{\it{}normal}());{\nwrbrace}
{\bf{}inline} {\bf{}cycle} {\it{}expand}() {\bf{}const} {\nwlbrace} {\bf{}return} {\bf{}cycle}({\it{}k}.{\it{}expand}(), {\it{}l}.{\it{}expand}(), {\it{}m}.{\it{}expand}(), {\it{}unit});{\nwrbrace}
\nwindexdefn{\nwixident{nops}}{nops}{NW3gGP3e-rnSJR-3}\nwindexdefn{\nwixident{op}}{op}{NW3gGP3e-rnSJR-3}\nwindexdefn{\nwixident{let{\_}op}}{let:unop}{NW3gGP3e-rnSJR-3}\nwindexdefn{\nwixident{is{\_}equal}}{is:unequal}{NW3gGP3e-rnSJR-3}\nwindexdefn{\nwixident{is{\_}zero}}{is:unzero}{NW3gGP3e-rnSJR-3}\nwindexdefn{\nwixident{subs}}{subs}{NW3gGP3e-rnSJR-3}\nwindexdefn{\nwixident{normal}}{normal}{NW3gGP3e-rnSJR-3}\nwindexdefn{\nwixident{expand}}{expand}{NW3gGP3e-rnSJR-3}\eatline
\nwused{\\{NW3gGP3e-2wwyff-1}}\nwidentdefs{\\{{\nwixident{expand}}{expand}}\\{{\nwixident{is{\_}equal}}{is:unequal}}\\{{\nwixident{is{\_}zero}}{is:unzero}}\\{{\nwixident{let{\_}op}}{let:unop}}\\{{\nwixident{nops}}{nops}}\\{{\nwixident{normal}}{normal}}\\{{\nwixident{op}}{op}}\\{{\nwixident{subs}}{subs}}}\nwidentuses{\\{{\nwixident{bool}}{bool}}\\{{\nwixident{cycle}}{cycle}}\\{{\nwixident{ex}}{ex}}\\{{\nwixident{k}}{k}}\\{{\nwixident{l}}{l}}\\{{\nwixident{m}}{m}}}\nwindexuse{\nwixident{bool}}{bool}{NW3gGP3e-rnSJR-3}\nwindexuse{\nwixident{cycle}}{cycle}{NW3gGP3e-rnSJR-3}\nwindexuse{\nwixident{ex}}{ex}{NW3gGP3e-rnSJR-3}\nwindexuse{\nwixident{k}}{k}{NW3gGP3e-rnSJR-3}\nwindexuse{\nwixident{l}}{l}{NW3gGP3e-rnSJR-3}\nwindexuse{\nwixident{m}}{m}{NW3gGP3e-rnSJR-3}\nwendcode{}\nwbegindocs{35} 
\nwenddocs{}\nwbegindocs{36}We also provide a method {\Tt{}\Rm{}{\it{}the\_same\_as}()\nwendquote} which return a {\Tt{}\Rm{}{\it{}GiNaC}::{\bf{}lst}\nwendquote}
of identities (i.e. {\Tt{}\Rm{}{\it{}GiNaC}::{\bf{}relational}\nwendquote}s), which defines that two
cycles are given by the same point of the projective space
\(\Space{P}{3}\).
\nwenddocs{}\nwbegincode{37}\sublabel{NW3gGP3e-rnSJR-4}\nwmargintag{{\nwtagstyle{}\subpageref{NW3gGP3e-rnSJR-4}}}\moddef{accessing the data of a cycle~{\nwtagstyle{}\subpageref{NW3gGP3e-rnSJR-1}}}\plusendmoddef\Rm{}\nwstartdeflinemarkup\nwusesondefline{\\{NW3gGP3e-2wwyff-1}}\nwprevnextdefs{NW3gGP3e-rnSJR-3}{\relax}\nwenddeflinemarkup
{\bf{}ex} {\it{}the\_same\_as}({\bf{}const} {\bf{}basic} & {\it{}other}) {\bf{}const};

\nwused{\\{NW3gGP3e-2wwyff-1}}\nwidentuses{\\{{\nwixident{ex}}{ex}}}\nwindexuse{\nwixident{ex}}{ex}{NW3gGP3e-rnSJR-4}\nwendcode{}\nwbegindocs{38}\nwdocspar
\subsection{Linear Operations on Cycles}
\label{sec:line-oper-cycl}
Cycles are represented by a points in a projective vector space, thus
we wish to have a full set of linear operation on them. The metric is
inherited from the first {\Tt{}\Rm{}{\bf{}cycle}\nwendquote} object. First we define it as an
methods of the {\Tt{}\Rm{}{\bf{}cycle}\nwendquote} class.
\nwenddocs{}\nwbegincode{39}\sublabel{NW3gGP3e-3Bzv5Q-1}\nwmargintag{{\nwtagstyle{}\subpageref{NW3gGP3e-3Bzv5Q-1}}}\moddef{Linear operation as cycle methods~{\nwtagstyle{}\subpageref{NW3gGP3e-3Bzv5Q-1}}}\endmoddef\Rm{}\nwstartdeflinemarkup\nwusesondefline{\\{NW3gGP3e-2wwyff-1}}\nwenddeflinemarkup
{\bf{}virtual} {\bf{}cycle} {\it{}add}({\bf{}const} {\bf{}cycle} & {\it{}rh}) {\bf{}const};
{\bf{}virtual} {\bf{}cycle} {\it{}sub}({\bf{}const} {\bf{}cycle} & {\it{}rh}) {\bf{}const};
{\bf{}virtual} {\bf{}cycle} {\it{}exmul}({\bf{}const} {\bf{}ex} & {\it{}rh}) {\bf{}const};
{\bf{}virtual} {\bf{}cycle} {\it{}div}({\bf{}const} {\bf{}ex} & {\it{}rh}) {\bf{}const};
\nwindexdefn{\nwixident{add}}{add}{NW3gGP3e-3Bzv5Q-1}\nwindexdefn{\nwixident{sub}}{sub}{NW3gGP3e-3Bzv5Q-1}\nwindexdefn{\nwixident{exmul}}{exmul}{NW3gGP3e-3Bzv5Q-1}\nwindexdefn{\nwixident{div}}{div}{NW3gGP3e-3Bzv5Q-1}\eatline
\nwused{\\{NW3gGP3e-2wwyff-1}}\nwidentdefs{\\{{\nwixident{add}}{add}}\\{{\nwixident{div}}{div}}\\{{\nwixident{exmul}}{exmul}}\\{{\nwixident{sub}}{sub}}}\nwidentuses{\\{{\nwixident{cycle}}{cycle}}\\{{\nwixident{ex}}{ex}}}\nwindexuse{\nwixident{cycle}}{cycle}{NW3gGP3e-3Bzv5Q-1}\nwindexuse{\nwixident{ex}}{ex}{NW3gGP3e-3Bzv5Q-1}\nwendcode{}\nwbegindocs{40}\nwdocspar
\nwenddocs{}\nwbegindocs{41}After that we overload standard binary operations for {\Tt{}\Rm{}{\bf{}cycle}\nwendquote}.
\nwenddocs{}\nwbegincode{42}\sublabel{NW3gGP3e-3LRIX-1}\nwmargintag{{\nwtagstyle{}\subpageref{NW3gGP3e-3LRIX-1}}}\moddef{Linear operation on cycles~{\nwtagstyle{}\subpageref{NW3gGP3e-3LRIX-1}}}\endmoddef\Rm{}\nwstartdeflinemarkup\nwusesondefline{\\{NW3gGP3e-2wwyff-1}}\nwprevnextdefs{\relax}{NW3gGP3e-3LRIX-2}\nwenddeflinemarkup
{\bf{}const} {\bf{}cycle} {\bf{}operator}+({\bf{}const} {\bf{}cycle} & {\it{}lh}, {\bf{}const} {\bf{}cycle} & {\it{}rh});\nwindexdefn{\nwixident{cycle}}{cycle}{NW3gGP3e-3LRIX-1}
{\bf{}const} {\bf{}cycle} {\bf{}operator}-({\bf{}const} {\bf{}cycle} & {\it{}lh}, {\bf{}const} {\bf{}cycle} & {\it{}rh});\nwindexdefn{\nwixident{cycle}}{cycle}{NW3gGP3e-3LRIX-1}
{\bf{}const} {\bf{}cycle} {\bf{}operator}\begin{math}\ast\end{math}({\bf{}const} {\bf{}cycle} & {\it{}lh}, {\bf{}const} {\bf{}ex} & {\it{}rh});\nwindexdefn{\nwixident{cycle}}{cycle}{NW3gGP3e-3LRIX-1}
{\bf{}const} {\bf{}cycle} {\bf{}operator}\begin{math}\ast\end{math}({\bf{}const} {\bf{}ex} & {\it{}lh}, {\bf{}const} {\bf{}cycle} & {\it{}rh});\nwindexdefn{\nwixident{cycle}}{cycle}{NW3gGP3e-3LRIX-1}
{\bf{}const} {\bf{}cycle} {\bf{}operator}\begin{math}\div\end{math}({\bf{}const} {\bf{}cycle} & {\it{}lh}, {\bf{}const} {\bf{}ex} & {\it{}rh});\nwindexdefn{\nwixident{cycle}}{cycle}{NW3gGP3e-3LRIX-1}
\nwindexdefn{\nwixident{operator+}}{operator+}{NW3gGP3e-3LRIX-1}\nwindexdefn{\nwixident{operator-}}{operator-}{NW3gGP3e-3LRIX-1}\nwindexdefn{\nwixident{operator*}}{operator*}{NW3gGP3e-3LRIX-1}\nwindexdefn{\nwixident{operator/}}{operator/}{NW3gGP3e-3LRIX-1}\eatline
\nwalsodefined{\\{NW3gGP3e-3LRIX-2}}\nwused{\\{NW3gGP3e-2wwyff-1}}\nwidentdefs{\\{{\nwixident{cycle}}{cycle}}\\{{\nwixident{operator*}}{operator*}}\\{{\nwixident{operator+}}{operator+}}\\{{\nwixident{operator-}}{operator-}}\\{{\nwixident{operator/}}{operator/}}}\nwidentuses{\\{{\nwixident{ex}}{ex}}}\nwindexuse{\nwixident{ex}}{ex}{NW3gGP3e-3LRIX-1}\nwendcode{}\nwbegindocs{43}\nwdocspar
\nwenddocs{}\nwbegindocs{44}We also define a product of two cycles through their matrix
representation~\eqref{eq:matrix-from-cycle}.
\nwenddocs{}\nwbegincode{45}\sublabel{NW3gGP3e-3LRIX-2}\nwmargintag{{\nwtagstyle{}\subpageref{NW3gGP3e-3LRIX-2}}}\moddef{Linear operation on cycles~{\nwtagstyle{}\subpageref{NW3gGP3e-3LRIX-1}}}\plusendmoddef\Rm{}\nwstartdeflinemarkup\nwusesondefline{\\{NW3gGP3e-2wwyff-1}}\nwprevnextdefs{NW3gGP3e-3LRIX-1}{\relax}\nwenddeflinemarkup
{\bf{}const} {\bf{}ex} {\bf{}operator}\begin{math}\ast\end{math}({\bf{}const} {\bf{}cycle} & {\it{}lh}, {\bf{}const} {\bf{}cycle} & {\it{}rh});\nwindexdefn{\nwixident{ex}}{ex}{NW3gGP3e-3LRIX-2}

\nwused{\\{NW3gGP3e-2wwyff-1}}\nwidentdefs{\\{{\nwixident{ex}}{ex}}}\nwidentuses{\\{{\nwixident{cycle}}{cycle}}\\{{\nwixident{operator*}}{operator*}}}\nwindexuse{\nwixident{cycle}}{cycle}{NW3gGP3e-3LRIX-2}\nwindexuse{\nwixident{operator*}}{operator*}{NW3gGP3e-3LRIX-2}\nwendcode{}\nwbegindocs{46}\nwdocspar
\subsection[Geometric methods in cycle]{Geometric methods in {\Tt{}\Rm{}{\bf{}cycle}\nwendquote}}
\label{sec:geom-meth-cycle}

We start from some general methods which deal with {\Tt{}\Rm{}{\bf{}cycle}\nwendquote}.
\nwenddocs{}\nwbegindocs{47}The next method is needed to get rid of the homogeneous ambiguity in the
projective space of cycles. If the cycle has non-zero determinant,
then it is scaled to have new determinant equal {\Tt{}\Rm{}{\it{}D}\nwendquote}, with \(1\) as
the default value.
\nwenddocs{}\nwbegincode{48}\sublabel{NW3gGP3e-mq5nH-1}\nwmargintag{{\nwtagstyle{}\subpageref{NW3gGP3e-mq5nH-1}}}\moddef{specific methods of the class cycle~{\nwtagstyle{}\subpageref{NW3gGP3e-mq5nH-1}}}\endmoddef\Rm{}\nwstartdeflinemarkup\nwusesondefline{\\{NW3gGP3e-2wwyff-1}}\nwprevnextdefs{\relax}{NW3gGP3e-mq5nH-2}\nwenddeflinemarkup
{\bf{}public}:
{\bf{}cycle} {\it{}normalize\_det}({\bf{}const} {\bf{}ex} & {\it{}e} = 0,
                    {\bf{}const} {\bf{}ex} & {\it{}sign} = ({\bf{}new} {\it{}tensdelta})\begin{math}\rightarrow\end{math}{\it{}setflag}({\it{}status\_flags}::{\it{}dynallocated}),
                    {\bf{}const} {\bf{}ex} & {\it{}D} = 1) {\bf{}const};
\nwindexdefn{\nwixident{normalize{\_}det}}{normalize:undet}{NW3gGP3e-mq5nH-1}\eatline
\nwalsodefined{\\{NW3gGP3e-mq5nH-2}\\{NW3gGP3e-mq5nH-3}\\{NW3gGP3e-mq5nH-4}\\{NW3gGP3e-mq5nH-5}\\{NW3gGP3e-mq5nH-6}\\{NW3gGP3e-mq5nH-7}\\{NW3gGP3e-mq5nH-8}\\{NW3gGP3e-mq5nH-9}\\{NW3gGP3e-mq5nH-A}\\{NW3gGP3e-mq5nH-B}\\{NW3gGP3e-mq5nH-C}\\{NW3gGP3e-mq5nH-D}\\{NW3gGP3e-mq5nH-E}\\{NW3gGP3e-mq5nH-F}\\{NW3gGP3e-mq5nH-G}\\{NW3gGP3e-mq5nH-H}\\{NW3gGP3e-mq5nH-I}\\{NW3gGP3e-mq5nH-J}\\{NW3gGP3e-mq5nH-K}}\nwused{\\{NW3gGP3e-2wwyff-1}}\nwidentdefs{\\{{\nwixident{normalize{\_}det}}{normalize:undet}}}\nwidentuses{\\{{\nwixident{cycle}}{cycle}}\\{{\nwixident{ex}}{ex}}}\nwindexuse{\nwixident{cycle}}{cycle}{NW3gGP3e-mq5nH-1}\nwindexuse{\nwixident{ex}}{ex}{NW3gGP3e-mq5nH-1}\nwendcode{}\nwbegindocs{49}\nwdocspar
\nwenddocs{}\nwbegindocs{50}The square \(\scalar{\cycle{}{}}{\cycle{}{}}\) of the norm of a cycle
\(\cycle{}{}\) is twice its determinant \(\det \cycle{}{}\), we provide a method to 
mormalize the norm as well.
\nwenddocs{}\nwbegincode{51}\sublabel{NW3gGP3e-mq5nH-2}\nwmargintag{{\nwtagstyle{}\subpageref{NW3gGP3e-mq5nH-2}}}\moddef{specific methods of the class cycle~{\nwtagstyle{}\subpageref{NW3gGP3e-mq5nH-1}}}\plusendmoddef\Rm{}\nwstartdeflinemarkup\nwusesondefline{\\{NW3gGP3e-2wwyff-1}}\nwprevnextdefs{NW3gGP3e-mq5nH-1}{NW3gGP3e-mq5nH-3}\nwenddeflinemarkup
{\bf{}inline} {\bf{}cycle} {\it{}normalize\_norm}({\bf{}const} {\bf{}ex} & {\it{}e} = 0,
                    {\bf{}const} {\bf{}ex} & {\it{}sign} = ({\bf{}new} {\it{}tensdelta})\begin{math}\rightarrow\end{math}{\it{}setflag}({\it{}status\_flags}::{\it{}dynallocated}),
                    {\bf{}const} {\bf{}ex} & {\it{}N} = 1) {\bf{}const}
{\nwlbrace}{\bf{}return} {\it{}normalize\_det}({\it{}e}, {\it{}sign}, {\it{}N}\begin{math}\ast\end{math}{\bf{}numeric}(1,2));{\nwrbrace}
\nwindexdefn{\nwixident{normalize{\_}norm}}{normalize:unnorm}{NW3gGP3e-mq5nH-2}\eatline
\nwused{\\{NW3gGP3e-2wwyff-1}}\nwidentdefs{\\{{\nwixident{normalize{\_}norm}}{normalize:unnorm}}}\nwidentuses{\\{{\nwixident{cycle}}{cycle}}\\{{\nwixident{ex}}{ex}}\\{{\nwixident{normalize{\_}det}}{normalize:undet}}\\{{\nwixident{numeric}}{numeric}}}\nwindexuse{\nwixident{cycle}}{cycle}{NW3gGP3e-mq5nH-2}\nwindexuse{\nwixident{ex}}{ex}{NW3gGP3e-mq5nH-2}\nwindexuse{\nwixident{normalize{\_}det}}{normalize:undet}{NW3gGP3e-mq5nH-2}\nwindexuse{\nwixident{numeric}}{numeric}{NW3gGP3e-mq5nH-2}\nwendcode{}\nwbegindocs{52}\nwdocspar
\nwenddocs{}\nwbegindocs{53}The next normalization acts as follows: if {\Tt{}\Rm{}{\it{}k\_new}=0\nwendquote} the {\Tt{}\Rm{}{\bf{}cycle}\nwendquote}
is normalised such that its \(\det\) becomes \(1\). Otherwise the
first non-zero coefficient among {\Tt{}\Rm{}{\it{}k}\nwendquote}, {\Tt{}\Rm{}{\it{}m}\nwendquote}, \(l_0\), \(l_1\),
\ldots\ is set to {\Tt{}\Rm{}{\it{}k\_new}\nwendquote}.
\nwenddocs{}\nwbegincode{54}\sublabel{NW3gGP3e-mq5nH-3}\nwmargintag{{\nwtagstyle{}\subpageref{NW3gGP3e-mq5nH-3}}}\moddef{specific methods of the class cycle~{\nwtagstyle{}\subpageref{NW3gGP3e-mq5nH-1}}}\plusendmoddef\Rm{}\nwstartdeflinemarkup\nwusesondefline{\\{NW3gGP3e-2wwyff-1}}\nwprevnextdefs{NW3gGP3e-mq5nH-2}{NW3gGP3e-mq5nH-4}\nwenddeflinemarkup
{\bf{}cycle} {\it{}normalize}({\bf{}const} {\bf{}ex} & {\it{}k\_new} = {\bf{}numeric}(1), {\bf{}const} {\bf{}ex} & {\it{}e} = 0) {\bf{}const};
\nwindexdefn{\nwixident{normalize}}{normalize}{NW3gGP3e-mq5nH-3}\eatline
\nwused{\\{NW3gGP3e-2wwyff-1}}\nwidentdefs{\\{{\nwixident{normalize}}{normalize}}}\nwidentuses{\\{{\nwixident{cycle}}{cycle}}\\{{\nwixident{ex}}{ex}}\\{{\nwixident{numeric}}{numeric}}}\nwindexuse{\nwixident{cycle}}{cycle}{NW3gGP3e-mq5nH-3}\nwindexuse{\nwixident{ex}}{ex}{NW3gGP3e-mq5nH-3}\nwindexuse{\nwixident{numeric}}{numeric}{NW3gGP3e-mq5nH-3}\nwendcode{}\nwbegindocs{55}\nwdocspar
\nwenddocs{}\nwbegindocs{56}The method {\Tt{}\Rm{}{\it{}center}()\nwendquote} returns a list of components of the cycle
centre or the corresponding vector ({\Tt{}\Rm{}{\it{}D}\nwendquote} matrix) if the dimension
is not symbolic. The metric, if not supplied is taken from the cycle.

\nwenddocs{}\nwbegincode{57}\sublabel{NW3gGP3e-mq5nH-4}\nwmargintag{{\nwtagstyle{}\subpageref{NW3gGP3e-mq5nH-4}}}\moddef{specific methods of the class cycle~{\nwtagstyle{}\subpageref{NW3gGP3e-mq5nH-1}}}\plusendmoddef\Rm{}\nwstartdeflinemarkup\nwusesondefline{\\{NW3gGP3e-2wwyff-1}}\nwprevnextdefs{NW3gGP3e-mq5nH-3}{NW3gGP3e-mq5nH-5}\nwenddeflinemarkup
{\bf{}virtual} {\bf{}ex} {\it{}center}({\bf{}const} {\bf{}ex} & {\it{}metr} = 0, {\bf{}bool} {\it{}return\_matrix} = {\bf{}false}) {\bf{}const};
\nwindexdefn{\nwixident{center}}{center}{NW3gGP3e-mq5nH-4}\eatline
\nwused{\\{NW3gGP3e-2wwyff-1}}\nwidentdefs{\\{{\nwixident{center}}{center}}}\nwidentuses{\\{{\nwixident{bool}}{bool}}\\{{\nwixident{ex}}{ex}}\\{{\nwixident{metr}}{metr}}}\nwindexuse{\nwixident{bool}}{bool}{NW3gGP3e-mq5nH-4}\nwindexuse{\nwixident{ex}}{ex}{NW3gGP3e-mq5nH-4}\nwindexuse{\nwixident{metr}}{metr}{NW3gGP3e-mq5nH-4}\nwendcode{}\nwbegindocs{58}\nwdocspar
\nwenddocs{}\nwbegindocs{59}The next method returns the value of the expression
\(-k\vecbf{x}^2-2\scalar{\vecbf{l}}{\vecbf{x}}+m\) for the given
cycle and point \(\vecbf{x}\). Obviously it should be \(0\) if \(\vecbf{x}\) belongs to the
cycle.

\nwenddocs{}\nwbegincode{60}\sublabel{NW3gGP3e-mq5nH-5}\nwmargintag{{\nwtagstyle{}\subpageref{NW3gGP3e-mq5nH-5}}}\moddef{specific methods of the class cycle~{\nwtagstyle{}\subpageref{NW3gGP3e-mq5nH-1}}}\plusendmoddef\Rm{}\nwstartdeflinemarkup\nwusesondefline{\\{NW3gGP3e-2wwyff-1}}\nwprevnextdefs{NW3gGP3e-mq5nH-4}{NW3gGP3e-mq5nH-6}\nwenddeflinemarkup
{\bf{}virtual} {\bf{}ex} {\it{}val}({\bf{}const} {\bf{}ex} & {\it{}y}) {\bf{}const};
\nwindexdefn{\nwixident{val}}{val}{NW3gGP3e-mq5nH-5}\eatline
\nwused{\\{NW3gGP3e-2wwyff-1}}\nwidentdefs{\\{{\nwixident{val}}{val}}}\nwidentuses{\\{{\nwixident{ex}}{ex}}}\nwindexuse{\nwixident{ex}}{ex}{NW3gGP3e-mq5nH-5}\nwendcode{}\nwbegindocs{61}\nwdocspar
\nwenddocs{}\nwbegindocs{62}Then method {\Tt{}\Rm{}{\it{}passing}()\nwendquote} returns a {\Tt{}\Rm{}{\bf{}relational}\nwendquote} defined by the identity
\(k\vecbf{x}^2-2\scalar{\vecbf{l}}{\vecbf{x}}+m\equiv 0\), i.e  this
relational describes incidence of point to a cycle.

\nwenddocs{}\nwbegincode{63}\sublabel{NW3gGP3e-mq5nH-6}\nwmargintag{{\nwtagstyle{}\subpageref{NW3gGP3e-mq5nH-6}}}\moddef{specific methods of the class cycle~{\nwtagstyle{}\subpageref{NW3gGP3e-mq5nH-1}}}\plusendmoddef\Rm{}\nwstartdeflinemarkup\nwusesondefline{\\{NW3gGP3e-2wwyff-1}}\nwprevnextdefs{NW3gGP3e-mq5nH-5}{NW3gGP3e-mq5nH-7}\nwenddeflinemarkup
{\bf{}inline} {\bf{}ex} {\it{}passing}({\bf{}const} {\bf{}ex} & {\it{}y}) {\bf{}const} {\nwlbrace}{\bf{}return}  {\it{}val}({\it{}y}).{\it{}numer}().{\it{}normal}() \begin{math}\equiv\end{math} 0;{\nwrbrace}
\nwindexdefn{\nwixident{passing}}{passing}{NW3gGP3e-mq5nH-6}\eatline
\nwused{\\{NW3gGP3e-2wwyff-1}}\nwidentdefs{\\{{\nwixident{passing}}{passing}}}\nwidentuses{\\{{\nwixident{ex}}{ex}}\\{{\nwixident{normal}}{normal}}\\{{\nwixident{val}}{val}}}\nwindexuse{\nwixident{ex}}{ex}{NW3gGP3e-mq5nH-6}\nwindexuse{\nwixident{normal}}{normal}{NW3gGP3e-mq5nH-6}\nwindexuse{\nwixident{val}}{val}{NW3gGP3e-mq5nH-6}\nwendcode{}\nwbegindocs{64}\nwdocspar
\nwenddocs{}\nwbegindocs{65}We oftenly need to consider a cycle which satisfies some additional
conditions, this can be done by the following method {\Tt{}\Rm{}{\it{}subject\_to}\nwendquote}. Its
typical application looks like:
\begin{webcode}{\Tt{}\Rm{}{\it{}C2} = {\it{}C}.{\it{}subject\_to}({\bf{}lst}({\it{}C}.{\it{}passing}({\it{}P}), {\it{}C}.{\it{}is\_orthogonal}({\it{}C1})));\nwendquote}\end{webcode}
The second parameters {\Tt{}\Rm{}{\it{}vars}\nwendquote} specifies which components of the
{\Tt{}\Rm{}{\bf{}cycle}\nwendquote} are considered as unknown. Its default value represents all
of them which are symbols.
\nwenddocs{}\nwbegincode{66}\sublabel{NW3gGP3e-mq5nH-7}\nwmargintag{{\nwtagstyle{}\subpageref{NW3gGP3e-mq5nH-7}}}\moddef{specific methods of the class cycle~{\nwtagstyle{}\subpageref{NW3gGP3e-mq5nH-1}}}\plusendmoddef\Rm{}\nwstartdeflinemarkup\nwusesondefline{\\{NW3gGP3e-2wwyff-1}}\nwprevnextdefs{NW3gGP3e-mq5nH-6}{NW3gGP3e-mq5nH-8}\nwenddeflinemarkup
{\bf{}cycle} {\it{}subject\_to}({\bf{}const} {\bf{}ex} & {\it{}condition}, {\bf{}const} {\bf{}ex} & {\it{}vars} = 0) {\bf{}const};
\nwindexdefn{\nwixident{subject{\_}to}}{subject:unto}{NW3gGP3e-mq5nH-7}\eatline
\nwused{\\{NW3gGP3e-2wwyff-1}}\nwidentdefs{\\{{\nwixident{subject{\_}to}}{subject:unto}}}\nwidentuses{\\{{\nwixident{cycle}}{cycle}}\\{{\nwixident{ex}}{ex}}}\nwindexuse{\nwixident{cycle}}{cycle}{NW3gGP3e-mq5nH-7}\nwindexuse{\nwixident{ex}}{ex}{NW3gGP3e-mq5nH-7}\nwendcode{}\nwbegindocs{67} 
\nwenddocs{}\nwbegindocs{68}\nwdocspar
\subsection{Methods representing SFSCc}
\label{sec:meth-repr-fscc}
There is a set of specific methods which represent mathematical side
of SFSCc.

\nwenddocs{}\nwbegindocs{69}The next method is the main gateway to the SFSCc, it generates the
\(2\times 2\) matrix
\begin{equation}
  \label{eq:matrix-from-cycle}
  \begin{pmatrix}
    \vecbf{l}_i \sigma^i_j\clifford[1]{e}^j & m\\
    k &  -\vecbf{l}_i  \sigma^i_j \clifford[1]{e}^j
  \end{pmatrix}
  \quad
  \textrm{ from the cycle }
  k\vecbf{x}^2-2\scalar{\vecbf{l}}{\vecbf{x}}+m=0.
  \quad
\end{equation}
Note, that the Clifford unit \(\clifford[1]{e}\) has an arbitrary
metric unrelated to the initial metric stored in the {\Tt{}\Rm{}{\it{}unit}\nwendquote} member variable.

\nwenddocs{}\nwbegincode{70}\sublabel{NW3gGP3e-mq5nH-8}\nwmargintag{{\nwtagstyle{}\subpageref{NW3gGP3e-mq5nH-8}}}\moddef{specific methods of the class cycle~{\nwtagstyle{}\subpageref{NW3gGP3e-mq5nH-1}}}\plusendmoddef\Rm{}\nwstartdeflinemarkup\nwusesondefline{\\{NW3gGP3e-2wwyff-1}}\nwprevnextdefs{NW3gGP3e-mq5nH-7}{NW3gGP3e-mq5nH-9}\nwenddeflinemarkup
{\bf{}virtual} {\bf{}matrix} {\it{}to\_matrix}({\bf{}const} {\bf{}ex} & {\it{}e} = 0,
     {\bf{}const} {\bf{}ex} & {\it{}sign} = ({\bf{}new} {\it{}tensdelta})\begin{math}\rightarrow\end{math}{\it{}setflag}({\it{}status\_flags}::{\it{}dynallocated})) {\bf{}const};
\nwindexdefn{\nwixident{to{\_}matrix}}{to:unmatrix}{NW3gGP3e-mq5nH-8}\eatline
\nwused{\\{NW3gGP3e-2wwyff-1}}\nwidentdefs{\\{{\nwixident{to{\_}matrix}}{to:unmatrix}}}\nwidentuses{\\{{\nwixident{ex}}{ex}}\\{{\nwixident{matrix}}{matrix}}}\nwindexuse{\nwixident{ex}}{ex}{NW3gGP3e-mq5nH-8}\nwindexuse{\nwixident{matrix}}{matrix}{NW3gGP3e-mq5nH-8}\nwendcode{}\nwbegindocs{71} 
\nwenddocs{}\nwbegindocs{72}The next method returns the value of determinant of the
matrix~\eqref{eq:matrix-from-cycle} corresponding to the {\Tt{}\Rm{}{\bf{}cycle}\nwendquote}. It has
explicit geometric meaning,
see~\cite[\S~\ref{E-sec:lengths-orth}]{Kisil05a}. Before calculation
the cycle is normalised by the condition {\Tt{}\Rm{}{\it{}k}\begin{math}\equiv\end{math}{\it{}k\_norm}\nwendquote}, if {\Tt{}\Rm{}{\it{}k\_norm}\nwendquote}
is zero then no normalisation is done.
\nwenddocs{}\nwbegincode{73}\sublabel{NW3gGP3e-mq5nH-9}\nwmargintag{{\nwtagstyle{}\subpageref{NW3gGP3e-mq5nH-9}}}\moddef{specific methods of the class cycle~{\nwtagstyle{}\subpageref{NW3gGP3e-mq5nH-1}}}\plusendmoddef\Rm{}\nwstartdeflinemarkup\nwusesondefline{\\{NW3gGP3e-2wwyff-1}}\nwprevnextdefs{NW3gGP3e-mq5nH-8}{NW3gGP3e-mq5nH-A}\nwenddeflinemarkup
{\bf{}virtual} {\bf{}ex} {\it{}det}({\bf{}const} {\bf{}ex} & {\it{}e} = 0,
  {\bf{}const} {\bf{}ex} & {\it{}sign} = ({\bf{}new} {\it{}tensdelta})\begin{math}\rightarrow\end{math}{\it{}setflag}({\it{}status\_flags}::{\it{}dynallocated}),
  {\bf{}const} {\bf{}ex} & {\it{}k\_norm} = 0) {\bf{}const};
\nwindexdefn{\nwixident{det}}{det}{NW3gGP3e-mq5nH-9}\eatline
\nwused{\\{NW3gGP3e-2wwyff-1}}\nwidentdefs{\\{{\nwixident{det}}{det}}}\nwidentuses{\\{{\nwixident{ex}}{ex}}}\nwindexuse{\nwixident{ex}}{ex}{NW3gGP3e-mq5nH-9}\nwendcode{}\nwbegindocs{74} 
\nwenddocs{}\nwbegindocs{75}The determinant of a k-normalised cycle can be treated as the square of its radius
\nwenddocs{}\nwbegincode{76}\sublabel{NW3gGP3e-mq5nH-A}\nwmargintag{{\nwtagstyle{}\subpageref{NW3gGP3e-mq5nH-A}}}\moddef{specific methods of the class cycle~{\nwtagstyle{}\subpageref{NW3gGP3e-mq5nH-1}}}\plusendmoddef\Rm{}\nwstartdeflinemarkup\nwusesondefline{\\{NW3gGP3e-2wwyff-1}}\nwprevnextdefs{NW3gGP3e-mq5nH-9}{NW3gGP3e-mq5nH-B}\nwenddeflinemarkup
{\bf{}virtual} {\bf{}inline} {\bf{}ex} {\it{}radius\_sq}({\bf{}const} {\bf{}ex} & {\it{}e} = 0,
                    {\bf{}const} {\bf{}ex} & {\it{}sign} = ({\bf{}new} {\it{}tensdelta})\begin{math}\rightarrow\end{math}{\it{}setflag}({\it{}status\_flags}::{\it{}dynallocated})) {\bf{}const}
    {\nwlbrace} {\bf{}return} {\it{}this}\begin{math}\rightarrow\end{math}{\it{}det}({\it{}e}, {\it{}sign}, {\bf{}numeric}(1));{\nwrbrace}
\nwindexdefn{\nwixident{radius{\_}sq}}{radius:unsq}{NW3gGP3e-mq5nH-A}\eatline
\nwused{\\{NW3gGP3e-2wwyff-1}}\nwidentdefs{\\{{\nwixident{radius{\_}sq}}{radius:unsq}}}\nwidentuses{\\{{\nwixident{det}}{det}}\\{{\nwixident{ex}}{ex}}\\{{\nwixident{numeric}}{numeric}}}\nwindexuse{\nwixident{det}}{det}{NW3gGP3e-mq5nH-A}\nwindexuse{\nwixident{ex}}{ex}{NW3gGP3e-mq5nH-A}\nwindexuse{\nwixident{numeric}}{numeric}{NW3gGP3e-mq5nH-A}\nwendcode{}\nwbegindocs{77}\nwdocspar
\nwenddocs{}\nwbegindocs{78}The matrix~\eqref{eq:matrix-from-cycle} corresponding to a cycle may
be multiplied by another matrix, which in turn may be either generated
by another cycle or be of a different origin. The next methods
multiplies a cycle by another cycle or matrix supplied in {\Tt{}\Rm{}{\it{}C}\nwendquote}.

\nwenddocs{}\nwbegincode{79}\sublabel{NW3gGP3e-mq5nH-B}\nwmargintag{{\nwtagstyle{}\subpageref{NW3gGP3e-mq5nH-B}}}\moddef{specific methods of the class cycle~{\nwtagstyle{}\subpageref{NW3gGP3e-mq5nH-1}}}\plusendmoddef\Rm{}\nwstartdeflinemarkup\nwusesondefline{\\{NW3gGP3e-2wwyff-1}}\nwprevnextdefs{NW3gGP3e-mq5nH-A}{NW3gGP3e-mq5nH-C}\nwenddeflinemarkup
{\bf{}virtual} {\bf{}ex} {\it{}mul}({\bf{}const} {\bf{}ex} & {\it{}C}, {\bf{}const} {\bf{}ex} & {\it{}e} = 0,
     {\bf{}const} {\bf{}ex} & {\it{}sign} = ({\bf{}new} {\it{}tensdelta})\begin{math}\rightarrow\end{math}{\it{}setflag}({\it{}status\_flags}::{\it{}dynallocated}),
     {\bf{}const} {\bf{}ex} & {\it{}sign1} =0) {\bf{}const};
\nwindexdefn{\nwixident{mul}}{mul}{NW3gGP3e-mq5nH-B}\eatline
\nwused{\\{NW3gGP3e-2wwyff-1}}\nwidentdefs{\\{{\nwixident{mul}}{mul}}}\nwidentuses{\\{{\nwixident{ex}}{ex}}}\nwindexuse{\nwixident{ex}}{ex}{NW3gGP3e-mq5nH-B}\nwendcode{}\nwbegindocs{80}\nwdocspar
\nwenddocs{}\nwbegindocs{81}Having a matrix \(C\) which represents a cycle and another matrix
\(M\) we can consider a similar matrix \(M^{-1}CM\).  The later matrix
will correspond to a cycle as well, which may be obtained by the
following three methods.  In the case then \(M\) belongs to the
\(\SL\) group the next two methods make a proper conversion of \(M\)
into Clifford-valued form.

\nwenddocs{}\nwbegincode{82}\sublabel{NW3gGP3e-mq5nH-C}\nwmargintag{{\nwtagstyle{}\subpageref{NW3gGP3e-mq5nH-C}}}\moddef{specific methods of the class cycle~{\nwtagstyle{}\subpageref{NW3gGP3e-mq5nH-1}}}\plusendmoddef\Rm{}\nwstartdeflinemarkup\nwusesondefline{\\{NW3gGP3e-2wwyff-1}}\nwprevnextdefs{NW3gGP3e-mq5nH-B}{NW3gGP3e-mq5nH-D}\nwenddeflinemarkup
{\bf{}virtual} {\bf{}cycle} {\it{}sl2\_similarity}({\bf{}const} {\bf{}ex} & {\it{}a}, {\bf{}const} {\bf{}ex} & {\it{}b}, {\bf{}const} {\bf{}ex} & {\it{}c}, {\bf{}const} {\bf{}ex} & {\it{}d},
    {\bf{}const} {\bf{}ex} & {\it{}e} = 0,
    {\bf{}const} {\bf{}ex} & {\it{}sign} = ({\bf{}new} {\it{}tensdelta})\begin{math}\rightarrow\end{math}{\it{}setflag}({\it{}status\_flags}::{\it{}dynallocated}),
    {\bf{}bool} {\it{}not\_inverse}={\bf{}true},
    {\bf{}const} {\bf{}ex} & {\it{}sign\_inv} = ({\bf{}new} {\it{}tensdelta})\begin{math}\rightarrow\end{math}{\it{}setflag}({\it{}status\_flags}::{\it{}dynallocated})) {\bf{}const};
{\bf{}virtual} {\bf{}cycle} {\it{}sl2\_similarity}({\bf{}const} {\bf{}ex} & {\it{}M}, {\bf{}const} {\bf{}ex} & {\it{}e} = 0,
      {\bf{}const} {\bf{}ex} & {\it{}sign} = ({\bf{}new} {\it{}tensdelta})\begin{math}\rightarrow\end{math}{\it{}setflag}({\it{}status\_flags}::{\it{}dynallocated}),
    {\bf{}bool} {\it{}not\_inverse}={\bf{}true},
      {\bf{}const} {\bf{}ex} & {\it{}sign\_inv} = ({\bf{}new} {\it{}tensdelta})\begin{math}\rightarrow\end{math}{\it{}setflag}({\it{}status\_flags}::{\it{}dynallocated})) {\bf{}const};
\nwindexdefn{\nwixident{sl2{\_}similarity}}{sl2:unsimilarity}{NW3gGP3e-mq5nH-C}\eatline
\nwused{\\{NW3gGP3e-2wwyff-1}}\nwidentdefs{\\{{\nwixident{sl2{\_}similarity}}{sl2:unsimilarity}}}\nwidentuses{\\{{\nwixident{bool}}{bool}}\\{{\nwixident{cycle}}{cycle}}\\{{\nwixident{ex}}{ex}}}\nwindexuse{\nwixident{bool}}{bool}{NW3gGP3e-mq5nH-C}\nwindexuse{\nwixident{cycle}}{cycle}{NW3gGP3e-mq5nH-C}\nwindexuse{\nwixident{ex}}{ex}{NW3gGP3e-mq5nH-C}\nwendcode{}\nwbegindocs{83} 
\nwenddocs{}\nwbegindocs{84}If \(M\) is a generic \(2\times 2\)-matrix of another sort then it is used in the
similarity in the unchanged form by the next method.
\nwenddocs{}\nwbegincode{85}\sublabel{NW3gGP3e-mq5nH-D}\nwmargintag{{\nwtagstyle{}\subpageref{NW3gGP3e-mq5nH-D}}}\moddef{specific methods of the class cycle~{\nwtagstyle{}\subpageref{NW3gGP3e-mq5nH-1}}}\plusendmoddef\Rm{}\nwstartdeflinemarkup\nwusesondefline{\\{NW3gGP3e-2wwyff-1}}\nwprevnextdefs{NW3gGP3e-mq5nH-C}{NW3gGP3e-mq5nH-E}\nwenddeflinemarkup
{\bf{}virtual} {\bf{}cycle} {\it{}matrix\_similarity}({\bf{}const} {\bf{}ex} & {\it{}M}, {\bf{}const} {\bf{}ex} & {\it{}e} = 0,
      {\bf{}const} {\bf{}ex} & {\it{}sign} = ({\bf{}new} {\it{}tensdelta})\begin{math}\rightarrow\end{math}{\it{}setflag}({\it{}status\_flags}::{\it{}dynallocated}),
    {\bf{}bool} {\it{}not\_inverse}={\bf{}true},
      {\bf{}const} {\bf{}ex} & {\it{}sign\_inv} = ({\bf{}new} {\it{}tensdelta})\begin{math}\rightarrow\end{math}{\it{}setflag}({\it{}status\_flags}::{\it{}dynallocated})) {\bf{}const};
\nwindexdefn{\nwixident{matrix{\_}similarity}}{matrix:unsimilarity}{NW3gGP3e-mq5nH-D}\eatline
\nwused{\\{NW3gGP3e-2wwyff-1}}\nwidentdefs{\\{{\nwixident{matrix{\_}similarity}}{matrix:unsimilarity}}}\nwidentuses{\\{{\nwixident{bool}}{bool}}\\{{\nwixident{cycle}}{cycle}}\\{{\nwixident{ex}}{ex}}}\nwindexuse{\nwixident{bool}}{bool}{NW3gGP3e-mq5nH-D}\nwindexuse{\nwixident{cycle}}{cycle}{NW3gGP3e-mq5nH-D}\nwindexuse{\nwixident{ex}}{ex}{NW3gGP3e-mq5nH-D}\nwendcode{}\nwbegindocs{86} 
\nwenddocs{}\nwbegindocs{87}The \(2\times 2\)-matrix \(M=
\begin{pmatrix}
  a&b\\c&d
\end{pmatrix}\) can be also defined by the collection of its elements.
\nwenddocs{}\nwbegincode{88}\sublabel{NW3gGP3e-mq5nH-E}\nwmargintag{{\nwtagstyle{}\subpageref{NW3gGP3e-mq5nH-E}}}\moddef{specific methods of the class cycle~{\nwtagstyle{}\subpageref{NW3gGP3e-mq5nH-1}}}\plusendmoddef\Rm{}\nwstartdeflinemarkup\nwusesondefline{\\{NW3gGP3e-2wwyff-1}}\nwprevnextdefs{NW3gGP3e-mq5nH-D}{NW3gGP3e-mq5nH-F}\nwenddeflinemarkup
{\bf{}virtual} {\bf{}cycle} {\it{}matrix\_similarity}({\bf{}const} {\bf{}ex} & {\it{}a}, {\bf{}const} {\bf{}ex} & {\it{}b}, {\bf{}const} {\bf{}ex} & {\it{}c}, {\bf{}const} {\bf{}ex} & {\it{}d},
      {\bf{}const} {\bf{}ex} & {\it{}e} = 0,
      {\bf{}const} {\bf{}ex} & {\it{}sign} = ({\bf{}new} {\it{}tensdelta})\begin{math}\rightarrow\end{math}{\it{}setflag}({\it{}status\_flags}::{\it{}dynallocated}),
      {\bf{}bool} {\it{}not\_inverse}={\bf{}true},
      {\bf{}const} {\bf{}ex} & {\it{}sign\_inv} = ({\bf{}new} {\it{}tensdelta})\begin{math}\rightarrow\end{math}{\it{}setflag}({\it{}status\_flags}::{\it{}dynallocated})) {\bf{}const};

\nwused{\\{NW3gGP3e-2wwyff-1}}\nwidentuses{\\{{\nwixident{bool}}{bool}}\\{{\nwixident{cycle}}{cycle}}\\{{\nwixident{ex}}{ex}}\\{{\nwixident{matrix{\_}similarity}}{matrix:unsimilarity}}}\nwindexuse{\nwixident{bool}}{bool}{NW3gGP3e-mq5nH-E}\nwindexuse{\nwixident{cycle}}{cycle}{NW3gGP3e-mq5nH-E}\nwindexuse{\nwixident{ex}}{ex}{NW3gGP3e-mq5nH-E}\nwindexuse{\nwixident{matrix{\_}similarity}}{matrix:unsimilarity}{NW3gGP3e-mq5nH-E}\nwendcode{}\nwbegindocs{89}Finally, we have a method for reflection of a cycle in another cycle
{\Tt{}\Rm{}{\it{}C}\nwendquote}, which is given by the similarity of the representing matrices:
\(C C_1 C\), see~\cite[\S~\ref{E-sec:invers-in-cycl}]{Kisil05a}.
\nwenddocs{}\nwbegincode{90}\sublabel{NW3gGP3e-mq5nH-F}\nwmargintag{{\nwtagstyle{}\subpageref{NW3gGP3e-mq5nH-F}}}\moddef{specific methods of the class cycle~{\nwtagstyle{}\subpageref{NW3gGP3e-mq5nH-1}}}\plusendmoddef\Rm{}\nwstartdeflinemarkup\nwusesondefline{\\{NW3gGP3e-2wwyff-1}}\nwprevnextdefs{NW3gGP3e-mq5nH-E}{NW3gGP3e-mq5nH-G}\nwenddeflinemarkup
{\bf{}virtual} {\bf{}cycle} {\it{}cycle\_similarity}({\bf{}const} {\bf{}cycle} & {\it{}C}, {\bf{}const} {\bf{}ex} & {\it{}e} = 0,
        {\bf{}const} {\bf{}ex} & {\it{}sign} = ({\bf{}new} {\it{}tensdelta})\begin{math}\rightarrow\end{math}{\it{}setflag}({\it{}status\_flags}::{\it{}dynallocated}),
        {\bf{}const} {\bf{}ex} & {\it{}sign1} = 0,
      {\bf{}const} {\bf{}ex} & {\it{}sign\_inv} = ({\bf{}new} {\it{}tensdelta})\begin{math}\rightarrow\end{math}{\it{}setflag}({\it{}status\_flags}::{\it{}dynallocated})) {\bf{}const};
\nwindexdefn{\nwixident{cycle{\_}similarity}}{cycle:unsimilarity}{NW3gGP3e-mq5nH-F}\eatline
\nwused{\\{NW3gGP3e-2wwyff-1}}\nwidentdefs{\\{{\nwixident{cycle{\_}similarity}}{cycle:unsimilarity}}}\nwidentuses{\\{{\nwixident{cycle}}{cycle}}\\{{\nwixident{ex}}{ex}}}\nwindexuse{\nwixident{cycle}}{cycle}{NW3gGP3e-mq5nH-F}\nwindexuse{\nwixident{ex}}{ex}{NW3gGP3e-mq5nH-F}\nwendcode{}\nwbegindocs{91} 
\nwenddocs{}\nwbegindocs{92}A cycle in the matrix form~\eqref{eq:matrix-from-cycle} naturally
defines a M\"obius transformations of the points:
\begin{equation}
  \label{eq:moebius-from-cycle}
  \begin{pmatrix}
    \vecbf{l}_i \sigma^i_j\clifford[1]{e}^j & m\\
    k &  -\vecbf{l}_i  \sigma^i_j \clifford[1]{e}^j
  \end{pmatrix} : \vecbf{x} \mapsto
    \frac{\vecbf{l}_i \sigma^i_j\clifford[1]{e}^j \vecbf{x} + m}
    {k \vecbf{x} -\vecbf{l}_i  \sigma^i_j \clifford[1]{e}^j}
\end{equation}
The following methods realised this transformations.

\nwenddocs{}\nwbegincode{93}\sublabel{NW3gGP3e-mq5nH-G}\nwmargintag{{\nwtagstyle{}\subpageref{NW3gGP3e-mq5nH-G}}}\moddef{specific methods of the class cycle~{\nwtagstyle{}\subpageref{NW3gGP3e-mq5nH-1}}}\plusendmoddef\Rm{}\nwstartdeflinemarkup\nwusesondefline{\\{NW3gGP3e-2wwyff-1}}\nwprevnextdefs{NW3gGP3e-mq5nH-F}{NW3gGP3e-mq5nH-H}\nwenddeflinemarkup
{\bf{}virtual} {\bf{}inline} {\bf{}ex} {\it{}moebius\_map}({\bf{}const} {\bf{}ex} & {\it{}P}, {\bf{}const} {\bf{}ex} & {\it{}e} = 0,
       {\bf{}const} {\bf{}ex} & {\it{}sign} = ({\bf{}new} {\it{}tensdelta})\begin{math}\rightarrow\end{math}{\it{}setflag}({\it{}status\_flags}::{\it{}dynallocated})) {\bf{}const}
{\nwlbrace}{\bf{}return} {\it{}clifford\_moebius\_map}({\it{}to\_matrix}({\it{}e}, {\it{}sign}), {\it{}P}, ({\it{}e}.{\it{}is\_zero}()?{\it{}unit}:{\it{}e}));{\nwrbrace}
\nwindexdefn{\nwixident{moebius{\_}map}}{moebius:unmap}{NW3gGP3e-mq5nH-G}\eatline
\nwused{\\{NW3gGP3e-2wwyff-1}}\nwidentdefs{\\{{\nwixident{moebius{\_}map}}{moebius:unmap}}}\nwidentuses{\\{{\nwixident{ex}}{ex}}\\{{\nwixident{is{\_}zero}}{is:unzero}}\\{{\nwixident{to{\_}matrix}}{to:unmatrix}}}\nwindexuse{\nwixident{ex}}{ex}{NW3gGP3e-mq5nH-G}\nwindexuse{\nwixident{is{\_}zero}}{is:unzero}{NW3gGP3e-mq5nH-G}\nwindexuse{\nwixident{to{\_}matrix}}{to:unmatrix}{NW3gGP3e-mq5nH-G}\nwendcode{}\nwbegindocs{94} 
\nwenddocs{}\nwbegindocs{95}For two matrices \(C_1\) and \(C_2\) obtained from cycles the expression
\begin{equation}
  \label{eq:cycle-inner-product}
   \scalar{C_1}{C_2}=-\Re\tr{(C_1C_2)}
\end{equation} 
naturally defines an inner product in the space of cycles. The follwong methods realised it.
\nwenddocs{}\nwbegincode{96}\sublabel{NW3gGP3e-mq5nH-H}\nwmargintag{{\nwtagstyle{}\subpageref{NW3gGP3e-mq5nH-H}}}\moddef{specific methods of the class cycle~{\nwtagstyle{}\subpageref{NW3gGP3e-mq5nH-1}}}\plusendmoddef\Rm{}\nwstartdeflinemarkup\nwusesondefline{\\{NW3gGP3e-2wwyff-1}}\nwprevnextdefs{NW3gGP3e-mq5nH-G}{NW3gGP3e-mq5nH-I}\nwenddeflinemarkup
{\bf{}virtual} {\bf{}inline} {\bf{}ex} {\it{}cycle\_product}({\bf{}const} {\bf{}cycle} & {\it{}C}, {\bf{}const} {\bf{}ex} & {\it{}e} = 0,
        {\bf{}const} {\bf{}ex} & {\it{}sign} = ({\bf{}new} {\it{}tensdelta})\begin{math}\rightarrow\end{math}{\it{}setflag}({\it{}status\_flags}::{\it{}dynallocated})) {\bf{}const}
{\nwlbrace}{\bf{}return} {\it{}scalar\_part}({\it{}ex\_to}\begin{math}<\end{math}{\bf{}matrix}\begin{math}>\end{math}({\it{}mul}({\it{}C}, {\it{}e}, {\it{}sign})).{\it{}trace}());{\nwrbrace}
\nwindexdefn{\nwixident{cycle{\_}product}}{cycle:unproduct}{NW3gGP3e-mq5nH-H}\eatline
\nwused{\\{NW3gGP3e-2wwyff-1}}\nwidentdefs{\\{{\nwixident{cycle{\_}product}}{cycle:unproduct}}}\nwidentuses{\\{{\nwixident{cycle}}{cycle}}\\{{\nwixident{ex}}{ex}}\\{{\nwixident{matrix}}{matrix}}\\{{\nwixident{mul}}{mul}}}\nwindexuse{\nwixident{cycle}}{cycle}{NW3gGP3e-mq5nH-H}\nwindexuse{\nwixident{ex}}{ex}{NW3gGP3e-mq5nH-H}\nwindexuse{\nwixident{matrix}}{matrix}{NW3gGP3e-mq5nH-H}\nwindexuse{\nwixident{mul}}{mul}{NW3gGP3e-mq5nH-H}\nwendcode{}\nwbegindocs{97} 
\nwenddocs{}\nwbegindocs{98}The inner product~\eqref{eq:cycle-inner-product} defines an
orthogonality relation \(\scalar{C_1}{C_2}\equiv 0\) in the space of
cycles which returned by the method {\Tt{}\Rm{}{\it{}is\_orthogonal}()\nwendquote}.
\nwenddocs{}\nwbegincode{99}\sublabel{NW3gGP3e-mq5nH-I}\nwmargintag{{\nwtagstyle{}\subpageref{NW3gGP3e-mq5nH-I}}}\moddef{specific methods of the class cycle~{\nwtagstyle{}\subpageref{NW3gGP3e-mq5nH-1}}}\plusendmoddef\Rm{}\nwstartdeflinemarkup\nwusesondefline{\\{NW3gGP3e-2wwyff-1}}\nwprevnextdefs{NW3gGP3e-mq5nH-H}{NW3gGP3e-mq5nH-J}\nwenddeflinemarkup
{\bf{}virtual} {\bf{}inline} {\bf{}ex} {\it{}is\_orthogonal}({\bf{}const} {\bf{}cycle} & {\it{}C}, {\bf{}const} {\bf{}ex} & {\it{}e} = 0,
   {\bf{}const} {\bf{}ex} & {\it{}sign} = ({\bf{}new} {\it{}tensdelta})\begin{math}\rightarrow\end{math}{\it{}setflag}({\it{}status\_flags}::{\it{}dynallocated})) {\bf{}const}
    {\nwlbrace}{\bf{}return} ({\it{}cycle\_product}({\it{}C}, {\it{}e}, {\it{}sign}) \begin{math}\equiv\end{math} 0);{\nwrbrace}
\nwindexdefn{\nwixident{is{\_}orthogonal}}{is:unorthogonal}{NW3gGP3e-mq5nH-I}\eatline
\nwused{\\{NW3gGP3e-2wwyff-1}}\nwidentdefs{\\{{\nwixident{is{\_}orthogonal}}{is:unorthogonal}}}\nwidentuses{\\{{\nwixident{cycle}}{cycle}}\\{{\nwixident{cycle{\_}product}}{cycle:unproduct}}\\{{\nwixident{ex}}{ex}}}\nwindexuse{\nwixident{cycle}}{cycle}{NW3gGP3e-mq5nH-I}\nwindexuse{\nwixident{cycle{\_}product}}{cycle:unproduct}{NW3gGP3e-mq5nH-I}\nwindexuse{\nwixident{ex}}{ex}{NW3gGP3e-mq5nH-I}\nwendcode{}\nwbegindocs{100} 
\nwenddocs{}\nwbegindocs{101}In many cases we need a higher order orthogonal relation between
cycles--- so called f-orthogonality, see
\cite[\S~\ref{E-sec:focal-orthogonality}]{Kisil05a}, which is given by
the relation:
\begin{displaymath}
      \Re\tr(\cycle{s}{\bs} \cycle[\tilde]{s}{\bs}\cycle{s}{\bs}\realline{s}{\bs})=0.
\end{displaymath}
\nwenddocs{}\nwbegincode{102}\sublabel{NW3gGP3e-mq5nH-J}\nwmargintag{{\nwtagstyle{}\subpageref{NW3gGP3e-mq5nH-J}}}\moddef{specific methods of the class cycle~{\nwtagstyle{}\subpageref{NW3gGP3e-mq5nH-1}}}\plusendmoddef\Rm{}\nwstartdeflinemarkup\nwusesondefline{\\{NW3gGP3e-2wwyff-1}}\nwprevnextdefs{NW3gGP3e-mq5nH-I}{NW3gGP3e-mq5nH-K}\nwenddeflinemarkup
{\bf{}ex} {\it{}is\_f\_orthogonal}({\bf{}const} {\bf{}cycle} & {\it{}C}, {\bf{}const} {\bf{}ex} & {\it{}e} = 0,
                   {\bf{}const} {\bf{}ex} & {\it{}sign} = ({\bf{}new} {\it{}tensdelta})\begin{math}\rightarrow\end{math}{\it{}setflag}({\it{}status\_flags}::{\it{}dynallocated}),
                   {\bf{}const} {\bf{}ex} & {\it{}sign1} = 0,
                   {\bf{}const} {\bf{}ex} & {\it{}sign\_inv} = ({\bf{}new} {\it{}tensdelta})\begin{math}\rightarrow\end{math}{\it{}setflag}({\it{}status\_flags}::{\it{}dynallocated})) {\bf{}const};
\nwindexdefn{\nwixident{is{\_}f{\_}orthogonal}}{is:unf:unorthogonal}{NW3gGP3e-mq5nH-J}\eatline
\nwused{\\{NW3gGP3e-2wwyff-1}}\nwidentdefs{\\{{\nwixident{is{\_}f{\_}orthogonal}}{is:unf:unorthogonal}}}\nwidentuses{\\{{\nwixident{cycle}}{cycle}}\\{{\nwixident{ex}}{ex}}}\nwindexuse{\nwixident{cycle}}{cycle}{NW3gGP3e-mq5nH-J}\nwindexuse{\nwixident{ex}}{ex}{NW3gGP3e-mq5nH-J}\nwendcode{}\nwbegindocs{103}\nwdocspar
\nwenddocs{}\nwbegindocs{104}The remaining to methods check if a cycle is a liner object and if
it is normalised to \(k=1\).
\nwenddocs{}\nwbegincode{105}\sublabel{NW3gGP3e-mq5nH-K}\nwmargintag{{\nwtagstyle{}\subpageref{NW3gGP3e-mq5nH-K}}}\moddef{specific methods of the class cycle~{\nwtagstyle{}\subpageref{NW3gGP3e-mq5nH-1}}}\plusendmoddef\Rm{}\nwstartdeflinemarkup\nwusesondefline{\\{NW3gGP3e-2wwyff-1}}\nwprevnextdefs{NW3gGP3e-mq5nH-J}{\relax}\nwenddeflinemarkup
 {\bf{}inline} {\bf{}ex} {\it{}is\_linear}() {\bf{}const} {\nwlbrace}{\bf{}return} ({\it{}k} \begin{math}\equiv\end{math} 0);{\nwrbrace}
 {\bf{}inline} {\bf{}ex} {\it{}is\_normalized}() {\bf{}const} {\nwlbrace}{\bf{}return} ({\it{}k} \begin{math}\equiv\end{math} 1);{\nwrbrace}
\nwindexdefn{\nwixident{is{\_}linear}}{is:unlinear}{NW3gGP3e-mq5nH-K}\nwindexdefn{\nwixident{is{\_}normalized}}{is:unnormalized}{NW3gGP3e-mq5nH-K}\eatline
\nwused{\\{NW3gGP3e-2wwyff-1}}\nwidentdefs{\\{{\nwixident{is{\_}linear}}{is:unlinear}}\\{{\nwixident{is{\_}normalized}}{is:unnormalized}}}\nwidentuses{\\{{\nwixident{ex}}{ex}}\\{{\nwixident{k}}{k}}}\nwindexuse{\nwixident{ex}}{ex}{NW3gGP3e-mq5nH-K}\nwindexuse{\nwixident{k}}{k}{NW3gGP3e-mq5nH-K}\nwendcode{}\nwbegindocs{106} 
\nwenddocs{}\nwbegindocs{107}\nwdocspar
\subsection{Two dimensional cycles}
\label{sec:two-dimens-cycl}
Two dimensional cycle {\Tt{}\Rm{}{\bf{}cycle2D}\nwendquote} is a derived class of {\Tt{}\Rm{}{\bf{}cycle}\nwendquote}. We
need to add only very few specific methods for two dimensions, notably
for the visualisation.

\nwenddocs{}\nwbegindocs{108}This a specialisation of the constructors from {\Tt{}\Rm{}{\bf{}cycle}\nwendquote} class to
{\Tt{}\Rm{}{\bf{}cycle2D}\nwendquote}. Here is the main constructor.

\nwenddocs{}\nwbegincode{109}\sublabel{NW3gGP3e-3e3TWq-1}\nwmargintag{{\nwtagstyle{}\subpageref{NW3gGP3e-3e3TWq-1}}}\moddef{constructors of the class cycle2D~{\nwtagstyle{}\subpageref{NW3gGP3e-3e3TWq-1}}}\endmoddef\Rm{}\nwstartdeflinemarkup\nwusesondefline{\\{NW3gGP3e-2ARAe1-1}}\nwprevnextdefs{\relax}{NW3gGP3e-3e3TWq-2}\nwenddeflinemarkup
{\bf{}public}:
    {\bf{}cycle2D}({\bf{}const} {\bf{}ex} & {\it{}k1}, {\bf{}const} {\bf{}ex} & {\it{}l1}, {\bf{}const} {\bf{}ex} & {\it{}m1},
            {\bf{}const} {\bf{}ex} & {\it{}metr} = -{\it{}unit\_matrix}(2));
\nwindexdefn{\nwixident{cycle2D}}{cycle2D}{NW3gGP3e-3e3TWq-1}\eatline
\nwalsodefined{\\{NW3gGP3e-3e3TWq-2}\\{NW3gGP3e-3e3TWq-3}\\{NW3gGP3e-3e3TWq-4}}\nwused{\\{NW3gGP3e-2ARAe1-1}}\nwidentdefs{\\{{\nwixident{cycle2D}}{cycle2D}}}\nwidentuses{\\{{\nwixident{ex}}{ex}}\\{{\nwixident{metr}}{metr}}}\nwindexuse{\nwixident{ex}}{ex}{NW3gGP3e-3e3TWq-1}\nwindexuse{\nwixident{metr}}{metr}{NW3gGP3e-3e3TWq-1}\nwendcode{}\nwbegindocs{110}\nwdocspar
\nwenddocs{}\nwbegindocs{111}Constructor for the {\Tt{}\Rm{}{\bf{}cycle2D}\nwendquote} from {\Tt{}\Rm{}{\it{}l}\nwendquote} and square of its radius.

\nwenddocs{}\nwbegincode{112}\sublabel{NW3gGP3e-3e3TWq-2}\nwmargintag{{\nwtagstyle{}\subpageref{NW3gGP3e-3e3TWq-2}}}\moddef{constructors of the class cycle2D~{\nwtagstyle{}\subpageref{NW3gGP3e-3e3TWq-1}}}\plusendmoddef\Rm{}\nwstartdeflinemarkup\nwusesondefline{\\{NW3gGP3e-2ARAe1-1}}\nwprevnextdefs{NW3gGP3e-3e3TWq-1}{NW3gGP3e-3e3TWq-3}\nwenddeflinemarkup
{\bf{}cycle2D}({\bf{}const} {\bf{}lst} & {\it{}l}, {\bf{}const} {\bf{}ex} & {\it{}metr} = -{\it{}unit\_matrix}(2), {\bf{}const} {\bf{}ex} & {\it{}r\_squared} =0,\nwindexdefn{\nwixident{cycle2D}}{cycle2D}{NW3gGP3e-3e3TWq-2}
  {\bf{}const} {\bf{}ex} & {\it{}e} =0, {\bf{}const} {\bf{}ex} & {\it{}sign} = {\it{}unit\_matrix}(2));

\nwused{\\{NW3gGP3e-2ARAe1-1}}\nwidentdefs{\\{{\nwixident{cycle2D}}{cycle2D}}}\nwidentuses{\\{{\nwixident{ex}}{ex}}\\{{\nwixident{l}}{l}}\\{{\nwixident{lst}}{lst}}\\{{\nwixident{metr}}{metr}}}\nwindexuse{\nwixident{ex}}{ex}{NW3gGP3e-3e3TWq-2}\nwindexuse{\nwixident{l}}{l}{NW3gGP3e-3e3TWq-2}\nwindexuse{\nwixident{lst}}{lst}{NW3gGP3e-3e3TWq-2}\nwindexuse{\nwixident{metr}}{metr}{NW3gGP3e-3e3TWq-2}\nwendcode{}\nwbegindocs{113}Construction of {\Tt{}\Rm{}{\bf{}cycle2D}\nwendquote} from its SFSCc matrix. 
\nwenddocs{}\nwbegincode{114}\sublabel{NW3gGP3e-3e3TWq-3}\nwmargintag{{\nwtagstyle{}\subpageref{NW3gGP3e-3e3TWq-3}}}\moddef{constructors of the class cycle2D~{\nwtagstyle{}\subpageref{NW3gGP3e-3e3TWq-1}}}\plusendmoddef\Rm{}\nwstartdeflinemarkup\nwusesondefline{\\{NW3gGP3e-2ARAe1-1}}\nwprevnextdefs{NW3gGP3e-3e3TWq-2}{NW3gGP3e-3e3TWq-4}\nwenddeflinemarkup
{\bf{}cycle2D}({\bf{}const} {\bf{}matrix} & {\it{}M}, {\bf{}const} {\bf{}ex} & {\it{}metr}, {\bf{}const} {\bf{}ex} & {\it{}e} = 0, {\bf{}const} {\bf{}ex} & {\it{}sign} = 0);

\nwused{\\{NW3gGP3e-2ARAe1-1}}\nwidentuses{\\{{\nwixident{cycle2D}}{cycle2D}}\\{{\nwixident{ex}}{ex}}\\{{\nwixident{matrix}}{matrix}}\\{{\nwixident{metr}}{metr}}}\nwindexuse{\nwixident{cycle2D}}{cycle2D}{NW3gGP3e-3e3TWq-3}\nwindexuse{\nwixident{ex}}{ex}{NW3gGP3e-3e3TWq-3}\nwindexuse{\nwixident{matrix}}{matrix}{NW3gGP3e-3e3TWq-3}\nwindexuse{\nwixident{metr}}{metr}{NW3gGP3e-3e3TWq-3}\nwendcode{}\nwbegindocs{115}Make a two dimensional cycle out of a general one, if the
dimensionality of the space permits. The metric of point space can
be replaced as well if a valid {\Tt{}\Rm{}{\it{}metr}\nwendquote} is supplied.
\nwenddocs{}\nwbegincode{116}\sublabel{NW3gGP3e-3e3TWq-4}\nwmargintag{{\nwtagstyle{}\subpageref{NW3gGP3e-3e3TWq-4}}}\moddef{constructors of the class cycle2D~{\nwtagstyle{}\subpageref{NW3gGP3e-3e3TWq-1}}}\plusendmoddef\Rm{}\nwstartdeflinemarkup\nwusesondefline{\\{NW3gGP3e-2ARAe1-1}}\nwprevnextdefs{NW3gGP3e-3e3TWq-3}{\relax}\nwenddeflinemarkup
{\bf{}cycle2D}({\bf{}const} {\bf{}cycle} & {\it{}C}, {\bf{}const} {\bf{}ex} & {\it{}metr} = 0);

\nwused{\\{NW3gGP3e-2ARAe1-1}}\nwidentuses{\\{{\nwixident{cycle}}{cycle}}\\{{\nwixident{cycle2D}}{cycle2D}}\\{{\nwixident{ex}}{ex}}\\{{\nwixident{metr}}{metr}}}\nwindexuse{\nwixident{cycle}}{cycle}{NW3gGP3e-3e3TWq-4}\nwindexuse{\nwixident{cycle2D}}{cycle2D}{NW3gGP3e-3e3TWq-4}\nwindexuse{\nwixident{ex}}{ex}{NW3gGP3e-3e3TWq-4}\nwindexuse{\nwixident{metr}}{metr}{NW3gGP3e-3e3TWq-4}\nwendcode{}\nwbegindocs{117}The realisation of 2D cycles through matrices with hypercomplex
numbers~\cites{Kisil06a,Kisil12a,Kisil11d} lead to some important differences
with this library using the Clifford algebras. One of them: the
determinant of a matrix change since. The next method return the
determinant as it will be calculated on those hypercomplex matrices. 
\nwenddocs{}\nwbegincode{118}\sublabel{NW3gGP3e-447MWQ-1}\nwmargintag{{\nwtagstyle{}\subpageref{NW3gGP3e-447MWQ-1}}}\moddef{methods specific for class cycle2D~{\nwtagstyle{}\subpageref{NW3gGP3e-447MWQ-1}}}\endmoddef\Rm{}\nwstartdeflinemarkup\nwusesondefline{\\{NW3gGP3e-2ARAe1-1}}\nwprevnextdefs{\relax}{NW3gGP3e-447MWQ-2}\nwenddeflinemarkup
{\bf{}public}:
{\bf{}virtual} {\bf{}inline} {\bf{}ex} {\it{}hdet}({\bf{}const} {\bf{}ex} & {\it{}e} = 0,
  {\bf{}const} {\bf{}ex} & {\it{}sign} = ({\bf{}new} {\it{}tensdelta})\begin{math}\rightarrow\end{math}{\it{}setflag}({\it{}status\_flags}::{\it{}dynallocated}),
  {\bf{}const} {\bf{}ex} & {\it{}k\_norm} = 0) {\bf{}const}
  {\nwlbrace}{\bf{}return} -{\it{}det}({\it{}e}, {\it{}sign}, {\it{}k\_norm});{\nwrbrace}
\nwindexdefn{\nwixident{hdet}}{hdet}{NW3gGP3e-447MWQ-1}\eatline
\nwalsodefined{\\{NW3gGP3e-447MWQ-2}\\{NW3gGP3e-447MWQ-3}\\{NW3gGP3e-447MWQ-4}\\{NW3gGP3e-447MWQ-5}\\{NW3gGP3e-447MWQ-6}\\{NW3gGP3e-447MWQ-7}}\nwused{\\{NW3gGP3e-2ARAe1-1}}\nwidentdefs{\\{{\nwixident{hdet}}{hdet}}}\nwidentuses{\\{{\nwixident{det}}{det}}\\{{\nwixident{ex}}{ex}}}\nwindexuse{\nwixident{det}}{det}{NW3gGP3e-447MWQ-1}\nwindexuse{\nwixident{ex}}{ex}{NW3gGP3e-447MWQ-1}\nwendcode{}\nwbegindocs{119}\nwdocspar
\nwenddocs{}\nwbegindocs{120}The method {\Tt{}\Rm{}{\it{}focus}()\nwendquote} returns list of the focus coordinates and the
focal length is provided by {\Tt{}\Rm{}{\it{}focal\_length}()\nwendquote}.
This turns to be meaningful not only for parabolas, see~\cite{Kisil05a}.

\nwenddocs{}\nwbegincode{121}\sublabel{NW3gGP3e-447MWQ-2}\nwmargintag{{\nwtagstyle{}\subpageref{NW3gGP3e-447MWQ-2}}}\moddef{methods specific for class cycle2D~{\nwtagstyle{}\subpageref{NW3gGP3e-447MWQ-1}}}\plusendmoddef\Rm{}\nwstartdeflinemarkup\nwusesondefline{\\{NW3gGP3e-2ARAe1-1}}\nwprevnextdefs{NW3gGP3e-447MWQ-1}{NW3gGP3e-447MWQ-3}\nwenddeflinemarkup
 {\bf{}ex} {\it{}focus}({\bf{}const} {\bf{}ex} & {\it{}e} = {\it{}diag\_matrix}({\bf{}lst}(-1, 1)), {\bf{}bool} {\it{}return\_matrix} = {\bf{}false}) {\bf{}const};
 {\bf{}inline} {\bf{}ex} {\it{}focal\_length}() {\bf{}const} {\nwlbrace}{\bf{}return} ({\it{}get\_l}(1)\begin{math}\div\end{math}2\begin{math}\div\end{math}{\it{}k});{\nwrbrace} // focal length of the cycle
\nwindexdefn{\nwixident{focus}}{focus}{NW3gGP3e-447MWQ-2}\nwindexdefn{\nwixident{focal{\_}length}}{focal:unlength}{NW3gGP3e-447MWQ-2}\eatline
\nwused{\\{NW3gGP3e-2ARAe1-1}}\nwidentdefs{\\{{\nwixident{focal{\_}length}}{focal:unlength}}\\{{\nwixident{focus}}{focus}}}\nwidentuses{\\{{\nwixident{bool}}{bool}}\\{{\nwixident{cycle}}{cycle}}\\{{\nwixident{ex}}{ex}}\\{{\nwixident{get{\_}l}}{get:unl}}\\{{\nwixident{k}}{k}}\\{{\nwixident{lst}}{lst}}}\nwindexuse{\nwixident{bool}}{bool}{NW3gGP3e-447MWQ-2}\nwindexuse{\nwixident{cycle}}{cycle}{NW3gGP3e-447MWQ-2}\nwindexuse{\nwixident{ex}}{ex}{NW3gGP3e-447MWQ-2}\nwindexuse{\nwixident{get{\_}l}}{get:unl}{NW3gGP3e-447MWQ-2}\nwindexuse{\nwixident{k}}{k}{NW3gGP3e-447MWQ-2}\nwindexuse{\nwixident{lst}}{lst}{NW3gGP3e-447MWQ-2}\nwendcode{}\nwbegindocs{122}\nwdocspar
\nwenddocs{}\nwbegindocs{123}The methods {\Tt{}\Rm{}{\it{}roots}()\nwendquote} returns values of \(u\) (if {\Tt{}\Rm{}{\it{}first} = {\bf{}true}\nwendquote})
such that \(k(u^2-\sigma y^2)-2l_1u-2l_2y+m=0\),
i.e. solves a quadratic equations. If {\Tt{}\Rm{}{\it{}first} = {\bf{}false}\nwendquote} then values
of \(v\) satisfying to \(k(y^2-\sigma v^2)-2l_1y-2l_2v+m=0\)
are returned.
\nwenddocs{}\nwbegincode{124}\sublabel{NW3gGP3e-447MWQ-3}\nwmargintag{{\nwtagstyle{}\subpageref{NW3gGP3e-447MWQ-3}}}\moddef{methods specific for class cycle2D~{\nwtagstyle{}\subpageref{NW3gGP3e-447MWQ-1}}}\plusendmoddef\Rm{}\nwstartdeflinemarkup\nwusesondefline{\\{NW3gGP3e-2ARAe1-1}}\nwprevnextdefs{NW3gGP3e-447MWQ-2}{NW3gGP3e-447MWQ-4}\nwenddeflinemarkup
 {\bf{}lst} {\it{}roots}({\bf{}const} {\bf{}ex} & {\it{}y} = 0, {\bf{}bool} {\it{}first} = {\bf{}true}) {\bf{}const};
\nwindexdefn{\nwixident{roots}}{roots}{NW3gGP3e-447MWQ-3}\eatline
\nwused{\\{NW3gGP3e-2ARAe1-1}}\nwidentdefs{\\{{\nwixident{roots}}{roots}}}\nwidentuses{\\{{\nwixident{bool}}{bool}}\\{{\nwixident{ex}}{ex}}\\{{\nwixident{lst}}{lst}}}\nwindexuse{\nwixident{bool}}{bool}{NW3gGP3e-447MWQ-3}\nwindexuse{\nwixident{ex}}{ex}{NW3gGP3e-447MWQ-3}\nwindexuse{\nwixident{lst}}{lst}{NW3gGP3e-447MWQ-3}\nwendcode{}\nwbegindocs{125} 
\nwenddocs{}\nwbegindocs{126}The next methods is a generalisation of the previous one: it returns intersection
points with the line \(ax+b\).
\nwenddocs{}\nwbegincode{127}\sublabel{NW3gGP3e-447MWQ-4}\nwmargintag{{\nwtagstyle{}\subpageref{NW3gGP3e-447MWQ-4}}}\moddef{methods specific for class cycle2D~{\nwtagstyle{}\subpageref{NW3gGP3e-447MWQ-1}}}\plusendmoddef\Rm{}\nwstartdeflinemarkup\nwusesondefline{\\{NW3gGP3e-2ARAe1-1}}\nwprevnextdefs{NW3gGP3e-447MWQ-3}{NW3gGP3e-447MWQ-5}\nwenddeflinemarkup
 {\bf{}lst} {\it{}line\_intersect}({\bf{}const} {\bf{}ex} & {\it{}a}, {\bf{}const} {\bf{}ex} & {\it{}b}) {\bf{}const};
\nwindexdefn{\nwixident{line{\_}intersect}}{line:unintersect}{NW3gGP3e-447MWQ-4}\eatline
\nwused{\\{NW3gGP3e-2ARAe1-1}}\nwidentdefs{\\{{\nwixident{line{\_}intersect}}{line:unintersect}}}\nwidentuses{\\{{\nwixident{ex}}{ex}}\\{{\nwixident{lst}}{lst}}}\nwindexuse{\nwixident{ex}}{ex}{NW3gGP3e-447MWQ-4}\nwindexuse{\nwixident{lst}}{lst}{NW3gGP3e-447MWQ-4}\nwendcode{}\nwbegindocs{128} 
\nwenddocs{}\nwbegindocs{129}The method {\Tt{}\Rm{}{\it{}metapost\_draw}()\nwendquote} outputs to the stream {\Tt{}\Rm{}{\it{}ost}\nwendquote} \MetaPost\ comands
to draw parts of two the {\Tt{}\Rm{}{\bf{}cycle2D}\nwendquote} within the rectangle with the
lower left vertex ({\Tt{}\Rm{}{\it{}xmin}\nwendquote}, {\Tt{}\Rm{}{\it{}ymin}\nwendquote}) and upper right  ({\Tt{}\Rm{}{\it{}xmax}\nwendquote},
{\Tt{}\Rm{}{\it{}ymax}\nwendquote}). The colour of drawing is specified by {\Tt{}\Rm{}{\it{}color}\nwendquote} (the
default is black) and any additional \MetaPost\ options can be provided
in the string {\Tt{}\Rm{}{\it{}more\_options}\nwendquote}. By default each set of the drawing
commands is preceded a comment line giving description of the cycle,
this can be suppressed by setting {\Tt{}\Rm{}{\it{}with\_header} = {\bf{}false}\nwendquote}. The default
number of points per arc is reasonable in most cases, however user can
override this with supplying a value to {\Tt{}\Rm{}{\it{}points\_per\_arc}\nwendquote}. The last
parameter is for internal use. If you do not want imaginary cycles to
be shown use the value {\Tt{}\Rm{}{\tt{}"invisible"}\nwendquote} for {\Tt{}\Rm{}{\it{}imaginary\_options}\nwendquote}.

\nwenddocs{}\nwbegincode{130}\sublabel{NW3gGP3e-447MWQ-5}\nwmargintag{{\nwtagstyle{}\subpageref{NW3gGP3e-447MWQ-5}}}\moddef{methods specific for class cycle2D~{\nwtagstyle{}\subpageref{NW3gGP3e-447MWQ-1}}}\plusendmoddef\Rm{}\nwstartdeflinemarkup\nwusesondefline{\\{NW3gGP3e-2ARAe1-1}}\nwprevnextdefs{NW3gGP3e-447MWQ-4}{NW3gGP3e-447MWQ-6}\nwenddeflinemarkup
 {\bf{}void} {\it{}metapost\_draw}({\it{}ostream} & {\it{}ost}, {\bf{}const} {\bf{}ex} & {\it{}xmin} = -5, {\bf{}const} {\bf{}ex} & {\it{}xmax} = 5,
                    {\bf{}const} {\bf{}ex} & {\it{}ymin} = -5, {\bf{}const} {\bf{}ex} & {\it{}ymax} = 5, {\bf{}const} {\bf{}lst} & {\it{}color} = {\bf{}lst}(),
                    {\bf{}const} {\it{}string} {\it{}more\_options} = {\tt{}""},
                    {\bf{}bool} {\it{}with\_header} = {\bf{}true}, {\bf{}int} {\it{}points\_per\_arc} = 0, {\bf{}bool} {\it{}asymptote} = {\bf{}false},
                    {\bf{}const} {\it{}string} {\it{}picture} = {\tt{}""}, {\bf{}bool} {\it{}only\_path}={\bf{}false}, {\bf{}bool} {\it{}is\_continuation}={\bf{}false},
                    {\bf{}const} {\it{}string} {\it{}imaginary\_options}={\tt{}"withcolor .9*green withpen pencircle scaled 4pt"}) {\bf{}const};
\nwindexdefn{\nwixident{metapost{\_}draw}}{metapost:undraw}{NW3gGP3e-447MWQ-5}\eatline
\nwused{\\{NW3gGP3e-2ARAe1-1}}\nwidentdefs{\\{{\nwixident{metapost{\_}draw}}{metapost:undraw}}}\nwidentuses{\\{{\nwixident{bool}}{bool}}\\{{\nwixident{ex}}{ex}}\\{{\nwixident{lst}}{lst}}\\{{\nwixident{string}}{string}}}\nwindexuse{\nwixident{bool}}{bool}{NW3gGP3e-447MWQ-5}\nwindexuse{\nwixident{ex}}{ex}{NW3gGP3e-447MWQ-5}\nwindexuse{\nwixident{lst}}{lst}{NW3gGP3e-447MWQ-5}\nwindexuse{\nwixident{string}}{string}{NW3gGP3e-447MWQ-5}\nwendcode{}\nwbegindocs{131} 
\nwenddocs{}\nwbegindocs{132} The similar method provides a drawing output for
\Asymptote~\cite{Asymptote} with the same meaning of
parameters. However, format of {\Tt{}\Rm{}{\it{}more\_options}\nwendquote} and
{\Tt{}\Rm{}{\it{}imaginary\_options}\nwendquote} should be adjusted
correspondingly. Currently {\Tt{}\Rm{}{\it{}asy\_draw}()\nwendquote} is realised as a
wrapper around  {\Tt{}\Rm{}{\it{}metapost\_draw}()\nwendquote} but this may be changed.

\nwenddocs{}\nwbegincode{133}\sublabel{NW3gGP3e-447MWQ-6}\nwmargintag{{\nwtagstyle{}\subpageref{NW3gGP3e-447MWQ-6}}}\moddef{methods specific for class cycle2D~{\nwtagstyle{}\subpageref{NW3gGP3e-447MWQ-1}}}\plusendmoddef\Rm{}\nwstartdeflinemarkup\nwusesondefline{\\{NW3gGP3e-2ARAe1-1}}\nwprevnextdefs{NW3gGP3e-447MWQ-5}{NW3gGP3e-447MWQ-7}\nwenddeflinemarkup
{\bf{}inline} {\bf{}void} {\it{}asy\_draw}({\it{}ostream} & {\it{}ost}, {\bf{}const} {\it{}string} {\it{}picture},
                     {\bf{}const} {\bf{}ex} & {\it{}xmin} = -5, {\bf{}const} {\bf{}ex} & {\it{}xmax} = 5,
                     {\bf{}const} {\bf{}ex} & {\it{}ymin} = -5, {\bf{}const} {\bf{}ex} & {\it{}ymax} = 5, {\bf{}const} {\bf{}lst} & {\it{}color} = {\bf{}lst}(),
                     {\bf{}const} {\it{}string} {\it{}more\_options} = {\tt{}""}, {\bf{}bool} {\it{}with\_header} = {\bf{}true},
                     {\bf{}int} {\it{}points\_per\_arc} = 0, {\bf{}const} {\it{}string} {\it{}imaginary\_options}={\tt{}"rgb(0,.9,0)+4pt"}) {\bf{}const}
{\nwlbrace}{\it{}metapost\_draw}({\it{}ost}, {\it{}xmin}, {\it{}xmax}, {\it{}ymin}, {\it{}ymax}, {\it{}color}, {\it{}more\_options}, {\it{}with\_header},
                {\it{}points\_per\_arc}, {\bf{}true}, {\it{}picture}, {\bf{}false}, {\bf{}false}, {\it{}imaginary\_options}); {\nwrbrace}

{\bf{}inline} {\bf{}void} {\it{}asy\_draw}({\it{}ostream} & {\it{}ost} = {\it{}std}::{\it{}cout},
                     {\bf{}const} {\bf{}ex} & {\it{}xmin} = -5, {\bf{}const} {\bf{}ex} & {\it{}xmax} = 5,
                     {\bf{}const} {\bf{}ex} & {\it{}ymin} = -5, {\bf{}const} {\bf{}ex} & {\it{}ymax} = 5, {\bf{}const} {\bf{}lst} & {\it{}color} = {\bf{}lst}(),
                     {\bf{}const} {\it{}string} {\it{}more\_options} = {\tt{}""},
                     {\bf{}bool} {\it{}with\_header} = {\bf{}true}, {\bf{}int} {\it{}points\_per\_arc} = 0, {\bf{}const} {\it{}string} {\it{}imaginary\_options}={\tt{}"rgb(0,.9,0)+4pt"}) {\bf{}const}
{\nwlbrace}{\it{}metapost\_draw}({\it{}ost}, {\it{}xmin}, {\it{}xmax}, {\it{}ymin}, {\it{}ymax}, {\it{}color}, {\it{}more\_options}, {\it{}with\_header},
                {\it{}points\_per\_arc}, {\bf{}true}, {\tt{}""}, {\bf{}false}, {\bf{}false}, {\it{}imaginary\_options}); {\nwrbrace}
\nwindexdefn{\nwixident{asy{\_}draw}}{asy:undraw}{NW3gGP3e-447MWQ-6}\eatline
\nwused{\\{NW3gGP3e-2ARAe1-1}}\nwidentdefs{\\{{\nwixident{asy{\_}draw}}{asy:undraw}}}\nwidentuses{\\{{\nwixident{bool}}{bool}}\\{{\nwixident{ex}}{ex}}\\{{\nwixident{lst}}{lst}}\\{{\nwixident{metapost{\_}draw}}{metapost:undraw}}\\{{\nwixident{string}}{string}}}\nwindexuse{\nwixident{bool}}{bool}{NW3gGP3e-447MWQ-6}\nwindexuse{\nwixident{ex}}{ex}{NW3gGP3e-447MWQ-6}\nwindexuse{\nwixident{lst}}{lst}{NW3gGP3e-447MWQ-6}\nwindexuse{\nwixident{metapost{\_}draw}}{metapost:undraw}{NW3gGP3e-447MWQ-6}\nwindexuse{\nwixident{string}}{string}{NW3gGP3e-447MWQ-6}\nwendcode{}\nwbegindocs{134} 
\nwenddocs{}\nwbegindocs{135}Finally, we have a similar method which does not issue drawing
command, instead it writes a definition for a (array of) path, which
may be manipulated later.
\nwenddocs{}\nwbegincode{136}\sublabel{NW3gGP3e-447MWQ-7}\nwmargintag{{\nwtagstyle{}\subpageref{NW3gGP3e-447MWQ-7}}}\moddef{methods specific for class cycle2D~{\nwtagstyle{}\subpageref{NW3gGP3e-447MWQ-1}}}\plusendmoddef\Rm{}\nwstartdeflinemarkup\nwusesondefline{\\{NW3gGP3e-2ARAe1-1}}\nwprevnextdefs{NW3gGP3e-447MWQ-6}{\relax}\nwenddeflinemarkup
{\bf{}inline} {\bf{}void} {\it{}asy\_path}({\it{}ostream} & {\it{}ost} = {\it{}std}::{\it{}cout},
                     {\bf{}const} {\bf{}ex} & {\it{}xmin} = -5, {\bf{}const} {\bf{}ex} & {\it{}xmax} = 5,
                     {\bf{}const} {\bf{}ex} & {\it{}ymin} = -5, {\bf{}const} {\bf{}ex} & {\it{}ymax} = 5,
                     {\bf{}int} {\it{}points\_per\_arc} = 0, {\bf{}bool} {\it{}is\_continuation} = {\bf{}false}) {\bf{}const}
{\nwlbrace}{\it{}metapost\_draw}({\it{}ost}, {\it{}xmin}, {\it{}xmax}, {\it{}ymin}, {\it{}ymax}, {\bf{}lst}(), {\tt{}""}, {\bf{}false},
                {\it{}points\_per\_arc}, {\bf{}true}, {\tt{}""}, {\bf{}true}, {\it{}is\_continuation}); {\nwrbrace}
\nwindexdefn{\nwixident{asy{\_}path}}{asy:unpath}{NW3gGP3e-447MWQ-7}\eatline
\nwused{\\{NW3gGP3e-2ARAe1-1}}\nwidentdefs{\\{{\nwixident{asy{\_}path}}{asy:unpath}}}\nwidentuses{\\{{\nwixident{bool}}{bool}}\\{{\nwixident{ex}}{ex}}\\{{\nwixident{lst}}{lst}}\\{{\nwixident{metapost{\_}draw}}{metapost:undraw}}}\nwindexuse{\nwixident{bool}}{bool}{NW3gGP3e-447MWQ-7}\nwindexuse{\nwixident{ex}}{ex}{NW3gGP3e-447MWQ-7}\nwindexuse{\nwixident{lst}}{lst}{NW3gGP3e-447MWQ-7}\nwindexuse{\nwixident{metapost{\_}draw}}{metapost:undraw}{NW3gGP3e-447MWQ-7}\nwendcode{}\nwbegindocs{137}\nwdocspar
\nwenddocs{}\nwbegindocs{138}\nwdocspar
\subsection[An Example: Moebius Invariance of cycles]{An Example: M\"obius Invariance of cycles}
\label{sec:mobi-invar-cycl}

A quick illustration of the library usage is the symbolic calculation
which proves the Lem.~\ref{E-pr:transform-cycles}
from~\cite{Kisil05b}: We check that a M\"obius transformation
\(g\in\SL\) acts on cycles by similarity \(g: C \rightarrow
gCg^{-1}\). We use the following predefined objects:
\begin{webcode}
  cycle2D C(k,lst(l,n),m,e);
  ex W=lst(u,v);
\end{webcode}

Firstly we define a {\Tt{}\Rm{}{\bf{}cycle2D}\nwendquote} {\Tt{}\Rm{}{\it{}C2}\nwendquote} by the condition
between {\Tt{}\Rm{}{\it{}k}\nwendquote}, {\Tt{}\Rm{}{\it{}l}\nwendquote} and {\Tt{}\Rm{}{\it{}m}\nwendquote} in the generic {\Tt{}\Rm{}{\bf{}cycle2D}\nwendquote} {\Tt{}\Rm{}{\it{}C}\nwendquote} that
{\Tt{}\Rm{}{\it{}C}\nwendquote} passes through some point {\Tt{}\Rm{}{\it{}W}\nwendquote}.

\nwenddocs{}\nwbegincode{139}\sublabel{NW3gGP3e-2em4FU-1}\nwmargintag{{\nwtagstyle{}\subpageref{NW3gGP3e-2em4FU-1}}}\moddef{Moebius transformation of cycles~{\nwtagstyle{}\subpageref{NW3gGP3e-2em4FU-1}}}\endmoddef\Rm{}\nwstartdeflinemarkup\nwusesondefline{\\{NW3gGP3e-1zkOI-1}}\nwprevnextdefs{\relax}{NW3gGP3e-2em4FU-2}\nwenddeflinemarkup
{\it{}C2} = {\it{}C}.{\it{}subject\_to}({\bf{}lst}({\it{}C}.{\it{}passing}({\it{}W})));

\nwalsodefined{\\{NW3gGP3e-2em4FU-2}}\nwused{\\{NW3gGP3e-1zkOI-1}}\nwidentuses{\\{{\nwixident{lst}}{lst}}\\{{\nwixident{passing}}{passing}}\\{{\nwixident{subject{\_}to}}{subject:unto}}}\nwindexuse{\nwixident{lst}}{lst}{NW3gGP3e-2em4FU-1}\nwindexuse{\nwixident{passing}}{passing}{NW3gGP3e-2em4FU-1}\nwindexuse{\nwixident{subject{\_}to}}{subject:unto}{NW3gGP3e-2em4FU-1}\nwendcode{}\nwbegindocs{140}The point {\Tt{}\Rm{}{\it{}gW}\nwendquote} is defined to be the M\"obius transform of
{\Tt{}\Rm{}{\it{}W}\nwendquote} by  an arbitrary \(g\).
\nwenddocs{}\nwbegincode{141}\sublabel{NW3gGP3e-XJsVP-1}\nwmargintag{{\nwtagstyle{}\subpageref{NW3gGP3e-XJsVP-1}}}\moddef{Moebius transforms of W~{\nwtagstyle{}\subpageref{NW3gGP3e-XJsVP-1}}}\endmoddef\Rm{}\nwstartdeflinemarkup\nwusesondefline{\\{NW3gGP3e-3hvAAH-9}}\nwenddeflinemarkup
{\bf{}const} {\bf{}matrix} {\it{}gW}={\it{}ex\_to}\begin{math}<\end{math}{\bf{}matrix}\begin{math}>\end{math}({\it{}clifford\_moebius\_map}({\it{}sl2\_clifford}({\it{}a}, {\it{}b}, {\it{}c}, {\it{}d}, {\it{}e}), {\it{}W}, {\it{}e}).{\it{}subs}({\it{}sl2\_relation1},\nwindexdefn{\nwixident{matrix}}{matrix}{NW3gGP3e-XJsVP-1}
  {\it{}subs\_options}::{\it{}algebraic} \begin{math}\mid\end{math} {\it{}subs\_options}::{\it{}no\_pattern}).{\it{}normal}());

\nwused{\\{NW3gGP3e-3hvAAH-9}}\nwidentdefs{\\{{\nwixident{matrix}}{matrix}}}\nwidentuses{\\{{\nwixident{normal}}{normal}}\\{{\nwixident{subs}}{subs}}}\nwindexuse{\nwixident{normal}}{normal}{NW3gGP3e-XJsVP-1}\nwindexuse{\nwixident{subs}}{subs}{NW3gGP3e-XJsVP-1}\nwendcode{}\nwbegindocs{142}Finally we verify that the new cycle \(gCg^{-1}\) passes through {\Tt{}\Rm{}{\it{}P}\nwendquote}. This
proves Lem.~\ref{E-pr:transform-cycles} from~\cite{Kisil05a}.
\nwenddocs{}\nwbegincode{143}\sublabel{NW3gGP3e-2em4FU-2}\nwmargintag{{\nwtagstyle{}\subpageref{NW3gGP3e-2em4FU-2}}}\moddef{Moebius transformation of cycles~{\nwtagstyle{}\subpageref{NW3gGP3e-2em4FU-1}}}\plusendmoddef\Rm{}\nwstartdeflinemarkup\nwusesondefline{\\{NW3gGP3e-1zkOI-1}}\nwprevnextdefs{NW3gGP3e-2em4FU-1}{\relax}\nwenddeflinemarkup
{\it{}cout} \begin{math}\ll\end{math} {\tt{}"Conjugation of a cycle comes through Moebius transformation: "}
        \begin{math}\ll\end{math} {\it{}C2}.{\it{}sl2\_similarity}({\it{}a}, {\it{}b}, {\it{}c}, {\it{}d}, {\it{}es}, {\it{}S2}, {\bf{}true}, {\it{}S2}).{\it{}val}({\it{}gW}).{\it{}subs}({\it{}sl2\_relation1},
                {\it{}subs\_options}::{\it{}algebraic} \begin{math}\mid\end{math} {\it{}subs\_options}::{\it{}no\_pattern}).{\it{}normal}().{\it{}is\_zero}()
 \begin{math}\ll\end{math} {\it{}endl} \begin{math}\ll\end{math} {\it{}endl};

\nwused{\\{NW3gGP3e-1zkOI-1}}\nwidentuses{\\{{\nwixident{cycle}}{cycle}}\\{{\nwixident{is{\_}zero}}{is:unzero}}\\{{\nwixident{normal}}{normal}}\\{{\nwixident{sl2{\_}similarity}}{sl2:unsimilarity}}\\{{\nwixident{subs}}{subs}}\\{{\nwixident{val}}{val}}}\nwindexuse{\nwixident{cycle}}{cycle}{NW3gGP3e-2em4FU-2}\nwindexuse{\nwixident{is{\_}zero}}{is:unzero}{NW3gGP3e-2em4FU-2}\nwindexuse{\nwixident{normal}}{normal}{NW3gGP3e-2em4FU-2}\nwindexuse{\nwixident{sl2{\_}similarity}}{sl2:unsimilarity}{NW3gGP3e-2em4FU-2}\nwindexuse{\nwixident{subs}}{subs}{NW3gGP3e-2em4FU-2}\nwindexuse{\nwixident{val}}{val}{NW3gGP3e-2em4FU-2}\nwendcode{}\nwbegindocs{144}\nwdocspar

\nwenddocs{}\nwbegindocs{145}\nwdocspar
\section{Demonstration through example}
\label{sec:main-procedure}
We illustrate the library usage by the complete program which was used
for computer-assisted proofs in the paper~\cite{Kisil05a}. The
numerous cross-references between these two papers are active
hyperlinks. It is recommended to obtain PDF files for both of them
from \url{http://arXiv.org} and put into the same local directory. In
this case clicking on a reference in a PDF reader will automatically
transfer to the appropriate place (even in the other paper).

\subsection[Outline of the main()]{Outline of the {\Tt{}\Rm{}{\it{}main}()\nwendquote}}
\label{sec:outline-main}

The {\Tt{}\Rm{}{\it{}main}()\nwendquote} procedure does several things:
\begin{enumerate}
\item Makes symbolic calculations related to M\"obius invariance;
\nwenddocs{}\nwbegincode{146}\sublabel{NW3gGP3e-1zkOI-1}\nwmargintag{{\nwtagstyle{}\subpageref{NW3gGP3e-1zkOI-1}}}\moddef{List of symbolic calculations~{\nwtagstyle{}\subpageref{NW3gGP3e-1zkOI-1}}}\endmoddef\Rm{}\nwstartdeflinemarkup\nwusesondefline{\\{NW3gGP3e-1p0Y9w-4}}\nwprevnextdefs{\relax}{NW3gGP3e-1zkOI-2}\nwenddeflinemarkup
  \LA{}Moebius transformation of cycles~{\nwtagstyle{}\subpageref{NW3gGP3e-2em4FU-1}}\RA{}
  \LA{}K-orbit invariance~{\nwtagstyle{}\subpageref{NW3gGP3e-3pbeiz-1}}\RA{}
  \LA{}Check Moebius transformations of zero cycles~{\nwtagstyle{}\subpageref{NW3gGP3e-19ZdJC-1}}\RA{}
  \LA{}Check transformations of zero cycles by conjugation~{\nwtagstyle{}\subpageref{NW3gGP3e-3O7JaU-1}}\RA{}
{\it{}cout} \begin{math}\ll\end{math} {\it{}endl};

\nwalsodefined{\\{NW3gGP3e-1zkOI-2}\\{NW3gGP3e-1zkOI-3}\\{NW3gGP3e-1zkOI-4}\\{NW3gGP3e-1zkOI-5}}\nwused{\\{NW3gGP3e-1p0Y9w-4}}\nwendcode{}\nwbegindocs{147}\nwdocspar
\item Calculates properties of orthogonality conditions and
  corresponding inversion in cycles;
\nwenddocs{}\nwbegincode{148}\sublabel{NW3gGP3e-1zkOI-2}\nwmargintag{{\nwtagstyle{}\subpageref{NW3gGP3e-1zkOI-2}}}\moddef{List of symbolic calculations~{\nwtagstyle{}\subpageref{NW3gGP3e-1zkOI-1}}}\plusendmoddef\Rm{}\nwstartdeflinemarkup\nwusesondefline{\\{NW3gGP3e-1p0Y9w-4}}\nwprevnextdefs{NW3gGP3e-1zkOI-1}{NW3gGP3e-1zkOI-3}\nwenddeflinemarkup
  \LA{}Orthogonality conditions~{\nwtagstyle{}\subpageref{NW3gGP3e-3aTNrh-1}}\RA{}
  \LA{}Two points and orthogonality~{\nwtagstyle{}\subpageref{NW3gGP3e-3nCzcM-1}}\RA{}
  \LA{}One point and orthogonality~{\nwtagstyle{}\subpageref{NW3gGP3e-mJ5pz-1}}\RA{}
  \LA{}Orthogonal line~{\nwtagstyle{}\subpageref{NW3gGP3e-3uJ41M-1}}\RA{}
  \LA{}Inversion in cycle~{\nwtagstyle{}\subpageref{NW3gGP3e-47k9wZ-1}}\RA{}
  \LA{}Reflection in cycle~{\nwtagstyle{}\subpageref{NW3gGP3e-2W1nIA-1}}\RA{}
  \LA{}Yaglom inversion~{\nwtagstyle{}\subpageref{NW3gGP3e-4Lsx4s-1}}\RA{}
{\it{}cout} \begin{math}\ll\end{math} {\it{}endl};

\nwused{\\{NW3gGP3e-1p0Y9w-4}}\nwendcode{}\nwbegindocs{149}\nwdocspar
\item Calculates properties of f-orthogonality conditions and second
  type of inversion;
\nwenddocs{}\nwbegincode{150}\sublabel{NW3gGP3e-1zkOI-3}\nwmargintag{{\nwtagstyle{}\subpageref{NW3gGP3e-1zkOI-3}}}\moddef{List of symbolic calculations~{\nwtagstyle{}\subpageref{NW3gGP3e-1zkOI-1}}}\plusendmoddef\Rm{}\nwstartdeflinemarkup\nwusesondefline{\\{NW3gGP3e-1p0Y9w-4}}\nwprevnextdefs{NW3gGP3e-1zkOI-2}{NW3gGP3e-1zkOI-4}\nwenddeflinemarkup
  \LA{}Focal orthogonality conditions~{\nwtagstyle{}\subpageref{NW3gGP3e-2oH953-1}}\RA{}
  \LA{}One point and f-orthogonality~{\nwtagstyle{}\subpageref{NW3gGP3e-P6T5o-1}}\RA{}
  \LA{}f-orthogonal line~{\nwtagstyle{}\subpageref{NW3gGP3e-3jX7bo-1}}\RA{}
  \LA{}f-inversion in cycle~{\nwtagstyle{}\subpageref{NW3gGP3e-22oRI1-1}}\RA{}
{\it{}cout} \begin{math}\ll\end{math} {\it{}endl};

\nwused{\\{NW3gGP3e-1p0Y9w-4}}\nwendcode{}\nwbegindocs{151}\nwdocspar
\item Calculates various length formulae;
\nwenddocs{}\nwbegincode{152}\sublabel{NW3gGP3e-1zkOI-4}\nwmargintag{{\nwtagstyle{}\subpageref{NW3gGP3e-1zkOI-4}}}\moddef{List of symbolic calculations~{\nwtagstyle{}\subpageref{NW3gGP3e-1zkOI-1}}}\plusendmoddef\Rm{}\nwstartdeflinemarkup\nwusesondefline{\\{NW3gGP3e-1p0Y9w-4}}\nwprevnextdefs{NW3gGP3e-1zkOI-3}{NW3gGP3e-1zkOI-5}\nwenddeflinemarkup
  \LA{}Distances from cycles~{\nwtagstyle{}\subpageref{NW3gGP3e-O1KCX-1}}\RA{}
  \LA{}Lengths from centre~{\nwtagstyle{}\subpageref{NW3gGP3e-20e6xZ-1}}\RA{}
  \LA{}Lengths from focus~{\nwtagstyle{}\subpageref{NW3gGP3e-Nadcx-1}}\RA{}
  \LA{}Infinitesimal cycle~{\nwtagstyle{}\subpageref{NW3gGP3e-2Xuors-1}}\RA{}
{\it{}cout} \begin{math}\ll\end{math} {\it{}endl};

\nwused{\\{NW3gGP3e-1p0Y9w-4}}\nwendcode{}\nwbegindocs{153}\nwdocspar
\item Generates \Asymptote\ output of the for illustrations.
\nwenddocs{}\nwbegincode{154}\sublabel{NW3gGP3e-1zkOI-5}\nwmargintag{{\nwtagstyle{}\subpageref{NW3gGP3e-1zkOI-5}}}\moddef{List of symbolic calculations~{\nwtagstyle{}\subpageref{NW3gGP3e-1zkOI-1}}}\plusendmoddef\Rm{}\nwstartdeflinemarkup\nwusesondefline{\\{NW3gGP3e-1p0Y9w-4}}\nwprevnextdefs{NW3gGP3e-1zkOI-4}{\relax}\nwenddeflinemarkup

\nwused{\\{NW3gGP3e-1p0Y9w-4}}\nwendcode{}\nwbegindocs{155}\nwdocspar
\end{enumerate}
Since we aiming into two targets simultaneously---validate our
software and use it for mathematical proofs---there are many double checks
and superfluous calculations. The positive aspect of this---a better
illustration of the library usage.

\nwenddocs{}\nwbegindocs{156}\nwdocspar
\subsubsection{The program outline}
\label{sec:programs-outline}

Here is the main entry into the program and its outline. We start from
some inclusions, note that \GiNaC\ is included through {\Tt{}\Rm{}\begin{math}<\end{math}{\bf{}cycle}.{\it{}h}\begin{math}>\end{math}\nwendquote}.
\nwenddocs{}\nwbegincode{157}\sublabel{NW3gGP3e-1p0Y9w-1}\nwmargintag{{\nwtagstyle{}\subpageref{NW3gGP3e-1p0Y9w-1}}}\moddef{*~{\nwtagstyle{}\subpageref{NW3gGP3e-1p0Y9w-1}}}\endmoddef\Rm{}\nwstartdeflinemarkup\nwprevnextdefs{\relax}{NW3gGP3e-1p0Y9w-2}\nwenddeflinemarkup
\LA{}license~{\nwtagstyle{}\subpageref{NW3gGP3e-ZXuKx-1}}\RA{}
{\bf{}\char35{}include}{\tt{} \begin{math}<\end{math}cycle.h\begin{math}>\end{math}}
{\bf{}\char35{}include}{\tt{} \begin{math}<\end{math}fstream\begin{math}>\end{math}}

{\bf{}\char35{}define}{\tt{} par\_matr diag\_matrix(lst(-1, 0))}\nwindexdefn{\nwixident{par{\_}matr}}{par:unmatr}{NW3gGP3e-1p0Y9w-1}
{\bf{}\char35{}define}{\tt{} hyp\_matr diag\_matrix(lst(-1, 1))}\nwindexdefn{\nwixident{hyp{\_}matr}}{hyp:unmatr}{NW3gGP3e-1p0Y9w-1}
{\bf{}using} {\bf{}namespace} {\it{}MoebInv};
{\bf{}using} {\bf{}namespace} {\it{}std};
{\bf{}using} {\bf{}namespace} {\it{}GiNaC};

\nwalsodefined{\\{NW3gGP3e-1p0Y9w-2}\\{NW3gGP3e-1p0Y9w-3}\\{NW3gGP3e-1p0Y9w-4}\\{NW3gGP3e-1p0Y9w-5}}\nwnotused{*}\nwidentdefs{\\{{\nwixident{hyp{\_}matr}}{hyp:unmatr}}\\{{\nwixident{par{\_}matr}}{par:unmatr}}}\nwidentuses{\\{{\nwixident{cycle}}{cycle}}\\{{\nwixident{lst}}{lst}}\\{{\nwixident{MoebInv}}{MoebInv}}}\nwindexuse{\nwixident{cycle}}{cycle}{NW3gGP3e-1p0Y9w-1}\nwindexuse{\nwixident{lst}}{lst}{NW3gGP3e-1p0Y9w-1}\nwindexuse{\nwixident{MoebInv}}{MoebInv}{NW3gGP3e-1p0Y9w-1}\nwendcode{}\nwbegindocs{158}We try to make the output more readable both in simple text
and \LaTeX\ modes.
\nwenddocs{}\nwbegincode{159}\sublabel{NW3gGP3e-1p0Y9w-2}\nwmargintag{{\nwtagstyle{}\subpageref{NW3gGP3e-1p0Y9w-2}}}\moddef{*~{\nwtagstyle{}\subpageref{NW3gGP3e-1p0Y9w-1}}}\plusendmoddef\Rm{}\nwstartdeflinemarkup\nwprevnextdefs{NW3gGP3e-1p0Y9w-1}{NW3gGP3e-1p0Y9w-3}\nwenddeflinemarkup
{\bf{}\char35{}define}{\tt{} math\_string \begin{math}<\end{math}\begin{math}<\end{math} (output\_latex?"$":"")//$ (this is to balance dollar signs for LaTeX highlights in Xemacs)}\nwindexdefn{\nwixident{math{\_}string}}{math:unstring}{NW3gGP3e-1p0Y9w-2}
{\bf{}\char35{}define}{\tt{} wspaces (output\_latex?"\begin{math}\backslash\end{math}\begin{math}\backslash\end{math}quad ":"  ")}\nwindexdefn{\nwixident{wspaces}}{wspaces}{NW3gGP3e-1p0Y9w-2}

\nwidentdefs{\\{{\nwixident{math{\_}string}}{math:unstring}}\\{{\nwixident{wspaces}}{wspaces}}}\nwendcode{}\nwbegindocs{160}The structure of the program is transparent. We declare all variables.
\nwenddocs{}\nwbegincode{161}\sublabel{NW3gGP3e-1p0Y9w-3}\nwmargintag{{\nwtagstyle{}\subpageref{NW3gGP3e-1p0Y9w-3}}}\moddef{*~{\nwtagstyle{}\subpageref{NW3gGP3e-1p0Y9w-1}}}\plusendmoddef\Rm{}\nwstartdeflinemarkup\nwprevnextdefs{NW3gGP3e-1p0Y9w-2}{NW3gGP3e-1p0Y9w-4}\nwenddeflinemarkup
\LA{}Declaration of variables~{\nwtagstyle{}\subpageref{NW3gGP3e-3hvAAH-1}}\RA{}
\LA{}Subroutines definitions~{\nwtagstyle{}\subpageref{NW3gGP3e-1Z2pUX-1}}\RA{}
{\bf{}int} {\it{}main}(){\nwlbrace}\nwindexdefn{\nwixident{main}}{main}{NW3gGP3e-1p0Y9w-3}
 {\it{}cout} \begin{math}\ll\end{math} {\it{}boolalpha};

 {\bf{}if} ({\it{}output\_latex}) {\it{}cout} \begin{math}\ll\end{math} {\it{}latex};

\nwidentdefs{\\{{\nwixident{main}}{main}}}\nwendcode{}\nwbegindocs{162}Then we make all symbolic calculations listed above. The exception
catcher helps to identify the possible problems.
\nwenddocs{}\nwbegincode{163}\sublabel{NW3gGP3e-1p0Y9w-4}\nwmargintag{{\nwtagstyle{}\subpageref{NW3gGP3e-1p0Y9w-4}}}\moddef{*~{\nwtagstyle{}\subpageref{NW3gGP3e-1p0Y9w-1}}}\plusendmoddef\Rm{}\nwstartdeflinemarkup\nwprevnextdefs{NW3gGP3e-1p0Y9w-3}{NW3gGP3e-1p0Y9w-5}\nwenddeflinemarkup
{\bf{}try} {\nwlbrace}
    \LA{}List of symbolic calculations~{\nwtagstyle{}\subpageref{NW3gGP3e-1zkOI-1}}\RA{}
        {\nwrbrace} {\bf{}catch} ({\it{}exception} &{\it{}p}) {\nwlbrace}
    {\it{}cerr} \begin{math}\ll\end{math} {\tt{}"*****       Got problem1: "} \begin{math}\ll\end{math}  {\it{}p}.{\it{}what}() \begin{math}\ll\end{math} {\it{}endl};
{\nwrbrace}

\nwidentuses{\\{{\nwixident{catch}}{catch}}}\nwindexuse{\nwixident{catch}}{catch}{NW3gGP3e-1p0Y9w-4}\nwendcode{}\nwbegindocs{164}We end up with drawing illustration to our paper~\cite{Kisil05a}.
\nwenddocs{}\nwbegincode{165}\sublabel{NW3gGP3e-1p0Y9w-5}\nwmargintag{{\nwtagstyle{}\subpageref{NW3gGP3e-1p0Y9w-5}}}\moddef{*~{\nwtagstyle{}\subpageref{NW3gGP3e-1p0Y9w-1}}}\plusendmoddef\Rm{}\nwstartdeflinemarkup\nwprevnextdefs{NW3gGP3e-1p0Y9w-4}{\relax}\nwenddeflinemarkup
 \LA{}Draw Asymptote pictures~{\nwtagstyle{}\subpageref{NW3gGP3e-Xmoi0-1}}\RA{}
{\nwrbrace}

\nwendcode{}\nwbegindocs{166}\nwdocspar
\subsubsection{Declaration of variables}
\label{sec:decl-vari}

First we declare all variables from the standard \GiNaC\ classes here.
\nwenddocs{}\nwbegincode{167}\sublabel{NW3gGP3e-3hvAAH-1}\nwmargintag{{\nwtagstyle{}\subpageref{NW3gGP3e-3hvAAH-1}}}\moddef{Declaration of variables~{\nwtagstyle{}\subpageref{NW3gGP3e-3hvAAH-1}}}\endmoddef\Rm{}\nwstartdeflinemarkup\nwusesondefline{\\{NW3gGP3e-1p0Y9w-3}}\nwprevnextdefs{\relax}{NW3gGP3e-3hvAAH-2}\nwenddeflinemarkup
{\bf{}const} {\it{}string} {\it{}eph\_names}={\tt{}"eph"};\nwindexdefn{\nwixident{string}}{string}{NW3gGP3e-3hvAAH-1}
{\bf{}const} {\bf{}numeric} {\it{}half}(1,2);\nwindexdefn{\nwixident{numeric}}{numeric}{NW3gGP3e-3hvAAH-1}

{\bf{}const} {\bf{}realsymbol} {\it{}a}({\tt{}"a"}), {\it{}b}({\tt{}"b"}), {\it{}c}({\tt{}"c"}), {\it{}d}({\tt{}"d"}), {\it{}x}({\tt{}"x"}), {\it{}y}({\tt{}"y"}), {\it{}z}({\tt{}"z"}), {\it{}t}({\tt{}"t"}),\nwindexdefn{\nwixident{realsymbol}}{realsymbol}{NW3gGP3e-3hvAAH-1}
  {\it{}k}({\tt{}"k"}), {\it{}l}({\tt{}"L"},{\tt{}"l"}), {\it{}m}({\tt{}"m"}), {\it{}n}({\tt{}"n"}), // Cycles parameters
  {\it{}k1}({\tt{}"k1"},{\tt{}"{\char92}{\char92}tilde{\char123}k{\char125}"}), {\it{}l1}({\tt{}"l1"},{\tt{}"{\char92}{\char92}tilde{\char123}l{\char125}"}), {\it{}m1}({\tt{}"m1"},{\tt{}"{\char92}{\char92}tilde{\char123}m{\char125}"}), {\it{}n1}({\tt{}"n1"},{\tt{}"{\char92}{\char92}tilde{\char123}n{\char125}"}),
  {\it{}u}({\tt{}"u"}), {\it{}v}({\tt{}"v"}), {\it{}u1}({\tt{}"u1"}), {\it{}v1}({\tt{}"v1"}), // Coordinates of points in \(\Space{R}{2}\)
  {\it{}epsilon}({\tt{}"eps"}, {\tt{}"{\char92}{\char92}epsilon"}); // The "infinitesimal" number

{\bf{}const} {\bf{}varidx} {\it{}nu}({\bf{}symbol}({\tt{}"nu"}, {\tt{}"{\char92}{\char92}nu"}), 2), {\it{}mu}({\bf{}symbol}({\tt{}"mu"}, {\tt{}"{\char92}{\char92}mu"}), 2);\nwindexdefn{\nwixident{varidx}}{varidx}{NW3gGP3e-3hvAAH-1}

\nwalsodefined{\\{NW3gGP3e-3hvAAH-2}\\{NW3gGP3e-3hvAAH-3}\\{NW3gGP3e-3hvAAH-4}\\{NW3gGP3e-3hvAAH-5}\\{NW3gGP3e-3hvAAH-6}\\{NW3gGP3e-3hvAAH-7}\\{NW3gGP3e-3hvAAH-8}\\{NW3gGP3e-3hvAAH-9}\\{NW3gGP3e-3hvAAH-A}\\{NW3gGP3e-3hvAAH-B}\\{NW3gGP3e-3hvAAH-C}}\nwused{\\{NW3gGP3e-1p0Y9w-3}}\nwidentdefs{\\{{\nwixident{numeric}}{numeric}}\\{{\nwixident{realsymbol}}{realsymbol}}\\{{\nwixident{string}}{string}}\\{{\nwixident{varidx}}{varidx}}}\nwidentuses{\\{{\nwixident{k}}{k}}\\{{\nwixident{l}}{l}}\\{{\nwixident{m}}{m}}\\{{\nwixident{points}}{points}}\\{{\nwixident{u}}{u}}\\{{\nwixident{v}}{v}}}\nwindexuse{\nwixident{k}}{k}{NW3gGP3e-3hvAAH-1}\nwindexuse{\nwixident{l}}{l}{NW3gGP3e-3hvAAH-1}\nwindexuse{\nwixident{m}}{m}{NW3gGP3e-3hvAAH-1}\nwindexuse{\nwixident{points}}{points}{NW3gGP3e-3hvAAH-1}\nwindexuse{\nwixident{u}}{u}{NW3gGP3e-3hvAAH-1}\nwindexuse{\nwixident{v}}{v}{NW3gGP3e-3hvAAH-1}\nwendcode{}\nwbegindocs{168}We need a plenty of symbols which will hold various parameters like
\(e_1^2\),  \(\breve{e}_1^2\), \(s\) for the SFSCc.
\nwenddocs{}\nwbegincode{169}\sublabel{NW3gGP3e-3hvAAH-2}\nwmargintag{{\nwtagstyle{}\subpageref{NW3gGP3e-3hvAAH-2}}}\moddef{Declaration of variables~{\nwtagstyle{}\subpageref{NW3gGP3e-3hvAAH-1}}}\plusendmoddef\Rm{}\nwstartdeflinemarkup\nwusesondefline{\\{NW3gGP3e-1p0Y9w-3}}\nwprevnextdefs{NW3gGP3e-3hvAAH-1}{NW3gGP3e-3hvAAH-3}\nwenddeflinemarkup
{\bf{}const} {\bf{}realsymbol} {\it{}sign}({\tt{}"s"}, {\tt{}"{\char92}{\char92}sigma"}), {\it{}sign1}({\tt{}"s1"}, {\tt{}"{\char92}{\char92}breve{\char123}{\char92}{\char92}sigma{\char125}"}), //Signs of \(e\sb1\sp2\) of  \(\breve{e}\sb1\sp2\)\nwindexdefn{\nwixident{realsymbol}}{realsymbol}{NW3gGP3e-3hvAAH-2}
            {\it{}sign2}({\tt{}"s2"}, {\tt{}"{\char92}{\char92}sigma\_2"}), {\it{}sign3}({\tt{}"s3"}, {\tt{}"{\char92}{\char92}sigma\_3"}), {\it{}sign4}({\tt{}"s4"}, {\tt{}"{\char92}{\char92}mathring{\char123}{\char92}{\char92}sigma{\char125}"});
{\bf{}int} {\it{}si}, {\it{}si1}; // Values of \(e\sb1\sp2\) and \(\breve{e}\sb1\sp2\) for substitutions\nwindexdefn{\nwixident{si}}{si}{NW3gGP3e-3hvAAH-2}\nwindexdefn{\nwixident{si1}}{si1}{NW3gGP3e-3hvAAH-2}

{\bf{}const} {\bf{}matrix} {\it{}S2}(2, 2, {\bf{}lst}(1, 0, 0, {\it{}jump\_fnct}({\it{}sign2}))),\nwindexdefn{\nwixident{matrix}}{matrix}{NW3gGP3e-3hvAAH-2}
            {\it{}S3}(2, 2, {\bf{}lst}(1, 0, 0, {\it{}jump\_fnct}({\it{}sign3}))),
            {\it{}S4}(2, 2, {\bf{}lst}(1, 0, 0, {\it{}jump\_fnct}({\it{}sign4}))); //Signs of {\it{}l} in the matrix representations of cycles

\nwused{\\{NW3gGP3e-1p0Y9w-3}}\nwidentdefs{\\{{\nwixident{matrix}}{matrix}}\\{{\nwixident{realsymbol}}{realsymbol}}\\{{\nwixident{si}}{si}}\\{{\nwixident{si1}}{si1}}}\nwidentuses{\\{{\nwixident{jump{\_}fnct}}{jump:unfnct}}\\{{\nwixident{l}}{l}}\\{{\nwixident{lst}}{lst}}}\nwindexuse{\nwixident{jump{\_}fnct}}{jump:unfnct}{NW3gGP3e-3hvAAH-2}\nwindexuse{\nwixident{l}}{l}{NW3gGP3e-3hvAAH-2}\nwindexuse{\nwixident{lst}}{lst}{NW3gGP3e-3hvAAH-2}\nwendcode{}\nwbegindocs{170}Here are several expressions which will keep results of calculations.
\nwenddocs{}\nwbegincode{171}\sublabel{NW3gGP3e-3hvAAH-3}\nwmargintag{{\nwtagstyle{}\subpageref{NW3gGP3e-3hvAAH-3}}}\moddef{Declaration of variables~{\nwtagstyle{}\subpageref{NW3gGP3e-3hvAAH-1}}}\plusendmoddef\Rm{}\nwstartdeflinemarkup\nwusesondefline{\\{NW3gGP3e-1p0Y9w-3}}\nwprevnextdefs{NW3gGP3e-3hvAAH-2}{NW3gGP3e-3hvAAH-4}\nwenddeflinemarkup
{\bf{}ex} {\it{}u2}, {\it{}v2}, // Coordinates of the Moebius transform of ({\it{}u}, {\it{}v})
            {\it{}u3}, {\it{}v3}, {\it{}u4}, {\it{}v4}, {\it{}u5}, {\it{}v5},
            {\it{}P}, {\it{}P1}, // points on the plain
            {\it{}K}, {\it{}L0}, {\it{}L1}, // Parameters of cycles
            {\it{}Len\_c}, // Expressions of Lengths
            {\it{}p};

\nwused{\\{NW3gGP3e-1p0Y9w-3}}\nwidentuses{\\{{\nwixident{ex}}{ex}}\\{{\nwixident{points}}{points}}\\{{\nwixident{u}}{u}}\\{{\nwixident{v}}{v}}}\nwindexuse{\nwixident{ex}}{ex}{NW3gGP3e-3hvAAH-3}\nwindexuse{\nwixident{points}}{points}{NW3gGP3e-3hvAAH-3}\nwindexuse{\nwixident{u}}{u}{NW3gGP3e-3hvAAH-3}\nwindexuse{\nwixident{v}}{v}{NW3gGP3e-3hvAAH-3}\nwendcode{}\nwbegindocs{172} Next we define metrics (through Clifford units) for the space of
points ({\Tt{}\Rm{}{\it{}M}\nwendquote}, {\Tt{}\Rm{}{\it{}e}\nwendquote}) and space of spheres ({\Tt{}\Rm{}{\it{}M1}\nwendquote}, {\Tt{}\Rm{}{\it{}es}\nwendquote}).
\nwenddocs{}\nwbegincode{173}\sublabel{NW3gGP3e-3hvAAH-4}\nwmargintag{{\nwtagstyle{}\subpageref{NW3gGP3e-3hvAAH-4}}}\moddef{Declaration of variables~{\nwtagstyle{}\subpageref{NW3gGP3e-3hvAAH-1}}}\plusendmoddef\Rm{}\nwstartdeflinemarkup\nwusesondefline{\\{NW3gGP3e-1p0Y9w-3}}\nwprevnextdefs{NW3gGP3e-3hvAAH-3}{NW3gGP3e-3hvAAH-5}\nwenddeflinemarkup
{\bf{}const} {\bf{}ex} {\it{}M} = {\it{}diag\_matrix}({\bf{}lst}(-1, {\it{}sign})), // Metrics of point spaces\nwindexdefn{\nwixident{ex}}{ex}{NW3gGP3e-3hvAAH-4}
            {\it{}e} =  {\it{}clifford\_unit}({\it{}mu}, {\it{}M}, 0), // Clifford algebra generators in the point space
            {\it{}M1} = {\it{}diag\_matrix}({\bf{}lst}(-1, {\it{}sign1})), // Metrics of cycles spaces
            {\it{}es} =  {\it{}clifford\_unit}({\it{}nu}, {\it{}M1}, 1);  // Clifford algebra generators in the sphere space

\nwused{\\{NW3gGP3e-1p0Y9w-3}}\nwidentdefs{\\{{\nwixident{ex}}{ex}}}\nwidentuses{\\{{\nwixident{lst}}{lst}}}\nwindexuse{\nwixident{lst}}{lst}{NW3gGP3e-3hvAAH-4}\nwendcode{}\nwbegindocs{174}Now we define instances of {\Tt{}\Rm{}{\bf{}cycle2D}\nwendquote} class. Some of them (like
{\Tt{}\Rm{}{\it{}real\_line}\nwendquote} or generic cycles {\Tt{}\Rm{}{\it{}C}\nwendquote} and {\Tt{}\Rm{}{\it{}C1}\nwendquote}) are constants.
\nwenddocs{}\nwbegincode{175}\sublabel{NW3gGP3e-3hvAAH-5}\nwmargintag{{\nwtagstyle{}\subpageref{NW3gGP3e-3hvAAH-5}}}\moddef{Declaration of variables~{\nwtagstyle{}\subpageref{NW3gGP3e-3hvAAH-1}}}\plusendmoddef\Rm{}\nwstartdeflinemarkup\nwusesondefline{\\{NW3gGP3e-1p0Y9w-3}}\nwprevnextdefs{NW3gGP3e-3hvAAH-4}{NW3gGP3e-3hvAAH-6}\nwenddeflinemarkup
{\bf{}cycle2D} {\it{}C2}, {\it{}C3}, {\it{}C4}, {\it{}C5}, {\it{}C6}, {\it{}C7}, {\it{}C8}, {\it{}C9}, {\it{}C10}, {\it{}C11};

{\bf{}const} {\bf{}cycle2D} {\it{}real\_line}(0, {\bf{}lst}(0, {\bf{}numeric}(1)), 0, {\it{}e}), // the real line\nwindexdefn{\nwixident{cycle2D}}{cycle2D}{NW3gGP3e-3hvAAH-5}
            {\it{}C}({\it{}k}, {\bf{}lst}({\it{}l}, {\it{}n}), {\it{}m}, {\it{}e}), {\it{}C1}({\it{}k1}, {\bf{}lst}({\it{}l1}, {\it{}n1}), {\it{}m1}, {\it{}e}); // two generic cycles
{\bf{}const} {\bf{}cycle2D} {\it{}Zinf}(0, {\bf{}lst}(0, 0), 1, {\it{}e}), // the zero-radius cycle at infinity\nwindexdefn{\nwixident{cycle2D}}{cycle2D}{NW3gGP3e-3hvAAH-5}
            {\it{}Z}({\bf{}lst}({\it{}u}, {\it{}v}), {\it{}e}), {\it{}Z1}({\bf{}lst}({\it{}u}, {\it{}v}), {\it{}e}, 0, {\it{}es}), // two generic cycles of zero-radius
            {\it{}Z2}({\bf{}lst}({\it{}u}, {\it{}v}), {\it{}e}, 0, {\it{}es}, {\it{}S2});

\nwused{\\{NW3gGP3e-1p0Y9w-3}}\nwidentdefs{\\{{\nwixident{cycle2D}}{cycle2D}}}\nwidentuses{\\{{\nwixident{cycle}}{cycle}}\\{{\nwixident{k}}{k}}\\{{\nwixident{l}}{l}}\\{{\nwixident{lst}}{lst}}\\{{\nwixident{m}}{m}}\\{{\nwixident{numeric}}{numeric}}\\{{\nwixident{u}}{u}}\\{{\nwixident{v}}{v}}}\nwindexuse{\nwixident{cycle}}{cycle}{NW3gGP3e-3hvAAH-5}\nwindexuse{\nwixident{k}}{k}{NW3gGP3e-3hvAAH-5}\nwindexuse{\nwixident{l}}{l}{NW3gGP3e-3hvAAH-5}\nwindexuse{\nwixident{lst}}{lst}{NW3gGP3e-3hvAAH-5}\nwindexuse{\nwixident{m}}{m}{NW3gGP3e-3hvAAH-5}\nwindexuse{\nwixident{numeric}}{numeric}{NW3gGP3e-3hvAAH-5}\nwindexuse{\nwixident{u}}{u}{NW3gGP3e-3hvAAH-5}\nwindexuse{\nwixident{v}}{v}{NW3gGP3e-3hvAAH-5}\nwendcode{}\nwbegindocs{176}For solution of various systems of linear equations we need the
followings {\Tt{}\Rm{}{\bf{}lst}\nwendquote}s.
\nwenddocs{}\nwbegincode{177}\sublabel{NW3gGP3e-3hvAAH-6}\nwmargintag{{\nwtagstyle{}\subpageref{NW3gGP3e-3hvAAH-6}}}\moddef{Declaration of variables~{\nwtagstyle{}\subpageref{NW3gGP3e-3hvAAH-1}}}\plusendmoddef\Rm{}\nwstartdeflinemarkup\nwusesondefline{\\{NW3gGP3e-1p0Y9w-3}}\nwprevnextdefs{NW3gGP3e-3hvAAH-5}{NW3gGP3e-3hvAAH-7}\nwenddeflinemarkup
{\bf{}lst} {\it{}eqns}, {\it{}eqns1},
            {\it{}vars}={\bf{}lst}({\it{}k1}, {\it{}l1}, {\it{}m1}, {\it{}n1}),
            {\it{}solns}, {\it{}solns1}, // Solutions of linear systems
            {\it{}sign\_val};

\nwused{\\{NW3gGP3e-1p0Y9w-3}}\nwidentuses{\\{{\nwixident{lst}}{lst}}}\nwindexuse{\nwixident{lst}}{lst}{NW3gGP3e-3hvAAH-6}\nwendcode{}\nwbegindocs{178}Here are {\Tt{}\Rm{}{\bf{}relational}\nwendquote}s and lists of {\Tt{}\Rm{}{\bf{}relational}\nwendquote}s which will be
used for automatic simplifications in calculations.
They are based on properties of \(\SL\) and values of the parameters.
\nwenddocs{}\nwbegincode{179}\sublabel{NW3gGP3e-3hvAAH-7}\nwmargintag{{\nwtagstyle{}\subpageref{NW3gGP3e-3hvAAH-7}}}\moddef{Declaration of variables~{\nwtagstyle{}\subpageref{NW3gGP3e-3hvAAH-1}}}\plusendmoddef\Rm{}\nwstartdeflinemarkup\nwusesondefline{\\{NW3gGP3e-1p0Y9w-3}}\nwprevnextdefs{NW3gGP3e-3hvAAH-6}{NW3gGP3e-3hvAAH-8}\nwenddeflinemarkup
{\bf{}const} {\bf{}ex} {\it{}sl2\_relation} = ({\it{}c}\begin{math}\ast\end{math}{\it{}b} \begin{math}\equiv\end{math} {\it{}a}\begin{math}\ast\end{math}{\it{}d}-1), {\it{}sl2\_relation1} = ({\it{}a} \begin{math}\equiv\end{math} (1+{\it{}b}\begin{math}\ast\end{math}{\it{}c})\begin{math}\div\end{math}{\it{}d}); // since \(ad-bc\equiv 1\)\nwindexdefn{\nwixident{ex}}{ex}{NW3gGP3e-3hvAAH-7}
{\bf{}const} {\bf{}lst} {\it{}signs\_cube} = {\bf{}lst}({\it{}pow}({\it{}sign}, 3) \begin{math}\equiv\end{math} {\it{}sign}, {\it{}pow}({\it{}sign1}, 3) \begin{math}\equiv\end{math} {\it{}sign1}); // \(s\sb{i}\sp3\equiv s\_{i}\) since \(s\sb{i}=-1, 0, 1\)\nwindexdefn{\nwixident{lst}}{lst}{NW3gGP3e-3hvAAH-7}

{\bf{}const} {\bf{}int} {\it{}debug} = 0;\nwindexdefn{\nwixident{debug}}{debug}{NW3gGP3e-3hvAAH-7}
{\bf{}const} {\bf{}bool} {\it{}output\_latex} = {\bf{}true};\nwindexdefn{\nwixident{bool}}{bool}{NW3gGP3e-3hvAAH-7}

\nwused{\\{NW3gGP3e-1p0Y9w-3}}\nwidentdefs{\\{{\nwixident{bool}}{bool}}\\{{\nwixident{debug}}{debug}}\\{{\nwixident{ex}}{ex}}\\{{\nwixident{lst}}{lst}}}\nwendcode{}\nwbegindocs{180}Two generic points on the plain are defined as constant vectors
(\(2\times 1\)matrices).
\nwenddocs{}\nwbegincode{181}\sublabel{NW3gGP3e-3hvAAH-8}\nwmargintag{{\nwtagstyle{}\subpageref{NW3gGP3e-3hvAAH-8}}}\moddef{Declaration of variables~{\nwtagstyle{}\subpageref{NW3gGP3e-3hvAAH-1}}}\plusendmoddef\Rm{}\nwstartdeflinemarkup\nwusesondefline{\\{NW3gGP3e-1p0Y9w-3}}\nwprevnextdefs{NW3gGP3e-3hvAAH-7}{NW3gGP3e-3hvAAH-9}\nwenddeflinemarkup
{\bf{}const} {\bf{}matrix} {\it{}W}(2,1, {\bf{}lst}({\it{}u}, {\it{}v})), {\it{}W1}(2,1, {\bf{}lst}({\it{}u1}, {\it{}v1}));\nwindexdefn{\nwixident{matrix}}{matrix}{NW3gGP3e-3hvAAH-8}

\nwused{\\{NW3gGP3e-1p0Y9w-3}}\nwidentdefs{\\{{\nwixident{matrix}}{matrix}}}\nwidentuses{\\{{\nwixident{lst}}{lst}}\\{{\nwixident{u}}{u}}\\{{\nwixident{v}}{v}}}\nwindexuse{\nwixident{lst}}{lst}{NW3gGP3e-3hvAAH-8}\nwindexuse{\nwixident{u}}{u}{NW3gGP3e-3hvAAH-8}\nwindexuse{\nwixident{v}}{v}{NW3gGP3e-3hvAAH-8}\nwendcode{}\nwbegindocs{182}We will also frequently use their M\"obius transforms.
\nwenddocs{}\nwbegincode{183}\sublabel{NW3gGP3e-3hvAAH-9}\nwmargintag{{\nwtagstyle{}\subpageref{NW3gGP3e-3hvAAH-9}}}\moddef{Declaration of variables~{\nwtagstyle{}\subpageref{NW3gGP3e-3hvAAH-1}}}\plusendmoddef\Rm{}\nwstartdeflinemarkup\nwusesondefline{\\{NW3gGP3e-1p0Y9w-3}}\nwprevnextdefs{NW3gGP3e-3hvAAH-8}{NW3gGP3e-3hvAAH-A}\nwenddeflinemarkup
{\bf{}const} {\bf{}matrix} {\it{}gW1}={\it{}ex\_to}\begin{math}<\end{math}{\bf{}matrix}\begin{math}>\end{math}({\it{}clifford\_moebius\_map}({\it{}sl2\_clifford}({\it{}a}, {\it{}b}, {\it{}c}, {\it{}d}, {\it{}e}), {\it{}W1}, {\it{}e}).{\it{}subs}({\it{}sl2\_relation1},\nwindexdefn{\nwixident{matrix}}{matrix}{NW3gGP3e-3hvAAH-9}
   {\it{}subs\_options}::{\it{}algebraic} \begin{math}\mid\end{math} {\it{}subs\_options}::{\it{}no\_pattern}).{\it{}normal}());
\LA{}Moebius transforms of W~{\nwtagstyle{}\subpageref{NW3gGP3e-XJsVP-1}}\RA{}

\nwused{\\{NW3gGP3e-1p0Y9w-3}}\nwidentdefs{\\{{\nwixident{matrix}}{matrix}}}\nwidentuses{\\{{\nwixident{normal}}{normal}}\\{{\nwixident{subs}}{subs}}}\nwindexuse{\nwixident{normal}}{normal}{NW3gGP3e-3hvAAH-9}\nwindexuse{\nwixident{subs}}{subs}{NW3gGP3e-3hvAAH-9}\nwendcode{}\nwbegindocs{184} We repeat some calculations several times for carious values of
parameters, such calculations are gathered here as subroutines.
\nwenddocs{}\nwbegincode{185}\sublabel{NW3gGP3e-1Z2pUX-1}\nwmargintag{{\nwtagstyle{}\subpageref{NW3gGP3e-1Z2pUX-1}}}\moddef{Subroutines definitions~{\nwtagstyle{}\subpageref{NW3gGP3e-1Z2pUX-1}}}\endmoddef\Rm{}\nwstartdeflinemarkup\nwusesondefline{\\{NW3gGP3e-1p0Y9w-3}}\nwenddeflinemarkup
\LA{}Parabolic Cayley transform of cycles~{\nwtagstyle{}\subpageref{NW3gGP3e-3vL8tu-1}}\RA{}
\LA{}Check conformal property~{\nwtagstyle{}\subpageref{NW3gGP3e-2eoTsH-1}}\RA{}
\LA{}Print perpendicular~{\nwtagstyle{}\subpageref{NW3gGP3e-4PKbLl-1}}\RA{}
\LA{}Focal length checks~{\nwtagstyle{}\subpageref{NW3gGP3e-2gYzHL-1}}\RA{}
\LA{}Infinitesimal cycle calculations~{\nwtagstyle{}\subpageref{NW3gGP3e-4WB4RW-1}}\RA{}

\nwused{\\{NW3gGP3e-1p0Y9w-3}}\nwendcode{}\nwbegindocs{186}\nwdocspar
\subsection[Moebius Transformation and Conjugation of Cycles]{M\"obius
  Transformation and Conjugation of Cycles}
\label{sec:symb-calc}

\nwenddocs{}\nwbegindocs{187}\nwdocspar

\subsubsection[Transformations of K-orbits]{Transformations of \(K\)-orbits}
\label{sec:transf-k-orbits}
As a simple check we verify that cycles given by the equation
\((u^2-\sigma v^2)-2v\frac{t^{-1}-\sigma t }{2}+1=0\),
see~\cite[Lem.~\ref{E-le:k-orbit-gen}]{Kisil05a} are \(K\)-invariant,
i.e. are \(K\)-orbits. To this end we make a similarity of a cycle
{\Tt{}\Rm{}{\it{}C2}\nwendquote} of this from with a matrix from \(K\) and check that the result
coincides with {\Tt{}\Rm{}{\it{}C2}\nwendquote}.
\nwenddocs{}\nwbegincode{188}\sublabel{NW3gGP3e-3pbeiz-1}\nwmargintag{{\nwtagstyle{}\subpageref{NW3gGP3e-3pbeiz-1}}}\moddef{K-orbit invariance~{\nwtagstyle{}\subpageref{NW3gGP3e-3pbeiz-1}}}\endmoddef\Rm{}\nwstartdeflinemarkup\nwusesondefline{\\{NW3gGP3e-1zkOI-1}}\nwprevnextdefs{\relax}{NW3gGP3e-3pbeiz-2}\nwenddeflinemarkup
{\it{}C2} = {\bf{}cycle2D}(1, {\bf{}lst}(0, ({\it{}pow}({\it{}t},-1)-{\it{}sign}\begin{math}\ast\end{math}{\it{}t})\begin{math}\div\end{math}2), 1, {\it{}e});
{\it{}cout} \begin{math}\ll\end{math} {\tt{}"A K-orbit is preserved: "} \begin{math}\ll\end{math} {\it{}C2}.{\it{}sl2\_similarity}({\it{}cos}({\it{}x}), {\it{}sin}({\it{}x}), -{\it{}sin}({\it{}x}), {\it{}cos}({\it{}x}), {\it{}e}).{\it{}is\_equal}({\it{}C2})

\nwalsodefined{\\{NW3gGP3e-3pbeiz-2}}\nwused{\\{NW3gGP3e-1zkOI-1}}\nwidentuses{\\{{\nwixident{cycle2D}}{cycle2D}}\\{{\nwixident{is{\_}equal}}{is:unequal}}\\{{\nwixident{lst}}{lst}}\\{{\nwixident{sl2{\_}similarity}}{sl2:unsimilarity}}}\nwindexuse{\nwixident{cycle2D}}{cycle2D}{NW3gGP3e-3pbeiz-1}\nwindexuse{\nwixident{is{\_}equal}}{is:unequal}{NW3gGP3e-3pbeiz-1}\nwindexuse{\nwixident{lst}}{lst}{NW3gGP3e-3pbeiz-1}\nwindexuse{\nwixident{sl2{\_}similarity}}{sl2:unsimilarity}{NW3gGP3e-3pbeiz-1}\nwendcode{}\nwbegindocs{189}We also check that {\Tt{}\Rm{}{\it{}C2}\nwendquote} passing the point \((0, t)\).
\nwenddocs{}\nwbegincode{190}\sublabel{NW3gGP3e-3pbeiz-2}\nwmargintag{{\nwtagstyle{}\subpageref{NW3gGP3e-3pbeiz-2}}}\moddef{K-orbit invariance~{\nwtagstyle{}\subpageref{NW3gGP3e-3pbeiz-1}}}\plusendmoddef\Rm{}\nwstartdeflinemarkup\nwusesondefline{\\{NW3gGP3e-1zkOI-1}}\nwprevnextdefs{NW3gGP3e-3pbeiz-1}{\relax}\nwenddeflinemarkup
    \begin{math}\ll\end{math} {\tt{}", and  passing (0, t): "} \begin{math}\ll\end{math} ({\bf{}bool}){\it{}ex\_to}\begin{math}<\end{math}{\bf{}relational}\begin{math}>\end{math}({\it{}C2}.{\it{}passing}({\bf{}lst}(0, {\it{}t}))) \begin{math}\ll\end{math} {\it{}endl};

\nwused{\\{NW3gGP3e-1zkOI-1}}\nwidentuses{\\{{\nwixident{bool}}{bool}}\\{{\nwixident{lst}}{lst}}\\{{\nwixident{passing}}{passing}}}\nwindexuse{\nwixident{bool}}{bool}{NW3gGP3e-3pbeiz-2}\nwindexuse{\nwixident{lst}}{lst}{NW3gGP3e-3pbeiz-2}\nwindexuse{\nwixident{passing}}{passing}{NW3gGP3e-3pbeiz-2}\nwendcode{}\nwbegindocs{191}\nwdocspar
\subsubsection{Transformation of Zero-Radius Cycles}
\label{sec:transf-zero-radi}

Firstly, we check some basic information about the zero-radius
cycles. This mainly done to verify our library.
\nwenddocs{}\nwbegincode{192}\sublabel{NW3gGP3e-19ZdJC-1}\nwmargintag{{\nwtagstyle{}\subpageref{NW3gGP3e-19ZdJC-1}}}\moddef{Check Moebius transformations of zero cycles~{\nwtagstyle{}\subpageref{NW3gGP3e-19ZdJC-1}}}\endmoddef\Rm{}\nwstartdeflinemarkup\nwusesondefline{\\{NW3gGP3e-1zkOI-1}}\nwprevnextdefs{\relax}{NW3gGP3e-19ZdJC-2}\nwenddeflinemarkup
{\it{}cout} \begin{math}\ll\end{math} {\it{}wspaces} \begin{math}\ll\end{math} {\tt{}"Determinant of zero-radius Z1 cycle in metric e is: "}
  {\it{}math\_string} \begin{math}\ll\end{math} {\it{}canonicalize\_clifford}({\it{}Z1}.{\it{}det}({\it{}e}, {\it{}S2})) {\it{}math\_string} \begin{math}\ll\end{math} {\it{}endl}
  \begin{math}\ll\end{math} {\it{}wspaces} \begin{math}\ll\end{math} {\tt{}"Focus of zero-radius cycle is: "} {\it{}math\_string} \begin{math}\ll\end{math} {\it{}Z1}.{\it{}focus}({\it{}e}) {\it{}math\_string} \begin{math}\ll\end{math} {\it{}endl}
  \begin{math}\ll\end{math} {\it{}wspaces} \begin{math}\ll\end{math} {\tt{}"Centre of zero-radius cycle is: "} {\it{}math\_string} \begin{math}\ll\end{math} {\it{}Z1}.{\it{}center}({\it{}e}) {\it{}math\_string} \begin{math}\ll\end{math} {\it{}endl}
  \begin{math}\ll\end{math} {\it{}wspaces} \begin{math}\ll\end{math} {\tt{}"Focal length of zero-radius cycle is: "} {\it{}math\_string} \begin{math}\ll\end{math} {\it{}Z1}.{\it{}focal\_length}() {\it{}math\_string} \begin{math}\ll\end{math} {\it{}endl};

\nwalsodefined{\\{NW3gGP3e-19ZdJC-2}\\{NW3gGP3e-19ZdJC-3}\\{NW3gGP3e-19ZdJC-4}\\{NW3gGP3e-19ZdJC-5}}\nwused{\\{NW3gGP3e-1zkOI-1}}\nwidentuses{\\{{\nwixident{center}}{center}}\\{{\nwixident{cycle}}{cycle}}\\{{\nwixident{det}}{det}}\\{{\nwixident{focal{\_}length}}{focal:unlength}}\\{{\nwixident{focus}}{focus}}\\{{\nwixident{math{\_}string}}{math:unstring}}\\{{\nwixident{wspaces}}{wspaces}}}\nwindexuse{\nwixident{center}}{center}{NW3gGP3e-19ZdJC-1}\nwindexuse{\nwixident{cycle}}{cycle}{NW3gGP3e-19ZdJC-1}\nwindexuse{\nwixident{det}}{det}{NW3gGP3e-19ZdJC-1}\nwindexuse{\nwixident{focal{\_}length}}{focal:unlength}{NW3gGP3e-19ZdJC-1}\nwindexuse{\nwixident{focus}}{focus}{NW3gGP3e-19ZdJC-1}\nwindexuse{\nwixident{math{\_}string}}{math:unstring}{NW3gGP3e-19ZdJC-1}\nwindexuse{\nwixident{wspaces}}{wspaces}{NW3gGP3e-19ZdJC-1}\nwendcode{}\nwbegindocs{193}This chunk checks that M\"obius transformation of a zero-radius
cycle is a zero-radius cycle with centre obtained from the first one
by the same M\"obius transformation.
\nwenddocs{}\nwbegincode{194}\sublabel{NW3gGP3e-19ZdJC-2}\nwmargintag{{\nwtagstyle{}\subpageref{NW3gGP3e-19ZdJC-2}}}\moddef{Check Moebius transformations of zero cycles~{\nwtagstyle{}\subpageref{NW3gGP3e-19ZdJC-1}}}\plusendmoddef\Rm{}\nwstartdeflinemarkup\nwusesondefline{\\{NW3gGP3e-1zkOI-1}}\nwprevnextdefs{NW3gGP3e-19ZdJC-1}{NW3gGP3e-19ZdJC-3}\nwenddeflinemarkup
{\it{}C2} = {\it{}Z1}.{\it{}sl2\_similarity}({\it{}a}, {\it{}b}, {\it{}c}, {\it{}d}, {\it{}e}, {\it{}S2});
{\it{}cout} \begin{math}\ll\end{math} {\tt{}"Image of the zero-radius cycle under Moebius transform has zero radius: "}
 \begin{math}\ll\end{math} {\it{}canonicalize\_clifford}({\it{}C2}.{\it{}det}({\it{}es}, {\it{}S2})).{\it{}subs}({\it{}sl2\_relation1},
              {\it{}subs\_options}::{\it{}algebraic} \begin{math}\mid\end{math} {\it{}subs\_options}::{\it{}no\_pattern}).{\it{}is\_zero}() \begin{math}\ll\end{math} {\it{}endl};

\nwused{\\{NW3gGP3e-1zkOI-1}}\nwidentuses{\\{{\nwixident{cycle}}{cycle}}\\{{\nwixident{det}}{det}}\\{{\nwixident{is{\_}zero}}{is:unzero}}\\{{\nwixident{sl2{\_}similarity}}{sl2:unsimilarity}}\\{{\nwixident{subs}}{subs}}}\nwindexuse{\nwixident{cycle}}{cycle}{NW3gGP3e-19ZdJC-2}\nwindexuse{\nwixident{det}}{det}{NW3gGP3e-19ZdJC-2}\nwindexuse{\nwixident{is{\_}zero}}{is:unzero}{NW3gGP3e-19ZdJC-2}\nwindexuse{\nwixident{sl2{\_}similarity}}{sl2:unsimilarity}{NW3gGP3e-19ZdJC-2}\nwindexuse{\nwixident{subs}}{subs}{NW3gGP3e-19ZdJC-2}\nwendcode{}\nwbegindocs{195}Here we find parameters of the transformed zero-radius cycle \(C_2=gZg^{-1}\).
\nwenddocs{}\nwbegincode{196}\sublabel{NW3gGP3e-19ZdJC-3}\nwmargintag{{\nwtagstyle{}\subpageref{NW3gGP3e-19ZdJC-3}}}\moddef{Check Moebius transformations of zero cycles~{\nwtagstyle{}\subpageref{NW3gGP3e-19ZdJC-1}}}\plusendmoddef\Rm{}\nwstartdeflinemarkup\nwusesondefline{\\{NW3gGP3e-1zkOI-1}}\nwprevnextdefs{NW3gGP3e-19ZdJC-2}{NW3gGP3e-19ZdJC-4}\nwenddeflinemarkup
{\it{}C2} = {\it{}Z}.{\it{}sl2\_similarity}({\it{}a}, {\it{}b}, {\it{}c}, {\it{}d}, {\it{}e}, {\it{}S2});
{\it{}K} = {\it{}C2}.{\it{}get\_k}();
{\it{}L0} = {\it{}C2}.{\it{}get\_l}(0);
{\it{}L1} = {\it{}C2}.{\it{}get\_l}(1);

\nwused{\\{NW3gGP3e-1zkOI-1}}\nwidentuses{\\{{\nwixident{get{\_}k}}{get:unk}}\\{{\nwixident{get{\_}l}}{get:unl}}\\{{\nwixident{sl2{\_}similarity}}{sl2:unsimilarity}}}\nwindexuse{\nwixident{get{\_}k}}{get:unk}{NW3gGP3e-19ZdJC-3}\nwindexuse{\nwixident{get{\_}l}}{get:unl}{NW3gGP3e-19ZdJC-3}\nwindexuse{\nwixident{sl2{\_}similarity}}{sl2:unsimilarity}{NW3gGP3e-19ZdJC-3}\nwendcode{}\nwbegindocs{197}Now we calculate the M\"obius transformation of the centre of {\Tt{}\Rm{}{\it{}Z}\nwendquote}
\nwenddocs{}\nwbegincode{198}\sublabel{NW3gGP3e-19ZdJC-4}\nwmargintag{{\nwtagstyle{}\subpageref{NW3gGP3e-19ZdJC-4}}}\moddef{Check Moebius transformations of zero cycles~{\nwtagstyle{}\subpageref{NW3gGP3e-19ZdJC-1}}}\plusendmoddef\Rm{}\nwstartdeflinemarkup\nwusesondefline{\\{NW3gGP3e-1zkOI-1}}\nwprevnextdefs{NW3gGP3e-19ZdJC-3}{NW3gGP3e-19ZdJC-5}\nwenddeflinemarkup
{\it{}u2} = {\it{}gW}.{\it{}op}(0);
{\it{}v2} = {\it{}gW}.{\it{}op}(1);

\nwused{\\{NW3gGP3e-1zkOI-1}}\nwidentuses{\\{{\nwixident{op}}{op}}}\nwindexuse{\nwixident{op}}{op}{NW3gGP3e-19ZdJC-4}\nwendcode{}\nwbegindocs{199} And we finally check that {\Tt{}\Rm{}{\it{}gW}\nwendquote} coincides with the centre of the transformed cycle
{\Tt{}\Rm{}{\it{}C2}\nwendquote}. This proves~\cite[Lem.~\ref{E-le:moeb-conj-z-cycle}]{Kisil05a}.
\nwenddocs{}\nwbegincode{200}\sublabel{NW3gGP3e-19ZdJC-5}\nwmargintag{{\nwtagstyle{}\subpageref{NW3gGP3e-19ZdJC-5}}}\moddef{Check Moebius transformations of zero cycles~{\nwtagstyle{}\subpageref{NW3gGP3e-19ZdJC-1}}}\plusendmoddef\Rm{}\nwstartdeflinemarkup\nwusesondefline{\\{NW3gGP3e-1zkOI-1}}\nwprevnextdefs{NW3gGP3e-19ZdJC-4}{\relax}\nwenddeflinemarkup
{\it{}cout} \begin{math}\ll\end{math}{\tt{}"The centre of the Moebius transformed zero-radius cycle is: "}
 \begin{math}\ll\end{math} {\it{}equality}(({\it{}u2}\begin{math}\ast\end{math}{\it{}K}-{\it{}L0}).{\it{}subs}({\it{}sl2\_relation}, {\it{}subs\_options}::{\it{}algebraic} \begin{math}\mid\end{math} {\it{}subs\_options}::{\it{}no\_pattern})) \begin{math}\ll\end{math} {\tt{}", "}
 \begin{math}\ll\end{math} {\it{}equality}(({\it{}v2}\begin{math}\ast\end{math}{\it{}K}-{\it{}L1}).{\it{}subs}({\it{}sl2\_relation}, {\it{}subs\_options}::{\it{}algebraic} \begin{math}\mid\end{math} {\it{}subs\_options}::{\it{}no\_pattern}))
 \begin{math}\ll\end{math} {\it{}endl} ;

 {\it{}cerr} \begin{math}\ll\end{math} ({\it{}v2}\begin{math}\ast\end{math}{\it{}K}-{\it{}L1}).{\it{}subs}({\it{}sl2\_relation}, {\it{}subs\_options}::{\it{}algebraic} \begin{math}\mid\end{math} {\it{}subs\_options}::{\it{}no\_pattern}).{\it{}expand}().{\it{}eval}() \begin{math}\ll\end{math} {\it{}endl}
      \begin{math}\ll\end{math} ({\it{}v2}\begin{math}\ast\end{math}{\it{}K}).{\it{}subs}({\it{}sl2\_relation}, {\it{}subs\_options}::{\it{}algebraic} \begin{math}\mid\end{math} {\it{}subs\_options}::{\it{}no\_pattern}).{\it{}expand}().{\it{}eval}()  \begin{math}\ll\end{math} {\it{}endl}
\begin{math}\ll\end{math} ({\it{}L1}).{\it{}subs}({\it{}sl2\_relation}, {\it{}subs\_options}::{\it{}algebraic} \begin{math}\mid\end{math} {\it{}subs\_options}::{\it{}no\_pattern}).{\it{}expand}().{\it{}eval}()  \begin{math}\ll\end{math} {\it{}endl};

\nwused{\\{NW3gGP3e-1zkOI-1}}\nwidentuses{\\{{\nwixident{cycle}}{cycle}}\\{{\nwixident{expand}}{expand}}\\{{\nwixident{subs}}{subs}}}\nwindexuse{\nwixident{cycle}}{cycle}{NW3gGP3e-19ZdJC-5}\nwindexuse{\nwixident{expand}}{expand}{NW3gGP3e-19ZdJC-5}\nwindexuse{\nwixident{subs}}{subs}{NW3gGP3e-19ZdJC-5}\nwendcode{}\nwbegindocs{201}\nwdocspar
\subsubsection{Cycles conjugation}
\label{sec:cycles-conjugation}
This chunk checks that transformation of a zero-radius cycle by
conjugation with a cycle is a zero-radius cycle with centre
obtained from the first one by the same transformation.

Firstly we calculate parameters of \(C_2=CZC\) .
\nwenddocs{}\nwbegincode{202}\sublabel{NW3gGP3e-3O7JaU-1}\nwmargintag{{\nwtagstyle{}\subpageref{NW3gGP3e-3O7JaU-1}}}\moddef{Check transformations of zero cycles by conjugation~{\nwtagstyle{}\subpageref{NW3gGP3e-3O7JaU-1}}}\endmoddef\Rm{}\nwstartdeflinemarkup\nwusesondefline{\\{NW3gGP3e-1zkOI-1}}\nwprevnextdefs{\relax}{NW3gGP3e-3O7JaU-2}\nwenddeflinemarkup
{\it{}C2} = {\it{}Z}.{\it{}cycle\_similarity}({\it{}C}, {\it{}e}, {\it{}S2}, {\it{}S3});
{\it{}cout} \begin{math}\ll\end{math} {\tt{}"Image of the zero-radius cycle under cycle similarity has zero radius: "}
 \begin{math}\ll\end{math} {\it{}canonicalize\_clifford}({\it{}C2}.{\it{}det}({\it{}e}, {\it{}S2})).{\it{}subs}({\it{}sl2\_relation1},
             {\it{}subs\_options}::{\it{}algebraic} \begin{math}\mid\end{math} {\it{}subs\_options}::{\it{}no\_pattern}).{\it{}normal}().{\it{}is\_zero}() \begin{math}\ll\end{math} {\it{}endl};

{\it{}K} = {\it{}C2}.{\it{}get\_k}();
{\it{}L0} = {\it{}C2}.{\it{}get\_l}(0);
{\it{}L1} = {\it{}C2}.{\it{}get\_l}(1);

\nwalsodefined{\\{NW3gGP3e-3O7JaU-2}}\nwused{\\{NW3gGP3e-1zkOI-1}}\nwidentuses{\\{{\nwixident{cycle}}{cycle}}\\{{\nwixident{cycle{\_}similarity}}{cycle:unsimilarity}}\\{{\nwixident{det}}{det}}\\{{\nwixident{get{\_}k}}{get:unk}}\\{{\nwixident{get{\_}l}}{get:unl}}\\{{\nwixident{is{\_}zero}}{is:unzero}}\\{{\nwixident{normal}}{normal}}\\{{\nwixident{subs}}{subs}}}\nwindexuse{\nwixident{cycle}}{cycle}{NW3gGP3e-3O7JaU-1}\nwindexuse{\nwixident{cycle{\_}similarity}}{cycle:unsimilarity}{NW3gGP3e-3O7JaU-1}\nwindexuse{\nwixident{det}}{det}{NW3gGP3e-3O7JaU-1}\nwindexuse{\nwixident{get{\_}k}}{get:unk}{NW3gGP3e-3O7JaU-1}\nwindexuse{\nwixident{get{\_}l}}{get:unl}{NW3gGP3e-3O7JaU-1}\nwindexuse{\nwixident{is{\_}zero}}{is:unzero}{NW3gGP3e-3O7JaU-1}\nwindexuse{\nwixident{normal}}{normal}{NW3gGP3e-3O7JaU-1}\nwindexuse{\nwixident{subs}}{subs}{NW3gGP3e-3O7JaU-1}\nwendcode{}\nwbegindocs{203}Then we check that it coincides with transformation point {\Tt{}\Rm{}{\it{}P}\nwendquote}
which is calculated in agreement with above used matrices {\Tt{}\Rm{}{\it{}S2}\nwendquote} and
{\Tt{}\Rm{}{\it{}S3}\nwendquote}. This proves the result~\cite[Lem.~\ref{E-le:cycle-conj-is-moeb}]{Kisil05a}
\nwenddocs{}\nwbegincode{204}\sublabel{NW3gGP3e-3O7JaU-2}\nwmargintag{{\nwtagstyle{}\subpageref{NW3gGP3e-3O7JaU-2}}}\moddef{Check transformations of zero cycles by conjugation~{\nwtagstyle{}\subpageref{NW3gGP3e-3O7JaU-1}}}\plusendmoddef\Rm{}\nwstartdeflinemarkup\nwusesondefline{\\{NW3gGP3e-1zkOI-1}}\nwprevnextdefs{NW3gGP3e-3O7JaU-1}{\relax}\nwenddeflinemarkup
{\it{}P} = {\it{}C}.{\it{}moebius\_map}({\it{}W}, {\it{}e}, {\it{}S2}.{\it{}mul}({\it{}S3}));
{\it{}u2} = {\it{}P}.{\it{}op}(0);
{\it{}v2} = {\it{}P}.{\it{}op}(1);

{\it{}cout} \begin{math}\ll\end{math}{\tt{}"The centre of the conjugated zero-radius cycle coinsides with Moebius tr: "}
 \begin{math}\ll\end{math}  {\it{}equality}({\it{}u2}\begin{math}\ast\end{math}{\it{}K}-{\it{}L0}) \begin{math}\ll\end{math} {\tt{}", "} \begin{math}\ll\end{math} {\it{}equality}({\it{}v2}\begin{math}\ast\end{math}{\it{}K}-{\it{}L1})  \begin{math}\ll\end{math} {\it{}endl} ;

\nwused{\\{NW3gGP3e-1zkOI-1}}\nwidentuses{\\{{\nwixident{cycle}}{cycle}}\\{{\nwixident{moebius{\_}map}}{moebius:unmap}}\\{{\nwixident{mul}}{mul}}\\{{\nwixident{op}}{op}}}\nwindexuse{\nwixident{cycle}}{cycle}{NW3gGP3e-3O7JaU-2}\nwindexuse{\nwixident{moebius{\_}map}}{moebius:unmap}{NW3gGP3e-3O7JaU-2}\nwindexuse{\nwixident{mul}}{mul}{NW3gGP3e-3O7JaU-2}\nwindexuse{\nwixident{op}}{op}{NW3gGP3e-3O7JaU-2}\nwendcode{}\nwbegindocs{205}\nwdocspar
\subsection{Orthogonality of Cycles}
\label{sec:orthogonality-cycles}

\subsubsection{Various orthogonality conditions}
\label{sec:vari-orth-cond}

We calculate orthogonality condition between two
{\Tt{}\Rm{}{\bf{}cycle2D}\nwendquote}s by the identity \(\Re \tr(C_1C_2)=0\). The expression
are stored in variables, which will be used later in our calculations.

\nwenddocs{}\nwbegindocs{206}Here is the orthogonality of two generic {\Tt{}\Rm{}{\bf{}cycle2D}\nwendquote}s\ldots
\nwenddocs{}\nwbegincode{207}\sublabel{NW3gGP3e-3aTNrh-1}\nwmargintag{{\nwtagstyle{}\subpageref{NW3gGP3e-3aTNrh-1}}}\moddef{Orthogonality conditions~{\nwtagstyle{}\subpageref{NW3gGP3e-3aTNrh-1}}}\endmoddef\Rm{}\nwstartdeflinemarkup\nwusesondefline{\\{NW3gGP3e-1zkOI-2}}\nwprevnextdefs{\relax}{NW3gGP3e-3aTNrh-2}\nwenddeflinemarkup
{\it{}cout} \begin{math}\ll\end{math} {\it{}wspaces} \begin{math}\ll\end{math} {\tt{}"The orthogonality is: "} {\it{}math\_string}
 \begin{math}\ll\end{math} ({\bf{}ex}){\it{}C}.{\it{}is\_orthogonal}({\it{}C1}, {\it{}es}, {\it{}S2}) {\it{}math\_string} \begin{math}\ll\end{math} {\it{}endl}

\nwalsodefined{\\{NW3gGP3e-3aTNrh-2}\\{NW3gGP3e-3aTNrh-3}\\{NW3gGP3e-3aTNrh-4}}\nwused{\\{NW3gGP3e-1zkOI-2}}\nwidentuses{\\{{\nwixident{ex}}{ex}}\\{{\nwixident{is{\_}orthogonal}}{is:unorthogonal}}\\{{\nwixident{math{\_}string}}{math:unstring}}\\{{\nwixident{wspaces}}{wspaces}}}\nwindexuse{\nwixident{ex}}{ex}{NW3gGP3e-3aTNrh-1}\nwindexuse{\nwixident{is{\_}orthogonal}}{is:unorthogonal}{NW3gGP3e-3aTNrh-1}\nwindexuse{\nwixident{math{\_}string}}{math:unstring}{NW3gGP3e-3aTNrh-1}\nwindexuse{\nwixident{wspaces}}{wspaces}{NW3gGP3e-3aTNrh-1}\nwendcode{}\nwbegindocs{208}\ldots and then its reduction to orthogonality of two straight lines.
\nwenddocs{}\nwbegincode{209}\sublabel{NW3gGP3e-3aTNrh-2}\nwmargintag{{\nwtagstyle{}\subpageref{NW3gGP3e-3aTNrh-2}}}\moddef{Orthogonality conditions~{\nwtagstyle{}\subpageref{NW3gGP3e-3aTNrh-1}}}\plusendmoddef\Rm{}\nwstartdeflinemarkup\nwusesondefline{\\{NW3gGP3e-1zkOI-2}}\nwprevnextdefs{NW3gGP3e-3aTNrh-1}{NW3gGP3e-3aTNrh-3}\nwenddeflinemarkup
 \begin{math}\ll\end{math} {\it{}wspaces} \begin{math}\ll\end{math} {\tt{}"The orthogonality of two lines is: "} {\it{}math\_string}
 \begin{math}\ll\end{math} ({\bf{}ex}){\it{}C}.{\it{}subs}({\it{}k} \begin{math}\equiv\end{math} 0).{\it{}is\_orthogonal}({\it{}C1}.{\it{}subs}({\it{}k1} \begin{math}\equiv\end{math} 0), {\it{}es}, {\it{}S2}) {\it{}math\_string} \begin{math}\ll\end{math} {\it{}endl};

\nwused{\\{NW3gGP3e-1zkOI-2}}\nwidentuses{\\{{\nwixident{ex}}{ex}}\\{{\nwixident{is{\_}orthogonal}}{is:unorthogonal}}\\{{\nwixident{k}}{k}}\\{{\nwixident{math{\_}string}}{math:unstring}}\\{{\nwixident{subs}}{subs}}\\{{\nwixident{wspaces}}{wspaces}}}\nwindexuse{\nwixident{ex}}{ex}{NW3gGP3e-3aTNrh-2}\nwindexuse{\nwixident{is{\_}orthogonal}}{is:unorthogonal}{NW3gGP3e-3aTNrh-2}\nwindexuse{\nwixident{k}}{k}{NW3gGP3e-3aTNrh-2}\nwindexuse{\nwixident{math{\_}string}}{math:unstring}{NW3gGP3e-3aTNrh-2}\nwindexuse{\nwixident{subs}}{subs}{NW3gGP3e-3aTNrh-2}\nwindexuse{\nwixident{wspaces}}{wspaces}{NW3gGP3e-3aTNrh-2}\nwendcode{}\nwbegindocs{210} Here is the orthogonality of a generic {\Tt{}\Rm{}{\bf{}cycle2D}\nwendquote} to a zero-radius
{\Tt{}\Rm{}{\bf{}cycle2D}\nwendquote}. This reduces to concurrence of the centre the zero-radius
and generic cycle.
\nwenddocs{}\nwbegincode{211}\sublabel{NW3gGP3e-3aTNrh-3}\nwmargintag{{\nwtagstyle{}\subpageref{NW3gGP3e-3aTNrh-3}}}\moddef{Orthogonality conditions~{\nwtagstyle{}\subpageref{NW3gGP3e-3aTNrh-1}}}\plusendmoddef\Rm{}\nwstartdeflinemarkup\nwusesondefline{\\{NW3gGP3e-1zkOI-2}}\nwprevnextdefs{NW3gGP3e-3aTNrh-2}{NW3gGP3e-3aTNrh-4}\nwenddeflinemarkup
{\it{}cout} \begin{math}\ll\end{math} {\it{}wspaces} \begin{math}\ll\end{math} {\tt{}"The orthogonality to z-r-cycle is: "} {\it{}math\_string}
 \begin{math}\ll\end{math} ({\bf{}ex}){\it{}C}.{\it{}is\_orthogonal}({\it{}Z}, {\it{}es})  {\it{}math\_string} \begin{math}\ll\end{math} {\it{}endl};

\nwused{\\{NW3gGP3e-1zkOI-2}}\nwidentuses{\\{{\nwixident{cycle}}{cycle}}\\{{\nwixident{ex}}{ex}}\\{{\nwixident{is{\_}orthogonal}}{is:unorthogonal}}\\{{\nwixident{math{\_}string}}{math:unstring}}\\{{\nwixident{wspaces}}{wspaces}}}\nwindexuse{\nwixident{cycle}}{cycle}{NW3gGP3e-3aTNrh-3}\nwindexuse{\nwixident{ex}}{ex}{NW3gGP3e-3aTNrh-3}\nwindexuse{\nwixident{is{\_}orthogonal}}{is:unorthogonal}{NW3gGP3e-3aTNrh-3}\nwindexuse{\nwixident{math{\_}string}}{math:unstring}{NW3gGP3e-3aTNrh-3}\nwindexuse{\nwixident{wspaces}}{wspaces}{NW3gGP3e-3aTNrh-3}\nwendcode{}\nwbegindocs{212} Here is the orthogonality of two zero-radius {\Tt{}\Rm{}{\bf{}cycle2D}\nwendquote}s.
\nwenddocs{}\nwbegincode{213}\sublabel{NW3gGP3e-3aTNrh-4}\nwmargintag{{\nwtagstyle{}\subpageref{NW3gGP3e-3aTNrh-4}}}\moddef{Orthogonality conditions~{\nwtagstyle{}\subpageref{NW3gGP3e-3aTNrh-1}}}\plusendmoddef\Rm{}\nwstartdeflinemarkup\nwusesondefline{\\{NW3gGP3e-1zkOI-2}}\nwprevnextdefs{NW3gGP3e-3aTNrh-3}{\relax}\nwenddeflinemarkup
{\it{}C2} = {\bf{}cycle2D}({\bf{}lst}({\it{}u1}, {\it{}v1}), {\it{}e}, 0, {\it{}S2});
{\it{}cout} \begin{math}\ll\end{math} {\it{}wspaces} \begin{math}\ll\end{math} {\tt{}"The orthogonality of two z-r-cycle is: "} {\it{}math\_string}
 \begin{math}\ll\end{math}  ({\bf{}ex}){\it{}C2}.{\it{}is\_orthogonal}({\it{}Z1}, {\it{}es}) {\it{}math\_string} \begin{math}\ll\end{math} {\it{}endl};

\nwused{\\{NW3gGP3e-1zkOI-2}}\nwidentuses{\\{{\nwixident{cycle}}{cycle}}\\{{\nwixident{cycle2D}}{cycle2D}}\\{{\nwixident{ex}}{ex}}\\{{\nwixident{is{\_}orthogonal}}{is:unorthogonal}}\\{{\nwixident{lst}}{lst}}\\{{\nwixident{math{\_}string}}{math:unstring}}\\{{\nwixident{wspaces}}{wspaces}}}\nwindexuse{\nwixident{cycle}}{cycle}{NW3gGP3e-3aTNrh-4}\nwindexuse{\nwixident{cycle2D}}{cycle2D}{NW3gGP3e-3aTNrh-4}\nwindexuse{\nwixident{ex}}{ex}{NW3gGP3e-3aTNrh-4}\nwindexuse{\nwixident{is{\_}orthogonal}}{is:unorthogonal}{NW3gGP3e-3aTNrh-4}\nwindexuse{\nwixident{lst}}{lst}{NW3gGP3e-3aTNrh-4}\nwindexuse{\nwixident{math{\_}string}}{math:unstring}{NW3gGP3e-3aTNrh-4}\nwindexuse{\nwixident{wspaces}}{wspaces}{NW3gGP3e-3aTNrh-4}\nwendcode{}\nwbegindocs{214}This chunk finds the parameters of a cycle {\Tt{}\Rm{}{\it{}C2}\nwendquote} passing through two
points \((u,v)\), \((u_1,v_1)\) and orthogonal to the given cycle
{\Tt{}\Rm{}{\it{}C}\nwendquote}. This gives three linear equations with four variables which are
consistent in a generic position.

\nwenddocs{}\nwbegincode{215}\sublabel{NW3gGP3e-3nCzcM-1}\nwmargintag{{\nwtagstyle{}\subpageref{NW3gGP3e-3nCzcM-1}}}\moddef{Two points and orthogonality~{\nwtagstyle{}\subpageref{NW3gGP3e-3nCzcM-1}}}\endmoddef\Rm{}\nwstartdeflinemarkup\nwusesondefline{\\{NW3gGP3e-1zkOI-2}}\nwprevnextdefs{\relax}{NW3gGP3e-3nCzcM-2}\nwenddeflinemarkup
{\it{}C2} = {\it{}C1}.{\it{}subject\_to}({\bf{}lst}({\it{}C1}.{\it{}passing}({\it{}W}),
        {\it{}C1}.{\it{}passing}({\it{}W1}),
        {\it{}C1}.{\it{}is\_orthogonal}({\it{}C}, {\it{}es})), {\it{}vars});

\nwalsodefined{\\{NW3gGP3e-3nCzcM-2}}\nwused{\\{NW3gGP3e-1zkOI-2}}\nwidentuses{\\{{\nwixident{is{\_}orthogonal}}{is:unorthogonal}}\\{{\nwixident{lst}}{lst}}\\{{\nwixident{passing}}{passing}}\\{{\nwixident{subject{\_}to}}{subject:unto}}}\nwindexuse{\nwixident{is{\_}orthogonal}}{is:unorthogonal}{NW3gGP3e-3nCzcM-1}\nwindexuse{\nwixident{lst}}{lst}{NW3gGP3e-3nCzcM-1}\nwindexuse{\nwixident{passing}}{passing}{NW3gGP3e-3nCzcM-1}\nwindexuse{\nwixident{subject{\_}to}}{subject:unto}{NW3gGP3e-3nCzcM-1}\nwendcode{}\nwbegindocs{216}To find the singularity condition of the above solution we
analyse the denominator of {\Tt{}\Rm{}{\it{}k}\nwendquote}, which calculated to
be:
\begin{displaymath}
  k = \frac{-2 (u^\prime (\sigma_1 n+v k)-v l+(-k v^\prime-\sigma_1 n) u+l v^\prime) n_1 }
  {-u^{\prime 2} l+u^{\prime 2} u k+\sigma l v^{\prime 2}-u^\prime
    u^2 k+u^\prime v^2 \sigma k+u^\prime m-u \sigma k v^{\prime 2}+u^2 l-v^2 \sigma l-u m}.
\end{displaymath}

\nwenddocs{}\nwbegincode{217}\sublabel{NW3gGP3e-3nCzcM-2}\nwmargintag{{\nwtagstyle{}\subpageref{NW3gGP3e-3nCzcM-2}}}\moddef{Two points and orthogonality~{\nwtagstyle{}\subpageref{NW3gGP3e-3nCzcM-1}}}\plusendmoddef\Rm{}\nwstartdeflinemarkup\nwusesondefline{\\{NW3gGP3e-1zkOI-2}}\nwprevnextdefs{NW3gGP3e-3nCzcM-1}{\relax}\nwenddeflinemarkup
{\bf{}if} ({\it{}debug} \begin{math}>\end{math} 0)
{\it{}cout} \begin{math}\ll\end{math} {\tt{}"Cycle through two point is possible and unique if denominator is not zero: "} \begin{math}\ll\end{math} {\it{}endl}
 {\it{}math\_string} \begin{math}\ll\end{math} {\it{}C2}.{\it{}get\_k}() {\it{}math\_string} \begin{math}\ll\end{math} {\it{}endl} \begin{math}\ll\end{math} {\it{}endl};

\nwused{\\{NW3gGP3e-1zkOI-2}}\nwidentuses{\\{{\nwixident{debug}}{debug}}\\{{\nwixident{get{\_}k}}{get:unk}}\\{{\nwixident{math{\_}string}}{math:unstring}}}\nwindexuse{\nwixident{debug}}{debug}{NW3gGP3e-3nCzcM-2}\nwindexuse{\nwixident{get{\_}k}}{get:unk}{NW3gGP3e-3nCzcM-2}\nwindexuse{\nwixident{math{\_}string}}{math:unstring}{NW3gGP3e-3nCzcM-2}\nwendcode{}\nwbegindocs{218}\nwdocspar
\subsubsection{Orthogonality and Inversion}
\label{sec:orth-invers}

Now we check that any orthogonal cycle comes through the inverse of
any its point. To this end we calculate a generic cycle {\Tt{}\Rm{}{\it{}C2}\nwendquote}
passing through a point \((u, v)\) and orthogonal to a cycle
{\Tt{}\Rm{}{\it{}C}\nwendquote}.
\nwenddocs{}\nwbegincode{219}\sublabel{NW3gGP3e-mJ5pz-1}\nwmargintag{{\nwtagstyle{}\subpageref{NW3gGP3e-mJ5pz-1}}}\moddef{One point and orthogonality~{\nwtagstyle{}\subpageref{NW3gGP3e-mJ5pz-1}}}\endmoddef\Rm{}\nwstartdeflinemarkup\nwusesondefline{\\{NW3gGP3e-1zkOI-2}}\nwprevnextdefs{\relax}{NW3gGP3e-mJ5pz-2}\nwenddeflinemarkup
{\it{}C2} = {\it{}C1}.{\it{}subject\_to}({\bf{}lst}({\it{}C1}.{\it{}passing}({\it{}W}),
        {\it{}C1}.{\it{}is\_orthogonal}({\it{}C}, {\it{}es})));

\nwalsodefined{\\{NW3gGP3e-mJ5pz-2}\\{NW3gGP3e-mJ5pz-3}\\{NW3gGP3e-mJ5pz-4}}\nwused{\\{NW3gGP3e-1zkOI-2}}\nwidentuses{\\{{\nwixident{is{\_}orthogonal}}{is:unorthogonal}}\\{{\nwixident{lst}}{lst}}\\{{\nwixident{passing}}{passing}}\\{{\nwixident{subject{\_}to}}{subject:unto}}}\nwindexuse{\nwixident{is{\_}orthogonal}}{is:unorthogonal}{NW3gGP3e-mJ5pz-1}\nwindexuse{\nwixident{lst}}{lst}{NW3gGP3e-mJ5pz-1}\nwindexuse{\nwixident{passing}}{passing}{NW3gGP3e-mJ5pz-1}\nwindexuse{\nwixident{subject{\_}to}}{subject:unto}{NW3gGP3e-mJ5pz-1}\nwendcode{}\nwbegindocs{220}\nwdocspar
 \sublabel{chu:orth-invers-2}
Then we calculate another cycle {\Tt{}\Rm{}{\it{}C3}\nwendquote} with an additional
condition that it passing through the M\"obius transform \(P\) of
\((u, v)\).
\nwenddocs{}\nwbegincode{221}\sublabel{NW3gGP3e-mJ5pz-2}\nwmargintag{{\nwtagstyle{}\subpageref{NW3gGP3e-mJ5pz-2}}}\moddef{One point and orthogonality~{\nwtagstyle{}\subpageref{NW3gGP3e-mJ5pz-1}}}\plusendmoddef\Rm{}\nwstartdeflinemarkup\nwusesondefline{\\{NW3gGP3e-1zkOI-2}}\nwprevnextdefs{NW3gGP3e-mJ5pz-1}{NW3gGP3e-mJ5pz-3}\nwenddeflinemarkup
{\it{}P} = {\it{}C}.{\it{}moebius\_map}({\it{}W}, {\it{}e}, -{\it{}M1});
{\it{}C3} = {\it{}C1}.{\it{}subject\_to}({\bf{}lst}({\it{}C1}.{\it{}passing}({\it{}P}),
        {\it{}C1}.{\it{}passing}({\it{}W}),
        {\it{}C1}.{\it{}is\_orthogonal}({\it{}C}, {\it{}es})));

\nwused{\\{NW3gGP3e-1zkOI-2}}\nwidentuses{\\{{\nwixident{is{\_}orthogonal}}{is:unorthogonal}}\\{{\nwixident{lst}}{lst}}\\{{\nwixident{moebius{\_}map}}{moebius:unmap}}\\{{\nwixident{passing}}{passing}}\\{{\nwixident{subject{\_}to}}{subject:unto}}}\nwindexuse{\nwixident{is{\_}orthogonal}}{is:unorthogonal}{NW3gGP3e-mJ5pz-2}\nwindexuse{\nwixident{lst}}{lst}{NW3gGP3e-mJ5pz-2}\nwindexuse{\nwixident{moebius{\_}map}}{moebius:unmap}{NW3gGP3e-mJ5pz-2}\nwindexuse{\nwixident{passing}}{passing}{NW3gGP3e-mJ5pz-2}\nwindexuse{\nwixident{subject{\_}to}}{subject:unto}{NW3gGP3e-mJ5pz-2}\nwendcode{}\nwbegindocs{222} Then we check twice in different ways the same mathematical statement:
\begin{enumerate}
\item that both cycles {\Tt{}\Rm{}{\it{}C2}\nwendquote} and {\Tt{}\Rm{}{\it{}C3}\nwendquote} are identical,
  i.e. the addition of inverse point does not put more restrictions;
\nwenddocs{}\nwbegincode{223}\sublabel{NW3gGP3e-mJ5pz-3}\nwmargintag{{\nwtagstyle{}\subpageref{NW3gGP3e-mJ5pz-3}}}\moddef{One point and orthogonality~{\nwtagstyle{}\subpageref{NW3gGP3e-mJ5pz-1}}}\plusendmoddef\Rm{}\nwstartdeflinemarkup\nwusesondefline{\\{NW3gGP3e-1zkOI-2}}\nwprevnextdefs{NW3gGP3e-mJ5pz-2}{NW3gGP3e-mJ5pz-4}\nwenddeflinemarkup
{\it{}cout} \begin{math}\ll\end{math} {\tt{}"Both orthogonal cycles (through one point and through its inverse) are the same: "}
        \begin{math}\ll\end{math} {\it{}C2}.{\it{}is\_equal}({\it{}C3}) \begin{math}\ll\end{math} {\it{}endl}

\nwused{\\{NW3gGP3e-1zkOI-2}}\nwidentuses{\\{{\nwixident{is{\_}equal}}{is:unequal}}}\nwindexuse{\nwixident{is{\_}equal}}{is:unequal}{NW3gGP3e-mJ5pz-3}\nwendcode{}\nwbegindocs{224}\item that cycle {\Tt{}\Rm{}{\it{}C2}\nwendquote} passes through the inversion {\Tt{}\Rm{}{\it{}P}\nwendquote} as well.
\nwenddocs{}\nwbegincode{225}\sublabel{NW3gGP3e-mJ5pz-4}\nwmargintag{{\nwtagstyle{}\subpageref{NW3gGP3e-mJ5pz-4}}}\moddef{One point and orthogonality~{\nwtagstyle{}\subpageref{NW3gGP3e-mJ5pz-1}}}\plusendmoddef\Rm{}\nwstartdeflinemarkup\nwusesondefline{\\{NW3gGP3e-1zkOI-2}}\nwprevnextdefs{NW3gGP3e-mJ5pz-3}{\relax}\nwenddeflinemarkup
        \begin{math}\ll\end{math} {\tt{}"Orthogonal cycle passes through the transformed point: "}
           \begin{math}\ll\end{math} {\it{}C2}.{\it{}val}({\it{}P}).{\it{}normal}().{\it{}is\_zero}() \begin{math}\ll\end{math} {\it{}endl} \begin{math}\ll\end{math} {\it{}endl};

\nwused{\\{NW3gGP3e-1zkOI-2}}\nwidentuses{\\{{\nwixident{cycle}}{cycle}}\\{{\nwixident{is{\_}zero}}{is:unzero}}\\{{\nwixident{normal}}{normal}}\\{{\nwixident{val}}{val}}}\nwindexuse{\nwixident{cycle}}{cycle}{NW3gGP3e-mJ5pz-4}\nwindexuse{\nwixident{is{\_}zero}}{is:unzero}{NW3gGP3e-mJ5pz-4}\nwindexuse{\nwixident{normal}}{normal}{NW3gGP3e-mJ5pz-4}\nwindexuse{\nwixident{val}}{val}{NW3gGP3e-mJ5pz-4}\nwendcode{}\nwbegindocs{226}\nwdocspar
\end{enumerate}

\nwenddocs{}\nwbegindocs{227}\nwdocspar
\subsubsection{Orthogonal Lines}
\label{sec:orthogonal-lines}

This chunk checks that the straight line {\Tt{}\Rm{}{\it{}C4}\nwendquote} passing through a point \((u,
v)\) and its inverse {\Tt{}\Rm{}{\it{}P}\nwendquote} in the cycle {\Tt{}\Rm{}{\it{}C}\nwendquote} is orthogonal to the
initial cycle {\Tt{}\Rm{}{\it{}C}\nwendquote}.
\nwenddocs{}\nwbegincode{228}\sublabel{NW3gGP3e-3uJ41M-1}\nwmargintag{{\nwtagstyle{}\subpageref{NW3gGP3e-3uJ41M-1}}}\moddef{Orthogonal line~{\nwtagstyle{}\subpageref{NW3gGP3e-3uJ41M-1}}}\endmoddef\Rm{}\nwstartdeflinemarkup\nwusesondefline{\\{NW3gGP3e-1zkOI-2}}\nwprevnextdefs{\relax}{NW3gGP3e-3uJ41M-2}\nwenddeflinemarkup
{\it{}C4} = {\it{}C1}.{\it{}subject\_to}({\bf{}lst}({\it{}C1}.{\it{}passing}({\it{}W}),
        {\it{}C1}.{\it{}passing}({\it{}P}),
        {\it{}C1}.{\it{}is\_linear}()));
{\it{}cout} \begin{math}\ll\end{math} {\tt{}"Line through point and its inverse is orthogonal: "} \begin{math}\ll\end{math} {\it{}C4}.{\it{}cycle\_product}({\it{}C}, {\it{}es}).{\it{}is\_zero}() \begin{math}\ll\end{math} {\it{}endl};

\nwalsodefined{\\{NW3gGP3e-3uJ41M-2}\\{NW3gGP3e-3uJ41M-3}}\nwused{\\{NW3gGP3e-1zkOI-2}}\nwidentuses{\\{{\nwixident{cycle{\_}product}}{cycle:unproduct}}\\{{\nwixident{is{\_}linear}}{is:unlinear}}\\{{\nwixident{is{\_}zero}}{is:unzero}}\\{{\nwixident{lst}}{lst}}\\{{\nwixident{passing}}{passing}}\\{{\nwixident{subject{\_}to}}{subject:unto}}}\nwindexuse{\nwixident{cycle{\_}product}}{cycle:unproduct}{NW3gGP3e-3uJ41M-1}\nwindexuse{\nwixident{is{\_}linear}}{is:unlinear}{NW3gGP3e-3uJ41M-1}\nwindexuse{\nwixident{is{\_}zero}}{is:unzero}{NW3gGP3e-3uJ41M-1}\nwindexuse{\nwixident{lst}}{lst}{NW3gGP3e-3uJ41M-1}\nwindexuse{\nwixident{passing}}{passing}{NW3gGP3e-3uJ41M-1}\nwindexuse{\nwixident{subject{\_}to}}{subject:unto}{NW3gGP3e-3uJ41M-1}\nwendcode{}\nwbegindocs{229}We also calculate that all such lines intersect in a single point
\((u_3, v_3)\), which is independent from \((u,v)\). This point will be
understood as centre of the cycle {\Tt{}\Rm{}{\it{}C5}\nwendquote} in \S~\ref{sec:ghost-cycle}.
\nwenddocs{}\nwbegincode{230}\sublabel{NW3gGP3e-3uJ41M-2}\nwmargintag{{\nwtagstyle{}\subpageref{NW3gGP3e-3uJ41M-2}}}\moddef{Orthogonal line~{\nwtagstyle{}\subpageref{NW3gGP3e-3uJ41M-1}}}\plusendmoddef\Rm{}\nwstartdeflinemarkup\nwusesondefline{\\{NW3gGP3e-1zkOI-2}}\nwprevnextdefs{NW3gGP3e-3uJ41M-1}{NW3gGP3e-3uJ41M-3}\nwenddeflinemarkup
{\it{}u3} = {\it{}C}.{\it{}center}().{\it{}op}(0);
{\it{}v3} = {\it{}C4}.{\it{}roots}({\it{}u3}, {\bf{}false}).{\it{}op}(0).{\it{}normal}();
{\it{}cout} \begin{math}\ll\end{math} {\tt{}"All lines come through the point "} {\it{}math\_string} \begin{math}\ll\end{math}{\tt{}"("} \begin{math}\ll\end{math} {\it{}u3} \begin{math}\ll\end{math} {\tt{}", "} \begin{math}\ll\end{math} {\it{}v3} \begin{math}\ll\end{math} {\tt{}")"} {\it{}math\_string} \begin{math}\ll\end{math} {\it{}endl};

\nwused{\\{NW3gGP3e-1zkOI-2}}\nwidentuses{\\{{\nwixident{center}}{center}}\\{{\nwixident{math{\_}string}}{math:unstring}}\\{{\nwixident{normal}}{normal}}\\{{\nwixident{op}}{op}}\\{{\nwixident{roots}}{roots}}}\nwindexuse{\nwixident{center}}{center}{NW3gGP3e-3uJ41M-2}\nwindexuse{\nwixident{math{\_}string}}{math:unstring}{NW3gGP3e-3uJ41M-2}\nwindexuse{\nwixident{normal}}{normal}{NW3gGP3e-3uJ41M-2}\nwindexuse{\nwixident{op}}{op}{NW3gGP3e-3uJ41M-2}\nwindexuse{\nwixident{roots}}{roots}{NW3gGP3e-3uJ41M-2}\nwendcode{}\nwbegindocs{231}The double check is done next: we calculate the inverse {\Tt{}\Rm{}{\it{}P1}\nwendquote} of
a vector {\Tt{}\Rm{}({\it{}u3}+{\it{}u}, {\it{}v3}+{\it{}v})\nwendquote} and check that {\Tt{}\Rm{}{\it{}P1}-({\it{}u3}, {\it{}v3})\nwendquote} is collinear
to {\Tt{}\Rm{}({\it{}u}, {\it{}v})\nwendquote}.

\nwenddocs{}\nwbegincode{232}\sublabel{NW3gGP3e-3uJ41M-3}\nwmargintag{{\nwtagstyle{}\subpageref{NW3gGP3e-3uJ41M-3}}}\moddef{Orthogonal line~{\nwtagstyle{}\subpageref{NW3gGP3e-3uJ41M-1}}}\plusendmoddef\Rm{}\nwstartdeflinemarkup\nwusesondefline{\\{NW3gGP3e-1zkOI-2}}\nwprevnextdefs{NW3gGP3e-3uJ41M-2}{\relax}\nwenddeflinemarkup
{\it{}P1} = {\it{}C}.{\it{}moebius\_map}({\bf{}lst}({\it{}u3}+{\it{}u}, {\it{}v3}+{\it{}v}), {\it{}e}, -{\it{}M1});
{\it{}cout} \begin{math}\ll\end{math} {\tt{}"Conjugated vector is parallel to (u,v): "}
        \begin{math}\ll\end{math} (({\it{}P1}.{\it{}op}(0)-{\it{}u3})\begin{math}\ast\end{math}{\it{}v}-({\it{}P1}.{\it{}op}(1)-{\it{}v3})\begin{math}\ast\end{math}{\it{}u}).{\it{}normal}().{\it{}is\_zero}() \begin{math}\ll\end{math} {\it{}endl};
{\bf{}if} ({\it{}debug} \begin{math}>\end{math} 1)
    {\it{}cout} \begin{math}\ll\end{math} {\tt{}"Conjugated vector to (u, v) is: "} {\it{}math\_string} \begin{math}\ll\end{math} {\tt{}"("} \begin{math}\ll\end{math} ({\it{}P1}.{\it{}op}(0)-{\it{}u3}).{\it{}normal}() \begin{math}\ll\end{math} {\tt{}", "}
         \begin{math}\ll\end{math} ({\it{}P1}.{\it{}op}(1)-{\it{}v3}).{\it{}normal}() \begin{math}\ll\end{math} {\tt{}")"} {\it{}math\_string} \begin{math}\ll\end{math} {\it{}endl};

\nwused{\\{NW3gGP3e-1zkOI-2}}\nwidentuses{\\{{\nwixident{debug}}{debug}}\\{{\nwixident{is{\_}zero}}{is:unzero}}\\{{\nwixident{lst}}{lst}}\\{{\nwixident{math{\_}string}}{math:unstring}}\\{{\nwixident{moebius{\_}map}}{moebius:unmap}}\\{{\nwixident{normal}}{normal}}\\{{\nwixident{op}}{op}}\\{{\nwixident{u}}{u}}\\{{\nwixident{v}}{v}}}\nwindexuse{\nwixident{debug}}{debug}{NW3gGP3e-3uJ41M-3}\nwindexuse{\nwixident{is{\_}zero}}{is:unzero}{NW3gGP3e-3uJ41M-3}\nwindexuse{\nwixident{lst}}{lst}{NW3gGP3e-3uJ41M-3}\nwindexuse{\nwixident{math{\_}string}}{math:unstring}{NW3gGP3e-3uJ41M-3}\nwindexuse{\nwixident{moebius{\_}map}}{moebius:unmap}{NW3gGP3e-3uJ41M-3}\nwindexuse{\nwixident{normal}}{normal}{NW3gGP3e-3uJ41M-3}\nwindexuse{\nwixident{op}}{op}{NW3gGP3e-3uJ41M-3}\nwindexuse{\nwixident{u}}{u}{NW3gGP3e-3uJ41M-3}\nwindexuse{\nwixident{v}}{v}{NW3gGP3e-3uJ41M-3}\nwendcode{}\nwbegindocs{233}\nwdocspar
\subsubsection{The Ghost Cycle}
\label{sec:ghost-cycle}
We build now the cycle {\Tt{}\Rm{}{\it{}C5}\nwendquote} which defines inversion. We build
it from two conditions:
\begin{enumerate}
\item {\Tt{}\Rm{}{\it{}C5}\nwendquote} has its centre in the point {\Tt{}\Rm{}({\it{}u3}, {\it{}v3})\nwendquote} which is the
  intersection of all orthogonal lines (see \S~\ref{sec:orthogonal-lines}).
\item The determinant of {\Tt{}\Rm{}{\it{}C5}\nwendquote} with delta-sign is equal to
  determinant of {\Tt{}\Rm{}{\it{}C}\nwendquote} with signs defined by {\Tt{}\Rm{}{\it{}M1}\nwendquote}.
\end{enumerate}

\nwenddocs{}\nwbegincode{234}\sublabel{NW3gGP3e-47k9wZ-1}\nwmargintag{{\nwtagstyle{}\subpageref{NW3gGP3e-47k9wZ-1}}}\moddef{Inversion in cycle~{\nwtagstyle{}\subpageref{NW3gGP3e-47k9wZ-1}}}\endmoddef\Rm{}\nwstartdeflinemarkup\nwusesondefline{\\{NW3gGP3e-1zkOI-2}}\nwprevnextdefs{\relax}{NW3gGP3e-47k9wZ-2}\nwenddeflinemarkup
      {\it{}C5} = {\bf{}cycle2D}({\bf{}lst}({\it{}u3}, -{\it{}v3}\begin{math}\ast\end{math}{\it{}jump\_fnct}({\it{}sign})), {\it{}e}, {\it{}C}.{\it{}radius\_sq}({\it{}e}, {\it{}M1})).{\it{}subs}({\it{}signs\_cube},
                 {\it{}subs\_options}::{\it{}algebraic} \begin{math}\mid\end{math} {\it{}subs\_options}::{\it{}no\_pattern});

\nwalsodefined{\\{NW3gGP3e-47k9wZ-2}\\{NW3gGP3e-47k9wZ-3}\\{NW3gGP3e-47k9wZ-4}}\nwused{\\{NW3gGP3e-1zkOI-2}}\nwidentuses{\\{{\nwixident{cycle2D}}{cycle2D}}\\{{\nwixident{jump{\_}fnct}}{jump:unfnct}}\\{{\nwixident{lst}}{lst}}\\{{\nwixident{radius{\_}sq}}{radius:unsq}}\\{{\nwixident{subs}}{subs}}}\nwindexuse{\nwixident{cycle2D}}{cycle2D}{NW3gGP3e-47k9wZ-1}\nwindexuse{\nwixident{jump{\_}fnct}}{jump:unfnct}{NW3gGP3e-47k9wZ-1}\nwindexuse{\nwixident{lst}}{lst}{NW3gGP3e-47k9wZ-1}\nwindexuse{\nwixident{radius{\_}sq}}{radius:unsq}{NW3gGP3e-47k9wZ-1}\nwindexuse{\nwixident{subs}}{subs}{NW3gGP3e-47k9wZ-1}\nwendcode{}\nwbegindocs{235}As a consequence we find out that {\Tt{}\Rm{}{\it{}C5}\nwendquote} has the same roots as {\Tt{}\Rm{}{\it{}C}\nwendquote}.
\nwenddocs{}\nwbegincode{236}\sublabel{NW3gGP3e-47k9wZ-2}\nwmargintag{{\nwtagstyle{}\subpageref{NW3gGP3e-47k9wZ-2}}}\moddef{Inversion in cycle~{\nwtagstyle{}\subpageref{NW3gGP3e-47k9wZ-1}}}\plusendmoddef\Rm{}\nwstartdeflinemarkup\nwusesondefline{\\{NW3gGP3e-1zkOI-2}}\nwprevnextdefs{NW3gGP3e-47k9wZ-1}{NW3gGP3e-47k9wZ-3}\nwenddeflinemarkup
{\it{}cout} \begin{math}\ll\end{math} {\tt{}"C5 has common roots with C : "} \begin{math}\ll\end{math} ({\it{}C5}.{\it{}val}({\bf{}lst}({\it{}C}.{\it{}roots}().{\it{}op}(0), 0)).{\it{}normal}().{\it{}is\_zero}()
                                            \begin{math}\wedge\end{math} {\it{}C5}.{\it{}val}({\bf{}lst}({\it{}C}.{\it{}roots}().{\it{}op}(1), 0)).{\it{}normal}().{\it{}is\_zero}()) \begin{math}\ll\end{math} {\it{}endl}
                                                                                                      \begin{math}\ll\end{math} {\tt{}"${\char92}{\char92}chi({\char92}{\char92}sigma)$-centre of C5 is equal to ${\char92}{\char92}breve{\char123}{\char92}{\char92}sigma{\char125}$-centre of C: "}
                                                                                                         \begin{math}\ll\end{math} ({\it{}C5}.{\it{}center}({\it{}diag\_matrix}({\bf{}lst}(-1,{\it{}jump\_fnct}({\it{}sign}))), {\bf{}true})-{\it{}C}.{\it{}center}({\it{}es}, {\bf{}true})).{\it{}normal}().{\it{}is\_zero}() \begin{math}\ll\end{math}  {\it{}endl};

\nwused{\\{NW3gGP3e-1zkOI-2}}\nwidentuses{\\{{\nwixident{center}}{center}}\\{{\nwixident{is{\_}zero}}{is:unzero}}\\{{\nwixident{jump{\_}fnct}}{jump:unfnct}}\\{{\nwixident{lst}}{lst}}\\{{\nwixident{normal}}{normal}}\\{{\nwixident{op}}{op}}\\{{\nwixident{roots}}{roots}}\\{{\nwixident{val}}{val}}}\nwindexuse{\nwixident{center}}{center}{NW3gGP3e-47k9wZ-2}\nwindexuse{\nwixident{is{\_}zero}}{is:unzero}{NW3gGP3e-47k9wZ-2}\nwindexuse{\nwixident{jump{\_}fnct}}{jump:unfnct}{NW3gGP3e-47k9wZ-2}\nwindexuse{\nwixident{lst}}{lst}{NW3gGP3e-47k9wZ-2}\nwindexuse{\nwixident{normal}}{normal}{NW3gGP3e-47k9wZ-2}\nwindexuse{\nwixident{op}}{op}{NW3gGP3e-47k9wZ-2}\nwindexuse{\nwixident{roots}}{roots}{NW3gGP3e-47k9wZ-2}\nwindexuse{\nwixident{val}}{val}{NW3gGP3e-47k9wZ-2}\nwendcode{}\nwbegindocs{237}Finally we calculate point {\Tt{}\Rm{}{\it{}P1}\nwendquote} which is the inverse of \((u_3,
v_3)\) in {\Tt{}\Rm{}{\it{}C5}\nwendquote}.
\nwenddocs{}\nwbegincode{238}\sublabel{NW3gGP3e-47k9wZ-3}\nwmargintag{{\nwtagstyle{}\subpageref{NW3gGP3e-47k9wZ-3}}}\moddef{Inversion in cycle~{\nwtagstyle{}\subpageref{NW3gGP3e-47k9wZ-1}}}\plusendmoddef\Rm{}\nwstartdeflinemarkup\nwusesondefline{\\{NW3gGP3e-1zkOI-2}}\nwprevnextdefs{NW3gGP3e-47k9wZ-2}{NW3gGP3e-47k9wZ-4}\nwenddeflinemarkup
{\it{}P1} = {\it{}C5}.{\it{}moebius\_map}({\it{}W}, {\it{}e}, {\it{}diag\_matrix}({\bf{}lst}(1, -{\it{}jump\_fnct}({\it{}sign}))));

\nwused{\\{NW3gGP3e-1zkOI-2}}\nwidentuses{\\{{\nwixident{jump{\_}fnct}}{jump:unfnct}}\\{{\nwixident{lst}}{lst}}\\{{\nwixident{moebius{\_}map}}{moebius:unmap}}}\nwindexuse{\nwixident{jump{\_}fnct}}{jump:unfnct}{NW3gGP3e-47k9wZ-3}\nwindexuse{\nwixident{lst}}{lst}{NW3gGP3e-47k9wZ-3}\nwindexuse{\nwixident{moebius{\_}map}}{moebius:unmap}{NW3gGP3e-47k9wZ-3}\nwendcode{}\nwbegindocs{239}The final check: {\Tt{}\Rm{}{\it{}P1}\nwendquote} (inversion in {\Tt{}\Rm{}{\it{}C5}\nwendquote} in terms of {\Tt{}\Rm{}{\it{}sign}\nwendquote})
coincides with {\Tt{}\Rm{}{\it{}P}\nwendquote}---the inversion  in {\Tt{}\Rm{}{\it{}C}\nwendquote} in terms of {\Tt{}\Rm{}{\it{}sign1}\nwendquote},
see chunk~\subpageref{chu:orth-invers-2}.
\nwenddocs{}\nwbegincode{240}\sublabel{NW3gGP3e-47k9wZ-4}\nwmargintag{{\nwtagstyle{}\subpageref{NW3gGP3e-47k9wZ-4}}}\moddef{Inversion in cycle~{\nwtagstyle{}\subpageref{NW3gGP3e-47k9wZ-1}}}\plusendmoddef\Rm{}\nwstartdeflinemarkup\nwusesondefline{\\{NW3gGP3e-1zkOI-2}}\nwprevnextdefs{NW3gGP3e-47k9wZ-3}{\relax}\nwenddeflinemarkup
{\it{}cout} \begin{math}\ll\end{math} {\tt{}"Inversion in (C5, sign)  coincides with inversion in (C, sign1): "}
 \begin{math}\ll\end{math} ({\it{}P1}-{\it{}P}).{\it{}subs}({\it{}signs\_cube}, {\it{}subs\_options}::{\it{}algebraic} \begin{math}\mid\end{math} {\it{}subs\_options}::{\it{}no\_pattern}).{\it{}normal}().{\it{}is\_zero}()
 \begin{math}\ll\end{math} {\it{}endl};

\nwused{\\{NW3gGP3e-1zkOI-2}}\nwidentuses{\\{{\nwixident{is{\_}zero}}{is:unzero}}\\{{\nwixident{normal}}{normal}}\\{{\nwixident{subs}}{subs}}}\nwindexuse{\nwixident{is{\_}zero}}{is:unzero}{NW3gGP3e-47k9wZ-4}\nwindexuse{\nwixident{normal}}{normal}{NW3gGP3e-47k9wZ-4}\nwindexuse{\nwixident{subs}}{subs}{NW3gGP3e-47k9wZ-4}\nwendcode{}\nwbegindocs{241}\nwdocspar
\subsubsection{The real line and reflection in cycles}
\label{sec:reflection-cycles}

We check that conjugation \(C_1 \Space{R}{} C_1\) maps the
{\Tt{}\Rm{}{\it{}real\_line}\nwendquote} to the cycle {\Tt{}\Rm{}{\it{}C}\nwendquote} and wise verse for the properly
chosen {\Tt{}\Rm{}{\it{}C1}\nwendquote},
see~\cite[Lem.~\ref{E-le:cycle-refl-real-line}]{Kisil05a}.
\nwenddocs{}\nwbegincode{242}\sublabel{NW3gGP3e-2W1nIA-1}\nwmargintag{{\nwtagstyle{}\subpageref{NW3gGP3e-2W1nIA-1}}}\moddef{Reflection in cycle~{\nwtagstyle{}\subpageref{NW3gGP3e-2W1nIA-1}}}\endmoddef\Rm{}\nwstartdeflinemarkup\nwusesondefline{\\{NW3gGP3e-1zkOI-2}}\nwprevnextdefs{\relax}{NW3gGP3e-2W1nIA-2}\nwenddeflinemarkup
{\bf{}for} ({\it{}si}=-1; {\it{}si}\begin{math}<\end{math}2; {\it{}si}+=2) {\nwlbrace}
    {\it{}C9} = {\bf{}cycle2D}({\it{}k}, {\bf{}lst}({\it{}l}, {\it{}n}+{\it{}si}\begin{math}\ast\end{math}{\it{}sqrt}(-{\it{}C}.{\it{}det}({\it{}es})\begin{math}\ast\end{math}{\it{}sign1})),{\it{}m},{\it{}es});
    {\it{}cout} \begin{math}\ll\end{math} {\tt{}"Inversion to the real line (with "} \begin{math}\ll\end{math} ({\it{}si}\begin{math}\equiv\end{math}-1? {\tt{}"-"} : {\tt{}"+"}) \begin{math}\ll\end{math} {\tt{}" sign): "} \begin{math}\ll\end{math} {\it{}endl}
         \begin{math}\ll\end{math} {\it{}wspaces} \begin{math}\ll\end{math} {\tt{}"Conjugation of the real line is the cycle C: "}
         \begin{math}\ll\end{math} {\it{}real\_line}.{\it{}cycle\_similarity}({\it{}C9}, {\it{}es}).{\it{}subs}({\it{}pow}({\it{}sign1},2)\begin{math}\equiv\end{math}1, {\it{}subs\_options}::{\it{}algebraic}).{\it{}is\_equal}({\it{}C}) \begin{math}\ll\end{math} {\it{}endl}
         \begin{math}\ll\end{math} {\it{}wspaces} \begin{math}\ll\end{math} {\tt{}"Conjugation of the cycle C is the real line: "}
         \begin{math}\ll\end{math} {\it{}C}.{\it{}cycle\_similarity}({\it{}C9}, {\it{}es}).{\it{}subs}({\it{}pow}({\it{}sign1},2)\begin{math}\equiv\end{math}1, {\it{}subs\_options}::{\it{}algebraic}).{\it{}is\_equal}({\it{}real\_line}) \begin{math}\ll\end{math} {\it{}endl}
        
\nwalsodefined{\\{NW3gGP3e-2W1nIA-2}}\nwused{\\{NW3gGP3e-1zkOI-2}}\nwidentuses{\\{{\nwixident{cycle}}{cycle}}\\{{\nwixident{cycle2D}}{cycle2D}}\\{{\nwixident{cycle{\_}similarity}}{cycle:unsimilarity}}\\{{\nwixident{det}}{det}}\\{{\nwixident{is{\_}equal}}{is:unequal}}\\{{\nwixident{k}}{k}}\\{{\nwixident{l}}{l}}\\{{\nwixident{lst}}{lst}}\\{{\nwixident{m}}{m}}\\{{\nwixident{si}}{si}}\\{{\nwixident{subs}}{subs}}\\{{\nwixident{wspaces}}{wspaces}}}\nwindexuse{\nwixident{cycle}}{cycle}{NW3gGP3e-2W1nIA-1}\nwindexuse{\nwixident{cycle2D}}{cycle2D}{NW3gGP3e-2W1nIA-1}\nwindexuse{\nwixident{cycle{\_}similarity}}{cycle:unsimilarity}{NW3gGP3e-2W1nIA-1}\nwindexuse{\nwixident{det}}{det}{NW3gGP3e-2W1nIA-1}\nwindexuse{\nwixident{is{\_}equal}}{is:unequal}{NW3gGP3e-2W1nIA-1}\nwindexuse{\nwixident{k}}{k}{NW3gGP3e-2W1nIA-1}\nwindexuse{\nwixident{l}}{l}{NW3gGP3e-2W1nIA-1}\nwindexuse{\nwixident{lst}}{lst}{NW3gGP3e-2W1nIA-1}\nwindexuse{\nwixident{m}}{m}{NW3gGP3e-2W1nIA-1}\nwindexuse{\nwixident{si}}{si}{NW3gGP3e-2W1nIA-1}\nwindexuse{\nwixident{subs}}{subs}{NW3gGP3e-2W1nIA-1}\nwindexuse{\nwixident{wspaces}}{wspaces}{NW3gGP3e-2W1nIA-1}\nwendcode{}\nwbegindocs{243}We also check two additional properties which caracterises the inversion
cycle {\Tt{}\Rm{}{\it{}C9}\nwendquote} in term of common roots of {\Tt{}\Rm{}{\it{}C}\nwendquote}
\cite[Lem.~\ref{E-it:inv-cycle-root}]{Kisil05a} and {\Tt{}\Rm{}{\it{}C}\nwendquote} passing
through {\Tt{}\Rm{}{\it{}C9}\nwendquote} centre \cite[Lem.~\ref{E-it:inv-cycle-center}]{Kisil05a}.
\nwenddocs{}\nwbegincode{244}\sublabel{NW3gGP3e-2W1nIA-2}\nwmargintag{{\nwtagstyle{}\subpageref{NW3gGP3e-2W1nIA-2}}}\moddef{Reflection in cycle~{\nwtagstyle{}\subpageref{NW3gGP3e-2W1nIA-1}}}\plusendmoddef\Rm{}\nwstartdeflinemarkup\nwusesondefline{\\{NW3gGP3e-1zkOI-2}}\nwprevnextdefs{NW3gGP3e-2W1nIA-1}{\relax}\nwenddeflinemarkup
      \begin{math}\ll\end{math} {\it{}wspaces} \begin{math}\ll\end{math} {\tt{}"Inversion cycle has common roots with C: "}
      \begin{math}\ll\end{math} ({\it{}C9}.{\it{}val}({\bf{}lst}({\it{}C}.{\it{}roots}().{\it{}op}(0), 0)).{\it{}numer}().{\it{}normal}().{\it{}is\_zero}()
          \begin{math}\wedge\end{math} {\it{}C9}.{\it{}val}({\bf{}lst}({\it{}C}.{\it{}roots}().{\it{}op}(1), 0)).{\it{}numer}().{\it{}normal}().{\it{}is\_zero}()) \begin{math}\ll\end{math} {\it{}endl}
      \begin{math}\ll\end{math} {\it{}wspaces} \begin{math}\ll\end{math} {\tt{}"C passing the centre of inversion cycle: "}
      \begin{math}\ll\end{math} {\bf{}cycle2D}({\it{}C}, {\it{}es}).{\it{}val}({\it{}C9}.{\it{}center}()).{\it{}numer}().{\it{}subs}({\it{}sign1}\begin{math}\equiv\end{math}{\it{}sign}, {\it{}subs\_options}::{\it{}no\_pattern}).{\it{}normal}()
      .{\it{}subs}({\it{}pow}({\it{}sign},2)\begin{math}\equiv\end{math}1, {\it{}subs\_options}::{\it{}algebraic} \begin{math}\mid\end{math} {\it{}subs\_options}::{\it{}no\_pattern}).{\it{}is\_zero}() \begin{math}\ll\end{math} {\it{}endl};
{\nwrbrace}

\nwused{\\{NW3gGP3e-1zkOI-2}}\nwidentuses{\\{{\nwixident{center}}{center}}\\{{\nwixident{cycle}}{cycle}}\\{{\nwixident{cycle2D}}{cycle2D}}\\{{\nwixident{is{\_}zero}}{is:unzero}}\\{{\nwixident{lst}}{lst}}\\{{\nwixident{normal}}{normal}}\\{{\nwixident{op}}{op}}\\{{\nwixident{passing}}{passing}}\\{{\nwixident{roots}}{roots}}\\{{\nwixident{subs}}{subs}}\\{{\nwixident{val}}{val}}\\{{\nwixident{wspaces}}{wspaces}}}\nwindexuse{\nwixident{center}}{center}{NW3gGP3e-2W1nIA-2}\nwindexuse{\nwixident{cycle}}{cycle}{NW3gGP3e-2W1nIA-2}\nwindexuse{\nwixident{cycle2D}}{cycle2D}{NW3gGP3e-2W1nIA-2}\nwindexuse{\nwixident{is{\_}zero}}{is:unzero}{NW3gGP3e-2W1nIA-2}\nwindexuse{\nwixident{lst}}{lst}{NW3gGP3e-2W1nIA-2}\nwindexuse{\nwixident{normal}}{normal}{NW3gGP3e-2W1nIA-2}\nwindexuse{\nwixident{op}}{op}{NW3gGP3e-2W1nIA-2}\nwindexuse{\nwixident{passing}}{passing}{NW3gGP3e-2W1nIA-2}\nwindexuse{\nwixident{roots}}{roots}{NW3gGP3e-2W1nIA-2}\nwindexuse{\nwixident{subs}}{subs}{NW3gGP3e-2W1nIA-2}\nwindexuse{\nwixident{val}}{val}{NW3gGP3e-2W1nIA-2}\nwindexuse{\nwixident{wspaces}}{wspaces}{NW3gGP3e-2W1nIA-2}\nwendcode{}\nwbegindocs{245}\nwdocspar
\subsubsection{Yaglom inversion of the second kind}
\label{sec:invers-second-kind}
In the book~\cite[\S~10]{Yaglom79} the inversion of second kind related to a
parabola \(v=k(u-l)^2+m\) is defined by the map:
\begin{displaymath}
  (u,v) \mapsto (u, 2(k(u-l)^2+m)-v).
\end{displaymath}
We shows here that this is a composition of three inversions in two
parabolas and the real line,
see~\cite[Prop.\ref{E-pr:inv-2nd-kind-decomp}]{Kisil05b}.
\nwenddocs{}\nwbegincode{246}\sublabel{NW3gGP3e-4Lsx4s-1}\nwmargintag{{\nwtagstyle{}\subpageref{NW3gGP3e-4Lsx4s-1}}}\moddef{Yaglom inversion~{\nwtagstyle{}\subpageref{NW3gGP3e-4Lsx4s-1}}}\endmoddef\Rm{}\nwstartdeflinemarkup\nwusesondefline{\\{NW3gGP3e-1zkOI-2}}\nwenddeflinemarkup
{\it{}cout} \begin{math}\ll\end{math} {\tt{}"Yaglom inversion of the second kind is three reflections in the cycles: "}
 \begin{math}\ll\end{math} ({\it{}real\_line}.{\it{}moebius\_map}({\bf{}cycle2D}({\bf{}lst}({\it{}l}, 0), {\it{}e}, -{\it{}m}\begin{math}\div\end{math}{\it{}k}).{\it{}moebius\_map}({\bf{}cycle2D}({\bf{}lst}({\it{}l}, 2\begin{math}\ast\end{math}{\it{}m}), {\it{}e}, -{\it{}m}\begin{math}\div\end{math}{\it{}k})
      .{\it{}moebius\_map}({\it{}W}))).{\it{}subs}({\it{}sign}\begin{math}\equiv\end{math}0)
  -{\bf{}matrix}(2,1,{\bf{}lst}({\it{}u}, 2\begin{math}\ast\end{math}({\it{}k}\begin{math}\ast\end{math}{\it{}pow}({\it{}u}-{\it{}l},2)+{\it{}m})-{\it{}v}))).{\it{}normal}().{\it{}is\_zero}() \begin{math}\ll\end{math} {\it{}endl};

\nwused{\\{NW3gGP3e-1zkOI-2}}\nwidentuses{\\{{\nwixident{cycle2D}}{cycle2D}}\\{{\nwixident{is{\_}zero}}{is:unzero}}\\{{\nwixident{k}}{k}}\\{{\nwixident{l}}{l}}\\{{\nwixident{lst}}{lst}}\\{{\nwixident{m}}{m}}\\{{\nwixident{matrix}}{matrix}}\\{{\nwixident{moebius{\_}map}}{moebius:unmap}}\\{{\nwixident{normal}}{normal}}\\{{\nwixident{subs}}{subs}}\\{{\nwixident{u}}{u}}\\{{\nwixident{v}}{v}}}\nwindexuse{\nwixident{cycle2D}}{cycle2D}{NW3gGP3e-4Lsx4s-1}\nwindexuse{\nwixident{is{\_}zero}}{is:unzero}{NW3gGP3e-4Lsx4s-1}\nwindexuse{\nwixident{k}}{k}{NW3gGP3e-4Lsx4s-1}\nwindexuse{\nwixident{l}}{l}{NW3gGP3e-4Lsx4s-1}\nwindexuse{\nwixident{lst}}{lst}{NW3gGP3e-4Lsx4s-1}\nwindexuse{\nwixident{m}}{m}{NW3gGP3e-4Lsx4s-1}\nwindexuse{\nwixident{matrix}}{matrix}{NW3gGP3e-4Lsx4s-1}\nwindexuse{\nwixident{moebius{\_}map}}{moebius:unmap}{NW3gGP3e-4Lsx4s-1}\nwindexuse{\nwixident{normal}}{normal}{NW3gGP3e-4Lsx4s-1}\nwindexuse{\nwixident{subs}}{subs}{NW3gGP3e-4Lsx4s-1}\nwindexuse{\nwixident{u}}{u}{NW3gGP3e-4Lsx4s-1}\nwindexuse{\nwixident{v}}{v}{NW3gGP3e-4Lsx4s-1}\nwendcode{}\nwbegindocs{247}\nwdocspar
\subsection{Focal Orthogonality}
\label{sec:focal-orth-1}

We study now the focal orthogonality condition
(f-orthogonality), \cite[\S~\ref{E-sec:focal-orthogonality}]{Kisil05a}.

\subsubsection{Expressions for f-orthogonality}
\label{sec:expr-orth-s.k}

\nwenddocs{}\nwbegindocs{248}One more simple consistency check: the {\Tt{}\Rm{}{\it{}real\_line}\nwendquote} is invariant
under all M\"obius transformations.
\nwenddocs{}\nwbegincode{249}\sublabel{NW3gGP3e-2oH953-1}\nwmargintag{{\nwtagstyle{}\subpageref{NW3gGP3e-2oH953-1}}}\moddef{Focal orthogonality conditions~{\nwtagstyle{}\subpageref{NW3gGP3e-2oH953-1}}}\endmoddef\Rm{}\nwstartdeflinemarkup\nwusesondefline{\\{NW3gGP3e-1zkOI-3}}\nwprevnextdefs{\relax}{NW3gGP3e-2oH953-2}\nwenddeflinemarkup
{\it{}cout} \begin{math}\ll\end{math} {\tt{}"The real line is Moebius invariant: "} \begin{math}\ll\end{math} {\it{}real\_line}.{\it{}is\_equal}({\it{}real\_line}.{\it{}sl2\_similarity}({\it{}a}, {\it{}b}, {\it{}c}, {\it{}d}, {\it{}es})) \begin{math}\ll\end{math} {\it{}endl}
  \begin{math}\ll\end{math} {\tt{}"Reflection in the real line: "}
  {\it{}math\_string} \begin{math}\ll\end{math} {\it{}Z}.{\it{}cycle\_similarity}({\it{}real\_line}, {\it{}es}).{\it{}normalize}() {\it{}math\_string} \begin{math}\ll\end{math} {\it{}endl}
  \begin{math}\ll\end{math} {\tt{}"Reflection of the real line in cycle C: "} \begin{math}\ll\end{math} {\it{}endl}
  {\it{}math\_string} \begin{math}\ll\end{math} {\it{}real\_line}.{\it{}cycle\_similarity}({\it{}C}, {\it{}es}, {\it{}S2}, {\it{}S3}) {\it{}math\_string} \begin{math}\ll\end{math} {\it{}endl};

\nwalsodefined{\\{NW3gGP3e-2oH953-2}\\{NW3gGP3e-2oH953-3}\\{NW3gGP3e-2oH953-4}\\{NW3gGP3e-2oH953-5}\\{NW3gGP3e-2oH953-6}}\nwused{\\{NW3gGP3e-1zkOI-3}}\nwidentuses{\\{{\nwixident{cycle}}{cycle}}\\{{\nwixident{cycle{\_}similarity}}{cycle:unsimilarity}}\\{{\nwixident{is{\_}equal}}{is:unequal}}\\{{\nwixident{math{\_}string}}{math:unstring}}\\{{\nwixident{normalize}}{normalize}}\\{{\nwixident{sl2{\_}similarity}}{sl2:unsimilarity}}}\nwindexuse{\nwixident{cycle}}{cycle}{NW3gGP3e-2oH953-1}\nwindexuse{\nwixident{cycle{\_}similarity}}{cycle:unsimilarity}{NW3gGP3e-2oH953-1}\nwindexuse{\nwixident{is{\_}equal}}{is:unequal}{NW3gGP3e-2oH953-1}\nwindexuse{\nwixident{math{\_}string}}{math:unstring}{NW3gGP3e-2oH953-1}\nwindexuse{\nwixident{normalize}}{normalize}{NW3gGP3e-2oH953-1}\nwindexuse{\nwixident{sl2{\_}similarity}}{sl2:unsimilarity}{NW3gGP3e-2oH953-1}\nwendcode{}\nwbegindocs{250}The focal orthogonality condition between two different
cycles is calculated by the identity~\cite[\S~\ref{E-sec:focal-orthogonality}]{Kisil05a}
\begin{displaymath}
  \Re \tr \scalar{C_1C_2C_1}{\Space{R}{}}=0.
\end{displaymath}
\nwenddocs{}\nwbegindocs{251}Here is f-orthogonality of two generic {\Tt{}\Rm{}{\bf{}cycle2D}\nwendquote}s\ldots
\nwenddocs{}\nwbegincode{252}\sublabel{NW3gGP3e-2oH953-2}\nwmargintag{{\nwtagstyle{}\subpageref{NW3gGP3e-2oH953-2}}}\moddef{Focal orthogonality conditions~{\nwtagstyle{}\subpageref{NW3gGP3e-2oH953-1}}}\plusendmoddef\Rm{}\nwstartdeflinemarkup\nwusesondefline{\\{NW3gGP3e-1zkOI-3}}\nwprevnextdefs{NW3gGP3e-2oH953-1}{NW3gGP3e-2oH953-3}\nwenddeflinemarkup
{\it{}cout} \begin{math}\ll\end{math} {\it{}wspaces} \begin{math}\ll\end{math} {\tt{}"The f-orthogonality is: "} {\it{}math\_string}
 \begin{math}\ll\end{math} ({\bf{}ex}){\it{}C}.{\it{}is\_f\_orthogonal}({\it{}C1}, {\it{}es}, {\it{}S2}) {\it{}math\_string} \begin{math}\ll\end{math} {\it{}endl}

\nwused{\\{NW3gGP3e-1zkOI-3}}\nwidentuses{\\{{\nwixident{ex}}{ex}}\\{{\nwixident{is{\_}f{\_}orthogonal}}{is:unf:unorthogonal}}\\{{\nwixident{math{\_}string}}{math:unstring}}\\{{\nwixident{wspaces}}{wspaces}}}\nwindexuse{\nwixident{ex}}{ex}{NW3gGP3e-2oH953-2}\nwindexuse{\nwixident{is{\_}f{\_}orthogonal}}{is:unf:unorthogonal}{NW3gGP3e-2oH953-2}\nwindexuse{\nwixident{math{\_}string}}{math:unstring}{NW3gGP3e-2oH953-2}\nwindexuse{\nwixident{wspaces}}{wspaces}{NW3gGP3e-2oH953-2}\nwendcode{}\nwbegindocs{253}\ldots and its reduction to the straight lines case.
\nwenddocs{}\nwbegincode{254}\sublabel{NW3gGP3e-2oH953-3}\nwmargintag{{\nwtagstyle{}\subpageref{NW3gGP3e-2oH953-3}}}\moddef{Focal orthogonality conditions~{\nwtagstyle{}\subpageref{NW3gGP3e-2oH953-1}}}\plusendmoddef\Rm{}\nwstartdeflinemarkup\nwusesondefline{\\{NW3gGP3e-1zkOI-3}}\nwprevnextdefs{NW3gGP3e-2oH953-2}{NW3gGP3e-2oH953-4}\nwenddeflinemarkup
 \begin{math}\ll\end{math} {\it{}wspaces} \begin{math}\ll\end{math} {\tt{}"The f-orthogonality of two lines is: "} {\it{}math\_string}
 \begin{math}\ll\end{math} ({\bf{}ex}){\it{}C}.{\it{}subs}({\it{}k} \begin{math}\equiv\end{math} 0).{\it{}is\_f\_orthogonal}({\it{}C1}.{\it{}subs}({\it{}k1}\begin{math}\equiv\end{math}0), {\it{}es}, {\it{}S2}) {\it{}math\_string} \begin{math}\ll\end{math} {\it{}endl};

\nwused{\\{NW3gGP3e-1zkOI-3}}\nwidentuses{\\{{\nwixident{ex}}{ex}}\\{{\nwixident{is{\_}f{\_}orthogonal}}{is:unf:unorthogonal}}\\{{\nwixident{k}}{k}}\\{{\nwixident{math{\_}string}}{math:unstring}}\\{{\nwixident{subs}}{subs}}\\{{\nwixident{wspaces}}{wspaces}}}\nwindexuse{\nwixident{ex}}{ex}{NW3gGP3e-2oH953-3}\nwindexuse{\nwixident{is{\_}f{\_}orthogonal}}{is:unf:unorthogonal}{NW3gGP3e-2oH953-3}\nwindexuse{\nwixident{k}}{k}{NW3gGP3e-2oH953-3}\nwindexuse{\nwixident{math{\_}string}}{math:unstring}{NW3gGP3e-2oH953-3}\nwindexuse{\nwixident{subs}}{subs}{NW3gGP3e-2oH953-3}\nwindexuse{\nwixident{wspaces}}{wspaces}{NW3gGP3e-2oH953-3}\nwendcode{}\nwbegindocs{255}Here is f-orthogonality of a generic {\Tt{}\Rm{}{\bf{}cycle2D}\nwendquote} to a zero-radius {\Tt{}\Rm{}{\bf{}cycle2D}\nwendquote}.
\nwenddocs{}\nwbegincode{256}\sublabel{NW3gGP3e-2oH953-4}\nwmargintag{{\nwtagstyle{}\subpageref{NW3gGP3e-2oH953-4}}}\moddef{Focal orthogonality conditions~{\nwtagstyle{}\subpageref{NW3gGP3e-2oH953-1}}}\plusendmoddef\Rm{}\nwstartdeflinemarkup\nwusesondefline{\\{NW3gGP3e-1zkOI-3}}\nwprevnextdefs{NW3gGP3e-2oH953-3}{NW3gGP3e-2oH953-5}\nwenddeflinemarkup
{\it{}cout} \begin{math}\ll\end{math} {\it{}wspaces} \begin{math}\ll\end{math} {\tt{}"The f-orthogonality to z-r-cycle is first way: "}  \begin{math}\ll\end{math} {\it{}endl}
 {\it{}math\_string} \begin{math}\ll\end{math} ({\bf{}ex}){\it{}C}.{\it{}is\_f\_orthogonal}({\it{}Z1}, {\it{}es}, {\it{}S2}) {\it{}math\_string} \begin{math}\ll\end{math} {\it{}endl};

\nwused{\\{NW3gGP3e-1zkOI-3}}\nwidentuses{\\{{\nwixident{cycle}}{cycle}}\\{{\nwixident{ex}}{ex}}\\{{\nwixident{is{\_}f{\_}orthogonal}}{is:unf:unorthogonal}}\\{{\nwixident{math{\_}string}}{math:unstring}}\\{{\nwixident{wspaces}}{wspaces}}}\nwindexuse{\nwixident{cycle}}{cycle}{NW3gGP3e-2oH953-4}\nwindexuse{\nwixident{ex}}{ex}{NW3gGP3e-2oH953-4}\nwindexuse{\nwixident{is{\_}f{\_}orthogonal}}{is:unf:unorthogonal}{NW3gGP3e-2oH953-4}\nwindexuse{\nwixident{math{\_}string}}{math:unstring}{NW3gGP3e-2oH953-4}\nwindexuse{\nwixident{wspaces}}{wspaces}{NW3gGP3e-2oH953-4}\nwendcode{}\nwbegindocs{257}Since f-orthogonality is not
symmetric~\cite[\S~\ref{E-sec:focal-orthogonality}]{Kisil05a},
we calculate separately f-orthogonality of a
zero-radius {\Tt{}\Rm{}{\bf{}cycle2D}\nwendquote} to a generic {\Tt{}\Rm{}{\bf{}cycle2D}\nwendquote}.
\nwenddocs{}\nwbegincode{258}\sublabel{NW3gGP3e-2oH953-5}\nwmargintag{{\nwtagstyle{}\subpageref{NW3gGP3e-2oH953-5}}}\moddef{Focal orthogonality conditions~{\nwtagstyle{}\subpageref{NW3gGP3e-2oH953-1}}}\plusendmoddef\Rm{}\nwstartdeflinemarkup\nwusesondefline{\\{NW3gGP3e-1zkOI-3}}\nwprevnextdefs{NW3gGP3e-2oH953-4}{NW3gGP3e-2oH953-6}\nwenddeflinemarkup
{\it{}cout} \begin{math}\ll\end{math} {\it{}wspaces} \begin{math}\ll\end{math} {\tt{}"The f-orthogonality to z-r-cycle in second way: "} \begin{math}\ll\end{math} {\it{}endl}
 {\it{}math\_string} \begin{math}\ll\end{math} ({\bf{}ex}){\it{}Z1}.{\it{}is\_f\_orthogonal}({\it{}C}, {\it{}es}, {\it{}S2}) {\it{}math\_string} \begin{math}\ll\end{math} {\it{}endl};

\nwused{\\{NW3gGP3e-1zkOI-3}}\nwidentuses{\\{{\nwixident{cycle}}{cycle}}\\{{\nwixident{ex}}{ex}}\\{{\nwixident{is{\_}f{\_}orthogonal}}{is:unf:unorthogonal}}\\{{\nwixident{math{\_}string}}{math:unstring}}\\{{\nwixident{wspaces}}{wspaces}}}\nwindexuse{\nwixident{cycle}}{cycle}{NW3gGP3e-2oH953-5}\nwindexuse{\nwixident{ex}}{ex}{NW3gGP3e-2oH953-5}\nwindexuse{\nwixident{is{\_}f{\_}orthogonal}}{is:unf:unorthogonal}{NW3gGP3e-2oH953-5}\nwindexuse{\nwixident{math{\_}string}}{math:unstring}{NW3gGP3e-2oH953-5}\nwindexuse{\nwixident{wspaces}}{wspaces}{NW3gGP3e-2oH953-5}\nwendcode{}\nwbegindocs{259}Here is f-orthogonality of two zero-radius {\Tt{}\Rm{}{\bf{}cycle2D}\nwendquote}s.
\nwenddocs{}\nwbegincode{260}\sublabel{NW3gGP3e-2oH953-6}\nwmargintag{{\nwtagstyle{}\subpageref{NW3gGP3e-2oH953-6}}}\moddef{Focal orthogonality conditions~{\nwtagstyle{}\subpageref{NW3gGP3e-2oH953-1}}}\plusendmoddef\Rm{}\nwstartdeflinemarkup\nwusesondefline{\\{NW3gGP3e-1zkOI-3}}\nwprevnextdefs{NW3gGP3e-2oH953-5}{\relax}\nwenddeflinemarkup
{\it{}C9} = {\bf{}cycle2D}({\bf{}lst}({\it{}u1}, {\it{}v1}), {\it{}e});
{\it{}cout} \begin{math}\ll\end{math} {\it{}wspaces} \begin{math}\ll\end{math} {\tt{}"The f-orthogonality of two z-r-cycle is: "} \begin{math}\ll\end{math} {\it{}endl}
 {\it{}math\_string} \begin{math}\ll\end{math} ({\bf{}ex}){\it{}Z1}.{\it{}is\_f\_orthogonal}({\it{}C9}, {\it{}es}, {\it{}S2}) {\it{}math\_string} \begin{math}\ll\end{math} {\it{}endl};

\nwused{\\{NW3gGP3e-1zkOI-3}}\nwidentuses{\\{{\nwixident{cycle}}{cycle}}\\{{\nwixident{cycle2D}}{cycle2D}}\\{{\nwixident{ex}}{ex}}\\{{\nwixident{is{\_}f{\_}orthogonal}}{is:unf:unorthogonal}}\\{{\nwixident{lst}}{lst}}\\{{\nwixident{math{\_}string}}{math:unstring}}\\{{\nwixident{wspaces}}{wspaces}}}\nwindexuse{\nwixident{cycle}}{cycle}{NW3gGP3e-2oH953-6}\nwindexuse{\nwixident{cycle2D}}{cycle2D}{NW3gGP3e-2oH953-6}\nwindexuse{\nwixident{ex}}{ex}{NW3gGP3e-2oH953-6}\nwindexuse{\nwixident{is{\_}f{\_}orthogonal}}{is:unf:unorthogonal}{NW3gGP3e-2oH953-6}\nwindexuse{\nwixident{lst}}{lst}{NW3gGP3e-2oH953-6}\nwindexuse{\nwixident{math{\_}string}}{math:unstring}{NW3gGP3e-2oH953-6}\nwindexuse{\nwixident{wspaces}}{wspaces}{NW3gGP3e-2oH953-6}\nwendcode{}\nwbegindocs{261}\nwdocspar
\subsubsection{Properies of f-orthogonality }
\label{sec:prop-orth-s.k}

Find the parameters of cycle passing through a point
and f-orthogonal to the given one
\nwenddocs{}\nwbegincode{262}\sublabel{NW3gGP3e-P6T5o-1}\nwmargintag{{\nwtagstyle{}\subpageref{NW3gGP3e-P6T5o-1}}}\moddef{One point and f-orthogonality~{\nwtagstyle{}\subpageref{NW3gGP3e-P6T5o-1}}}\endmoddef\Rm{}\nwstartdeflinemarkup\nwusesondefline{\\{NW3gGP3e-1zkOI-3}}\nwenddeflinemarkup
{\it{}C6} = {\it{}C1}.{\it{}subject\_to}({\bf{}lst}({\it{}C1}.{\it{}passing}({\it{}W}), {\it{}C}.{\it{}is\_f\_orthogonal}({\it{}C1}, {\it{}es})));
{\bf{}if} ({\it{}debug} \begin{math}>\end{math} 1)
 {\it{}cout} \begin{math}\ll\end{math} {\tt{}"Cycle f-orthogonal to (k, (l, n), m) is: "} \begin{math}\ll\end{math} {\it{}endl}
 {\it{}math\_string} \begin{math}\ll\end{math} {\it{}C6} {\it{}math\_string} \begin{math}\ll\end{math} {\it{}endl};

\nwused{\\{NW3gGP3e-1zkOI-3}}\nwidentuses{\\{{\nwixident{debug}}{debug}}\\{{\nwixident{is{\_}f{\_}orthogonal}}{is:unf:unorthogonal}}\\{{\nwixident{k}}{k}}\\{{\nwixident{l}}{l}}\\{{\nwixident{lst}}{lst}}\\{{\nwixident{m}}{m}}\\{{\nwixident{math{\_}string}}{math:unstring}}\\{{\nwixident{passing}}{passing}}\\{{\nwixident{subject{\_}to}}{subject:unto}}}\nwindexuse{\nwixident{debug}}{debug}{NW3gGP3e-P6T5o-1}\nwindexuse{\nwixident{is{\_}f{\_}orthogonal}}{is:unf:unorthogonal}{NW3gGP3e-P6T5o-1}\nwindexuse{\nwixident{k}}{k}{NW3gGP3e-P6T5o-1}\nwindexuse{\nwixident{l}}{l}{NW3gGP3e-P6T5o-1}\nwindexuse{\nwixident{lst}}{lst}{NW3gGP3e-P6T5o-1}\nwindexuse{\nwixident{m}}{m}{NW3gGP3e-P6T5o-1}\nwindexuse{\nwixident{math{\_}string}}{math:unstring}{NW3gGP3e-P6T5o-1}\nwindexuse{\nwixident{passing}}{passing}{NW3gGP3e-P6T5o-1}\nwindexuse{\nwixident{subject{\_}to}}{subject:unto}{NW3gGP3e-P6T5o-1}\nwendcode{}\nwbegindocs{263}Check the orthogonality of the line through a point to the
cycle.
\nwenddocs{}\nwbegincode{264}\sublabel{NW3gGP3e-3jX7bo-1}\nwmargintag{{\nwtagstyle{}\subpageref{NW3gGP3e-3jX7bo-1}}}\moddef{f-orthogonal line~{\nwtagstyle{}\subpageref{NW3gGP3e-3jX7bo-1}}}\endmoddef\Rm{}\nwstartdeflinemarkup\nwusesondefline{\\{NW3gGP3e-1zkOI-3}}\nwprevnextdefs{\relax}{NW3gGP3e-3jX7bo-2}\nwenddeflinemarkup
{\it{}C7} = {\it{}C6}.{\it{}subject\_to}({\bf{}lst}({\it{}C6}.{\it{}is\_linear}()));
{\it{}u4} = {\it{}C}.{\it{}center}().{\it{}op}(0);
{\it{}v4} = {\it{}C7}.{\it{}roots}({\it{}u4}, {\bf{}false}).{\it{}op}(0).{\it{}normal}();

\nwalsodefined{\\{NW3gGP3e-3jX7bo-2}}\nwused{\\{NW3gGP3e-1zkOI-3}}\nwidentuses{\\{{\nwixident{center}}{center}}\\{{\nwixident{is{\_}linear}}{is:unlinear}}\\{{\nwixident{lst}}{lst}}\\{{\nwixident{normal}}{normal}}\\{{\nwixident{op}}{op}}\\{{\nwixident{roots}}{roots}}\\{{\nwixident{subject{\_}to}}{subject:unto}}}\nwindexuse{\nwixident{center}}{center}{NW3gGP3e-3jX7bo-1}\nwindexuse{\nwixident{is{\_}linear}}{is:unlinear}{NW3gGP3e-3jX7bo-1}\nwindexuse{\nwixident{lst}}{lst}{NW3gGP3e-3jX7bo-1}\nwindexuse{\nwixident{normal}}{normal}{NW3gGP3e-3jX7bo-1}\nwindexuse{\nwixident{op}}{op}{NW3gGP3e-3jX7bo-1}\nwindexuse{\nwixident{roots}}{roots}{NW3gGP3e-3jX7bo-1}\nwindexuse{\nwixident{subject{\_}to}}{subject:unto}{NW3gGP3e-3jX7bo-1}\nwendcode{}\nwbegindocs{265}All orthogonal lines come through the same point, which the
focus of the cycle {\Tt{}\Rm{}{\it{}C}\nwendquote} with respect to metric {\Tt{}\Rm{}(-1, -{\it{}sign1})\nwendquote}.
\nwenddocs{}\nwbegincode{266}\sublabel{NW3gGP3e-3jX7bo-2}\nwmargintag{{\nwtagstyle{}\subpageref{NW3gGP3e-3jX7bo-2}}}\moddef{f-orthogonal line~{\nwtagstyle{}\subpageref{NW3gGP3e-3jX7bo-1}}}\plusendmoddef\Rm{}\nwstartdeflinemarkup\nwusesondefline{\\{NW3gGP3e-1zkOI-3}}\nwprevnextdefs{NW3gGP3e-3jX7bo-1}{\relax}\nwenddeflinemarkup
{\it{}cout} \begin{math}\ll\end{math} {\tt{}"All lines come through the focus related ${\char92}{\char92}breve{\char123}e{\char125}$: "}
  \begin{math}\ll\end{math} ({\it{}C}.{\it{}focus}({\it{}diag\_matrix}({\bf{}lst}(-1, -{\it{}sign1})), {\bf{}true})-{\bf{}matrix}(2, 1, {\bf{}lst}({\it{}u4}, {\it{}v4}))).{\it{}normal}().{\it{}is\_zero}() \begin{math}\ll\end{math} {\it{}endl};

\nwused{\\{NW3gGP3e-1zkOI-3}}\nwidentuses{\\{{\nwixident{focus}}{focus}}\\{{\nwixident{is{\_}zero}}{is:unzero}}\\{{\nwixident{lst}}{lst}}\\{{\nwixident{matrix}}{matrix}}\\{{\nwixident{normal}}{normal}}}\nwindexuse{\nwixident{focus}}{focus}{NW3gGP3e-3jX7bo-2}\nwindexuse{\nwixident{is{\_}zero}}{is:unzero}{NW3gGP3e-3jX7bo-2}\nwindexuse{\nwixident{lst}}{lst}{NW3gGP3e-3jX7bo-2}\nwindexuse{\nwixident{matrix}}{matrix}{NW3gGP3e-3jX7bo-2}\nwindexuse{\nwixident{normal}}{normal}{NW3gGP3e-3jX7bo-2}\nwendcode{}\nwbegindocs{267}\nwdocspar
\subsubsection{Inversion from the f-orthogonality}
\label{sec:invers-from-orth}
We express f-orthogonality to a cycle {\Tt{}\Rm{}{\it{}C}\nwendquote} through the usual
orthogonality to another cycle {\Tt{}\Rm{}{\it{}C8}\nwendquote}. This cycle is the reflection of
the real line in {\Tt{}\Rm{}{\it{}C}\nwendquote}, see~\ref{sec:reflection-cycles}.

\nwenddocs{}\nwbegincode{268}\sublabel{NW3gGP3e-22oRI1-1}\nwmargintag{{\nwtagstyle{}\subpageref{NW3gGP3e-22oRI1-1}}}\moddef{f-inversion in cycle~{\nwtagstyle{}\subpageref{NW3gGP3e-22oRI1-1}}}\endmoddef\Rm{}\nwstartdeflinemarkup\nwusesondefline{\\{NW3gGP3e-1zkOI-3}}\nwprevnextdefs{\relax}{NW3gGP3e-22oRI1-2}\nwenddeflinemarkup
{\it{}C8} = {\it{}real\_line}.{\it{}cycle\_similarity}({\it{}C}, {\it{}es}, {\it{}diag\_matrix}({\bf{}lst}(1, {\it{}sign1})), {\it{}diag\_matrix}({\bf{}lst}(1, {\it{}jump\_fnct}({\it{}sign}))), {\it{}diag\_matrix}({\bf{}lst}(1, {\it{}sign1}))).{\it{}normalize}({\it{}n}\begin{math}\ast\end{math}{\it{}k});
{\bf{}if} ({\it{}debug} \begin{math}>\end{math} 1)
 {\it{}cout} \begin{math}\ll\end{math} {\tt{}"f-ghost cycleis : "} {\it{}math\_string} \begin{math}\ll\end{math} {\it{}C8} {\it{}math\_string} \begin{math}\ll\end{math} {\it{}endl};

\nwalsodefined{\\{NW3gGP3e-22oRI1-2}\\{NW3gGP3e-22oRI1-3}\\{NW3gGP3e-22oRI1-4}}\nwused{\\{NW3gGP3e-1zkOI-3}}\nwidentuses{\\{{\nwixident{cycle{\_}similarity}}{cycle:unsimilarity}}\\{{\nwixident{debug}}{debug}}\\{{\nwixident{jump{\_}fnct}}{jump:unfnct}}\\{{\nwixident{k}}{k}}\\{{\nwixident{lst}}{lst}}\\{{\nwixident{math{\_}string}}{math:unstring}}\\{{\nwixident{normalize}}{normalize}}}\nwindexuse{\nwixident{cycle{\_}similarity}}{cycle:unsimilarity}{NW3gGP3e-22oRI1-1}\nwindexuse{\nwixident{debug}}{debug}{NW3gGP3e-22oRI1-1}\nwindexuse{\nwixident{jump{\_}fnct}}{jump:unfnct}{NW3gGP3e-22oRI1-1}\nwindexuse{\nwixident{k}}{k}{NW3gGP3e-22oRI1-1}\nwindexuse{\nwixident{lst}}{lst}{NW3gGP3e-22oRI1-1}\nwindexuse{\nwixident{math{\_}string}}{math:unstring}{NW3gGP3e-22oRI1-1}\nwindexuse{\nwixident{normalize}}{normalize}{NW3gGP3e-22oRI1-1}\nwendcode{}\nwbegindocs{269}We check that {\Tt{}\Rm{}{\it{}C8}\nwendquote} has common roots with {\Tt{}\Rm{}{\it{}C}\nwendquote}.
\nwenddocs{}\nwbegincode{270}\sublabel{NW3gGP3e-22oRI1-2}\nwmargintag{{\nwtagstyle{}\subpageref{NW3gGP3e-22oRI1-2}}}\moddef{f-inversion in cycle~{\nwtagstyle{}\subpageref{NW3gGP3e-22oRI1-1}}}\plusendmoddef\Rm{}\nwstartdeflinemarkup\nwusesondefline{\\{NW3gGP3e-1zkOI-3}}\nwprevnextdefs{NW3gGP3e-22oRI1-1}{NW3gGP3e-22oRI1-3}\nwenddeflinemarkup
{\it{}cout} \begin{math}\ll\end{math} {\tt{}"f-ghost cycle has common roots with C: "} \begin{math}\ll\end{math} ({\it{}C8}.{\it{}val}({\bf{}lst}({\it{}C}.{\it{}roots}().{\it{}op}(0), 0)).{\it{}numer}().{\it{}normal}().{\it{}is\_zero}()
 \begin{math}\wedge\end{math} {\it{}C8}.{\it{}val}({\bf{}lst}({\it{}C}.{\it{}roots}().{\it{}op}(1), 0)).{\it{}numer}().{\it{}normal}().{\it{}is\_zero}()) \begin{math}\ll\end{math} {\it{}endl};

\nwused{\\{NW3gGP3e-1zkOI-3}}\nwidentuses{\\{{\nwixident{cycle}}{cycle}}\\{{\nwixident{is{\_}zero}}{is:unzero}}\\{{\nwixident{lst}}{lst}}\\{{\nwixident{normal}}{normal}}\\{{\nwixident{op}}{op}}\\{{\nwixident{roots}}{roots}}\\{{\nwixident{val}}{val}}}\nwindexuse{\nwixident{cycle}}{cycle}{NW3gGP3e-22oRI1-2}\nwindexuse{\nwixident{is{\_}zero}}{is:unzero}{NW3gGP3e-22oRI1-2}\nwindexuse{\nwixident{lst}}{lst}{NW3gGP3e-22oRI1-2}\nwindexuse{\nwixident{normal}}{normal}{NW3gGP3e-22oRI1-2}\nwindexuse{\nwixident{op}}{op}{NW3gGP3e-22oRI1-2}\nwindexuse{\nwixident{roots}}{roots}{NW3gGP3e-22oRI1-2}\nwindexuse{\nwixident{val}}{val}{NW3gGP3e-22oRI1-2}\nwendcode{}\nwbegindocs{271}This chunk checks that centre of {\Tt{}\Rm{}{\it{}C8}\nwendquote} coincides with focus of {\Tt{}\Rm{}{\it{}C}\nwendquote}.
\nwenddocs{}\nwbegincode{272}\sublabel{NW3gGP3e-22oRI1-3}\nwmargintag{{\nwtagstyle{}\subpageref{NW3gGP3e-22oRI1-3}}}\moddef{f-inversion in cycle~{\nwtagstyle{}\subpageref{NW3gGP3e-22oRI1-1}}}\plusendmoddef\Rm{}\nwstartdeflinemarkup\nwusesondefline{\\{NW3gGP3e-1zkOI-3}}\nwprevnextdefs{NW3gGP3e-22oRI1-2}{NW3gGP3e-22oRI1-4}\nwenddeflinemarkup
{\it{}cout}  \begin{math}\ll\end{math} {\tt{}"${\char92}{\char92}chi({\char92}{\char92}sigma)$-center of f-ghost cycle coincides with ${\char92}{\char92}breve{\char123}{\char92}{\char92}sigma{\char125}$-focus of C : "}
   \begin{math}\ll\end{math} ({\it{}C8}.{\it{}center}({\it{}diag\_matrix}({\bf{}lst}(-1,{\it{}jump\_fnct}({\it{}sign}))), {\bf{}true})
    -{\it{}C}.{\it{}focus}({\it{}diag\_matrix}({\bf{}lst}(-1, -{\it{}sign1})), {\bf{}true})).{\it{}evalm}().{\it{}normal}().{\it{}is\_zero\_matrix}()
 \begin{math}\ll\end{math} {\it{}endl};

\nwused{\\{NW3gGP3e-1zkOI-3}}\nwidentuses{\\{{\nwixident{center}}{center}}\\{{\nwixident{cycle}}{cycle}}\\{{\nwixident{focus}}{focus}}\\{{\nwixident{jump{\_}fnct}}{jump:unfnct}}\\{{\nwixident{lst}}{lst}}\\{{\nwixident{normal}}{normal}}}\nwindexuse{\nwixident{center}}{center}{NW3gGP3e-22oRI1-3}\nwindexuse{\nwixident{cycle}}{cycle}{NW3gGP3e-22oRI1-3}\nwindexuse{\nwixident{focus}}{focus}{NW3gGP3e-22oRI1-3}\nwindexuse{\nwixident{jump{\_}fnct}}{jump:unfnct}{NW3gGP3e-22oRI1-3}\nwindexuse{\nwixident{lst}}{lst}{NW3gGP3e-22oRI1-3}\nwindexuse{\nwixident{normal}}{normal}{NW3gGP3e-22oRI1-3}\nwendcode{}\nwbegindocs{273}Finally we check that f-inversion in {\Tt{}\Rm{}{\it{}C}\nwendquote} defined through
f-orthogonality coincides with inversion in {\Tt{}\Rm{}{\it{}C8}\nwendquote}.
\nwenddocs{}\nwbegincode{274}\sublabel{NW3gGP3e-22oRI1-4}\nwmargintag{{\nwtagstyle{}\subpageref{NW3gGP3e-22oRI1-4}}}\moddef{f-inversion in cycle~{\nwtagstyle{}\subpageref{NW3gGP3e-22oRI1-1}}}\plusendmoddef\Rm{}\nwstartdeflinemarkup\nwusesondefline{\\{NW3gGP3e-1zkOI-3}}\nwprevnextdefs{NW3gGP3e-22oRI1-3}{\relax}\nwenddeflinemarkup
{\it{}P1} = {\it{}C8}.{\it{}moebius\_map}({\it{}W}, {\it{}e}, {\it{}diag\_matrix}({\bf{}lst}(1, -{\it{}jump\_fnct}({\it{}sign})))).{\it{}subs}({\it{}signs\_cube}, {\it{}subs\_options}::{\it{}algebraic} \begin{math}\mid\end{math}
    {\it{}subs\_options}::{\it{}no\_pattern}).{\it{}normal}();
{\it{}cout} \begin{math}\ll\end{math} {\tt{}"f-inversion in C coincides with inversion in f-ghost cycle: "}
  \begin{math}\ll\end{math} {\it{}C6}.{\it{}val}({\it{}P1}).{\it{}normal}().{\it{}subs}({\it{}signs\_cube}, {\it{}subs\_options}::{\it{}algebraic} \begin{math}\mid\end{math} {\it{}subs\_options}::{\it{}no\_pattern}).{\it{}normal}().{\it{}is\_zero}()
  \begin{math}\ll\end{math} {\it{}endl};

\nwused{\\{NW3gGP3e-1zkOI-3}}\nwidentuses{\\{{\nwixident{cycle}}{cycle}}\\{{\nwixident{is{\_}zero}}{is:unzero}}\\{{\nwixident{jump{\_}fnct}}{jump:unfnct}}\\{{\nwixident{lst}}{lst}}\\{{\nwixident{moebius{\_}map}}{moebius:unmap}}\\{{\nwixident{normal}}{normal}}\\{{\nwixident{subs}}{subs}}\\{{\nwixident{val}}{val}}}\nwindexuse{\nwixident{cycle}}{cycle}{NW3gGP3e-22oRI1-4}\nwindexuse{\nwixident{is{\_}zero}}{is:unzero}{NW3gGP3e-22oRI1-4}\nwindexuse{\nwixident{jump{\_}fnct}}{jump:unfnct}{NW3gGP3e-22oRI1-4}\nwindexuse{\nwixident{lst}}{lst}{NW3gGP3e-22oRI1-4}\nwindexuse{\nwixident{moebius{\_}map}}{moebius:unmap}{NW3gGP3e-22oRI1-4}\nwindexuse{\nwixident{normal}}{normal}{NW3gGP3e-22oRI1-4}\nwindexuse{\nwixident{subs}}{subs}{NW3gGP3e-22oRI1-4}\nwindexuse{\nwixident{val}}{val}{NW3gGP3e-22oRI1-4}\nwendcode{}\nwbegindocs{275}\nwdocspar
\subsection{Distances and Lengths}
\label{sec:distances-lengths}

\subsubsection{Distances between points}
\label{sec:dist-betw-points}

We calculate several distances from the cycles.

The distance is given by the extremal value of diameters for all
possible cycles passing through the both
points~\cite[Defn.~\ref{E-de:distance}]{Kisil05b}. Thus we first
construct a generic {\Tt{}\Rm{}{\it{}cycle2d}\nwendquote} {\Tt{}\Rm{}{\it{}C10}\nwendquote} passing through two points
\((u, v)\) and \((u^\prime,v^\prime)\).

\nwenddocs{}\nwbegincode{276}\sublabel{NW3gGP3e-O1KCX-1}\nwmargintag{{\nwtagstyle{}\subpageref{NW3gGP3e-O1KCX-1}}}\moddef{Distances from cycles~{\nwtagstyle{}\subpageref{NW3gGP3e-O1KCX-1}}}\endmoddef\Rm{}\nwstartdeflinemarkup\nwusesondefline{\\{NW3gGP3e-1zkOI-4}}\nwprevnextdefs{\relax}{NW3gGP3e-O1KCX-2}\nwenddeflinemarkup
{\it{}C10} = {\bf{}cycle2D}({\bf{}numeric}(1), {\bf{}lst}({\it{}l}, {\it{}n}), {\it{}m}, {\it{}e});
{\it{}C10} = {\it{}C10}.{\it{}subject\_to}({\bf{}lst}({\it{}C10}.{\it{}passing}({\it{}W}), {\it{}C10}.{\it{}passing}({\it{}W1})), {\bf{}lst}({\it{}m}, {\it{}n}, {\it{}l}));
{\bf{}if} ({\it{}debug} \begin{math}>\end{math} 0) {\it{}cout} \begin{math}\ll\end{math} {\it{}wspaces} \begin{math}\ll\end{math} {\tt{}"C10 is:   "}  \begin{math}\ll\end{math} {\it{}C10} \begin{math}\ll\end{math} {\it{}endl};

\nwalsodefined{\\{NW3gGP3e-O1KCX-2}\\{NW3gGP3e-O1KCX-3}\\{NW3gGP3e-O1KCX-4}\\{NW3gGP3e-O1KCX-5}\\{NW3gGP3e-O1KCX-6}\\{NW3gGP3e-O1KCX-7}\\{NW3gGP3e-O1KCX-8}}\nwused{\\{NW3gGP3e-1zkOI-4}}\nwidentuses{\\{{\nwixident{cycle2D}}{cycle2D}}\\{{\nwixident{debug}}{debug}}\\{{\nwixident{l}}{l}}\\{{\nwixident{lst}}{lst}}\\{{\nwixident{m}}{m}}\\{{\nwixident{numeric}}{numeric}}\\{{\nwixident{passing}}{passing}}\\{{\nwixident{subject{\_}to}}{subject:unto}}\\{{\nwixident{wspaces}}{wspaces}}}\nwindexuse{\nwixident{cycle2D}}{cycle2D}{NW3gGP3e-O1KCX-1}\nwindexuse{\nwixident{debug}}{debug}{NW3gGP3e-O1KCX-1}\nwindexuse{\nwixident{l}}{l}{NW3gGP3e-O1KCX-1}\nwindexuse{\nwixident{lst}}{lst}{NW3gGP3e-O1KCX-1}\nwindexuse{\nwixident{m}}{m}{NW3gGP3e-O1KCX-1}\nwindexuse{\nwixident{numeric}}{numeric}{NW3gGP3e-O1KCX-1}\nwindexuse{\nwixident{passing}}{passing}{NW3gGP3e-O1KCX-1}\nwindexuse{\nwixident{subject{\_}to}}{subject:unto}{NW3gGP3e-O1KCX-1}\nwindexuse{\nwixident{wspaces}}{wspaces}{NW3gGP3e-O1KCX-1}\nwendcode{}\nwbegindocs{277}Then we calculate the square of its radius as the value of the
determinant {\Tt{}\Rm{}{\it{}D}\nwendquote}. The point {\Tt{}\Rm{}{\it{}l}\nwendquote} of extremum {\Tt{}\Rm{}{\it{}Len\_c}\nwendquote} is calculated from the
condition \(D^\prime_l=0\).

\nwenddocs{}\nwbegincode{278}\sublabel{NW3gGP3e-O1KCX-2}\nwmargintag{{\nwtagstyle{}\subpageref{NW3gGP3e-O1KCX-2}}}\moddef{Distances from cycles~{\nwtagstyle{}\subpageref{NW3gGP3e-O1KCX-1}}}\plusendmoddef\Rm{}\nwstartdeflinemarkup\nwusesondefline{\\{NW3gGP3e-1zkOI-4}}\nwprevnextdefs{NW3gGP3e-O1KCX-1}{NW3gGP3e-O1KCX-3}\nwenddeflinemarkup

{\bf{}ex} {\it{}D} = 4\begin{math}\ast\end{math}{\it{}C10}.{\it{}radius\_sq}({\it{}es});
{\it{}Len\_c} = {\it{}D}.{\it{}subs}({\it{}lsolve}({\bf{}lst}({\it{}D}.{\it{}diff}({\it{}l}) \begin{math}\equiv\end{math} 0), {\bf{}lst}({\it{}l}))).{\it{}normal}();

\nwused{\\{NW3gGP3e-1zkOI-4}}\nwidentuses{\\{{\nwixident{ex}}{ex}}\\{{\nwixident{l}}{l}}\\{{\nwixident{lst}}{lst}}\\{{\nwixident{normal}}{normal}}\\{{\nwixident{radius{\_}sq}}{radius:unsq}}\\{{\nwixident{subs}}{subs}}}\nwindexuse{\nwixident{ex}}{ex}{NW3gGP3e-O1KCX-2}\nwindexuse{\nwixident{l}}{l}{NW3gGP3e-O1KCX-2}\nwindexuse{\nwixident{lst}}{lst}{NW3gGP3e-O1KCX-2}\nwindexuse{\nwixident{normal}}{normal}{NW3gGP3e-O1KCX-2}\nwindexuse{\nwixident{radius{\_}sq}}{radius:unsq}{NW3gGP3e-O1KCX-2}\nwindexuse{\nwixident{subs}}{subs}{NW3gGP3e-O1KCX-2}\nwendcode{}\nwbegindocs{279}Now we check that {\Tt{}\Rm{}{\it{}Len\_c}\nwendquote} is equal to~\cite[Lem.~\ref{E-le:distance-first}]{Kisil05a}
\begin{displaymath}
        d^2(y, y^\prime) = \frac{ \bs ((u-u^\prime)^2-\sigma(v- v^\prime)^2) +4(1-\sigma\bs) v v^\prime}
      {(u- u^\prime)^2 \bs-(v-v^\prime)^2} ((u-u^\prime)^2 -\sigma(v- v^\prime)^2),
\end{displaymath}
\nwenddocs{}\nwbegincode{280}\sublabel{NW3gGP3e-O1KCX-3}\nwmargintag{{\nwtagstyle{}\subpageref{NW3gGP3e-O1KCX-3}}}\moddef{Distances from cycles~{\nwtagstyle{}\subpageref{NW3gGP3e-O1KCX-1}}}\plusendmoddef\Rm{}\nwstartdeflinemarkup\nwusesondefline{\\{NW3gGP3e-1zkOI-4}}\nwprevnextdefs{NW3gGP3e-O1KCX-2}{NW3gGP3e-O1KCX-4}\nwenddeflinemarkup
{\it{}cout} \begin{math}\ll\end{math} {\tt{}"Distance between (u,v) and (u{\char92}',v{\char92}') in elliptic and hyperbolic spaces is "}  \begin{math}\ll\end{math} {\it{}endl};

{\bf{}if} ({\it{}output\_latex}) {\nwlbrace}
 {\bf{}ex} {\it{}dist} = ({\it{}sign1}\begin{math}\ast\end{math}({\it{}pow}({\it{}u}-{\it{}u1},2)-{\it{}sign}\begin{math}\ast\end{math}{\it{}pow}({\it{}v}-{\it{}v1},2))+4\begin{math}\ast\end{math}(1-{\it{}sign}\begin{math}\ast\end{math}{\it{}sign1})\begin{math}\ast\end{math}{\it{}v}\begin{math}\ast\end{math}{\it{}v1})\begin{math}\ast\end{math}({\it{}pow}({\it{}u}-{\it{}u1},2)
   -{\it{}sign}\begin{math}\ast\end{math}{\it{}pow}({\it{}v}-{\it{}v1},2))\begin{math}\div\end{math}({\it{}pow}({\it{}u}-{\it{}u1},2)\begin{math}\ast\end{math}{\it{}sign1}-{\it{}pow}({\it{}v}-{\it{}v1},2));
 {\it{}cout} \begin{math}\ll\end{math} {\tt{}"{\char92}{\char92}({\char92}{\char92}displaystyle "} \begin{math}\ll\end{math} {\it{}dist} \begin{math}\ll\end{math} {\tt{}"{\char92}{\char92}): "} \begin{math}\ll\end{math}  ({\it{}Len\_c}-{\it{}dist}).{\it{}normal}().{\it{}is\_zero}() \begin{math}\ll\end{math} {\it{}endl};
{\nwrbrace} {\bf{}else}
 {\it{}cout} \begin{math}\ll\end{math} {\it{}endl}
  \begin{math}\ll\end{math} {\tt{}"  s1*((u-u{\char92}')^2-s*(v-v{\char92}')^2)+4*(1-s*s1)*v*v{\char92}')*((u-u{\char92}')^2-s*(v-v{\char92}')^2)"}
  \begin{math}\ll\end{math} {\it{}endl}
  \begin{math}\ll\end{math} {\tt{}"  ---------------------------------------------------------------      : "}
  \begin{math}\ll\end{math}  ({\it{}Len\_c}-({\it{}sign1}\begin{math}\ast\end{math}({\it{}pow}({\it{}u}-{\it{}u1},2)-{\it{}sign}\begin{math}\ast\end{math}{\it{}pow}({\it{}v}-{\it{}v1},2))+4\begin{math}\ast\end{math}(1-{\it{}sign}\begin{math}\ast\end{math}{\it{}sign1})\begin{math}\ast\end{math}{\it{}v}\begin{math}\ast\end{math}{\it{}v1})\begin{math}\ast\end{math}({\it{}pow}({\it{}u}-{\it{}u1},2)
   -{\it{}sign}\begin{math}\ast\end{math}{\it{}pow}({\it{}v}-{\it{}v1},2))\begin{math}\div\end{math}({\it{}pow}({\it{}u}-{\it{}u1},2)\begin{math}\ast\end{math}{\it{}sign1}-{\it{}pow}({\it{}v}-{\it{}v1},2))).{\it{}normal}().{\it{}is\_zero}() \begin{math}\ll\end{math} {\it{}endl}
  \begin{math}\ll\end{math}{\tt{}"               (u-u{\char92}')^2*s1-(v-v{\char92}')^2"} \begin{math}\ll\end{math} {\it{}endl} \begin{math}\ll\end{math} {\it{}endl};

\nwused{\\{NW3gGP3e-1zkOI-4}}\nwidentuses{\\{{\nwixident{ex}}{ex}}\\{{\nwixident{is{\_}zero}}{is:unzero}}\\{{\nwixident{normal}}{normal}}\\{{\nwixident{u}}{u}}\\{{\nwixident{v}}{v}}}\nwindexuse{\nwixident{ex}}{ex}{NW3gGP3e-O1KCX-3}\nwindexuse{\nwixident{is{\_}zero}}{is:unzero}{NW3gGP3e-O1KCX-3}\nwindexuse{\nwixident{normal}}{normal}{NW3gGP3e-O1KCX-3}\nwindexuse{\nwixident{u}}{u}{NW3gGP3e-O1KCX-3}\nwindexuse{\nwixident{v}}{v}{NW3gGP3e-O1KCX-3}\nwendcode{}\nwbegindocs{281}Conformity is verified in the same chunk (see
\S~\ref{sec:check-conformity}) for this and all subsequent
distances and lengths. Value {\Tt{}\Rm{}{\it{}si} = -1\nwendquote} initiates
conformality checks only
in elliptic and hyperbolic point spaces.
\nwenddocs{}\nwbegincode{282}\sublabel{NW3gGP3e-O1KCX-4}\nwmargintag{{\nwtagstyle{}\subpageref{NW3gGP3e-O1KCX-4}}}\moddef{Distances from cycles~{\nwtagstyle{}\subpageref{NW3gGP3e-O1KCX-1}}}\plusendmoddef\Rm{}\nwstartdeflinemarkup\nwusesondefline{\\{NW3gGP3e-1zkOI-4}}\nwprevnextdefs{NW3gGP3e-O1KCX-3}{NW3gGP3e-O1KCX-5}\nwenddeflinemarkup
{\it{}check\_conformality}({\it{}Len\_c}, -1);
{\it{}C11} = {\it{}C10}.{\it{}subs}({\it{}lsolve}({\bf{}lst}({\it{}D}.{\it{}diff}({\it{}l}) \begin{math}\equiv\end{math} 0), {\bf{}lst}({\it{}l})));
{\it{}print\_perpendicular}({\it{}C11});

\nwused{\\{NW3gGP3e-1zkOI-4}}\nwidentuses{\\{{\nwixident{check{\_}conformality}}{check:unconformality}}\\{{\nwixident{l}}{l}}\\{{\nwixident{lst}}{lst}}\\{{\nwixident{print{\_}perpendicular}}{print:unperpendicular}}\\{{\nwixident{subs}}{subs}}}\nwindexuse{\nwixident{check{\_}conformality}}{check:unconformality}{NW3gGP3e-O1KCX-4}\nwindexuse{\nwixident{l}}{l}{NW3gGP3e-O1KCX-4}\nwindexuse{\nwixident{lst}}{lst}{NW3gGP3e-O1KCX-4}\nwindexuse{\nwixident{print{\_}perpendicular}}{print:unperpendicular}{NW3gGP3e-O1KCX-4}\nwindexuse{\nwixident{subs}}{subs}{NW3gGP3e-O1KCX-4}\nwendcode{}\nwbegindocs{283}In parabolic space the extremal value is attained in the point
\(\frac{1}{2}(u+u1)\), since it separates upward-branched parabolas
from down-branched.
\nwenddocs{}\nwbegincode{284}\sublabel{NW3gGP3e-O1KCX-5}\nwmargintag{{\nwtagstyle{}\subpageref{NW3gGP3e-O1KCX-5}}}\moddef{Distances from cycles~{\nwtagstyle{}\subpageref{NW3gGP3e-O1KCX-1}}}\plusendmoddef\Rm{}\nwstartdeflinemarkup\nwusesondefline{\\{NW3gGP3e-1zkOI-4}}\nwprevnextdefs{NW3gGP3e-O1KCX-4}{NW3gGP3e-O1KCX-6}\nwenddeflinemarkup
{\it{}Len\_c} = {\it{}D}.{\it{}subs}({\bf{}lst}({\it{}sign} \begin{math}\equiv\end{math}0, {\it{}l} \begin{math}\equiv\end{math} ({\it{}u}+{\it{}u1})\begin{math}\ast\end{math}{\it{}half})).{\it{}normal}();
{\it{}cout} \begin{math}\ll\end{math} {\tt{}"Value at the middle point (parabolic point space):"} \begin{math}\ll\end{math} {\it{}endl} \begin{math}\ll\end{math} {\it{}wspaces}
 {\it{}math\_string} \begin{math}\ll\end{math} {\it{}Len\_c} {\it{}math\_string} \begin{math}\ll\end{math} {\it{}endl};

\nwused{\\{NW3gGP3e-1zkOI-4}}\nwidentuses{\\{{\nwixident{l}}{l}}\\{{\nwixident{lst}}{lst}}\\{{\nwixident{math{\_}string}}{math:unstring}}\\{{\nwixident{normal}}{normal}}\\{{\nwixident{subs}}{subs}}\\{{\nwixident{u}}{u}}\\{{\nwixident{wspaces}}{wspaces}}}\nwindexuse{\nwixident{l}}{l}{NW3gGP3e-O1KCX-5}\nwindexuse{\nwixident{lst}}{lst}{NW3gGP3e-O1KCX-5}\nwindexuse{\nwixident{math{\_}string}}{math:unstring}{NW3gGP3e-O1KCX-5}\nwindexuse{\nwixident{normal}}{normal}{NW3gGP3e-O1KCX-5}\nwindexuse{\nwixident{subs}}{subs}{NW3gGP3e-O1KCX-5}\nwindexuse{\nwixident{u}}{u}{NW3gGP3e-O1KCX-5}\nwindexuse{\nwixident{wspaces}}{wspaces}{NW3gGP3e-O1KCX-5}\nwendcode{}\nwbegindocs{285}Value {\Tt{}\Rm{}{\it{}si} = 0\nwendquote} initiates conformality checks only in the parabolic point space.
\nwenddocs{}\nwbegincode{286}\sublabel{NW3gGP3e-O1KCX-6}\nwmargintag{{\nwtagstyle{}\subpageref{NW3gGP3e-O1KCX-6}}}\moddef{Distances from cycles~{\nwtagstyle{}\subpageref{NW3gGP3e-O1KCX-1}}}\plusendmoddef\Rm{}\nwstartdeflinemarkup\nwusesondefline{\\{NW3gGP3e-1zkOI-4}}\nwprevnextdefs{NW3gGP3e-O1KCX-5}{NW3gGP3e-O1KCX-7}\nwenddeflinemarkup
{\it{}check\_conformality}({\it{}Len\_c}, 0);
{\it{}C11} = {\it{}C10}.{\it{}subs}({\bf{}lst}({\it{}sign} \begin{math}\equiv\end{math}0, {\it{}l} \begin{math}\equiv\end{math} ({\it{}u}+{\it{}u1})\begin{math}\ast\end{math}{\it{}half}));
{\it{}print\_perpendicular}({\it{}C11});

\nwused{\\{NW3gGP3e-1zkOI-4}}\nwidentuses{\\{{\nwixident{check{\_}conformality}}{check:unconformality}}\\{{\nwixident{l}}{l}}\\{{\nwixident{lst}}{lst}}\\{{\nwixident{print{\_}perpendicular}}{print:unperpendicular}}\\{{\nwixident{subs}}{subs}}\\{{\nwixident{u}}{u}}}\nwindexuse{\nwixident{check{\_}conformality}}{check:unconformality}{NW3gGP3e-O1KCX-6}\nwindexuse{\nwixident{l}}{l}{NW3gGP3e-O1KCX-6}\nwindexuse{\nwixident{lst}}{lst}{NW3gGP3e-O1KCX-6}\nwindexuse{\nwixident{print{\_}perpendicular}}{print:unperpendicular}{NW3gGP3e-O1KCX-6}\nwindexuse{\nwixident{subs}}{subs}{NW3gGP3e-O1KCX-6}\nwindexuse{\nwixident{u}}{u}{NW3gGP3e-O1KCX-6}\nwendcode{}\nwbegindocs{287} We need to check the case \(v=v^\prime\) separately, since it is not covered
by the above chunk. This is done almost identically to the previous
case, with replacement of \(l\) by \(n\), since the value of \(l\) is
now fixed.
\nwenddocs{}\nwbegincode{288}\sublabel{NW3gGP3e-O1KCX-7}\nwmargintag{{\nwtagstyle{}\subpageref{NW3gGP3e-O1KCX-7}}}\moddef{Distances from cycles~{\nwtagstyle{}\subpageref{NW3gGP3e-O1KCX-1}}}\plusendmoddef\Rm{}\nwstartdeflinemarkup\nwusesondefline{\\{NW3gGP3e-1zkOI-4}}\nwprevnextdefs{NW3gGP3e-O1KCX-6}{NW3gGP3e-O1KCX-8}\nwenddeflinemarkup
{\it{}C10} = {\bf{}cycle2D}({\bf{}numeric}(1), {\bf{}lst}({\it{}l}, {\it{}n}), {\it{}m}, {\it{}e});
{\it{}C10} = {\it{}C10}.{\it{}subject\_to}({\bf{}lst}({\it{}C10}.{\it{}passing}({\it{}W}),
 {\it{}C10}.{\it{}passing}({\bf{}lst}({\it{}u1}, {\it{}v}))));
{\bf{}if} ({\it{}debug} \begin{math}>\end{math} 1)
  {\it{}cout} \begin{math}\ll\end{math} {\it{}wspaces}  \begin{math}\ll\end{math} {\it{}C10} \begin{math}\ll\end{math} {\it{}endl};

\nwused{\\{NW3gGP3e-1zkOI-4}}\nwidentuses{\\{{\nwixident{cycle2D}}{cycle2D}}\\{{\nwixident{debug}}{debug}}\\{{\nwixident{l}}{l}}\\{{\nwixident{lst}}{lst}}\\{{\nwixident{m}}{m}}\\{{\nwixident{numeric}}{numeric}}\\{{\nwixident{passing}}{passing}}\\{{\nwixident{subject{\_}to}}{subject:unto}}\\{{\nwixident{v}}{v}}\\{{\nwixident{wspaces}}{wspaces}}}\nwindexuse{\nwixident{cycle2D}}{cycle2D}{NW3gGP3e-O1KCX-7}\nwindexuse{\nwixident{debug}}{debug}{NW3gGP3e-O1KCX-7}\nwindexuse{\nwixident{l}}{l}{NW3gGP3e-O1KCX-7}\nwindexuse{\nwixident{lst}}{lst}{NW3gGP3e-O1KCX-7}\nwindexuse{\nwixident{m}}{m}{NW3gGP3e-O1KCX-7}\nwindexuse{\nwixident{numeric}}{numeric}{NW3gGP3e-O1KCX-7}\nwindexuse{\nwixident{passing}}{passing}{NW3gGP3e-O1KCX-7}\nwindexuse{\nwixident{subject{\_}to}}{subject:unto}{NW3gGP3e-O1KCX-7}\nwindexuse{\nwixident{v}}{v}{NW3gGP3e-O1KCX-7}\nwindexuse{\nwixident{wspaces}}{wspaces}{NW3gGP3e-O1KCX-7}\nwendcode{}\nwbegindocs{289}This time the extremal point {\Tt{}\Rm{}{\it{}n}\nwendquote} is found from the condition \(D^\prime_n=0\).
\nwenddocs{}\nwbegincode{290}\sublabel{NW3gGP3e-O1KCX-8}\nwmargintag{{\nwtagstyle{}\subpageref{NW3gGP3e-O1KCX-8}}}\moddef{Distances from cycles~{\nwtagstyle{}\subpageref{NW3gGP3e-O1KCX-1}}}\plusendmoddef\Rm{}\nwstartdeflinemarkup\nwusesondefline{\\{NW3gGP3e-1zkOI-4}}\nwprevnextdefs{NW3gGP3e-O1KCX-7}{\relax}\nwenddeflinemarkup
{\it{}D} = 4\begin{math}\ast\end{math}{\it{}C10}.{\it{}radius\_sq}({\it{}es});
{\it{}Len\_c} = {\it{}D}.{\it{}subs}({\it{}lsolve}({\bf{}lst}({\it{}D}.{\it{}diff}({\it{}n}) \begin{math}\equiv\end{math} 0), {\bf{}lst}({\it{}n}))).{\it{}normal}();
{\it{}cout} \begin{math}\ll\end{math} {\tt{}"Distance between (u,v) and (u{\char92}',v{\char92}'): "} \begin{math}\ll\end{math} {\it{}endl}
 \begin{math}\ll\end{math} {\it{}wspaces} \begin{math}\ll\end{math} {\tt{}"Value at critical point:"} \begin{math}\ll\end{math} {\it{}endl} \begin{math}\ll\end{math} {\it{}wspaces} {\it{}math\_string} \begin{math}\ll\end{math} {\it{}Len\_c} {\it{}math\_string}
 \begin{math}\ll\end{math} {\it{}endl} \begin{math}\ll\end{math} {\it{}endl};

\nwused{\\{NW3gGP3e-1zkOI-4}}\nwidentuses{\\{{\nwixident{lst}}{lst}}\\{{\nwixident{math{\_}string}}{math:unstring}}\\{{\nwixident{normal}}{normal}}\\{{\nwixident{radius{\_}sq}}{radius:unsq}}\\{{\nwixident{subs}}{subs}}\\{{\nwixident{u}}{u}}\\{{\nwixident{v}}{v}}\\{{\nwixident{wspaces}}{wspaces}}}\nwindexuse{\nwixident{lst}}{lst}{NW3gGP3e-O1KCX-8}\nwindexuse{\nwixident{math{\_}string}}{math:unstring}{NW3gGP3e-O1KCX-8}\nwindexuse{\nwixident{normal}}{normal}{NW3gGP3e-O1KCX-8}\nwindexuse{\nwixident{radius{\_}sq}}{radius:unsq}{NW3gGP3e-O1KCX-8}\nwindexuse{\nwixident{subs}}{subs}{NW3gGP3e-O1KCX-8}\nwindexuse{\nwixident{u}}{u}{NW3gGP3e-O1KCX-8}\nwindexuse{\nwixident{v}}{v}{NW3gGP3e-O1KCX-8}\nwindexuse{\nwixident{wspaces}}{wspaces}{NW3gGP3e-O1KCX-8}\nwendcode{}\nwbegindocs{291}\nwdocspar
\subsubsection{Check of the conformal property}
\label{sec:check-conformity}

We check conformal property of all distances and lengths. This is most
time-consuming portion of the program and it took few minutes
on my computer. The rest is calculated within twenty seconds.
\nwenddocs{}\nwbegincode{292}\sublabel{NW3gGP3e-2eoTsH-1}\nwmargintag{{\nwtagstyle{}\subpageref{NW3gGP3e-2eoTsH-1}}}\moddef{Check conformal property~{\nwtagstyle{}\subpageref{NW3gGP3e-2eoTsH-1}}}\endmoddef\Rm{}\nwstartdeflinemarkup\nwusesondefline{\\{NW3gGP3e-1Z2pUX-1}}\nwprevnextdefs{\relax}{NW3gGP3e-2eoTsH-2}\nwenddeflinemarkup
{\bf{}void} {\it{}check\_conformality}({\bf{}const} {\bf{}ex} & {\it{}Len\_c}, {\bf{}int} {\it{}si} = 3) {\nwlbrace}\nwindexdefn{\nwixident{check{\_}conformality}}{check:unconformality}{NW3gGP3e-2eoTsH-1}
 \LA{}Evaluate the fraction~{\nwtagstyle{}\subpageref{NW3gGP3e-5o8iC-1}}\RA{}

\nwalsodefined{\\{NW3gGP3e-2eoTsH-2}\\{NW3gGP3e-2eoTsH-3}\\{NW3gGP3e-2eoTsH-4}\\{NW3gGP3e-2eoTsH-5}\\{NW3gGP3e-2eoTsH-6}\\{NW3gGP3e-2eoTsH-7}\\{NW3gGP3e-2eoTsH-8}}\nwused{\\{NW3gGP3e-1Z2pUX-1}}\nwidentdefs{\\{{\nwixident{check{\_}conformality}}{check:unconformality}}}\nwidentuses{\\{{\nwixident{ex}}{ex}}\\{{\nwixident{si}}{si}}}\nwindexuse{\nwixident{ex}}{ex}{NW3gGP3e-2eoTsH-1}\nwindexuse{\nwixident{si}}{si}{NW3gGP3e-2eoTsH-1}\nwendcode{}\nwbegindocs{293}Several times we fork for two cases: the first one if the check is done for all
signs combinations simultaneously.
\nwenddocs{}\nwbegincode{294}\sublabel{NW3gGP3e-2eoTsH-2}\nwmargintag{{\nwtagstyle{}\subpageref{NW3gGP3e-2eoTsH-2}}}\moddef{Check conformal property~{\nwtagstyle{}\subpageref{NW3gGP3e-2eoTsH-1}}}\plusendmoddef\Rm{}\nwstartdeflinemarkup\nwusesondefline{\\{NW3gGP3e-1Z2pUX-1}}\nwprevnextdefs{NW3gGP3e-2eoTsH-1}{NW3gGP3e-2eoTsH-3}\nwenddeflinemarkup
{\bf{}if} ({\it{}si} \begin{math}>\end{math} 2)
 {\it{}cout} \begin{math}\ll\end{math} {\it{}wspaces} \begin{math}\ll\end{math} {\tt{}"This distance/length is conformal:"} ;

\nwused{\\{NW3gGP3e-1Z2pUX-1}}\nwidentuses{\\{{\nwixident{si}}{si}}\\{{\nwixident{wspaces}}{wspaces}}}\nwindexuse{\nwixident{si}}{si}{NW3gGP3e-2eoTsH-2}\nwindexuse{\nwixident{wspaces}}{wspaces}{NW3gGP3e-2eoTsH-2}\nwendcode{}\nwbegindocs{295}The second case is we output coresponding results for different metric signs.
\nwenddocs{}\nwbegincode{296}\sublabel{NW3gGP3e-2eoTsH-3}\nwmargintag{{\nwtagstyle{}\subpageref{NW3gGP3e-2eoTsH-3}}}\moddef{Check conformal property~{\nwtagstyle{}\subpageref{NW3gGP3e-2eoTsH-1}}}\plusendmoddef\Rm{}\nwstartdeflinemarkup\nwusesondefline{\\{NW3gGP3e-1Z2pUX-1}}\nwprevnextdefs{NW3gGP3e-2eoTsH-2}{NW3gGP3e-2eoTsH-4}\nwenddeflinemarkup
{\bf{}else}
 {\it{}cout} \begin{math}\ll\end{math} {\it{}wspaces} \begin{math}\ll\end{math} {\tt{}"Conformity in a cycle space with metric:   E      P      H "} \begin{math}\ll\end{math} {\it{}endl};

\nwused{\\{NW3gGP3e-1Z2pUX-1}}\nwidentuses{\\{{\nwixident{cycle}}{cycle}}\\{{\nwixident{wspaces}}{wspaces}}}\nwindexuse{\nwixident{cycle}}{cycle}{NW3gGP3e-2eoTsH-3}\nwindexuse{\nwixident{wspaces}}{wspaces}{NW3gGP3e-2eoTsH-3}\nwendcode{}\nwbegindocs{297}However we make the substitution of all possible combinations of {\Tt{}\Rm{}{\it{}sign}\nwendquote} and
{\Tt{}\Rm{}{\it{}sign1}\nwendquote} (an initial value of {\Tt{}\Rm{}{\it{}si}\nwendquote} should be set  before in
order to separate parabolic case from others). The first loop is for point space
metric sign.
\nwenddocs{}\nwbegincode{298}\sublabel{NW3gGP3e-2eoTsH-4}\nwmargintag{{\nwtagstyle{}\subpageref{NW3gGP3e-2eoTsH-4}}}\moddef{Check conformal property~{\nwtagstyle{}\subpageref{NW3gGP3e-2eoTsH-1}}}\plusendmoddef\Rm{}\nwstartdeflinemarkup\nwusesondefline{\\{NW3gGP3e-1Z2pUX-1}}\nwprevnextdefs{NW3gGP3e-2eoTsH-3}{NW3gGP3e-2eoTsH-5}\nwenddeflinemarkup
{\bf{}do}  {\nwlbrace}
 {\bf{}if} ({\it{}si} \begin{math}>\end{math} 1)
  {\it{}si1} = 2;
 {\bf{}else} {\nwlbrace}
  {\it{}cout} \begin{math}\ll\end{math} {\it{}wspaces} \begin{math}\ll\end{math} {\tt{}"Point space is "} \begin{math}\ll\end{math} {\it{}eph\_case}({\it{}si}) \begin{math}\ll\end{math} {\tt{}": "};
  {\it{}si1} = -1;
 {\nwrbrace}

\nwused{\\{NW3gGP3e-1Z2pUX-1}}\nwidentuses{\\{{\nwixident{si}}{si}}\\{{\nwixident{si1}}{si1}}\\{{\nwixident{wspaces}}{wspaces}}}\nwindexuse{\nwixident{si}}{si}{NW3gGP3e-2eoTsH-4}\nwindexuse{\nwixident{si1}}{si1}{NW3gGP3e-2eoTsH-4}\nwindexuse{\nwixident{wspaces}}{wspaces}{NW3gGP3e-2eoTsH-4}\nwendcode{}\nwbegindocs{299}The second loop is for cycle space metric sign.
\nwenddocs{}\nwbegincode{300}\sublabel{NW3gGP3e-2eoTsH-5}\nwmargintag{{\nwtagstyle{}\subpageref{NW3gGP3e-2eoTsH-5}}}\moddef{Check conformal property~{\nwtagstyle{}\subpageref{NW3gGP3e-2eoTsH-1}}}\plusendmoddef\Rm{}\nwstartdeflinemarkup\nwusesondefline{\\{NW3gGP3e-1Z2pUX-1}}\nwprevnextdefs{NW3gGP3e-2eoTsH-4}{NW3gGP3e-2eoTsH-6}\nwenddeflinemarkup
 {\bf{}do} {\nwlbrace}
  {\bf{}if} ({\it{}si} \begin{math}<\end{math} 2)

\nwused{\\{NW3gGP3e-1Z2pUX-1}}\nwidentuses{\\{{\nwixident{si}}{si}}}\nwindexuse{\nwixident{si}}{si}{NW3gGP3e-2eoTsH-5}\nwendcode{}\nwbegindocs{301} However the substition of signs is not done for dummy loops.
\nwenddocs{}\nwbegincode{302}\sublabel{NW3gGP3e-2eoTsH-6}\nwmargintag{{\nwtagstyle{}\subpageref{NW3gGP3e-2eoTsH-6}}}\moddef{Check conformal property~{\nwtagstyle{}\subpageref{NW3gGP3e-2eoTsH-1}}}\plusendmoddef\Rm{}\nwstartdeflinemarkup\nwusesondefline{\\{NW3gGP3e-1Z2pUX-1}}\nwprevnextdefs{NW3gGP3e-2eoTsH-5}{NW3gGP3e-2eoTsH-7}\nwenddeflinemarkup
   {\it{}Len\_cD} = {\it{}Len\_fD}.{\it{}subs}({\bf{}lst}({\it{}sign} \begin{math}\equiv\end{math} {\bf{}numeric}({\it{}si}), {\it{}sign1} \begin{math}\equiv\end{math} {\bf{}numeric}({\it{}si1})),
         {\it{}subs\_options}::{\it{}algebraic} \begin{math}\mid\end{math} {\it{}subs\_options}::{\it{}no\_pattern}).{\it{}normal}();

\nwused{\\{NW3gGP3e-1Z2pUX-1}}\nwidentuses{\\{{\nwixident{lst}}{lst}}\\{{\nwixident{normal}}{normal}}\\{{\nwixident{numeric}}{numeric}}\\{{\nwixident{si}}{si}}\\{{\nwixident{si1}}{si1}}\\{{\nwixident{subs}}{subs}}}\nwindexuse{\nwixident{lst}}{lst}{NW3gGP3e-2eoTsH-6}\nwindexuse{\nwixident{normal}}{normal}{NW3gGP3e-2eoTsH-6}\nwindexuse{\nwixident{numeric}}{numeric}{NW3gGP3e-2eoTsH-6}\nwindexuse{\nwixident{si}}{si}{NW3gGP3e-2eoTsH-6}\nwindexuse{\nwixident{si1}}{si1}{NW3gGP3e-2eoTsH-6}\nwindexuse{\nwixident{subs}}{subs}{NW3gGP3e-2eoTsH-6}\nwendcode{}\nwbegindocs{303}But even for dummy loops we make a check the conformity.
\nwenddocs{}\nwbegincode{304}\sublabel{NW3gGP3e-2eoTsH-7}\nwmargintag{{\nwtagstyle{}\subpageref{NW3gGP3e-2eoTsH-7}}}\moddef{Check conformal property~{\nwtagstyle{}\subpageref{NW3gGP3e-2eoTsH-1}}}\plusendmoddef\Rm{}\nwstartdeflinemarkup\nwusesondefline{\\{NW3gGP3e-1Z2pUX-1}}\nwprevnextdefs{NW3gGP3e-2eoTsH-6}{NW3gGP3e-2eoTsH-8}\nwenddeflinemarkup
  \LA{}Find the limit~{\nwtagstyle{}\subpageref{NW3gGP3e-2wjYgT-1}}\RA{}
  \LA{}Check independence~{\nwtagstyle{}\subpageref{NW3gGP3e-3N1PZV-1}}\RA{}

\nwused{\\{NW3gGP3e-1Z2pUX-1}}\nwendcode{}\nwbegindocs{305}and then finalise all loops.
\nwenddocs{}\nwbegincode{306}\sublabel{NW3gGP3e-2eoTsH-8}\nwmargintag{{\nwtagstyle{}\subpageref{NW3gGP3e-2eoTsH-8}}}\moddef{Check conformal property~{\nwtagstyle{}\subpageref{NW3gGP3e-2eoTsH-1}}}\plusendmoddef\Rm{}\nwstartdeflinemarkup\nwusesondefline{\\{NW3gGP3e-1Z2pUX-1}}\nwprevnextdefs{NW3gGP3e-2eoTsH-7}{\relax}\nwenddeflinemarkup
  {\it{}si1}\protect\PP;
 {\nwrbrace} {\bf{}while} ({\it{}si1} \begin{math}<\end{math} 2);
 {\it{}cout}  \begin{math}\ll\end{math} {\it{}endl};
 {\it{}si}+=2;
{\nwrbrace} {\bf{}while} ({\it{}si} \begin{math}<\end{math} 2);
{\nwrbrace}

\nwused{\\{NW3gGP3e-1Z2pUX-1}}\nwidentuses{\\{{\nwixident{si}}{si}}\\{{\nwixident{si1}}{si1}}}\nwindexuse{\nwixident{si}}{si}{NW3gGP3e-2eoTsH-8}\nwindexuse{\nwixident{si1}}{si1}{NW3gGP3e-2eoTsH-8}\nwendcode{}\nwbegindocs{307}To this end we consider the ratio of distances between \((u,v)\) and
\((u+tx,v+ty)\) and between their images {\Tt{}\Rm{}{\it{}gW}\nwendquote} and {\Tt{}\Rm{}{\it{}gW1}\nwendquote} under the
generic M\"obius transform.
\nwenddocs{}\nwbegincode{308}\sublabel{NW3gGP3e-5o8iC-1}\nwmargintag{{\nwtagstyle{}\subpageref{NW3gGP3e-5o8iC-1}}}\moddef{Evaluate the fraction~{\nwtagstyle{}\subpageref{NW3gGP3e-5o8iC-1}}}\endmoddef\Rm{}\nwstartdeflinemarkup\nwusesondefline{\\{NW3gGP3e-2eoTsH-1}}\nwenddeflinemarkup
{\bf{}ex} {\it{}Len\_cD}= (({\it{}Len\_c}.{\it{}subs}({\bf{}lst}({\it{}u} \begin{math}\equiv\end{math} {\it{}gW}.{\it{}op}(0), {\it{}v}\begin{math}\equiv\end{math}{\it{}gW}.{\it{}op}(1), {\it{}u1} \begin{math}\equiv\end{math} {\it{}gW1}.{\it{}op}(0),
  {\it{}v1}\begin{math}\equiv\end{math}{\it{}gW1}.{\it{}op}(1)), {\it{}subs\_options}::{\it{}algebraic} \begin{math}\mid\end{math} {\it{}subs\_options}::{\it{}no\_pattern})
  \begin{math}\div\end{math}{\it{}Len\_c}).{\it{}subs}({\bf{}lst}({\it{}u1}\begin{math}\equiv\end{math}{\it{}u}+{\it{}t}\begin{math}\ast\end{math}{\it{}x}, {\it{}v1}\begin{math}\equiv\end{math}{\it{}v}+{\it{}t}\begin{math}\ast\end{math}{\it{}y}), {\it{}subs\_options}::{\it{}algebraic} \begin{math}\mid\end{math} {\it{}subs\_options}::{\it{}no\_pattern}));
{\bf{}ex} {\it{}Len\_fD} = {\it{}Len\_cD};

\nwused{\\{NW3gGP3e-2eoTsH-1}}\nwidentuses{\\{{\nwixident{ex}}{ex}}\\{{\nwixident{lst}}{lst}}\\{{\nwixident{op}}{op}}\\{{\nwixident{subs}}{subs}}\\{{\nwixident{u}}{u}}\\{{\nwixident{v}}{v}}}\nwindexuse{\nwixident{ex}}{ex}{NW3gGP3e-5o8iC-1}\nwindexuse{\nwixident{lst}}{lst}{NW3gGP3e-5o8iC-1}\nwindexuse{\nwixident{op}}{op}{NW3gGP3e-5o8iC-1}\nwindexuse{\nwixident{subs}}{subs}{NW3gGP3e-5o8iC-1}\nwindexuse{\nwixident{u}}{u}{NW3gGP3e-5o8iC-1}\nwindexuse{\nwixident{v}}{v}{NW3gGP3e-5o8iC-1}\nwendcode{}\nwbegindocs{309}If {\Tt{}\Rm{}{\it{}Len\_cD}\nwendquote} has the variable {\Tt{}\Rm{}{\it{}t}\nwendquote}, we take the limit
\(t\rightarrow 0\) using the power series expansions.
\nwenddocs{}\nwbegincode{310}\sublabel{NW3gGP3e-2wjYgT-1}\nwmargintag{{\nwtagstyle{}\subpageref{NW3gGP3e-2wjYgT-1}}}\moddef{Find the limit~{\nwtagstyle{}\subpageref{NW3gGP3e-2wjYgT-1}}}\endmoddef\Rm{}\nwstartdeflinemarkup\nwusesondefline{\\{NW3gGP3e-2eoTsH-7}}\nwenddeflinemarkup
{\bf{}if} ({\it{}Len\_cD}.{\it{}has}({\it{}t}))
 {\it{}Len\_cD} = {\it{}Len\_cD}.{\it{}series}({\it{}t}\begin{math}\equiv\end{math}0,1).{\it{}op}(0).{\it{}normal}();

\nwused{\\{NW3gGP3e-2eoTsH-7}}\nwidentuses{\\{{\nwixident{normal}}{normal}}\\{{\nwixident{op}}{op}}}\nwindexuse{\nwixident{normal}}{normal}{NW3gGP3e-2wjYgT-1}\nwindexuse{\nwixident{op}}{op}{NW3gGP3e-2wjYgT-1}\nwendcode{}\nwbegindocs{311}The limit of this ratio for \(t\rightarrow 0\) should be independent
from \((x,y)\) (see \cite[Defn.~\ref{E-de:conformal}]{Kisil05a}).
\nwenddocs{}\nwbegincode{312}\sublabel{NW3gGP3e-3N1PZV-1}\nwmargintag{{\nwtagstyle{}\subpageref{NW3gGP3e-3N1PZV-1}}}\moddef{Check independence~{\nwtagstyle{}\subpageref{NW3gGP3e-3N1PZV-1}}}\endmoddef\Rm{}\nwstartdeflinemarkup\nwusesondefline{\\{NW3gGP3e-2eoTsH-7}}\nwenddeflinemarkup
{\bf{}bool} {\it{}is\_conformal} = \begin{math}\neg\end{math}({\it{}Len\_cD}.{\it{}is\_zero}() \begin{math}\vee\end{math} {\it{}Len\_cD}.{\it{}has}({\it{}t})
     \begin{math}\vee\end{math} {\it{}Len\_cD}.{\it{}has}({\it{}x}) \begin{math}\vee\end{math} {\it{}Len\_cD}.{\it{}has}({\it{}y}));
{\it{}cout} \begin{math}\ll\end{math} {\tt{}" "} \begin{math}\ll\end{math} {\it{}is\_conformal};
{\bf{}if} ({\it{}debug} \begin{math}>\end{math} 0 \begin{math}\vee\end{math} (\begin{math}\neg\end{math}{\it{}is\_conformal} \begin{math}\wedge\end{math} ({\it{}si} \begin{math}>\end{math} 2))) {\nwlbrace}
    {\it{}cout} \begin{math}\ll\end{math} {\tt{}". The factor is: "} \begin{math}\ll\end{math} {\it{}endl} \begin{math}\ll\end{math} {\it{}wspaces} {\it{}math\_string} \begin{math}\ll\end{math} {\it{}Len\_cD}.{\it{}normal}() {\it{}math\_string} ;

{\nwrbrace}

\nwused{\\{NW3gGP3e-2eoTsH-7}}\nwidentuses{\\{{\nwixident{bool}}{bool}}\\{{\nwixident{debug}}{debug}}\\{{\nwixident{is{\_}zero}}{is:unzero}}\\{{\nwixident{math{\_}string}}{math:unstring}}\\{{\nwixident{normal}}{normal}}\\{{\nwixident{si}}{si}}\\{{\nwixident{wspaces}}{wspaces}}}\nwindexuse{\nwixident{bool}}{bool}{NW3gGP3e-3N1PZV-1}\nwindexuse{\nwixident{debug}}{debug}{NW3gGP3e-3N1PZV-1}\nwindexuse{\nwixident{is{\_}zero}}{is:unzero}{NW3gGP3e-3N1PZV-1}\nwindexuse{\nwixident{math{\_}string}}{math:unstring}{NW3gGP3e-3N1PZV-1}\nwindexuse{\nwixident{normal}}{normal}{NW3gGP3e-3N1PZV-1}\nwindexuse{\nwixident{si}}{si}{NW3gGP3e-3N1PZV-1}\nwindexuse{\nwixident{wspaces}}{wspaces}{NW3gGP3e-3N1PZV-1}\nwendcode{}\nwbegindocs{313}\nwdocspar
\subsubsection{Calculation of Perpendiculars}
\label{sec:calc-perp}

Lengths define corresponding perpendicular conditions in terms of
shortest routes, see~\cite[Defn.~\ref{E-de:perpendicular}]{Kisil05a}.
\nwenddocs{}\nwbegincode{314}\sublabel{NW3gGP3e-4PKbLl-1}\nwmargintag{{\nwtagstyle{}\subpageref{NW3gGP3e-4PKbLl-1}}}\moddef{Print perpendicular~{\nwtagstyle{}\subpageref{NW3gGP3e-4PKbLl-1}}}\endmoddef\Rm{}\nwstartdeflinemarkup\nwusesondefline{\\{NW3gGP3e-1Z2pUX-1}}\nwenddeflinemarkup
{\bf{}void} {\it{}print\_perpendicular}({\bf{}const} {\bf{}cycle2D} & {\it{}C}) {\nwlbrace}\nwindexdefn{\nwixident{print{\_}perpendicular}}{print:unperpendicular}{NW3gGP3e-4PKbLl-1}
 {\it{}cout} \begin{math}\ll\end{math} {\it{}wspaces} \begin{math}\ll\end{math} {\tt{}"Perpendicular to ((u,v); (u{\char92}',v{\char92}')) is: "}
  {\it{}math\_string} \begin{math}\ll\end{math} ({\it{}C}.{\it{}get\_l}(1)+{\it{}sign}\begin{math}\ast\end{math}{\it{}C}.{\it{}get\_k}()\begin{math}\ast\end{math}{\it{}v1}).{\it{}normal}() {\it{}math\_string} \begin{math}\ll\end{math} {\tt{}"; "}
  {\it{}math\_string} \begin{math}\ll\end{math} ({\it{}C}.{\it{}get\_l}(0)-{\it{}C}.{\it{}get\_k}()\begin{math}\ast\end{math}{\it{}u1}).{\it{}normal}() {\it{}math\_string} \begin{math}\ll\end{math} {\it{}endl} \begin{math}\ll\end{math} {\it{}endl};
{\nwrbrace}

\nwused{\\{NW3gGP3e-1Z2pUX-1}}\nwidentdefs{\\{{\nwixident{print{\_}perpendicular}}{print:unperpendicular}}}\nwidentuses{\\{{\nwixident{cycle2D}}{cycle2D}}\\{{\nwixident{get{\_}k}}{get:unk}}\\{{\nwixident{get{\_}l}}{get:unl}}\\{{\nwixident{math{\_}string}}{math:unstring}}\\{{\nwixident{normal}}{normal}}\\{{\nwixident{u}}{u}}\\{{\nwixident{v}}{v}}\\{{\nwixident{wspaces}}{wspaces}}}\nwindexuse{\nwixident{cycle2D}}{cycle2D}{NW3gGP3e-4PKbLl-1}\nwindexuse{\nwixident{get{\_}k}}{get:unk}{NW3gGP3e-4PKbLl-1}\nwindexuse{\nwixident{get{\_}l}}{get:unl}{NW3gGP3e-4PKbLl-1}\nwindexuse{\nwixident{math{\_}string}}{math:unstring}{NW3gGP3e-4PKbLl-1}\nwindexuse{\nwixident{normal}}{normal}{NW3gGP3e-4PKbLl-1}\nwindexuse{\nwixident{u}}{u}{NW3gGP3e-4PKbLl-1}\nwindexuse{\nwixident{v}}{v}{NW3gGP3e-4PKbLl-1}\nwindexuse{\nwixident{wspaces}}{wspaces}{NW3gGP3e-4PKbLl-1}\nwendcode{}\nwbegindocs{315}\nwdocspar
\subsubsection{Length of intervals from centre}
\label{sec:length-direct-interv}

We calculate the lengths derived from the cycle
with a \emph{centre} at one point and passing through the second,
see~\cite[Defn.~\ref{E-de:length}]{Kisil05a}.

\nwenddocs{}\nwbegindocs{316}Firstly we need some more imaginary units, to accommodate different
types of centres (foci).
\nwenddocs{}\nwbegincode{317}\sublabel{NW3gGP3e-3hvAAH-A}\nwmargintag{{\nwtagstyle{}\subpageref{NW3gGP3e-3hvAAH-A}}}\moddef{Declaration of variables~{\nwtagstyle{}\subpageref{NW3gGP3e-3hvAAH-1}}}\plusendmoddef\Rm{}\nwstartdeflinemarkup\nwusesondefline{\\{NW3gGP3e-1p0Y9w-3}}\nwprevnextdefs{NW3gGP3e-3hvAAH-9}{NW3gGP3e-3hvAAH-B}\nwenddeflinemarkup
{\bf{}ex} {\it{}sign5}={\it{}sign4};
{\bf{}ex} {\it{}e4} = {\it{}clifford\_unit}({\it{}nu}, {\it{}diag\_matrix}({\bf{}lst}(-1, {\it{}sign4})), 2);

\nwused{\\{NW3gGP3e-1p0Y9w-3}}\nwidentuses{\\{{\nwixident{ex}}{ex}}\\{{\nwixident{lst}}{lst}}}\nwindexuse{\nwixident{ex}}{ex}{NW3gGP3e-3hvAAH-A}\nwindexuse{\nwixident{lst}}{lst}{NW3gGP3e-3hvAAH-A}\nwendcode{}\nwbegindocs{318}Then we build a {\Tt{}\Rm{}{\bf{}cycle2D}\nwendquote} {\Tt{}\Rm{}{\it{}C11}\nwendquote} which passes through \((u^\prime, v^\prime)\) and
has its centre at \((u,v)\).
\nwenddocs{}\nwbegincode{319}\sublabel{NW3gGP3e-20e6xZ-1}\nwmargintag{{\nwtagstyle{}\subpageref{NW3gGP3e-20e6xZ-1}}}\moddef{Lengths from centre~{\nwtagstyle{}\subpageref{NW3gGP3e-20e6xZ-1}}}\endmoddef\Rm{}\nwstartdeflinemarkup\nwusesondefline{\\{NW3gGP3e-1zkOI-4}}\nwprevnextdefs{\relax}{NW3gGP3e-20e6xZ-2}\nwenddeflinemarkup
{\it{}C11} = {\it{}C}.{\it{}subject\_to}({\bf{}lst}({\it{}C}.{\it{}passing}({\it{}W1}), {\it{}C}.{\it{}is\_normalized}()));
{\it{}C11} = {\it{}C11}.{\it{}subject\_to}({\bf{}lst}({\it{}C11}.{\it{}center}().{\it{}op}(0) \begin{math}\equiv\end{math} {\it{}u}, {\it{}C11}.{\it{}center}({\it{}e4}).{\it{}op}(1)\begin{math}\equiv\end{math} {\it{}v}));

\nwalsodefined{\\{NW3gGP3e-20e6xZ-2}}\nwused{\\{NW3gGP3e-1zkOI-4}}\nwidentuses{\\{{\nwixident{center}}{center}}\\{{\nwixident{is{\_}normalized}}{is:unnormalized}}\\{{\nwixident{lst}}{lst}}\\{{\nwixident{op}}{op}}\\{{\nwixident{passing}}{passing}}\\{{\nwixident{subject{\_}to}}{subject:unto}}\\{{\nwixident{u}}{u}}\\{{\nwixident{v}}{v}}}\nwindexuse{\nwixident{center}}{center}{NW3gGP3e-20e6xZ-1}\nwindexuse{\nwixident{is{\_}normalized}}{is:unnormalized}{NW3gGP3e-20e6xZ-1}\nwindexuse{\nwixident{lst}}{lst}{NW3gGP3e-20e6xZ-1}\nwindexuse{\nwixident{op}}{op}{NW3gGP3e-20e6xZ-1}\nwindexuse{\nwixident{passing}}{passing}{NW3gGP3e-20e6xZ-1}\nwindexuse{\nwixident{subject{\_}to}}{subject:unto}{NW3gGP3e-20e6xZ-1}\nwindexuse{\nwixident{u}}{u}{NW3gGP3e-20e6xZ-1}\nwindexuse{\nwixident{v}}{v}{NW3gGP3e-20e6xZ-1}\nwendcode{}\nwbegindocs{320}Then the distance is  radius the {\Tt{}\Rm{}{\it{}C11}\nwendquote},
see~\cite[Lem.~\ref{E-it:length-centre}]{Kisil05a}. We check conformity
and calculate the perpendicular at the end.
\nwenddocs{}\nwbegincode{321}\sublabel{NW3gGP3e-20e6xZ-2}\nwmargintag{{\nwtagstyle{}\subpageref{NW3gGP3e-20e6xZ-2}}}\moddef{Lengths from centre~{\nwtagstyle{}\subpageref{NW3gGP3e-20e6xZ-1}}}\plusendmoddef\Rm{}\nwstartdeflinemarkup\nwusesondefline{\\{NW3gGP3e-1zkOI-4}}\nwprevnextdefs{NW3gGP3e-20e6xZ-1}{\relax}\nwenddeflinemarkup
{\it{}Len\_c} = {\it{}C11}.{\it{}radius\_sq}({\it{}es}).{\it{}normal}();
{\it{}cout} \begin{math}\ll\end{math} {\tt{}"Length from *center* between (u,v) and (u^{\char92}{\char92}prime,v^{\char92}{\char92}prime):"} \begin{math}\ll\end{math} {\it{}endl}  \begin{math}\ll\end{math} {\it{}wspaces}
 {\it{}math\_string} \begin{math}\ll\end{math} {\it{}Len\_c} {\it{}math\_string} \begin{math}\ll\end{math} {\it{}endl} ;
{\it{}check\_conformality}({\it{}Len\_c});
{\it{}print\_perpendicular}({\it{}C11});

\nwused{\\{NW3gGP3e-1zkOI-4}}\nwidentuses{\\{{\nwixident{center}}{center}}\\{{\nwixident{check{\_}conformality}}{check:unconformality}}\\{{\nwixident{math{\_}string}}{math:unstring}}\\{{\nwixident{normal}}{normal}}\\{{\nwixident{print{\_}perpendicular}}{print:unperpendicular}}\\{{\nwixident{radius{\_}sq}}{radius:unsq}}\\{{\nwixident{u}}{u}}\\{{\nwixident{v}}{v}}\\{{\nwixident{wspaces}}{wspaces}}}\nwindexuse{\nwixident{center}}{center}{NW3gGP3e-20e6xZ-2}\nwindexuse{\nwixident{check{\_}conformality}}{check:unconformality}{NW3gGP3e-20e6xZ-2}\nwindexuse{\nwixident{math{\_}string}}{math:unstring}{NW3gGP3e-20e6xZ-2}\nwindexuse{\nwixident{normal}}{normal}{NW3gGP3e-20e6xZ-2}\nwindexuse{\nwixident{print{\_}perpendicular}}{print:unperpendicular}{NW3gGP3e-20e6xZ-2}\nwindexuse{\nwixident{radius{\_}sq}}{radius:unsq}{NW3gGP3e-20e6xZ-2}\nwindexuse{\nwixident{u}}{u}{NW3gGP3e-20e6xZ-2}\nwindexuse{\nwixident{v}}{v}{NW3gGP3e-20e6xZ-2}\nwindexuse{\nwixident{wspaces}}{wspaces}{NW3gGP3e-20e6xZ-2}\nwendcode{}\nwbegindocs{322}\nwdocspar
\subsubsection{Length of intervals from focus}
\label{sec:length-interv-from}

We calculate the length derived from the cycle with a
\emph{focus} at one point. To use the linear solver in \GiNaC\ we need
to replace the condition {\Tt{}\Rm{}{\it{}C10}.{\it{}focus}().{\it{}op}(1) \begin{math}\equiv\end{math} {\it{}v}\nwendquote} by hand-made value
for the parameter {\Tt{}\Rm{}{\it{}n}\nwendquote}.

There are two suitable values of {\Tt{}\Rm{}{\it{}n}\nwendquote} which correspond upward and
downward parabolas, which are expressed by plus or minus before the
square root.  After the value of length was found we master
a simpler expression for it which utilises the focal length {\Tt{}\Rm{}{\it{}p}\nwendquote} of
the parabola.
\nwenddocs{}\nwbegincode{323}\sublabel{NW3gGP3e-Nadcx-1}\nwmargintag{{\nwtagstyle{}\subpageref{NW3gGP3e-Nadcx-1}}}\moddef{Lengths from focus~{\nwtagstyle{}\subpageref{NW3gGP3e-Nadcx-1}}}\endmoddef\Rm{}\nwstartdeflinemarkup\nwusesondefline{\\{NW3gGP3e-1zkOI-4}}\nwprevnextdefs{\relax}{NW3gGP3e-Nadcx-2}\nwenddeflinemarkup
{\it{}focal\_length\_check}({\it{}sign5}\begin{math}\ast\end{math}(-({\it{}v1}-{\it{}v})+{\it{}sqrt}({\it{}sign5}\begin{math}\ast\end{math}{\it{}pow}(({\it{}u1}-{\it{}u}), 2)+{\it{}pow}(({\it{}v1}-{\it{}v}), 2) -{\it{}sign5}\begin{math}\ast\end{math}{\it{}sign}\begin{math}\ast\end{math}{\it{}pow}({\it{}v1}, 2))));

\nwalsodefined{\\{NW3gGP3e-Nadcx-2}\\{NW3gGP3e-Nadcx-3}}\nwused{\\{NW3gGP3e-1zkOI-4}}\nwidentuses{\\{{\nwixident{focal{\_}length{\_}check}}{focal:unlength:uncheck}}\\{{\nwixident{u}}{u}}\\{{\nwixident{v}}{v}}}\nwindexuse{\nwixident{focal{\_}length{\_}check}}{focal:unlength:uncheck}{NW3gGP3e-Nadcx-1}\nwindexuse{\nwixident{u}}{u}{NW3gGP3e-Nadcx-1}\nwindexuse{\nwixident{v}}{v}{NW3gGP3e-Nadcx-1}\nwendcode{}\nwbegindocs{324}This chunk is similar to an above one but checks the second
parabola (the minus sign before the square root).
\nwenddocs{}\nwbegincode{325}\sublabel{NW3gGP3e-Nadcx-2}\nwmargintag{{\nwtagstyle{}\subpageref{NW3gGP3e-Nadcx-2}}}\moddef{Lengths from focus~{\nwtagstyle{}\subpageref{NW3gGP3e-Nadcx-1}}}\plusendmoddef\Rm{}\nwstartdeflinemarkup\nwusesondefline{\\{NW3gGP3e-1zkOI-4}}\nwprevnextdefs{NW3gGP3e-Nadcx-1}{NW3gGP3e-Nadcx-3}\nwenddeflinemarkup
{\it{}focal\_length\_check}({\it{}sign5}\begin{math}\ast\end{math}(-({\it{}v1}-{\it{}v})-{\it{}sqrt}({\it{}sign5}\begin{math}\ast\end{math}{\it{}pow}({\it{}u1}-{\it{}u}, 2)+{\it{}pow}(({\it{}v1}-{\it{}v}), 2) -{\it{}sign5}\begin{math}\ast\end{math}{\it{}sign}\begin{math}\ast\end{math}{\it{}pow}({\it{}v1}, 2))));

\nwused{\\{NW3gGP3e-1zkOI-4}}\nwidentuses{\\{{\nwixident{focal{\_}length{\_}check}}{focal:unlength:uncheck}}\\{{\nwixident{u}}{u}}\\{{\nwixident{v}}{v}}}\nwindexuse{\nwixident{focal{\_}length{\_}check}}{focal:unlength:uncheck}{NW3gGP3e-Nadcx-2}\nwindexuse{\nwixident{u}}{u}{NW3gGP3e-Nadcx-2}\nwindexuse{\nwixident{v}}{v}{NW3gGP3e-Nadcx-2}\nwendcode{}\nwbegindocs{326}We need to verify separately the case of {\Tt{}\Rm{}{\it{}sign5}=0\nwendquote}, in this case
{\Tt{}\Rm{}{\it{}p}\nwendquote} has a rational value.
\nwenddocs{}\nwbegincode{327}\sublabel{NW3gGP3e-Nadcx-3}\nwmargintag{{\nwtagstyle{}\subpageref{NW3gGP3e-Nadcx-3}}}\moddef{Lengths from focus~{\nwtagstyle{}\subpageref{NW3gGP3e-Nadcx-1}}}\plusendmoddef\Rm{}\nwstartdeflinemarkup\nwusesondefline{\\{NW3gGP3e-1zkOI-4}}\nwprevnextdefs{NW3gGP3e-Nadcx-2}{\relax}\nwenddeflinemarkup
{\it{}sign5}=0;
{\it{}focal\_length\_check}(({\it{}pow}({\it{}u1}-{\it{}u},2)-{\it{}sign}\begin{math}\ast\end{math}{\it{}pow}({\it{}v1},2))\begin{math}\div\end{math}({\it{}v1}-{\it{}v})\begin{math}\div\end{math}2);

\nwused{\\{NW3gGP3e-1zkOI-4}}\nwidentuses{\\{{\nwixident{focal{\_}length{\_}check}}{focal:unlength:uncheck}}\\{{\nwixident{u}}{u}}\\{{\nwixident{v}}{v}}}\nwindexuse{\nwixident{focal{\_}length{\_}check}}{focal:unlength:uncheck}{NW3gGP3e-Nadcx-3}\nwindexuse{\nwixident{u}}{u}{NW3gGP3e-Nadcx-3}\nwindexuse{\nwixident{v}}{v}{NW3gGP3e-Nadcx-3}\nwendcode{}\nwbegindocs{328} Again to avoid non-linearity of equation, we first
construct a desired cycle.
\nwenddocs{}\nwbegincode{329}\sublabel{NW3gGP3e-2gYzHL-1}\nwmargintag{{\nwtagstyle{}\subpageref{NW3gGP3e-2gYzHL-1}}}\moddef{Focal length checks~{\nwtagstyle{}\subpageref{NW3gGP3e-2gYzHL-1}}}\endmoddef\Rm{}\nwstartdeflinemarkup\nwusesondefline{\\{NW3gGP3e-1Z2pUX-1}}\nwprevnextdefs{\relax}{NW3gGP3e-2gYzHL-2}\nwenddeflinemarkup
{\bf{}void} {\it{}focal\_length\_check}({\bf{}const} {\bf{}ex} & {\it{}p}) {\nwlbrace}\nwindexdefn{\nwixident{focal{\_}length{\_}check}}{focal:unlength:uncheck}{NW3gGP3e-2gYzHL-1}
{\it{}cout} \begin{math}\ll\end{math} {\tt{}"Length from *focus* check for "} {\it{}math\_string} \begin{math}\ll\end{math} {\tt{}"p = "} \begin{math}\ll\end{math} {\it{}p} {\it{}math\_string} \begin{math}\ll\end{math} {\it{}endl};
{\it{}C11} = {\it{}C}.{\it{}subject\_to}({\bf{}lst}({\it{}C}.{\it{}passing}({\it{}W1}), {\it{}k}\begin{math}\equiv\end{math}1, {\it{}l} \begin{math}\equiv\end{math} {\it{}u}, {\it{}n} \begin{math}\equiv\end{math} {\it{}p}));

\nwalsodefined{\\{NW3gGP3e-2gYzHL-2}\\{NW3gGP3e-2gYzHL-3}\\{NW3gGP3e-2gYzHL-4}}\nwused{\\{NW3gGP3e-1Z2pUX-1}}\nwidentdefs{\\{{\nwixident{focal{\_}length{\_}check}}{focal:unlength:uncheck}}}\nwidentuses{\\{{\nwixident{ex}}{ex}}\\{{\nwixident{focus}}{focus}}\\{{\nwixident{k}}{k}}\\{{\nwixident{l}}{l}}\\{{\nwixident{lst}}{lst}}\\{{\nwixident{math{\_}string}}{math:unstring}}\\{{\nwixident{passing}}{passing}}\\{{\nwixident{subject{\_}to}}{subject:unto}}\\{{\nwixident{u}}{u}}}\nwindexuse{\nwixident{ex}}{ex}{NW3gGP3e-2gYzHL-1}\nwindexuse{\nwixident{focus}}{focus}{NW3gGP3e-2gYzHL-1}\nwindexuse{\nwixident{k}}{k}{NW3gGP3e-2gYzHL-1}\nwindexuse{\nwixident{l}}{l}{NW3gGP3e-2gYzHL-1}\nwindexuse{\nwixident{lst}}{lst}{NW3gGP3e-2gYzHL-1}\nwindexuse{\nwixident{math{\_}string}}{math:unstring}{NW3gGP3e-2gYzHL-1}\nwindexuse{\nwixident{passing}}{passing}{NW3gGP3e-2gYzHL-1}\nwindexuse{\nwixident{subject{\_}to}}{subject:unto}{NW3gGP3e-2gYzHL-1}\nwindexuse{\nwixident{u}}{u}{NW3gGP3e-2gYzHL-1}\nwendcode{}\nwbegindocs{330}And now we verify that the length is equal
to \((1-\sigma_1)p^2-2vp\), see~\cite[Lem.~\ref{E-it:length-focus}]{Kisil05a}.
\nwenddocs{}\nwbegincode{331}\sublabel{NW3gGP3e-2gYzHL-2}\nwmargintag{{\nwtagstyle{}\subpageref{NW3gGP3e-2gYzHL-2}}}\moddef{Focal length checks~{\nwtagstyle{}\subpageref{NW3gGP3e-2gYzHL-1}}}\plusendmoddef\Rm{}\nwstartdeflinemarkup\nwusesondefline{\\{NW3gGP3e-1Z2pUX-1}}\nwprevnextdefs{NW3gGP3e-2gYzHL-1}{NW3gGP3e-2gYzHL-3}\nwenddeflinemarkup
{\it{}Len\_c} = {\it{}C11}.{\it{}radius\_sq}({\it{}es}).{\it{}subs}({\it{}pow}({\it{}sign4},2)\begin{math}\equiv\end{math}1,{\it{}subs\_options}::{\it{}algebraic} \begin{math}\mid\end{math} {\it{}subs\_options}::{\it{}no\_pattern}).{\it{}normal}();
{\it{}cout} \begin{math}\ll\end{math} {\it{}wspaces} \begin{math}\ll\end{math} {\tt{}"Length between (u,v) and (u{\char92}', v{\char92}') is equal to "}
 \begin{math}\ll\end{math} ({\it{}output\_latex} ?  {\tt{}"{\char92}{\char92}(({\char92}{\char92}mathring({\char92}{\char92}sigma)-{\char92}{\char92}breve{\char123}{\char92}{\char92}sigma{\char125})p^2-2vp{\char92}{\char92}): "}: {\tt{}"(s4-s1)*p^2-2vp: "})
 \begin{math}\ll\end{math} ({\it{}Len\_c} - (({\it{}sign5}-{\it{}sign1})\begin{math}\ast\end{math}{\it{}pow}({\it{}p}, 2) - 2\begin{math}\ast\end{math}{\it{}v}\begin{math}\ast\end{math}{\it{}p})).{\it{}subs}({\it{}signs\_cube}, {\it{}subs\_options}::{\it{}algebraic} \begin{math}\mid\end{math} {\it{}subs\_options}::{\it{}no\_pattern}).{\it{}expand}()
  .{\it{}subs}({\it{}pow}({\it{}sign4},2)\begin{math}\equiv\end{math}1,{\it{}subs\_options}::{\it{}algebraic} \begin{math}\mid\end{math} {\it{}subs\_options}::{\it{}no\_pattern}).{\it{}normal}().{\it{}is\_zero}() \begin{math}\ll\end{math} {\it{}endl};

\nwused{\\{NW3gGP3e-1Z2pUX-1}}\nwidentuses{\\{{\nwixident{expand}}{expand}}\\{{\nwixident{is{\_}zero}}{is:unzero}}\\{{\nwixident{normal}}{normal}}\\{{\nwixident{radius{\_}sq}}{radius:unsq}}\\{{\nwixident{subs}}{subs}}\\{{\nwixident{u}}{u}}\\{{\nwixident{v}}{v}}\\{{\nwixident{wspaces}}{wspaces}}}\nwindexuse{\nwixident{expand}}{expand}{NW3gGP3e-2gYzHL-2}\nwindexuse{\nwixident{is{\_}zero}}{is:unzero}{NW3gGP3e-2gYzHL-2}\nwindexuse{\nwixident{normal}}{normal}{NW3gGP3e-2gYzHL-2}\nwindexuse{\nwixident{radius{\_}sq}}{radius:unsq}{NW3gGP3e-2gYzHL-2}\nwindexuse{\nwixident{subs}}{subs}{NW3gGP3e-2gYzHL-2}\nwindexuse{\nwixident{u}}{u}{NW3gGP3e-2gYzHL-2}\nwindexuse{\nwixident{v}}{v}{NW3gGP3e-2gYzHL-2}\nwindexuse{\nwixident{wspaces}}{wspaces}{NW3gGP3e-2gYzHL-2}\nwendcode{}\nwbegindocs{332}and we check all requested properties for {\Tt{}\Rm{}{\it{}C11}\nwendquote}: it passes
{\Tt{}\Rm{}({\it{}u1}, {\it{}v1})\nwendquote} and has focus at {\Tt{}\Rm{}({\it{}u}, {\it{}v})\nwendquote}.
\nwenddocs{}\nwbegincode{333}\sublabel{NW3gGP3e-2gYzHL-3}\nwmargintag{{\nwtagstyle{}\subpageref{NW3gGP3e-2gYzHL-3}}}\moddef{Focal length checks~{\nwtagstyle{}\subpageref{NW3gGP3e-2gYzHL-1}}}\plusendmoddef\Rm{}\nwstartdeflinemarkup\nwusesondefline{\\{NW3gGP3e-1Z2pUX-1}}\nwprevnextdefs{NW3gGP3e-2gYzHL-2}{NW3gGP3e-2gYzHL-4}\nwenddeflinemarkup
{\it{}cout} \begin{math}\ll\end{math} {\it{}wspaces} \begin{math}\ll\end{math} {\tt{}"checks: C11 passes through (u{\char92}', v{\char92}'): "} \begin{math}\ll\end{math} {\it{}C11}.{\it{}val}({\it{}W1}).{\it{}normal}().{\it{}is\_zero}()
 \begin{math}\ll\end{math} {\tt{}"; C11 focus is at (u, v): "} 
    \begin{math}\ll\end{math} ({\it{}C11}.{\it{}focus}({\it{}diag\_matrix}({\bf{}lst}(-1,{\it{}sign5})),{\bf{}true}).{\it{}subs}({\it{}pow}({\it{}sign4},2)\begin{math}\equiv\end{math}1,{\it{}subs\_options}::{\it{}algebraic})-{\bf{}matrix}(2,1,{\bf{}lst}({\it{}u},{\it{}v}))).{\it{}evalm}().{\it{}normal}().{\it{}is\_zero\_matrix}() \begin{math}\ll\end{math} {\it{}endl};
{\it{}check\_conformality}({\it{}Len\_c});

\nwused{\\{NW3gGP3e-1Z2pUX-1}}\nwidentuses{\\{{\nwixident{check{\_}conformality}}{check:unconformality}}\\{{\nwixident{focus}}{focus}}\\{{\nwixident{is{\_}zero}}{is:unzero}}\\{{\nwixident{lst}}{lst}}\\{{\nwixident{matrix}}{matrix}}\\{{\nwixident{normal}}{normal}}\\{{\nwixident{subs}}{subs}}\\{{\nwixident{u}}{u}}\\{{\nwixident{v}}{v}}\\{{\nwixident{val}}{val}}\\{{\nwixident{wspaces}}{wspaces}}}\nwindexuse{\nwixident{check{\_}conformality}}{check:unconformality}{NW3gGP3e-2gYzHL-3}\nwindexuse{\nwixident{focus}}{focus}{NW3gGP3e-2gYzHL-3}\nwindexuse{\nwixident{is{\_}zero}}{is:unzero}{NW3gGP3e-2gYzHL-3}\nwindexuse{\nwixident{lst}}{lst}{NW3gGP3e-2gYzHL-3}\nwindexuse{\nwixident{matrix}}{matrix}{NW3gGP3e-2gYzHL-3}\nwindexuse{\nwixident{normal}}{normal}{NW3gGP3e-2gYzHL-3}\nwindexuse{\nwixident{subs}}{subs}{NW3gGP3e-2gYzHL-3}\nwindexuse{\nwixident{u}}{u}{NW3gGP3e-2gYzHL-3}\nwindexuse{\nwixident{v}}{v}{NW3gGP3e-2gYzHL-3}\nwindexuse{\nwixident{val}}{val}{NW3gGP3e-2gYzHL-3}\nwindexuse{\nwixident{wspaces}}{wspaces}{NW3gGP3e-2gYzHL-3}\nwendcode{}\nwbegindocs{334}We finally verify that focal perpendiculars are multiples of the
vector \((\sigma v^\prime+p, u-u^\prime)\), see~\cite[E-it:focal-perpendicularity]{Kisil05a}.
\nwenddocs{}\nwbegincode{335}\sublabel{NW3gGP3e-2gYzHL-4}\nwmargintag{{\nwtagstyle{}\subpageref{NW3gGP3e-2gYzHL-4}}}\moddef{Focal length checks~{\nwtagstyle{}\subpageref{NW3gGP3e-2gYzHL-1}}}\plusendmoddef\Rm{}\nwstartdeflinemarkup\nwusesondefline{\\{NW3gGP3e-1Z2pUX-1}}\nwprevnextdefs{NW3gGP3e-2gYzHL-3}{\relax}\nwenddeflinemarkup
{\it{}cout} \begin{math}\ll\end{math} {\it{}wspaces} \begin{math}\ll\end{math} {\tt{}"Perpendicular to ((u,v); (u{\char92}',v{\char92}')) is "}
 \begin{math}\ll\end{math} ({\it{}output\_latex} ? {\tt{}"{\char92}{\char92}(({\char92}{\char92}sigma v{\char92}'+p, u-u{\char92}'){\char92}{\char92}): "} : {\tt{}"(s*v{\char92}'+p, u-u{\char92}'): "})
  \begin{math}\ll\end{math} (({\it{}C11}.{\it{}get\_l}(1)+{\it{}sign}\begin{math}\ast\end{math}{\it{}C11}.{\it{}get\_k}()\begin{math}\ast\end{math}{\it{}v1}-({\it{}sign}\begin{math}\ast\end{math}{\it{}v1}+{\it{}p})).{\it{}normal}().{\it{}is\_zero}()
    \begin{math}\wedge\end{math} ({\it{}C11}.{\it{}get\_l}(0)-{\it{}C11}.{\it{}get\_k}()\begin{math}\ast\end{math}{\it{}u1}-({\it{}u}-{\it{}u1})).{\it{}normal}().{\it{}is\_zero}())
 \begin{math}\ll\end{math} {\it{}endl} \begin{math}\ll\end{math} {\it{}endl};
{\nwrbrace}

\nwused{\\{NW3gGP3e-1Z2pUX-1}}\nwidentuses{\\{{\nwixident{get{\_}k}}{get:unk}}\\{{\nwixident{get{\_}l}}{get:unl}}\\{{\nwixident{is{\_}zero}}{is:unzero}}\\{{\nwixident{normal}}{normal}}\\{{\nwixident{u}}{u}}\\{{\nwixident{v}}{v}}\\{{\nwixident{wspaces}}{wspaces}}}\nwindexuse{\nwixident{get{\_}k}}{get:unk}{NW3gGP3e-2gYzHL-4}\nwindexuse{\nwixident{get{\_}l}}{get:unl}{NW3gGP3e-2gYzHL-4}\nwindexuse{\nwixident{is{\_}zero}}{is:unzero}{NW3gGP3e-2gYzHL-4}\nwindexuse{\nwixident{normal}}{normal}{NW3gGP3e-2gYzHL-4}\nwindexuse{\nwixident{u}}{u}{NW3gGP3e-2gYzHL-4}\nwindexuse{\nwixident{v}}{v}{NW3gGP3e-2gYzHL-4}\nwindexuse{\nwixident{wspaces}}{wspaces}{NW3gGP3e-2gYzHL-4}\nwendcode{}\nwbegindocs{336}\nwdocspar
\subsection{Infinitesimal Cycles}
\label{sec:infinites-cycles}
The final bit of our calculation is related with the infinitesimal
radius cycles,
see~\cite[\S~\ref{E-sec:zero-length-cycles}]{Kisil05a}.

Some additional parameters.
\nwenddocs{}\nwbegincode{337}\sublabel{NW3gGP3e-3hvAAH-B}\nwmargintag{{\nwtagstyle{}\subpageref{NW3gGP3e-3hvAAH-B}}}\moddef{Declaration of variables~{\nwtagstyle{}\subpageref{NW3gGP3e-3hvAAH-1}}}\plusendmoddef\Rm{}\nwstartdeflinemarkup\nwusesondefline{\\{NW3gGP3e-1p0Y9w-3}}\nwprevnextdefs{NW3gGP3e-3hvAAH-A}{NW3gGP3e-3hvAAH-C}\nwenddeflinemarkup
{\it{}possymbol} {\it{}vp}({\tt{}"vp"},{\tt{}"v\_p"}); //the positive instance of symbol {\it{}v}
{\bf{}ex} {\it{}displ}; //displacement of the focus

\nwused{\\{NW3gGP3e-1p0Y9w-3}}\nwidentuses{\\{{\nwixident{ex}}{ex}}\\{{\nwixident{focus}}{focus}}\\{{\nwixident{v}}{v}}}\nwindexuse{\nwixident{ex}}{ex}{NW3gGP3e-3hvAAH-B}\nwindexuse{\nwixident{focus}}{focus}{NW3gGP3e-3hvAAH-B}\nwindexuse{\nwixident{v}}{v}{NW3gGP3e-3hvAAH-B}\nwendcode{}\nwbegindocs{338}\nwdocspar
\subsubsection{Basic properties of infinitesimal cycles}
\label{sec:basic-prop-infin}

@We define an infinitesimal cycle {\Tt{}\Rm{}{\it{}C10}\nwendquote} such that its squared radius
(\(\det\)) is an infinitesimal number \(\varepsilon^2\) and focus is
at \((u,v)\). This defined by the cycle \((1, u_0, n, u_0^2+2 n v_0
    -\rs n^2)\) where \(n\) satisfies to the equation
\begin{equation}
  \label{eq:inf-cycle-cond}
  (\rs-\bs)n^2-2v_0n+\varepsilon^2=0.
\end{equation}
Only one root of the quadratic case produces a cycle with an
infinitesimal focal length, and we consider it here:
\nwenddocs{}\nwbegincode{339}\sublabel{NW3gGP3e-2Xuors-1}\nwmargintag{{\nwtagstyle{}\subpageref{NW3gGP3e-2Xuors-1}}}\moddef{Infinitesimal cycle~{\nwtagstyle{}\subpageref{NW3gGP3e-2Xuors-1}}}\endmoddef\Rm{}\nwstartdeflinemarkup\nwusesondefline{\\{NW3gGP3e-1zkOI-4}}\nwprevnextdefs{\relax}{NW3gGP3e-2Xuors-2}\nwenddeflinemarkup
{\it{}infinitesimal\_calculations}({\it{}n}\begin{math}\equiv\end{math}({\it{}vp}-{\it{}sqrt}({\it{}pow}({\it{}vp},2)+{\it{}pow}({\it{}epsilon},2)\begin{math}\ast\end{math}({\it{}sign4}-{\it{}sign1})))\begin{math}\div\end{math}({\it{}sign4}-{\it{}sign1}));
//infinitesimal\_calculations(n==(vp-abs(pow(pow(vp,2)-pow(epsilon,2)*(sign4-sign1),half)))/(sign4-sign1));

\nwalsodefined{\\{NW3gGP3e-2Xuors-2}}\nwused{\\{NW3gGP3e-1zkOI-4}}\nwidentuses{\\{{\nwixident{infinitesimal{\_}calculations}}{infinitesimal:uncalculations}}}\nwindexuse{\nwixident{infinitesimal{\_}calculations}}{infinitesimal:uncalculations}{NW3gGP3e-2Xuors-1}\nwendcode{}\nwbegindocs{340}The second expression for an infinitesimal cycle for the case
\(\rs=\bs\) is given by the substitution
\(n=-\frac{\varepsilon^2}{2v}\), which the root
of~\eqref{eq:inf-cycle-cond} in this case.
\nwenddocs{}\nwbegincode{341}\sublabel{NW3gGP3e-2Xuors-2}\nwmargintag{{\nwtagstyle{}\subpageref{NW3gGP3e-2Xuors-2}}}\moddef{Infinitesimal cycle~{\nwtagstyle{}\subpageref{NW3gGP3e-2Xuors-1}}}\plusendmoddef\Rm{}\nwstartdeflinemarkup\nwusesondefline{\\{NW3gGP3e-1zkOI-4}}\nwprevnextdefs{NW3gGP3e-2Xuors-1}{\relax}\nwenddeflinemarkup
{\it{}infinitesimal\_calculations}({\bf{}lst}({\it{}n}\begin{math}\equiv\end{math}{\it{}pow}({\it{}epsilon},2)\begin{math}\div\end{math}2\begin{math}\div\end{math}{\it{}vp}, {\it{}sign4}\begin{math}\equiv\end{math}{\it{}sign1}));

\nwused{\\{NW3gGP3e-1zkOI-4}}\nwidentuses{\\{{\nwixident{infinitesimal{\_}calculations}}{infinitesimal:uncalculations}}\\{{\nwixident{lst}}{lst}}}\nwindexuse{\nwixident{infinitesimal{\_}calculations}}{infinitesimal:uncalculations}{NW3gGP3e-2Xuors-2}\nwindexuse{\nwixident{lst}}{lst}{NW3gGP3e-2Xuors-2}\nwendcode{}\nwbegindocs{342}We organise the infinitesimal cycles check as a separate subroutine
and start it from several local variables definition.
\nwenddocs{}\nwbegincode{343}\sublabel{NW3gGP3e-4WB4RW-1}\nwmargintag{{\nwtagstyle{}\subpageref{NW3gGP3e-4WB4RW-1}}}\moddef{Infinitesimal cycle calculations~{\nwtagstyle{}\subpageref{NW3gGP3e-4WB4RW-1}}}\endmoddef\Rm{}\nwstartdeflinemarkup\nwusesondefline{\\{NW3gGP3e-1Z2pUX-1}}\nwprevnextdefs{\relax}{NW3gGP3e-4WB4RW-2}\nwenddeflinemarkup
{\bf{}void} {\it{}infinitesimal\_calculations}({\bf{}const} {\bf{}ex} & {\it{}nval}) {\nwlbrace}\nwindexdefn{\nwixident{infinitesimal{\_}calculations}}{infinitesimal:uncalculations}{NW3gGP3e-4WB4RW-1}
{\it{}exmap} {\it{}smap};
{\it{}smap}[{\it{}v}]={\it{}vp};

\nwalsodefined{\\{NW3gGP3e-4WB4RW-2}\\{NW3gGP3e-4WB4RW-3}\\{NW3gGP3e-4WB4RW-4}\\{NW3gGP3e-4WB4RW-5}\\{NW3gGP3e-4WB4RW-6}\\{NW3gGP3e-4WB4RW-7}\\{NW3gGP3e-4WB4RW-8}\\{NW3gGP3e-4WB4RW-9}\\{NW3gGP3e-4WB4RW-A}}\nwused{\\{NW3gGP3e-1Z2pUX-1}}\nwidentdefs{\\{{\nwixident{infinitesimal{\_}calculations}}{infinitesimal:uncalculations}}}\nwidentuses{\\{{\nwixident{ex}}{ex}}\\{{\nwixident{v}}{v}}}\nwindexuse{\nwixident{ex}}{ex}{NW3gGP3e-4WB4RW-1}\nwindexuse{\nwixident{v}}{v}{NW3gGP3e-4WB4RW-1}\nwendcode{}\nwbegindocs{344}\nwdocspar
\nwenddocs{}\nwbegincode{345}\sublabel{NW3gGP3e-4WB4RW-2}\nwmargintag{{\nwtagstyle{}\subpageref{NW3gGP3e-4WB4RW-2}}}\moddef{Infinitesimal cycle calculations~{\nwtagstyle{}\subpageref{NW3gGP3e-4WB4RW-1}}}\plusendmoddef\Rm{}\nwstartdeflinemarkup\nwusesondefline{\\{NW3gGP3e-1Z2pUX-1}}\nwprevnextdefs{NW3gGP3e-4WB4RW-1}{NW3gGP3e-4WB4RW-3}\nwenddeflinemarkup
{\it{}C10} = {\bf{}cycle2D}(1, {\bf{}lst}({\it{}u}, {\it{}n}),  {\it{}pow}({\it{}u},2)-{\it{}pow}({\it{}n},2)\begin{math}\ast\end{math}{\it{}sign1}-{\it{}pow}({\it{}epsilon},2), {\it{}e}).{\it{}subs}({\it{}nval});
{\it{}cout} \begin{math}\ll\end{math} {\tt{}"Inf cycle is: "} {\it{}math\_string} \begin{math}\ll\end{math} {\it{}C10} {\it{}math\_string} \begin{math}\ll\end{math} {\it{}endl};
{\it{}cout} \begin{math}\ll\end{math} {\tt{}"Square of radius of the infinitesimal cycle is: "}
 {\it{}math\_string} \begin{math}\ll\end{math} {\it{}C10}.{\it{}radius\_sq}({\it{}es}).{\it{}subs}({\it{}signs\_cube}, {\it{}subs\_options}::{\it{}algebraic}
   \begin{math}\mid\end{math} {\it{}subs\_options}::{\it{}no\_pattern}).{\it{}normal}() {\it{}math\_string} \begin{math}\ll\end{math} {\it{}endl};

\nwused{\\{NW3gGP3e-1Z2pUX-1}}\nwidentuses{\\{{\nwixident{cycle}}{cycle}}\\{{\nwixident{cycle2D}}{cycle2D}}\\{{\nwixident{lst}}{lst}}\\{{\nwixident{math{\_}string}}{math:unstring}}\\{{\nwixident{normal}}{normal}}\\{{\nwixident{radius{\_}sq}}{radius:unsq}}\\{{\nwixident{subs}}{subs}}\\{{\nwixident{u}}{u}}}\nwindexuse{\nwixident{cycle}}{cycle}{NW3gGP3e-4WB4RW-2}\nwindexuse{\nwixident{cycle2D}}{cycle2D}{NW3gGP3e-4WB4RW-2}\nwindexuse{\nwixident{lst}}{lst}{NW3gGP3e-4WB4RW-2}\nwindexuse{\nwixident{math{\_}string}}{math:unstring}{NW3gGP3e-4WB4RW-2}\nwindexuse{\nwixident{normal}}{normal}{NW3gGP3e-4WB4RW-2}\nwindexuse{\nwixident{radius{\_}sq}}{radius:unsq}{NW3gGP3e-4WB4RW-2}\nwindexuse{\nwixident{subs}}{subs}{NW3gGP3e-4WB4RW-2}\nwindexuse{\nwixident{u}}{u}{NW3gGP3e-4WB4RW-2}\nwendcode{}\nwbegindocs{346}Then we verify that in parabolic space it focus is in the point
\((u,v)\) and the focal length is an infinitesimal.
\nwenddocs{}\nwbegincode{347}\sublabel{NW3gGP3e-4WB4RW-3}\nwmargintag{{\nwtagstyle{}\subpageref{NW3gGP3e-4WB4RW-3}}}\moddef{Infinitesimal cycle calculations~{\nwtagstyle{}\subpageref{NW3gGP3e-4WB4RW-1}}}\plusendmoddef\Rm{}\nwstartdeflinemarkup\nwusesondefline{\\{NW3gGP3e-1Z2pUX-1}}\nwprevnextdefs{NW3gGP3e-4WB4RW-2}{NW3gGP3e-4WB4RW-4}\nwenddeflinemarkup
{\it{}cout} \begin{math}\ll\end{math} {\tt{}"Focus of infinitesimal cycle is: "} {\it{}math\_string} \begin{math}\ll\end{math} {\it{}C10}.{\it{}focus}({\it{}e4}).{\it{}subs}({\it{}nval}) {\it{}math\_string} \begin{math}\ll\end{math} {\it{}endl}
 \begin{math}\ll\end{math} {\tt{}"Focal length is: "} {\it{}math\_string} \begin{math}\ll\end{math} {\it{}C10}.{\it{}focal\_length}().{\it{}series}({\it{}epsilon}\begin{math}\equiv\end{math}0,3).{\it{}normal}() {\it{}math\_string} \begin{math}\ll\end{math} {\it{}endl};

{\it{}cout} \begin{math}\ll\end{math} {\tt{}"Infinitesimal cycle passing points"} {\it{}math\_string} \begin{math}\ll\end{math} {\tt{}"(u+"} \begin{math}\ll\end{math} {\it{}epsilon}\begin{math}\ast\end{math}{\it{}x} \begin{math}\ll\end{math}{\tt{}", vp+"}
  \begin{math}\ll\end{math} {\it{}lsolve}({\it{}C10}.{\it{}subs}({\it{}sign}\begin{math}\equiv\end{math}0).{\it{}passing}({\bf{}lst}({\it{}u}+{\it{}epsilon}\begin{math}\ast\end{math}{\it{}x},{\it{}vp}+{\it{}y})),{\it{}y}).{\it{}series}({\it{}epsilon}\begin{math}\equiv\end{math}0,3).{\it{}normal}()
 \begin{math}\ll\end{math} {\tt{}"), "} {\it{}math\_string} \begin{math}\ll\end{math} {\it{}endl};

\nwused{\\{NW3gGP3e-1Z2pUX-1}}\nwidentuses{\\{{\nwixident{cycle}}{cycle}}\\{{\nwixident{focal{\_}length}}{focal:unlength}}\\{{\nwixident{focus}}{focus}}\\{{\nwixident{lst}}{lst}}\\{{\nwixident{math{\_}string}}{math:unstring}}\\{{\nwixident{normal}}{normal}}\\{{\nwixident{passing}}{passing}}\\{{\nwixident{points}}{points}}\\{{\nwixident{subs}}{subs}}\\{{\nwixident{u}}{u}}}\nwindexuse{\nwixident{cycle}}{cycle}{NW3gGP3e-4WB4RW-3}\nwindexuse{\nwixident{focal{\_}length}}{focal:unlength}{NW3gGP3e-4WB4RW-3}\nwindexuse{\nwixident{focus}}{focus}{NW3gGP3e-4WB4RW-3}\nwindexuse{\nwixident{lst}}{lst}{NW3gGP3e-4WB4RW-3}\nwindexuse{\nwixident{math{\_}string}}{math:unstring}{NW3gGP3e-4WB4RW-3}\nwindexuse{\nwixident{normal}}{normal}{NW3gGP3e-4WB4RW-3}\nwindexuse{\nwixident{passing}}{passing}{NW3gGP3e-4WB4RW-3}\nwindexuse{\nwixident{points}}{points}{NW3gGP3e-4WB4RW-3}\nwindexuse{\nwixident{subs}}{subs}{NW3gGP3e-4WB4RW-3}\nwindexuse{\nwixident{u}}{u}{NW3gGP3e-4WB4RW-3}\nwendcode{}\nwbegindocs{348}\nwdocspar
\subsubsection{M\"obius transformations of infinitesimal cycles }
\label{sec:mobi-transf-infin}

Now we check that transformation of an infinitesimal cycle is an
infinitesimal cycle again\ldots
\nwenddocs{}\nwbegincode{349}\sublabel{NW3gGP3e-4WB4RW-4}\nwmargintag{{\nwtagstyle{}\subpageref{NW3gGP3e-4WB4RW-4}}}\moddef{Infinitesimal cycle calculations~{\nwtagstyle{}\subpageref{NW3gGP3e-4WB4RW-1}}}\plusendmoddef\Rm{}\nwstartdeflinemarkup\nwusesondefline{\\{NW3gGP3e-1Z2pUX-1}}\nwprevnextdefs{NW3gGP3e-4WB4RW-3}{NW3gGP3e-4WB4RW-5}\nwenddeflinemarkup
{\it{}C11}={\it{}C10}.{\it{}sl2\_similarity}({\it{}a}, {\it{}b}, {\it{}c}, {\it{}d}, {\it{}es});
{\it{}cout} \begin{math}\ll\end{math} {\tt{}"Image under SL2(R) of infinitesimal cycle has radius squared: "} \begin{math}\ll\end{math} {\it{}endl}
 {\it{}math\_string} \begin{math}\ll\end{math} {\it{}C11}.{\it{}radius\_sq}({\it{}es}).{\it{}subs}({\it{}sl2\_relation1},
  {\it{}subs\_options}::{\it{}algebraic} \begin{math}\mid\end{math} {\it{}subs\_options}::{\it{}no\_pattern}).{\it{}subs}({\it{}signs\_cube},
  {\it{}subs\_options}::{\it{}algebraic} \begin{math}\mid\end{math} {\it{}subs\_options}::{\it{}no\_pattern}).{\it{}series}({\it{}epsilon}\begin{math}\equiv\end{math}0,3).{\it{}normal}()
 {\it{}math\_string} \begin{math}\ll\end{math} {\it{}endl}
 \begin{math}\ll\end{math} {\tt{}"Image under cycle similarity of infinitesimal cycle has radius squared: "} \begin{math}\ll\end{math} {\it{}endl}
 {\it{}math\_string} \begin{math}\ll\end{math} {\it{}C10}.{\it{}cycle\_similarity}({\it{}C}, {\it{}es}).{\it{}radius\_sq}({\it{}es}).{\it{}subs}({\it{}signs\_cube}, {\it{}subs\_options}::{\it{}algebraic}
  \begin{math}\mid\end{math} {\it{}subs\_options}::{\it{}no\_pattern}).{\it{}series}({\it{}epsilon}\begin{math}\equiv\end{math}0,3).{\it{}normal}() {\it{}math\_string} \begin{math}\ll\end{math} {\it{}endl};

\nwused{\\{NW3gGP3e-1Z2pUX-1}}\nwidentuses{\\{{\nwixident{cycle}}{cycle}}\\{{\nwixident{cycle{\_}similarity}}{cycle:unsimilarity}}\\{{\nwixident{math{\_}string}}{math:unstring}}\\{{\nwixident{normal}}{normal}}\\{{\nwixident{radius{\_}sq}}{radius:unsq}}\\{{\nwixident{sl2{\_}similarity}}{sl2:unsimilarity}}\\{{\nwixident{subs}}{subs}}}\nwindexuse{\nwixident{cycle}}{cycle}{NW3gGP3e-4WB4RW-4}\nwindexuse{\nwixident{cycle{\_}similarity}}{cycle:unsimilarity}{NW3gGP3e-4WB4RW-4}\nwindexuse{\nwixident{math{\_}string}}{math:unstring}{NW3gGP3e-4WB4RW-4}\nwindexuse{\nwixident{normal}}{normal}{NW3gGP3e-4WB4RW-4}\nwindexuse{\nwixident{radius{\_}sq}}{radius:unsq}{NW3gGP3e-4WB4RW-4}\nwindexuse{\nwixident{sl2{\_}similarity}}{sl2:unsimilarity}{NW3gGP3e-4WB4RW-4}\nwindexuse{\nwixident{subs}}{subs}{NW3gGP3e-4WB4RW-4}\nwendcode{}\nwbegindocs{350}\ldots and focus of the transformed cycle is (up to infinitesimals)
obtained from the focus of initial cycle by the same transformation.
\nwenddocs{}\nwbegincode{351}\sublabel{NW3gGP3e-4WB4RW-5}\nwmargintag{{\nwtagstyle{}\subpageref{NW3gGP3e-4WB4RW-5}}}\moddef{Infinitesimal cycle calculations~{\nwtagstyle{}\subpageref{NW3gGP3e-4WB4RW-1}}}\plusendmoddef\Rm{}\nwstartdeflinemarkup\nwusesondefline{\\{NW3gGP3e-1Z2pUX-1}}\nwprevnextdefs{NW3gGP3e-4WB4RW-4}{NW3gGP3e-4WB4RW-6}\nwenddeflinemarkup
{\it{}displ} = ({\it{}C11}.{\it{}focus}({\it{}e4}, {\bf{}true}).{\it{}subs}({\it{}nval}) - {\it{}gW}.{\it{}subs}({\it{}smap}, {\it{}subs\_options}::{\it{}no\_pattern})).{\it{}evalm}();
{\it{}cout} \begin{math}\ll\end{math} {\tt{}"Focus of the transormed cycle is from transformation of focus by: "}
  {\it{}math\_string} \begin{math}\ll\end{math} {\it{}displ}.{\it{}subs}({\it{}sl2\_relation}, {\it{}subs\_options}::{\it{}algebraic}
       \begin{math}\mid\end{math} {\it{}subs\_options}::{\it{}no\_pattern}).{\it{}subs}({\bf{}lst}({\it{}sign}\begin{math}\equiv\end{math}0,{\it{}a}\begin{math}\equiv\end{math}(1+{\it{}b}\begin{math}\ast\end{math}{\it{}c})\begin{math}\div\end{math}{\it{}d})).{\it{}series}({\it{}epsilon}\begin{math}\equiv\end{math}0,2).{\it{}normal}()
 {\it{}math\_string} \begin{math}\ll\end{math} {\it{}endl};

\nwused{\\{NW3gGP3e-1Z2pUX-1}}\nwidentuses{\\{{\nwixident{cycle}}{cycle}}\\{{\nwixident{focus}}{focus}}\\{{\nwixident{lst}}{lst}}\\{{\nwixident{math{\_}string}}{math:unstring}}\\{{\nwixident{normal}}{normal}}\\{{\nwixident{subs}}{subs}}}\nwindexuse{\nwixident{cycle}}{cycle}{NW3gGP3e-4WB4RW-5}\nwindexuse{\nwixident{focus}}{focus}{NW3gGP3e-4WB4RW-5}\nwindexuse{\nwixident{lst}}{lst}{NW3gGP3e-4WB4RW-5}\nwindexuse{\nwixident{math{\_}string}}{math:unstring}{NW3gGP3e-4WB4RW-5}\nwindexuse{\nwixident{normal}}{normal}{NW3gGP3e-4WB4RW-5}\nwindexuse{\nwixident{subs}}{subs}{NW3gGP3e-4WB4RW-5}\nwendcode{}\nwbegindocs{352}\nwdocspar
\subsubsection{Orthogonality with infinitesimal cycles}
\label{sec:orth-with-infin}

We also find expressions for the orthogonality (see
\S~\ref{sec:orthogonality-cycles}) with the infinitesimal radius
cycle.
\nwenddocs{}\nwbegincode{353}\sublabel{NW3gGP3e-4WB4RW-6}\nwmargintag{{\nwtagstyle{}\subpageref{NW3gGP3e-4WB4RW-6}}}\moddef{Infinitesimal cycle calculations~{\nwtagstyle{}\subpageref{NW3gGP3e-4WB4RW-1}}}\plusendmoddef\Rm{}\nwstartdeflinemarkup\nwusesondefline{\\{NW3gGP3e-1Z2pUX-1}}\nwprevnextdefs{NW3gGP3e-4WB4RW-5}{NW3gGP3e-4WB4RW-7}\nwenddeflinemarkup
{\it{}cout} \begin{math}\ll\end{math} {\tt{}"Orthogonality (leading term) to infinitesimal cycle is:"} \begin{math}\ll\end{math} {\it{}endl} \begin{math}\ll\end{math} {\it{}wspaces}
  {\it{}math\_string} \begin{math}\ll\end{math} {\bf{}ex}({\it{}C}.{\it{}is\_orthogonal}({\it{}C10}, {\it{}es})).{\it{}series}({\it{}epsilon}\begin{math}\equiv\end{math}0,1).{\it{}normal}() {\it{}math\_string} \begin{math}\ll\end{math} {\it{}endl};

\nwused{\\{NW3gGP3e-1Z2pUX-1}}\nwidentuses{\\{{\nwixident{cycle}}{cycle}}\\{{\nwixident{ex}}{ex}}\\{{\nwixident{is{\_}orthogonal}}{is:unorthogonal}}\\{{\nwixident{math{\_}string}}{math:unstring}}\\{{\nwixident{normal}}{normal}}\\{{\nwixident{wspaces}}{wspaces}}}\nwindexuse{\nwixident{cycle}}{cycle}{NW3gGP3e-4WB4RW-6}\nwindexuse{\nwixident{ex}}{ex}{NW3gGP3e-4WB4RW-6}\nwindexuse{\nwixident{is{\_}orthogonal}}{is:unorthogonal}{NW3gGP3e-4WB4RW-6}\nwindexuse{\nwixident{math{\_}string}}{math:unstring}{NW3gGP3e-4WB4RW-6}\nwindexuse{\nwixident{normal}}{normal}{NW3gGP3e-4WB4RW-6}\nwindexuse{\nwixident{wspaces}}{wspaces}{NW3gGP3e-4WB4RW-6}\nwendcode{}\nwbegindocs{354}And the both expressions for the f-orthogonality (see
\S~\ref{sec:focal-orth-1}) conditions with the infinitesimal
radius cycle.  The second relation verifies the
Lem.~\ref{E-le:infinitesimal-ortho} from~\cite{Kisil05a}.
\nwenddocs{}\nwbegincode{355}\sublabel{NW3gGP3e-4WB4RW-7}\nwmargintag{{\nwtagstyle{}\subpageref{NW3gGP3e-4WB4RW-7}}}\moddef{Infinitesimal cycle calculations~{\nwtagstyle{}\subpageref{NW3gGP3e-4WB4RW-1}}}\plusendmoddef\Rm{}\nwstartdeflinemarkup\nwusesondefline{\\{NW3gGP3e-1Z2pUX-1}}\nwprevnextdefs{NW3gGP3e-4WB4RW-6}{NW3gGP3e-4WB4RW-8}\nwenddeflinemarkup
{\it{}cout} \begin{math}\ll\end{math} {\tt{}"f-orthogonality of other cycle to infinitesimal:"} \begin{math}\ll\end{math} {\it{}endl} \begin{math}\ll\end{math} {\it{}wspaces}
  {\it{}math\_string} \begin{math}\ll\end{math} {\it{}C}.{\it{}is\_f\_orthogonal}({\it{}C10}, {\it{}es}).{\it{}series}({\it{}epsilon}\begin{math}\equiv\end{math}0,1).{\it{}normal}() {\it{}math\_string} \begin{math}\ll\end{math} {\it{}endl}
  \begin{math}\ll\end{math} {\tt{}"f-orthogonality of infinitesimal cycle to other:"} \begin{math}\ll\end{math} {\it{}endl} \begin{math}\ll\end{math} {\it{}wspaces}
  {\it{}math\_string} \begin{math}\ll\end{math} {\it{}C10}.{\it{}is\_f\_orthogonal}({\it{}C}, {\it{}es}).{\it{}series}({\it{}epsilon}\begin{math}\equiv\end{math}0,3).{\it{}normal}() {\it{}math\_string} \begin{math}\ll\end{math} {\it{}endl};

\nwused{\\{NW3gGP3e-1Z2pUX-1}}\nwidentuses{\\{{\nwixident{cycle}}{cycle}}\\{{\nwixident{is{\_}f{\_}orthogonal}}{is:unf:unorthogonal}}\\{{\nwixident{math{\_}string}}{math:unstring}}\\{{\nwixident{normal}}{normal}}\\{{\nwixident{wspaces}}{wspaces}}}\nwindexuse{\nwixident{cycle}}{cycle}{NW3gGP3e-4WB4RW-7}\nwindexuse{\nwixident{is{\_}f{\_}orthogonal}}{is:unf:unorthogonal}{NW3gGP3e-4WB4RW-7}\nwindexuse{\nwixident{math{\_}string}}{math:unstring}{NW3gGP3e-4WB4RW-7}\nwindexuse{\nwixident{normal}}{normal}{NW3gGP3e-4WB4RW-7}\nwindexuse{\nwixident{wspaces}}{wspaces}{NW3gGP3e-4WB4RW-7}\nwendcode{}\nwbegindocs{356}\nwdocspar
\subsubsection{Cayley transform of infinitesimal cycles}
\label{sec:cayl-transf-infin}

\nwenddocs{}\nwbegindocs{357}Here is two matrices which defines the Cayley transform and its inverses:
\nwenddocs{}\nwbegincode{358}\sublabel{NW3gGP3e-3hvAAH-C}\nwmargintag{{\nwtagstyle{}\subpageref{NW3gGP3e-3hvAAH-C}}}\moddef{Declaration of variables~{\nwtagstyle{}\subpageref{NW3gGP3e-3hvAAH-1}}}\plusendmoddef\Rm{}\nwstartdeflinemarkup\nwusesondefline{\\{NW3gGP3e-1p0Y9w-3}}\nwprevnextdefs{NW3gGP3e-3hvAAH-B}{\relax}\nwenddeflinemarkup
{\bf{}const} {\bf{}matrix} {\it{}TC}(2,2, {\bf{}lst}({\it{}dirac\_ONE}(), -{\it{}e}.{\it{}subs}({\it{}mu}\begin{math}\equiv\end{math}1), {\it{}sign1}\begin{math}\ast\end{math}{\it{}e}.{\it{}subs}({\it{}mu}\begin{math}\equiv\end{math}1), {\it{}dirac\_ONE}()));\nwindexdefn{\nwixident{matrix}}{matrix}{NW3gGP3e-3hvAAH-C}
// the inverse is TCI(2,2, lst(dirac\_ONE(), e.subs(mu==1), -sign1*e.subs(mu==1), dirac\_ONE()));

\nwused{\\{NW3gGP3e-1p0Y9w-3}}\nwidentdefs{\\{{\nwixident{matrix}}{matrix}}}\nwidentuses{\\{{\nwixident{lst}}{lst}}\\{{\nwixident{subs}}{subs}}}\nwindexuse{\nwixident{lst}}{lst}{NW3gGP3e-3hvAAH-C}\nwindexuse{\nwixident{subs}}{subs}{NW3gGP3e-3hvAAH-C}\nwendcode{}\nwbegindocs{359}\nwdocspar
We conclude with calculations of the parabolic Cayley
transform~\cite[\S~\ref{E-sec:cayl-transf-cycl}]{Kisil05a} on
infinitesimal radius cycles. The parabolic Cayley transform on cycles  is defined by the following transformation.
\nwenddocs{}\nwbegincode{360}\sublabel{NW3gGP3e-3vL8tu-1}\nwmargintag{{\nwtagstyle{}\subpageref{NW3gGP3e-3vL8tu-1}}}\moddef{Parabolic Cayley transform of cycles~{\nwtagstyle{}\subpageref{NW3gGP3e-3vL8tu-1}}}\endmoddef\Rm{}\nwstartdeflinemarkup\nwusesondefline{\\{NW3gGP3e-1Z2pUX-1}}\nwenddeflinemarkup
{\bf{}cycle2D} {\it{}cayley\_parab}({\bf{}const} {\bf{}cycle2D} & {\it{}C}, {\bf{}const} {\bf{}ex} & {\it{}sign} = -1)
{\nwlbrace}
 {\bf{}return} {\bf{}cycle2D}({\it{}C}.{\it{}get\_k}()-2\begin{math}\ast\end{math}{\it{}sign}\begin{math}\ast\end{math}{\it{}C}.{\it{}get\_l}(1), {\it{}C}.{\it{}get\_l}(), {\it{}C}.{\it{}get\_m}()-2\begin{math}\ast\end{math}{\it{}C}.{\it{}get\_l}(1), {\it{}C}.{\it{}get\_unit}());
{\nwrbrace}

\nwused{\\{NW3gGP3e-1Z2pUX-1}}\nwidentuses{\\{{\nwixident{cycle2D}}{cycle2D}}\\{{\nwixident{ex}}{ex}}\\{{\nwixident{get{\_}k}}{get:unk}}\\{{\nwixident{get{\_}l}}{get:unl}}\\{{\nwixident{get{\_}m}}{get:unm}}\\{{\nwixident{get{\_}unit}}{get:ununit}}}\nwindexuse{\nwixident{cycle2D}}{cycle2D}{NW3gGP3e-3vL8tu-1}\nwindexuse{\nwixident{ex}}{ex}{NW3gGP3e-3vL8tu-1}\nwindexuse{\nwixident{get{\_}k}}{get:unk}{NW3gGP3e-3vL8tu-1}\nwindexuse{\nwixident{get{\_}l}}{get:unl}{NW3gGP3e-3vL8tu-1}\nwindexuse{\nwixident{get{\_}m}}{get:unm}{NW3gGP3e-3vL8tu-1}\nwindexuse{\nwixident{get{\_}unit}}{get:ununit}{NW3gGP3e-3vL8tu-1}\nwendcode{}\nwbegindocs{361}The image of an infinitesimal cycle is another infinitesimal radius
cycle\ldots
\nwenddocs{}\nwbegincode{362}\sublabel{NW3gGP3e-4WB4RW-8}\nwmargintag{{\nwtagstyle{}\subpageref{NW3gGP3e-4WB4RW-8}}}\moddef{Infinitesimal cycle calculations~{\nwtagstyle{}\subpageref{NW3gGP3e-4WB4RW-1}}}\plusendmoddef\Rm{}\nwstartdeflinemarkup\nwusesondefline{\\{NW3gGP3e-1Z2pUX-1}}\nwprevnextdefs{NW3gGP3e-4WB4RW-7}{NW3gGP3e-4WB4RW-9}\nwenddeflinemarkup
{\it{}C11} = {\it{}cayley\_parab}({\it{}C10}, {\it{}sign1});
{\it{}cout} \begin{math}\ll\end{math} {\tt{}"Det of Cayley-transformed infinitesimal cycle: "}
  {\it{}math\_string} \begin{math}\ll\end{math} {\it{}C11}.{\it{}radius\_sq}({\it{}es}).{\it{}subs}({\bf{}lst}({\it{}sign} \begin{math}\equiv\end{math} 0),
   {\it{}subs\_options}::{\it{}algebraic} \begin{math}\mid\end{math} {\it{}subs\_options}::{\it{}no\_pattern}).{\it{}series}({\it{}epsilon}\begin{math}\equiv\end{math}0,3).{\it{}normal}()
  {\it{}math\_string} \begin{math}\ll\end{math} {\it{}endl};

\nwused{\\{NW3gGP3e-1Z2pUX-1}}\nwidentuses{\\{{\nwixident{cycle}}{cycle}}\\{{\nwixident{lst}}{lst}}\\{{\nwixident{math{\_}string}}{math:unstring}}\\{{\nwixident{normal}}{normal}}\\{{\nwixident{radius{\_}sq}}{radius:unsq}}\\{{\nwixident{subs}}{subs}}}\nwindexuse{\nwixident{cycle}}{cycle}{NW3gGP3e-4WB4RW-8}\nwindexuse{\nwixident{lst}}{lst}{NW3gGP3e-4WB4RW-8}\nwindexuse{\nwixident{math{\_}string}}{math:unstring}{NW3gGP3e-4WB4RW-8}\nwindexuse{\nwixident{normal}}{normal}{NW3gGP3e-4WB4RW-8}\nwindexuse{\nwixident{radius{\_}sq}}{radius:unsq}{NW3gGP3e-4WB4RW-8}\nwindexuse{\nwixident{subs}}{subs}{NW3gGP3e-4WB4RW-8}\nwendcode{}\nwbegindocs{363}\ldots with its focus mapped by the Cayley transform.
\nwenddocs{}\nwbegincode{364}\sublabel{NW3gGP3e-4WB4RW-9}\nwmargintag{{\nwtagstyle{}\subpageref{NW3gGP3e-4WB4RW-9}}}\moddef{Infinitesimal cycle calculations~{\nwtagstyle{}\subpageref{NW3gGP3e-4WB4RW-1}}}\plusendmoddef\Rm{}\nwstartdeflinemarkup\nwusesondefline{\\{NW3gGP3e-1Z2pUX-1}}\nwprevnextdefs{NW3gGP3e-4WB4RW-8}{NW3gGP3e-4WB4RW-A}\nwenddeflinemarkup
{\it{}displ} = ({\it{}C11}.{\it{}focus}({\it{}e4}, {\bf{}true}).{\it{}subs}({\it{}nval})
   - {\it{}clifford\_moebius\_map}({\it{}TC}, {\bf{}matrix}(2,1,{\bf{}lst}({\it{}u},{\it{}vp})), {\it{}e})).{\it{}evalm}().{\it{}normal}();
{\it{}cout} \begin{math}\ll\end{math} {\tt{}"Focus of the Cayley-transformed infinitesimal cycle displaced by: "} {\it{}math\_string} ;
{\bf{}try}{\nwlbrace}
 {\it{}cout} \begin{math}\ll\end{math} {\it{}displ}.{\it{}subs}({\bf{}lst}({\it{}sign} \begin{math}\equiv\end{math} 0),
        {\it{}subs\_options}::{\it{}algebraic} \begin{math}\mid\end{math} {\it{}subs\_options}::{\it{}no\_pattern}).{\it{}series}({\it{}epsilon}\begin{math}\equiv\end{math}0, 2).{\it{}normal}();
{\nwrbrace} {\bf{}catch} ({\it{}exception} &{\it{}p}) {\nwlbrace}\nwindexdefn{\nwixident{catch}}{catch}{NW3gGP3e-4WB4RW-9}
 {\it{}cout} \begin{math}\ll\end{math} {\tt{}"("} \begin{math}\ll\end{math} {\it{}displ}.{\it{}op}(0).{\it{}subs}({\bf{}lst}({\it{}sign} \begin{math}\equiv\end{math} 0),
           {\it{}subs\_options}::{\it{}algebraic} \begin{math}\mid\end{math} {\it{}subs\_options}::{\it{}no\_pattern}).{\it{}series}({\it{}epsilon}\begin{math}\equiv\end{math}0, 2).{\it{}normal}()
   \begin{math}\ll\end{math} {\tt{}", "} \begin{math}\ll\end{math} {\it{}displ}.{\it{}op}(1).{\it{}subs}({\bf{}lst}({\it{}sign} \begin{math}\equiv\end{math} 0),
         {\it{}subs\_options}::{\it{}algebraic} \begin{math}\mid\end{math} {\it{}subs\_options}::{\it{}no\_pattern}).{\it{}series}({\it{}epsilon}\begin{math}\equiv\end{math}0, 2).{\it{}normal}()
   \begin{math}\ll\end{math} {\tt{}")"};
{\nwrbrace}

\nwused{\\{NW3gGP3e-1Z2pUX-1}}\nwidentdefs{\\{{\nwixident{catch}}{catch}}}\nwidentuses{\\{{\nwixident{cycle}}{cycle}}\\{{\nwixident{focus}}{focus}}\\{{\nwixident{lst}}{lst}}\\{{\nwixident{math{\_}string}}{math:unstring}}\\{{\nwixident{matrix}}{matrix}}\\{{\nwixident{normal}}{normal}}\\{{\nwixident{op}}{op}}\\{{\nwixident{subs}}{subs}}\\{{\nwixident{u}}{u}}}\nwindexuse{\nwixident{cycle}}{cycle}{NW3gGP3e-4WB4RW-9}\nwindexuse{\nwixident{focus}}{focus}{NW3gGP3e-4WB4RW-9}\nwindexuse{\nwixident{lst}}{lst}{NW3gGP3e-4WB4RW-9}\nwindexuse{\nwixident{math{\_}string}}{math:unstring}{NW3gGP3e-4WB4RW-9}\nwindexuse{\nwixident{matrix}}{matrix}{NW3gGP3e-4WB4RW-9}\nwindexuse{\nwixident{normal}}{normal}{NW3gGP3e-4WB4RW-9}\nwindexuse{\nwixident{op}}{op}{NW3gGP3e-4WB4RW-9}\nwindexuse{\nwixident{subs}}{subs}{NW3gGP3e-4WB4RW-9}\nwindexuse{\nwixident{u}}{u}{NW3gGP3e-4WB4RW-9}\nwendcode{}\nwbegindocs{365}f-orthogonality of
\nwenddocs{}\nwbegincode{366}\sublabel{NW3gGP3e-4WB4RW-A}\nwmargintag{{\nwtagstyle{}\subpageref{NW3gGP3e-4WB4RW-A}}}\moddef{Infinitesimal cycle calculations~{\nwtagstyle{}\subpageref{NW3gGP3e-4WB4RW-1}}}\plusendmoddef\Rm{}\nwstartdeflinemarkup\nwusesondefline{\\{NW3gGP3e-1Z2pUX-1}}\nwprevnextdefs{NW3gGP3e-4WB4RW-9}{\relax}\nwenddeflinemarkup
{\it{}cout}  {\it{}math\_string} \begin{math}\ll\end{math} {\it{}endl}
 \begin{math}\ll\end{math} {\tt{}"f-orthogonality of Cayley transforms of infinitesimal cycle to other:"} \begin{math}\ll\end{math} {\it{}endl} \begin{math}\ll\end{math} {\it{}wspaces}
 {\it{}math\_string} \begin{math}\ll\end{math} {\it{}C11}.{\it{}is\_f\_orthogonal}({\it{}cayley\_parab}({\it{}C},{\it{}sign1}), {\it{}es}).{\it{}series}({\it{}epsilon}\begin{math}\equiv\end{math}0,3).{\it{}normal}()
 {\it{}math\_string} \begin{math}\ll\end{math} {\it{}endl} \begin{math}\ll\end{math} {\it{}endl};
{\nwrbrace}

\nwused{\\{NW3gGP3e-1Z2pUX-1}}\nwidentuses{\\{{\nwixident{cycle}}{cycle}}\\{{\nwixident{is{\_}f{\_}orthogonal}}{is:unf:unorthogonal}}\\{{\nwixident{math{\_}string}}{math:unstring}}\\{{\nwixident{normal}}{normal}}\\{{\nwixident{wspaces}}{wspaces}}}\nwindexuse{\nwixident{cycle}}{cycle}{NW3gGP3e-4WB4RW-A}\nwindexuse{\nwixident{is{\_}f{\_}orthogonal}}{is:unf:unorthogonal}{NW3gGP3e-4WB4RW-A}\nwindexuse{\nwixident{math{\_}string}}{math:unstring}{NW3gGP3e-4WB4RW-A}\nwindexuse{\nwixident{normal}}{normal}{NW3gGP3e-4WB4RW-A}\nwindexuse{\nwixident{wspaces}}{wspaces}{NW3gGP3e-4WB4RW-A}\nwendcode{}\nwbegindocs{367}\nwdocspar
\subsection[Drawing the Asymptote output]{Drawing the \Asymptote\ output}
\label{sec:draw-metap-outp}
Although we use every possibility above to make double and cross
checks one may still wish to see ``by his own eyes'' that the
all calculations are correct. This may be done as follows.

We draw some \Asymptote\ pictures which are included
in~\cite{Kisil05a}, see also Fig.~\ref{fig:example}. We start from
illustration of the both orthogonality relations, see
\S~\ref{sec:orthogonality-cycles} and~\ref{sec:focal-orth-1}.
They are done for nine (\(=3\times 3\)) possible combinations of
metrics (elliptic, parabolic and hyperbolic) for the space of points
and space of cycles.

\nwenddocs{}\nwbegincode{368}\sublabel{NW3gGP3e-Xmoi0-1}\nwmargintag{{\nwtagstyle{}\subpageref{NW3gGP3e-Xmoi0-1}}}\moddef{Draw Asymptote pictures~{\nwtagstyle{}\subpageref{NW3gGP3e-Xmoi0-1}}}\endmoddef\Rm{}\nwstartdeflinemarkup\nwusesondefline{\\{NW3gGP3e-1p0Y9w-5}}\nwprevnextdefs{\relax}{NW3gGP3e-Xmoi0-2}\nwenddeflinemarkup
{\it{}ofstream} {\it{}asymptote}({\tt{}"parab-ortho1.asy"});
{\it{}asymptote} \begin{math}\ll\end{math} {\it{}setprecision}(2);
{\bf{}for} ({\it{}si} = -1; {\it{}si} \begin{math}<\end{math} 2; {\it{}si}\protect\PP) {\nwlbrace}
 {\bf{}for} ({\it{}si1} = -1; {\it{}si1} \begin{math}<\end{math} 2; {\it{}si1}\protect\PP) {\nwlbrace}
  {\it{}sign\_val} = {\bf{}lst}({\it{}sign} \begin{math}\equiv\end{math} {\it{}si}, {\it{}sign1} \begin{math}\equiv\end{math} {\it{}si1});

\nwalsodefined{\\{NW3gGP3e-Xmoi0-2}\\{NW3gGP3e-Xmoi0-3}}\nwused{\\{NW3gGP3e-1p0Y9w-5}}\nwidentuses{\\{{\nwixident{lst}}{lst}}\\{{\nwixident{si}}{si}}\\{{\nwixident{si1}}{si1}}}\nwindexuse{\nwixident{lst}}{lst}{NW3gGP3e-Xmoi0-1}\nwindexuse{\nwixident{si}}{si}{NW3gGP3e-Xmoi0-1}\nwindexuse{\nwixident{si1}}{si1}{NW3gGP3e-Xmoi0-1}\nwendcode{}\nwbegindocs{369}For each of those combinations we produce pictures from the set of
data which is almost identical. This help to see the influence of
{\Tt{}\Rm{}{\it{}sign}\nwendquote} and {\Tt{}\Rm{}{\it{}sign1}\nwendquote} parameters with constant other ones. All those
graphics are mainly application of {\Tt{}\Rm{}{\it{}asy\_draw}()\nwendquote} method (see
\S~\ref{sec:two-dimens-cycl} mixed with some \Asymptote\ drawing
instructions. Since this is rather technical issue we put it
separately in Appendix~\ref{sec:deta-metap-draw}.
\nwenddocs{}\nwbegincode{370}\sublabel{NW3gGP3e-Xmoi0-2}\nwmargintag{{\nwtagstyle{}\subpageref{NW3gGP3e-Xmoi0-2}}}\moddef{Draw Asymptote pictures~{\nwtagstyle{}\subpageref{NW3gGP3e-Xmoi0-1}}}\plusendmoddef\Rm{}\nwstartdeflinemarkup\nwusesondefline{\\{NW3gGP3e-1p0Y9w-5}}\nwprevnextdefs{NW3gGP3e-Xmoi0-1}{NW3gGP3e-Xmoi0-3}\nwenddeflinemarkup
{\bf{}try} {\nwlbrace}
 {\nwlbrace}\LA{}Drawing first orthogonality~{\nwtagstyle{}\subpageref{NW3gGP3e-1VpO8V-1}}\RA{}{\nwrbrace}
 {\nwlbrace}\LA{}Drawing focal orthogonality~{\nwtagstyle{}\subpageref{NW3gGP3e-384dQC-1}}\RA{}{\nwrbrace}
{\nwrbrace} {\bf{}catch}  ({\it{}exception} &{\it{}p}) {\nwlbrace}\nwindexdefn{\nwixident{catch}}{catch}{NW3gGP3e-Xmoi0-2}
 {\it{}cerr} \begin{math}\ll\end{math} {\tt{}"*****       Got problem2: "} \begin{math}\ll\end{math}  {\it{}p}.{\it{}what}() \begin{math}\ll\end{math} {\it{}endl};
{\nwrbrace}
{\nwrbrace}
{\nwrbrace}

\nwused{\\{NW3gGP3e-1p0Y9w-5}}\nwidentdefs{\\{{\nwixident{catch}}{catch}}}\nwendcode{}\nwbegindocs{371}We finish the code with generation of some additional pictures for
the paper~\cite{Kisil05a}.
\nwenddocs{}\nwbegincode{372}\sublabel{NW3gGP3e-Xmoi0-3}\nwmargintag{{\nwtagstyle{}\subpageref{NW3gGP3e-Xmoi0-3}}}\moddef{Draw Asymptote pictures~{\nwtagstyle{}\subpageref{NW3gGP3e-Xmoi0-1}}}\plusendmoddef\Rm{}\nwstartdeflinemarkup\nwusesondefline{\\{NW3gGP3e-1p0Y9w-5}}\nwprevnextdefs{NW3gGP3e-Xmoi0-2}{\relax}\nwenddeflinemarkup
{\bf{}try} {\nwlbrace}
 \LA{}Extra pictures from Asymptote~{\nwtagstyle{}\subpageref{NW3gGP3e-3gefqu-1}}\RA{}
{\nwrbrace} {\bf{}catch}  ({\it{}exception} &{\it{}p}) {\nwlbrace}\nwindexdefn{\nwixident{catch}}{catch}{NW3gGP3e-Xmoi0-3}
 {\it{}cerr} \begin{math}\ll\end{math} {\tt{}"*****       Got problem3: "} \begin{math}\ll\end{math}  {\it{}p}.{\it{}what}() \begin{math}\ll\end{math} {\it{}endl};
{\nwrbrace}
{\it{}asymptote}.{\it{}close}();

\nwused{\\{NW3gGP3e-1p0Y9w-5}}\nwidentdefs{\\{{\nwixident{catch}}{catch}}}\nwendcode{}\nwbegindocs{373}\nwdocspar

\nwenddocs{}\nwbegindocs{374}\nwdocspar
\appendix

\nwenddocs{}\nwbegindocs{375}\nwdocspar
{
\let\oldsection=\section
\let\oldsubsection=\subsection
\let\chapter=\oldsection
\let\section=\oldsubsection
\let\subsection=\subsubsection

}
\nwenddocs{}\nwbegindocs{376}\nwdocspar

\nwenddocs{}\nwbegindocs{377}\nwdocspar
\section{Textual output of the program}
\label{sec:text-outp-progr}

{\obeylines
Conjugation of a cycle comes through Moebius transformation: true

A K-orbit is preserved: true, and  passing (0, t): true
\quad Determinant of zero-radius Z1 cycle in metric e is: $- \sigma v^{2}+ v^{2} \breve{\sigma}$
\quad Focus of zero-radius cycle is: ${u,-\frac{1}{2}  v \breve{\sigma}+\frac{1}{2}  \sigma v}$
\quad Centre of zero-radius cycle is: ${u,- \sigma v}$
\quad Focal length of zero-radius cycle is: $\frac{1}{2}  v$
Image of the zero-radius cycle under Moebius transform has zero radius: true
The centre of the Moebius transformed zero-radius cycle is: -equal-, -equal-
Image of the zero-radius cycle under cycle similarity has zero radius: true
The centre of the conjugated zero-radius cycle coinsides with Moebius tr: -equal-, -equal-

\quad The orthogonality is: $ \tilde{n} n \breve{\sigma}+\frac{1}{2}  \tilde{k} m- l \tilde{l}+\frac{1}{2}  \tilde{m} k==0$
\quad The orthogonality of two lines is: $ \tilde{n} n \breve{\sigma}- l \tilde{l}==0$
\quad The orthogonality to z-r-cycle is: $-\frac{1}{2}  \sigma v^{2} k+ v n \breve{\sigma}+\frac{1}{2} m+\frac{1}{2}  u^{2} k- l u==0$
\quad The orthogonality of two z-r-cycle is: $- u u'-\frac{1}{2}  \chi(\sigma_2) v'^{2}+\frac{1}{2} u^{2}-\frac{1}{2} u'^{2}+ v' v \breve{\sigma}-\frac{1}{2}  v^{2} \breve{\sigma}==0$
Both orthogonal cycles (through one point and through its inverse) are the same: true
Orthogonal cycle passes through the transformed point: true

Line through point and its inverse is orthogonal: true
All lines come through the point $(\frac{l}{k}, -\frac{ n \breve{\sigma}}{k})$
Conjugated vector is parallel to (u,v): true
C5 has common roots with C : true
$\chi(\sigma)$-centre of C5 is equal to $\breve{\sigma}$-centre of C: true
Inversion in (C5, sign)  coincides with inversion in (C, sign1): true
Inversion to the real line (with - sign): 
\quad Conjugation of the real line is the cycle C: true
\quad Conjugation of the cycle C is the real line: true
\quad Inversion cycle has common roots with C: true
\quad C passing the centre of inversion cycle: true
Inversion to the real line (with + sign): 
\quad Conjugation of the real line is the cycle C: true
\quad Conjugation of the cycle C is the real line: true
\quad Inversion cycle has common roots with C: true
\quad C passing the centre of inversion cycle: true
Yaglom inversion of the second kind is three reflections in the cycles: true

The real line is Moebius invariant: true
Reflection in the real line: $(1, {\left(\begin{array}{cc}u&- v\end{array}\right)}_{{symbol2640} }, - \sigma v^{2}+u^{2})$
Reflection of the real line in cycle C: 
$(2  \chi(\sigma_2) \chi(\sigma_3) n k \breve{\sigma}, {\left(\begin{array}{cc}2  l \chi(\sigma_2) \chi(\sigma_3) n \breve{\sigma}& \chi(\sigma_2) {( \chi(\sigma_2) n^{2} \breve{\sigma}- \chi(\sigma_2) m k+ l^{2} \chi(\sigma_2))}\end{array}\right)}_{{symbol2602} }, 2  \chi(\sigma_2) m \chi(\sigma_3) n \breve{\sigma})$
\quad The f-orthogonality is: $ \chi(\sigma_2) {( \chi(\sigma_2) n \tilde{m} k-2  l \chi(\sigma_2) \tilde{l} n+ \tilde{n} \chi(\sigma_2) n^{2} \breve{\sigma}- \tilde{n} \chi(\sigma_2) m k+ \tilde{n} l^{2} \chi(\sigma_2)+ \tilde{k} \chi(\sigma_2) m n)}==0$
\quad The f-orthogonality of two lines is: $- \chi(\sigma_2) {(2  l \chi(\sigma_2) \tilde{l} n- \tilde{n} \chi(\sigma_2) n^{2} \breve{\sigma}- \tilde{n} l^{2} \chi(\sigma_2))}==0$
\quad The f-orthogonality to z-r-cycle is first way: 
$ \chi(\sigma_2) {( l^{2} \chi(\sigma_2) v- \chi(\sigma_2) m v k- \chi(\sigma_2) v^{2} n k \breve{\sigma}+ \chi(\sigma_2) m n+ \chi(\sigma_2) v n^{2} \breve{\sigma}-2  l \chi(\sigma_2) u n+ \chi(\sigma_2) u^{2} n k)}==0$
\quad The f-orthogonality to z-r-cycle in second way: 
$- \chi(\sigma_2) {( \chi(\sigma_2) v^{3} k \breve{\sigma}-2  \chi(\sigma_2) v^{2} n \breve{\sigma}- \chi(\sigma_2) m v- \chi(\sigma_2) u^{2} v k+2  l \chi(\sigma_2) u v)}==0$
\quad The f-orthogonality of two z-r-cycle is: 
$- \chi(\sigma_2) {(2  \chi(\sigma_2) u v u'+ \chi(\sigma_2) v^{3} \breve{\sigma}-2  \chi(\sigma_2) v' v^{2} \breve{\sigma}- \chi(\sigma_2) u^{2} v+ \chi(\sigma_2) v'^{2} \sigma v- \chi(\sigma_2) v u'^{2})}==0$
All lines come through the focus related $\breve{e}$: true
C8 has common roots with C : true
$\chi(\sigma)$-center of C8  coincides with $\breve{\sigma}$-focus of C : true
f-inversion in C coincides with inversion in C8 : true

Distance between (u,v) and (u',v') in elliptic and hyperbolic spaces is 
\(\displaystyle \frac{ {( {(v'-v)}^{2} \sigma-{(u-u')}^{2})} {(4  v' {(-1+ \sigma \breve{\sigma})} v+ {( {(v'-v)}^{2} \sigma-{(u-u')}^{2})} \breve{\sigma})}}{ {(u-u')}^{2} \breve{\sigma}-{(v'-v)}^{2}}\): true
\quad Conformity in a cycle space with metric:   E      P      H 
\quad Point space is Elliptic case (sign = -1):  true false false
\quad Point space is Hyperbolic case (sign = 1):  false false true
\quad Perpendicular to ((u,v); (u',v')) is: $\frac{1}{2} \frac{3  v'^{2} \sigma v- v u'^{2}- u^{2} v+2  v' u^{2} \sigma \breve{\sigma}+ \sigma v^{3}+2  v' \sigma \breve{\sigma} u'^{2}-4  v' u \sigma \breve{\sigma} u'-3  v' \sigma v^{2}+2  u v u'- v' u'^{2}- v' u^{2}+2  v' u u'- v'^{3} \sigma}{2  v' v-v^{2}-2  u \breve{\sigma} u'+ \breve{\sigma} u'^{2}+ u^{2} \breve{\sigma}-v'^{2}}$; $\frac{1}{2} \frac{3  u \breve{\sigma} u'^{2}- u \sigma v^{2} \breve{\sigma}+ v'^{2} u \sigma \breve{\sigma}+2  v'^{2} u'+2  v' u v-2  v'^{2} u- v'^{2} \sigma \breve{\sigma} u'+ \sigma v^{2} \breve{\sigma} u'- \breve{\sigma} u'^{3}-3  u^{2} \breve{\sigma} u'-2  v' v u'+ u^{3} \breve{\sigma}}{2  v' v-v^{2}-2  u \breve{\sigma} u'+ \breve{\sigma} u'^{2}+ u^{2} \breve{\sigma}-v'^{2}}$

Value at the middle point (parabolic point space):
\quad $-2  u u'+u^{2}+u'^{2}$
\quad Conformity in a cycle space with metric:   E      P      H 
\quad Point space is Parabolic case (sign = 0):  true true true
\quad Perpendicular to ((u,v); (u',v')) is: $ v' \sigma$; $\frac{1}{2} u-\frac{1}{2} u'$

Distance between (u,v) and (u',v): 
\quad Value at critical point:
\quad $-\frac{4  \sigma v^{2} \breve{\sigma}-4 v^{2}+2  u \breve{\sigma} u'- \breve{\sigma} u'^{2}- u^{2} \breve{\sigma}}{\breve{\sigma}}$

Length from *center* between (u,v) and (u',v'):
\quad $\frac{ \mathring{\sigma}^{2} u'^{2}- v'^{2} \sigma \mathring{\sigma}^{2}+2  v' v \mathring{\sigma}+ u^{2} \mathring{\sigma}^{2}-2  u \mathring{\sigma}^{2} u'- v^{2} \breve{\sigma}}{\mathring{\sigma}^{2}}$
\quad This distance/length is conformal: true
\quad Perpendicular to ((u,v); (u',v')) is: $\frac{ v' \sigma \mathring{\sigma}-v}{\mathring{\sigma}}$; $u-u'$

Length from *focus* check for $p = - {(v'-v-\sqrt{ {(u-u')}^{2} \mathring{\sigma}+{(v'-v)}^{2}- v'^{2} \sigma \mathring{\sigma}})} \mathring{\sigma}$
\quad Length between (u,v) and (u', v') is equal to \((\mathring(\sigma)-\breve{\sigma})p^2-2vp\): true
\quad checks: C11 passes through (u', v'): true; C11 focus is at (u, v): true
\quad This distance/length is conformal: true
\quad Perpendicular to ((u,v); (u',v')) is \((\sigma v'+p, u-u')\): true

Length from *focus* check for $p = - \mathring{\sigma} {(v'-v+\sqrt{ {(u-u')}^{2} \mathring{\sigma}+{(v'-v)}^{2}- v'^{2} \sigma \mathring{\sigma}})}$
\quad Length between (u,v) and (u', v') is equal to \((\mathring(\sigma)-\breve{\sigma})p^2-2vp\): true
\quad checks: C11 passes through (u', v'): true; C11 focus is at (u, v): true
\quad This distance/length is conformal: true
\quad Perpendicular to ((u,v); (u',v')) is \((\sigma v'+p, u-u')\): true

Length from *focus* check for $p = \frac{1}{2} \frac{{(u-u')}^{2}- v'^{2} \sigma}{v'-v}$
\quad Length between (u,v) and (u', v') is equal to \((\mathring(\sigma)-\breve{\sigma})p^2-2vp\): true
\quad checks: C11 passes through (u', v'): true; C11 focus is at (u, v): true
\quad This distance/length is conformal: false. The factor is: 
\quad $\frac{y^{2}}{{( d^{2} y+2  d u y c-2  d x v c+ u^{2} y c^{2}+ \sigma v^{2} y c^{2}-2  u x v c^{2})}^{2}}$
\quad Perpendicular to ((u,v); (u',v')) is \((\sigma v'+p, u-u')\): true

Inf cycle is: $(1, {\left(\begin{array}{cc}u&-\frac{\sqrt{- \epsilon^{2} {(\mathring{\sigma}-\breve{\sigma})}+v_p^{2}}-v_p}{\mathring{\sigma}-\breve{\sigma}}\end{array}\right)}^{{symbol3675} }, -\frac{ {(\sqrt{- \epsilon^{2} {(\mathring{\sigma}-\breve{\sigma})}+v_p^{2}}-v_p)}^{2} \mathring{\sigma}}{{(\mathring{\sigma}-\breve{\sigma})}^{2}}+u^{2}-2 \frac{ v_p {(\sqrt{- \epsilon^{2} {(\mathring{\sigma}-\breve{\sigma})}+v_p^{2}}-v_p)}}{\mathring{\sigma}-\breve{\sigma}})$
Square of radius of the infinitesimal cycle is: $- \epsilon^{2}$
Focus of infinitesimal cycle is: ${u,v_p}$
Focal length is: ${(\frac{1}{4} \frac{1}{v_p})} \epsilon^{2}+\mathcal{O}(\epsilon^{3})$
Infinitesimal cycle passing points$(u+ x \epsilon, vp+{( v_p x^{2})}+{(\frac{1}{4} \frac{ x^{2} \breve{\sigma}- x^{2} \mathring{\sigma}-\mathring{\sigma}}{v_p})} \epsilon^{2}+\mathcal{O}(\epsilon^{3})), $
Image under SL2(R) of infinitesimal cycle has radius squared: 
${(-\frac{\mathring{\sigma}^{4}+6  \mathring{\sigma}^{2} \breve{\sigma}^{2}-4  \mathring{\sigma}^{3} \breve{\sigma}-4  \mathring{\sigma} \breve{\sigma}+\breve{\sigma}^{2}}{{(4  d u \mathring{\sigma} \breve{\sigma} c- u^{2} \breve{\sigma}^{2} c^{2}- d^{2} \breve{\sigma}^{2}- u^{2} \mathring{\sigma}^{2} c^{2}- d^{2} \mathring{\sigma}^{2}+2  u^{2} \mathring{\sigma} \breve{\sigma} c^{2}+2  d^{2} \mathring{\sigma} \breve{\sigma}-2  d u \breve{\sigma}^{2} c-2  d u \mathring{\sigma}^{2} c)}^{2}})} \epsilon^{2}+\mathcal{O}(\epsilon^{3})$
Image under cycle similarity of infinitesimal cycle has radius squared: 
${(\frac{2  l^{2} n^{2} \mathring{\sigma}^{4} \breve{\sigma}+4  l^{4} \mathring{\sigma} \breve{\sigma}- n^{4} \breve{\sigma}^{2}-6  m^{2} \mathring{\sigma}^{2} k^{2} \breve{\sigma}^{2}+4  n^{4} \mathring{\sigma} \breve{\sigma}- n^{4} \mathring{\sigma}^{4} \breve{\sigma}^{2}+8  m n^{2} \mathring{\sigma} k \breve{\sigma}^{2}-8  l^{2} n^{2} \mathring{\sigma} \breve{\sigma}^{2}+2  l^{2} m k \breve{\sigma}^{2}+12  l^{2} m \mathring{\sigma}^{2} k \breve{\sigma}^{2}+8  m n^{2} \mathring{\sigma}^{3} k \breve{\sigma}^{2}+2  l^{2} n^{2} \breve{\sigma}- l^{4} \mathring{\sigma}^{4}+2  l^{2} m \mathring{\sigma}^{4} k- m^{2} k^{2} \breve{\sigma}^{2}- l^{4} \breve{\sigma}^{2}-12  m n^{2} \mathring{\sigma}^{2} k \breve{\sigma}-8  l^{2} n^{2} \mathring{\sigma}^{3} \breve{\sigma}^{2}-8  l^{2} m \mathring{\sigma}^{3} k \breve{\sigma}-6  n^{4} \mathring{\sigma}^{2} \breve{\sigma}^{2}-6  l^{4} \mathring{\sigma}^{2} \breve{\sigma}^{2}+4  n^{4} \mathring{\sigma}^{3} \breve{\sigma}+4  m^{2} \mathring{\sigma} k^{2} \breve{\sigma}+12  l^{2} n^{2} \mathring{\sigma}^{2} \breve{\sigma}+4  l^{4} \mathring{\sigma}^{3} \breve{\sigma}-2  m n^{2} \mathring{\sigma}^{4} k \breve{\sigma}+4  m^{2} \mathring{\sigma}^{3} k^{2} \breve{\sigma}- m^{2} \mathring{\sigma}^{4} k^{2}-2  m n^{2} k \breve{\sigma}-8  l^{2} m \mathring{\sigma} k \breve{\sigma}}{{(2  l u \mathring{\sigma}^{2} k+2  u^{2} \mathring{\sigma} k^{2} \breve{\sigma}- l^{2} \mathring{\sigma}^{2}- u^{2} \mathring{\sigma}^{2} k^{2}- l^{2} \breve{\sigma}^{2}- u^{2} k^{2} \breve{\sigma}^{2}-2  n^{2} \mathring{\sigma} \breve{\sigma}^{2}+2  l^{2} \mathring{\sigma} \breve{\sigma}-4  l u \mathring{\sigma} k \breve{\sigma}+2  l u k \breve{\sigma}^{2}+ n^{2} \mathring{\sigma}^{2} \breve{\sigma}+ n^{2} \breve{\sigma})}^{2}})} \epsilon^{2}+\mathcal{O}(\epsilon^{3})$
Focus of the transormed cycle is from transformation of focus by: ${(\left(\begin{array}{c}0\\0\end{array}\right))}+{(\left(\begin{array}{c}0\\0\end{array}\right))} \epsilon+\mathcal{O}(\epsilon^{2})$
Orthogonality (leading term) to infinitesimal cycle is:
\quad ${(\frac{1}{2} m+\frac{1}{2}  u^{2} k- l u==0)}+\mathcal{O}(\epsilon)$
f-orthogonality of other cycle to infinitesimal:
\quad ${( u^{2} n k+ m n-2  l u n==0)}+\mathcal{O}(\epsilon)$
f-orthogonality of infinitesimal cycle to other:
\quad ${(0==0)}+{(0==0)} \epsilon+{(\frac{1}{2}  (\frac{m+ u^{2} k-2  l u-2  v_p n}{v_p}==0))} \epsilon^{2}+\mathcal{O}(\epsilon^{3})$
Det of Cayley-transformed infinitesimal cycle: ${(-\frac{-1+v_p- u^{2} \breve{\sigma}}{v_p})} \epsilon^{2}+\mathcal{O}(\epsilon^{3})$
Focus of the Cayley-transformed infinitesimal cycle displaced by: $(\mathcal{O}(\epsilon^{2}), \mathcal{O}(\epsilon^{2}))$
f-orthogonality of Cayley transforms of infinitesimal cycle to other:
\quad ${(0==0)}+{(0==0)} \epsilon+{(\frac{1}{2}  (\frac{m+ u^{2} k-2  l u-2  v_p n}{v_p}==0))} \epsilon^{2}+\mathcal{O}(\epsilon^{3})$

Inf cycle is: $(1, {\left(\begin{array}{cc}u&\frac{1}{2} \frac{\epsilon^{2}}{v_p}\end{array}\right)}^{{symbol9532} }, -\frac{1}{4} \frac{ \epsilon^{4} \breve{\sigma}}{v_p^{2}}+\epsilon^{2}+u^{2})$
Square of radius of the infinitesimal cycle is: $- \epsilon^{2}$
Focus of infinitesimal cycle is: ${u,v_p}$
Focal length is: ${(\frac{1}{4} \frac{1}{v_p})} \epsilon^{2}$
Infinitesimal cycle passing points$(u+ x \epsilon, vp+{( v_p x^{2})}+{(-\frac{1}{4} \frac{\breve{\sigma}}{v_p})} \epsilon^{2}), $
Image under SL2(R) of infinitesimal cycle has radius squared: 
${(-\frac{1}{{(2  d u c+ u^{2} c^{2}+d^{2})}^{2}})} \epsilon^{2}+\mathcal{O}(\epsilon^{3})$
Image under cycle similarity of infinitesimal cycle has radius squared: 
${(-\frac{ n^{4} \breve{\sigma}^{2}-2  l^{2} n^{2} \breve{\sigma}+l^{4}+ m^{2} k^{2}-2  l^{2} m k+2  m n^{2} k \breve{\sigma}}{{(2  l u k-l^{2}+ n^{2} \breve{\sigma}- u^{2} k^{2})}^{2}})} \epsilon^{2}+\mathcal{O}(\epsilon^{3})$
Focus of the transormed cycle is from transformation of focus by: ${(\left(\begin{array}{c}0\\0\end{array}\right))}+{(\left(\begin{array}{c}0\\0\end{array}\right))} \epsilon+\mathcal{O}(\epsilon^{2})$
Orthogonality (leading term) to infinitesimal cycle is:
\quad ${(\frac{1}{2} m+\frac{1}{2}  u^{2} k- l u==0)}+\mathcal{O}(\epsilon)$
f-orthogonality of other cycle to infinitesimal:
\quad ${( u^{2} n k+ m n-2  l u n==0)}+\mathcal{O}(\epsilon)$
f-orthogonality of infinitesimal cycle to other:
\quad ${(0==0)}+{(0==0)} \epsilon+{(\frac{1}{2}  (\frac{m+ u^{2} k-2  l u-2  v_p n}{v_p}==0))} \epsilon^{2}+\mathcal{O}(\epsilon^{3})$
Det of Cayley-transformed infinitesimal cycle: ${(-\frac{-1+v_p- u^{2} \breve{\sigma}}{v_p})} \epsilon^{2}+\mathcal{O}(\epsilon^{3})$
Focus of the Cayley-transformed infinitesimal cycle displaced by: ${(\left(\begin{array}{c}0\\0\end{array}\right))}+{(\left(\begin{array}{c}0\\0\end{array}\right))} \epsilon+\mathcal{O}(\epsilon^{2})$
f-orthogonality of Cayley transforms of infinitesimal cycle to other:
\quad ${(0==0)}+{(0==0)} \epsilon+{(\frac{1}{2}  (\frac{m+ u^{2} k-2  l u-2  v_p n}{v_p}==0))} \epsilon^{2}+\mathcal{O}(\epsilon^{3})$

}

\nwenddocs{}\nwbegindocs{378}\nwdocspar
\nwenddocs{}\nwbegindocs{379}\nwdocspar
\section{Example of the produced graphics}
\label{sec:example-prod-graph}
\begin{figure}[htbp]
  \includegraphics[scale=.9]{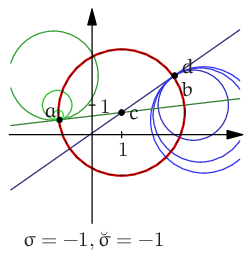}\hfill
  \includegraphics[scale=.9]{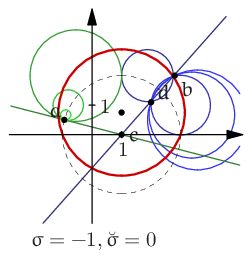}\hfill
  \includegraphics[scale=.9]{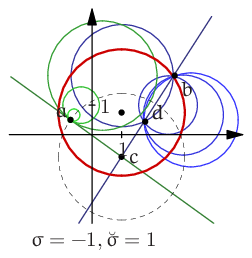}\\[4mm]
  \includegraphics[scale=.9]{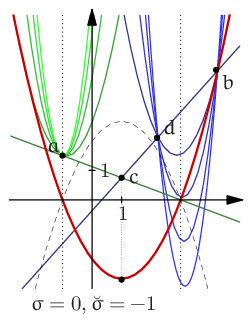}\hfill
  \includegraphics[scale=.9]{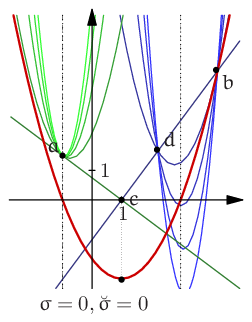}\hfill
  \includegraphics[scale=.9]{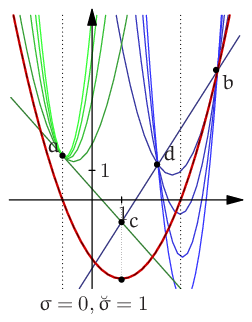}\\[4mm]
  \includegraphics[scale=.9]{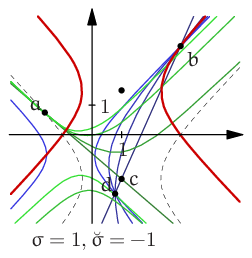}\hfill
  \includegraphics[scale=.9]{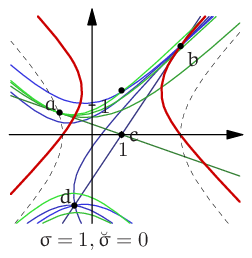}\hfill
  \includegraphics[scale=.9]{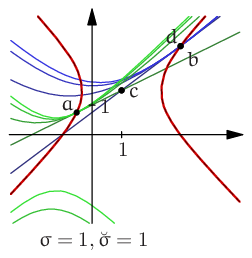}
  \caption[Orthogonality of the first kind]{Orthogonality of the first
    kind in nine combinations.}
\label{fig:example}
\end{figure}

An example of graphics generated by the program is given in
Figure~\ref{fig:example}.
This was produced by the part of program
from the Section~\ref{sec:first-orth-cond}.

\nwenddocs{}\nwbegindocs{380}\nwdocspar

\nwenddocs{}\nwbegindocs{381}\nwdocspar
\section[Details of the Asymptote Drawing]{Details of the \Asymptote\ Drawing}
\label{sec:deta-metap-draw}

\subsection{Drawing Orthogonality Conditions}
\label{sec:draw-orth-cond}

\subsubsection{First Orthogonality Condition}
\label{sec:first-orth-cond}

We define numeric values of all involved parameters first.
\nwenddocs{}\nwbegincode{382}\sublabel{NW3gGP3e-1VpO8V-1}\nwmargintag{{\nwtagstyle{}\subpageref{NW3gGP3e-1VpO8V-1}}}\moddef{Drawing first orthogonality~{\nwtagstyle{}\subpageref{NW3gGP3e-1VpO8V-1}}}\endmoddef\Rm{}\nwstartdeflinemarkup\nwusesondefline{\\{NW3gGP3e-Xmoi0-2}}\nwenddeflinemarkup
{\bf{}numeric} {\it{}xmin}(-11,4), {\it{}xmax}(5), {\it{}ymin}(-3), {\it{}ymax} = ({\it{}si} \begin{math}\equiv\end{math} 0?{\bf{}numeric}(25, 4): 4);
{\bf{}lst} {\it{}cycle\_val} = {\bf{}lst}({\it{}sign} \begin{math}\equiv\end{math} {\bf{}numeric}({\it{}si}), {\it{}sign1} \begin{math}\equiv\end{math} {\bf{}numeric}({\it{}si1}),
     {\it{}k} \begin{math}\equiv\end{math} {\bf{}numeric}(2,3), {\it{}l} \begin{math}\equiv\end{math} {\bf{}numeric}(2,3), {\it{}n} \begin{math}\equiv\end{math} ({\it{}si} \begin{math}\equiv\end{math} 1?{\bf{}numeric}(-1):{\bf{}numeric}(1,2)), {\it{}m} \begin{math}\equiv\end{math}{\bf{}numeric}(-2));
{\bf{}cycle2D} {\it{}Cf} = {\it{}C}.{\it{}subs}({\it{}cycle\_val}), {\it{}Cg} = {\it{}C5}.{\it{}subs}({\it{}cycle\_val}), {\it{}Cp} ={\it{}C2};
{\bf{}lst} {\it{}U}, {\it{}V};
{\bf{}switch} ({\it{}si}) {\nwlbrace}
{\bf{}case} -1: // points b, a, center, c, d
 {\it{}U} = {\bf{}numeric}(11,4), {\it{}Cg}.{\it{}roots}({\it{}half}).{\it{}op}(0), {\it{}Cf}.{\it{}center}().{\it{}op}(0).{\it{}subs}({\it{}cycle\_val}), ({\it{}l}\begin{math}\div\end{math}{\it{}k}).{\it{}subs}({\it{}cycle\_val});
 {\it{}V} = {\it{}Cf}.{\it{}roots}({\it{}U}.{\it{}op}(0), {\bf{}false}).{\it{}op}(1), {\it{}half}, {\it{}Cf}.{\it{}center}().{\it{}op}(1).{\it{}subs}({\it{}cycle\_val}),
  {\it{}C4}.{\it{}roots}({\it{}l}\begin{math}\div\end{math}{\it{}k}, {\bf{}false}).{\it{}op}(0).{\it{}normal}().{\it{}subs}({\it{}cycle\_val});
 {\bf{}break};
{\bf{}case} 0:
 {\it{}U} = {\bf{}numeric}(17,4), {\it{}Cg}.{\it{}roots}().{\it{}op}(0), {\it{}Cf}.{\it{}center}().{\it{}op}(0).{\it{}subs}({\it{}cycle\_val}), ({\it{}l}\begin{math}\div\end{math}{\it{}k}).{\it{}subs}({\it{}cycle\_val});
 {\it{}V} = {\it{}Cf}.{\it{}roots}({\it{}U}.{\it{}op}(0), {\bf{}false}).{\it{}op}(0), {\bf{}numeric}(3,2), {\it{}Cf}.{\it{}roots}({\it{}l}\begin{math}\div\end{math}{\it{}k}, {\bf{}false}).{\it{}op}(0).{\it{}subs}({\it{}cycle\_val}),
  {\it{}C4}.{\it{}roots}({\it{}l}\begin{math}\div\end{math}{\it{}k}, {\bf{}false}).{\it{}op}(0).{\it{}normal}().{\it{}subs}({\it{}cycle\_val});
 {\bf{}break};
{\bf{}case} 1:
 {\it{}U} = {\bf{}numeric}(12,4), {\it{}Cg}.{\it{}roots}({\bf{}numeric}(3,4)).{\it{}op}(0), {\it{}Cf}.{\it{}center}().{\it{}op}(0).{\it{}subs}({\it{}cycle\_val}), ({\it{}l}\begin{math}\div\end{math}{\it{}k}).{\it{}subs}({\it{}cycle\_val});
 {\it{}V} = {\it{}Cf}.{\it{}roots}({\it{}U}.{\it{}op}(0), {\bf{}false}).{\it{}op}(0), {\bf{}numeric}(3,4), {\it{}Cf}.{\it{}center}().{\it{}op}(1).{\it{}subs}({\it{}cycle\_val}),
  {\it{}C4}.{\it{}roots}({\it{}l}\begin{math}\div\end{math}{\it{}k}, {\bf{}false}).{\it{}op}(0).{\it{}normal}().{\it{}subs}({\it{}cycle\_val});
 {\bf{}break};
{\nwrbrace}
{\it{}U}.{\it{}append}({\it{}P}.{\it{}op}(0).{\it{}subs}({\it{}cycle\_val}).{\it{}subs}({\bf{}lst}({\it{}u} \begin{math}\equiv\end{math} {\it{}U}.{\it{}op}(0), {\it{}v} \begin{math}\equiv\end{math} {\it{}V}.{\it{}op}(0))).{\it{}normal}()); // Moebius transform of the first point
{\it{}V}.{\it{}append}({\it{}P}.{\it{}op}(1).{\it{}subs}({\it{}cycle\_val}).{\it{}subs}({\bf{}lst}({\it{}u} \begin{math}\equiv\end{math} {\it{}U}.{\it{}op}(0), {\it{}v} \begin{math}\equiv\end{math} {\it{}V}.{\it{}op}(0))).{\it{}normal}());

{\it{}asymptote} \begin{math}\ll\end{math} {\it{}endl} \begin{math}\ll\end{math}  {\tt{}"erase();"}  \begin{math}\ll\end{math} {\it{}endl};
\LA{}Drawing orthogonal cycles~{\nwtagstyle{}\subpageref{NW3gGP3e-3iFOPx-1}}\RA{}
{\it{}asymptote} \begin{math}\ll\end{math} {\tt{}"shipout({\char92}"first-ort-"} \begin{math}\ll\end{math} {\it{}eph\_names}[{\it{}si}+1] \begin{math}\ll\end{math} {\it{}eph\_names}[{\it{}si1}+1] \begin{math}\ll\end{math} {\tt{}"{\char92}");"} \begin{math}\ll\end{math} {\it{}endl};

\nwused{\\{NW3gGP3e-Xmoi0-2}}\nwidentuses{\\{{\nwixident{center}}{center}}\\{{\nwixident{cycle2D}}{cycle2D}}\\{{\nwixident{k}}{k}}\\{{\nwixident{l}}{l}}\\{{\nwixident{lst}}{lst}}\\{{\nwixident{m}}{m}}\\{{\nwixident{normal}}{normal}}\\{{\nwixident{numeric}}{numeric}}\\{{\nwixident{op}}{op}}\\{{\nwixident{points}}{points}}\\{{\nwixident{roots}}{roots}}\\{{\nwixident{si}}{si}}\\{{\nwixident{si1}}{si1}}\\{{\nwixident{subs}}{subs}}\\{{\nwixident{u}}{u}}\\{{\nwixident{v}}{v}}}\nwindexuse{\nwixident{center}}{center}{NW3gGP3e-1VpO8V-1}\nwindexuse{\nwixident{cycle2D}}{cycle2D}{NW3gGP3e-1VpO8V-1}\nwindexuse{\nwixident{k}}{k}{NW3gGP3e-1VpO8V-1}\nwindexuse{\nwixident{l}}{l}{NW3gGP3e-1VpO8V-1}\nwindexuse{\nwixident{lst}}{lst}{NW3gGP3e-1VpO8V-1}\nwindexuse{\nwixident{m}}{m}{NW3gGP3e-1VpO8V-1}\nwindexuse{\nwixident{normal}}{normal}{NW3gGP3e-1VpO8V-1}\nwindexuse{\nwixident{numeric}}{numeric}{NW3gGP3e-1VpO8V-1}\nwindexuse{\nwixident{op}}{op}{NW3gGP3e-1VpO8V-1}\nwindexuse{\nwixident{points}}{points}{NW3gGP3e-1VpO8V-1}\nwindexuse{\nwixident{roots}}{roots}{NW3gGP3e-1VpO8V-1}\nwindexuse{\nwixident{si}}{si}{NW3gGP3e-1VpO8V-1}\nwindexuse{\nwixident{si1}}{si1}{NW3gGP3e-1VpO8V-1}\nwindexuse{\nwixident{subs}}{subs}{NW3gGP3e-1VpO8V-1}\nwindexuse{\nwixident{u}}{u}{NW3gGP3e-1VpO8V-1}\nwindexuse{\nwixident{v}}{v}{NW3gGP3e-1VpO8V-1}\nwendcode{}\nwbegindocs{383}We start drawing from cycles.
\nwenddocs{}\nwbegincode{384}\sublabel{NW3gGP3e-3iFOPx-1}\nwmargintag{{\nwtagstyle{}\subpageref{NW3gGP3e-3iFOPx-1}}}\moddef{Drawing orthogonal cycles~{\nwtagstyle{}\subpageref{NW3gGP3e-3iFOPx-1}}}\endmoddef\Rm{}\nwstartdeflinemarkup\nwusesondefline{\\{NW3gGP3e-1VpO8V-1}\\{NW3gGP3e-384dQC-1}}\nwprevnextdefs{\relax}{NW3gGP3e-3iFOPx-2}\nwenddeflinemarkup
{\bf{}for} ({\bf{}int} {\it{}j} = 0; {\it{}j}\begin{math}<\end{math}2; {\it{}j}\protect\PP)
 {\bf{}for} ({\bf{}int} {\it{}i}=0; {\it{}i}\begin{math}<\end{math}({\it{}si}\begin{math}\equiv\end{math}1?4:5); {\it{}i}\protect\PP)
  {\it{}Cp}.{\it{}subs}({\bf{}lst}({\it{}k1} \begin{math}\equiv\end{math} ({\it{}si} \begin{math}\equiv\end{math} 0? {\bf{}numeric}(3\begin{math}\ast\end{math}{\it{}i},2): {\bf{}numeric}({\it{}i}, 4)), {\it{}n1} \begin{math}\equiv\end{math} {\it{}half}, {\it{}u} \begin{math}\equiv\end{math} {\it{}U}.{\it{}op}({\it{}j}),
     {\it{}v} \begin{math}\equiv\end{math} {\it{}V}.{\it{}op}({\it{}j}))).{\it{}subs}({\it{}cycle\_val}).{\it{}asy\_draw}({\it{}asymptote}, {\it{}xmin}, {\it{}xmax}, {\it{}ymin}, {\it{}ymax},
               {\bf{}lst}(0.2, 0.2+{\it{}j}\begin{math}\ast\end{math}(0.3+{\it{}i}\begin{math}\div\end{math}8.0), 0.2+(1-{\it{}j})\begin{math}\ast\end{math}(0.3+{\it{}i}\begin{math}\div\end{math}8.0)));

{\it{}Cf}.{\it{}asy\_draw}({\it{}asymptote}, {\it{}xmin}, {\it{}xmax}, {\it{}ymin}, {\it{}ymax}, {\bf{}lst}(0.8, 0, 0), {\tt{}"+1"});
{\it{}Cg}.{\it{}asy\_draw}({\it{}asymptote}, {\it{}xmin}, {\it{}xmax}, {\it{}ymin}, {\it{}ymax}, {\bf{}lst}(0, 0, 0), {\tt{}"+0.3+dashed"});
{\bf{}if} ({\it{}si} \begin{math}\equiv\end{math} 0)
 {\it{}C5}.{\it{}subs}({\bf{}lst}({\it{}sign} \begin{math}\equiv\end{math}0, {\it{}sign1}\begin{math}\equiv\end{math}0)).{\it{}subs}({\it{}cycle\_val}).{\it{}asy\_draw}({\it{}asymptote}, {\it{}xmin}, {\it{}xmax}, {\it{}ymin}, {\it{}ymax}, {\bf{}lst}(0, 0, 0),
                 {\tt{}"+dotted"});

\nwalsodefined{\\{NW3gGP3e-3iFOPx-2}}\nwused{\\{NW3gGP3e-1VpO8V-1}\\{NW3gGP3e-384dQC-1}}\nwidentuses{\\{{\nwixident{asy{\_}draw}}{asy:undraw}}\\{{\nwixident{lst}}{lst}}\\{{\nwixident{numeric}}{numeric}}\\{{\nwixident{op}}{op}}\\{{\nwixident{si}}{si}}\\{{\nwixident{subs}}{subs}}\\{{\nwixident{u}}{u}}\\{{\nwixident{v}}{v}}}\nwindexuse{\nwixident{asy{\_}draw}}{asy:undraw}{NW3gGP3e-3iFOPx-1}\nwindexuse{\nwixident{lst}}{lst}{NW3gGP3e-3iFOPx-1}\nwindexuse{\nwixident{numeric}}{numeric}{NW3gGP3e-3iFOPx-1}\nwindexuse{\nwixident{op}}{op}{NW3gGP3e-3iFOPx-1}\nwindexuse{\nwixident{si}}{si}{NW3gGP3e-3iFOPx-1}\nwindexuse{\nwixident{subs}}{subs}{NW3gGP3e-3iFOPx-1}\nwindexuse{\nwixident{u}}{u}{NW3gGP3e-3iFOPx-1}\nwindexuse{\nwixident{v}}{v}{NW3gGP3e-3iFOPx-1}\nwendcode{}\nwbegindocs{385}To finish we add some additional drawing explaining the picture.
\nwenddocs{}\nwbegincode{386}\sublabel{NW3gGP3e-3iFOPx-2}\nwmargintag{{\nwtagstyle{}\subpageref{NW3gGP3e-3iFOPx-2}}}\moddef{Drawing orthogonal cycles~{\nwtagstyle{}\subpageref{NW3gGP3e-3iFOPx-1}}}\plusendmoddef\Rm{}\nwstartdeflinemarkup\nwusesondefline{\\{NW3gGP3e-1VpO8V-1}\\{NW3gGP3e-384dQC-1}}\nwprevnextdefs{NW3gGP3e-3iFOPx-1}{\relax}\nwenddeflinemarkup
{\it{}asymptote} \begin{math}\ll\end{math} {\tt{}"pair[] z={\char123}("} \begin{math}\ll\end{math} {\it{}ex\_to}\begin{math}<\end{math}{\bf{}numeric}\begin{math}>\end{math}({\it{}U}.{\it{}op}(0).{\it{}evalf}()).{\it{}to\_double}() \begin{math}\ll\end{math} {\tt{}", "}
 \begin{math}\ll\end{math} {\it{}ex\_to}\begin{math}<\end{math}{\bf{}numeric}\begin{math}>\end{math}({\it{}V}.{\it{}op}(0).{\it{}evalf}()).{\it{}to\_double}() \begin{math}\ll\end{math} {\tt{}")"};
{\bf{}for} ({\bf{}int} {\it{}j} = 1; {\it{}j}\begin{math}<\end{math}5; {\it{}j}\protect\PP)
 {\it{}asymptote} \begin{math}\ll\end{math} {\tt{}", ("} \begin{math}\ll\end{math} {\it{}ex\_to}\begin{math}<\end{math}{\bf{}numeric}\begin{math}>\end{math}({\it{}U}.{\it{}op}({\it{}j}).{\it{}evalf}()).{\it{}to\_double}() \begin{math}\ll\end{math} {\tt{}", "}
     \begin{math}\ll\end{math} {\it{}ex\_to}\begin{math}<\end{math}{\bf{}numeric}\begin{math}>\end{math}({\it{}V}.{\it{}op}({\it{}j}).{\it{}evalf}()).{\it{}to\_double}() \begin{math}\ll\end{math} {\tt{}")"} ;

{\it{}asymptote} \begin{math}\ll\end{math} {\tt{}"{\char125};"} \begin{math}\ll\end{math} {\it{}endl}    \begin{math}\ll\end{math} {\tt{}"  dot(z);"} \begin{math}\ll\end{math} {\it{}endl}
 \begin{math}\ll\end{math} ({\it{}si} \begin{math}\equiv\end{math} 0? {\tt{}"  draw((z[2].x,0)--z[2], 0.3+dotted);"} : {\tt{}""}) \begin{math}\ll\end{math} {\it{}endl}
 \begin{math}\ll\end{math} ({\it{}si} \begin{math}\equiv\end{math} 0? {\tt{}"  draw((z[3].x,0)--z[3], 0.3+dotted);"} : {\tt{}""}) \begin{math}\ll\end{math} {\it{}endl}
 \begin{math}\ll\end{math} {\tt{}"  label({\char92}"$a${\char92}", z[1], NW);"} \begin{math}\ll\end{math} {\it{}endl}
   \begin{math}\ll\end{math} {\tt{}"  label({\char92}"$b${\char92}", z[0], SE);"} \begin{math}\ll\end{math} {\it{}endl}
  \begin{math}\ll\end{math} {\tt{}"  label({\char92}"$c${\char92}", z[3], E);"} \begin{math}\ll\end{math} {\it{}endl}
   \begin{math}\ll\end{math} {\tt{}"  label"} \begin{math}\ll\end{math} {\tt{}"({\char92}"$d${\char92}", z[4], "} \begin{math}\ll\end{math} ({\it{}si} \begin{math}\equiv\end{math}1?{\tt{}"NW);"}:{\tt{}"NE);"}) \begin{math}\ll\end{math} {\it{}endl};

\LA{}Put units~{\nwtagstyle{}\subpageref{NW3gGP3e-1r99tr-1}}\RA{}
\LA{}Draw axes~{\nwtagstyle{}\subpageref{NW3gGP3e-ph2PF-1}}\RA{}

\nwused{\\{NW3gGP3e-1VpO8V-1}\\{NW3gGP3e-384dQC-1}}\nwidentuses{\\{{\nwixident{numeric}}{numeric}}\\{{\nwixident{op}}{op}}\\{{\nwixident{si}}{si}}}\nwindexuse{\nwixident{numeric}}{numeric}{NW3gGP3e-3iFOPx-2}\nwindexuse{\nwixident{op}}{op}{NW3gGP3e-3iFOPx-2}\nwindexuse{\nwixident{si}}{si}{NW3gGP3e-3iFOPx-2}\nwendcode{}\nwbegindocs{387}This chunk draws the standard coordinat axes.
\nwenddocs{}\nwbegincode{388}\sublabel{NW3gGP3e-ph2PF-1}\nwmargintag{{\nwtagstyle{}\subpageref{NW3gGP3e-ph2PF-1}}}\moddef{Draw axes~{\nwtagstyle{}\subpageref{NW3gGP3e-ph2PF-1}}}\endmoddef\Rm{}\nwstartdeflinemarkup\nwusesondefline{\\{NW3gGP3e-3iFOPx-2}\\{NW3gGP3e-4CF7DS-1}\\{NW3gGP3e-3gnOun-1}\\{NW3gGP3e-QUZd0-1}\\{NW3gGP3e-Y6OSa-1}\\{NW3gGP3e-1g9SsP-1}\\{NW3gGP3e-FS4xg-1}\\{NW3gGP3e-q6Aul-5}\\{NW3gGP3e-3xVrpR-1}\\{NW3gGP3e-3xVrpR-3}}\nwenddeflinemarkup
{\it{}asymptote} \begin{math}\ll\end{math} {\tt{}"  draw\_axes(("} \begin{math}\ll\end{math} {\it{}xmin}.{\it{}to\_double}() \begin{math}\ll\end{math} {\tt{}", "} \begin{math}\ll\end{math} {\it{}ymin}.{\it{}to\_double}()
 \begin{math}\ll\end{math} {\tt{}"), ( "} \begin{math}\ll\end{math} {\it{}xmax}.{\it{}to\_double}() \begin{math}\ll\end{math} {\tt{}", "} \begin{math}\ll\end{math} {\it{}ymax}.{\it{}to\_double}() \begin{math}\ll\end{math} {\tt{}"));"} \begin{math}\ll\end{math} {\it{}endl};

\nwused{\\{NW3gGP3e-3iFOPx-2}\\{NW3gGP3e-4CF7DS-1}\\{NW3gGP3e-3gnOun-1}\\{NW3gGP3e-QUZd0-1}\\{NW3gGP3e-Y6OSa-1}\\{NW3gGP3e-1g9SsP-1}\\{NW3gGP3e-FS4xg-1}\\{NW3gGP3e-q6Aul-5}\\{NW3gGP3e-3xVrpR-1}\\{NW3gGP3e-3xVrpR-3}}\nwendcode{}\nwbegindocs{389}\nwdocspar
\nwenddocs{}\nwbegincode{390}\sublabel{NW3gGP3e-1r99tr-1}\nwmargintag{{\nwtagstyle{}\subpageref{NW3gGP3e-1r99tr-1}}}\moddef{Put units~{\nwtagstyle{}\subpageref{NW3gGP3e-1r99tr-1}}}\endmoddef\Rm{}\nwstartdeflinemarkup\nwusesondefline{\\{NW3gGP3e-3iFOPx-2}\\{NW3gGP3e-q6Aul-5}}\nwenddeflinemarkup
{\it{}asymptote}  \begin{math}\ll\end{math} {\tt{}"  label({\char92}"${\char92}{\char92}sigma="} \begin{math}\ll\end{math} {\it{}si} \begin{math}\ll\end{math} {\tt{}", {\char92}{\char92}breve{\char123}{\char92}{\char92}sigma{\char125}="} \begin{math}\ll\end{math} {\it{}si1}
  \begin{math}\ll\end{math} {\tt{}"${\char92}", (0, "} \begin{math}\ll\end{math} {\it{}ymin}.{\it{}to\_double}() \begin{math}\ll\end{math} {\tt{}"), S);"} \begin{math}\ll\end{math} {\it{}endl} \begin{math}\ll\end{math} {\tt{}"draw((1,-0.1)--(1,0.1));"} \begin{math}\ll\end{math} {\it{}endl}
  \begin{math}\ll\end{math} {\tt{}"draw((-0.1,1)--(0.1,1));"} \begin{math}\ll\end{math} {\it{}endl}
  \begin{math}\ll\end{math} {\tt{}"label({\char92}"$1${\char92}", (1,0), S);"} \begin{math}\ll\end{math} {\it{}endl}
  \begin{math}\ll\end{math} {\tt{}"label({\char92}"$1${\char92}", (0,1), E);"} \begin{math}\ll\end{math} {\it{}endl};

\nwused{\\{NW3gGP3e-3iFOPx-2}\\{NW3gGP3e-q6Aul-5}}\nwidentuses{\\{{\nwixident{si}}{si}}\\{{\nwixident{si1}}{si1}}}\nwindexuse{\nwixident{si}}{si}{NW3gGP3e-1r99tr-1}\nwindexuse{\nwixident{si1}}{si1}{NW3gGP3e-1r99tr-1}\nwendcode{}\nwbegindocs{391}\nwdocspar
\subsubsection{Focal Orthogonality Condition}
\label{sec:focal-orth-cond}

We draw some \Asymptote\ pictures to illustrate the focal orthogonality
relation. We define numeric values of all involved parameters first.

\nwenddocs{}\nwbegincode{392}\sublabel{NW3gGP3e-384dQC-1}\nwmargintag{{\nwtagstyle{}\subpageref{NW3gGP3e-384dQC-1}}}\moddef{Drawing focal orthogonality~{\nwtagstyle{}\subpageref{NW3gGP3e-384dQC-1}}}\endmoddef\Rm{}\nwstartdeflinemarkup\nwusesondefline{\\{NW3gGP3e-Xmoi0-2}}\nwenddeflinemarkup
{\bf{}numeric} {\it{}xmin}(-11,4), {\it{}xmax}(5), {\it{}ymin}(-13,4), {\it{}ymax} = ({\it{}si} \begin{math}\equiv\end{math} 0?{\bf{}numeric}(6): {\bf{}numeric}(15,4));
{\bf{}lst} {\it{}cycle\_val} = {\bf{}lst}({\it{}sign} \begin{math}\equiv\end{math} {\bf{}numeric}({\it{}si}), {\it{}sign1} \begin{math}\equiv\end{math} {\bf{}numeric}({\it{}si1}), {\it{}sign2} \begin{math}\equiv\end{math} {\bf{}numeric}(1), //sign3 == jump\_fnct(-si), //sign3 == (si \begin{math}>\end{math} 0?numeric(-1):numeric(1)),
     {\it{}k} \begin{math}\equiv\end{math} {\bf{}numeric}(2,3), {\it{}l} \begin{math}\equiv\end{math} {\bf{}numeric}(2,3), {\it{}n} \begin{math}\equiv\end{math} ({\it{}si} \begin{math}\equiv\end{math} 1?{\bf{}numeric}(-4,3):{\it{}half}), {\it{}m} \begin{math}\equiv\end{math}({\it{}si} \begin{math}\equiv\end{math} 1?{\bf{}numeric}(-9,3):{\bf{}numeric}(-2)));
{\bf{}cycle2D} {\it{}Cf} = {\it{}C}.{\it{}subs}({\it{}cycle\_val}), {\it{}Cg} = {\it{}C8}.{\it{}subs}({\it{}cycle\_val}), {\it{}Cp} ={\it{}C6};
{\bf{}lst} {\it{}U}, {\it{}V};
{\bf{}switch} ({\it{}si}) {\nwlbrace}
{\bf{}case} -1: // points b, a, center, c, d
 {\it{}U} = {\bf{}numeric}(11,4), {\it{}Cg}.{\it{}roots}({\it{}half}).{\it{}op}(0), {\it{}Cf}.{\it{}focus}().{\it{}op}(0).{\it{}subs}({\it{}cycle\_val}), ({\it{}l}\begin{math}\div\end{math}{\it{}k}).{\it{}subs}({\it{}cycle\_val});
 {\it{}V} = {\it{}Cf}.{\it{}roots}({\it{}U}.{\it{}op}(0), {\bf{}false}).{\it{}op}(1), {\it{}half}, {\it{}Cf}.{\it{}focus}().{\it{}op}(1).{\it{}subs}({\it{}cycle\_val}),
  {\it{}C7}.{\it{}roots}({\it{}l}\begin{math}\div\end{math}{\it{}k}, {\bf{}false}).{\it{}op}(0).{\it{}normal}().{\it{}subs}({\it{}cycle\_val});
 {\bf{}break};
{\bf{}case} 0:
 {\it{}U} = {\bf{}numeric}(4), {\it{}Cf}.{\it{}roots}().{\it{}op}(0), {\it{}Cf}.{\it{}focus}().{\it{}op}(0).{\it{}subs}({\it{}cycle\_val}), ({\it{}l}\begin{math}\div\end{math}{\it{}k}).{\it{}subs}({\it{}cycle\_val});
 {\it{}V} = {\it{}Cf}.{\it{}roots}({\it{}U}.{\it{}op}(0), {\bf{}false}).{\it{}op}(0), {\bf{}numeric}(3,2), {\it{}Cf}.{\it{}focus}().{\it{}op}(0).{\it{}subs}({\it{}cycle\_val}),
  {\it{}C7}.{\it{}roots}({\it{}l}\begin{math}\div\end{math}{\it{}k}, {\bf{}false}).{\it{}op}(0).{\it{}normal}().{\it{}subs}({\it{}cycle\_val});
 {\bf{}break};
{\bf{}case} 1:
 {\it{}U} = {\it{}Cf}.{\it{}roots}({\bf{}numeric}(1)).{\it{}op}(1), {\it{}Cg}.{\it{}roots}({\bf{}numeric}(6, 4)).{\it{}op}(1),
  {\it{}Cf}.{\it{}focus}().{\it{}op}(0).{\it{}subs}({\it{}cycle\_val}), ({\it{}l}\begin{math}\div\end{math}{\it{}k}).{\it{}subs}({\it{}cycle\_val});
 {\it{}V} = {\bf{}numeric}(1), {\bf{}numeric}(6, 4), {\it{}Cf}.{\it{}focus}().{\it{}op}(1).{\it{}subs}({\it{}cycle\_val}),
  {\it{}C7}.{\it{}roots}({\it{}l}\begin{math}\div\end{math}{\it{}k}, {\bf{}false}).{\it{}op}(0).{\it{}normal}().{\it{}subs}({\it{}cycle\_val});
 {\bf{}break};
{\nwrbrace}
{\it{}U}.{\it{}append}({\it{}P1}.{\it{}op}(0).{\it{}subs}({\it{}cycle\_val}).{\it{}subs}({\bf{}lst}({\it{}u} \begin{math}\equiv\end{math} {\it{}U}.{\it{}op}(0), {\it{}v} \begin{math}\equiv\end{math} {\it{}V}.{\it{}op}(0))).{\it{}normal}()); // Moebius transform of {\it{}U}.{\it{}op}(0)
{\it{}V}.{\it{}append}({\it{}P1}.{\it{}op}(1).{\it{}subs}({\it{}cycle\_val}).{\it{}subs}({\bf{}lst}({\it{}u} \begin{math}\equiv\end{math} {\it{}U}.{\it{}op}(0), {\it{}v} \begin{math}\equiv\end{math} {\it{}V}.{\it{}op}(0))).{\it{}normal}());

{\it{}asymptote} \begin{math}\ll\end{math} {\it{}endl} \begin{math}\ll\end{math} {\tt{}"erase();"} //\begin{math}<\end{math}\begin{math}<\end{math} endl \begin{math}<\end{math}\begin{math}<\end{math} "size(250);"
  \begin{math}\ll\end{math} {\it{}endl};
\LA{}Drawing orthogonal cycles~{\nwtagstyle{}\subpageref{NW3gGP3e-3iFOPx-1}}\RA{}
{\it{}asymptote} \begin{math}\ll\end{math} {\tt{}"shipout({\char92}"sec-ort-"} \begin{math}\ll\end{math} {\it{}eph\_names}[{\it{}si}+1] \begin{math}\ll\end{math} {\it{}eph\_names}[{\it{}si1}+1] \begin{math}\ll\end{math} {\tt{}"{\char92}");"} \begin{math}\ll\end{math} {\it{}endl};

\nwused{\\{NW3gGP3e-Xmoi0-2}}\nwidentuses{\\{{\nwixident{center}}{center}}\\{{\nwixident{cycle2D}}{cycle2D}}\\{{\nwixident{focus}}{focus}}\\{{\nwixident{jump{\_}fnct}}{jump:unfnct}}\\{{\nwixident{k}}{k}}\\{{\nwixident{l}}{l}}\\{{\nwixident{lst}}{lst}}\\{{\nwixident{m}}{m}}\\{{\nwixident{normal}}{normal}}\\{{\nwixident{numeric}}{numeric}}\\{{\nwixident{op}}{op}}\\{{\nwixident{points}}{points}}\\{{\nwixident{roots}}{roots}}\\{{\nwixident{si}}{si}}\\{{\nwixident{si1}}{si1}}\\{{\nwixident{subs}}{subs}}\\{{\nwixident{u}}{u}}\\{{\nwixident{v}}{v}}}\nwindexuse{\nwixident{center}}{center}{NW3gGP3e-384dQC-1}\nwindexuse{\nwixident{cycle2D}}{cycle2D}{NW3gGP3e-384dQC-1}\nwindexuse{\nwixident{focus}}{focus}{NW3gGP3e-384dQC-1}\nwindexuse{\nwixident{jump{\_}fnct}}{jump:unfnct}{NW3gGP3e-384dQC-1}\nwindexuse{\nwixident{k}}{k}{NW3gGP3e-384dQC-1}\nwindexuse{\nwixident{l}}{l}{NW3gGP3e-384dQC-1}\nwindexuse{\nwixident{lst}}{lst}{NW3gGP3e-384dQC-1}\nwindexuse{\nwixident{m}}{m}{NW3gGP3e-384dQC-1}\nwindexuse{\nwixident{normal}}{normal}{NW3gGP3e-384dQC-1}\nwindexuse{\nwixident{numeric}}{numeric}{NW3gGP3e-384dQC-1}\nwindexuse{\nwixident{op}}{op}{NW3gGP3e-384dQC-1}\nwindexuse{\nwixident{points}}{points}{NW3gGP3e-384dQC-1}\nwindexuse{\nwixident{roots}}{roots}{NW3gGP3e-384dQC-1}\nwindexuse{\nwixident{si}}{si}{NW3gGP3e-384dQC-1}\nwindexuse{\nwixident{si1}}{si1}{NW3gGP3e-384dQC-1}\nwindexuse{\nwixident{subs}}{subs}{NW3gGP3e-384dQC-1}\nwindexuse{\nwixident{u}}{u}{NW3gGP3e-384dQC-1}\nwindexuse{\nwixident{v}}{v}{NW3gGP3e-384dQC-1}\nwendcode{}\nwbegindocs{393}\nwdocspar
\subsection[Extra pictures from Asymptote]{Extra pictures from \Asymptote}
\label{sec:extra-pictures-from}
We draw few more pictures in \Asymptote.
\nwenddocs{}\nwbegincode{394}\sublabel{NW3gGP3e-3gefqu-1}\nwmargintag{{\nwtagstyle{}\subpageref{NW3gGP3e-3gefqu-1}}}\moddef{Extra pictures from Asymptote~{\nwtagstyle{}\subpageref{NW3gGP3e-3gefqu-1}}}\endmoddef\Rm{}\nwstartdeflinemarkup\nwusesondefline{\\{NW3gGP3e-Xmoi0-3}}\nwenddeflinemarkup
 {\bf{}numeric} {\it{}xmin}(-5), {\it{}xmax}(5), {\it{}ymin}(-13,4), {\it{}ymax} = {\bf{}numeric}(6);
 \LA{}Three images of the same cycle~{\nwtagstyle{}\subpageref{NW3gGP3e-4CF7DS-1}}\RA{}
 \LA{}Centres and foci of parabolas~{\nwtagstyle{}\subpageref{NW3gGP3e-3gnOun-1}}\RA{}
 \LA{}Zero-radius cycle implementations~{\nwtagstyle{}\subpageref{NW3gGP3e-QUZd0-1}}\RA{}
 \LA{}Parabolic diameters~{\nwtagstyle{}\subpageref{NW3gGP3e-Y6OSa-1}}\RA{}
 \LA{}Distance as an extremum~{\nwtagstyle{}\subpageref{NW3gGP3e-1g9SsP-1}}\RA{}
 \LA{}Infinitesimal cycles draw~{\nwtagstyle{}\subpageref{NW3gGP3e-FS4xg-1}}\RA{}
 \LA{}Cayley transform pictures~{\nwtagstyle{}\subpageref{NW3gGP3e-q6Aul-1}}\RA{}
 \LA{}Three inversions~{\nwtagstyle{}\subpageref{NW3gGP3e-3xVrpR-1}}\RA{}
 \LA{}Hyperbolic inversion of a ball~{\nwtagstyle{}\subpageref{NW3gGP3e-2DcUMV-1}}\RA{}

\nwused{\\{NW3gGP3e-Xmoi0-3}}\nwidentuses{\\{{\nwixident{numeric}}{numeric}}}\nwindexuse{\nwixident{numeric}}{numeric}{NW3gGP3e-3gefqu-1}\nwendcode{}\nwbegindocs{395}\nwdocspar
\subsubsection{Different implementations of the same cycle}
\label{sec:diff-impl-same}

A cycle represented by a four numbers \((k, l, n, m\) looks
different in three spaces with different metrics.
\nwenddocs{}\nwbegincode{396}\sublabel{NW3gGP3e-4CF7DS-1}\nwmargintag{{\nwtagstyle{}\subpageref{NW3gGP3e-4CF7DS-1}}}\moddef{Three images of the same cycle~{\nwtagstyle{}\subpageref{NW3gGP3e-4CF7DS-1}}}\endmoddef\Rm{}\nwstartdeflinemarkup\nwusesondefline{\\{NW3gGP3e-3gefqu-1}}\nwenddeflinemarkup
{\it{}asymptote} \begin{math}\ll\end{math} {\it{}endl} \begin{math}\ll\end{math} {\tt{}"erase();"} \begin{math}\ll\end{math} {\it{}endl};
{\bf{}cycle2D} {\it{}C1f}, {\it{}C2f};
{\it{}asymptote} \begin{math}\ll\end{math} {\tt{}"pair[] z;"};
{\bf{}for} ({\bf{}int} {\it{}j} = -1; {\it{}j}\begin{math}<\end{math}2; {\it{}j}\protect\PP) {\nwlbrace}
 {\it{}C1f} = {\bf{}cycle2D}(1, {\bf{}lst}(-2.5, 1), 3.75, {\it{}diag\_matrix}({\bf{}lst}(-1, {\it{}j})));
 {\it{}C2f} = {\bf{}cycle2D}(1, {\bf{}lst}(2.75, 3), 14.0625, {\it{}diag\_matrix}({\bf{}lst}(-1, {\it{}j})));
 {\it{}C1f}.{\it{}asy\_draw}({\it{}asymptote}, {\it{}xmin}, {\it{}xmax}, {\it{}ymin}, {\it{}ymax}, {\bf{}lst}(0, 1.0-0.4\begin{math}\ast\end{math}({\it{}j}+1), 0.4\begin{math}\ast\end{math}({\it{}j}+1)), {\tt{}"+.75"}, {\bf{}true}, 7);
 {\it{}C2f}.{\it{}asy\_draw}({\it{}asymptote}, {\it{}xmin}, {\it{}xmax}, {\it{}ymin}, {\it{}ymax}, {\bf{}lst}(0, 1.0-0.4\begin{math}\ast\end{math}({\it{}j}+1), 0.4\begin{math}\ast\end{math}({\it{}j}+1)), {\tt{}"+.75"}, {\bf{}true}, 7);
 {\it{}asymptote} \begin{math}\ll\end{math} {\tt{}"z.push(("} \begin{math}\ll\end{math} {\it{}C1f}.{\it{}center}().{\it{}op}(0) \begin{math}\ll\end{math} {\tt{}", "}  \begin{math}\ll\end{math} {\it{}C1f}.{\it{}center}().{\it{}op}(1) \begin{math}\ll\end{math} {\tt{}")); z.push(("}
     \begin{math}\ll\end{math} {\it{}C2f}.{\it{}center}().{\it{}op}(0) \begin{math}\ll\end{math} {\tt{}", "}  \begin{math}\ll\end{math} {\it{}C2f}.{\it{}center}().{\it{}op}(1) \begin{math}\ll\end{math} {\tt{}"));"} \begin{math}\ll\end{math} {\it{}endl};
{\nwrbrace}
{\it{}asymptote} \begin{math}\ll\end{math} {\tt{}"z.push(("} \begin{math}\ll\end{math} {\it{}C1f}.{\it{}roots}().{\it{}op}(0) \begin{math}\ll\end{math} {\tt{}", 0));  z.push(("} \begin{math}\ll\end{math} {\it{}C1f}.{\it{}roots}().{\it{}op}(1) \begin{math}\ll\end{math} {\tt{}", 0));"} \begin{math}\ll\end{math} {\it{}endl}
  \begin{math}\ll\end{math} {\tt{}" dot(z);"} \begin{math}\ll\end{math} {\it{}endl}
  \begin{math}\ll\end{math} {\tt{}"  for (int j = 0; j<2; ++j) {\char123}"}
  \begin{math}\ll\end{math} {\tt{}"    label({\char92}"$c\_e${\char92}", z[j], E);"} \begin{math}\ll\end{math} {\it{}endl}
  \begin{math}\ll\end{math} {\tt{}"    label({\char92}"$c\_p${\char92}", z[j+2], SE);"} \begin{math}\ll\end{math} {\it{}endl}
  \begin{math}\ll\end{math} {\tt{}"    label({\char92}"$c\_h${\char92}", z[j+4], E);"} \begin{math}\ll\end{math} {\it{}endl}
  \begin{math}\ll\end{math} {\tt{}"    label((j==0?{\char92}"$r\_0${\char92}":{\char92}"$r\_1${\char92}"), z[j+6], (j==0? SW: SE));"} \begin{math}\ll\end{math} {\it{}endl}
  \begin{math}\ll\end{math} {\tt{}"    draw(z[j]--z[j+4], .3+dashed);"} \begin{math}\ll\end{math} {\it{}endl}
  \begin{math}\ll\end{math} {\tt{}"  {\char125}"} \begin{math}\ll\end{math} {\it{}endl};
\LA{}Draw axes~{\nwtagstyle{}\subpageref{NW3gGP3e-ph2PF-1}}\RA{}
{\it{}asymptote} \begin{math}\ll\end{math} {\tt{}"shipout({\char92}"same-cycle{\char92}");"} \begin{math}\ll\end{math} {\it{}endl};

\nwused{\\{NW3gGP3e-3gefqu-1}}\nwidentuses{\\{{\nwixident{asy{\_}draw}}{asy:undraw}}\\{{\nwixident{center}}{center}}\\{{\nwixident{cycle}}{cycle}}\\{{\nwixident{cycle2D}}{cycle2D}}\\{{\nwixident{lst}}{lst}}\\{{\nwixident{op}}{op}}\\{{\nwixident{roots}}{roots}}}\nwindexuse{\nwixident{asy{\_}draw}}{asy:undraw}{NW3gGP3e-4CF7DS-1}\nwindexuse{\nwixident{center}}{center}{NW3gGP3e-4CF7DS-1}\nwindexuse{\nwixident{cycle}}{cycle}{NW3gGP3e-4CF7DS-1}\nwindexuse{\nwixident{cycle2D}}{cycle2D}{NW3gGP3e-4CF7DS-1}\nwindexuse{\nwixident{lst}}{lst}{NW3gGP3e-4CF7DS-1}\nwindexuse{\nwixident{op}}{op}{NW3gGP3e-4CF7DS-1}\nwindexuse{\nwixident{roots}}{roots}{NW3gGP3e-4CF7DS-1}\nwendcode{}\nwbegindocs{397}\nwdocspar
\subsubsection{Centres and foci of cycles}
\label{sec:centres-foci-cycles}

We draw two parabolas and their centres with three type of foci.
\nwenddocs{}\nwbegincode{398}\sublabel{NW3gGP3e-3gnOun-1}\nwmargintag{{\nwtagstyle{}\subpageref{NW3gGP3e-3gnOun-1}}}\moddef{Centres and foci of parabolas~{\nwtagstyle{}\subpageref{NW3gGP3e-3gnOun-1}}}\endmoddef\Rm{}\nwstartdeflinemarkup\nwusesondefline{\\{NW3gGP3e-3gefqu-1}}\nwenddeflinemarkup
{\it{}asymptote} \begin{math}\ll\end{math} {\it{}endl} \begin{math}\ll\end{math} {\tt{}"erase();"} \begin{math}\ll\end{math} {\it{}endl};
{\it{}C1f} = {\bf{}cycle2D}(1, {\bf{}lst}(-1.5, 2), 3.75, {\it{}par\_matr});
{\it{}C2f} = {\bf{}cycle2D}(1, {\bf{}lst}(2, 2), -3.5, {\it{}par\_matr});
{\it{}C1f}.{\it{}asy\_draw}({\it{}asymptote}, {\it{}xmin}, {\it{}xmax}, {\it{}ymin}, {\it{}ymax}, {\bf{}lst}(0, 1.0-0.4, 0.4), {\tt{}"+.75"}, {\bf{}true}, 7);
{\it{}C2f}.{\it{}asy\_draw}({\it{}asymptote}, {\it{}xmin}, {\it{}xmax}, {\it{}ymin}, {\it{}ymax}, {\bf{}lst}(0, 1.0-0.4, 0.4), {\tt{}"+.75"}, {\bf{}true}, 7);

{\it{}asymptote} \begin{math}\ll\end{math} {\tt{}"pair[] z= {\char123}("} \begin{math}\ll\end{math} {\it{}C1f}.{\it{}center}(-{\it{}unit\_matrix}(2)).{\it{}op}(0) \begin{math}\ll\end{math} {\tt{}", "}  \begin{math}\ll\end{math} {\it{}C1f}.{\it{}center}(-{\it{}unit\_matrix}(2)).{\it{}op}(1)
 \begin{math}\ll\end{math} {\tt{}"), ("} \begin{math}\ll\end{math} {\it{}C2f}.{\it{}center}(-{\it{}unit\_matrix}(2)).{\it{}op}(0) \begin{math}\ll\end{math} {\tt{}", "}  \begin{math}\ll\end{math} {\it{}C2f}.{\it{}center}(-{\it{}unit\_matrix}(2)).{\it{}op}(1) \begin{math}\ll\end{math} {\tt{}"), "};
{\bf{}for} ({\bf{}int} {\it{}j} = -1; {\it{}j}\begin{math}<\end{math}2; {\it{}j}\protect\PP) {\nwlbrace}
 {\bf{}ex} {\it{}MS} = {\it{}diag\_matrix}({\bf{}lst}(-1, {\it{}j}));
 {\bf{}lst} {\it{}F1} =  {\it{}ex\_to}\begin{math}<\end{math}{\bf{}lst}\begin{math}>\end{math}({\it{}C1f}.{\it{}focus}({\it{}MS})),   {\it{}F2} = {\it{}ex\_to}\begin{math}<\end{math}{\bf{}lst}\begin{math}>\end{math}({\it{}C2f}.{\it{}focus}({\it{}MS}));
 {\it{}asymptote} \begin{math}\ll\end{math} {\tt{}"   ("} \begin{math}\ll\end{math} {\it{}F1}.{\it{}op}(0) \begin{math}\ll\end{math} {\tt{}", "}  \begin{math}\ll\end{math} {\it{}F1}.{\it{}op}(1) \begin{math}\ll\end{math} {\tt{}"), ("}
     \begin{math}\ll\end{math} {\it{}F2}.{\it{}op}(0) \begin{math}\ll\end{math} {\tt{}", "}  \begin{math}\ll\end{math} {\it{}F2}.{\it{}op}(1) \begin{math}\ll\end{math} {\tt{}")"} \begin{math}\ll\end{math} ({\it{}j}\begin{math}\equiv\end{math}1? {\tt{}"{\char125};"} : {\tt{}","} ) \begin{math}\ll\end{math} {\it{}endl};
{\nwrbrace}
{\it{}asymptote} \begin{math}\ll\end{math} {\tt{}" dot (z);"} \begin{math}\ll\end{math} {\it{}endl}
  \begin{math}\ll\end{math} {\tt{}" draw(z[0]--z[1], dashed);"} \begin{math}\ll\end{math} {\it{}endl};

{\it{}asymptote} \begin{math}\ll\end{math} {\tt{}"for (int j=1; j<3; ++j) {\char123}"} \begin{math}\ll\end{math} {\it{}endl}
 \begin{math}\ll\end{math} {\tt{}"  label({\char92}"$c\_e${\char92}", z[j-1], N);"} \begin{math}\ll\end{math} {\it{}endl}
 \begin{math}\ll\end{math} {\tt{}"  label({\char92}"$f\_e${\char92}", z[j+1], E);"} \begin{math}\ll\end{math} {\it{}endl}
 \begin{math}\ll\end{math} {\tt{}"  label({\char92}"$f\_p${\char92}", z[j+3], E);"} \begin{math}\ll\end{math} {\it{}endl}
 \begin{math}\ll\end{math} {\tt{}"  label({\char92}"$f\_h${\char92}", z[j+5], E);"} \begin{math}\ll\end{math} {\it{}endl}
 \begin{math}\ll\end{math} {\tt{}" draw(z[j+1]--z[j+5], dotted+0.5);"} \begin{math}\ll\end{math} {\it{}endl}
 \begin{math}\ll\end{math} {\tt{}"{\char125}"} \begin{math}\ll\end{math} {\it{}endl};
\LA{}Draw axes~{\nwtagstyle{}\subpageref{NW3gGP3e-ph2PF-1}}\RA{}
{\it{}asymptote} \begin{math}\ll\end{math} {\tt{}"shipout({\char92}"parab-cent{\char92}");"} \begin{math}\ll\end{math} {\it{}endl};

\nwused{\\{NW3gGP3e-3gefqu-1}}\nwidentuses{\\{{\nwixident{asy{\_}draw}}{asy:undraw}}\\{{\nwixident{center}}{center}}\\{{\nwixident{cycle2D}}{cycle2D}}\\{{\nwixident{ex}}{ex}}\\{{\nwixident{focus}}{focus}}\\{{\nwixident{lst}}{lst}}\\{{\nwixident{op}}{op}}\\{{\nwixident{par{\_}matr}}{par:unmatr}}}\nwindexuse{\nwixident{asy{\_}draw}}{asy:undraw}{NW3gGP3e-3gnOun-1}\nwindexuse{\nwixident{center}}{center}{NW3gGP3e-3gnOun-1}\nwindexuse{\nwixident{cycle2D}}{cycle2D}{NW3gGP3e-3gnOun-1}\nwindexuse{\nwixident{ex}}{ex}{NW3gGP3e-3gnOun-1}\nwindexuse{\nwixident{focus}}{focus}{NW3gGP3e-3gnOun-1}\nwindexuse{\nwixident{lst}}{lst}{NW3gGP3e-3gnOun-1}\nwindexuse{\nwixident{op}}{op}{NW3gGP3e-3gnOun-1}\nwindexuse{\nwixident{par{\_}matr}}{par:unmatr}{NW3gGP3e-3gnOun-1}\nwendcode{}\nwbegindocs{399}\nwdocspar
\subsubsection{Zero-radius cycles}
\label{sec:zer-radius-cycles}

Zero-radius cycles can look different in different EPH realisations,
here is an illustration.
\nwenddocs{}\nwbegincode{400}\sublabel{NW3gGP3e-QUZd0-1}\nwmargintag{{\nwtagstyle{}\subpageref{NW3gGP3e-QUZd0-1}}}\moddef{Zero-radius cycle implementations~{\nwtagstyle{}\subpageref{NW3gGP3e-QUZd0-1}}}\endmoddef\Rm{}\nwstartdeflinemarkup\nwusesondefline{\\{NW3gGP3e-3gefqu-1}}\nwenddeflinemarkup
{\it{}asymptote} \begin{math}\ll\end{math} {\it{}endl} \begin{math}\ll\end{math} {\tt{}"erase();"} \begin{math}\ll\end{math} {\it{}endl}
 \begin{math}\ll\end{math} {\tt{}"pair[] z;"} \begin{math}\ll\end{math} {\it{}endl};
{\nwlbrace}
 {\bf{}numeric} {\it{}xmin}(-5), {\it{}xmax}(15), {\it{}ymin}(-5), {\it{}ymax}(5);
 {\bf{}for} ({\bf{}int} {\it{}i1}=-1; {\it{}i1}\begin{math}<\end{math}2; {\it{}i1}\protect\PP) {\nwlbrace}
  {\bf{}for}({\bf{}int} {\it{}i2}=-1; {\it{}i2}\begin{math}<\end{math}2; {\it{}i2}\protect\PP) {\nwlbrace}
   {\bf{}lst} {\it{}val}({\it{}sign}\begin{math}\equiv\end{math}{\it{}i1}, {\it{}sign1}\begin{math}\equiv\end{math}{\it{}i2}, {\it{}u}\begin{math}\equiv\end{math}6\begin{math}\ast\end{math}{\it{}i1}+4, {\it{}v}\begin{math}\equiv\end{math}1.7);
   {\it{}Z1}.{\it{}subs}({\it{}val}).{\it{}asy\_draw}({\it{}asymptote}, {\it{}xmin}, {\it{}xmax}, {\it{}ymin}, {\it{}ymax}, {\bf{}lst}(0.5+0.4\begin{math}\ast\end{math}{\it{}i1}, .5-0.3\begin{math}\ast\end{math}{\it{}i2}, 0.5+0.3\begin{math}\ast\end{math}{\it{}i2}),{\tt{}""}, {\bf{}true}, 7);
   {\it{}asymptote} \begin{math}\ll\end{math} {\tt{}"dot(("} \begin{math}\ll\end{math} {\it{}ex\_to}\begin{math}<\end{math}{\bf{}numeric}\begin{math}>\end{math}({\it{}Z1}.{\it{}focus}({\it{}e}).{\it{}op}(0).{\it{}subs}({\it{}val})).{\it{}to\_double}()
       \begin{math}\ll\end{math} {\tt{}", "}\begin{math}\ll\end{math} {\it{}ex\_to}\begin{math}<\end{math}{\bf{}numeric}\begin{math}>\end{math}({\it{}Z1}.{\it{}focus}({\it{}e}).{\it{}op}(1).{\it{}subs}({\it{}val})).{\it{}to\_double}()
       \begin{math}\ll\end{math} {\tt{}"), "} \begin{math}\ll\end{math} 0.4+0.4\begin{math}\ast\end{math}{\it{}i1} \begin{math}\ll\end{math} {\tt{}"red+"}
       \begin{math}\ll\end{math}  .4-0.3\begin{math}\ast\end{math}{\it{}i2} \begin{math}\ll\end{math} {\tt{}"green+"}
       \begin{math}\ll\end{math} 0.6+0.3\begin{math}\ast\end{math}{\it{}i2} \begin{math}\ll\end{math} {\tt{}"blue);"} \begin{math}\ll\end{math} {\it{}endl};
  {\nwrbrace}
 {\nwrbrace}
 \LA{}Draw axes~{\nwtagstyle{}\subpageref{NW3gGP3e-ph2PF-1}}\RA{}
{\nwrbrace}
{\it{}asymptote} \begin{math}\ll\end{math} {\tt{}"shipout({\char92}"zero-cycles{\char92}");"} \begin{math}\ll\end{math} {\it{}endl};

\nwused{\\{NW3gGP3e-3gefqu-1}}\nwidentuses{\\{{\nwixident{asy{\_}draw}}{asy:undraw}}\\{{\nwixident{focus}}{focus}}\\{{\nwixident{lst}}{lst}}\\{{\nwixident{numeric}}{numeric}}\\{{\nwixident{op}}{op}}\\{{\nwixident{subs}}{subs}}\\{{\nwixident{u}}{u}}\\{{\nwixident{v}}{v}}\\{{\nwixident{val}}{val}}}\nwindexuse{\nwixident{asy{\_}draw}}{asy:undraw}{NW3gGP3e-QUZd0-1}\nwindexuse{\nwixident{focus}}{focus}{NW3gGP3e-QUZd0-1}\nwindexuse{\nwixident{lst}}{lst}{NW3gGP3e-QUZd0-1}\nwindexuse{\nwixident{numeric}}{numeric}{NW3gGP3e-QUZd0-1}\nwindexuse{\nwixident{op}}{op}{NW3gGP3e-QUZd0-1}\nwindexuse{\nwixident{subs}}{subs}{NW3gGP3e-QUZd0-1}\nwindexuse{\nwixident{u}}{u}{NW3gGP3e-QUZd0-1}\nwindexuse{\nwixident{v}}{v}{NW3gGP3e-QUZd0-1}\nwindexuse{\nwixident{val}}{val}{NW3gGP3e-QUZd0-1}\nwendcode{}\nwbegindocs{401}\nwdocspar
\subsubsection{Diameters of cycles}
\label{sec:diameters-cycles}

The notion of diameter and related distance became strange in parabolic case.
\nwenddocs{}\nwbegincode{402}\sublabel{NW3gGP3e-Y6OSa-1}\nwmargintag{{\nwtagstyle{}\subpageref{NW3gGP3e-Y6OSa-1}}}\moddef{Parabolic diameters~{\nwtagstyle{}\subpageref{NW3gGP3e-Y6OSa-1}}}\endmoddef\Rm{}\nwstartdeflinemarkup\nwusesondefline{\\{NW3gGP3e-3gefqu-1}}\nwenddeflinemarkup
{\it{}asymptote} \begin{math}\ll\end{math} {\it{}endl} \begin{math}\ll\end{math} {\tt{}"erase();"} \begin{math}\ll\end{math} {\it{}endl};
{\it{}C10} = {\bf{}cycle2D}(1, {\bf{}lst}((-4-1)\begin{math}\div\end{math}2.0, 0.5), 4,{\it{}par\_matr});
{\it{}C10}.{\it{}asy\_draw}({\it{}asymptote}, {\it{}xmin}, {\it{}xmax}, {\it{}ymin}, {\it{}ymax}, {\bf{}lst}(0.1, 0, 0.6));
{\it{}asymptote} \begin{math}\ll\end{math} {\tt{}"pair[] z = {\char123}("} \begin{math}\ll\end{math} {\it{}C10}.{\it{}roots}().{\it{}op}(0) \begin{math}\ll\end{math} {\tt{}", 0), ("} \begin{math}\ll\end{math} {\it{}C10}.{\it{}roots}().{\it{}op}(1) \begin{math}\ll\end{math} {\tt{}", 0){\char125};"} \begin{math}\ll\end{math} {\it{}endl};
{\bf{}cycle2D}(1, {\bf{}lst}(5\begin{math}\div\end{math}2.0, 0.5), 8,{\it{}par\_matr}).{\it{}asy\_draw}({\it{}asymptote}, {\it{}xmin}, {\it{}xmax}, {\it{}ymin}, {\it{}ymax},\nwindexdefn{\nwixident{cycle2D}}{cycle2D}{NW3gGP3e-Y6OSa-1}
                  {\bf{}lst}(0.1, 0.6, 0), {\tt{}""}, {\bf{}true}, 7);
{\it{}C10} ={\bf{}cycle2D}(-1, {\bf{}lst}(-5\begin{math}\div\end{math}2.0, 0.5), 8-5.0\begin{math}\ast\end{math}5\begin{math}\div\end{math}2.0,{\it{}par\_matr});
{\it{}C10}.{\it{}asy\_draw}({\it{}asymptote}, {\it{}xmin}, {\it{}xmax}, {\it{}ymin}, {\it{}ymax}, {\bf{}lst}(0.1, 0.6, 0),
{\tt{}"+dashed "}, {\bf{}true}, 7);
{\it{}asymptote} \begin{math}\ll\end{math} {\tt{}"z.push(("} \begin{math}\ll\end{math} {\it{}C10}.{\it{}roots}().{\it{}op}(1) \begin{math}\ll\end{math} {\tt{}", 0)); z.push(("} \begin{math}\ll\end{math} {\it{}C10}.{\it{}roots}().{\it{}op}(0) \begin{math}\ll\end{math} {\tt{}", 0));"} \begin{math}\ll\end{math} {\it{}endl};
\LA{}Put labels on 22-23~{\nwtagstyle{}\subpageref{NW3gGP3e-3wMlZE-1}}\RA{}
\LA{}Draw axes~{\nwtagstyle{}\subpageref{NW3gGP3e-ph2PF-1}}\RA{}
{\it{}asymptote} \begin{math}\ll\end{math} {\tt{}"shipout({\char92}"parab-diam{\char92}");"} \begin{math}\ll\end{math} {\it{}endl};

\nwused{\\{NW3gGP3e-3gefqu-1}}\nwidentdefs{\\{{\nwixident{cycle2D}}{cycle2D}}}\nwidentuses{\\{{\nwixident{asy{\_}draw}}{asy:undraw}}\\{{\nwixident{lst}}{lst}}\\{{\nwixident{op}}{op}}\\{{\nwixident{par{\_}matr}}{par:unmatr}}\\{{\nwixident{roots}}{roots}}}\nwindexuse{\nwixident{asy{\_}draw}}{asy:undraw}{NW3gGP3e-Y6OSa-1}\nwindexuse{\nwixident{lst}}{lst}{NW3gGP3e-Y6OSa-1}\nwindexuse{\nwixident{op}}{op}{NW3gGP3e-Y6OSa-1}\nwindexuse{\nwixident{par{\_}matr}}{par:unmatr}{NW3gGP3e-Y6OSa-1}\nwindexuse{\nwixident{roots}}{roots}{NW3gGP3e-Y6OSa-1}\nwendcode{}\nwbegindocs{403}Here is the common part of drawing points and labels on the figures 22-23.
\nwenddocs{}\nwbegincode{404}\sublabel{NW3gGP3e-3wMlZE-1}\nwmargintag{{\nwtagstyle{}\subpageref{NW3gGP3e-3wMlZE-1}}}\moddef{Put labels on 22-23~{\nwtagstyle{}\subpageref{NW3gGP3e-3wMlZE-1}}}\endmoddef\Rm{}\nwstartdeflinemarkup\nwusesondefline{\\{NW3gGP3e-Y6OSa-1}\\{NW3gGP3e-1g9SsP-1}}\nwenddeflinemarkup
{\it{}asymptote}  \begin{math}\ll\end{math} {\tt{}"z.push((z[2].x,0)); z.push((z[3].x,0));"} \begin{math}\ll\end{math} {\it{}endl}
  \begin{math}\ll\end{math} {\tt{}" dot(z);"} \begin{math}\ll\end{math} {\it{}endl}
  \begin{math}\ll\end{math} {\tt{}" draw(z[2]--z[3], black+.3);"} \begin{math}\ll\end{math} {\it{}endl}
  \begin{math}\ll\end{math} {\tt{}" draw(z[0]--z[1], black+1.2);"} \begin{math}\ll\end{math} {\it{}endl}
  \begin{math}\ll\end{math} {\tt{}" draw(z[4]--z[5], black+1.2);"} \begin{math}\ll\end{math} {\it{}endl}
  \begin{math}\ll\end{math} {\tt{}"  label({\char92}"$z\_1${\char92}", z[0], NW);"} \begin{math}\ll\end{math} {\it{}endl}
  \begin{math}\ll\end{math} {\tt{}"  label({\char92}"$z\_2${\char92}", z[1], SE);"} \begin{math}\ll\end{math} {\it{}endl}
  \begin{math}\ll\end{math} {\tt{}"  label({\char92}"$z\_3${\char92}", z[2], SW);"} \begin{math}\ll\end{math} {\it{}endl}
  \begin{math}\ll\end{math} {\tt{}"  label({\char92}"$z\_4${\char92}", z[3], SE);"} \begin{math}\ll\end{math} {\it{}endl};

\nwused{\\{NW3gGP3e-Y6OSa-1}\\{NW3gGP3e-1g9SsP-1}}\nwendcode{}\nwbegindocs{405}\nwdocspar
\subsubsection{Extremal property of the distance}
\label{sec:extr-prop-dist}

To illustrate the variational definition of the
distance~\cite[Defn.\ref{E-de:distance}]{Kisil05b}
we draw several cycles  which passes
two given points. The cycles with the extremal value of diameter
is highlighted in bold.
\nwenddocs{}\nwbegincode{406}\sublabel{NW3gGP3e-1g9SsP-1}\nwmargintag{{\nwtagstyle{}\subpageref{NW3gGP3e-1g9SsP-1}}}\moddef{Distance as an extremum~{\nwtagstyle{}\subpageref{NW3gGP3e-1g9SsP-1}}}\endmoddef\Rm{}\nwstartdeflinemarkup\nwusesondefline{\\{NW3gGP3e-3gefqu-1}}\nwenddeflinemarkup
{\it{}asymptote} \begin{math}\ll\end{math} {\it{}endl} \begin{math}\ll\end{math} {\tt{}"erase();"} \begin{math}\ll\end{math} {\it{}endl};
{\bf{}for} ({\bf{}int} {\it{}j}=-2; {\it{}j} \begin{math}<\end{math} 3; {\it{}j}\protect\PP) {\nwlbrace}
 {\it{}ex\_to}\begin{math}<\end{math}{\bf{}cycle2D}\begin{math}>\end{math}({\it{}C}.{\it{}subject\_to}({\bf{}lst}({\it{}C}.{\it{}passing}({\bf{}lst}({\it{}xmin}+1, {\it{}ymax}-5)), {\it{}C}.{\it{}passing}({\bf{}lst}({\it{}xmin}+3, {\it{}ymax}-6.5)), {\it{}k} \begin{math}\equiv\end{math} 1,
         {\it{}l} \begin{math}\equiv\end{math} {\it{}xmin}+2+0.5\begin{math}\ast\end{math}{\it{}j})).{\it{}subs}({\it{}sign} \begin{math}\equiv\end{math} -1)).{\it{}asy\_draw}({\it{}asymptote}, {\it{}xmin}, {\it{}xmax}, {\it{}ymin}, {\it{}ymax},
                      {\bf{}lst}(0, 0.4\begin{math}\ast\end{math}{\it{}abs}({\it{}j}), 1.0-0.4\begin{math}\ast\end{math}{\it{}abs}({\it{}j})), ({\it{}j} \begin{math}\equiv\end{math} 0 ? {\tt{}"+1"} : {\tt{}"+.3"}));
 {\it{}ex\_to}\begin{math}<\end{math}{\bf{}cycle2D}\begin{math}>\end{math}({\it{}C}.{\it{}subject\_to}({\bf{}lst}({\it{}C}.{\it{}passing}({\bf{}lst}({\it{}xmax}-4, {\it{}ymax}-5)), {\it{}C}.{\it{}passing}({\bf{}lst}({\it{}xmax}-1, {\it{}ymax}-2)), {\it{}k} \begin{math}\equiv\end{math} 1,
         {\it{}l} \begin{math}\equiv\end{math} {\it{}xmax}-2.5-0.2\begin{math}\ast\end{math}({\it{}j}+2))).{\it{}subs}({\it{}sign} \begin{math}\equiv\end{math} 0)).{\it{}asy\_draw}({\it{}asymptote}, {\it{}xmin}, {\it{}xmax}, {\it{}ymin}, {\it{}ymax},
                        {\bf{}lst}(0.2\begin{math}\ast\end{math}({\it{}j}+2), 0, 1.0-0.2\begin{math}\ast\end{math}({\it{}j}+2)), ({\it{}j} \begin{math}\equiv\end{math} -2 ? {\tt{}"+1"} : {\tt{}"+.3"}), {\bf{}true}, 7);
{\nwrbrace}
{\it{}asymptote} \begin{math}\ll\end{math} {\tt{}"pair[] z ={\char123} ("} \begin{math}\ll\end{math} {\it{}xmin}+1 \begin{math}\ll\end{math} {\tt{}", "} \begin{math}\ll\end{math} {\it{}ymax}-5 \begin{math}\ll\end{math} {\tt{}"),  ("} \begin{math}\ll\end{math} {\it{}xmin}+3 \begin{math}\ll\end{math} {\tt{}", "}
  \begin{math}\ll\end{math} {\it{}ymax}-6.5 \begin{math}\ll\end{math} {\tt{}"),  ("} \begin{math}\ll\end{math} {\it{}xmax}-4 \begin{math}\ll\end{math} {\tt{}", "} \begin{math}\ll\end{math} {\it{}ymax}-5 \begin{math}\ll\end{math} {\tt{}"),  ("} \begin{math}\ll\end{math} {\it{}xmax}-1
  \begin{math}\ll\end{math} {\tt{}", "} \begin{math}\ll\end{math} {\it{}ymax}-2 \begin{math}\ll\end{math} {\tt{}"){\char125};"} \begin{math}\ll\end{math} {\it{}endl};
\LA{}Put labels on 22-23~{\nwtagstyle{}\subpageref{NW3gGP3e-3wMlZE-1}}\RA{}
{\it{}asymptote} \begin{math}\ll\end{math} {\tt{}"  label({\char92}"$d\_e${\char92}", .5z[0]+.5z[1], NE);"} \begin{math}\ll\end{math} {\it{}endl}
   \begin{math}\ll\end{math} {\tt{}"  label({\char92}"$d\_p${\char92}", .5z[4]+.5z[5], S);"} \begin{math}\ll\end{math} {\it{}endl};
\LA{}Draw axes~{\nwtagstyle{}\subpageref{NW3gGP3e-ph2PF-1}}\RA{}
{\it{}asymptote} \begin{math}\ll\end{math} {\tt{}"shipout({\char92}"dist-extr{\char92}");"} \begin{math}\ll\end{math} {\it{}endl};

\nwused{\\{NW3gGP3e-3gefqu-1}}\nwidentuses{\\{{\nwixident{asy{\_}draw}}{asy:undraw}}\\{{\nwixident{cycle2D}}{cycle2D}}\\{{\nwixident{k}}{k}}\\{{\nwixident{l}}{l}}\\{{\nwixident{lst}}{lst}}\\{{\nwixident{passing}}{passing}}\\{{\nwixident{subject{\_}to}}{subject:unto}}\\{{\nwixident{subs}}{subs}}}\nwindexuse{\nwixident{asy{\_}draw}}{asy:undraw}{NW3gGP3e-1g9SsP-1}\nwindexuse{\nwixident{cycle2D}}{cycle2D}{NW3gGP3e-1g9SsP-1}\nwindexuse{\nwixident{k}}{k}{NW3gGP3e-1g9SsP-1}\nwindexuse{\nwixident{l}}{l}{NW3gGP3e-1g9SsP-1}\nwindexuse{\nwixident{lst}}{lst}{NW3gGP3e-1g9SsP-1}\nwindexuse{\nwixident{passing}}{passing}{NW3gGP3e-1g9SsP-1}\nwindexuse{\nwixident{subject{\_}to}}{subject:unto}{NW3gGP3e-1g9SsP-1}\nwindexuse{\nwixident{subs}}{subs}{NW3gGP3e-1g9SsP-1}\nwendcode{}\nwbegindocs{407}\nwdocspar
\subsubsection{Infinitesimal cycles}
\label{sec:infinitesimal-cycles}

Here we draw a set of parabola with the same focus and the focal length
tensing to zero.
\nwenddocs{}\nwbegincode{408}\sublabel{NW3gGP3e-FS4xg-1}\nwmargintag{{\nwtagstyle{}\subpageref{NW3gGP3e-FS4xg-1}}}\moddef{Infinitesimal cycles draw~{\nwtagstyle{}\subpageref{NW3gGP3e-FS4xg-1}}}\endmoddef\Rm{}\nwstartdeflinemarkup\nwusesondefline{\\{NW3gGP3e-3gefqu-1}}\nwenddeflinemarkup
{\it{}asymptote} \begin{math}\ll\end{math} {\it{}endl} \begin{math}\ll\end{math} {\tt{}"erase();"}             \begin{math}\ll\end{math} {\it{}endl};
{\bf{}for} ({\bf{}int} {\it{}j}=1; {\it{}j} \begin{math}<\end{math} 5; {\it{}j}\protect\PP) {\nwlbrace}
 {\bf{}cycle2D}({\bf{}lst}(-2.5, 4.5), -{\it{}unit\_matrix}(2), 16.0\begin{math}\ast\end{math}{\it{}GiNaC}::{\it{}pow}(2, -2\begin{math}\ast\end{math}{\it{}j})).{\it{}asy\_draw}({\it{}asymptote}, {\it{}xmin}, {\it{}xmax}, {\it{}ymin}, {\it{}ymax},
                     {\bf{}lst}(0, 0.2\begin{math}\ast\end{math}{\it{}abs}({\it{}j}), 1.0-0.2\begin{math}\ast\end{math}{\it{}abs}({\it{}j})), {\tt{}"+.3"});
 {\bf{}cycle2D}({\bf{}lst}(1, 1.25), {\it{}hyp\_matr}, 25\begin{math}\ast\end{math}{\it{}GiNaC}::{\it{}pow}(1.8, -2\begin{math}\ast\end{math}{\it{}j})).{\it{}asy\_draw}({\it{}asymptote}, {\it{}xmin}, {\it{}xmax}, {\it{}ymin}, {\it{}ymax}\begin{math}\div\end{math}3,
                     {\bf{}lst}(0.2\begin{math}\ast\end{math}{\it{}abs}({\it{}j}), 1.0-0.2\begin{math}\ast\end{math}{\it{}abs}({\it{}j}), 0), {\tt{}"+.3"}, {\bf{}true}, 5+{\it{}j});
 {\bf{}cycle2D}(1, {\bf{}lst}(2, {\it{}GiNaC}::{\it{}pow}(3,-{\it{}j})), 2\begin{math}\ast\end{math}2+2.0\begin{math}\ast\end{math}{\it{}GiNaC}::{\it{}pow}(3,-{\it{}j})-{\it{}GiNaC}::{\it{}pow}(3,-2\begin{math}\ast\end{math}{\it{}j}), {\it{}par\_matr}).{\it{}asy\_draw}({\it{}asymptote}, {\it{}xmin},
                          {\it{}xmax}, {\it{}ymin}, {\it{}ymax}, {\bf{}lst}(1.0-0.17\begin{math}\ast\end{math}{\it{}j}, 0, 0.17\begin{math}\ast\end{math}{\it{}j}), {\tt{}"+.3"}, {\bf{}true}, 7);
{\nwrbrace}
{\it{}asymptote}  \begin{math}\ll\end{math} {\tt{}" draw((2,1)--(2,"} \begin{math}\ll\end{math} {\it{}ymax} \begin{math}\ll\end{math} {\tt{}"), blue+1);"} \begin{math}\ll\end{math} {\it{}endl};
{\bf{}cycle2D}({\bf{}lst}(1, 1.25), {\it{}hyp\_matr}).{\it{}asy\_draw}({\it{}asymptote}, {\it{}xmin}, {\it{}xmax}, {\it{}ymin}, {\it{}ymax}\begin{math}\div\end{math}3, {\bf{}lst}(1, 0, 0), {\tt{}"+1"});\nwindexdefn{\nwixident{cycle2D}}{cycle2D}{NW3gGP3e-FS4xg-1}
{\it{}asymptote}  \begin{math}\ll\end{math} {\tt{}" dot((-2.5,4.5));"} \begin{math}\ll\end{math} {\it{}endl}
  \begin{math}\ll\end{math} {\tt{}" dot((2,1));"} \begin{math}\ll\end{math} {\it{}endl};
\LA{}Draw axes~{\nwtagstyle{}\subpageref{NW3gGP3e-ph2PF-1}}\RA{}
{\it{}asymptote} \begin{math}\ll\end{math} {\tt{}"shipout({\char92}"infinites{\char92}");"} \begin{math}\ll\end{math} {\it{}endl};

\nwused{\\{NW3gGP3e-3gefqu-1}}\nwidentdefs{\\{{\nwixident{cycle2D}}{cycle2D}}}\nwidentuses{\\{{\nwixident{asy{\_}draw}}{asy:undraw}}\\{{\nwixident{hyp{\_}matr}}{hyp:unmatr}}\\{{\nwixident{lst}}{lst}}\\{{\nwixident{par{\_}matr}}{par:unmatr}}}\nwindexuse{\nwixident{asy{\_}draw}}{asy:undraw}{NW3gGP3e-FS4xg-1}\nwindexuse{\nwixident{hyp{\_}matr}}{hyp:unmatr}{NW3gGP3e-FS4xg-1}\nwindexuse{\nwixident{lst}}{lst}{NW3gGP3e-FS4xg-1}\nwindexuse{\nwixident{par{\_}matr}}{par:unmatr}{NW3gGP3e-FS4xg-1}\nwendcode{}\nwbegindocs{409}\nwdocspar
\subsubsection{Pictures of the Cayley transform}
\label{sec:pict-cayl-transf}

We draw now pictures of Cayley transform, which shows that the unit
cycle {\Tt{}\Rm{}{\it{}UC}\nwendquote} may be obtained as a reflection of the real line into the cycle {\Tt{}\Rm{}{\it{}C10f}\nwendquote}.
\nwenddocs{}\nwbegincode{410}\sublabel{NW3gGP3e-q6Aul-1}\nwmargintag{{\nwtagstyle{}\subpageref{NW3gGP3e-q6Aul-1}}}\moddef{Cayley transform pictures~{\nwtagstyle{}\subpageref{NW3gGP3e-q6Aul-1}}}\endmoddef\Rm{}\nwstartdeflinemarkup\nwusesondefline{\\{NW3gGP3e-3gefqu-1}}\nwprevnextdefs{\relax}{NW3gGP3e-q6Aul-2}\nwenddeflinemarkup
{\it{}xmin} = -{\bf{}numeric}(4,2); {\it{}xmax}={\bf{}numeric}(4,2); {\it{}ymin}=-{\bf{}numeric}(7,2); {\it{}ymax}={\bf{}numeric}(3);
{\bf{}cycle2D} {\it{}C10f}, {\it{}UC};
{\it{}C10f}  = {\bf{}cycle2D}(1, {\bf{}lst}(0, {\it{}sign2}), {\it{}sign}, {\it{}e});
{\it{}UC}={\it{}real\_line}.{\it{}cycle\_similarity}({\it{}C10f}, {\it{}es}).{\it{}normalize}();

\nwalsodefined{\\{NW3gGP3e-q6Aul-2}\\{NW3gGP3e-q6Aul-3}\\{NW3gGP3e-q6Aul-4}\\{NW3gGP3e-q6Aul-5}}\nwused{\\{NW3gGP3e-3gefqu-1}}\nwidentuses{\\{{\nwixident{cycle2D}}{cycle2D}}\\{{\nwixident{cycle{\_}similarity}}{cycle:unsimilarity}}\\{{\nwixident{lst}}{lst}}\\{{\nwixident{normalize}}{normalize}}\\{{\nwixident{numeric}}{numeric}}}\nwindexuse{\nwixident{cycle2D}}{cycle2D}{NW3gGP3e-q6Aul-1}\nwindexuse{\nwixident{cycle{\_}similarity}}{cycle:unsimilarity}{NW3gGP3e-q6Aul-1}\nwindexuse{\nwixident{lst}}{lst}{NW3gGP3e-q6Aul-1}\nwindexuse{\nwixident{normalize}}{normalize}{NW3gGP3e-q6Aul-1}\nwindexuse{\nwixident{numeric}}{numeric}{NW3gGP3e-q6Aul-1}\nwendcode{}\nwbegindocs{411}Now we run cycles over signatures of point and cycle spaces and sign of {\Tt{}\Rm{}{\it{}sign2}\nwendquote}.
\nwenddocs{}\nwbegincode{412}\sublabel{NW3gGP3e-q6Aul-2}\nwmargintag{{\nwtagstyle{}\subpageref{NW3gGP3e-q6Aul-2}}}\moddef{Cayley transform pictures~{\nwtagstyle{}\subpageref{NW3gGP3e-q6Aul-1}}}\plusendmoddef\Rm{}\nwstartdeflinemarkup\nwusesondefline{\\{NW3gGP3e-3gefqu-1}}\nwprevnextdefs{NW3gGP3e-q6Aul-1}{NW3gGP3e-q6Aul-3}\nwenddeflinemarkup
{\bf{}for} ({\it{}si}=-1; {\it{}si}\begin{math}<\end{math}2; {\it{}si}\protect\PP) {\nwlbrace}
 {\bf{}for} ({\it{}si1}=-1; {\it{}si1}\begin{math}<\end{math}2; {\it{}si1}\protect\PP)
  {\bf{}if} (({\it{}si} \begin{math}\equiv\end{math}0 ) \begin{math}\vee\end{math} ({\it{}si} \begin{math}\equiv\end{math} {\it{}si1})) {\nwlbrace}
   {\it{}asymptote} \begin{math}\ll\end{math} {\it{}endl} \begin{math}\ll\end{math} {\tt{}"erase();"} \begin{math}\ll\end{math} {\it{}endl};
   {\bf{}for} ({\bf{}int} {\it{}si2}=-1; {\it{}si2}\begin{math}<\end{math}2; {\it{}si2}={\it{}si2}+2) {\nwlbrace}
    {\bf{}lst} {\it{}cycle\_val} = {\bf{}lst}({\it{}sign} \begin{math}\equiv\end{math} {\it{}si}, {\it{}sign1} \begin{math}\equiv\end{math} {\it{}si1}, {\it{}sign2}\begin{math}\equiv\end{math}{\it{}si2});

\nwused{\\{NW3gGP3e-3gefqu-1}}\nwidentuses{\\{{\nwixident{lst}}{lst}}\\{{\nwixident{si}}{si}}\\{{\nwixident{si1}}{si1}}}\nwindexuse{\nwixident{lst}}{lst}{NW3gGP3e-q6Aul-2}\nwindexuse{\nwixident{si}}{si}{NW3gGP3e-q6Aul-2}\nwindexuse{\nwixident{si1}}{si1}{NW3gGP3e-q6Aul-2}\nwendcode{}\nwbegindocs{413}If point space is not parabolic,  the unit cycle {\Tt{}\Rm{}{\it{}UC}\nwendquote} is the reflection of real line
in  {\Tt{}\Rm{}{\it{}C10f}\nwendquote} and we draw both of them.
\nwenddocs{}\nwbegincode{414}\sublabel{NW3gGP3e-q6Aul-3}\nwmargintag{{\nwtagstyle{}\subpageref{NW3gGP3e-q6Aul-3}}}\moddef{Cayley transform pictures~{\nwtagstyle{}\subpageref{NW3gGP3e-q6Aul-1}}}\plusendmoddef\Rm{}\nwstartdeflinemarkup\nwusesondefline{\\{NW3gGP3e-3gefqu-1}}\nwprevnextdefs{NW3gGP3e-q6Aul-2}{NW3gGP3e-q6Aul-4}\nwenddeflinemarkup
    {\bf{}if} ({\it{}si} \begin{math}\neq\end{math} 0 ) {\nwlbrace}
     {\it{}ex\_to}\begin{math}<\end{math}{\bf{}cycle2D}\begin{math}>\end{math}({\it{}UC}.{\it{}subs}({\it{}cycle\_val}, {\it{}subs\_options}::{\it{}algebraic} \begin{math}\mid\end{math} {\it{}subs\_options}::{\it{}no\_pattern}))
      .{\it{}asy\_draw}({\it{}asymptote}, {\it{}xmin}, {\it{}xmax}, {\it{}ymin}, {\it{}ymax}, {\bf{}lst}(0, 0, 0.7), {\tt{}"+1.5"}, {\bf{}true}, 7);
     {\it{}C10f}.{\it{}subs}({\it{}cycle\_val}, {\it{}subs\_options}::{\it{}algebraic} \begin{math}\mid\end{math} {\it{}subs\_options}::{\it{}no\_pattern}).{\it{}normalize}()
      .{\it{}asy\_draw}({\it{}asymptote}, {\it{}xmin}, {\it{}xmax}, {\it{}ymin}, {\it{}ymax}, {\bf{}lst}(0, 0.7, 0), ({\it{}si2} \begin{math}\equiv\end{math}{\it{}si1} ? {\tt{}"+1"} : {\tt{}"+Dotted "}), {\bf{}true}, 7);

\nwused{\\{NW3gGP3e-3gefqu-1}}\nwidentuses{\\{{\nwixident{asy{\_}draw}}{asy:undraw}}\\{{\nwixident{cycle2D}}{cycle2D}}\\{{\nwixident{lst}}{lst}}\\{{\nwixident{normalize}}{normalize}}\\{{\nwixident{si}}{si}}\\{{\nwixident{si1}}{si1}}\\{{\nwixident{subs}}{subs}}}\nwindexuse{\nwixident{asy{\_}draw}}{asy:undraw}{NW3gGP3e-q6Aul-3}\nwindexuse{\nwixident{cycle2D}}{cycle2D}{NW3gGP3e-q6Aul-3}\nwindexuse{\nwixident{lst}}{lst}{NW3gGP3e-q6Aul-3}\nwindexuse{\nwixident{normalize}}{normalize}{NW3gGP3e-q6Aul-3}\nwindexuse{\nwixident{si}}{si}{NW3gGP3e-q6Aul-3}\nwindexuse{\nwixident{si1}}{si1}{NW3gGP3e-q6Aul-3}\nwindexuse{\nwixident{subs}}{subs}{NW3gGP3e-q6Aul-3}\nwendcode{}\nwbegindocs{415}In the parabolic space unit cycle obtained from the real line by {\Tt{}\Rm{}{\it{}cayley\_parab}()\nwendquote} procedure.
\nwenddocs{}\nwbegincode{416}\sublabel{NW3gGP3e-q6Aul-4}\nwmargintag{{\nwtagstyle{}\subpageref{NW3gGP3e-q6Aul-4}}}\moddef{Cayley transform pictures~{\nwtagstyle{}\subpageref{NW3gGP3e-q6Aul-1}}}\plusendmoddef\Rm{}\nwstartdeflinemarkup\nwusesondefline{\\{NW3gGP3e-3gefqu-1}}\nwprevnextdefs{NW3gGP3e-q6Aul-3}{NW3gGP3e-q6Aul-5}\nwenddeflinemarkup
    {\nwrbrace} {\bf{}else}
     {\it{}ex\_to}\begin{math}<\end{math}{\bf{}cycle2D}\begin{math}>\end{math}({\it{}cayley\_parab}({\it{}real\_line},{\it{}sign1}).{\it{}subs}({\it{}cycle\_val}, {\it{}subs\_options}::{\it{}algebraic} \begin{math}\mid\end{math} {\it{}subs\_options}::{\it{}no\_pattern}))
      .{\it{}asy\_draw}({\it{}asymptote}, {\it{}xmin}, {\it{}xmax}, {\it{}ymin}, {\it{}ymax}, {\bf{}lst}(0, 0, 0.7), {\tt{}"+1.5"}, {\bf{}true}, 7);
   {\nwrbrace}

\nwused{\\{NW3gGP3e-3gefqu-1}}\nwidentuses{\\{{\nwixident{asy{\_}draw}}{asy:undraw}}\\{{\nwixident{cycle2D}}{cycle2D}}\\{{\nwixident{lst}}{lst}}\\{{\nwixident{subs}}{subs}}}\nwindexuse{\nwixident{asy{\_}draw}}{asy:undraw}{NW3gGP3e-q6Aul-4}\nwindexuse{\nwixident{cycle2D}}{cycle2D}{NW3gGP3e-q6Aul-4}\nwindexuse{\nwixident{lst}}{lst}{NW3gGP3e-q6Aul-4}\nwindexuse{\nwixident{subs}}{subs}{NW3gGP3e-q6Aul-4}\nwendcode{}\nwbegindocs{417}The pictures are finished with standard stuff.
\nwenddocs{}\nwbegincode{418}\sublabel{NW3gGP3e-q6Aul-5}\nwmargintag{{\nwtagstyle{}\subpageref{NW3gGP3e-q6Aul-5}}}\moddef{Cayley transform pictures~{\nwtagstyle{}\subpageref{NW3gGP3e-q6Aul-1}}}\plusendmoddef\Rm{}\nwstartdeflinemarkup\nwusesondefline{\\{NW3gGP3e-3gefqu-1}}\nwprevnextdefs{NW3gGP3e-q6Aul-4}{\relax}\nwenddeflinemarkup
   \LA{}Put units~{\nwtagstyle{}\subpageref{NW3gGP3e-1r99tr-1}}\RA{}
      \LA{}Draw axes~{\nwtagstyle{}\subpageref{NW3gGP3e-ph2PF-1}}\RA{}
   {\it{}asymptote} \begin{math}\ll\end{math} {\tt{}"shipout({\char92}"cayley-"}\begin{math}\ll\end{math} {\it{}eph\_names}[{\it{}si}+1] \begin{math}\ll\end{math} {\it{}eph\_names}[{\it{}si1}+1]\begin{math}\ll\end{math}{\tt{}"{\char92}");"} \begin{math}\ll\end{math} {\it{}endl};
  {\nwrbrace}
{\nwrbrace}

\nwused{\\{NW3gGP3e-3gefqu-1}}\nwidentuses{\\{{\nwixident{si}}{si}}\\{{\nwixident{si1}}{si1}}}\nwindexuse{\nwixident{si}}{si}{NW3gGP3e-q6Aul-5}\nwindexuse{\nwixident{si1}}{si1}{NW3gGP3e-q6Aul-5}\nwendcode{}\nwbegindocs{419}\nwdocspar
\subsubsection{Three types of inversions}
\label{sec:three-types-invers}

We draw here pictures for three types of the inversions. First we
make a rectangular grid.
\nwenddocs{}\nwbegincode{420}\sublabel{NW3gGP3e-3xVrpR-1}\nwmargintag{{\nwtagstyle{}\subpageref{NW3gGP3e-3xVrpR-1}}}\moddef{Three inversions~{\nwtagstyle{}\subpageref{NW3gGP3e-3xVrpR-1}}}\endmoddef\Rm{}\nwstartdeflinemarkup\nwusesondefline{\\{NW3gGP3e-3gefqu-1}}\nwprevnextdefs{\relax}{NW3gGP3e-3xVrpR-2}\nwenddeflinemarkup
{\it{}xmin}=-2; {\it{}xmax}=2; {\it{}ymin}=-2; {\it{}ymax}=2;
{\it{}C2}={\bf{}cycle2D}({\bf{}lst}(0,(1-{\it{}abs}({\it{}sign}))\begin{math}\div\end{math}2),{\it{}e}, 1);
{\it{}C3}={\bf{}cycle2D}(0,{\bf{}lst}({\it{}l},{\it{}n}),{\it{}m},{\it{}e});
{\it{}asymptote} \begin{math}\ll\end{math} {\it{}endl} \begin{math}\ll\end{math} {\tt{}"erase(); u=1cm;"} \begin{math}\ll\end{math} {\it{}endl};
{\bf{}for}({\bf{}double} {\it{}i}=-4; {\it{}i}\begin{math}\leq\end{math}4; {\it{}i}+=.4) {\nwlbrace}
 {\it{}C3}.{\it{}subs}({\bf{}lst}({\it{}sign}\begin{math}\equiv\end{math}-1, {\it{}l}\begin{math}\equiv\end{math}0, {\it{}n}\begin{math}\equiv\end{math}1, {\it{}m}\begin{math}\equiv\end{math}{\it{}i})).{\it{}asy\_draw}(
  {\it{}asymptote}, {\it{}xmin}, {\it{}xmax}, {\it{}ymin}, {\it{}ymax}, {\bf{}lst}(0.5, .75, 0.5),{\tt{}"+0.25pt"}, {\bf{}true}, 7);
 {\it{}C3}.{\it{}subs}({\bf{}lst}({\it{}sign}\begin{math}\equiv\end{math}-1, {\it{}l}\begin{math}\equiv\end{math}1, {\it{}n}\begin{math}\equiv\end{math}0, {\it{}m}\begin{math}\equiv\end{math}{\it{}i})).{\it{}asy\_draw}(
  {\it{}asymptote}, {\it{}xmin}, {\it{}xmax}, {\it{}ymin}, {\it{}ymax}, {\bf{}lst}(0.5, .5, 0.75),{\tt{}"+0.25pt"}, {\bf{}true}, 7);
{\nwrbrace}
{\it{}C2}.{\it{}subs}({\it{}sign}\begin{math}\equiv\end{math}-1).{\it{}asy\_draw}({\it{}asymptote}, {\it{}xmin}, {\it{}xmax}, {\it{}ymin}, {\it{}ymax}, {\bf{}lst}(1,0,0),{\tt{}"+.75pt"}, {\bf{}true}, 7);
\LA{}Draw axes~{\nwtagstyle{}\subpageref{NW3gGP3e-ph2PF-1}}\RA{}
{\it{}asymptote} \begin{math}\ll\end{math} {\tt{}"shipout({\char92}"pre-invers{\char92}");"} \begin{math}\ll\end{math} {\it{}endl};

\nwalsodefined{\\{NW3gGP3e-3xVrpR-2}\\{NW3gGP3e-3xVrpR-3}}\nwused{\\{NW3gGP3e-3gefqu-1}}\nwidentuses{\\{{\nwixident{asy{\_}draw}}{asy:undraw}}\\{{\nwixident{cycle2D}}{cycle2D}}\\{{\nwixident{l}}{l}}\\{{\nwixident{lst}}{lst}}\\{{\nwixident{m}}{m}}\\{{\nwixident{subs}}{subs}}\\{{\nwixident{u}}{u}}}\nwindexuse{\nwixident{asy{\_}draw}}{asy:undraw}{NW3gGP3e-3xVrpR-1}\nwindexuse{\nwixident{cycle2D}}{cycle2D}{NW3gGP3e-3xVrpR-1}\nwindexuse{\nwixident{l}}{l}{NW3gGP3e-3xVrpR-1}\nwindexuse{\nwixident{lst}}{lst}{NW3gGP3e-3xVrpR-1}\nwindexuse{\nwixident{m}}{m}{NW3gGP3e-3xVrpR-1}\nwindexuse{\nwixident{subs}}{subs}{NW3gGP3e-3xVrpR-1}\nwindexuse{\nwixident{u}}{u}{NW3gGP3e-3xVrpR-1}\nwendcode{}\nwbegindocs{421}Now we define inversions of the grid lines in the unit cycle and
draw them for three different metrics.
\nwenddocs{}\nwbegincode{422}\sublabel{NW3gGP3e-3xVrpR-2}\nwmargintag{{\nwtagstyle{}\subpageref{NW3gGP3e-3xVrpR-2}}}\moddef{Three inversions~{\nwtagstyle{}\subpageref{NW3gGP3e-3xVrpR-1}}}\plusendmoddef\Rm{}\nwstartdeflinemarkup\nwusesondefline{\\{NW3gGP3e-3gefqu-1}}\nwprevnextdefs{NW3gGP3e-3xVrpR-1}{NW3gGP3e-3xVrpR-3}\nwenddeflinemarkup
{\it{}C4}={\it{}C3}.{\it{}cycle\_similarity}({\it{}C2});
{\bf{}for}({\bf{}int} {\it{}si}=-1; {\it{}si}\begin{math}<\end{math}2; {\it{}si}\protect\PP) {\nwlbrace}
    {\it{}asymptote} \begin{math}\ll\end{math} {\it{}endl} \begin{math}\ll\end{math} {\tt{}"erase();"} \begin{math}\ll\end{math} {\it{}endl};
    {\bf{}for}({\bf{}double} {\it{}i}=-4; {\it{}i}\begin{math}\leq\end{math}4; {\it{}i}+=.4) {\nwlbrace}
        {\it{}C4}.{\it{}subs}({\bf{}lst}({\it{}sign}\begin{math}\equiv\end{math}{\it{}si}, {\it{}l}\begin{math}\equiv\end{math}0, {\it{}n}\begin{math}\equiv\end{math}1, {\it{}m}\begin{math}\equiv\end{math}{\it{}i})).{\it{}asy\_draw}(
            {\it{}asymptote}, {\it{}xmin}, {\it{}xmax}, {\it{}ymin}, {\it{}ymax}, {\bf{}lst}(0.5, .75, 0.5),{\tt{}"+0.25pt"}, {\bf{}true}, 9);
        {\it{}C4}.{\it{}subs}({\bf{}lst}({\it{}sign}\begin{math}\equiv\end{math}{\it{}si}, {\it{}l}\begin{math}\equiv\end{math}1, {\it{}n}\begin{math}\equiv\end{math}0, {\it{}m}\begin{math}\equiv\end{math}{\it{}i})).{\it{}asy\_draw}(
            {\it{}asymptote}, {\it{}xmin}, {\it{}xmax}, {\it{}ymin}, {\it{}ymax}, {\bf{}lst}(0.5, .5, 0.75),{\tt{}"+0.25pt"}, {\bf{}true}, 9);
    {\nwrbrace}
    {\it{}C2}.{\it{}subs}({\it{}sign}\begin{math}\equiv\end{math}{\it{}si}).{\it{}asy\_draw}({\it{}asymptote}, {\it{}xmin}, {\it{}xmax}, {\it{}ymin}, {\it{}ymax}, {\bf{}lst}(1,0,0),{\tt{}"+.75pt"}, {\bf{}true}, 7);

\nwused{\\{NW3gGP3e-3gefqu-1}}\nwidentuses{\\{{\nwixident{asy{\_}draw}}{asy:undraw}}\\{{\nwixident{cycle{\_}similarity}}{cycle:unsimilarity}}\\{{\nwixident{l}}{l}}\\{{\nwixident{lst}}{lst}}\\{{\nwixident{m}}{m}}\\{{\nwixident{si}}{si}}\\{{\nwixident{subs}}{subs}}}\nwindexuse{\nwixident{asy{\_}draw}}{asy:undraw}{NW3gGP3e-3xVrpR-2}\nwindexuse{\nwixident{cycle{\_}similarity}}{cycle:unsimilarity}{NW3gGP3e-3xVrpR-2}\nwindexuse{\nwixident{l}}{l}{NW3gGP3e-3xVrpR-2}\nwindexuse{\nwixident{lst}}{lst}{NW3gGP3e-3xVrpR-2}\nwindexuse{\nwixident{m}}{m}{NW3gGP3e-3xVrpR-2}\nwindexuse{\nwixident{si}}{si}{NW3gGP3e-3xVrpR-2}\nwindexuse{\nwixident{subs}}{subs}{NW3gGP3e-3xVrpR-2}\nwendcode{}\nwbegindocs{423}We conclude by drawing the image of the cycle at infinity {\Tt{}\Rm{}{\it{}Zinf}\nwendquote}.
\nwenddocs{}\nwbegincode{424}\sublabel{NW3gGP3e-3xVrpR-3}\nwmargintag{{\nwtagstyle{}\subpageref{NW3gGP3e-3xVrpR-3}}}\moddef{Three inversions~{\nwtagstyle{}\subpageref{NW3gGP3e-3xVrpR-1}}}\plusendmoddef\Rm{}\nwstartdeflinemarkup\nwusesondefline{\\{NW3gGP3e-3gefqu-1}}\nwprevnextdefs{NW3gGP3e-3xVrpR-2}{\relax}\nwenddeflinemarkup
    {\it{}ex\_to}\begin{math}<\end{math}{\bf{}cycle2D}\begin{math}>\end{math}({\it{}Zinf}.{\it{}cycle\_similarity}({\it{}C2})).{\it{}subs}({\it{}sign}\begin{math}\equiv\end{math}{\it{}si}).{\it{}asy\_draw}(
        {\it{}asymptote}, {\it{}xmin}, {\it{}xmax}, {\it{}ymin}, {\it{}ymax}, {\bf{}lst}(0,0,1), ({\it{}si}\begin{math}\equiv\end{math}-1? {\tt{}"+3pt"}: {\tt{}"+.75pt"}));
    \LA{}Draw axes~{\nwtagstyle{}\subpageref{NW3gGP3e-ph2PF-1}}\RA{}
    {\it{}asymptote} \begin{math}\ll\end{math} {\tt{}"shipout({\char92}"inversion-"} \begin{math}\ll\end{math} {\it{}eph\_names}[{\it{}si}+1] \begin{math}\ll\end{math} {\tt{}"{\char92}");"} \begin{math}\ll\end{math} {\it{}endl};
{\nwrbrace}

\nwused{\\{NW3gGP3e-3gefqu-1}}\nwidentuses{\\{{\nwixident{asy{\_}draw}}{asy:undraw}}\\{{\nwixident{cycle2D}}{cycle2D}}\\{{\nwixident{cycle{\_}similarity}}{cycle:unsimilarity}}\\{{\nwixident{lst}}{lst}}\\{{\nwixident{si}}{si}}\\{{\nwixident{subs}}{subs}}}\nwindexuse{\nwixident{asy{\_}draw}}{asy:undraw}{NW3gGP3e-3xVrpR-3}\nwindexuse{\nwixident{cycle2D}}{cycle2D}{NW3gGP3e-3xVrpR-3}\nwindexuse{\nwixident{cycle{\_}similarity}}{cycle:unsimilarity}{NW3gGP3e-3xVrpR-3}\nwindexuse{\nwixident{lst}}{lst}{NW3gGP3e-3xVrpR-3}\nwindexuse{\nwixident{si}}{si}{NW3gGP3e-3xVrpR-3}\nwindexuse{\nwixident{subs}}{subs}{NW3gGP3e-3xVrpR-3}\nwendcode{}\nwbegindocs{425}\nwdocspar
\subsubsection{Drawing inversion of the hyperbolic ball}
\label{sec:draw-invers-hyperb}

A hyperbolic ball can be inverted without self-intersection. We
produce here an illustration of this.

Firstly we define some parameters
\nwenddocs{}\nwbegincode{426}\sublabel{NW3gGP3e-2DcUMV-1}\nwmargintag{{\nwtagstyle{}\subpageref{NW3gGP3e-2DcUMV-1}}}\moddef{Hyperbolic inversion of a ball~{\nwtagstyle{}\subpageref{NW3gGP3e-2DcUMV-1}}}\endmoddef\Rm{}\nwstartdeflinemarkup\nwusesondefline{\\{NW3gGP3e-3gefqu-1}}\nwprevnextdefs{\relax}{NW3gGP3e-2DcUMV-2}\nwenddeflinemarkup
{\bf{}const} {\bf{}int} {\it{}frames}=20, {\it{}balls}=10; // number of frames and balls\nwindexdefn{\nwixident{frames}}{frames}{NW3gGP3e-2DcUMV-1}
{\bf{}const} {\bf{}double} {\it{}r1}=.1, {\it{}r2}=1, {\it{}tmin}=-3, {\it{}tmax}=3, // limits of balls' filling and inversions\nwindexdefn{\nwixident{r1}}{r1}{NW3gGP3e-2DcUMV-1}
    {\it{}step2}=({\it{}r2}-{\it{}r1})\begin{math}\div\end{math}({\it{}balls}-1); // steps between balls

\nwalsodefined{\\{NW3gGP3e-2DcUMV-2}\\{NW3gGP3e-2DcUMV-3}\\{NW3gGP3e-2DcUMV-4}\\{NW3gGP3e-2DcUMV-5}\\{NW3gGP3e-2DcUMV-6}\\{NW3gGP3e-2DcUMV-7}}\nwused{\\{NW3gGP3e-3gefqu-1}}\nwidentdefs{\\{{\nwixident{frames}}{frames}}\\{{\nwixident{r1}}{r1}}}\nwendcode{}\nwbegindocs{427}Then we open the file and put initialisation into it.
\nwenddocs{}\nwbegincode{428}\sublabel{NW3gGP3e-2DcUMV-2}\nwmargintag{{\nwtagstyle{}\subpageref{NW3gGP3e-2DcUMV-2}}}\moddef{Hyperbolic inversion of a ball~{\nwtagstyle{}\subpageref{NW3gGP3e-2DcUMV-1}}}\plusendmoddef\Rm{}\nwstartdeflinemarkup\nwusesondefline{\\{NW3gGP3e-3gefqu-1}}\nwprevnextdefs{NW3gGP3e-2DcUMV-1}{NW3gGP3e-2DcUMV-3}\nwenddeflinemarkup
{\it{}ofstream} {\it{}asymptote}({\tt{}"ball-inv-d.asy"});
{\it{}asymptote} \begin{math}\ll\end{math} {\it{}setprecision}(2);
{\bf{}const} {\bf{}numeric} {\it{}scale}=2.5; //size of the picture\nwindexdefn{\nwixident{numeric}}{numeric}{NW3gGP3e-2DcUMV-2}
{\it{}asymptote} \begin{math}\ll\end{math} {\tt{}"scale = "} \begin{math}\ll\end{math} {\it{}scale} \begin{math}\ll\end{math} {\tt{}";"} \begin{math}\ll\end{math} {\it{}endl};

\nwused{\\{NW3gGP3e-3gefqu-1}}\nwidentdefs{\\{{\nwixident{numeric}}{numeric}}}\nwendcode{}\nwbegindocs{429}We have one cycle which will inverted by the matrix {\Tt{}\Rm{}{\it{}T}\nwendquote}.
\nwenddocs{}\nwbegincode{430}\sublabel{NW3gGP3e-2DcUMV-3}\nwmargintag{{\nwtagstyle{}\subpageref{NW3gGP3e-2DcUMV-3}}}\moddef{Hyperbolic inversion of a ball~{\nwtagstyle{}\subpageref{NW3gGP3e-2DcUMV-1}}}\plusendmoddef\Rm{}\nwstartdeflinemarkup\nwusesondefline{\\{NW3gGP3e-3gefqu-1}}\nwprevnextdefs{NW3gGP3e-2DcUMV-2}{NW3gGP3e-2DcUMV-4}\nwenddeflinemarkup
{\bf{}matrix} {\it{}T}={\bf{}matrix}(2, 2, {\bf{}lst}({\it{}dirac\_ONE}(), -{\it{}t}\begin{math}\ast\end{math}{\it{}e}.{\it{}subs}({\it{}mu}\begin{math}\equiv\end{math}1), {\it{}t}\begin{math}\ast\end{math}{\it{}e}.{\it{}subs}({\it{}mu}\begin{math}\equiv\end{math}1), {\it{}dirac\_ONE}()));
{\bf{}const} {\bf{}cycle2D} {\it{}Hyp}={\bf{}cycle2D}({\bf{}lst}(0,0),{\it{}e}, {\it{}a}).{\it{}matrix\_similarity}({\it{}T});\nwindexdefn{\nwixident{cycle2D}}{cycle2D}{NW3gGP3e-2DcUMV-3}

\nwused{\\{NW3gGP3e-3gefqu-1}}\nwidentdefs{\\{{\nwixident{cycle2D}}{cycle2D}}}\nwidentuses{\\{{\nwixident{lst}}{lst}}\\{{\nwixident{matrix}}{matrix}}\\{{\nwixident{matrix{\_}similarity}}{matrix:unsimilarity}}\\{{\nwixident{subs}}{subs}}}\nwindexuse{\nwixident{lst}}{lst}{NW3gGP3e-2DcUMV-3}\nwindexuse{\nwixident{matrix}}{matrix}{NW3gGP3e-2DcUMV-3}\nwindexuse{\nwixident{matrix{\_}similarity}}{matrix:unsimilarity}{NW3gGP3e-2DcUMV-3}\nwindexuse{\nwixident{subs}}{subs}{NW3gGP3e-2DcUMV-3}\nwendcode{}\nwbegindocs{431}We run a cycle for different frames, the parameter {\Tt{}\Rm{}{\it{}t}\nwendquote} from the
matrix {\Tt{}\Rm{}{\it{}T}\nwendquote} get specific values.
\nwenddocs{}\nwbegincode{432}\sublabel{NW3gGP3e-2DcUMV-4}\nwmargintag{{\nwtagstyle{}\subpageref{NW3gGP3e-2DcUMV-4}}}\moddef{Hyperbolic inversion of a ball~{\nwtagstyle{}\subpageref{NW3gGP3e-2DcUMV-1}}}\plusendmoddef\Rm{}\nwstartdeflinemarkup\nwusesondefline{\\{NW3gGP3e-3gefqu-1}}\nwprevnextdefs{NW3gGP3e-2DcUMV-3}{NW3gGP3e-2DcUMV-5}\nwenddeflinemarkup
{\bf{}for} ({\bf{}int} {\it{}j}=0; {\it{}j}\begin{math}\leq\end{math}2\begin{math}\ast\end{math}{\it{}frames} ;{\it{}j}\protect\PP ) {\nwlbrace}
 {\bf{}double} {\it{}tval}=({\it{}j}\begin{math}\equiv\end{math}0 \begin{math}\wedge\end{math} {\it{}j}\begin{math}\equiv\end{math}2\begin{math}\ast\end{math}{\it{}frames} ? 0 :
     ({\it{}j}\begin{math}\equiv\end{math}{\it{}frames} ? 10000000 :
      {\it{}ex\_to}\begin{math}<\end{math}{\bf{}numeric}\begin{math}>\end{math}(({\it{}j}\begin{math}<\end{math}{\it{}frames} ? {\it{}exp}({\it{}tmin}+{\it{}j}\begin{math}\ast\end{math}({\it{}tmax}-{\it{}tmin})\begin{math}\div\end{math}({\it{}frames}-2)) :
                      -{\it{}GiNaC}::{\it{}exp}({\it{}tmin}+(2\begin{math}\ast\end{math}{\it{}frames}-{\it{}j})\begin{math}\ast\end{math}({\it{}tmax}-{\it{}tmin})\begin{math}\div\end{math}({\it{}frames}-2))).{\it{}evalf}()).{\it{}to\_double}()));

\nwused{\\{NW3gGP3e-3gefqu-1}}\nwidentuses{\\{{\nwixident{frames}}{frames}}\\{{\nwixident{numeric}}{numeric}}}\nwindexuse{\nwixident{frames}}{frames}{NW3gGP3e-2DcUMV-4}\nwindexuse{\nwixident{numeric}}{numeric}{NW3gGP3e-2DcUMV-4}\nwendcode{}\nwbegindocs{433}Then we run a cycle over different hyperbolas filling up the
ball. Two copies are drown for GIF and PDF images.
\nwenddocs{}\nwbegincode{434}\sublabel{NW3gGP3e-2DcUMV-5}\nwmargintag{{\nwtagstyle{}\subpageref{NW3gGP3e-2DcUMV-5}}}\moddef{Hyperbolic inversion of a ball~{\nwtagstyle{}\subpageref{NW3gGP3e-2DcUMV-1}}}\plusendmoddef\Rm{}\nwstartdeflinemarkup\nwusesondefline{\\{NW3gGP3e-3gefqu-1}}\nwprevnextdefs{NW3gGP3e-2DcUMV-4}{NW3gGP3e-2DcUMV-6}\nwenddeflinemarkup
 {\bf{}for} ({\bf{}int} {\it{}i}=0; {\it{}i} \begin{math}<\end{math}{\it{}balls}; {\it{}i}\protect\PP) {\nwlbrace}
     {\it{}Hyp}.{\it{}subs}({\bf{}lst}({\it{}sign}\begin{math}\equiv\end{math}1, {\it{}a}\begin{math}\equiv\end{math}{\it{}GiNaC}::{\it{}pow}({\it{}r1}+{\it{}i}\begin{math}\ast\end{math}{\it{}step2},2), {\it{}t}\begin{math}\equiv\end{math}{\it{}tval})).{\it{}asy\_draw}({\it{}asymptote}, {\tt{}"pa"},
   -{\it{}scale}, {\it{}scale}, -{\it{}scale}, {\it{}scale}, {\bf{}lst}(0.1+0.8\begin{math}\ast\end{math}{\it{}i}\begin{math}\div\end{math}{\it{}balls}, 0, 0.9-0.8\begin{math}\ast\end{math}{\it{}i}\begin{math}\div\end{math}{\it{}balls}));
  {\it{}Hyp}.{\it{}subs}({\bf{}lst}({\it{}sign}\begin{math}\equiv\end{math}1, {\it{}a}\begin{math}\equiv\end{math}{\it{}GiNaC}::{\it{}pow}({\it{}r1}+{\it{}i}\begin{math}\ast\end{math}{\it{}step2},2), {\it{}t}\begin{math}\equiv\end{math}{\it{}tval})).{\it{}asy\_draw}({\it{}asymptote}, {\tt{}"pb"},
   -{\it{}scale}, {\it{}scale}, -{\it{}scale}, {\it{}scale}, {\bf{}lst}(0.1+0.8\begin{math}\ast\end{math}{\it{}i}\begin{math}\div\end{math}{\it{}balls}, 0, 0.9-0.8\begin{math}\ast\end{math}{\it{}i}\begin{math}\div\end{math}{\it{}balls}));
 {\nwrbrace}

\nwused{\\{NW3gGP3e-3gefqu-1}}\nwidentuses{\\{{\nwixident{asy{\_}draw}}{asy:undraw}}\\{{\nwixident{lst}}{lst}}\\{{\nwixident{r1}}{r1}}\\{{\nwixident{subs}}{subs}}}\nwindexuse{\nwixident{asy{\_}draw}}{asy:undraw}{NW3gGP3e-2DcUMV-5}\nwindexuse{\nwixident{lst}}{lst}{NW3gGP3e-2DcUMV-5}\nwindexuse{\nwixident{r1}}{r1}{NW3gGP3e-2DcUMV-5}\nwindexuse{\nwixident{subs}}{subs}{NW3gGP3e-2DcUMV-5}\nwendcode{}\nwbegindocs{435}The boundary of the ball is drown in a highlighted way.
\nwenddocs{}\nwbegincode{436}\sublabel{NW3gGP3e-2DcUMV-6}\nwmargintag{{\nwtagstyle{}\subpageref{NW3gGP3e-2DcUMV-6}}}\moddef{Hyperbolic inversion of a ball~{\nwtagstyle{}\subpageref{NW3gGP3e-2DcUMV-1}}}\plusendmoddef\Rm{}\nwstartdeflinemarkup\nwusesondefline{\\{NW3gGP3e-3gefqu-1}}\nwprevnextdefs{NW3gGP3e-2DcUMV-5}{NW3gGP3e-2DcUMV-7}\nwenddeflinemarkup
 {\it{}Hyp}.{\it{}subs}({\bf{}lst}({\it{}sign}\begin{math}\equiv\end{math}1, {\it{}a}\begin{math}\equiv\end{math}1, {\it{}t}\begin{math}\equiv\end{math}{\it{}tval})).{\it{}asy\_draw}({\it{}asymptote}, {\tt{}"pa"},
  -{\it{}scale}, {\it{}scale}, -{\it{}scale}, {\it{}scale}, {\bf{}lst}(1,0,0),{\tt{}"+2pt"});
 {\it{}Hyp}.{\it{}subs}({\bf{}lst}({\it{}sign}\begin{math}\equiv\end{math}1, {\it{}a}\begin{math}\equiv\end{math}1, {\it{}t}\begin{math}\equiv\end{math}{\it{}tval})).{\it{}asy\_draw}({\it{}asymptote}, {\tt{}"pb"},
  -{\it{}scale}, {\it{}scale}, -{\it{}scale}, {\it{}scale}, {\bf{}lst}(1,0,0),{\tt{}"+2pt"});
 {\it{}asymptote} \begin{math}\ll\end{math} {\tt{}"newpic();"} \begin{math}\ll\end{math} {\it{}endl}  \begin{math}\ll\end{math} {\it{}endl} ;
{\nwrbrace}

\nwused{\\{NW3gGP3e-3gefqu-1}}\nwidentuses{\\{{\nwixident{asy{\_}draw}}{asy:undraw}}\\{{\nwixident{lst}}{lst}}\\{{\nwixident{subs}}{subs}}}\nwindexuse{\nwixident{asy{\_}draw}}{asy:undraw}{NW3gGP3e-2DcUMV-6}\nwindexuse{\nwixident{lst}}{lst}{NW3gGP3e-2DcUMV-6}\nwindexuse{\nwixident{subs}}{subs}{NW3gGP3e-2DcUMV-6}\nwendcode{}\nwbegindocs{437}Finally we close the file.
\nwenddocs{}\nwbegincode{438}\sublabel{NW3gGP3e-2DcUMV-7}\nwmargintag{{\nwtagstyle{}\subpageref{NW3gGP3e-2DcUMV-7}}}\moddef{Hyperbolic inversion of a ball~{\nwtagstyle{}\subpageref{NW3gGP3e-2DcUMV-1}}}\plusendmoddef\Rm{}\nwstartdeflinemarkup\nwusesondefline{\\{NW3gGP3e-3gefqu-1}}\nwprevnextdefs{NW3gGP3e-2DcUMV-6}{\relax}\nwenddeflinemarkup
{\it{}asymptote}.{\it{}close}();

\nwused{\\{NW3gGP3e-3gefqu-1}}\nwendcode{}\nwbegindocs{439}\nwdocspar
\section[The Implementation the Classes cycle and cycle2D]{The Implementation the Classes {\Tt{}\Rm{}{\bf{}cycle}\nwendquote} and {\Tt{}\Rm{}{\bf{}cycle2D}\nwendquote}}
\label{sec:main-file-class}

This is the main file providing implementation the Classes {\Tt{}\Rm{}{\bf{}cycle}\nwendquote}
and {\Tt{}\Rm{}{\bf{}cycle2D}\nwendquote}. It is not well documented yet.

\nwenddocs{}\nwbegindocs{440}\nwdocspar
\subsection{Cycle and cycle2D classes header files}
\label{sec:cycle-class}

\subsubsection{Cycle header file}
\label{sec:header-file}

This the header file describing the classes {\Tt{}\Rm{}{\bf{}cycle}\nwendquote} and
{\Tt{}\Rm{}{\it{}cycle2d}\nwendquote}. We start from the general inclusions and definitions and
then defining those two classes.
\nwenddocs{}\nwbegincode{441}\sublabel{NW3gGP3e-4Ef0r4-1}\nwmargintag{{\nwtagstyle{}\subpageref{NW3gGP3e-4Ef0r4-1}}}\moddef{cycle.h~{\nwtagstyle{}\subpageref{NW3gGP3e-4Ef0r4-1}}}\endmoddef\Rm{}\nwstartdeflinemarkup\nwprevnextdefs{\relax}{NW3gGP3e-4Ef0r4-2}\nwenddeflinemarkup
\LA{}license~{\nwtagstyle{}\subpageref{NW3gGP3e-ZXuKx-1}}\RA{}
{\bf{}\char35{}include}{\tt{} \begin{math}<\end{math}stdexcept\begin{math}>\end{math}}
{\bf{}\char35{}include}{\tt{} \begin{math}<\end{math}ostream\begin{math}>\end{math}}
{\bf{}\char35{}include}{\tt{} \begin{math}<\end{math}sstream\begin{math}>\end{math}}

{\bf{}\char35{}include}{\tt{} \begin{math}<\end{math}ginac/ginac.h\begin{math}>\end{math}}

{\bf{}namespace} {\it{}MoebInv} {\nwlbrace}
{\bf{}using} {\bf{}namespace} {\it{}std};
{\bf{}using} {\bf{}namespace} {\it{}GiNaC};
\nwindexdefn{\nwixident{MoebInv}}{MoebInv}{NW3gGP3e-4Ef0r4-1}\eatline
\nwalsodefined{\\{NW3gGP3e-4Ef0r4-2}\\{NW3gGP3e-4Ef0r4-3}\\{NW3gGP3e-4Ef0r4-4}}\nwnotused{cycle.h}\nwidentdefs{\\{{\nwixident{MoebInv}}{MoebInv}}}\nwendcode{}\nwbegindocs{442}\nwdocspar
\nwenddocs{}\nwbegindocs{443} We need to know if \GiNaC\ is too old or not.
\nwenddocs{}\nwbegincode{444}\sublabel{NW3gGP3e-4Ef0r4-2}\nwmargintag{{\nwtagstyle{}\subpageref{NW3gGP3e-4Ef0r4-2}}}\moddef{cycle.h~{\nwtagstyle{}\subpageref{NW3gGP3e-4Ef0r4-1}}}\plusendmoddef\Rm{}\nwstartdeflinemarkup\nwprevnextdefs{NW3gGP3e-4Ef0r4-1}{NW3gGP3e-4Ef0r4-3}\nwenddeflinemarkup
{\bf{}\char35{}define}{\tt{} GINAC\_VERSION\_ATLEAST( major, minor) \begin{math}\backslash\end{math}}\nwindexdefn{\nwixident{GINAC{\_}VERSION{\_}ATLEAST}}{GINAC:unVERSION:unATLEAST}{NW3gGP3e-4Ef0r4-2}
        ({\it{}GINACLIB\_MAJOR\_VERSION} \begin{math}>\end{math} {\it{}major} \begin{math}\backslash\end{math}
        \begin{math}\vee\end{math} ({\it{}GINACLIB\_MAJOR\_VERSION} \begin{math}\equiv\end{math} {\it{}major} \begin{math}\wedge\end{math} {\it{}GINACLIB\_MINOR\_VERSION} \begin{math}\geq\end{math} {\it{}minor}))

\nwidentdefs{\\{{\nwixident{GINAC{\_}VERSION{\_}ATLEAST}}{GINAC:unVERSION:unATLEAST}}}\nwendcode{}\nwbegindocs{445}We define version number for our own library.
\nwenddocs{}\nwbegincode{446}\sublabel{NW3gGP3e-4Ef0r4-3}\nwmargintag{{\nwtagstyle{}\subpageref{NW3gGP3e-4Ef0r4-3}}}\moddef{cycle.h~{\nwtagstyle{}\subpageref{NW3gGP3e-4Ef0r4-1}}}\plusendmoddef\Rm{}\nwstartdeflinemarkup\nwprevnextdefs{NW3gGP3e-4Ef0r4-2}{NW3gGP3e-4Ef0r4-4}\nwenddeflinemarkup
{\bf{}\char35{}define}{\tt{} MOEBINV\_MAJOR\_VERSION 2}\nwindexdefn{\nwixident{MOEBINV{\_}MAJOR{\_}VERSION}}{MOEBINV:unMAJOR:unVERSION}{NW3gGP3e-4Ef0r4-3}
{\bf{}\char35{}define}{\tt{} MOEBINV\_MINOR\_VERSION 3}\nwindexdefn{\nwixident{MOEBINV{\_}MINOR{\_}VERSION}}{MOEBINV:unMINOR:unVERSION}{NW3gGP3e-4Ef0r4-3}
\nwindexdefn{\nwixident{MOEBINV{\_}MAJOR{\_}VERSION}}{MOEBINV:unMAJOR:unVERSION}{NW3gGP3e-4Ef0r4-3}\nwindexdefn{\nwixident{MOEBINV{\_}MINOR{\_}VERSION}}{MOEBINV:unMINOR:unVERSION}{NW3gGP3e-4Ef0r4-3}\eatline
\nwidentdefs{\\{{\nwixident{MOEBINV{\_}MAJOR{\_}VERSION}}{MOEBINV:unMAJOR:unVERSION}}\\{{\nwixident{MOEBINV{\_}MINOR{\_}VERSION}}{MOEBINV:unMINOR:unVERSION}}}\nwendcode{}\nwbegindocs{447}\nwdocspar
\nwenddocs{}\nwbegindocs{448}The brief outline of the header file.
\nwenddocs{}\nwbegincode{449}\sublabel{NW3gGP3e-4Ef0r4-4}\nwmargintag{{\nwtagstyle{}\subpageref{NW3gGP3e-4Ef0r4-4}}}\moddef{cycle.h~{\nwtagstyle{}\subpageref{NW3gGP3e-4Ef0r4-1}}}\plusendmoddef\Rm{}\nwstartdeflinemarkup\nwprevnextdefs{NW3gGP3e-4Ef0r4-3}{\relax}\nwenddeflinemarkup
\LA{}Auxiliary functions headers~{\nwtagstyle{}\subpageref{NW3gGP3e-ugGZb-1}}\RA{}
\LA{}cycle class~{\nwtagstyle{}\subpageref{NW3gGP3e-2wwyff-1}}\RA{}
\LA{}cycle2D class~{\nwtagstyle{}\subpageref{NW3gGP3e-2ARAe1-1}}\RA{}

{\nwrbrace} // namespace MoebInv

\nwidentuses{\\{{\nwixident{MoebInv}}{MoebInv}}}\nwindexuse{\nwixident{MoebInv}}{MoebInv}{NW3gGP3e-4Ef0r4-4}\nwendcode{}\nwbegindocs{450}\nwdocspar
\subsubsection{Some auxillary functions}
\label{sec:some-auxill-funct}

 Here is the list of some auxiliary functions which are defined and
used in the \texttt{cycle.h}.

\nwenddocs{}\nwbegindocs{451}\nwdocspar
\nwenddocs{}\nwbegincode{452}\sublabel{NW3gGP3e-ugGZb-1}\nwmargintag{{\nwtagstyle{}\subpageref{NW3gGP3e-ugGZb-1}}}\moddef{Auxiliary functions headers~{\nwtagstyle{}\subpageref{NW3gGP3e-ugGZb-1}}}\endmoddef\Rm{}\nwstartdeflinemarkup\nwusesondefline{\\{NW3gGP3e-4Ef0r4-4}}\nwenddeflinemarkup
{\commopen}* Check of equality of two expression and report the string {\commclose}
{\it{}DECLARE\_FUNCTION\_1P}({\it{}jump\_fnct})\nwindexdefn{\nwixident{jump{\_}fnct}}{jump:unfnct}{NW3gGP3e-ugGZb-1}

{\bf{}const} {\it{}string} {\it{}equality}({\bf{}const} {\bf{}ex} & {\it{}E});\nwindexdefn{\nwixident{string}}{string}{NW3gGP3e-ugGZb-1}
{\bf{}inline} {\bf{}const} {\it{}string} {\it{}equality}({\bf{}const} {\bf{}ex} & {\it{}E1}, {\bf{}const} {\bf{}ex} & {\it{}E2}) {\nwlbrace} {\bf{}return} {\it{}equality}({\it{}E1}-{\it{}E2});{\nwrbrace}
{\bf{}inline} {\bf{}const} {\it{}string} {\it{}equality}({\bf{}const} {\bf{}ex} & {\it{}E}, {\bf{}const} {\bf{}ex} & {\it{}solns1}, {\bf{}const} {\bf{}ex} & {\it{}solns2})
{\nwlbrace} {\bf{}ex} {\it{}e} = {\it{}E}; {\bf{}return} {\it{}equality}({\it{}e}.{\it{}subs}({\it{}solns1}), {\it{}e}.{\it{}subs}({\it{}solns2}));{\nwrbrace}

{\commopen}* Return the string describing the case (elliptic, parabolic or hyperbolic)  {\commclose}
{\bf{}const} {\it{}string} {\it{}eph\_case}({\bf{}const} {\bf{}numeric} & {\it{}sign});\nwindexdefn{\nwixident{string}}{string}{NW3gGP3e-ugGZb-1}

{\commopen}* Return even (real) part of a Clifford number {\commclose}
{\bf{}ex} {\it{}scalar\_part}({\bf{}const} {\bf{}ex} & {\it{}e});

///** Return odd part of a Clifford number */
//inline ex clifford\_part(const ex & e) {\nwlbrace} return normal(canonicalize\_clifford(e - clifford\_bar(e)))/numeric(2);{\nwrbrace}

{\commopen}* Produces a Clifford matrix form of element of SL2 {\commclose}
{\bf{}matrix} {\it{}sl2\_clifford}({\bf{}const} {\bf{}ex} & {\it{}a}, {\bf{}const} {\bf{}ex} & {\it{}b}, {\bf{}const} {\bf{}ex} & {\it{}c}, {\bf{}const} {\bf{}ex} & {\it{}d}, {\bf{}const} {\bf{}ex} & {\it{}e}, {\bf{}bool} {\it{}not\_inverse}={\bf{}true});

{\bf{}matrix} {\it{}sl2\_clifford}({\bf{}const} {\bf{}ex} & {\it{}M}, {\bf{}const} {\bf{}ex} & {\it{}e}, {\bf{}bool} {\it{}not\_inverse}={\bf{}true});

\nwused{\\{NW3gGP3e-4Ef0r4-4}}\nwidentdefs{\\{{\nwixident{jump{\_}fnct}}{jump:unfnct}}\\{{\nwixident{string}}{string}}}\nwidentuses{\\{{\nwixident{bool}}{bool}}\\{{\nwixident{ex}}{ex}}\\{{\nwixident{matrix}}{matrix}}\\{{\nwixident{normal}}{normal}}\\{{\nwixident{numeric}}{numeric}}\\{{\nwixident{subs}}{subs}}}\nwindexuse{\nwixident{bool}}{bool}{NW3gGP3e-ugGZb-1}\nwindexuse{\nwixident{ex}}{ex}{NW3gGP3e-ugGZb-1}\nwindexuse{\nwixident{matrix}}{matrix}{NW3gGP3e-ugGZb-1}\nwindexuse{\nwixident{normal}}{normal}{NW3gGP3e-ugGZb-1}\nwindexuse{\nwixident{numeric}}{numeric}{NW3gGP3e-ugGZb-1}\nwindexuse{\nwixident{subs}}{subs}{NW3gGP3e-ugGZb-1}\nwendcode{}\nwbegindocs{453}\nwdocspar
\subsubsection[Members and methods in class cycle]{Members and methods in class {\Tt{}\Rm{}{\bf{}cycle}\nwendquote}}
\label{sec:class-cycle-header}

The class {\Tt{}\Rm{}{\bf{}cycle}\nwendquote} is derived from class {\Tt{}\Rm{}{\bf{}basic}\nwendquote} in \GiNaC\
according to the general guidelines given in the \GiNaC\ tutorial. is defined through the general s
\nwenddocs{}\nwbegincode{454}\sublabel{NW3gGP3e-2wwyff-1}\nwmargintag{{\nwtagstyle{}\subpageref{NW3gGP3e-2wwyff-1}}}\moddef{cycle class~{\nwtagstyle{}\subpageref{NW3gGP3e-2wwyff-1}}}\endmoddef\Rm{}\nwstartdeflinemarkup\nwusesondefline{\\{NW3gGP3e-4Ef0r4-4}}\nwenddeflinemarkup
{\commopen}* The class holding cycles kx^2-2\begin{math}<\end{math}l,x\begin{math}>\end{math}+m=0 {\commclose}
{\bf{}class} {\bf{}cycle} : {\bf{}public} {\bf{}basic}
{\nwlbrace}
 {\it{}GINAC\_DECLARE\_REGISTERED\_CLASS}({\bf{}cycle}, {\bf{}basic})

 \LA{}cycle class constructors~{\nwtagstyle{}\subpageref{NW3gGP3e-2EuhSt-1}}\RA{}
 \LA{}service functions for class cycle~{\nwtagstyle{}\subpageref{NW3gGP3e-2AZq7I-1}}\RA{}
 \LA{}accessing the data of a cycle~{\nwtagstyle{}\subpageref{NW3gGP3e-rnSJR-1}}\RA{}
 \LA{}specific methods of the class cycle~{\nwtagstyle{}\subpageref{NW3gGP3e-mq5nH-1}}\RA{}
 \LA{}Linear operation as cycle methods~{\nwtagstyle{}\subpageref{NW3gGP3e-3Bzv5Q-1}}\RA{}

{\bf{}protected}:
 {\bf{}ex} {\it{}unit}; // A Clifford unit to store the dimensionality and metric of the point space
 {\bf{}ex} {\it{}k};
 {\bf{}ex} {\it{}l};
 {\bf{}ex} {\it{}m};
{\nwrbrace};
 {\it{}GINAC\_DECLARE\_UNARCHIVER}({\bf{}cycle});

 \LA{}Linear operation on cycles~{\nwtagstyle{}\subpageref{NW3gGP3e-3LRIX-1}}\RA{}

\nwused{\\{NW3gGP3e-4Ef0r4-4}}\nwidentuses{\\{{\nwixident{cycle}}{cycle}}\\{{\nwixident{ex}}{ex}}\\{{\nwixident{k}}{k}}\\{{\nwixident{l}}{l}}\\{{\nwixident{m}}{m}}}\nwindexuse{\nwixident{cycle}}{cycle}{NW3gGP3e-2wwyff-1}\nwindexuse{\nwixident{ex}}{ex}{NW3gGP3e-2wwyff-1}\nwindexuse{\nwixident{k}}{k}{NW3gGP3e-2wwyff-1}\nwindexuse{\nwixident{l}}{l}{NW3gGP3e-2wwyff-1}\nwindexuse{\nwixident{m}}{m}{NW3gGP3e-2wwyff-1}\nwendcode{}\nwbegindocs{455}This is a set of the service functions which is required that a
{\Tt{}\Rm{}{\bf{}cycle}\nwendquote} is properly archived or printed to a stream.
\nwenddocs{}\nwbegincode{456}\sublabel{NW3gGP3e-2AZq7I-1}\nwmargintag{{\nwtagstyle{}\subpageref{NW3gGP3e-2AZq7I-1}}}\moddef{service functions for class cycle~{\nwtagstyle{}\subpageref{NW3gGP3e-2AZq7I-1}}}\endmoddef\Rm{}\nwstartdeflinemarkup\nwusesondefline{\\{NW3gGP3e-2wwyff-1}}\nwprevnextdefs{\relax}{NW3gGP3e-2AZq7I-2}\nwenddeflinemarkup
{\bf{}\char35{}if}{\tt{} GINAC\_VERSION\_ATLEAST(1,5)}
 {\bf{}void} {\it{}archive}({\it{}archive\_node} &{\it{}n}) {\bf{}const};
 {\bf{}void} {\it{}read\_archive}({\bf{}const} {\it{}archive\_node} &{\it{}n}, {\bf{}lst} &{\it{}sym\_lst});
 {\it{}return\_type\_t} {\it{}return\_type\_tinfo}() {\bf{}const};
{\bf{}\char35{}endif}{\tt{}}

\nwalsodefined{\\{NW3gGP3e-2AZq7I-2}\\{NW3gGP3e-2AZq7I-3}}\nwused{\\{NW3gGP3e-2wwyff-1}}\nwidentuses{\\{{\nwixident{GINAC{\_}VERSION{\_}ATLEAST}}{GINAC:unVERSION:unATLEAST}}\\{{\nwixident{lst}}{lst}}}\nwindexuse{\nwixident{GINAC{\_}VERSION{\_}ATLEAST}}{GINAC:unVERSION:unATLEAST}{NW3gGP3e-2AZq7I-1}\nwindexuse{\nwixident{lst}}{lst}{NW3gGP3e-2AZq7I-1}\nwendcode{}\nwbegindocs{457} Real and imaginary part of the representing vector.
\nwenddocs{}\nwbegincode{458}\sublabel{NW3gGP3e-2AZq7I-2}\nwmargintag{{\nwtagstyle{}\subpageref{NW3gGP3e-2AZq7I-2}}}\moddef{service functions for class cycle~{\nwtagstyle{}\subpageref{NW3gGP3e-2AZq7I-1}}}\plusendmoddef\Rm{}\nwstartdeflinemarkup\nwusesondefline{\\{NW3gGP3e-2wwyff-1}}\nwprevnextdefs{NW3gGP3e-2AZq7I-1}{NW3gGP3e-2AZq7I-3}\nwenddeflinemarkup
{\bf{}ex} {\it{}real\_part}() {\bf{}const};
{\bf{}ex} {\it{}imag\_part}() {\bf{}const};
{\bf{}inline} {\bf{}ex} {\it{}evalf}() {\bf{}const} {\nwlbrace} {\bf{}return} {\bf{}cycle}({\it{}k}.{\it{}evalf}(), {\it{}l}.{\it{}evalf}(), {\it{}m}.{\it{}evalf}(), {\it{}unit});{\nwrbrace}

\nwused{\\{NW3gGP3e-2wwyff-1}}\nwidentuses{\\{{\nwixident{cycle}}{cycle}}\\{{\nwixident{ex}}{ex}}\\{{\nwixident{k}}{k}}\\{{\nwixident{l}}{l}}\\{{\nwixident{m}}{m}}}\nwindexuse{\nwixident{cycle}}{cycle}{NW3gGP3e-2AZq7I-2}\nwindexuse{\nwixident{ex}}{ex}{NW3gGP3e-2AZq7I-2}\nwindexuse{\nwixident{k}}{k}{NW3gGP3e-2AZq7I-2}\nwindexuse{\nwixident{l}}{l}{NW3gGP3e-2AZq7I-2}\nwindexuse{\nwixident{m}}{m}{NW3gGP3e-2AZq7I-2}\nwendcode{}\nwbegindocs{459}Printing of cycles.
\nwenddocs{}\nwbegincode{460}\sublabel{NW3gGP3e-2AZq7I-3}\nwmargintag{{\nwtagstyle{}\subpageref{NW3gGP3e-2AZq7I-3}}}\moddef{service functions for class cycle~{\nwtagstyle{}\subpageref{NW3gGP3e-2AZq7I-1}}}\plusendmoddef\Rm{}\nwstartdeflinemarkup\nwusesondefline{\\{NW3gGP3e-2wwyff-1}}\nwprevnextdefs{NW3gGP3e-2AZq7I-2}{\relax}\nwenddeflinemarkup
{\bf{}protected}:
 {\bf{}void} {\it{}do\_print}({\bf{}const} {\it{}print\_dflt} & {\it{}c}, {\bf{}unsigned} {\it{}level}) {\bf{}const};
// void do\_print\_python(const print\_dflt & c, unsigned level) const;
 {\bf{}void} {\it{}do\_print\_dflt}({\bf{}const} {\it{}print\_dflt} & {\it{}c}, {\bf{}unsigned} {\it{}level}) {\bf{}const};
 {\bf{}void} {\it{}do\_print\_latex}({\bf{}const} {\it{}print\_latex} & {\it{}c}, {\bf{}unsigned} {\it{}level}) {\bf{}const};

\nwused{\\{NW3gGP3e-2wwyff-1}}\nwendcode{}\nwbegindocs{461}\nwdocspar
\subsubsection[The derived class cycle2D for two dimensional cycles]{The derived class {\Tt{}\Rm{}{\bf{}cycle2D}\nwendquote} for two dimensional cycles}
\label{sec:derived-class-two}

We derive a derived class {\Tt{}\Rm{}{\bf{}cycle2D}\nwendquote} from {\Tt{}\Rm{}{\bf{}cycle}\nwendquote} in order to add some
more methods which only make sense in two dimensions.
\nwenddocs{}\nwbegincode{462}\sublabel{NW3gGP3e-2ARAe1-1}\nwmargintag{{\nwtagstyle{}\subpageref{NW3gGP3e-2ARAe1-1}}}\moddef{cycle2D class~{\nwtagstyle{}\subpageref{NW3gGP3e-2ARAe1-1}}}\endmoddef\Rm{}\nwstartdeflinemarkup\nwusesondefline{\\{NW3gGP3e-4Ef0r4-4}}\nwenddeflinemarkup
{\bf{}class} {\bf{}cycle2D} : {\bf{}public} {\bf{}cycle}
{\nwlbrace}
 {\it{}GINAC\_DECLARE\_REGISTERED\_CLASS}({\bf{}cycle2D}, {\bf{}cycle})

 \LA{}constructors of the class cycle2D~{\nwtagstyle{}\subpageref{NW3gGP3e-3e3TWq-1}}\RA{}
 \LA{}methods specific for class cycle2D~{\nwtagstyle{}\subpageref{NW3gGP3e-447MWQ-1}}\RA{}
 \LA{}duplicated methods for class cycle2D~{\nwtagstyle{}\subpageref{NW3gGP3e-xs0zU-1}}\RA{}
{\nwrbrace};
{\it{}GINAC\_DECLARE\_UNARCHIVER}({\bf{}cycle2D});\nwindexdefn{\nwixident{cycle2D}}{cycle2D}{NW3gGP3e-2ARAe1-1}

 \LA{}duplicated linear operation on cycle2D~{\nwtagstyle{}\subpageref{NW3gGP3e-1Ihica-1}}\RA{}

\nwused{\\{NW3gGP3e-4Ef0r4-4}}\nwidentdefs{\\{{\nwixident{cycle2D}}{cycle2D}}}\nwidentuses{\\{{\nwixident{cycle}}{cycle}}}\nwindexuse{\nwixident{cycle}}{cycle}{NW3gGP3e-2ARAe1-1}\nwendcode{}\nwbegindocs{463}The general framework developed in the {\Tt{}\Rm{}{\bf{}cycle}\nwendquote} class have some
duplicates for two dimensions.
\nwenddocs{}\nwbegincode{464}\sublabel{NW3gGP3e-xs0zU-1}\nwmargintag{{\nwtagstyle{}\subpageref{NW3gGP3e-xs0zU-1}}}\moddef{duplicated methods for class cycle2D~{\nwtagstyle{}\subpageref{NW3gGP3e-xs0zU-1}}}\endmoddef\Rm{}\nwstartdeflinemarkup\nwusesondefline{\\{NW3gGP3e-2ARAe1-1}}\nwprevnextdefs{\relax}{NW3gGP3e-xs0zU-2}\nwenddeflinemarkup
{\bf{}inline} {\bf{}cycle2D} {\it{}subs}({\bf{}const} {\bf{}ex} & {\it{}e}, {\bf{}unsigned} {\it{}options} = 0) {\bf{}const} {\nwlbrace}
 {\bf{}return} {\it{}ex\_to}\begin{math}<\end{math}{\bf{}cycle2D}\begin{math}>\end{math}({\it{}inherited}::{\it{}subs}({\it{}e}, {\it{}options})); {\nwrbrace}
{\bf{}inline} {\bf{}cycle2D} {\it{}normalize}({\bf{}const} {\bf{}ex} & {\it{}k\_new} = {\bf{}numeric}(1), {\bf{}const} {\bf{}ex} & {\it{}e} = 0) {\bf{}const} {\nwlbrace}
 {\bf{}return} {\it{}ex\_to}\begin{math}<\end{math}{\bf{}cycle2D}\begin{math}>\end{math}({\it{}inherited}::{\it{}normalize}({\it{}k\_new}, {\it{}e})); {\nwrbrace}
 {\bf{}inline} {\bf{}cycle2D} {\it{}normalize\_det}({\bf{}const} {\bf{}ex} & {\it{}e} = 0, {\bf{}const} {\bf{}ex} & {\it{}sign} = ({\bf{}new} {\it{}tensdelta})\begin{math}\rightarrow\end{math}{\it{}setflag}({\it{}status\_flags}::{\it{}dynallocated}), {\bf{}const} {\bf{}ex} & {\it{}D} = 1) {\bf{}const} {\nwlbrace}
     {\bf{}return} {\it{}ex\_to}\begin{math}<\end{math}{\bf{}cycle2D}\begin{math}>\end{math}({\it{}inherited}::{\it{}normalize\_det}({\it{}e}, {\it{}sign}, {\it{}D})); {\nwrbrace}
 {\bf{}inline} {\bf{}cycle2D} {\it{}normalize\_norm}({\bf{}const} {\bf{}ex} & {\it{}e} = 0, {\bf{}const} {\bf{}ex} & {\it{}sign} = ({\bf{}new} {\it{}tensdelta})\begin{math}\rightarrow\end{math}{\it{}setflag}({\it{}status\_flags}::{\it{}dynallocated}), {\bf{}const} {\bf{}ex} & {\it{}N} = 1) {\bf{}const} {\nwlbrace}
     {\bf{}return} {\it{}ex\_to}\begin{math}<\end{math}{\bf{}cycle2D}\begin{math}>\end{math}({\it{}inherited}::{\it{}normalize\_norm}({\it{}e}, {\it{}sign}, {\it{}N})); {\nwrbrace}
{\bf{}inline} {\bf{}cycle2D} {\it{}normal}() {\bf{}const} {\nwlbrace} {\bf{}return} {\bf{}cycle2D}({\it{}k}.{\it{}normal}(), {\it{}l}.{\it{}normal}(), {\it{}m}.{\it{}normal}(), {\it{}unit}.{\it{}normal}());{\nwrbrace}
{\bf{}inline} {\bf{}cycle2D} {\it{}expand}() {\bf{}const} {\nwlbrace} {\bf{}return} {\bf{}cycle2D}({\it{}k}.{\it{}expand}(), {\it{}l}.{\it{}expand}(), {\it{}m}.{\it{}expand}(), {\it{}unit});{\nwrbrace}
{\bf{}inline} {\bf{}cycle2D} {\it{}evalf}() {\bf{}const} {\nwlbrace} {\bf{}return} {\it{}ex\_to}\begin{math}<\end{math}{\bf{}cycle2D}\begin{math}>\end{math}({\it{}inherited}::{\it{}evalf}());{\nwrbrace}
{\bf{}inline} {\bf{}cycle2D} {\it{}subject\_to}({\bf{}const} {\bf{}ex} & {\it{}condition}, {\bf{}const} {\bf{}ex} & {\it{}vars} = 0) {\bf{}const} {\nwlbrace}
 {\bf{}return} {\it{}ex\_to}\begin{math}<\end{math}{\bf{}cycle2D}\begin{math}>\end{math}({\it{}inherited}::{\it{}subject\_to}({\it{}condition}, {\it{}vars})); {\nwrbrace}

// cycle2D(const archive\_node &n, lst &sym\_lst);
 {\bf{}void} {\it{}archive}({\it{}archive\_node} &{\it{}n}) {\bf{}const};
 // ex unarchive(const archive\_node &n, lst &sym\_lst);
 {\bf{}void} {\it{}read\_archive}({\bf{}const} {\it{}archive\_node} &{\it{}n}, {\bf{}lst} &{\it{}sym\_lst});

\nwalsodefined{\\{NW3gGP3e-xs0zU-2}}\nwused{\\{NW3gGP3e-2ARAe1-1}}\nwidentuses{\\{{\nwixident{cycle2D}}{cycle2D}}\\{{\nwixident{ex}}{ex}}\\{{\nwixident{expand}}{expand}}\\{{\nwixident{k}}{k}}\\{{\nwixident{l}}{l}}\\{{\nwixident{lst}}{lst}}\\{{\nwixident{m}}{m}}\\{{\nwixident{normal}}{normal}}\\{{\nwixident{normalize}}{normalize}}\\{{\nwixident{normalize{\_}det}}{normalize:undet}}\\{{\nwixident{normalize{\_}norm}}{normalize:unnorm}}\\{{\nwixident{numeric}}{numeric}}\\{{\nwixident{subject{\_}to}}{subject:unto}}\\{{\nwixident{subs}}{subs}}}\nwindexuse{\nwixident{cycle2D}}{cycle2D}{NW3gGP3e-xs0zU-1}\nwindexuse{\nwixident{ex}}{ex}{NW3gGP3e-xs0zU-1}\nwindexuse{\nwixident{expand}}{expand}{NW3gGP3e-xs0zU-1}\nwindexuse{\nwixident{k}}{k}{NW3gGP3e-xs0zU-1}\nwindexuse{\nwixident{l}}{l}{NW3gGP3e-xs0zU-1}\nwindexuse{\nwixident{lst}}{lst}{NW3gGP3e-xs0zU-1}\nwindexuse{\nwixident{m}}{m}{NW3gGP3e-xs0zU-1}\nwindexuse{\nwixident{normal}}{normal}{NW3gGP3e-xs0zU-1}\nwindexuse{\nwixident{normalize}}{normalize}{NW3gGP3e-xs0zU-1}\nwindexuse{\nwixident{normalize{\_}det}}{normalize:undet}{NW3gGP3e-xs0zU-1}\nwindexuse{\nwixident{normalize{\_}norm}}{normalize:unnorm}{NW3gGP3e-xs0zU-1}\nwindexuse{\nwixident{numeric}}{numeric}{NW3gGP3e-xs0zU-1}\nwindexuse{\nwixident{subject{\_}to}}{subject:unto}{NW3gGP3e-xs0zU-1}\nwindexuse{\nwixident{subs}}{subs}{NW3gGP3e-xs0zU-1}\nwendcode{}\nwbegindocs{465} Real and imaginary part of the representing vector.
\nwenddocs{}\nwbegincode{466}\sublabel{NW3gGP3e-xs0zU-2}\nwmargintag{{\nwtagstyle{}\subpageref{NW3gGP3e-xs0zU-2}}}\moddef{duplicated methods for class cycle2D~{\nwtagstyle{}\subpageref{NW3gGP3e-xs0zU-1}}}\plusendmoddef\Rm{}\nwstartdeflinemarkup\nwusesondefline{\\{NW3gGP3e-2ARAe1-1}}\nwprevnextdefs{NW3gGP3e-xs0zU-1}{\relax}\nwenddeflinemarkup
{\bf{}ex} {\it{}real\_part}() {\bf{}const};
{\bf{}ex} {\it{}imag\_part}() {\bf{}const};

\nwused{\\{NW3gGP3e-2ARAe1-1}}\nwidentuses{\\{{\nwixident{ex}}{ex}}}\nwindexuse{\nwixident{ex}}{ex}{NW3gGP3e-xs0zU-2}\nwendcode{}\nwbegindocs{467}We also specialise for the derived class {\Tt{}\Rm{}{\bf{}cycle2D}\nwendquote} all operations
defined in \S~\ref{sec:line-oper-cycl}
\nwenddocs{}\nwbegincode{468}\sublabel{NW3gGP3e-1Ihica-1}\nwmargintag{{\nwtagstyle{}\subpageref{NW3gGP3e-1Ihica-1}}}\moddef{duplicated linear operation on cycle2D~{\nwtagstyle{}\subpageref{NW3gGP3e-1Ihica-1}}}\endmoddef\Rm{}\nwstartdeflinemarkup\nwusesondefline{\\{NW3gGP3e-2ARAe1-1}}\nwenddeflinemarkup
{\bf{}const} {\bf{}cycle2D} {\bf{}operator}+({\bf{}const} {\bf{}cycle2D} & {\it{}lh}, {\bf{}const} {\bf{}cycle2D} & {\it{}rh});\nwindexdefn{\nwixident{cycle2D}}{cycle2D}{NW3gGP3e-1Ihica-1}
{\bf{}const} {\bf{}cycle2D} {\bf{}operator}-({\bf{}const} {\bf{}cycle2D} & {\it{}lh}, {\bf{}const} {\bf{}cycle2D} & {\it{}rh});\nwindexdefn{\nwixident{cycle2D}}{cycle2D}{NW3gGP3e-1Ihica-1}
{\bf{}const} {\bf{}cycle2D} {\bf{}operator}\begin{math}\ast\end{math}({\bf{}const} {\bf{}cycle2D} & {\it{}lh}, {\bf{}const} {\bf{}ex} & {\it{}rh});\nwindexdefn{\nwixident{cycle2D}}{cycle2D}{NW3gGP3e-1Ihica-1}
{\bf{}const} {\bf{}cycle2D} {\bf{}operator}\begin{math}\ast\end{math}({\bf{}const} {\bf{}ex} & {\it{}lh}, {\bf{}const} {\bf{}cycle2D} & {\it{}rh});\nwindexdefn{\nwixident{cycle2D}}{cycle2D}{NW3gGP3e-1Ihica-1}
{\bf{}const} {\bf{}cycle2D} {\bf{}operator}\begin{math}\div\end{math}({\bf{}const} {\bf{}cycle2D} & {\it{}lh}, {\bf{}const} {\bf{}ex} & {\it{}rh});\nwindexdefn{\nwixident{cycle2D}}{cycle2D}{NW3gGP3e-1Ihica-1}
{\bf{}const} {\bf{}ex} {\bf{}operator}\begin{math}\ast\end{math}({\bf{}const} {\bf{}cycle2D} & {\it{}lh}, {\bf{}const} {\bf{}cycle2D} & {\it{}rh});\nwindexdefn{\nwixident{ex}}{ex}{NW3gGP3e-1Ihica-1}

\nwused{\\{NW3gGP3e-2ARAe1-1}}\nwidentdefs{\\{{\nwixident{cycle2D}}{cycle2D}}\\{{\nwixident{ex}}{ex}}}\nwidentuses{\\{{\nwixident{operator*}}{operator*}}\\{{\nwixident{operator+}}{operator+}}\\{{\nwixident{operator-}}{operator-}}\\{{\nwixident{operator/}}{operator/}}}\nwindexuse{\nwixident{operator*}}{operator*}{NW3gGP3e-1Ihica-1}\nwindexuse{\nwixident{operator+}}{operator+}{NW3gGP3e-1Ihica-1}\nwindexuse{\nwixident{operator-}}{operator-}{NW3gGP3e-1Ihica-1}\nwindexuse{\nwixident{operator/}}{operator/}{NW3gGP3e-1Ihica-1}\nwendcode{}\nwbegindocs{469}\nwdocspar
\subsection[Implementation of the cycle class]{Implementation of the {\Tt{}\Rm{}{\bf{}cycle}\nwendquote} class}
\label{sec:impl-cycle-class}

We start from definitions of constructors in {\Tt{}\Rm{}{\bf{}cycle}\nwendquote} class
\nwenddocs{}\nwbegincode{470}\sublabel{NW3gGP3e-1Oxbp0-1}\nwmargintag{{\nwtagstyle{}\subpageref{NW3gGP3e-1Oxbp0-1}}}\moddef{cycle.cpp~{\nwtagstyle{}\subpageref{NW3gGP3e-1Oxbp0-1}}}\endmoddef\Rm{}\nwstartdeflinemarkup\nwprevnextdefs{\relax}{NW3gGP3e-1Oxbp0-2}\nwenddeflinemarkup
\LA{}license~{\nwtagstyle{}\subpageref{NW3gGP3e-ZXuKx-1}}\RA{}
{\bf{}\char35{}include}{\tt{} \begin{math}<\end{math}cycle.h\begin{math}>\end{math}}
{\bf{}namespace} {\it{}MoebInv} {\nwlbrace}
{\bf{}using} {\bf{}namespace} {\it{}std};
{\bf{}using} {\bf{}namespace} {\it{}GiNaC};

{\bf{}\char35{}define}{\tt{} PRINT\_CYCLE  c.s \begin{math}<\end{math}\begin{math}<\end{math} "("; \begin{math}\backslash\end{math}}\nwindexdefn{\nwixident{PRINT{\_}CYCLE}}{PRINT:unCYCLE}{NW3gGP3e-1Oxbp0-1}
 {\it{}k}.{\it{}print}({\it{}c}, {\it{}level}); \begin{math}\backslash\end{math}
 {\it{}c}.{\it{}s} \begin{math}\ll\end{math} {\tt{}", "}; \begin{math}\backslash\end{math}
 {\it{}l}.{\it{}print}({\it{}c}, {\it{}level}); \begin{math}\backslash\end{math}
 {\it{}c}.{\it{}s} \begin{math}\ll\end{math} {\tt{}", "}; \begin{math}\backslash\end{math}
 {\it{}m}.{\it{}print}({\it{}c}, {\it{}level}); \begin{math}\backslash\end{math}
 {\it{}c}.{\it{}s} \begin{math}\ll\end{math} {\tt{}")"};

 {\it{}GINAC\_IMPLEMENT\_REGISTERED\_CLASS\_OPT}({\bf{}cycle}, {\bf{}basic},
           {\it{}print\_func}\begin{math}<\end{math}{\it{}print\_dflt}\begin{math}>\end{math}(&{\bf{}cycle}::{\it{}do\_print}).
                                      //           print\_func\begin{math}<\end{math}print\_python\begin{math}>\end{math}(&cycle::do\_print\_python).
           {\it{}print\_func}\begin{math}<\end{math}{\it{}print\_latex}\begin{math}>\end{math}(&{\bf{}cycle}::{\it{}do\_print\_latex}))

 {\it{}GINAC\_IMPLEMENT\_REGISTERED\_CLASS}({\bf{}cycle2D}, {\bf{}cycle})
//,    print\_func\begin{math}<\end{math}print\_dflt\begin{math}>\end{math}(&cycle2D::do\_print)

{\bf{}\char35{}if}{\tt{} GINAC\_VERSION\_ATLEAST(1,5)}
{\it{}return\_type\_t} {\bf{}cycle}::{\it{}return\_type\_tinfo}() {\bf{}const}
{\nwlbrace}
    {\bf{}if} ({\it{}is\_a}\begin{math}<\end{math}{\bf{}numeric}\begin{math}>\end{math}({\it{}get\_dim}()))
        {\bf{}switch} ({\it{}ex\_to}\begin{math}<\end{math}{\bf{}numeric}\begin{math}>\end{math}({\it{}get\_dim}()).{\it{}to\_int}()) {\nwlbrace}
        {\bf{}case} 2:
            {\bf{}return} {\it{}make\_return\_type\_t}\begin{math}<\end{math}{\bf{}cycle2D}\begin{math}>\end{math}();
        {\bf{}default}:
        {\bf{}return} {\it{}make\_return\_type\_t}\begin{math}<\end{math}{\bf{}cycle}\begin{math}>\end{math}();
        {\nwrbrace}
    {\bf{}else}
        {\bf{}return} {\it{}make\_return\_type\_t}\begin{math}<\end{math}{\bf{}cycle}\begin{math}>\end{math}();
{\nwrbrace}
{\bf{}\char35{}endif}{\tt{}}
{\bf{}cycle}::{\bf{}cycle}() : {\it{}unit}(), {\it{}k}(), {\it{}l}(), {\it{}m}()
{\nwlbrace}
{\bf{}\char35{}if}{\tt{} GINAC\_VERSION\_ATLEAST(1,5)}
{\bf{}\char35{}else}{\tt{}}
 {\it{}tinfo\_key} = &{\bf{}cycle}::{\it{}tinfo\_static};
{\bf{}\char35{}endif}{\tt{}}
{\nwrbrace}

\nwalsodefined{\\{NW3gGP3e-1Oxbp0-2}\\{NW3gGP3e-1Oxbp0-3}\\{NW3gGP3e-1Oxbp0-4}\\{NW3gGP3e-1Oxbp0-5}\\{NW3gGP3e-1Oxbp0-6}\\{NW3gGP3e-1Oxbp0-7}\\{NW3gGP3e-1Oxbp0-8}\\{NW3gGP3e-1Oxbp0-9}\\{NW3gGP3e-1Oxbp0-A}\\{NW3gGP3e-1Oxbp0-B}\\{NW3gGP3e-1Oxbp0-C}\\{NW3gGP3e-1Oxbp0-D}\\{NW3gGP3e-1Oxbp0-E}\\{NW3gGP3e-1Oxbp0-F}\\{NW3gGP3e-1Oxbp0-G}\\{NW3gGP3e-1Oxbp0-H}\\{NW3gGP3e-1Oxbp0-I}\\{NW3gGP3e-1Oxbp0-J}\\{NW3gGP3e-1Oxbp0-K}\\{NW3gGP3e-1Oxbp0-L}\\{NW3gGP3e-1Oxbp0-M}\\{NW3gGP3e-1Oxbp0-N}\\{NW3gGP3e-1Oxbp0-O}\\{NW3gGP3e-1Oxbp0-P}\\{NW3gGP3e-1Oxbp0-Q}\\{NW3gGP3e-1Oxbp0-R}\\{NW3gGP3e-1Oxbp0-S}\\{NW3gGP3e-1Oxbp0-T}\\{NW3gGP3e-1Oxbp0-U}\\{NW3gGP3e-1Oxbp0-V}\\{NW3gGP3e-1Oxbp0-W}\\{NW3gGP3e-1Oxbp0-X}\\{NW3gGP3e-1Oxbp0-Y}\\{NW3gGP3e-1Oxbp0-Z}\\{NW3gGP3e-1Oxbp0-a}\\{NW3gGP3e-1Oxbp0-b}\\{NW3gGP3e-1Oxbp0-c}\\{NW3gGP3e-1Oxbp0-d}\\{NW3gGP3e-1Oxbp0-e}\\{NW3gGP3e-1Oxbp0-f}\\{NW3gGP3e-1Oxbp0-g}\\{NW3gGP3e-1Oxbp0-h}\\{NW3gGP3e-1Oxbp0-i}\\{NW3gGP3e-1Oxbp0-j}\\{NW3gGP3e-1Oxbp0-k}\\{NW3gGP3e-1Oxbp0-l}\\{NW3gGP3e-1Oxbp0-m}\\{NW3gGP3e-1Oxbp0-n}\\{NW3gGP3e-1Oxbp0-o}\\{NW3gGP3e-1Oxbp0-p}\\{NW3gGP3e-1Oxbp0-q}\\{NW3gGP3e-1Oxbp0-r}\\{NW3gGP3e-1Oxbp0-s}\\{NW3gGP3e-1Oxbp0-t}\\{NW3gGP3e-1Oxbp0-u}\\{NW3gGP3e-1Oxbp0-v}\\{NW3gGP3e-1Oxbp0-w}\\{NW3gGP3e-1Oxbp0-x}\\{NW3gGP3e-1Oxbp0-y}\\{NW3gGP3e-1Oxbp0-z}\\{NW3gGP3e-1Oxbp0-10}\\{NW3gGP3e-1Oxbp0-11}\\{NW3gGP3e-1Oxbp0-12}\\{NW3gGP3e-1Oxbp0-13}\\{NW3gGP3e-1Oxbp0-14}\\{NW3gGP3e-1Oxbp0-15}\\{NW3gGP3e-1Oxbp0-16}\\{NW3gGP3e-1Oxbp0-17}\\{NW3gGP3e-1Oxbp0-18}\\{NW3gGP3e-1Oxbp0-19}\\{NW3gGP3e-1Oxbp0-1A}\\{NW3gGP3e-1Oxbp0-1B}\\{NW3gGP3e-1Oxbp0-1C}\\{NW3gGP3e-1Oxbp0-1D}\\{NW3gGP3e-1Oxbp0-1E}\\{NW3gGP3e-1Oxbp0-1F}\\{NW3gGP3e-1Oxbp0-1G}\\{NW3gGP3e-1Oxbp0-1H}\\{NW3gGP3e-1Oxbp0-1I}\\{NW3gGP3e-1Oxbp0-1J}\\{NW3gGP3e-1Oxbp0-1K}\\{NW3gGP3e-1Oxbp0-1L}\\{NW3gGP3e-1Oxbp0-1M}}\nwnotused{cycle.cpp}\nwidentdefs{\\{{\nwixident{PRINT{\_}CYCLE}}{PRINT:unCYCLE}}}\nwidentuses{\\{{\nwixident{cycle}}{cycle}}\\{{\nwixident{cycle2D}}{cycle2D}}\\{{\nwixident{get{\_}dim}}{get:undim}}\\{{\nwixident{GINAC{\_}VERSION{\_}ATLEAST}}{GINAC:unVERSION:unATLEAST}}\\{{\nwixident{k}}{k}}\\{{\nwixident{l}}{l}}\\{{\nwixident{m}}{m}}\\{{\nwixident{MoebInv}}{MoebInv}}\\{{\nwixident{numeric}}{numeric}}}\nwindexuse{\nwixident{cycle}}{cycle}{NW3gGP3e-1Oxbp0-1}\nwindexuse{\nwixident{cycle2D}}{cycle2D}{NW3gGP3e-1Oxbp0-1}\nwindexuse{\nwixident{get{\_}dim}}{get:undim}{NW3gGP3e-1Oxbp0-1}\nwindexuse{\nwixident{GINAC{\_}VERSION{\_}ATLEAST}}{GINAC:unVERSION:unATLEAST}{NW3gGP3e-1Oxbp0-1}\nwindexuse{\nwixident{k}}{k}{NW3gGP3e-1Oxbp0-1}\nwindexuse{\nwixident{l}}{l}{NW3gGP3e-1Oxbp0-1}\nwindexuse{\nwixident{m}}{m}{NW3gGP3e-1Oxbp0-1}\nwindexuse{\nwixident{MoebInv}}{MoebInv}{NW3gGP3e-1Oxbp0-1}\nwindexuse{\nwixident{numeric}}{numeric}{NW3gGP3e-1Oxbp0-1}\nwendcode{}\nwbegindocs{471}\nwdocspar
\subsubsection{Main constructor of cycle from all parameters given}
\label{sec:main-constr-cycle}

If all parameters of the cycle are given this constructor is used.
\nwenddocs{}\nwbegincode{472}\sublabel{NW3gGP3e-1Oxbp0-2}\nwmargintag{{\nwtagstyle{}\subpageref{NW3gGP3e-1Oxbp0-2}}}\moddef{cycle.cpp~{\nwtagstyle{}\subpageref{NW3gGP3e-1Oxbp0-1}}}\plusendmoddef\Rm{}\nwstartdeflinemarkup\nwprevnextdefs{NW3gGP3e-1Oxbp0-1}{NW3gGP3e-1Oxbp0-3}\nwenddeflinemarkup
{\bf{}cycle}::{\bf{}cycle}({\bf{}const} {\bf{}ex} & {\it{}k1}, {\bf{}const} {\bf{}ex} & {\it{}l1}, {\bf{}const} {\bf{}ex} & {\it{}m1}, {\bf{}const} {\bf{}ex} & {\it{}metr}) // Main constructor
 : {\it{}k}({\it{}k1}), {\it{}m}({\it{}m1})
{\nwlbrace}
    {\bf{}ex} {\it{}D}, {\it{}metric};

\nwidentuses{\\{{\nwixident{cycle}}{cycle}}\\{{\nwixident{ex}}{ex}}\\{{\nwixident{k}}{k}}\\{{\nwixident{m}}{m}}\\{{\nwixident{metr}}{metr}}}\nwindexuse{\nwixident{cycle}}{cycle}{NW3gGP3e-1Oxbp0-2}\nwindexuse{\nwixident{ex}}{ex}{NW3gGP3e-1Oxbp0-2}\nwindexuse{\nwixident{k}}{k}{NW3gGP3e-1Oxbp0-2}\nwindexuse{\nwixident{m}}{m}{NW3gGP3e-1Oxbp0-2}\nwindexuse{\nwixident{metr}}{metr}{NW3gGP3e-1Oxbp0-2}\nwendcode{}\nwbegindocs{473} The first portion of the code processes various form of
presentation for {\Tt{}\Rm{}{\it{}l}\nwendquote}.
\nwenddocs{}\nwbegincode{474}\sublabel{NW3gGP3e-1Oxbp0-3}\nwmargintag{{\nwtagstyle{}\subpageref{NW3gGP3e-1Oxbp0-3}}}\moddef{cycle.cpp~{\nwtagstyle{}\subpageref{NW3gGP3e-1Oxbp0-1}}}\plusendmoddef\Rm{}\nwstartdeflinemarkup\nwprevnextdefs{NW3gGP3e-1Oxbp0-2}{NW3gGP3e-1Oxbp0-4}\nwenddeflinemarkup
      {\bf{}if} ({\it{}is\_a}\begin{math}<\end{math}{\bf{}indexed}\begin{math}>\end{math}({\it{}l1}.{\it{}simplify\_indexed}())) {\nwlbrace}
          {\it{}l} = {\it{}ex\_to}\begin{math}<\end{math}{\bf{}indexed}\begin{math}>\end{math}({\it{}l1}.{\it{}simplify\_indexed}());
          {\bf{}if} ({\it{}ex\_to}\begin{math}<\end{math}{\bf{}indexed}\begin{math}>\end{math}({\it{}l}).{\it{}get\_indices}().{\it{}size}() \begin{math}\equiv\end{math} 1) {\nwlbrace}
              {\it{}D} = {\it{}ex\_to}\begin{math}<\end{math}{\bf{}varidx}\begin{math}>\end{math}({\it{}ex\_to}\begin{math}<\end{math}{\bf{}indexed}\begin{math}>\end{math}({\it{}l}).{\it{}get\_indices}()[0]).{\it{}get\_dim}();
          {\nwrbrace} {\bf{}else}
              {\bf{}throw}({\it{}std}::{\it{}invalid\_argument}({\tt{}"cycle::cycle(): the second parameter should be an indexed object"}
                                          {\tt{}"with one varindex"}));
      {\nwrbrace} {\bf{}else} {\bf{}if} ({\it{}is\_a}\begin{math}<\end{math}{\bf{}matrix}\begin{math}>\end{math}({\it{}l1}) \begin{math}\wedge\end{math} ({\it{}min}({\it{}ex\_to}\begin{math}<\end{math}{\bf{}matrix}\begin{math}>\end{math}({\it{}l1}).{\it{}rows}(), {\it{}ex\_to}\begin{math}<\end{math}{\bf{}matrix}\begin{math}>\end{math}({\it{}l1}).{\it{}cols}()) \begin{math}\equiv\end{math}1)) {\nwlbrace}
          {\it{}D} = {\it{}max}({\it{}ex\_to}\begin{math}<\end{math}{\bf{}matrix}\begin{math}>\end{math}({\it{}l1}).{\it{}rows}(), {\it{}ex\_to}\begin{math}<\end{math}{\bf{}matrix}\begin{math}>\end{math}({\it{}l1}).{\it{}cols}());
          {\it{}l} = {\bf{}indexed}({\it{}l1}, {\bf{}varidx}(({\bf{}new} {\bf{}symbol})\begin{math}\rightarrow\end{math}{\it{}setflag}({\it{}status\_flags}::{\it{}dynallocated}), {\it{}D}));
      {\nwrbrace} {\bf{}else} {\bf{}if} ({\it{}l1}.{\it{}info}({\it{}info\_flags}::{\it{}list}) \begin{math}\wedge\end{math} ({\it{}l1}.{\it{}nops}() \begin{math}>\end{math} 0)) {\nwlbrace}
          {\it{}D} = {\it{}l1}.{\it{}nops}();
          {\it{}l} = {\bf{}indexed}({\bf{}matrix}(1, {\it{}l1}.{\it{}nops}(), {\it{}ex\_to}\begin{math}<\end{math}{\bf{}lst}\begin{math}>\end{math}({\it{}l1})), {\bf{}varidx}(({\bf{}new} {\bf{}symbol})\begin{math}\rightarrow\end{math}{\it{}setflag}({\it{}status\_flags}::{\it{}dynallocated}), {\it{}D}));

\nwidentuses{\\{{\nwixident{cycle}}{cycle}}\\{{\nwixident{get{\_}dim}}{get:undim}}\\{{\nwixident{l}}{l}}\\{{\nwixident{lst}}{lst}}\\{{\nwixident{matrix}}{matrix}}\\{{\nwixident{nops}}{nops}}\\{{\nwixident{varidx}}{varidx}}}\nwindexuse{\nwixident{cycle}}{cycle}{NW3gGP3e-1Oxbp0-3}\nwindexuse{\nwixident{get{\_}dim}}{get:undim}{NW3gGP3e-1Oxbp0-3}\nwindexuse{\nwixident{l}}{l}{NW3gGP3e-1Oxbp0-3}\nwindexuse{\nwixident{lst}}{lst}{NW3gGP3e-1Oxbp0-3}\nwindexuse{\nwixident{matrix}}{matrix}{NW3gGP3e-1Oxbp0-3}\nwindexuse{\nwixident{nops}}{nops}{NW3gGP3e-1Oxbp0-3}\nwindexuse{\nwixident{varidx}}{varidx}{NW3gGP3e-1Oxbp0-3}\nwendcode{}\nwbegindocs{475}If {\Tt{}\Rm{}{\it{}l1}\nwendquote} is zero we will try to get missing information from the matrix in
the next chunk, otherwise throw an exception.
\nwenddocs{}\nwbegincode{476}\sublabel{NW3gGP3e-1Oxbp0-4}\nwmargintag{{\nwtagstyle{}\subpageref{NW3gGP3e-1Oxbp0-4}}}\moddef{cycle.cpp~{\nwtagstyle{}\subpageref{NW3gGP3e-1Oxbp0-1}}}\plusendmoddef\Rm{}\nwstartdeflinemarkup\nwprevnextdefs{NW3gGP3e-1Oxbp0-3}{NW3gGP3e-1Oxbp0-5}\nwenddeflinemarkup
    {\nwrbrace} {\bf{}else} {\bf{}if} ({\it{}not} {\it{}l1}.{\it{}simplify\_indexed}().{\it{}is\_zero}()) {\nwlbrace}
          {\bf{}throw}({\it{}std}::{\it{}invalid\_argument}({\tt{}"cycle::cycle(): the second parameter should be an indexed object, "}
                                      {\tt{}"matrix or list"}));
      {\nwrbrace}

\nwidentuses{\\{{\nwixident{cycle}}{cycle}}\\{{\nwixident{is{\_}zero}}{is:unzero}}\\{{\nwixident{matrix}}{matrix}}}\nwindexuse{\nwixident{cycle}}{cycle}{NW3gGP3e-1Oxbp0-4}\nwindexuse{\nwixident{is{\_}zero}}{is:unzero}{NW3gGP3e-1Oxbp0-4}\nwindexuse{\nwixident{matrix}}{matrix}{NW3gGP3e-1Oxbp0-4}\nwendcode{}\nwbegindocs{477}Now we process the metric parameter, in case {\Tt{}\Rm{}{\it{}l1}\nwendquote} did not provide
information on the dimensionality we try to get it here.
\nwenddocs{}\nwbegincode{478}\sublabel{NW3gGP3e-1Oxbp0-5}\nwmargintag{{\nwtagstyle{}\subpageref{NW3gGP3e-1Oxbp0-5}}}\moddef{cycle.cpp~{\nwtagstyle{}\subpageref{NW3gGP3e-1Oxbp0-1}}}\plusendmoddef\Rm{}\nwstartdeflinemarkup\nwprevnextdefs{NW3gGP3e-1Oxbp0-4}{NW3gGP3e-1Oxbp0-6}\nwenddeflinemarkup
    {\bf{}if} ({\it{}is\_a}\begin{math}<\end{math}{\bf{}clifford}\begin{math}>\end{math}({\it{}metr})) {\nwlbrace}
        {\bf{}if} ({\it{}D}.{\it{}is\_zero}())
            {\it{}D} = {\it{}ex\_to}\begin{math}<\end{math}{\bf{}varidx}\begin{math}>\end{math}({\it{}metr}.{\it{}op}(1)).{\it{}get\_dim}();
        {\it{}unit} ={\it{}metr};
    {\nwrbrace} {\bf{}else} {\nwlbrace}
        {\bf{}if} ({\it{}D}.{\it{}is\_zero}()) {\nwlbrace}
            {\bf{}if} ({\it{}is\_a}\begin{math}<\end{math}{\bf{}indexed}\begin{math}>\end{math}({\it{}metr}))
                {\it{}D} = {\it{}ex\_to}\begin{math}<\end{math}{\bf{}varidx}\begin{math}>\end{math}({\it{}metr}.{\it{}op}(1)).{\it{}get\_dim}();
            {\bf{}else} {\bf{}if} ({\it{}is\_a}\begin{math}<\end{math}{\bf{}matrix}\begin{math}>\end{math}({\it{}metr}))
                {\it{}D} = {\it{}ex\_to}\begin{math}<\end{math}{\bf{}matrix}\begin{math}>\end{math}({\it{}metr}).{\it{}rows}();
            {\bf{}else} {\nwlbrace}
                {\it{}exvector} {\it{}indices} = {\it{}metr}.{\it{}get\_free\_indices}();
                {\bf{}if} ({\it{}indices}.{\it{}size}() \begin{math}\equiv\end{math} 2)
                    {\it{}D} = {\it{}ex\_to}\begin{math}<\end{math}{\bf{}varidx}\begin{math}>\end{math}({\it{}indices}[0]).{\it{}get\_dim}();
            {\nwrbrace}
        {\nwrbrace}

\nwidentuses{\\{{\nwixident{get{\_}dim}}{get:undim}}\\{{\nwixident{is{\_}zero}}{is:unzero}}\\{{\nwixident{matrix}}{matrix}}\\{{\nwixident{metr}}{metr}}\\{{\nwixident{op}}{op}}\\{{\nwixident{varidx}}{varidx}}}\nwindexuse{\nwixident{get{\_}dim}}{get:undim}{NW3gGP3e-1Oxbp0-5}\nwindexuse{\nwixident{is{\_}zero}}{is:unzero}{NW3gGP3e-1Oxbp0-5}\nwindexuse{\nwixident{matrix}}{matrix}{NW3gGP3e-1Oxbp0-5}\nwindexuse{\nwixident{metr}}{metr}{NW3gGP3e-1Oxbp0-5}\nwindexuse{\nwixident{op}}{op}{NW3gGP3e-1Oxbp0-5}\nwindexuse{\nwixident{varidx}}{varidx}{NW3gGP3e-1Oxbp0-5}\nwendcode{}\nwbegindocs{479}For metric of unknown type we throw an exception.
\nwenddocs{}\nwbegincode{480}\sublabel{NW3gGP3e-1Oxbp0-6}\nwmargintag{{\nwtagstyle{}\subpageref{NW3gGP3e-1Oxbp0-6}}}\moddef{cycle.cpp~{\nwtagstyle{}\subpageref{NW3gGP3e-1Oxbp0-1}}}\plusendmoddef\Rm{}\nwstartdeflinemarkup\nwprevnextdefs{NW3gGP3e-1Oxbp0-5}{NW3gGP3e-1Oxbp0-7}\nwenddeflinemarkup
    {\bf{}if} ({\it{}D}.{\it{}is\_zero}())
        {\bf{}throw}({\it{}std}::{\it{}invalid\_argument}({\tt{}"cycle::cycle(): the metric should be either tensor, "}
                                    {\tt{}"matrix, Clifford unit or indexed by two indices. "}
                                    {\tt{}"Otherwise supply the through the second parameter."}));

{\it{}unit} = {\it{}clifford\_unit}({\bf{}varidx}(0, {\it{}D}), {\it{}metr});
    {\nwrbrace}

\nwidentuses{\\{{\nwixident{cycle}}{cycle}}\\{{\nwixident{is{\_}zero}}{is:unzero}}\\{{\nwixident{matrix}}{matrix}}\\{{\nwixident{metr}}{metr}}\\{{\nwixident{varidx}}{varidx}}}\nwindexuse{\nwixident{cycle}}{cycle}{NW3gGP3e-1Oxbp0-6}\nwindexuse{\nwixident{is{\_}zero}}{is:unzero}{NW3gGP3e-1Oxbp0-6}\nwindexuse{\nwixident{matrix}}{matrix}{NW3gGP3e-1Oxbp0-6}\nwindexuse{\nwixident{metr}}{metr}{NW3gGP3e-1Oxbp0-6}\nwindexuse{\nwixident{varidx}}{varidx}{NW3gGP3e-1Oxbp0-6}\nwendcode{}\nwbegindocs{481}Now we come back to the case {\Tt{}\Rm{}{\it{}l1}\nwendquote} is zero and try to resolve it
with new info on {\Tt{}\Rm{}{\it{}D}\nwendquote}.
\nwenddocs{}\nwbegincode{482}\sublabel{NW3gGP3e-1Oxbp0-7}\nwmargintag{{\nwtagstyle{}\subpageref{NW3gGP3e-1Oxbp0-7}}}\moddef{cycle.cpp~{\nwtagstyle{}\subpageref{NW3gGP3e-1Oxbp0-1}}}\plusendmoddef\Rm{}\nwstartdeflinemarkup\nwprevnextdefs{NW3gGP3e-1Oxbp0-6}{NW3gGP3e-1Oxbp0-8}\nwenddeflinemarkup
{\bf{}\char35{}if}{\tt{} GINAC\_VERSION\_ATLEAST(1,5)}
{\bf{}\char35{}else}{\tt{}}
 \LA{}Set tinfo to dimension~{\nwtagstyle{}\subpageref{NW3gGP3e-MkLdY-1}}\RA{}
{\bf{}\char35{}endif}{\tt{}}
{\nwrbrace}

\nwidentuses{\\{{\nwixident{GINAC{\_}VERSION{\_}ATLEAST}}{GINAC:unVERSION:unATLEAST}}}\nwindexuse{\nwixident{GINAC{\_}VERSION{\_}ATLEAST}}{GINAC:unVERSION:unATLEAST}{NW3gGP3e-1Oxbp0-7}\nwendcode{}\nwbegindocs{483}We set tinfo key for cycle according to its dimension
\nwenddocs{}\nwbegincode{484}\sublabel{NW3gGP3e-MkLdY-1}\nwmargintag{{\nwtagstyle{}\subpageref{NW3gGP3e-MkLdY-1}}}\moddef{Set tinfo to dimension~{\nwtagstyle{}\subpageref{NW3gGP3e-MkLdY-1}}}\endmoddef\Rm{}\nwstartdeflinemarkup\nwusesondefline{\\{NW3gGP3e-1Oxbp0-7}}\nwenddeflinemarkup
{\bf{}if} ({\it{}is\_a}\begin{math}<\end{math}{\bf{}numeric}\begin{math}>\end{math}({\it{}D}))
    {\bf{}switch} ({\it{}ex\_to}\begin{math}<\end{math}{\bf{}numeric}\begin{math}>\end{math}({\it{}D}).{\it{}to\_int}()) {\nwlbrace}
    {\bf{}case} 2:
        {\it{}tinfo\_key} = &{\bf{}cycle2D}::{\it{}tinfo\_static};
        {\bf{}break};
    {\bf{}default}:
        {\it{}tinfo\_key} = &{\bf{}cycle}::{\it{}tinfo\_static};
        {\bf{}break};
    {\nwrbrace}
{\bf{}else}
    {\it{}tinfo\_key} = &{\bf{}cycle}::{\it{}tinfo\_static};

\nwused{\\{NW3gGP3e-1Oxbp0-7}}\nwidentuses{\\{{\nwixident{cycle}}{cycle}}\\{{\nwixident{cycle2D}}{cycle2D}}\\{{\nwixident{numeric}}{numeric}}}\nwindexuse{\nwixident{cycle}}{cycle}{NW3gGP3e-MkLdY-1}\nwindexuse{\nwixident{cycle2D}}{cycle2D}{NW3gGP3e-MkLdY-1}\nwindexuse{\nwixident{numeric}}{numeric}{NW3gGP3e-MkLdY-1}\nwendcode{}\nwbegindocs{485}\nwdocspar
\subsubsection{Specific cycle constructors}
\label{sec:spec-cycle-constr}

\nwenddocs{}\nwbegindocs{486}Constructor for cycle with the given determinant {\Tt{}\Rm{}{\it{}r\_squared}\nwendquote}, e.g.
zero-radius cycle by default.
\nwenddocs{}\nwbegincode{487}\sublabel{NW3gGP3e-1Oxbp0-8}\nwmargintag{{\nwtagstyle{}\subpageref{NW3gGP3e-1Oxbp0-8}}}\moddef{cycle.cpp~{\nwtagstyle{}\subpageref{NW3gGP3e-1Oxbp0-1}}}\plusendmoddef\Rm{}\nwstartdeflinemarkup\nwprevnextdefs{NW3gGP3e-1Oxbp0-7}{NW3gGP3e-1Oxbp0-9}\nwenddeflinemarkup
{\bf{}cycle}::{\bf{}cycle}({\bf{}const} {\bf{}lst} & {\it{}l}, {\bf{}const} {\bf{}ex} & {\it{}metr}, {\bf{}const} {\bf{}ex} & {\it{}r\_squared}, {\bf{}const} {\bf{}ex} & {\it{}e}, {\bf{}const} {\bf{}ex} & {\it{}sign})
{\nwlbrace}
    {\bf{}symbol} {\it{}m\_temp};
    {\bf{}cycle} {\it{}C}({\bf{}numeric}(1), {\it{}l}, {\it{}m\_temp}, {\it{}metr});
    (\begin{math}\ast\end{math}{\it{}this}) = {\it{}C}.{\it{}subject\_to}({\bf{}lst}({\it{}C}.{\it{}radius\_sq}({\it{}e}, {\it{}sign}) \begin{math}\equiv\end{math} {\it{}r\_squared}), {\bf{}lst}({\it{}m\_temp}));
{\nwrbrace}

\nwidentuses{\\{{\nwixident{cycle}}{cycle}}\\{{\nwixident{ex}}{ex}}\\{{\nwixident{l}}{l}}\\{{\nwixident{lst}}{lst}}\\{{\nwixident{metr}}{metr}}\\{{\nwixident{numeric}}{numeric}}\\{{\nwixident{radius{\_}sq}}{radius:unsq}}\\{{\nwixident{subject{\_}to}}{subject:unto}}}\nwindexuse{\nwixident{cycle}}{cycle}{NW3gGP3e-1Oxbp0-8}\nwindexuse{\nwixident{ex}}{ex}{NW3gGP3e-1Oxbp0-8}\nwindexuse{\nwixident{l}}{l}{NW3gGP3e-1Oxbp0-8}\nwindexuse{\nwixident{lst}}{lst}{NW3gGP3e-1Oxbp0-8}\nwindexuse{\nwixident{metr}}{metr}{NW3gGP3e-1Oxbp0-8}\nwindexuse{\nwixident{numeric}}{numeric}{NW3gGP3e-1Oxbp0-8}\nwindexuse{\nwixident{radius{\_}sq}}{radius:unsq}{NW3gGP3e-1Oxbp0-8}\nwindexuse{\nwixident{subject{\_}to}}{subject:unto}{NW3gGP3e-1Oxbp0-8}\nwendcode{}\nwbegindocs{488}This is the constructor of a cycle identical to the given one with replaced metric
in the point space.
\nwenddocs{}\nwbegincode{489}\sublabel{NW3gGP3e-1Oxbp0-9}\nwmargintag{{\nwtagstyle{}\subpageref{NW3gGP3e-1Oxbp0-9}}}\moddef{cycle.cpp~{\nwtagstyle{}\subpageref{NW3gGP3e-1Oxbp0-1}}}\plusendmoddef\Rm{}\nwstartdeflinemarkup\nwprevnextdefs{NW3gGP3e-1Oxbp0-8}{NW3gGP3e-1Oxbp0-A}\nwenddeflinemarkup
{\bf{}cycle}::{\bf{}cycle}({\bf{}const} {\bf{}cycle} & {\it{}C}, {\bf{}const} {\bf{}ex} & {\it{}metr})
{\nwlbrace}
 (\begin{math}\ast\end{math}{\it{}this}) = {\it{}metr}.{\it{}is\_zero}()? {\it{}C} : {\bf{}cycle}({\it{}C}.{\it{}get\_k}(), {\it{}C}.{\it{}get\_l}(), {\it{}C}.{\it{}get\_m}(), {\it{}metr});
{\nwrbrace}

\nwidentuses{\\{{\nwixident{cycle}}{cycle}}\\{{\nwixident{ex}}{ex}}\\{{\nwixident{get{\_}k}}{get:unk}}\\{{\nwixident{get{\_}l}}{get:unl}}\\{{\nwixident{get{\_}m}}{get:unm}}\\{{\nwixident{is{\_}zero}}{is:unzero}}\\{{\nwixident{metr}}{metr}}}\nwindexuse{\nwixident{cycle}}{cycle}{NW3gGP3e-1Oxbp0-9}\nwindexuse{\nwixident{ex}}{ex}{NW3gGP3e-1Oxbp0-9}\nwindexuse{\nwixident{get{\_}k}}{get:unk}{NW3gGP3e-1Oxbp0-9}\nwindexuse{\nwixident{get{\_}l}}{get:unl}{NW3gGP3e-1Oxbp0-9}\nwindexuse{\nwixident{get{\_}m}}{get:unm}{NW3gGP3e-1Oxbp0-9}\nwindexuse{\nwixident{is{\_}zero}}{is:unzero}{NW3gGP3e-1Oxbp0-9}\nwindexuse{\nwixident{metr}}{metr}{NW3gGP3e-1Oxbp0-9}\nwendcode{}\nwbegindocs{490}Constructor of a cycle from a matrix representations. First we check
that matrix is in a proper form.
\nwenddocs{}\nwbegincode{491}\sublabel{NW3gGP3e-1Oxbp0-A}\nwmargintag{{\nwtagstyle{}\subpageref{NW3gGP3e-1Oxbp0-A}}}\moddef{cycle.cpp~{\nwtagstyle{}\subpageref{NW3gGP3e-1Oxbp0-1}}}\plusendmoddef\Rm{}\nwstartdeflinemarkup\nwprevnextdefs{NW3gGP3e-1Oxbp0-9}{NW3gGP3e-1Oxbp0-B}\nwenddeflinemarkup
{\bf{}cycle}::{\bf{}cycle}({\bf{}const} {\bf{}matrix} & {\it{}M}, {\bf{}const} {\bf{}ex} & {\it{}metr}, {\bf{}const} {\bf{}ex} & {\it{}e}, {\bf{}const} {\bf{}ex} & {\it{}sign})
{\nwlbrace}
    {\bf{}if} ({\it{}not} ({\it{}M}.{\it{}rows}() \begin{math}\equiv\end{math} 2 \begin{math}\wedge\end{math} {\it{}M}.{\it{}cols}() \begin{math}\equiv\end{math} 2 \begin{math}\wedge\end{math} ({\it{}M}.{\it{}op}(0)+{\it{}M}.{\it{}op}(3)).{\it{}normal}().{\it{}is\_zero}()))
        {\bf{}throw}({\it{}std}::{\it{}invalid\_argument}({\tt{}"cycle::cycle(): the second argument should be square 2x2 matrix "}
                                    {\tt{}"with M(1,1)=-M(2,2)"}));

    \LA{}Create a Clifford unit ~{\nwtagstyle{}\subpageref{NW3gGP3e-4RDj8q-1}}\RA{}
    {\bf{}varidx} {\it{}i0}(({\bf{}new} {\bf{}symbol})\begin{math}\rightarrow\end{math}{\it{}setflag}({\it{}status\_flags}::{\it{}dynallocated}), {\it{}D}),
    {\it{}i1}(({\bf{}new} {\bf{}symbol})\begin{math}\rightarrow\end{math}{\it{}setflag}({\it{}status\_flags}::{\it{}dynallocated}), {\it{}D}, {\bf{}true});

\nwidentuses{\\{{\nwixident{cycle}}{cycle}}\\{{\nwixident{ex}}{ex}}\\{{\nwixident{is{\_}zero}}{is:unzero}}\\{{\nwixident{matrix}}{matrix}}\\{{\nwixident{metr}}{metr}}\\{{\nwixident{normal}}{normal}}\\{{\nwixident{op}}{op}}\\{{\nwixident{varidx}}{varidx}}}\nwindexuse{\nwixident{cycle}}{cycle}{NW3gGP3e-1Oxbp0-A}\nwindexuse{\nwixident{ex}}{ex}{NW3gGP3e-1Oxbp0-A}\nwindexuse{\nwixident{is{\_}zero}}{is:unzero}{NW3gGP3e-1Oxbp0-A}\nwindexuse{\nwixident{matrix}}{matrix}{NW3gGP3e-1Oxbp0-A}\nwindexuse{\nwixident{metr}}{metr}{NW3gGP3e-1Oxbp0-A}\nwindexuse{\nwixident{normal}}{normal}{NW3gGP3e-1Oxbp0-A}\nwindexuse{\nwixident{op}}{op}{NW3gGP3e-1Oxbp0-A}\nwindexuse{\nwixident{varidx}}{varidx}{NW3gGP3e-1Oxbp0-A}\nwendcode{}\nwbegindocs{492}There are different options for {\Tt{}\Rm{}{\it{}sign}\nwendquote}, which should be
checked. First we verify is it zero and use the default value in this case.
\nwenddocs{}\nwbegincode{493}\sublabel{NW3gGP3e-1Oxbp0-B}\nwmargintag{{\nwtagstyle{}\subpageref{NW3gGP3e-1Oxbp0-B}}}\moddef{cycle.cpp~{\nwtagstyle{}\subpageref{NW3gGP3e-1Oxbp0-1}}}\plusendmoddef\Rm{}\nwstartdeflinemarkup\nwprevnextdefs{NW3gGP3e-1Oxbp0-A}{NW3gGP3e-1Oxbp0-C}\nwenddeflinemarkup
{\bf{}if} ({\it{}sign}.{\it{}is\_zero}()) {\nwlbrace}
    {\bf{}try} {\nwlbrace}
        (\begin{math}\ast\end{math}{\it{}this}) = {\bf{}cycle}({\it{}remove\_dirac\_ONE}({\it{}M}.{\it{}op}(2)), {\it{}clifford\_to\_lst}({\it{}M}.{\it{}op}(0), {\it{}e1}), {\it{}remove\_dirac\_ONE}({\it{}M}.{\it{}op}(1)), {\it{}metr});
    {\nwrbrace} {\bf{}catch}  ({\it{}std}::{\it{}exception} &{\it{}p}) {\nwlbrace}
        (\begin{math}\ast\end{math}{\it{}this}) = {\bf{}cycle}({\bf{}numeric}(1), {\it{}clifford\_to\_lst}({\it{}M}.{\it{}op}(0)\begin{math}\ast\end{math}{\it{}clifford\_inverse}({\it{}M}.{\it{}op}(2)), {\it{}e1}),
                        {\it{}canonicalize\_clifford}({\it{}M}.{\it{}op}(1)\begin{math}\ast\end{math}{\it{}clifford\_inverse}({\it{}M}.{\it{}op}(2))), {\it{}metr});
    {\nwrbrace}
{\nwrbrace} {\bf{}else} {\nwlbrace}
    {\bf{}ex} {\it{}sign\_m}, {\it{}conv};
    {\it{}sign\_m} = {\it{}sign}.{\it{}evalm}();

\nwidentuses{\\{{\nwixident{catch}}{catch}}\\{{\nwixident{cycle}}{cycle}}\\{{\nwixident{ex}}{ex}}\\{{\nwixident{is{\_}zero}}{is:unzero}}\\{{\nwixident{metr}}{metr}}\\{{\nwixident{numeric}}{numeric}}\\{{\nwixident{op}}{op}}}\nwindexuse{\nwixident{catch}}{catch}{NW3gGP3e-1Oxbp0-B}\nwindexuse{\nwixident{cycle}}{cycle}{NW3gGP3e-1Oxbp0-B}\nwindexuse{\nwixident{ex}}{ex}{NW3gGP3e-1Oxbp0-B}\nwindexuse{\nwixident{is{\_}zero}}{is:unzero}{NW3gGP3e-1Oxbp0-B}\nwindexuse{\nwixident{metr}}{metr}{NW3gGP3e-1Oxbp0-B}\nwindexuse{\nwixident{numeric}}{numeric}{NW3gGP3e-1Oxbp0-B}\nwindexuse{\nwixident{op}}{op}{NW3gGP3e-1Oxbp0-B}\nwendcode{}\nwbegindocs{494}If {\Tt{}\Rm{}{\it{}sign}\nwendquote} is not zero we process different types which can supply it.
\nwenddocs{}\nwbegincode{495}\sublabel{NW3gGP3e-1Oxbp0-C}\nwmargintag{{\nwtagstyle{}\subpageref{NW3gGP3e-1Oxbp0-C}}}\moddef{cycle.cpp~{\nwtagstyle{}\subpageref{NW3gGP3e-1Oxbp0-1}}}\plusendmoddef\Rm{}\nwstartdeflinemarkup\nwprevnextdefs{NW3gGP3e-1Oxbp0-B}{NW3gGP3e-1Oxbp0-D}\nwenddeflinemarkup
    {\bf{}if} ({\it{}is\_a}\begin{math}<\end{math}{\bf{}tensor}\begin{math}>\end{math}({\it{}sign\_m}))
        {\it{}conv} = {\bf{}indexed}({\it{}ex\_to}\begin{math}<\end{math}{\bf{}tensor}\begin{math}>\end{math}({\it{}sign\_m}), {\it{}i0}, {\it{}i1});
    {\bf{}else} {\bf{}if} ({\it{}is\_a}\begin{math}<\end{math}{\bf{}clifford}\begin{math}>\end{math}({\it{}sign\_m})) {\nwlbrace}
        {\bf{}if} ({\it{}ex\_to}\begin{math}<\end{math}{\bf{}varidx}\begin{math}>\end{math}({\it{}sign\_m}.{\it{}op}(1)).{\it{}get\_dim}() \begin{math}\equiv\end{math} {\it{}D})
            {\it{}conv} = {\it{}ex\_to}\begin{math}<\end{math}{\bf{}clifford}\begin{math}>\end{math}({\it{}sign\_m}).{\it{}get\_metric}({\it{}i0}, {\it{}i1});
        {\bf{}else}
            {\bf{}throw}({\it{}std}::{\it{}invalid\_argument}({\tt{}"cycle::cycle(): the sign should be a Clifford unit with "}
                                        {\tt{}"the dimensionality matching to the second parameter"}));
    {\nwrbrace} {\bf{}else} {\bf{}if} ({\it{}is\_a}\begin{math}<\end{math}{\bf{}indexed}\begin{math}>\end{math}({\it{}sign\_m})) {\nwlbrace}
        {\it{}exvector} {\it{}ind} = {\it{}ex\_to}\begin{math}<\end{math}{\bf{}indexed}\begin{math}>\end{math}({\it{}sign\_m}).{\it{}get\_indices}();
        {\bf{}if} (({\it{}ind}.{\it{}size}() \begin{math}\equiv\end{math} 2) \begin{math}\wedge\end{math} ({\it{}ex\_to}\begin{math}<\end{math}{\bf{}varidx}\begin{math}>\end{math}({\it{}ind}[0]).{\it{}get\_dim}() \begin{math}\equiv\end{math} {\it{}D}) \begin{math}\wedge\end{math} ({\it{}ex\_to}\begin{math}<\end{math}{\bf{}varidx}\begin{math}>\end{math}({\it{}ind}[1]).{\it{}get\_dim}() \begin{math}\equiv\end{math} {\it{}D}))
            {\it{}conv} = {\it{}sign\_m}.{\it{}subs}({\bf{}lst}({\it{}ind}[0] \begin{math}\equiv\end{math} {\it{}i0}, {\it{}ind}[1] \begin{math}\equiv\end{math} {\it{}i1}));
        {\bf{}else}
            {\bf{}throw}({\it{}std}::{\it{}invalid\_argument}({\tt{}"cycle::cycle(): the sign should be an indexed object with two "}
                                        {\tt{}"indices and their dimensionality matching to the second parameter"}));

\nwidentuses{\\{{\nwixident{cycle}}{cycle}}\\{{\nwixident{get{\_}dim}}{get:undim}}\\{{\nwixident{get{\_}metric}}{get:unmetric}}\\{{\nwixident{lst}}{lst}}\\{{\nwixident{op}}{op}}\\{{\nwixident{subs}}{subs}}\\{{\nwixident{varidx}}{varidx}}}\nwindexuse{\nwixident{cycle}}{cycle}{NW3gGP3e-1Oxbp0-C}\nwindexuse{\nwixident{get{\_}dim}}{get:undim}{NW3gGP3e-1Oxbp0-C}\nwindexuse{\nwixident{get{\_}metric}}{get:unmetric}{NW3gGP3e-1Oxbp0-C}\nwindexuse{\nwixident{lst}}{lst}{NW3gGP3e-1Oxbp0-C}\nwindexuse{\nwixident{op}}{op}{NW3gGP3e-1Oxbp0-C}\nwindexuse{\nwixident{subs}}{subs}{NW3gGP3e-1Oxbp0-C}\nwindexuse{\nwixident{varidx}}{varidx}{NW3gGP3e-1Oxbp0-C}\nwendcode{}\nwbegindocs{496} The sign given as a matrix is oftenly used.
\nwenddocs{}\nwbegincode{497}\sublabel{NW3gGP3e-1Oxbp0-D}\nwmargintag{{\nwtagstyle{}\subpageref{NW3gGP3e-1Oxbp0-D}}}\moddef{cycle.cpp~{\nwtagstyle{}\subpageref{NW3gGP3e-1Oxbp0-1}}}\plusendmoddef\Rm{}\nwstartdeflinemarkup\nwprevnextdefs{NW3gGP3e-1Oxbp0-C}{NW3gGP3e-1Oxbp0-E}\nwenddeflinemarkup
    {\nwrbrace} {\bf{}else} {\bf{}if} ({\it{}is\_a}\begin{math}<\end{math}{\bf{}matrix}\begin{math}>\end{math}({\it{}sign\_m})) {\nwlbrace}
        {\bf{}if} (({\it{}ex\_to}\begin{math}<\end{math}{\bf{}matrix}\begin{math}>\end{math}({\it{}sign\_m}).{\it{}cols}() \begin{math}\equiv\end{math} {\it{}D}) \begin{math}\wedge\end{math} ({\it{}ex\_to}\begin{math}<\end{math}{\bf{}matrix}\begin{math}>\end{math}({\it{}sign\_m}).{\it{}rows}() \begin{math}\equiv\end{math} {\it{}D}))
            {\it{}conv} = {\bf{}indexed}({\it{}ex\_to}\begin{math}<\end{math}{\bf{}matrix}\begin{math}>\end{math}({\it{}sign\_m}), {\it{}i0}, {\it{}i1});
        {\bf{}else}
            {\bf{}throw}({\it{}std}::{\it{}invalid\_argument}({\tt{}"cycle::cycle(): the sign should be a square matrix with the "}
                                        {\tt{}"dimensionality matching to the second parameter"}));
    {\nwrbrace} {\bf{}else}
        {\bf{}throw}({\it{}std}::{\it{}invalid\_argument}({\tt{}"cycle::cycle(): the sign should be either tensor, indexed, matrix "}
                                    {\tt{}"or Clifford unit"}));

\nwidentuses{\\{{\nwixident{cycle}}{cycle}}\\{{\nwixident{matrix}}{matrix}}}\nwindexuse{\nwixident{cycle}}{cycle}{NW3gGP3e-1Oxbp0-D}\nwindexuse{\nwixident{matrix}}{matrix}{NW3gGP3e-1Oxbp0-D}\nwendcode{}\nwbegindocs{498}Then all blocks of the matrix are used to construct the cycle in
main constructor.
\nwenddocs{}\nwbegincode{499}\sublabel{NW3gGP3e-1Oxbp0-E}\nwmargintag{{\nwtagstyle{}\subpageref{NW3gGP3e-1Oxbp0-E}}}\moddef{cycle.cpp~{\nwtagstyle{}\subpageref{NW3gGP3e-1Oxbp0-1}}}\plusendmoddef\Rm{}\nwstartdeflinemarkup\nwprevnextdefs{NW3gGP3e-1Oxbp0-D}{NW3gGP3e-1Oxbp0-F}\nwenddeflinemarkup
    {\bf{}try} {\nwlbrace}
        (\begin{math}\ast\end{math}{\it{}this}) = {\bf{}cycle}({\it{}remove\_dirac\_ONE}({\it{}M}.{\it{}op}(2)), {\bf{}indexed}({\bf{}matrix}(1, {\it{}ex\_to}\begin{math}<\end{math}{\bf{}numeric}\begin{math}>\end{math}({\it{}D}).{\it{}to\_int}(),
                                                                   {\it{}clifford\_to\_lst}({\it{}M}.{\it{}op}(0), {\it{}e1})), {\it{}i0}.{\it{}toggle\_variance}())\begin{math}\ast\end{math}{\it{}conv}, {\it{}remove\_dirac\_ONE}({\it{}M}.{\it{}op}(1)), {\it{}metr});
    {\nwrbrace} {\bf{}catch}  ({\it{}std}::{\it{}exception} &{\it{}p}) {\nwlbrace}
        (\begin{math}\ast\end{math}{\it{}this}) = {\bf{}cycle}({\bf{}numeric}(1), {\bf{}indexed}({\bf{}matrix}(1, {\it{}ex\_to}\begin{math}<\end{math}{\bf{}numeric}\begin{math}>\end{math}({\it{}D}).{\it{}to\_int}(), {\it{}clifford\_to\_lst}({\it{}M}.{\it{}op}(0)
                                                                                                  \begin{math}\ast\end{math}{\it{}clifford\_inverse}({\it{}M}.{\it{}op}(2)), {\it{}e1})), {\it{}i0}.{\it{}toggle\_variance}())\begin{math}\ast\end{math}{\it{}conv},
                        {\it{}canonicalize\_clifford}({\it{}M}.{\it{}op}(1)\begin{math}\ast\end{math}{\it{}clifford\_inverse}({\it{}M}.{\it{}op}(2))), {\it{}metr});
    {\nwrbrace}
{\nwrbrace}
{\nwrbrace}

\nwidentuses{\\{{\nwixident{catch}}{catch}}\\{{\nwixident{cycle}}{cycle}}\\{{\nwixident{matrix}}{matrix}}\\{{\nwixident{metr}}{metr}}\\{{\nwixident{numeric}}{numeric}}\\{{\nwixident{op}}{op}}}\nwindexuse{\nwixident{catch}}{catch}{NW3gGP3e-1Oxbp0-E}\nwindexuse{\nwixident{cycle}}{cycle}{NW3gGP3e-1Oxbp0-E}\nwindexuse{\nwixident{matrix}}{matrix}{NW3gGP3e-1Oxbp0-E}\nwindexuse{\nwixident{metr}}{metr}{NW3gGP3e-1Oxbp0-E}\nwindexuse{\nwixident{numeric}}{numeric}{NW3gGP3e-1Oxbp0-E}\nwindexuse{\nwixident{op}}{op}{NW3gGP3e-1Oxbp0-E}\nwendcode{}\nwbegindocs{500}We need the proper Clifford unit to decompose M(0,0) element into
vector for {\Tt{}\Rm{}{\it{}l}\nwendquote}.
\nwenddocs{}\nwbegincode{501}\sublabel{NW3gGP3e-4RDj8q-1}\nwmargintag{{\nwtagstyle{}\subpageref{NW3gGP3e-4RDj8q-1}}}\moddef{Create a Clifford unit ~{\nwtagstyle{}\subpageref{NW3gGP3e-4RDj8q-1}}}\endmoddef\Rm{}\nwstartdeflinemarkup\nwusesondefline{\\{NW3gGP3e-1Oxbp0-A}}\nwenddeflinemarkup
{\bf{}ex} {\it{}e1}, {\it{}D};
{\bf{}if} ({\it{}e}.{\it{}is\_zero}()) {\nwlbrace}
    {\bf{}if} ({\it{}is\_a}\begin{math}<\end{math}{\bf{}clifford}\begin{math}>\end{math}({\it{}metr})) {\nwlbrace}
        {\it{}D}={\it{}ex\_to}\begin{math}<\end{math}{\bf{}varidx}\begin{math}>\end{math}({\it{}metr}.{\it{}op}(1)).{\it{}get\_dim}();
        {\it{}e1}={\it{}metr};
    {\nwrbrace} {\bf{}else} {\nwlbrace}
        {\bf{}ex} {\it{}metr1};
        {\bf{}if} ({\it{}is\_a}\begin{math}<\end{math}{\bf{}matrix}\begin{math}>\end{math}({\it{}metr})) {\nwlbrace}
            {\it{}D} = {\it{}ex\_to}\begin{math}<\end{math}{\bf{}matrix}\begin{math}>\end{math}({\it{}metr}).{\it{}cols}();
            {\it{}metr1} = {\it{}metr};
        {\nwrbrace} {\bf{}else} {\bf{}if} ({\it{}is\_a}\begin{math}<\end{math}{\bf{}indexed}\begin{math}>\end{math}({\it{}metr})) {\nwlbrace}
            {\it{}D} = {\it{}ex\_to}\begin{math}<\end{math}{\bf{}varidx}\begin{math}>\end{math}({\it{}ex\_to}\begin{math}<\end{math}{\bf{}indexed}\begin{math}>\end{math}({\it{}metr}).{\it{}get\_indices}()[0]).{\it{}get\_dim}();
            {\it{}metr1} = {\it{}metr};
        {\nwrbrace} {\bf{}else}
            {\bf{}throw}({\it{}std}::{\it{}invalid\_argument}({\tt{}"cycle(): Could not determine the dimensionality of point space "}
                                        {\tt{}"from the supplied metric or Clifford unit"}));

        {\it{}e1} = {\it{}clifford\_unit}({\bf{}varidx}(({\bf{}new} {\bf{}symbol})\begin{math}\rightarrow\end{math}{\it{}setflag}({\it{}status\_flags}::{\it{}dynallocated}), {\it{}D}), {\it{}metr1});
    {\nwrbrace}
{\nwrbrace} {\bf{}else} {\nwlbrace}
    {\bf{}if} (\begin{math}\neg\end{math} {\it{}is\_a}\begin{math}<\end{math}{\bf{}clifford}\begin{math}>\end{math}({\it{}e}))
        {\bf{}throw}({\it{}std}::{\it{}invalid\_argument}({\tt{}"cycle(): if e is supplied, it shall be a Clifford unit"}));
    {\it{}e1} = {\it{}e};
    {\it{}D} = {\it{}ex\_to}\begin{math}<\end{math}{\bf{}varidx}\begin{math}>\end{math}({\it{}e}.{\it{}op}(1)).{\it{}get\_dim}();
 {\nwrbrace}

\nwused{\\{NW3gGP3e-1Oxbp0-A}}\nwidentuses{\\{{\nwixident{cycle}}{cycle}}\\{{\nwixident{ex}}{ex}}\\{{\nwixident{get{\_}dim}}{get:undim}}\\{{\nwixident{is{\_}zero}}{is:unzero}}\\{{\nwixident{matrix}}{matrix}}\\{{\nwixident{metr}}{metr}}\\{{\nwixident{op}}{op}}\\{{\nwixident{varidx}}{varidx}}}\nwindexuse{\nwixident{cycle}}{cycle}{NW3gGP3e-4RDj8q-1}\nwindexuse{\nwixident{ex}}{ex}{NW3gGP3e-4RDj8q-1}\nwindexuse{\nwixident{get{\_}dim}}{get:undim}{NW3gGP3e-4RDj8q-1}\nwindexuse{\nwixident{is{\_}zero}}{is:unzero}{NW3gGP3e-4RDj8q-1}\nwindexuse{\nwixident{matrix}}{matrix}{NW3gGP3e-4RDj8q-1}\nwindexuse{\nwixident{metr}}{metr}{NW3gGP3e-4RDj8q-1}\nwindexuse{\nwixident{op}}{op}{NW3gGP3e-4RDj8q-1}\nwindexuse{\nwixident{varidx}}{varidx}{NW3gGP3e-4RDj8q-1}\nwendcode{}\nwbegindocs{502}\nwdocspar
\subsubsection[Class cycle members access]{Class {\Tt{}\Rm{}{\bf{}cycle}\nwendquote} members access}
\label{sec:class-cycle-members}

Class {\Tt{}\Rm{}{\bf{}cycle}\nwendquote} has four operands.
\nwenddocs{}\nwbegincode{503}\sublabel{NW3gGP3e-1Oxbp0-F}\nwmargintag{{\nwtagstyle{}\subpageref{NW3gGP3e-1Oxbp0-F}}}\moddef{cycle.cpp~{\nwtagstyle{}\subpageref{NW3gGP3e-1Oxbp0-1}}}\plusendmoddef\Rm{}\nwstartdeflinemarkup\nwprevnextdefs{NW3gGP3e-1Oxbp0-E}{NW3gGP3e-1Oxbp0-G}\nwenddeflinemarkup
{\bf{}ex} {\bf{}cycle}::{\it{}op}({\it{}size\_t} {\it{}i}) {\bf{}const}
{\nwlbrace}
 {\it{}GINAC\_ASSERT}({\it{}i}\begin{math}<\end{math}{\it{}nops}());

 {\bf{}switch} ({\it{}i}) {\nwlbrace}
 {\bf{}case} 0:
  {\bf{}return} {\it{}k};
 {\bf{}case} 1:
  {\bf{}return} {\it{}l};
 {\bf{}case} 2:
  {\bf{}return} {\it{}m};
 {\bf{}case} 3:
  {\bf{}return} {\it{}unit};
 {\bf{}default}:
  {\bf{}throw}({\it{}std}::{\it{}invalid\_argument}({\tt{}"cycle::op(): requested operand out of the range (4)"}));
 {\nwrbrace}
{\nwrbrace}

\nwidentuses{\\{{\nwixident{cycle}}{cycle}}\\{{\nwixident{ex}}{ex}}\\{{\nwixident{k}}{k}}\\{{\nwixident{l}}{l}}\\{{\nwixident{m}}{m}}\\{{\nwixident{nops}}{nops}}\\{{\nwixident{op}}{op}}}\nwindexuse{\nwixident{cycle}}{cycle}{NW3gGP3e-1Oxbp0-F}\nwindexuse{\nwixident{ex}}{ex}{NW3gGP3e-1Oxbp0-F}\nwindexuse{\nwixident{k}}{k}{NW3gGP3e-1Oxbp0-F}\nwindexuse{\nwixident{l}}{l}{NW3gGP3e-1Oxbp0-F}\nwindexuse{\nwixident{m}}{m}{NW3gGP3e-1Oxbp0-F}\nwindexuse{\nwixident{nops}}{nops}{NW3gGP3e-1Oxbp0-F}\nwindexuse{\nwixident{op}}{op}{NW3gGP3e-1Oxbp0-F}\nwendcode{}\nwbegindocs{504}Operands may be set through this method.
\nwenddocs{}\nwbegincode{505}\sublabel{NW3gGP3e-1Oxbp0-G}\nwmargintag{{\nwtagstyle{}\subpageref{NW3gGP3e-1Oxbp0-G}}}\moddef{cycle.cpp~{\nwtagstyle{}\subpageref{NW3gGP3e-1Oxbp0-1}}}\plusendmoddef\Rm{}\nwstartdeflinemarkup\nwprevnextdefs{NW3gGP3e-1Oxbp0-F}{NW3gGP3e-1Oxbp0-H}\nwenddeflinemarkup
{\bf{}ex} & {\bf{}cycle}::{\it{}let\_op}({\it{}size\_t} {\it{}i})
{\nwlbrace}
 {\it{}GINAC\_ASSERT}({\it{}i}\begin{math}<\end{math}{\it{}nops}());

 {\it{}ensure\_if\_modifiable}();
 {\bf{}switch} ({\it{}i}) {\nwlbrace}
 {\bf{}case} 0:
  {\bf{}return} {\it{}k};
 {\bf{}case} 1:
  {\bf{}return} {\it{}l};
 {\bf{}case} 2:
  {\bf{}return} {\it{}m};
 {\bf{}case} 3:
  {\bf{}return} {\it{}unit};
 {\bf{}default}:
  {\bf{}throw}({\it{}std}::{\it{}invalid\_argument}({\tt{}"cycle::let\_op(): requested operand out of the range (4)"}));
 {\nwrbrace}
{\nwrbrace}

\nwidentuses{\\{{\nwixident{cycle}}{cycle}}\\{{\nwixident{ex}}{ex}}\\{{\nwixident{k}}{k}}\\{{\nwixident{l}}{l}}\\{{\nwixident{let{\_}op}}{let:unop}}\\{{\nwixident{m}}{m}}\\{{\nwixident{nops}}{nops}}}\nwindexuse{\nwixident{cycle}}{cycle}{NW3gGP3e-1Oxbp0-G}\nwindexuse{\nwixident{ex}}{ex}{NW3gGP3e-1Oxbp0-G}\nwindexuse{\nwixident{k}}{k}{NW3gGP3e-1Oxbp0-G}\nwindexuse{\nwixident{l}}{l}{NW3gGP3e-1Oxbp0-G}\nwindexuse{\nwixident{let{\_}op}}{let:unop}{NW3gGP3e-1Oxbp0-G}\nwindexuse{\nwixident{m}}{m}{NW3gGP3e-1Oxbp0-G}\nwindexuse{\nwixident{nops}}{nops}{NW3gGP3e-1Oxbp0-G}\nwendcode{}\nwbegindocs{506}Substitutions works as usual in \GiNaC.
\nwenddocs{}\nwbegincode{507}\sublabel{NW3gGP3e-1Oxbp0-H}\nwmargintag{{\nwtagstyle{}\subpageref{NW3gGP3e-1Oxbp0-H}}}\moddef{cycle.cpp~{\nwtagstyle{}\subpageref{NW3gGP3e-1Oxbp0-1}}}\plusendmoddef\Rm{}\nwstartdeflinemarkup\nwprevnextdefs{NW3gGP3e-1Oxbp0-G}{NW3gGP3e-1Oxbp0-I}\nwenddeflinemarkup
{\bf{}cycle} {\bf{}cycle}::{\it{}subs}({\bf{}const} {\bf{}ex} & {\it{}e}, {\bf{}unsigned} {\it{}options}) {\bf{}const}
{\nwlbrace}
    {\it{}exmap} {\it{}m};
    {\bf{}if} ({\it{}e}.{\it{}info}({\it{}info\_flags}::{\it{}list})) {\nwlbrace}
        {\bf{}lst} {\it{}l} = {\it{}ex\_to}\begin{math}<\end{math}{\bf{}lst}\begin{math}>\end{math}({\it{}e});
        {\bf{}for} ({\bf{}const} {\bf{}auto} & {\it{}i} : {\it{}l})
            {\it{}m}.{\it{}insert}({\it{}std}::{\it{}make\_pair}({\it{}i}.{\it{}op}(0), {\it{}i}.{\it{}op}(1)));
    {\nwrbrace} {\bf{}else} {\bf{}if} ({\it{}is\_a}\begin{math}<\end{math}{\bf{}relational}\begin{math}>\end{math}({\it{}e})) {\nwlbrace}
        {\it{}m}.{\it{}insert}({\it{}std}::{\it{}make\_pair}({\it{}e}.{\it{}op}(0), {\it{}e}.{\it{}op}(1)));
    {\nwrbrace} {\bf{}else}
        {\bf{}throw}({\it{}std}::{\it{}invalid\_argument}({\tt{}"cycle::subs(): the parameter should be a relational or a lst"}));

    {\bf{}return} {\it{}ex\_to}\begin{math}<\end{math}{\bf{}cycle}\begin{math}>\end{math}({\it{}inherited}::{\it{}subs}({\it{}m}, {\it{}options}));
{\nwrbrace}

\nwidentuses{\\{{\nwixident{cycle}}{cycle}}\\{{\nwixident{ex}}{ex}}\\{{\nwixident{l}}{l}}\\{{\nwixident{lst}}{lst}}\\{{\nwixident{m}}{m}}\\{{\nwixident{op}}{op}}\\{{\nwixident{subs}}{subs}}}\nwindexuse{\nwixident{cycle}}{cycle}{NW3gGP3e-1Oxbp0-H}\nwindexuse{\nwixident{ex}}{ex}{NW3gGP3e-1Oxbp0-H}\nwindexuse{\nwixident{l}}{l}{NW3gGP3e-1Oxbp0-H}\nwindexuse{\nwixident{lst}}{lst}{NW3gGP3e-1Oxbp0-H}\nwindexuse{\nwixident{m}}{m}{NW3gGP3e-1Oxbp0-H}\nwindexuse{\nwixident{op}}{op}{NW3gGP3e-1Oxbp0-H}\nwindexuse{\nwixident{subs}}{subs}{NW3gGP3e-1Oxbp0-H}\nwendcode{}\nwbegindocs{508}\nwdocspar
\subsubsection[Service methods for the GiNaC infrastructure]{Service methods for the {\GiNaC} infrastructure}
\label{sec:serv-meth-ginac}

Standard parts involving archiving, comparison and printing of the {\Tt{}\Rm{}{\bf{}cycle}\nwendquote} class
\nwenddocs{}\nwbegincode{509}\sublabel{NW3gGP3e-1Oxbp0-I}\nwmargintag{{\nwtagstyle{}\subpageref{NW3gGP3e-1Oxbp0-I}}}\moddef{cycle.cpp~{\nwtagstyle{}\subpageref{NW3gGP3e-1Oxbp0-1}}}\plusendmoddef\Rm{}\nwstartdeflinemarkup\nwprevnextdefs{NW3gGP3e-1Oxbp0-H}{NW3gGP3e-1Oxbp0-J}\nwenddeflinemarkup

\nwendcode{}\nwbegindocs{510}Archiving routine.
\nwenddocs{}\nwbegincode{511}\sublabel{NW3gGP3e-1Oxbp0-J}\nwmargintag{{\nwtagstyle{}\subpageref{NW3gGP3e-1Oxbp0-J}}}\moddef{cycle.cpp~{\nwtagstyle{}\subpageref{NW3gGP3e-1Oxbp0-1}}}\plusendmoddef\Rm{}\nwstartdeflinemarkup\nwprevnextdefs{NW3gGP3e-1Oxbp0-I}{NW3gGP3e-1Oxbp0-K}\nwenddeflinemarkup
{\bf{}void} {\bf{}cycle}::{\it{}archive}({\it{}archive\_node} &{\it{}n}) {\bf{}const}\nwindexdefn{\nwixident{cycle}}{cycle}{NW3gGP3e-1Oxbp0-J}
{\nwlbrace}
    {\it{}inherited}::{\it{}archive}({\it{}n});
    {\it{}n}.{\it{}add\_ex}({\tt{}"k-param"}, {\it{}k});
    {\it{}n}.{\it{}add\_ex}({\tt{}"l-param"}, {\it{}l});
    {\it{}n}.{\it{}add\_ex}({\tt{}"m-param"}, {\it{}m});
    {\it{}n}.{\it{}add\_ex}({\tt{}"unit"}, {\it{}unit});
{\nwrbrace}

\nwidentdefs{\\{{\nwixident{cycle}}{cycle}}}\nwidentuses{\\{{\nwixident{k}}{k}}\\{{\nwixident{l}}{l}}\\{{\nwixident{m}}{m}}}\nwindexuse{\nwixident{k}}{k}{NW3gGP3e-1Oxbp0-J}\nwindexuse{\nwixident{l}}{l}{NW3gGP3e-1Oxbp0-J}\nwindexuse{\nwixident{m}}{m}{NW3gGP3e-1Oxbp0-J}\nwendcode{}\nwbegindocs{512}Un-archiving routine.
\nwenddocs{}\nwbegincode{513}\sublabel{NW3gGP3e-1Oxbp0-K}\nwmargintag{{\nwtagstyle{}\subpageref{NW3gGP3e-1Oxbp0-K}}}\moddef{cycle.cpp~{\nwtagstyle{}\subpageref{NW3gGP3e-1Oxbp0-1}}}\plusendmoddef\Rm{}\nwstartdeflinemarkup\nwprevnextdefs{NW3gGP3e-1Oxbp0-J}{NW3gGP3e-1Oxbp0-L}\nwenddeflinemarkup
{\bf{}void} {\bf{}cycle}::{\it{}read\_archive}({\bf{}const} {\it{}archive\_node} &{\it{}n}, {\bf{}lst} &{\it{}sym\_lst}) \nwindexdefn{\nwixident{cycle}}{cycle}{NW3gGP3e-1Oxbp0-K}
{\nwlbrace}
    {\it{}inherited}::{\it{}read\_archive}({\it{}n}, {\it{}sym\_lst});
    {\it{}n}.{\it{}find\_ex}({\tt{}"k-param"}, {\it{}k}, {\it{}sym\_lst});
    {\it{}n}.{\it{}find\_ex}({\tt{}"l-param"}, {\it{}l}, {\it{}sym\_lst});
    {\it{}n}.{\it{}find\_ex}({\tt{}"m-param"}, {\it{}m}, {\it{}sym\_lst});
    {\it{}n}.{\it{}find\_ex}({\tt{}"unit"}, {\it{}unit}, {\it{}sym\_lst});
{\nwrbrace}
{\it{}GINAC\_BIND\_UNARCHIVER}({\bf{}cycle});\nwindexdefn{\nwixident{cycle}}{cycle}{NW3gGP3e-1Oxbp0-K}

//const char *cycle::get\_class\_name() {\nwlbrace} return "cycle"; {\nwrbrace}

\nwidentdefs{\\{{\nwixident{cycle}}{cycle}}}\nwidentuses{\\{{\nwixident{k}}{k}}\\{{\nwixident{l}}{l}}\\{{\nwixident{lst}}{lst}}\\{{\nwixident{m}}{m}}}\nwindexuse{\nwixident{k}}{k}{NW3gGP3e-1Oxbp0-K}\nwindexuse{\nwixident{l}}{l}{NW3gGP3e-1Oxbp0-K}\nwindexuse{\nwixident{lst}}{lst}{NW3gGP3e-1Oxbp0-K}\nwindexuse{\nwixident{m}}{m}{NW3gGP3e-1Oxbp0-K}\nwendcode{}\nwbegindocs{514}Comparison of {\Tt{}\Rm{}{\bf{}cycle}\nwendquote}s.
\nwenddocs{}\nwbegincode{515}\sublabel{NW3gGP3e-1Oxbp0-L}\nwmargintag{{\nwtagstyle{}\subpageref{NW3gGP3e-1Oxbp0-L}}}\moddef{cycle.cpp~{\nwtagstyle{}\subpageref{NW3gGP3e-1Oxbp0-1}}}\plusendmoddef\Rm{}\nwstartdeflinemarkup\nwprevnextdefs{NW3gGP3e-1Oxbp0-K}{NW3gGP3e-1Oxbp0-M}\nwenddeflinemarkup
{\bf{}int} {\bf{}cycle}::{\it{}compare\_same\_type}({\bf{}const} {\bf{}basic} &{\it{}other}) {\bf{}const}\nwindexdefn{\nwixident{cycle}}{cycle}{NW3gGP3e-1Oxbp0-L}
{\nwlbrace}
       {\it{}GINAC\_ASSERT}({\it{}is\_a}\begin{math}<\end{math}{\bf{}cycle}\begin{math}>\end{math}({\it{}other}));
       {\bf{}return} {\it{}inherited}::{\it{}compare\_same\_type}({\it{}other});
\begin{math}\div\end{math}\begin{math}\ast\end{math}
    {\bf{}const} {\bf{}cycle} &{\it{}o} = {\bf{}static\_cast}\begin{math}<\end{math}{\bf{}const} {\bf{}cycle} &\begin{math}>\end{math}({\it{}other});
    {\bf{}if} (({\it{}unit} \begin{math}\equiv\end{math} {\it{}o}.{\it{}unit}) \begin{math}\wedge\end{math} ({\it{}l}\begin{math}\ast\end{math}{\it{}o}.{\it{}get\_k}() - {\it{}o}.{\it{}get\_l}()\begin{math}\ast\end{math}{\it{}k}).{\it{}is\_zero}() \begin{math}\wedge\end{math} ({\it{}m}\begin{math}\ast\end{math}{\it{}o}.{\it{}get\_k}() - {\it{}o}.{\it{}get\_m}()\begin{math}\ast\end{math}{\it{}k}).{\it{}is\_zero}())
        {\bf{}return} 0;
    {\bf{}else} {\bf{}if} (({\it{}unit} \begin{math}<\end{math} {\it{}o}.{\it{}unit})
             \begin{math}\vee\end{math} ({\it{}l}\begin{math}\ast\end{math}{\it{}o}.{\it{}get\_k}() \begin{math}<\end{math} {\it{}o}.{\it{}get\_l}()\begin{math}\ast\end{math}{\it{}k}) \begin{math}\vee\end{math} ({\it{}m}\begin{math}\ast\end{math}{\it{}o}.{\it{}get\_k}() \begin{math}<\end{math} {\it{}o}.{\it{}get\_m}()\begin{math}\ast\end{math}{\it{}k}))
        {\bf{}return} -1;
    {\bf{}else}
        {\bf{}return} 1;\begin{math}\ast\end{math}\begin{math}\div\end{math}
{\nwrbrace}

\nwidentdefs{\\{{\nwixident{cycle}}{cycle}}}\nwidentuses{\\{{\nwixident{get{\_}k}}{get:unk}}\\{{\nwixident{get{\_}l}}{get:unl}}\\{{\nwixident{get{\_}m}}{get:unm}}\\{{\nwixident{is{\_}zero}}{is:unzero}}\\{{\nwixident{k}}{k}}\\{{\nwixident{l}}{l}}\\{{\nwixident{m}}{m}}}\nwindexuse{\nwixident{get{\_}k}}{get:unk}{NW3gGP3e-1Oxbp0-L}\nwindexuse{\nwixident{get{\_}l}}{get:unl}{NW3gGP3e-1Oxbp0-L}\nwindexuse{\nwixident{get{\_}m}}{get:unm}{NW3gGP3e-1Oxbp0-L}\nwindexuse{\nwixident{is{\_}zero}}{is:unzero}{NW3gGP3e-1Oxbp0-L}\nwindexuse{\nwixident{k}}{k}{NW3gGP3e-1Oxbp0-L}\nwindexuse{\nwixident{l}}{l}{NW3gGP3e-1Oxbp0-L}\nwindexuse{\nwixident{m}}{m}{NW3gGP3e-1Oxbp0-L}\nwendcode{}\nwbegindocs{516}Equality of {\Tt{}\Rm{}{\bf{}cycle}\nwendquote}s.
\nwenddocs{}\nwbegincode{517}\sublabel{NW3gGP3e-1Oxbp0-M}\nwmargintag{{\nwtagstyle{}\subpageref{NW3gGP3e-1Oxbp0-M}}}\moddef{cycle.cpp~{\nwtagstyle{}\subpageref{NW3gGP3e-1Oxbp0-1}}}\plusendmoddef\Rm{}\nwstartdeflinemarkup\nwprevnextdefs{NW3gGP3e-1Oxbp0-L}{NW3gGP3e-1Oxbp0-N}\nwenddeflinemarkup
{\bf{}bool} {\bf{}cycle}::{\it{}is\_equal}({\bf{}const} {\bf{}basic} & {\it{}other}, {\bf{}bool} {\it{}projectively}) {\bf{}const}
{\nwlbrace}
    {\bf{}if} ({\it{}not} {\it{}is\_a}\begin{math}<\end{math}{\bf{}cycle}\begin{math}>\end{math}({\it{}other}))
        {\bf{}return} {\bf{}false};
    {\bf{}const} {\bf{}cycle} {\it{}o} = {\it{}ex\_to}\begin{math}<\end{math}{\bf{}cycle}\begin{math}>\end{math}({\it{}other});
    {\bf{}ex} {\it{}factor}=0, {\it{}ofactor}=0;

    {\bf{}if} ({\it{}projectively}) {\nwlbrace}
        // Check that coefficients are scalar multiples of other 
        {\bf{}if} ({\it{}not} ({\it{}unit}.{\it{}is\_equal}({\it{}o}.{\it{}unit}) \begin{math}\wedge\end{math} ({\it{}m}\begin{math}\ast\end{math}{\it{}o}.{\it{}get\_k}()-{\it{}o}.{\it{}get\_m}()\begin{math}\ast\end{math}{\it{}k}).{\it{}normal}().{\it{}is\_zero}()))
            {\bf{}return} {\bf{}false};
        // Set up coefficients for proportionality
        {\bf{}if} ({\it{}get\_k}().{\it{}normal}().{\it{}is\_zero}()) {\nwlbrace}
            {\it{}factor}={\it{}get\_m}();
            {\it{}ofactor}={\it{}o}.{\it{}get\_m}();
        {\nwrbrace} {\bf{}else} {\nwlbrace}
            {\it{}factor}={\it{}get\_k}();
            {\it{}ofactor}={\it{}o}.{\it{}get\_k}();
        {\nwrbrace}
            
    {\nwrbrace} {\bf{}else}
        // Check the exact equality of coefficients
        {\bf{}if} ({\it{}not} ({\it{}unit}.{\it{}is\_equal}({\it{}o}.{\it{}unit}) \begin{math}\wedge\end{math} ({\it{}get\_k}()-{\it{}o}.{\it{}get\_k}()).{\it{}normal}().{\it{}is\_zero}()
                 \begin{math}\wedge\end{math} ({\it{}get\_m}()-{\it{}o}.{\it{}get\_m}()).{\it{}normal}().{\it{}is\_zero}()))
            {\bf{}return} {\bf{}false};

\nwidentuses{\\{{\nwixident{bool}}{bool}}\\{{\nwixident{cycle}}{cycle}}\\{{\nwixident{ex}}{ex}}\\{{\nwixident{get{\_}k}}{get:unk}}\\{{\nwixident{get{\_}m}}{get:unm}}\\{{\nwixident{is{\_}equal}}{is:unequal}}\\{{\nwixident{is{\_}zero}}{is:unzero}}\\{{\nwixident{k}}{k}}\\{{\nwixident{m}}{m}}\\{{\nwixident{normal}}{normal}}}\nwindexuse{\nwixident{bool}}{bool}{NW3gGP3e-1Oxbp0-M}\nwindexuse{\nwixident{cycle}}{cycle}{NW3gGP3e-1Oxbp0-M}\nwindexuse{\nwixident{ex}}{ex}{NW3gGP3e-1Oxbp0-M}\nwindexuse{\nwixident{get{\_}k}}{get:unk}{NW3gGP3e-1Oxbp0-M}\nwindexuse{\nwixident{get{\_}m}}{get:unm}{NW3gGP3e-1Oxbp0-M}\nwindexuse{\nwixident{is{\_}equal}}{is:unequal}{NW3gGP3e-1Oxbp0-M}\nwindexuse{\nwixident{is{\_}zero}}{is:unzero}{NW3gGP3e-1Oxbp0-M}\nwindexuse{\nwixident{k}}{k}{NW3gGP3e-1Oxbp0-M}\nwindexuse{\nwixident{m}}{m}{NW3gGP3e-1Oxbp0-M}\nwindexuse{\nwixident{normal}}{normal}{NW3gGP3e-1Oxbp0-M}\nwendcode{}\nwbegindocs{518}Now we iterate through the coefficients of {\Tt{}\Rm{}{\it{}l}\nwendquote}.
\nwenddocs{}\nwbegincode{519}\sublabel{NW3gGP3e-1Oxbp0-N}\nwmargintag{{\nwtagstyle{}\subpageref{NW3gGP3e-1Oxbp0-N}}}\moddef{cycle.cpp~{\nwtagstyle{}\subpageref{NW3gGP3e-1Oxbp0-1}}}\plusendmoddef\Rm{}\nwstartdeflinemarkup\nwprevnextdefs{NW3gGP3e-1Oxbp0-M}{NW3gGP3e-1Oxbp0-O}\nwenddeflinemarkup
    {\bf{}if} ({\it{}is\_a}\begin{math}<\end{math}{\bf{}numeric}\begin{math}>\end{math}({\it{}get\_dim}())) {\nwlbrace}
        {\bf{}int} {\it{}D} = {\it{}ex\_to}\begin{math}<\end{math}{\bf{}numeric}\begin{math}>\end{math}({\it{}get\_dim}()).{\it{}to\_int}();
        {\bf{}if} (\begin{math}\neg\end{math} ({\it{}is\_a}\begin{math}<\end{math}{\bf{}numeric}\begin{math}>\end{math}({\it{}o}.{\it{}get\_dim}()) \begin{math}\wedge\end{math} {\it{}D} \begin{math}\equiv\end{math}{\it{}ex\_to}\begin{math}<\end{math}{\bf{}numeric}\begin{math}>\end{math}({\it{}o}.{\it{}get\_dim}()).{\it{}to\_int}()))
            {\bf{}return} {\bf{}false};
        
        {\bf{}for} ({\bf{}int} {\it{}i}=0; {\it{}i}\begin{math}<\end{math}{\it{}D}; {\it{}i}\protect\PP)
            {\bf{}if} ({\it{}projectively}) {\nwlbrace}
                // search the the first non-zero coefficient 
                {\bf{}if} ({\it{}factor}.{\it{}is\_zero}()) {\nwlbrace}
                    {\it{}factor}={\it{}get\_l}({\it{}i});
                    {\it{}ofactor}={\it{}o}.{\it{}get\_l}({\it{}i});
                {\nwrbrace} {\bf{}else}
                    {\bf{}if} (\begin{math}\neg\end{math} ({\it{}get\_l}({\it{}i})\begin{math}\ast\end{math}{\it{}ofactor}-{\it{}o}.{\it{}get\_l}({\it{}i})\begin{math}\ast\end{math}{\it{}factor}).{\it{}normal}().{\it{}is\_zero}())
                        {\bf{}return} {\bf{}false};
            {\nwrbrace} {\bf{}else} 
                {\bf{}if} (\begin{math}\neg\end{math} ({\it{}get\_l}({\it{}i})-{\it{}o}.{\it{}get\_l}({\it{}i})).{\it{}normal}().{\it{}is\_zero}())
                    {\bf{}return} {\bf{}false};
        
        {\bf{}return} {\bf{}true};
    {\nwrbrace} {\bf{}else}
        {\bf{}return} ({\it{}l}\begin{math}\ast\end{math}{\it{}ofactor}-{\it{}o}.{\it{}get\_l}()\begin{math}\ast\end{math}{\it{}factor}).{\it{}normal}().{\it{}is\_zero}();
{\nwrbrace}

\nwidentuses{\\{{\nwixident{get{\_}dim}}{get:undim}}\\{{\nwixident{get{\_}l}}{get:unl}}\\{{\nwixident{is{\_}zero}}{is:unzero}}\\{{\nwixident{l}}{l}}\\{{\nwixident{normal}}{normal}}\\{{\nwixident{numeric}}{numeric}}}\nwindexuse{\nwixident{get{\_}dim}}{get:undim}{NW3gGP3e-1Oxbp0-N}\nwindexuse{\nwixident{get{\_}l}}{get:unl}{NW3gGP3e-1Oxbp0-N}\nwindexuse{\nwixident{is{\_}zero}}{is:unzero}{NW3gGP3e-1Oxbp0-N}\nwindexuse{\nwixident{l}}{l}{NW3gGP3e-1Oxbp0-N}\nwindexuse{\nwixident{normal}}{normal}{NW3gGP3e-1Oxbp0-N}\nwindexuse{\nwixident{numeric}}{numeric}{NW3gGP3e-1Oxbp0-N}\nwendcode{}\nwbegindocs{520} We return a {\Tt{}\Rm{}{\bf{}lst}\nwendquote} of equations, which describes the condition of
the given {\Tt{}\Rm{}{\bf{}cycle}\nwendquote} to be given by the same point of the projective space
as {\Tt{}\Rm{}{\it{}other}\nwendquote}.
\nwenddocs{}\nwbegincode{521}\sublabel{NW3gGP3e-1Oxbp0-O}\nwmargintag{{\nwtagstyle{}\subpageref{NW3gGP3e-1Oxbp0-O}}}\moddef{cycle.cpp~{\nwtagstyle{}\subpageref{NW3gGP3e-1Oxbp0-1}}}\plusendmoddef\Rm{}\nwstartdeflinemarkup\nwprevnextdefs{NW3gGP3e-1Oxbp0-N}{NW3gGP3e-1Oxbp0-P}\nwenddeflinemarkup
{\bf{}ex} {\bf{}cycle}::{\it{}the\_same\_as}({\bf{}const} {\bf{}basic} & {\it{}other}) {\bf{}const}
{\nwlbrace}
 {\bf{}if} (\begin{math}\neg\end{math} ({\it{}is\_a}\begin{math}<\end{math}{\bf{}cycle}\begin{math}>\end{math}({\it{}other}) \begin{math}\wedge\end{math} ({\it{}get\_dim}() \begin{math}\equiv\end{math} {\it{}ex\_to}\begin{math}<\end{math}{\bf{}cycle}\begin{math}>\end{math}({\it{}other}).{\it{}get\_dim}())))
  {\bf{}return} {\bf{}lst}(1\begin{math}\equiv\end{math}0);
 {\bf{}ex} {\it{}f}=1, {\it{}f1}=1;
 {\bf{}lst} {\it{}res};

\nwidentuses{\\{{\nwixident{cycle}}{cycle}}\\{{\nwixident{ex}}{ex}}\\{{\nwixident{get{\_}dim}}{get:undim}}\\{{\nwixident{lst}}{lst}}}\nwindexuse{\nwixident{cycle}}{cycle}{NW3gGP3e-1Oxbp0-O}\nwindexuse{\nwixident{ex}}{ex}{NW3gGP3e-1Oxbp0-O}\nwindexuse{\nwixident{get{\_}dim}}{get:undim}{NW3gGP3e-1Oxbp0-O}\nwindexuse{\nwixident{lst}}{lst}{NW3gGP3e-1Oxbp0-O}\nwendcode{}\nwbegindocs{522}If {\Tt{}\Rm{}{\it{}k}\nwendquote} is non-zero than we chose it as a normalizing factor.
\nwenddocs{}\nwbegincode{523}\sublabel{NW3gGP3e-1Oxbp0-P}\nwmargintag{{\nwtagstyle{}\subpageref{NW3gGP3e-1Oxbp0-P}}}\moddef{cycle.cpp~{\nwtagstyle{}\subpageref{NW3gGP3e-1Oxbp0-1}}}\plusendmoddef\Rm{}\nwstartdeflinemarkup\nwprevnextdefs{NW3gGP3e-1Oxbp0-O}{NW3gGP3e-1Oxbp0-Q}\nwenddeflinemarkup
 {\bf{}if} ({\it{}not} {\it{}k}.{\it{}is\_zero}()) {\nwlbrace}
  {\it{}f} = {\it{}k};
  {\it{}f1} = {\it{}ex\_to}\begin{math}<\end{math}{\bf{}cycle}\begin{math}>\end{math}({\it{}other}).{\it{}get\_k}();
  {\it{}res}.{\it{}append}({\it{}f1}\begin{math}\ast\end{math}{\it{}m} \begin{math}\equiv\end{math}{\it{}f}\begin{math}\ast\end{math}{\it{}ex\_to}\begin{math}<\end{math}{\bf{}cycle}\begin{math}>\end{math}({\it{}other}).{\it{}get\_m}());

\nwidentuses{\\{{\nwixident{cycle}}{cycle}}\\{{\nwixident{get{\_}k}}{get:unk}}\\{{\nwixident{get{\_}m}}{get:unm}}\\{{\nwixident{is{\_}zero}}{is:unzero}}\\{{\nwixident{k}}{k}}\\{{\nwixident{m}}{m}}}\nwindexuse{\nwixident{cycle}}{cycle}{NW3gGP3e-1Oxbp0-P}\nwindexuse{\nwixident{get{\_}k}}{get:unk}{NW3gGP3e-1Oxbp0-P}\nwindexuse{\nwixident{get{\_}m}}{get:unm}{NW3gGP3e-1Oxbp0-P}\nwindexuse{\nwixident{is{\_}zero}}{is:unzero}{NW3gGP3e-1Oxbp0-P}\nwindexuse{\nwixident{k}}{k}{NW3gGP3e-1Oxbp0-P}\nwindexuse{\nwixident{m}}{m}{NW3gGP3e-1Oxbp0-P}\nwendcode{}\nwbegindocs{524}Otherwise we try {\Tt{}\Rm{}{\it{}m}\nwendquote} for this.
\nwenddocs{}\nwbegincode{525}\sublabel{NW3gGP3e-1Oxbp0-Q}\nwmargintag{{\nwtagstyle{}\subpageref{NW3gGP3e-1Oxbp0-Q}}}\moddef{cycle.cpp~{\nwtagstyle{}\subpageref{NW3gGP3e-1Oxbp0-1}}}\plusendmoddef\Rm{}\nwstartdeflinemarkup\nwprevnextdefs{NW3gGP3e-1Oxbp0-P}{NW3gGP3e-1Oxbp0-R}\nwenddeflinemarkup
 {\nwrbrace} {\bf{}else} {\bf{}if} ({\it{}not} {\it{}m}.{\it{}is\_zero}()) {\nwlbrace}
  {\it{}f} = {\it{}m};
  {\it{}f1} = {\it{}ex\_to}\begin{math}<\end{math}{\bf{}cycle}\begin{math}>\end{math}({\it{}other}).{\it{}get\_m}();
 {\nwrbrace}

\nwidentuses{\\{{\nwixident{cycle}}{cycle}}\\{{\nwixident{get{\_}m}}{get:unm}}\\{{\nwixident{is{\_}zero}}{is:unzero}}\\{{\nwixident{m}}{m}}}\nwindexuse{\nwixident{cycle}}{cycle}{NW3gGP3e-1Oxbp0-Q}\nwindexuse{\nwixident{get{\_}m}}{get:unm}{NW3gGP3e-1Oxbp0-Q}\nwindexuse{\nwixident{is{\_}zero}}{is:unzero}{NW3gGP3e-1Oxbp0-Q}\nwindexuse{\nwixident{m}}{m}{NW3gGP3e-1Oxbp0-Q}\nwendcode{}\nwbegindocs{526}And then we build equations equating corresopnding {\Tt{}\Rm{}{\it{}l}\nwendquote}s.
\nwenddocs{}\nwbegincode{527}\sublabel{NW3gGP3e-1Oxbp0-R}\nwmargintag{{\nwtagstyle{}\subpageref{NW3gGP3e-1Oxbp0-R}}}\moddef{cycle.cpp~{\nwtagstyle{}\subpageref{NW3gGP3e-1Oxbp0-1}}}\plusendmoddef\Rm{}\nwstartdeflinemarkup\nwprevnextdefs{NW3gGP3e-1Oxbp0-Q}{NW3gGP3e-1Oxbp0-S}\nwenddeflinemarkup
 {\bf{}if} ({\it{}ex\_to}\begin{math}<\end{math}{\bf{}varidx}\begin{math}>\end{math}({\it{}unit}.{\it{}op}(1)).{\it{}is\_numeric}()) {\nwlbrace}
  {\bf{}int} {\it{}D} = {\it{}ex\_to}\begin{math}<\end{math}{\bf{}numeric}\begin{math}>\end{math}({\it{}get\_dim}()).{\it{}to\_int}();
  {\bf{}for} ({\bf{}int} {\it{}i}=0; {\it{}i} \begin{math}<\end{math} {\it{}D}; \protect\PP{\it{}i})
   {\it{}res}.{\it{}append}({\it{}f1}\begin{math}\ast\end{math}{\it{}get\_l}({\it{}i})\begin{math}\equiv\end{math}{\it{}f}\begin{math}\ast\end{math}{\it{}ex\_to}\begin{math}<\end{math}{\bf{}cycle}\begin{math}>\end{math}({\it{}other}).{\it{}get\_l}({\it{}i}));
 {\nwrbrace} {\bf{}else}
  {\it{}res}.{\it{}append}({\it{}f1}\begin{math}\ast\end{math}{\it{}l}\begin{math}\equiv\end{math}{\it{}f}\begin{math}\ast\end{math}{\it{}ex\_to}\begin{math}<\end{math}{\bf{}cycle}\begin{math}>\end{math}({\it{}other}).{\it{}get\_l}());
 {\bf{}return} {\it{}res};
{\nwrbrace}

\nwidentuses{\\{{\nwixident{cycle}}{cycle}}\\{{\nwixident{get{\_}dim}}{get:undim}}\\{{\nwixident{get{\_}l}}{get:unl}}\\{{\nwixident{l}}{l}}\\{{\nwixident{numeric}}{numeric}}\\{{\nwixident{op}}{op}}\\{{\nwixident{varidx}}{varidx}}}\nwindexuse{\nwixident{cycle}}{cycle}{NW3gGP3e-1Oxbp0-R}\nwindexuse{\nwixident{get{\_}dim}}{get:undim}{NW3gGP3e-1Oxbp0-R}\nwindexuse{\nwixident{get{\_}l}}{get:unl}{NW3gGP3e-1Oxbp0-R}\nwindexuse{\nwixident{l}}{l}{NW3gGP3e-1Oxbp0-R}\nwindexuse{\nwixident{numeric}}{numeric}{NW3gGP3e-1Oxbp0-R}\nwindexuse{\nwixident{op}}{op}{NW3gGP3e-1Oxbp0-R}\nwindexuse{\nwixident{varidx}}{varidx}{NW3gGP3e-1Oxbp0-R}\nwendcode{}\nwbegindocs{528}A {\Tt{}\Rm{}{\bf{}cycle}\nwendquote} is zero if and only if its all components are zero
\nwenddocs{}\nwbegincode{529}\sublabel{NW3gGP3e-1Oxbp0-S}\nwmargintag{{\nwtagstyle{}\subpageref{NW3gGP3e-1Oxbp0-S}}}\moddef{cycle.cpp~{\nwtagstyle{}\subpageref{NW3gGP3e-1Oxbp0-1}}}\plusendmoddef\Rm{}\nwstartdeflinemarkup\nwprevnextdefs{NW3gGP3e-1Oxbp0-R}{NW3gGP3e-1Oxbp0-T}\nwenddeflinemarkup
{\bf{}bool} {\bf{}cycle}::{\it{}is\_zero}() {\bf{}const}
{\nwlbrace}
 {\bf{}return} ({\it{}k}.{\it{}is\_zero}() \begin{math}\wedge\end{math} {\it{}l}.{\it{}is\_zero}() \begin{math}\wedge\end{math} {\it{}m}.{\it{}is\_zero}());
{\nwrbrace}

\nwidentuses{\\{{\nwixident{bool}}{bool}}\\{{\nwixident{cycle}}{cycle}}\\{{\nwixident{is{\_}zero}}{is:unzero}}\\{{\nwixident{k}}{k}}\\{{\nwixident{l}}{l}}\\{{\nwixident{m}}{m}}}\nwindexuse{\nwixident{bool}}{bool}{NW3gGP3e-1Oxbp0-S}\nwindexuse{\nwixident{cycle}}{cycle}{NW3gGP3e-1Oxbp0-S}\nwindexuse{\nwixident{is{\_}zero}}{is:unzero}{NW3gGP3e-1Oxbp0-S}\nwindexuse{\nwixident{k}}{k}{NW3gGP3e-1Oxbp0-S}\nwindexuse{\nwixident{l}}{l}{NW3gGP3e-1Oxbp0-S}\nwindexuse{\nwixident{m}}{m}{NW3gGP3e-1Oxbp0-S}\nwendcode{}\nwbegindocs{530}Real and imaginary part of the representing vector.
\nwenddocs{}\nwbegincode{531}\sublabel{NW3gGP3e-1Oxbp0-T}\nwmargintag{{\nwtagstyle{}\subpageref{NW3gGP3e-1Oxbp0-T}}}\moddef{cycle.cpp~{\nwtagstyle{}\subpageref{NW3gGP3e-1Oxbp0-1}}}\plusendmoddef\Rm{}\nwstartdeflinemarkup\nwprevnextdefs{NW3gGP3e-1Oxbp0-S}{NW3gGP3e-1Oxbp0-U}\nwenddeflinemarkup
{\bf{}ex} {\bf{}cycle}::{\it{}real\_part}() {\bf{}const}
{\nwlbrace}
    {\bf{}return} {\bf{}cycle}({\it{}k}.{\it{}real\_part}(),{\bf{}indexed}({\it{}l}.{\it{}op}(0).{\it{}real\_part}(),{\it{}l}.{\it{}op}(1)),{\it{}m}.{\it{}real\_part}(),{\it{}unit});
{\nwrbrace}

{\bf{}ex} {\bf{}cycle}::{\it{}imag\_part}() {\bf{}const}
{\nwlbrace}
    {\bf{}return} {\bf{}cycle}({\it{}k}.{\it{}imag\_part}(),{\bf{}indexed}({\it{}l}.{\it{}op}(0).{\it{}imag\_part}(),{\it{}l}.{\it{}op}(1)),{\it{}m}.{\it{}imag\_part}(),{\it{}unit});
{\nwrbrace}

\nwidentuses{\\{{\nwixident{cycle}}{cycle}}\\{{\nwixident{ex}}{ex}}\\{{\nwixident{k}}{k}}\\{{\nwixident{l}}{l}}\\{{\nwixident{m}}{m}}\\{{\nwixident{op}}{op}}}\nwindexuse{\nwixident{cycle}}{cycle}{NW3gGP3e-1Oxbp0-T}\nwindexuse{\nwixident{ex}}{ex}{NW3gGP3e-1Oxbp0-T}\nwindexuse{\nwixident{k}}{k}{NW3gGP3e-1Oxbp0-T}\nwindexuse{\nwixident{l}}{l}{NW3gGP3e-1Oxbp0-T}\nwindexuse{\nwixident{m}}{m}{NW3gGP3e-1Oxbp0-T}\nwindexuse{\nwixident{op}}{op}{NW3gGP3e-1Oxbp0-T}\nwendcode{}\nwbegindocs{532}Printing of {\Tt{}\Rm{}{\bf{}cycle}\nwendquote}s.
\nwenddocs{}\nwbegincode{533}\sublabel{NW3gGP3e-1Oxbp0-U}\nwmargintag{{\nwtagstyle{}\subpageref{NW3gGP3e-1Oxbp0-U}}}\moddef{cycle.cpp~{\nwtagstyle{}\subpageref{NW3gGP3e-1Oxbp0-1}}}\plusendmoddef\Rm{}\nwstartdeflinemarkup\nwprevnextdefs{NW3gGP3e-1Oxbp0-T}{NW3gGP3e-1Oxbp0-V}\nwenddeflinemarkup
{\bf{}void} {\bf{}cycle}::{\it{}do\_print}({\bf{}const} {\it{}print\_dflt} & {\it{}c}, {\bf{}unsigned} {\it{}level}) {\bf{}const}\nwindexdefn{\nwixident{cycle}}{cycle}{NW3gGP3e-1Oxbp0-U}
{\nwlbrace}
 {\it{}PRINT\_CYCLE}
{\nwrbrace}

\begin{math}\div\end{math}\begin{math}\ast\end{math}{\bf{}void} {\bf{}cycle}::{\it{}do\_print\_python}({\bf{}const} {\it{}print\_dflt} & {\it{}c}, {\bf{}unsigned} {\it{}level}) {\bf{}const}
{\nwlbrace}
 {\it{}PRINT\_CYCLE}
 {\nwrbrace}\begin{math}\ast\end{math}\begin{math}\div\end{math}

{\bf{}void} {\bf{}cycle}::{\it{}do\_print\_latex}({\bf{}const} {\it{}print\_latex} & {\it{}c}, {\bf{}unsigned} {\it{}level}) {\bf{}const}\nwindexdefn{\nwixident{cycle}}{cycle}{NW3gGP3e-1Oxbp0-U}
{\nwlbrace}
 {\it{}PRINT\_CYCLE}
{\nwrbrace}

\nwidentdefs{\\{{\nwixident{cycle}}{cycle}}}\nwidentuses{\\{{\nwixident{PRINT{\_}CYCLE}}{PRINT:unCYCLE}}}\nwindexuse{\nwixident{PRINT{\_}CYCLE}}{PRINT:unCYCLE}{NW3gGP3e-1Oxbp0-U}\nwendcode{}\nwbegindocs{534}\nwdocspar
\subsubsection{Linear operation on cycles}
\label{sec:line-oper-cycl-1}
Here are linear operations on {\Tt{}\Rm{}{\bf{}cycle}\nwendquote} defined as methods.
\nwenddocs{}\nwbegincode{535}\sublabel{NW3gGP3e-1Oxbp0-V}\nwmargintag{{\nwtagstyle{}\subpageref{NW3gGP3e-1Oxbp0-V}}}\moddef{cycle.cpp~{\nwtagstyle{}\subpageref{NW3gGP3e-1Oxbp0-1}}}\plusendmoddef\Rm{}\nwstartdeflinemarkup\nwprevnextdefs{NW3gGP3e-1Oxbp0-U}{NW3gGP3e-1Oxbp0-W}\nwenddeflinemarkup
{\bf{}cycle} {\bf{}cycle}::{\it{}add}({\bf{}const} {\bf{}cycle} & {\it{}rh}) {\bf{}const}
{\nwlbrace}
    {\bf{}if} ({\it{}get\_dim}() \begin{math}\neq\end{math} {\it{}rh}.{\it{}get\_dim}())
        {\bf{}throw}({\it{}std}::{\it{}invalid\_argument}({\tt{}"cycle::add(): cannot add two cycles from diferent dimensions"}));

    {\bf{}ex} {\it{}ln}={\bf{}indexed}((({\it{}get\_l}().{\it{}is\_zero}()?0:{\it{}get\_l}().{\it{}op}(0))+({\it{}rh}.{\it{}get\_l}().{\it{}is\_zero}()?0:{\it{}rh}.{\it{}get\_l}().{\it{}op}(0))).{\it{}evalm}(), 
                  {\bf{}varidx}(({\bf{}new} {\bf{}symbol})\begin{math}\rightarrow\end{math}{\it{}setflag}({\it{}status\_flags}::{\it{}dynallocated}), {\it{}get\_dim}()));
    {\bf{}return} {\bf{}cycle}({\it{}get\_k}()+{\it{}rh}.{\it{}get\_k}(), {\it{}ln}, {\it{}get\_m}()+{\it{}rh}.{\it{}get\_m}(), {\it{}unit});
{\nwrbrace}
{\bf{}cycle} {\bf{}cycle}::{\it{}sub}({\bf{}const} {\bf{}cycle} & {\it{}rh}) {\bf{}const}
{\nwlbrace}
    {\bf{}if} ({\it{}get\_dim}() \begin{math}\neq\end{math} {\it{}rh}.{\it{}get\_dim}())
        {\bf{}throw}({\it{}std}::{\it{}invalid\_argument}({\tt{}"cycle::add(): cannot subtract two cycles from diferent dimensions"}));

    {\bf{}ex} {\it{}ln}={\bf{}indexed}((({\it{}get\_l}().{\it{}is\_zero}()?0:{\it{}get\_l}().{\it{}op}(0))-({\it{}rh}.{\it{}get\_l}().{\it{}is\_zero}()?0:{\it{}rh}.{\it{}get\_l}().{\it{}op}(0))).{\it{}evalm}(), 
                  {\bf{}varidx}(({\bf{}new} {\bf{}symbol})\begin{math}\rightarrow\end{math}{\it{}setflag}({\it{}status\_flags}::{\it{}dynallocated}), {\it{}get\_dim}()));
    {\bf{}return} {\bf{}cycle}({\it{}get\_k}()-{\it{}rh}.{\it{}get\_k}(), {\it{}ln}, {\it{}get\_m}()-{\it{}rh}.{\it{}get\_m}(), {\it{}unit});
{\nwrbrace}
{\bf{}cycle} {\bf{}cycle}::{\it{}exmul}({\bf{}const} {\bf{}ex} & {\it{}rh}) {\bf{}const}
{\nwlbrace}
    {\bf{}return} {\bf{}cycle}({\it{}get\_k}()\begin{math}\ast\end{math}{\it{}rh}, {\bf{}indexed}({\it{}get\_l}().{\it{}is\_zero}() ? 0 : ({\it{}get\_l}().{\it{}op}(0)\begin{math}\ast\end{math}{\it{}rh}).{\it{}evalm}(), 
                                     {\bf{}varidx}(({\bf{}new} {\bf{}symbol})\begin{math}\rightarrow\end{math}{\it{}setflag}({\it{}status\_flags}::{\it{}dynallocated}), {\it{}get\_dim}())),
                 {\it{}get\_m}()\begin{math}\ast\end{math}{\it{}rh}, {\it{}unit});
{\nwrbrace}
{\bf{}cycle} {\bf{}cycle}::{\it{}div}({\bf{}const} {\bf{}ex} & {\it{}rh}) {\bf{}const}
{\nwlbrace}
    {\bf{}return} {\it{}exmul}({\it{}pow}({\it{}rh}, {\bf{}numeric}(-1)));
{\nwrbrace}

\nwidentuses{\\{{\nwixident{add}}{add}}\\{{\nwixident{cycle}}{cycle}}\\{{\nwixident{div}}{div}}\\{{\nwixident{ex}}{ex}}\\{{\nwixident{exmul}}{exmul}}\\{{\nwixident{get{\_}dim}}{get:undim}}\\{{\nwixident{get{\_}k}}{get:unk}}\\{{\nwixident{get{\_}l}}{get:unl}}\\{{\nwixident{get{\_}m}}{get:unm}}\\{{\nwixident{is{\_}zero}}{is:unzero}}\\{{\nwixident{numeric}}{numeric}}\\{{\nwixident{op}}{op}}\\{{\nwixident{sub}}{sub}}\\{{\nwixident{varidx}}{varidx}}}\nwindexuse{\nwixident{add}}{add}{NW3gGP3e-1Oxbp0-V}\nwindexuse{\nwixident{cycle}}{cycle}{NW3gGP3e-1Oxbp0-V}\nwindexuse{\nwixident{div}}{div}{NW3gGP3e-1Oxbp0-V}\nwindexuse{\nwixident{ex}}{ex}{NW3gGP3e-1Oxbp0-V}\nwindexuse{\nwixident{exmul}}{exmul}{NW3gGP3e-1Oxbp0-V}\nwindexuse{\nwixident{get{\_}dim}}{get:undim}{NW3gGP3e-1Oxbp0-V}\nwindexuse{\nwixident{get{\_}k}}{get:unk}{NW3gGP3e-1Oxbp0-V}\nwindexuse{\nwixident{get{\_}l}}{get:unl}{NW3gGP3e-1Oxbp0-V}\nwindexuse{\nwixident{get{\_}m}}{get:unm}{NW3gGP3e-1Oxbp0-V}\nwindexuse{\nwixident{is{\_}zero}}{is:unzero}{NW3gGP3e-1Oxbp0-V}\nwindexuse{\nwixident{numeric}}{numeric}{NW3gGP3e-1Oxbp0-V}\nwindexuse{\nwixident{op}}{op}{NW3gGP3e-1Oxbp0-V}\nwindexuse{\nwixident{sub}}{sub}{NW3gGP3e-1Oxbp0-V}\nwindexuse{\nwixident{varidx}}{varidx}{NW3gGP3e-1Oxbp0-V}\nwendcode{}\nwbegindocs{536}The same linear structure is represented in operators overloading.
\nwenddocs{}\nwbegincode{537}\sublabel{NW3gGP3e-1Oxbp0-W}\nwmargintag{{\nwtagstyle{}\subpageref{NW3gGP3e-1Oxbp0-W}}}\moddef{cycle.cpp~{\nwtagstyle{}\subpageref{NW3gGP3e-1Oxbp0-1}}}\plusendmoddef\Rm{}\nwstartdeflinemarkup\nwprevnextdefs{NW3gGP3e-1Oxbp0-V}{NW3gGP3e-1Oxbp0-X}\nwenddeflinemarkup
{\bf{}const} {\bf{}cycle} {\bf{}operator}+({\bf{}const} {\bf{}cycle} & {\it{}lh}, {\bf{}const} {\bf{}cycle} & {\it{}rh})\nwindexdefn{\nwixident{cycle}}{cycle}{NW3gGP3e-1Oxbp0-W}
{\nwlbrace}
 {\bf{}return} {\it{}lh}.{\it{}add}({\it{}rh});
{\nwrbrace}
{\bf{}const} {\bf{}cycle} {\bf{}operator}-({\bf{}const} {\bf{}cycle} & {\it{}lh}, {\bf{}const} {\bf{}cycle} & {\it{}rh})\nwindexdefn{\nwixident{cycle}}{cycle}{NW3gGP3e-1Oxbp0-W}
{\nwlbrace}
 {\bf{}return} {\it{}lh}.{\it{}sub}({\it{}rh});
{\nwrbrace}
{\bf{}const} {\bf{}cycle} {\bf{}operator}\begin{math}\ast\end{math}({\bf{}const} {\bf{}cycle} & {\it{}lh}, {\bf{}const} {\bf{}ex} & {\it{}rh})\nwindexdefn{\nwixident{cycle}}{cycle}{NW3gGP3e-1Oxbp0-W}
{\nwlbrace}
 {\bf{}return} {\it{}lh}.{\it{}exmul}({\it{}rh});
{\nwrbrace}
{\bf{}const} {\bf{}cycle} {\bf{}operator}\begin{math}\ast\end{math}({\bf{}const} {\bf{}ex} & {\it{}lh}, {\bf{}const} {\bf{}cycle} & {\it{}rh})\nwindexdefn{\nwixident{cycle}}{cycle}{NW3gGP3e-1Oxbp0-W}
{\nwlbrace}
 {\bf{}return} {\it{}rh}.{\it{}exmul}({\it{}lh});
{\nwrbrace}
{\bf{}const} {\bf{}cycle} {\bf{}operator}\begin{math}\div\end{math}({\bf{}const} {\bf{}cycle} & {\it{}lh}, {\bf{}const} {\bf{}ex} & {\it{}rh})\nwindexdefn{\nwixident{cycle}}{cycle}{NW3gGP3e-1Oxbp0-W}
{\nwlbrace}
 {\bf{}return} {\it{}lh}.{\it{}div}({\it{}rh});
{\nwrbrace}
{\bf{}const} {\bf{}ex} {\bf{}operator}\begin{math}\ast\end{math}({\bf{}const} {\bf{}cycle} & {\it{}lh}, {\bf{}const} {\bf{}cycle} & {\it{}rh})\nwindexdefn{\nwixident{ex}}{ex}{NW3gGP3e-1Oxbp0-W}
{\nwlbrace}
 {\bf{}return} {\it{}lh}.{\it{}mul}({\it{}rh});
{\nwrbrace}

\nwidentdefs{\\{{\nwixident{cycle}}{cycle}}\\{{\nwixident{ex}}{ex}}}\nwidentuses{\\{{\nwixident{add}}{add}}\\{{\nwixident{div}}{div}}\\{{\nwixident{exmul}}{exmul}}\\{{\nwixident{mul}}{mul}}\\{{\nwixident{operator*}}{operator*}}\\{{\nwixident{operator+}}{operator+}}\\{{\nwixident{operator-}}{operator-}}\\{{\nwixident{operator/}}{operator/}}\\{{\nwixident{sub}}{sub}}}\nwindexuse{\nwixident{add}}{add}{NW3gGP3e-1Oxbp0-W}\nwindexuse{\nwixident{div}}{div}{NW3gGP3e-1Oxbp0-W}\nwindexuse{\nwixident{exmul}}{exmul}{NW3gGP3e-1Oxbp0-W}\nwindexuse{\nwixident{mul}}{mul}{NW3gGP3e-1Oxbp0-W}\nwindexuse{\nwixident{operator*}}{operator*}{NW3gGP3e-1Oxbp0-W}\nwindexuse{\nwixident{operator+}}{operator+}{NW3gGP3e-1Oxbp0-W}\nwindexuse{\nwixident{operator-}}{operator-}{NW3gGP3e-1Oxbp0-W}\nwindexuse{\nwixident{operator/}}{operator/}{NW3gGP3e-1Oxbp0-W}\nwindexuse{\nwixident{sub}}{sub}{NW3gGP3e-1Oxbp0-W}\nwendcode{}\nwbegindocs{538}We make a specialisation of these operation for {\Tt{}\Rm{}{\bf{}cycle2D}\nwendquote} class as well.
\nwenddocs{}\nwbegincode{539}\sublabel{NW3gGP3e-1Oxbp0-X}\nwmargintag{{\nwtagstyle{}\subpageref{NW3gGP3e-1Oxbp0-X}}}\moddef{cycle.cpp~{\nwtagstyle{}\subpageref{NW3gGP3e-1Oxbp0-1}}}\plusendmoddef\Rm{}\nwstartdeflinemarkup\nwprevnextdefs{NW3gGP3e-1Oxbp0-W}{NW3gGP3e-1Oxbp0-Y}\nwenddeflinemarkup
{\bf{}const} {\bf{}cycle2D} {\bf{}operator}+({\bf{}const} {\bf{}cycle2D} & {\it{}lh}, {\bf{}const} {\bf{}cycle2D} & {\it{}rh})\nwindexdefn{\nwixident{cycle2D}}{cycle2D}{NW3gGP3e-1Oxbp0-X}
{\nwlbrace}
 {\bf{}return} {\it{}ex\_to}\begin{math}<\end{math}{\bf{}cycle2D}\begin{math}>\end{math}({\it{}lh}.{\it{}add}({\it{}rh}));
{\nwrbrace}
{\bf{}const} {\bf{}cycle2D} {\bf{}operator}-({\bf{}const} {\bf{}cycle2D} & {\it{}lh}, {\bf{}const} {\bf{}cycle2D} & {\it{}rh})\nwindexdefn{\nwixident{cycle2D}}{cycle2D}{NW3gGP3e-1Oxbp0-X}
{\nwlbrace}
 {\bf{}return} {\it{}ex\_to}\begin{math}<\end{math}{\bf{}cycle2D}\begin{math}>\end{math}({\it{}lh}.{\it{}sub}({\it{}rh}));
{\nwrbrace}
{\bf{}const} {\bf{}cycle2D} {\bf{}operator}\begin{math}\ast\end{math}({\bf{}const} {\bf{}cycle2D} & {\it{}lh}, {\bf{}const} {\bf{}ex} & {\it{}rh})\nwindexdefn{\nwixident{cycle2D}}{cycle2D}{NW3gGP3e-1Oxbp0-X}
{\nwlbrace}
 {\bf{}return} {\it{}ex\_to}\begin{math}<\end{math}{\bf{}cycle2D}\begin{math}>\end{math}({\it{}lh}.{\it{}exmul}({\it{}rh}));
{\nwrbrace}
{\bf{}const} {\bf{}cycle2D} {\bf{}operator}\begin{math}\ast\end{math}({\bf{}const} {\bf{}ex} & {\it{}lh}, {\bf{}const} {\bf{}cycle2D} & {\it{}rh})\nwindexdefn{\nwixident{cycle2D}}{cycle2D}{NW3gGP3e-1Oxbp0-X}
{\nwlbrace}
 {\bf{}return} {\it{}ex\_to}\begin{math}<\end{math}{\bf{}cycle2D}\begin{math}>\end{math}({\it{}rh}.{\it{}exmul}({\it{}lh}));
{\nwrbrace}
{\bf{}const} {\bf{}cycle2D} {\bf{}operator}\begin{math}\div\end{math}({\bf{}const} {\bf{}cycle2D} & {\it{}lh}, {\bf{}const} {\bf{}ex} & {\it{}rh})\nwindexdefn{\nwixident{cycle2D}}{cycle2D}{NW3gGP3e-1Oxbp0-X}
{\nwlbrace}
 {\bf{}return} {\it{}ex\_to}\begin{math}<\end{math}{\bf{}cycle2D}\begin{math}>\end{math}({\it{}lh}.{\it{}div}({\it{}rh}));
{\nwrbrace}
{\bf{}const} {\bf{}ex} {\bf{}operator}\begin{math}\ast\end{math}({\bf{}const} {\bf{}cycle2D} & {\it{}lh}, {\bf{}const} {\bf{}cycle2D} & {\it{}rh})\nwindexdefn{\nwixident{ex}}{ex}{NW3gGP3e-1Oxbp0-X}
{\nwlbrace}
 {\bf{}return} {\it{}ex\_to}\begin{math}<\end{math}{\bf{}cycle2D}\begin{math}>\end{math}({\it{}lh}.{\it{}mul}({\it{}rh}));
{\nwrbrace}

\nwidentdefs{\\{{\nwixident{cycle2D}}{cycle2D}}\\{{\nwixident{ex}}{ex}}}\nwidentuses{\\{{\nwixident{add}}{add}}\\{{\nwixident{div}}{div}}\\{{\nwixident{exmul}}{exmul}}\\{{\nwixident{mul}}{mul}}\\{{\nwixident{operator*}}{operator*}}\\{{\nwixident{operator+}}{operator+}}\\{{\nwixident{operator-}}{operator-}}\\{{\nwixident{operator/}}{operator/}}\\{{\nwixident{sub}}{sub}}}\nwindexuse{\nwixident{add}}{add}{NW3gGP3e-1Oxbp0-X}\nwindexuse{\nwixident{div}}{div}{NW3gGP3e-1Oxbp0-X}\nwindexuse{\nwixident{exmul}}{exmul}{NW3gGP3e-1Oxbp0-X}\nwindexuse{\nwixident{mul}}{mul}{NW3gGP3e-1Oxbp0-X}\nwindexuse{\nwixident{operator*}}{operator*}{NW3gGP3e-1Oxbp0-X}\nwindexuse{\nwixident{operator+}}{operator+}{NW3gGP3e-1Oxbp0-X}\nwindexuse{\nwixident{operator-}}{operator-}{NW3gGP3e-1Oxbp0-X}\nwindexuse{\nwixident{operator/}}{operator/}{NW3gGP3e-1Oxbp0-X}\nwindexuse{\nwixident{sub}}{sub}{NW3gGP3e-1Oxbp0-X}\nwendcode{}\nwbegindocs{540}\nwdocspar
\subsubsection[Specific methods for cycle]{Specific methods for {\Tt{}\Rm{}{\bf{}cycle}\nwendquote}}
\label{sec:specific-methods}

\nwenddocs{}\nwbegindocs{541}\nwdocspar
We oftenly need to normalise cycles to get rid of ambiguity in their
definition. This is typically by prescribing a value to {\Tt{}\Rm{}{\it{}k}\nwendquote}.
\nwenddocs{}\nwbegincode{542}\sublabel{NW3gGP3e-1Oxbp0-Y}\nwmargintag{{\nwtagstyle{}\subpageref{NW3gGP3e-1Oxbp0-Y}}}\moddef{cycle.cpp~{\nwtagstyle{}\subpageref{NW3gGP3e-1Oxbp0-1}}}\plusendmoddef\Rm{}\nwstartdeflinemarkup\nwprevnextdefs{NW3gGP3e-1Oxbp0-X}{NW3gGP3e-1Oxbp0-Z}\nwenddeflinemarkup
{\bf{}cycle} {\bf{}cycle}::{\it{}normalize}({\bf{}const} {\bf{}ex} & {\it{}k\_new}, {\bf{}const} {\bf{}ex} & {\it{}e}) {\bf{}const}
{\nwlbrace}
    {\bf{}ex} {\it{}ratio} = 0;
    {\bf{}if} ({\it{}k\_new}.{\it{}is\_zero}()) // Make the determinant equal 1
        {\it{}ratio} = {\it{}sqrt}({\it{}radius\_sq}({\it{}e}));
    {\bf{}else} {\nwlbrace} // First non-zero coefficient among k, m, l\_0, l\_1, ... is set to k\_new
        {\bf{}if} (\begin{math}\neg\end{math}{\it{}k}.{\it{}is\_zero}())
            {\it{}ratio} = {\it{}k}\begin{math}\div\end{math}{\it{}k\_new};
        {\bf{}else} {\bf{}if} (\begin{math}\neg\end{math}{\it{}m}.{\it{}is\_zero}())
            {\it{}ratio} = {\it{}m}\begin{math}\div\end{math}{\it{}k\_new};
        {\bf{}else} {\nwlbrace}
            {\bf{}int} {\it{}D} = {\it{}ex\_to}\begin{math}<\end{math}{\bf{}numeric}\begin{math}>\end{math}({\it{}get\_dim}()).{\it{}to\_int}();
            {\bf{}for} ({\bf{}int} {\it{}i}=0; {\it{}i}\begin{math}<\end{math}{\it{}D}; {\it{}i}\protect\PP)
                {\bf{}if} (\begin{math}\neg\end{math}{\it{}l}.{\it{}subs}({\it{}l}.{\it{}op}(1) \begin{math}\equiv\end{math} {\it{}i}).{\it{}is\_zero}()) {\nwlbrace}
                    {\it{}ratio} = {\it{}l}.{\it{}subs}({\it{}l}.{\it{}op}(1) \begin{math}\equiv\end{math} {\it{}i})\begin{math}\div\end{math}{\it{}k\_new};
                    {\bf{}break};
                {\nwrbrace}
        {\nwrbrace}
    {\nwrbrace}
    {\bf{}if} ({\it{}ratio}.{\it{}is\_zero}()) // No normalisation is possible
        {\bf{}return} (\begin{math}\ast\end{math}{\it{}this});
    
    {\bf{}return} {\bf{}cycle}(({\it{}k}\begin{math}\div\end{math}{\it{}ratio}).{\it{}normal}(), {\bf{}indexed}(({\it{}l}.{\it{}op}(0)\begin{math}\div\end{math}{\it{}ratio}).{\it{}evalm}().{\it{}normal}(), {\it{}l}.{\it{}op}(1)), ({\it{}m}\begin{math}\div\end{math}{\it{}ratio}).{\it{}normal}(), {\it{}unit});
{\nwrbrace}

\nwidentuses{\\{{\nwixident{cycle}}{cycle}}\\{{\nwixident{ex}}{ex}}\\{{\nwixident{get{\_}dim}}{get:undim}}\\{{\nwixident{is{\_}zero}}{is:unzero}}\\{{\nwixident{k}}{k}}\\{{\nwixident{l}}{l}}\\{{\nwixident{m}}{m}}\\{{\nwixident{normal}}{normal}}\\{{\nwixident{normalize}}{normalize}}\\{{\nwixident{numeric}}{numeric}}\\{{\nwixident{op}}{op}}\\{{\nwixident{radius{\_}sq}}{radius:unsq}}\\{{\nwixident{subs}}{subs}}}\nwindexuse{\nwixident{cycle}}{cycle}{NW3gGP3e-1Oxbp0-Y}\nwindexuse{\nwixident{ex}}{ex}{NW3gGP3e-1Oxbp0-Y}\nwindexuse{\nwixident{get{\_}dim}}{get:undim}{NW3gGP3e-1Oxbp0-Y}\nwindexuse{\nwixident{is{\_}zero}}{is:unzero}{NW3gGP3e-1Oxbp0-Y}\nwindexuse{\nwixident{k}}{k}{NW3gGP3e-1Oxbp0-Y}\nwindexuse{\nwixident{l}}{l}{NW3gGP3e-1Oxbp0-Y}\nwindexuse{\nwixident{m}}{m}{NW3gGP3e-1Oxbp0-Y}\nwindexuse{\nwixident{normal}}{normal}{NW3gGP3e-1Oxbp0-Y}\nwindexuse{\nwixident{normalize}}{normalize}{NW3gGP3e-1Oxbp0-Y}\nwindexuse{\nwixident{numeric}}{numeric}{NW3gGP3e-1Oxbp0-Y}\nwindexuse{\nwixident{op}}{op}{NW3gGP3e-1Oxbp0-Y}\nwindexuse{\nwixident{radius{\_}sq}}{radius:unsq}{NW3gGP3e-1Oxbp0-Y}\nwindexuse{\nwixident{subs}}{subs}{NW3gGP3e-1Oxbp0-Y}\nwendcode{}\nwbegindocs{543}The normalisation to determinant \(\pm 1\). We try to avoid
imaginary numbers, thus if {\Tt{}\Rm{}-{\it{}d}\begin{math}\div\end{math}{\it{}D}\nwendquote} is known to be nonnegative, then
we use it for square root.
\nwenddocs{}\nwbegincode{544}\sublabel{NW3gGP3e-1Oxbp0-Z}\nwmargintag{{\nwtagstyle{}\subpageref{NW3gGP3e-1Oxbp0-Z}}}\moddef{cycle.cpp~{\nwtagstyle{}\subpageref{NW3gGP3e-1Oxbp0-1}}}\plusendmoddef\Rm{}\nwstartdeflinemarkup\nwprevnextdefs{NW3gGP3e-1Oxbp0-Y}{NW3gGP3e-1Oxbp0-a}\nwenddeflinemarkup
{\bf{}cycle} {\bf{}cycle}::{\it{}normalize\_det}({\bf{}const} {\bf{}ex} & {\it{}e}, {\bf{}const} {\bf{}ex} & {\it{}sign}, {\bf{}const} {\bf{}ex} & {\it{}D}) {\bf{}const}
{\nwlbrace}
    {\bf{}ex} {\it{}d} = {\it{}det}({\it{}e}, {\it{}sign}), {\it{}k\_new};
    {\bf{}if} ((-{\it{}d}\begin{math}\div\end{math}{\it{}D}).{\it{}info}({\it{}info\_flags}::{\it{}nonnegative}))
        {\it{}k\_new}={\it{}k}\begin{math}\div\end{math}{\it{}sqrt}(-{\it{}d}\begin{math}\div\end{math}{\it{}D});
    {\bf{}else}
        {\it{}k\_new}={\it{}k}\begin{math}\div\end{math}{\it{}sqrt}({\it{}d}\begin{math}\div\end{math}{\it{}D});

    {\bf{}return} ({\it{}d}.{\it{}is\_zero}()? \begin{math}\ast\end{math}{\it{}this}: {\it{}normalize}({\it{}k\_new}, {\it{}e}));
{\nwrbrace}

\nwidentuses{\\{{\nwixident{cycle}}{cycle}}\\{{\nwixident{det}}{det}}\\{{\nwixident{ex}}{ex}}\\{{\nwixident{is{\_}zero}}{is:unzero}}\\{{\nwixident{k}}{k}}\\{{\nwixident{normalize}}{normalize}}\\{{\nwixident{normalize{\_}det}}{normalize:undet}}}\nwindexuse{\nwixident{cycle}}{cycle}{NW3gGP3e-1Oxbp0-Z}\nwindexuse{\nwixident{det}}{det}{NW3gGP3e-1Oxbp0-Z}\nwindexuse{\nwixident{ex}}{ex}{NW3gGP3e-1Oxbp0-Z}\nwindexuse{\nwixident{is{\_}zero}}{is:unzero}{NW3gGP3e-1Oxbp0-Z}\nwindexuse{\nwixident{k}}{k}{NW3gGP3e-1Oxbp0-Z}\nwindexuse{\nwixident{normalize}}{normalize}{NW3gGP3e-1Oxbp0-Z}\nwindexuse{\nwixident{normalize{\_}det}}{normalize:undet}{NW3gGP3e-1Oxbp0-Z}\nwendcode{}\nwbegindocs{545}This methods returns a centre of the {\Tt{}\Rm{}{\bf{}cycle}\nwendquote} depending from the
provided metric.
\nwenddocs{}\nwbegincode{546}\sublabel{NW3gGP3e-1Oxbp0-a}\nwmargintag{{\nwtagstyle{}\subpageref{NW3gGP3e-1Oxbp0-a}}}\moddef{cycle.cpp~{\nwtagstyle{}\subpageref{NW3gGP3e-1Oxbp0-1}}}\plusendmoddef\Rm{}\nwstartdeflinemarkup\nwprevnextdefs{NW3gGP3e-1Oxbp0-Z}{NW3gGP3e-1Oxbp0-b}\nwenddeflinemarkup
{\bf{}ex} {\bf{}cycle}::{\it{}center}({\bf{}const} {\bf{}ex} & {\it{}metr}, {\bf{}bool} {\it{}return\_matrix}) {\bf{}const}
{\nwlbrace}
    {\bf{}if} ({\it{}is\_a}\begin{math}<\end{math}{\bf{}numeric}\begin{math}>\end{math}({\it{}get\_dim}())) {\nwlbrace}
        {\bf{}ex} {\it{}e1}, {\it{}D} = {\it{}get\_dim}();
        {\bf{}if} ({\it{}metr}.{\it{}is\_zero}())
            {\it{}e1} = {\it{}unit};
        {\bf{}else} {\nwlbrace}
            {\bf{}if} ({\it{}is\_a}\begin{math}<\end{math}{\bf{}clifford}\begin{math}>\end{math}({\it{}metr}))
                {\it{}e1}={\it{}metr};
            {\bf{}else}
                {\bf{}try} {\nwlbrace}
                    {\it{}e1} = {\it{}clifford\_unit}({\bf{}varidx}(0, {\it{}D}), {\it{}metr});
                {\nwrbrace} {\bf{}catch} ({\it{}exception} &{\it{}p}) {\nwlbrace}
                    {\bf{}throw}({\it{}std}::{\it{}invalid\_argument}({\tt{}"cycle::center(): supplied metric"}
                                                {\tt{}" is not suitable for Clifford unit"}));
                {\nwrbrace}
        {\nwrbrace}
        
\nwidentuses{\\{{\nwixident{bool}}{bool}}\\{{\nwixident{catch}}{catch}}\\{{\nwixident{center}}{center}}\\{{\nwixident{cycle}}{cycle}}\\{{\nwixident{ex}}{ex}}\\{{\nwixident{get{\_}dim}}{get:undim}}\\{{\nwixident{is{\_}zero}}{is:unzero}}\\{{\nwixident{metr}}{metr}}\\{{\nwixident{numeric}}{numeric}}\\{{\nwixident{varidx}}{varidx}}}\nwindexuse{\nwixident{bool}}{bool}{NW3gGP3e-1Oxbp0-a}\nwindexuse{\nwixident{catch}}{catch}{NW3gGP3e-1Oxbp0-a}\nwindexuse{\nwixident{center}}{center}{NW3gGP3e-1Oxbp0-a}\nwindexuse{\nwixident{cycle}}{cycle}{NW3gGP3e-1Oxbp0-a}\nwindexuse{\nwixident{ex}}{ex}{NW3gGP3e-1Oxbp0-a}\nwindexuse{\nwixident{get{\_}dim}}{get:undim}{NW3gGP3e-1Oxbp0-a}\nwindexuse{\nwixident{is{\_}zero}}{is:unzero}{NW3gGP3e-1Oxbp0-a}\nwindexuse{\nwixident{metr}}{metr}{NW3gGP3e-1Oxbp0-a}\nwindexuse{\nwixident{numeric}}{numeric}{NW3gGP3e-1Oxbp0-a}\nwindexuse{\nwixident{varidx}}{varidx}{NW3gGP3e-1Oxbp0-a}\nwendcode{}\nwbegindocs{547}Finally, the centre is constructed for the cycle and given metric by
the formula~\cite[Defn.~\ref{E-de:center-first}]{Kisil05a}:
\begin{displaymath}
  \left(-e_0^2\frac{l_0}{k}, -e_1^2\frac{l_1}{k}, \ldots, -e_{D-1}^2\frac{l_{D-1}}{k}\right)
\end{displaymath}
\nwenddocs{}\nwbegincode{548}\sublabel{NW3gGP3e-1Oxbp0-b}\nwmargintag{{\nwtagstyle{}\subpageref{NW3gGP3e-1Oxbp0-b}}}\moddef{cycle.cpp~{\nwtagstyle{}\subpageref{NW3gGP3e-1Oxbp0-1}}}\plusendmoddef\Rm{}\nwstartdeflinemarkup\nwprevnextdefs{NW3gGP3e-1Oxbp0-a}{NW3gGP3e-1Oxbp0-c}\nwenddeflinemarkup
        {\bf{}lst} {\it{}c};
        {\bf{}for}({\bf{}int} {\it{}i}=0; {\it{}i}\begin{math}<\end{math}{\it{}D}; {\it{}i}\protect\PP)
            {\bf{}if} ({\it{}k}.{\it{}is\_zero}())
                {\it{}c}.{\it{}append}({\it{}get\_l}({\it{}i}));
            {\bf{}else}
                //c.append(jump\_fnct(-ex\_to\begin{math}<\end{math}clifford\begin{math}>\end{math}(e1).get\_metric(varidx(i, D), varidx(i, D)))*get\_l(i)/k);
                {\it{}c}.{\it{}append}(-{\it{}ex\_to}\begin{math}<\end{math}{\bf{}clifford}\begin{math}>\end{math}({\it{}e1}).{\it{}get\_metric}({\bf{}varidx}({\it{}i}, {\it{}D}), {\bf{}varidx}({\it{}i}, {\it{}D}))\begin{math}\ast\end{math}{\it{}get\_l}({\it{}i})\begin{math}\div\end{math}{\it{}k});
        {\bf{}return} ({\it{}return\_matrix}? ({\bf{}ex}){\bf{}matrix}({\it{}ex\_to}\begin{math}<\end{math}{\bf{}numeric}\begin{math}>\end{math}({\it{}D}).{\it{}to\_int}(), 1, {\it{}c}) : ({\bf{}ex}){\it{}c});
    {\nwrbrace} {\bf{}else} {\nwlbrace}
        {\bf{}return} {\it{}l}\begin{math}\div\end{math}{\it{}k};
    {\nwrbrace}
{\nwrbrace}

\nwidentuses{\\{{\nwixident{ex}}{ex}}\\{{\nwixident{get{\_}l}}{get:unl}}\\{{\nwixident{get{\_}metric}}{get:unmetric}}\\{{\nwixident{is{\_}zero}}{is:unzero}}\\{{\nwixident{jump{\_}fnct}}{jump:unfnct}}\\{{\nwixident{k}}{k}}\\{{\nwixident{l}}{l}}\\{{\nwixident{lst}}{lst}}\\{{\nwixident{matrix}}{matrix}}\\{{\nwixident{numeric}}{numeric}}\\{{\nwixident{varidx}}{varidx}}}\nwindexuse{\nwixident{ex}}{ex}{NW3gGP3e-1Oxbp0-b}\nwindexuse{\nwixident{get{\_}l}}{get:unl}{NW3gGP3e-1Oxbp0-b}\nwindexuse{\nwixident{get{\_}metric}}{get:unmetric}{NW3gGP3e-1Oxbp0-b}\nwindexuse{\nwixident{is{\_}zero}}{is:unzero}{NW3gGP3e-1Oxbp0-b}\nwindexuse{\nwixident{jump{\_}fnct}}{jump:unfnct}{NW3gGP3e-1Oxbp0-b}\nwindexuse{\nwixident{k}}{k}{NW3gGP3e-1Oxbp0-b}\nwindexuse{\nwixident{l}}{l}{NW3gGP3e-1Oxbp0-b}\nwindexuse{\nwixident{lst}}{lst}{NW3gGP3e-1Oxbp0-b}\nwindexuse{\nwixident{matrix}}{matrix}{NW3gGP3e-1Oxbp0-b}\nwindexuse{\nwixident{numeric}}{numeric}{NW3gGP3e-1Oxbp0-b}\nwindexuse{\nwixident{varidx}}{varidx}{NW3gGP3e-1Oxbp0-b}\nwendcode{}\nwbegindocs{549}\nwdocspar
\subsubsection{Build cycle with given properties}
\label{sec:build-cycle-with}

We oftenly need {\Tt{}\Rm{}{\bf{}cycle}\nwendquote}s with prescribed properties, e.g. when
converting of {\Tt{}\Rm{}{\bf{}cycle}\nwendquote}s to normalised form or matrix. This routine
takes a system of linear equations with the {\Tt{}\Rm{}{\bf{}cycle}\nwendquote} parameters and
try to resolve it. The list of unknown parameters is either supplied
or build automatically in a way suitable for most applications.
\nwenddocs{}\nwbegincode{550}\sublabel{NW3gGP3e-1Oxbp0-c}\nwmargintag{{\nwtagstyle{}\subpageref{NW3gGP3e-1Oxbp0-c}}}\moddef{cycle.cpp~{\nwtagstyle{}\subpageref{NW3gGP3e-1Oxbp0-1}}}\plusendmoddef\Rm{}\nwstartdeflinemarkup\nwprevnextdefs{NW3gGP3e-1Oxbp0-b}{NW3gGP3e-1Oxbp0-d}\nwenddeflinemarkup
{\bf{}cycle} {\bf{}cycle}::{\it{}subject\_to}({\bf{}const} {\bf{}ex} & {\it{}condition}, {\bf{}const} {\bf{}ex} & {\it{}vars}) {\bf{}const}
{\nwlbrace}
 {\bf{}lst} {\it{}vars1};
 {\bf{}if} ({\it{}vars}.{\it{}info}({\it{}info\_flags}::{\it{}list}) \begin{math}\wedge\end{math}({\it{}vars}.{\it{}nops}() \begin{math}\neq\end{math} 0))
   {\it{}vars1} = {\it{}ex\_to}\begin{math}<\end{math}{\bf{}lst}\begin{math}>\end{math}({\it{}vars});
 {\bf{}else} {\bf{}if} ({\it{}is\_a}\begin{math}<\end{math}{\bf{}symbol}\begin{math}>\end{math}({\it{}vars}))
   {\it{}vars1} = {\bf{}lst}({\it{}vars});
 {\bf{}else} {\bf{}if} (({\it{}vars} \begin{math}\equiv\end{math} 0) \begin{math}\vee\end{math} ({\it{}vars}.{\it{}nops}() \begin{math}\equiv\end{math} 0)) {\nwlbrace}
  {\bf{}if} ({\it{}is\_a}\begin{math}<\end{math}{\bf{}symbol}\begin{math}>\end{math}({\it{}m}))
   {\it{}vars1}.{\it{}append}({\it{}m});
  {\bf{}if} ({\it{}is\_a}\begin{math}<\end{math}{\bf{}numeric}\begin{math}>\end{math}({\it{}get\_dim}()))
   {\bf{}for} ({\bf{}int} {\it{}i} = 0; {\it{}i} \begin{math}<\end{math} {\it{}ex\_to}\begin{math}<\end{math}{\bf{}numeric}\begin{math}>\end{math}({\it{}get\_dim}()).{\it{}to\_double}(); {\it{}i}\protect\PP)
    {\bf{}if} ({\it{}is\_a}\begin{math}<\end{math}{\bf{}symbol}\begin{math}>\end{math}({\it{}get\_l}({\it{}i})))
     {\it{}vars1}.{\it{}append}({\it{}get\_l}({\it{}i}));
  {\bf{}if} ({\it{}is\_a}\begin{math}<\end{math}{\bf{}symbol}\begin{math}>\end{math}({\it{}k}))
   {\it{}vars1}.{\it{}append}({\it{}k});
  {\bf{}if} ({\it{}vars1}.{\it{}nops}() \begin{math}\equiv\end{math} 0)
   {\bf{}throw}({\it{}std}::{\it{}invalid\_argument}({\tt{}"cycle::subject\_to(): could not construct the default list of "}
          {\tt{}"parameters"}));
 {\nwrbrace} {\bf{}else}
  {\bf{}throw}({\it{}std}::{\it{}invalid\_argument}({\tt{}"cycle::subject\_to(): second parameter should be a list of symbols"}
         {\tt{}" or a single symbol"}));

 {\bf{}return} {\it{}subs}({\it{}lsolve}({\it{}condition}.{\it{}info}({\it{}info\_flags}::{\it{}relation\_equal})? {\bf{}lst}({\it{}condition}) : {\it{}condition},
        {\it{}vars1}), {\it{}subs\_options}::{\it{}algebraic} \begin{math}\mid\end{math} {\it{}subs\_options}::{\it{}no\_pattern});
{\nwrbrace}

\nwidentuses{\\{{\nwixident{cycle}}{cycle}}\\{{\nwixident{ex}}{ex}}\\{{\nwixident{get{\_}dim}}{get:undim}}\\{{\nwixident{get{\_}l}}{get:unl}}\\{{\nwixident{k}}{k}}\\{{\nwixident{lst}}{lst}}\\{{\nwixident{m}}{m}}\\{{\nwixident{nops}}{nops}}\\{{\nwixident{numeric}}{numeric}}\\{{\nwixident{subject{\_}to}}{subject:unto}}\\{{\nwixident{subs}}{subs}}}\nwindexuse{\nwixident{cycle}}{cycle}{NW3gGP3e-1Oxbp0-c}\nwindexuse{\nwixident{ex}}{ex}{NW3gGP3e-1Oxbp0-c}\nwindexuse{\nwixident{get{\_}dim}}{get:undim}{NW3gGP3e-1Oxbp0-c}\nwindexuse{\nwixident{get{\_}l}}{get:unl}{NW3gGP3e-1Oxbp0-c}\nwindexuse{\nwixident{k}}{k}{NW3gGP3e-1Oxbp0-c}\nwindexuse{\nwixident{lst}}{lst}{NW3gGP3e-1Oxbp0-c}\nwindexuse{\nwixident{m}}{m}{NW3gGP3e-1Oxbp0-c}\nwindexuse{\nwixident{nops}}{nops}{NW3gGP3e-1Oxbp0-c}\nwindexuse{\nwixident{numeric}}{numeric}{NW3gGP3e-1Oxbp0-c}\nwindexuse{\nwixident{subject{\_}to}}{subject:unto}{NW3gGP3e-1Oxbp0-c}\nwindexuse{\nwixident{subs}}{subs}{NW3gGP3e-1Oxbp0-c}\nwendcode{}\nwbegindocs{551}\nwdocspar
\subsubsection[Conversion of the cycle to the matrix form]{Conversion of the {\Tt{}\Rm{}{\bf{}cycle}\nwendquote} to the matrix form}
\label{sec:conv-cycle-matr}

This method is inverse to the constructor of the {\Tt{}\Rm{}{\bf{}cycle}\nwendquote} from its
matrix, see~\eqref{eq:matrix-from-cycle}
and~\cite[\S~\ref{E-sec:fillm-spring-cnops}]{Kisil05a}. First, we
process the supplied {\Tt{}\Rm{}{\it{}e}\nwendquote} to the standard form of the Clifford unit.
\nwenddocs{}\nwbegincode{552}\sublabel{NW3gGP3e-1Oxbp0-d}\nwmargintag{{\nwtagstyle{}\subpageref{NW3gGP3e-1Oxbp0-d}}}\moddef{cycle.cpp~{\nwtagstyle{}\subpageref{NW3gGP3e-1Oxbp0-1}}}\plusendmoddef\Rm{}\nwstartdeflinemarkup\nwprevnextdefs{NW3gGP3e-1Oxbp0-c}{NW3gGP3e-1Oxbp0-e}\nwenddeflinemarkup
{\bf{}matrix} {\bf{}cycle}::{\it{}to\_matrix}({\bf{}const} {\bf{}ex} & {\it{}e}, {\bf{}const} {\bf{}ex} & {\it{}sign}) {\bf{}const}
{\nwlbrace}
    {\bf{}ex} {\it{}one}, {\it{}es}, {\it{}conv}, {\it{}D} = {\it{}get\_dim}();
    {\bf{}varidx} {\it{}i0}(({\bf{}new} {\bf{}symbol})\begin{math}\rightarrow\end{math}{\it{}setflag}({\it{}status\_flags}::{\it{}dynallocated}), {\it{}D}),
        {\it{}i1}(({\bf{}new} {\bf{}symbol})\begin{math}\rightarrow\end{math}{\it{}setflag}({\it{}status\_flags}::{\it{}dynallocated}), {\it{}D}, {\bf{}true});
    {\bf{}if} ({\it{}e}.{\it{}is\_zero}()) {\nwlbrace}
        {\it{}one} = {\it{}dirac\_ONE}({\it{}ex\_to}\begin{math}<\end{math}{\bf{}clifford}\begin{math}>\end{math}({\it{}unit}).{\it{}get\_representation\_label}());
        {\it{}es} = {\it{}unit}.{\it{}subs}({\it{}unit}.{\it{}op}(1) \begin{math}\equiv\end{math} {\it{}i1}.{\it{}toggle\_variance}());
    {\nwrbrace} {\bf{}else} {\bf{}if} ({\it{}is\_a}\begin{math}<\end{math}{\bf{}clifford}\begin{math}>\end{math}({\it{}e})) {\nwlbrace}
        {\it{}one} = {\it{}dirac\_ONE}({\it{}ex\_to}\begin{math}<\end{math}{\bf{}clifford}\begin{math}>\end{math}({\it{}e}).{\it{}get\_representation\_label}());
        {\it{}es} = {\it{}e}.{\it{}subs}({\it{}e}.{\it{}op}(1) \begin{math}\equiv\end{math} {\it{}i1}.{\it{}toggle\_variance}());
    {\nwrbrace} {\bf{}else} {\bf{}if} ({\it{}is\_a}\begin{math}<\end{math}{\bf{}tensor}\begin{math}>\end{math}({\it{}e}) \begin{math}\vee\end{math} {\it{}is\_a}\begin{math}<\end{math}{\bf{}indexed}\begin{math}>\end{math}({\it{}e}) \begin{math}\vee\end{math} {\it{}is\_a}\begin{math}<\end{math}{\bf{}matrix}\begin{math}>\end{math}({\it{}e})) {\nwlbrace}
        {\it{}one} = {\it{}dirac\_ONE}();
        {\it{}es} = {\it{}clifford\_unit}({\it{}i1}.{\it{}toggle\_variance}(), {\it{}e});
    {\nwrbrace} {\bf{}else}
        {\bf{}throw}({\it{}std}::{\it{}invalid\_argument}({\tt{}"cycle::to\_matrix(): expect a clifford number, matrix, tensor or "}
                                    {\tt{}"indexed as the first parameter"}));

\nwidentuses{\\{{\nwixident{cycle}}{cycle}}\\{{\nwixident{ex}}{ex}}\\{{\nwixident{get{\_}dim}}{get:undim}}\\{{\nwixident{is{\_}zero}}{is:unzero}}\\{{\nwixident{matrix}}{matrix}}\\{{\nwixident{op}}{op}}\\{{\nwixident{subs}}{subs}}\\{{\nwixident{to{\_}matrix}}{to:unmatrix}}\\{{\nwixident{varidx}}{varidx}}}\nwindexuse{\nwixident{cycle}}{cycle}{NW3gGP3e-1Oxbp0-d}\nwindexuse{\nwixident{ex}}{ex}{NW3gGP3e-1Oxbp0-d}\nwindexuse{\nwixident{get{\_}dim}}{get:undim}{NW3gGP3e-1Oxbp0-d}\nwindexuse{\nwixident{is{\_}zero}}{is:unzero}{NW3gGP3e-1Oxbp0-d}\nwindexuse{\nwixident{matrix}}{matrix}{NW3gGP3e-1Oxbp0-d}\nwindexuse{\nwixident{op}}{op}{NW3gGP3e-1Oxbp0-d}\nwindexuse{\nwixident{subs}}{subs}{NW3gGP3e-1Oxbp0-d}\nwindexuse{\nwixident{to{\_}matrix}}{to:unmatrix}{NW3gGP3e-1Oxbp0-d}\nwindexuse{\nwixident{varidx}}{varidx}{NW3gGP3e-1Oxbp0-d}\nwendcode{}\nwbegindocs{553}Then we work out the sign, which should be used.
\nwenddocs{}\nwbegincode{554}\sublabel{NW3gGP3e-1Oxbp0-e}\nwmargintag{{\nwtagstyle{}\subpageref{NW3gGP3e-1Oxbp0-e}}}\moddef{cycle.cpp~{\nwtagstyle{}\subpageref{NW3gGP3e-1Oxbp0-1}}}\plusendmoddef\Rm{}\nwstartdeflinemarkup\nwprevnextdefs{NW3gGP3e-1Oxbp0-d}{NW3gGP3e-1Oxbp0-f}\nwenddeflinemarkup
    {\bf{}ex} {\it{}sign\_m} = {\it{}sign}.{\it{}evalm}();

{\bf{}if} ({\it{}is\_a}\begin{math}<\end{math}{\bf{}tensor}\begin{math}>\end{math}({\it{}sign\_m}))
    {\it{}conv} = {\bf{}indexed}({\it{}ex\_to}\begin{math}<\end{math}{\bf{}tensor}\begin{math}>\end{math}({\it{}sign\_m}), {\it{}i0}, {\it{}i1});
{\bf{}else} {\bf{}if} ({\it{}is\_a}\begin{math}<\end{math}{\bf{}clifford}\begin{math}>\end{math}({\it{}sign\_m})) {\nwlbrace}
    {\bf{}if} ({\it{}ex\_to}\begin{math}<\end{math}{\bf{}varidx}\begin{math}>\end{math}({\it{}sign\_m}.{\it{}op}(1)).{\it{}get\_dim}() \begin{math}\equiv\end{math} {\it{}D})
        {\it{}conv} = {\it{}ex\_to}\begin{math}<\end{math}{\bf{}clifford}\begin{math}>\end{math}({\it{}sign\_m}).{\it{}get\_metric}({\it{}i0}, {\it{}i1});
    {\bf{}else}
        {\bf{}throw}({\it{}std}::{\it{}invalid\_argument}({\tt{}"cycle::to\_matrix(): the sign should be a Clifford unit with the "}
                                    {\tt{}"dimensionality matching to the second parameter"}));
{\nwrbrace} {\bf{}else} {\bf{}if} ({\it{}is\_a}\begin{math}<\end{math}{\bf{}indexed}\begin{math}>\end{math}({\it{}sign\_m})) {\nwlbrace}
    {\it{}exvector} {\it{}ind} = {\it{}ex\_to}\begin{math}<\end{math}{\bf{}indexed}\begin{math}>\end{math}({\it{}sign\_m}).{\it{}get\_indices}();
    {\bf{}if} (({\it{}ind}.{\it{}size}() \begin{math}\equiv\end{math} 2) \begin{math}\wedge\end{math} ({\it{}ex\_to}\begin{math}<\end{math}{\bf{}varidx}\begin{math}>\end{math}({\it{}ind}[0]).{\it{}get\_dim}() \begin{math}\equiv\end{math} {\it{}D}) \begin{math}\wedge\end{math} ({\it{}ex\_to}\begin{math}<\end{math}{\bf{}varidx}\begin{math}>\end{math}({\it{}ind}[1]).{\it{}get\_dim}() \begin{math}\equiv\end{math} {\it{}D}))
        {\it{}conv} = {\it{}sign\_m}.{\it{}subs}({\bf{}lst}({\it{}ind}[0] \begin{math}\equiv\end{math} {\it{}i0}, {\it{}ind}[1] \begin{math}\equiv\end{math} {\it{}i1}));
    {\bf{}else}
        {\bf{}throw}({\it{}std}::{\it{}invalid\_argument}({\tt{}"cycle::to\_matrix(): the sign should be an indexed object with two "}
                                    {\tt{}"indices and their dimensionality matching to the second parameter"}));
{\nwrbrace} {\bf{}else} {\bf{}if} ({\it{}is\_a}\begin{math}<\end{math}{\bf{}matrix}\begin{math}>\end{math}({\it{}sign\_m})) {\nwlbrace}
    {\bf{}if} (({\it{}ex\_to}\begin{math}<\end{math}{\bf{}matrix}\begin{math}>\end{math}({\it{}sign\_m}).{\it{}cols}() \begin{math}\equiv\end{math} {\it{}D}) \begin{math}\wedge\end{math} ({\it{}ex\_to}\begin{math}<\end{math}{\bf{}matrix}\begin{math}>\end{math}({\it{}sign\_m}).{\it{}rows}() \begin{math}\equiv\end{math} {\it{}D}))
        {\it{}conv} = {\bf{}indexed}({\it{}ex\_to}\begin{math}<\end{math}{\bf{}matrix}\begin{math}>\end{math}({\it{}sign\_m}), {\it{}i0}, {\it{}i1});
    {\bf{}else}
        {\bf{}throw}({\it{}std}::{\it{}invalid\_argument}({\tt{}"cycle::to\_matrix(): the sign should be a square matrix with the "}
                                    {\tt{}"dimensionality matching to the second parameter"}));
{\nwrbrace} {\bf{}else}
    {\bf{}throw}({\it{}std}::{\it{}invalid\_argument}({\tt{}"cycle::to\_matrix(): the sign should be either tensor, indexed, "}
                                {\tt{}"matrix or Clifford unit"}));

\nwidentuses{\\{{\nwixident{cycle}}{cycle}}\\{{\nwixident{ex}}{ex}}\\{{\nwixident{get{\_}dim}}{get:undim}}\\{{\nwixident{get{\_}metric}}{get:unmetric}}\\{{\nwixident{lst}}{lst}}\\{{\nwixident{matrix}}{matrix}}\\{{\nwixident{op}}{op}}\\{{\nwixident{subs}}{subs}}\\{{\nwixident{to{\_}matrix}}{to:unmatrix}}\\{{\nwixident{varidx}}{varidx}}}\nwindexuse{\nwixident{cycle}}{cycle}{NW3gGP3e-1Oxbp0-e}\nwindexuse{\nwixident{ex}}{ex}{NW3gGP3e-1Oxbp0-e}\nwindexuse{\nwixident{get{\_}dim}}{get:undim}{NW3gGP3e-1Oxbp0-e}\nwindexuse{\nwixident{get{\_}metric}}{get:unmetric}{NW3gGP3e-1Oxbp0-e}\nwindexuse{\nwixident{lst}}{lst}{NW3gGP3e-1Oxbp0-e}\nwindexuse{\nwixident{matrix}}{matrix}{NW3gGP3e-1Oxbp0-e}\nwindexuse{\nwixident{op}}{op}{NW3gGP3e-1Oxbp0-e}\nwindexuse{\nwixident{subs}}{subs}{NW3gGP3e-1Oxbp0-e}\nwindexuse{\nwixident{to{\_}matrix}}{to:unmatrix}{NW3gGP3e-1Oxbp0-e}\nwindexuse{\nwixident{varidx}}{varidx}{NW3gGP3e-1Oxbp0-e}\nwendcode{}\nwbegindocs{555}When all components are ready the matrix is build in few lines.
\nwenddocs{}\nwbegincode{556}\sublabel{NW3gGP3e-1Oxbp0-f}\nwmargintag{{\nwtagstyle{}\subpageref{NW3gGP3e-1Oxbp0-f}}}\moddef{cycle.cpp~{\nwtagstyle{}\subpageref{NW3gGP3e-1Oxbp0-1}}}\plusendmoddef\Rm{}\nwstartdeflinemarkup\nwprevnextdefs{NW3gGP3e-1Oxbp0-e}{NW3gGP3e-1Oxbp0-g}\nwenddeflinemarkup
    {\bf{}ex} {\it{}a00} = {\it{}expand\_dummy\_sum}({\it{}l}.{\it{}subs}({\it{}ex\_to}\begin{math}<\end{math}{\bf{}indexed}\begin{math}>\end{math}({\it{}l}).{\it{}get\_indices}()[0] \begin{math}\equiv\end{math} {\it{}i0}.{\it{}toggle\_variance}()) \begin{math}\ast\end{math} {\it{}conv} \begin{math}\ast\end{math} {\it{}es});

{\bf{}return} {\bf{}matrix}(2, 2, {\bf{}lst}({\it{}a00}, {\it{}m} \begin{math}\ast\end{math} {\it{}one}, {\it{}k} \begin{math}\ast\end{math} {\it{}one}, -{\it{}a00}));
{\nwrbrace}

\nwidentuses{\\{{\nwixident{ex}}{ex}}\\{{\nwixident{k}}{k}}\\{{\nwixident{l}}{l}}\\{{\nwixident{lst}}{lst}}\\{{\nwixident{m}}{m}}\\{{\nwixident{matrix}}{matrix}}\\{{\nwixident{subs}}{subs}}}\nwindexuse{\nwixident{ex}}{ex}{NW3gGP3e-1Oxbp0-f}\nwindexuse{\nwixident{k}}{k}{NW3gGP3e-1Oxbp0-f}\nwindexuse{\nwixident{l}}{l}{NW3gGP3e-1Oxbp0-f}\nwindexuse{\nwixident{lst}}{lst}{NW3gGP3e-1Oxbp0-f}\nwindexuse{\nwixident{m}}{m}{NW3gGP3e-1Oxbp0-f}\nwindexuse{\nwixident{matrix}}{matrix}{NW3gGP3e-1Oxbp0-f}\nwindexuse{\nwixident{subs}}{subs}{NW3gGP3e-1Oxbp0-f}\nwendcode{}\nwbegindocs{557}\nwdocspar
\subsubsection{Calculation of a value of cycle at a point}
\label{sec:calc-value-cycle}

This is used in the construction of a
relational {\Tt{}\Rm{}{\bf{}cycle}::{\it{}passing}\nwendquote} describing incidence of a point  to cycle.
\nwenddocs{}\nwbegincode{558}\sublabel{NW3gGP3e-1Oxbp0-g}\nwmargintag{{\nwtagstyle{}\subpageref{NW3gGP3e-1Oxbp0-g}}}\moddef{cycle.cpp~{\nwtagstyle{}\subpageref{NW3gGP3e-1Oxbp0-1}}}\plusendmoddef\Rm{}\nwstartdeflinemarkup\nwprevnextdefs{NW3gGP3e-1Oxbp0-f}{NW3gGP3e-1Oxbp0-h}\nwenddeflinemarkup
{\bf{}ex} {\bf{}cycle}::{\it{}val}({\bf{}const} {\bf{}ex} & {\it{}y}) {\bf{}const}
{\nwlbrace}
    {\bf{}ex} {\it{}y0}, {\it{}D} = {\it{}get\_dim}();
    {\bf{}varidx} {\it{}i0}, {\it{}i1};
    {\bf{}if} ({\it{}is\_a}\begin{math}<\end{math}{\bf{}indexed}\begin{math}>\end{math}({\it{}y})) {\nwlbrace}
        {\it{}i0} = {\it{}ex\_to}\begin{math}<\end{math}{\bf{}varidx}\begin{math}>\end{math}({\it{}ex\_to}\begin{math}<\end{math}{\bf{}indexed}\begin{math}>\end{math}({\it{}y}).{\it{}get\_indices}()[0]);
        {\bf{}if} (({\it{}ex\_to}\begin{math}<\end{math}{\bf{}indexed}\begin{math}>\end{math}({\it{}y}).{\it{}get\_indices}().{\it{}size}() \begin{math}\equiv\end{math} 1) \begin{math}\wedge\end{math} ({\it{}i0}.{\it{}get\_dim}() \begin{math}\equiv\end{math} {\it{}D})) {\nwlbrace}
            {\it{}y0} = {\it{}ex\_to}\begin{math}<\end{math}{\bf{}indexed}\begin{math}>\end{math}({\it{}y});
            {\it{}i1} = {\bf{}varidx}(({\bf{}new} {\bf{}symbol})\begin{math}\rightarrow\end{math}{\it{}setflag}({\it{}status\_flags}::{\it{}dynallocated}), {\it{}D});
        {\nwrbrace} {\bf{}else}
            {\bf{}throw}({\it{}std}::{\it{}invalid\_argument}({\tt{}"cycle::val(): the second parameter should be a indexed object with "}
                                        {\tt{}"one varindex"}));
    {\nwrbrace} {\bf{}else} {\bf{}if} ({\it{}y}.{\it{}info}({\it{}info\_flags}::{\it{}list}) \begin{math}\wedge\end{math} ({\it{}y}.{\it{}nops}() \begin{math}\equiv\end{math} {\it{}D})) {\nwlbrace}
        {\it{}i0} = {\bf{}varidx}(({\bf{}new} {\bf{}symbol})\begin{math}\rightarrow\end{math}{\it{}setflag}({\it{}status\_flags}::{\it{}dynallocated}), {\it{}D});
        {\it{}i1} = {\bf{}varidx}(({\bf{}new} {\bf{}symbol})\begin{math}\rightarrow\end{math}{\it{}setflag}({\it{}status\_flags}::{\it{}dynallocated}), {\it{}D});
        {\it{}y0} = {\bf{}indexed}({\bf{}matrix}(1, {\it{}y}.{\it{}nops}(), {\it{}ex\_to}\begin{math}<\end{math}{\bf{}lst}\begin{math}>\end{math}({\it{}y})), {\it{}i0});
    {\nwrbrace} {\bf{}else} {\bf{}if} ({\it{}is\_a}\begin{math}<\end{math}{\bf{}matrix}\begin{math}>\end{math}({\it{}y}) \begin{math}\wedge\end{math} ({\it{}min}({\it{}ex\_to}\begin{math}<\end{math}{\bf{}matrix}\begin{math}>\end{math}({\it{}y}).{\it{}rows}(), {\it{}ex\_to}\begin{math}<\end{math}{\bf{}matrix}\begin{math}>\end{math}({\it{}y}).{\it{}cols}()) \begin{math}\equiv\end{math}1)
               \begin{math}\wedge\end{math} ({\it{}D} \begin{math}\equiv\end{math} {\it{}max}({\it{}ex\_to}\begin{math}<\end{math}{\bf{}matrix}\begin{math}>\end{math}({\it{}y}).{\it{}rows}(), {\it{}ex\_to}\begin{math}<\end{math}{\bf{}matrix}\begin{math}>\end{math}({\it{}y}).{\it{}cols}()))) {\nwlbrace}
        {\it{}i0} = {\bf{}varidx}(({\bf{}new} {\bf{}symbol})\begin{math}\rightarrow\end{math}{\it{}setflag}({\it{}status\_flags}::{\it{}dynallocated}), {\it{}D});
        {\it{}i1} = {\bf{}varidx}(({\bf{}new} {\bf{}symbol})\begin{math}\rightarrow\end{math}{\it{}setflag}({\it{}status\_flags}::{\it{}dynallocated}), {\it{}D});
        {\it{}y0} = {\bf{}indexed}({\it{}y}, {\it{}i0});
    {\nwrbrace} {\bf{}else}
        {\bf{}throw}({\it{}std}::{\it{}invalid\_argument}({\tt{}"cycle::val(): the second parameter should be a indexed object, "}
                                    {\tt{}"matrix or list"}));
    
    {\bf{}return} {\it{}expand\_dummy\_sum}(-{\it{}k}\begin{math}\ast\end{math}{\it{}y0}\begin{math}\ast\end{math}{\it{}y0}.{\it{}subs}({\it{}i0} \begin{math}\equiv\end{math} {\it{}i1})\begin{math}\ast\end{math}{\it{}get\_metric}({\it{}i0}.{\it{}toggle\_variance}(), {\it{}i1}.{\it{}toggle\_variance}())
                            - {\bf{}numeric}(2)\begin{math}\ast\end{math} {\it{}l}\begin{math}\ast\end{math}{\it{}y0}.{\it{}subs}({\it{}i0} \begin{math}\equiv\end{math} {\it{}ex\_to}\begin{math}<\end{math}{\bf{}varidx}\begin{math}>\end{math}({\it{}ex\_to}\begin{math}<\end{math}{\bf{}indexed}\begin{math}>\end{math}({\it{}l}).{\it{}get\_indices}()[0]).{\it{}toggle\_variance}()) +{\it{}m});
{\nwrbrace}

\nwidentuses{\\{{\nwixident{cycle}}{cycle}}\\{{\nwixident{ex}}{ex}}\\{{\nwixident{get{\_}dim}}{get:undim}}\\{{\nwixident{get{\_}metric}}{get:unmetric}}\\{{\nwixident{k}}{k}}\\{{\nwixident{l}}{l}}\\{{\nwixident{lst}}{lst}}\\{{\nwixident{m}}{m}}\\{{\nwixident{matrix}}{matrix}}\\{{\nwixident{nops}}{nops}}\\{{\nwixident{numeric}}{numeric}}\\{{\nwixident{subs}}{subs}}\\{{\nwixident{val}}{val}}\\{{\nwixident{varidx}}{varidx}}}\nwindexuse{\nwixident{cycle}}{cycle}{NW3gGP3e-1Oxbp0-g}\nwindexuse{\nwixident{ex}}{ex}{NW3gGP3e-1Oxbp0-g}\nwindexuse{\nwixident{get{\_}dim}}{get:undim}{NW3gGP3e-1Oxbp0-g}\nwindexuse{\nwixident{get{\_}metric}}{get:unmetric}{NW3gGP3e-1Oxbp0-g}\nwindexuse{\nwixident{k}}{k}{NW3gGP3e-1Oxbp0-g}\nwindexuse{\nwixident{l}}{l}{NW3gGP3e-1Oxbp0-g}\nwindexuse{\nwixident{lst}}{lst}{NW3gGP3e-1Oxbp0-g}\nwindexuse{\nwixident{m}}{m}{NW3gGP3e-1Oxbp0-g}\nwindexuse{\nwixident{matrix}}{matrix}{NW3gGP3e-1Oxbp0-g}\nwindexuse{\nwixident{nops}}{nops}{NW3gGP3e-1Oxbp0-g}\nwindexuse{\nwixident{numeric}}{numeric}{NW3gGP3e-1Oxbp0-g}\nwindexuse{\nwixident{subs}}{subs}{NW3gGP3e-1Oxbp0-g}\nwindexuse{\nwixident{val}}{val}{NW3gGP3e-1Oxbp0-g}\nwindexuse{\nwixident{varidx}}{varidx}{NW3gGP3e-1Oxbp0-g}\nwendcode{}\nwbegindocs{559}\nwdocspar
\subsubsection[Matrix methods for cycle]{Matrix methods for {\Tt{}\Rm{}{\bf{}cycle}\nwendquote}}
\label{sec:matrix-methods-cycle}

The method {\Tt{}\Rm{}{\it{}det}()\nwendquote} may be defined in several ways.  An alternative
to the present definition is \emph{pseudodeterminant}~\cite[(4.9)]{Cnops02a}
\begin{webcode}{\Tt{}\Rm{}{\bf{}ex} {\bf{}cycle}::{\it{}det}({\bf{}const} {\bf{}ex} & {\it{}e} = 0, {\bf{}const} {\bf{}ex} & {\it{}sign} = ({\bf{}new} {\it{}tensdelta})\begin{math}\rightarrow\end{math}{\it{}setflag}({\it{}status\_flags}::{\it{}dynallocated})) {\bf{}const}\nwendquote}
{\Tt{}\Rm{}{\nwlbrace}{\bf{}ex} {\it{}M} = {\it{}normalize}().{\it{}to\_matrix}({\it{}e}, {\it{}sign});\nwendquote}
{\Tt{}\Rm{}    {\bf{}return} {\it{}remove\_dirac\_ONE}({\it{}M}.{\it{}op}(0)\begin{math}\ast\end{math}{\it{}clifford\_star}({\it{}M}.{\it{}op}(3))-{\it{}M}.{\it{}op}(1)\begin{math}\ast\end{math}{\it{}clifford\_star}({\it{}M}.{\it{}op}(2))) ; {\nwrbrace}\nwendquote}
\end{webcode}
However due to the structure of matrix this coincides with the usual
determinant of the matrix.
\nwenddocs{}\nwbegincode{560}\sublabel{NW3gGP3e-1Oxbp0-h}\nwmargintag{{\nwtagstyle{}\subpageref{NW3gGP3e-1Oxbp0-h}}}\moddef{cycle.cpp~{\nwtagstyle{}\subpageref{NW3gGP3e-1Oxbp0-1}}}\plusendmoddef\Rm{}\nwstartdeflinemarkup\nwprevnextdefs{NW3gGP3e-1Oxbp0-g}{NW3gGP3e-1Oxbp0-i}\nwenddeflinemarkup
{\bf{}ex} {\bf{}cycle}::{\it{}det}({\bf{}const} {\bf{}ex} & {\it{}e}, {\bf{}const} {\bf{}ex} & {\it{}sign}, {\bf{}const} {\bf{}ex} & {\it{}k\_norm}) {\bf{}const}
{\nwlbrace}
    {\bf{}return} {\it{}remove\_dirac\_ONE}(({\it{}k\_norm}.{\it{}is\_zero}()?\begin{math}\ast\end{math}{\it{}this}:{\it{}normalize}({\it{}k\_norm})).{\it{}to\_matrix}({\it{}e}, {\it{}sign}).{\it{}determinant}());
{\nwrbrace}

\nwidentuses{\\{{\nwixident{cycle}}{cycle}}\\{{\nwixident{det}}{det}}\\{{\nwixident{ex}}{ex}}\\{{\nwixident{is{\_}zero}}{is:unzero}}\\{{\nwixident{normalize}}{normalize}}\\{{\nwixident{to{\_}matrix}}{to:unmatrix}}}\nwindexuse{\nwixident{cycle}}{cycle}{NW3gGP3e-1Oxbp0-h}\nwindexuse{\nwixident{det}}{det}{NW3gGP3e-1Oxbp0-h}\nwindexuse{\nwixident{ex}}{ex}{NW3gGP3e-1Oxbp0-h}\nwindexuse{\nwixident{is{\_}zero}}{is:unzero}{NW3gGP3e-1Oxbp0-h}\nwindexuse{\nwixident{normalize}}{normalize}{NW3gGP3e-1Oxbp0-h}\nwindexuse{\nwixident{to{\_}matrix}}{to:unmatrix}{NW3gGP3e-1Oxbp0-h}\nwendcode{}\nwbegindocs{561}Multiplication of cycles in the matrix representations and their
similarity with respect to elements of \(\SL\) and other cycles.
\nwenddocs{}\nwbegincode{562}\sublabel{NW3gGP3e-1Oxbp0-i}\nwmargintag{{\nwtagstyle{}\subpageref{NW3gGP3e-1Oxbp0-i}}}\moddef{cycle.cpp~{\nwtagstyle{}\subpageref{NW3gGP3e-1Oxbp0-1}}}\plusendmoddef\Rm{}\nwstartdeflinemarkup\nwprevnextdefs{NW3gGP3e-1Oxbp0-h}{NW3gGP3e-1Oxbp0-j}\nwenddeflinemarkup
{\bf{}ex} {\bf{}cycle}::{\it{}mul}({\bf{}const} {\bf{}ex} & {\it{}C}, {\bf{}const} {\bf{}ex} & {\it{}e}, {\bf{}const} {\bf{}ex} & {\it{}sign}, {\bf{}const} {\bf{}ex} & {\it{}sign1}) {\bf{}const}
{\nwlbrace}
    {\bf{}if} ({\it{}is\_a}\begin{math}<\end{math}{\bf{}cycle}\begin{math}>\end{math}({\it{}C})) {\nwlbrace}
        {\bf{}return} {\it{}canonicalize\_clifford}({\it{}to\_matrix}({\it{}e}, {\it{}sign}).{\it{}mul}(
                {\it{}ex\_to}\begin{math}<\end{math}{\bf{}cycle}\begin{math}>\end{math}({\it{}C}).{\it{}to\_matrix}({\it{}e}.{\it{}is\_zero}()?{\it{}unit}:{\it{}e}, {\it{}sign1}.{\it{}is\_zero}()?{\it{}sign}:{\it{}sign1})));
    {\nwrbrace} {\bf{}else} {\bf{}if} ({\it{}is\_a}\begin{math}<\end{math}{\bf{}matrix}\begin{math}>\end{math}({\it{}C}) \begin{math}\wedge\end{math} ({\it{}ex\_to}\begin{math}<\end{math}{\bf{}matrix}\begin{math}>\end{math}({\it{}C}).{\it{}rows}() \begin{math}\equiv\end{math} 2)  \begin{math}\wedge\end{math} ({\it{}ex\_to}\begin{math}<\end{math}{\bf{}matrix}\begin{math}>\end{math}({\it{}C}).{\it{}cols}() \begin{math}\equiv\end{math} 2)) {\nwlbrace}
        {\bf{}return} {\it{}canonicalize\_clifford}({\it{}to\_matrix}({\it{}e}, {\it{}sign}).{\it{}mul}({\it{}ex\_to}\begin{math}<\end{math}{\bf{}matrix}\begin{math}>\end{math}({\it{}C})));
    {\nwrbrace} {\bf{}else}
        {\bf{}throw}({\it{}std}::{\it{}invalid\_argument}({\tt{}"cycle::mul(): cannot multiply a cycle by anything but a cycle "}
                                    {\tt{}"or 2x2 matrix"}));
{\nwrbrace}

\nwidentuses{\\{{\nwixident{cycle}}{cycle}}\\{{\nwixident{ex}}{ex}}\\{{\nwixident{is{\_}zero}}{is:unzero}}\\{{\nwixident{matrix}}{matrix}}\\{{\nwixident{mul}}{mul}}\\{{\nwixident{to{\_}matrix}}{to:unmatrix}}}\nwindexuse{\nwixident{cycle}}{cycle}{NW3gGP3e-1Oxbp0-i}\nwindexuse{\nwixident{ex}}{ex}{NW3gGP3e-1Oxbp0-i}\nwindexuse{\nwixident{is{\_}zero}}{is:unzero}{NW3gGP3e-1Oxbp0-i}\nwindexuse{\nwixident{matrix}}{matrix}{NW3gGP3e-1Oxbp0-i}\nwindexuse{\nwixident{mul}}{mul}{NW3gGP3e-1Oxbp0-i}\nwindexuse{\nwixident{to{\_}matrix}}{to:unmatrix}{NW3gGP3e-1Oxbp0-i}\nwendcode{}\nwbegindocs{563}\nwdocspar
\subsubsection[Actions of cycle as matrix]{Actions of {\Tt{}\Rm{}{\bf{}cycle}\nwendquote} as matrix}
\label{sec:actions-cycle-as}

{\Tt{}\Rm{}{\bf{}cycle}\nwendquote} in the matrix form can act on other objects, or matrices can
acts on {\Tt{}\Rm{}{\bf{}cycle}\nwendquote}.

\nwenddocs{}\nwbegindocs{564}Any \(2\times 2\)-matrix acts on a {\Tt{}\Rm{}{\bf{}cycle}\nwendquote} by the similarity: \(M:
C\mapsto MCM^{-1}\).
\nwenddocs{}\nwbegincode{565}\sublabel{NW3gGP3e-1Oxbp0-j}\nwmargintag{{\nwtagstyle{}\subpageref{NW3gGP3e-1Oxbp0-j}}}\moddef{cycle.cpp~{\nwtagstyle{}\subpageref{NW3gGP3e-1Oxbp0-1}}}\plusendmoddef\Rm{}\nwstartdeflinemarkup\nwprevnextdefs{NW3gGP3e-1Oxbp0-i}{NW3gGP3e-1Oxbp0-k}\nwenddeflinemarkup
{\bf{}cycle} {\bf{}cycle}::{\it{}matrix\_similarity}({\bf{}const} {\bf{}ex} & {\it{}M}, {\bf{}const} {\bf{}ex} & {\it{}e}, {\bf{}const} {\bf{}ex} & {\it{}sign}, {\bf{}bool} {\it{}not\_inverse}, {\bf{}const} {\bf{}ex} & {\it{}sign\_inv}) {\bf{}const}
{\nwlbrace}
    {\bf{}if} ({\it{}not} ({\it{}is\_a}\begin{math}<\end{math}{\bf{}matrix}\begin{math}>\end{math}({\it{}M}) \begin{math}\wedge\end{math} {\it{}ex\_to}\begin{math}<\end{math}{\bf{}matrix}\begin{math}>\end{math}({\it{}M}).{\it{}rows}()\begin{math}\equiv\end{math}2 \begin{math}\wedge\end{math} {\it{}ex\_to}\begin{math}<\end{math}{\bf{}matrix}\begin{math}>\end{math}({\it{}M}).{\it{}cols}()\begin{math}\equiv\end{math}2))
        {\bf{}throw}({\it{}std}::{\it{}invalid\_argument}({\tt{}"cycle::matrix\_similarity(): the first parameter sgould be a 2x2 matrix"}));
    {\bf{}return} {\it{}matrix\_similarity}({\it{}M}.{\it{}op}(0), {\it{}M}.{\it{}op}(1), {\it{}M}.{\it{}op}(2), {\it{}M}.{\it{}op}(3), {\it{}e}, {\it{}sign}, {\it{}not\_inverse}, {\it{}sign\_inv});
{\nwrbrace}

\nwidentuses{\\{{\nwixident{bool}}{bool}}\\{{\nwixident{cycle}}{cycle}}\\{{\nwixident{ex}}{ex}}\\{{\nwixident{matrix}}{matrix}}\\{{\nwixident{matrix{\_}similarity}}{matrix:unsimilarity}}\\{{\nwixident{op}}{op}}}\nwindexuse{\nwixident{bool}}{bool}{NW3gGP3e-1Oxbp0-j}\nwindexuse{\nwixident{cycle}}{cycle}{NW3gGP3e-1Oxbp0-j}\nwindexuse{\nwixident{ex}}{ex}{NW3gGP3e-1Oxbp0-j}\nwindexuse{\nwixident{matrix}}{matrix}{NW3gGP3e-1Oxbp0-j}\nwindexuse{\nwixident{matrix{\_}similarity}}{matrix:unsimilarity}{NW3gGP3e-1Oxbp0-j}\nwindexuse{\nwixident{op}}{op}{NW3gGP3e-1Oxbp0-j}\nwendcode{}\nwbegindocs{566}The same method works if the matrix is provided by its four
elements.
\nwenddocs{}\nwbegincode{567}\sublabel{NW3gGP3e-1Oxbp0-k}\nwmargintag{{\nwtagstyle{}\subpageref{NW3gGP3e-1Oxbp0-k}}}\moddef{cycle.cpp~{\nwtagstyle{}\subpageref{NW3gGP3e-1Oxbp0-1}}}\plusendmoddef\Rm{}\nwstartdeflinemarkup\nwprevnextdefs{NW3gGP3e-1Oxbp0-j}{NW3gGP3e-1Oxbp0-l}\nwenddeflinemarkup
{\bf{}cycle} {\bf{}cycle}::{\it{}matrix\_similarity}({\bf{}const} {\bf{}ex} & {\it{}a}, {\bf{}const} {\bf{}ex} & {\it{}b}, {\bf{}const} {\bf{}ex} & {\it{}c}, {\bf{}const} {\bf{}ex} & {\it{}d}, {\bf{}const} {\bf{}ex} & {\it{}e},
          {\bf{}const} {\bf{}ex} & {\it{}sign}, {\bf{}bool} {\it{}not\_inverse}, {\bf{}const} {\bf{}ex} & {\it{}sign\_inv}) {\bf{}const}
{\nwlbrace}
    {\bf{}matrix} {\it{}R}={\it{}ex\_to}\begin{math}<\end{math}{\bf{}matrix}\begin{math}>\end{math}({\it{}canonicalize\_clifford}({\bf{}matrix}(2,2,{\it{}not\_inverse}?{\bf{}lst}({\it{}a}, {\it{}b}, {\it{}c}, {\it{}d}):{\bf{}lst}({\it{}clifford\_star}({\it{}d}), -{\it{}clifford\_star}({\it{}b}), -{\it{}clifford\_star}({\it{}c}), {\it{}clifford\_star}({\it{}a})))
                                                 .{\it{}mul}({\it{}ex\_to}\begin{math}<\end{math}{\bf{}matrix}\begin{math}>\end{math}({\it{}mul}({\bf{}matrix}(2,2,{\it{}not\_inverse}?{\bf{}lst}({\it{}clifford\_star}({\it{}d}), -{\it{}clifford\_star}({\it{}b}), -{\it{}clifford\_star}({\it{}c}), {\it{}clifford\_star}({\it{}a})):{\bf{}lst}({\it{}a}, {\it{}b}, {\it{}c}, {\it{}d})), {\it{}e}.{\it{}is\_zero}()?{\it{}unit}:{\it{}e}, {\it{}sign})))
                                                 .{\it{}evalm}()).{\it{}normal}());

\nwidentuses{\\{{\nwixident{bool}}{bool}}\\{{\nwixident{cycle}}{cycle}}\\{{\nwixident{ex}}{ex}}\\{{\nwixident{is{\_}zero}}{is:unzero}}\\{{\nwixident{lst}}{lst}}\\{{\nwixident{matrix}}{matrix}}\\{{\nwixident{matrix{\_}similarity}}{matrix:unsimilarity}}\\{{\nwixident{mul}}{mul}}\\{{\nwixident{normal}}{normal}}}\nwindexuse{\nwixident{bool}}{bool}{NW3gGP3e-1Oxbp0-k}\nwindexuse{\nwixident{cycle}}{cycle}{NW3gGP3e-1Oxbp0-k}\nwindexuse{\nwixident{ex}}{ex}{NW3gGP3e-1Oxbp0-k}\nwindexuse{\nwixident{is{\_}zero}}{is:unzero}{NW3gGP3e-1Oxbp0-k}\nwindexuse{\nwixident{lst}}{lst}{NW3gGP3e-1Oxbp0-k}\nwindexuse{\nwixident{matrix}}{matrix}{NW3gGP3e-1Oxbp0-k}\nwindexuse{\nwixident{matrix{\_}similarity}}{matrix:unsimilarity}{NW3gGP3e-1Oxbp0-k}\nwindexuse{\nwixident{mul}}{mul}{NW3gGP3e-1Oxbp0-k}\nwindexuse{\nwixident{normal}}{normal}{NW3gGP3e-1Oxbp0-k}\nwendcode{}\nwbegindocs{568}We do some anti-symmetrisation of the matrix before the call of
{\Tt{}\Rm{}{\bf{}cycle}()\nwendquote} constructor since matrix should posses it anyway but it
may not be apparent to \GiNaC.
\nwenddocs{}\nwbegincode{569}\sublabel{NW3gGP3e-1Oxbp0-l}\nwmargintag{{\nwtagstyle{}\subpageref{NW3gGP3e-1Oxbp0-l}}}\moddef{cycle.cpp~{\nwtagstyle{}\subpageref{NW3gGP3e-1Oxbp0-1}}}\plusendmoddef\Rm{}\nwstartdeflinemarkup\nwprevnextdefs{NW3gGP3e-1Oxbp0-k}{NW3gGP3e-1Oxbp0-m}\nwenddeflinemarkup
            {\bf{}return} {\bf{}cycle}({\bf{}matrix}(2,2,{\bf{}lst}(({\it{}R}.{\it{}op}(0)-{\it{}R}.{\it{}op}(3))\begin{math}\div\end{math}{\bf{}numeric}(2),{\it{}R}.{\it{}op}(1),{\it{}R}.{\it{}op}(2),(-{\it{}R}.{\it{}op}(0)+{\it{}R}.{\it{}op}(3))\begin{math}\div\end{math}{\bf{}numeric}(2))), {\it{}unit}, {\it{}e}.{\it{}is\_zero}()?{\it{}unit}:{\it{}e}, {\it{}sign\_inv});
{\nwrbrace}

\nwidentuses{\\{{\nwixident{cycle}}{cycle}}\\{{\nwixident{is{\_}zero}}{is:unzero}}\\{{\nwixident{lst}}{lst}}\\{{\nwixident{matrix}}{matrix}}\\{{\nwixident{numeric}}{numeric}}\\{{\nwixident{op}}{op}}}\nwindexuse{\nwixident{cycle}}{cycle}{NW3gGP3e-1Oxbp0-l}\nwindexuse{\nwixident{is{\_}zero}}{is:unzero}{NW3gGP3e-1Oxbp0-l}\nwindexuse{\nwixident{lst}}{lst}{NW3gGP3e-1Oxbp0-l}\nwindexuse{\nwixident{matrix}}{matrix}{NW3gGP3e-1Oxbp0-l}\nwindexuse{\nwixident{numeric}}{numeric}{NW3gGP3e-1Oxbp0-l}\nwindexuse{\nwixident{op}}{op}{NW3gGP3e-1Oxbp0-l}\nwendcode{}\nwbegindocs{570} For elements of \(\SL\) we have a specific method which make the
proper ``cliffordization'' of the matrix first.
\nwenddocs{}\nwbegincode{571}\sublabel{NW3gGP3e-1Oxbp0-m}\nwmargintag{{\nwtagstyle{}\subpageref{NW3gGP3e-1Oxbp0-m}}}\moddef{cycle.cpp~{\nwtagstyle{}\subpageref{NW3gGP3e-1Oxbp0-1}}}\plusendmoddef\Rm{}\nwstartdeflinemarkup\nwprevnextdefs{NW3gGP3e-1Oxbp0-l}{NW3gGP3e-1Oxbp0-n}\nwenddeflinemarkup
{\bf{}cycle} {\bf{}cycle}::{\it{}sl2\_similarity}({\bf{}const} {\bf{}ex} & {\it{}a}, {\bf{}const} {\bf{}ex} & {\it{}b}, {\bf{}const} {\bf{}ex} & {\it{}c}, {\bf{}const} {\bf{}ex} & {\it{}d}, {\bf{}const} {\bf{}ex} & {\it{}e}, 
                            {\bf{}const} {\bf{}ex} & {\it{}sign}, {\bf{}bool} {\it{}not\_inverse}, {\bf{}const} {\bf{}ex} & {\it{}sign\_inv}) {\bf{}const}
{\nwlbrace}
//  ex sign\_inv=is\_a\begin{math}<\end{math}matrix\begin{math}>\end{math}(sign)?pow(sign,-1):sign;
    {\bf{}relational} {\it{}sl2\_rel} = ({\it{}c}\begin{math}\ast\end{math}{\it{}b} \begin{math}\equiv\end{math} ({\it{}d}\begin{math}\ast\end{math}{\it{}a}-1));

\nwidentuses{\\{{\nwixident{bool}}{bool}}\\{{\nwixident{cycle}}{cycle}}\\{{\nwixident{ex}}{ex}}\\{{\nwixident{matrix}}{matrix}}\\{{\nwixident{sl2{\_}similarity}}{sl2:unsimilarity}}}\nwindexuse{\nwixident{bool}}{bool}{NW3gGP3e-1Oxbp0-m}\nwindexuse{\nwixident{cycle}}{cycle}{NW3gGP3e-1Oxbp0-m}\nwindexuse{\nwixident{ex}}{ex}{NW3gGP3e-1Oxbp0-m}\nwindexuse{\nwixident{matrix}}{matrix}{NW3gGP3e-1Oxbp0-m}\nwindexuse{\nwixident{sl2{\_}similarity}}{sl2:unsimilarity}{NW3gGP3e-1Oxbp0-m}\nwendcode{}\nwbegindocs{572} We check either the condition \(ad-bc=1\) can be used for
substitution later.
\nwenddocs{}\nwbegincode{573}\sublabel{NW3gGP3e-1Oxbp0-n}\nwmargintag{{\nwtagstyle{}\subpageref{NW3gGP3e-1Oxbp0-n}}}\moddef{cycle.cpp~{\nwtagstyle{}\subpageref{NW3gGP3e-1Oxbp0-1}}}\plusendmoddef\Rm{}\nwstartdeflinemarkup\nwprevnextdefs{NW3gGP3e-1Oxbp0-m}{NW3gGP3e-1Oxbp0-o}\nwenddeflinemarkup
    {\bf{}ex} {\it{}det}=({\it{}a}\begin{math}\ast\end{math}{\it{}d}-{\it{}b}\begin{math}\ast\end{math}{\it{}c}).{\it{}eval}();
    {\bf{}if} ({\it{}is\_a}\begin{math}<\end{math}{\bf{}numeric}\begin{math}>\end{math}({\it{}det}) \begin{math}\wedge\end{math} ({\it{}ex\_to}\begin{math}<\end{math}{\bf{}numeric}\begin{math}>\end{math}({\it{}det}).{\it{}evalf}() \begin{math}\neq\end{math}1)) 
        {\it{}sl2\_rel} = ({\it{}c}\begin{math}\ast\end{math}{\it{}b}\begin{math}\equiv\end{math}{\it{}c}\begin{math}\ast\end{math}{\it{}b});

\nwidentuses{\\{{\nwixident{det}}{det}}\\{{\nwixident{ex}}{ex}}\\{{\nwixident{numeric}}{numeric}}}\nwindexuse{\nwixident{det}}{det}{NW3gGP3e-1Oxbp0-n}\nwindexuse{\nwixident{ex}}{ex}{NW3gGP3e-1Oxbp0-n}\nwindexuse{\nwixident{numeric}}{numeric}{NW3gGP3e-1Oxbp0-n}\nwendcode{}\nwbegindocs{574}Evaluation of the matrix corresponding to the cycle.
\nwenddocs{}\nwbegincode{575}\sublabel{NW3gGP3e-1Oxbp0-o}\nwmargintag{{\nwtagstyle{}\subpageref{NW3gGP3e-1Oxbp0-o}}}\moddef{cycle.cpp~{\nwtagstyle{}\subpageref{NW3gGP3e-1Oxbp0-1}}}\plusendmoddef\Rm{}\nwstartdeflinemarkup\nwprevnextdefs{NW3gGP3e-1Oxbp0-n}{NW3gGP3e-1Oxbp0-p}\nwenddeflinemarkup
    {\bf{}matrix} {\it{}R}={\it{}ex\_to}\begin{math}<\end{math}{\bf{}matrix}\begin{math}>\end{math}({\it{}canonicalize\_clifford}(
                               {\it{}sl2\_clifford}({\it{}a}, {\it{}b}, {\it{}c}, {\it{}d}, {\it{}e}.{\it{}is\_zero}()?{\it{}unit}:{\it{}e}, {\it{}not\_inverse})
                               .{\it{}mul}({\it{}ex\_to}\begin{math}<\end{math}{\bf{}matrix}\begin{math}>\end{math}({\it{}mul}({\it{}sl2\_clifford}({\it{}a}, {\it{}b}, {\it{}c}, {\it{}d}, {\it{}e}.{\it{}is\_zero}()?{\it{}unit}:{\it{}e}, \begin{math}\neg\end{math}{\it{}not\_inverse}), {\it{}e}, {\it{}sign\_inv})))
                               .{\it{}evalm}().{\it{}subs}({\it{}sl2\_rel}, {\it{}subs\_options}::{\it{}algebraic} \begin{math}\mid\end{math} {\it{}subs\_options}::{\it{}no\_pattern})).{\it{}normal}());

\nwidentuses{\\{{\nwixident{is{\_}zero}}{is:unzero}}\\{{\nwixident{matrix}}{matrix}}\\{{\nwixident{mul}}{mul}}\\{{\nwixident{normal}}{normal}}\\{{\nwixident{subs}}{subs}}}\nwindexuse{\nwixident{is{\_}zero}}{is:unzero}{NW3gGP3e-1Oxbp0-o}\nwindexuse{\nwixident{matrix}}{matrix}{NW3gGP3e-1Oxbp0-o}\nwindexuse{\nwixident{mul}}{mul}{NW3gGP3e-1Oxbp0-o}\nwindexuse{\nwixident{normal}}{normal}{NW3gGP3e-1Oxbp0-o}\nwindexuse{\nwixident{subs}}{subs}{NW3gGP3e-1Oxbp0-o}\nwendcode{}\nwbegindocs{576}The same anti-symmetrisation of the matrix.
\nwenddocs{}\nwbegincode{577}\sublabel{NW3gGP3e-1Oxbp0-p}\nwmargintag{{\nwtagstyle{}\subpageref{NW3gGP3e-1Oxbp0-p}}}\moddef{cycle.cpp~{\nwtagstyle{}\subpageref{NW3gGP3e-1Oxbp0-1}}}\plusendmoddef\Rm{}\nwstartdeflinemarkup\nwprevnextdefs{NW3gGP3e-1Oxbp0-o}{NW3gGP3e-1Oxbp0-q}\nwenddeflinemarkup
    {\bf{}return} {\bf{}cycle}({\bf{}matrix}(2,2,{\bf{}lst}(({\it{}R}.{\it{}op}(0)-{\it{}R}.{\it{}op}(3))\begin{math}\div\end{math}{\bf{}numeric}(2),{\it{}R}.{\it{}op}(1),{\it{}R}.{\it{}op}(2),(-{\it{}R}.{\it{}op}(0)+{\it{}R}.{\it{}op}(3))\begin{math}\div\end{math}{\bf{}numeric}(2))), {\it{}unit}, {\it{}e}, {\it{}sign});
{\nwrbrace}

\nwidentuses{\\{{\nwixident{cycle}}{cycle}}\\{{\nwixident{lst}}{lst}}\\{{\nwixident{matrix}}{matrix}}\\{{\nwixident{numeric}}{numeric}}\\{{\nwixident{op}}{op}}}\nwindexuse{\nwixident{cycle}}{cycle}{NW3gGP3e-1Oxbp0-p}\nwindexuse{\nwixident{lst}}{lst}{NW3gGP3e-1Oxbp0-p}\nwindexuse{\nwixident{matrix}}{matrix}{NW3gGP3e-1Oxbp0-p}\nwindexuse{\nwixident{numeric}}{numeric}{NW3gGP3e-1Oxbp0-p}\nwindexuse{\nwixident{op}}{op}{NW3gGP3e-1Oxbp0-p}\nwendcode{}\nwbegindocs{578}\nwdocspar
\nwenddocs{}\nwbegincode{579}\sublabel{NW3gGP3e-1Oxbp0-q}\nwmargintag{{\nwtagstyle{}\subpageref{NW3gGP3e-1Oxbp0-q}}}\moddef{cycle.cpp~{\nwtagstyle{}\subpageref{NW3gGP3e-1Oxbp0-1}}}\plusendmoddef\Rm{}\nwstartdeflinemarkup\nwprevnextdefs{NW3gGP3e-1Oxbp0-p}{NW3gGP3e-1Oxbp0-r}\nwenddeflinemarkup
{\bf{}cycle} {\bf{}cycle}::{\it{}sl2\_similarity}({\bf{}const} {\bf{}ex} & {\it{}M}, {\bf{}const} {\bf{}ex} & {\it{}e}, {\bf{}const} {\bf{}ex} & {\it{}sign}, {\bf{}bool} {\it{}not\_inverse}, {\bf{}const} {\bf{}ex} & {\it{}sign\_inv}) {\bf{}const}
{\nwlbrace}
    {\bf{}if} ({\it{}is\_a}\begin{math}<\end{math}{\bf{}matrix}\begin{math}>\end{math}({\it{}M}) \begin{math}\vee\end{math} {\it{}M}.{\it{}info}({\it{}info\_flags}::{\it{}list}))
        {\bf{}return} {\it{}sl2\_similarity}({\it{}M}.{\it{}op}(0), {\it{}M}.{\it{}op}(1), {\it{}M}.{\it{}op}(2), {\it{}M}.{\it{}op}(3), {\it{}e}, {\it{}sign}, {\it{}not\_inverse}, {\it{}sign\_inv});
    {\bf{}else}
        {\bf{}throw}({\it{}std}::{\it{}invalid\_argument}({\tt{}"sl2\_clifford(): expect a list or matrix as the first parameter"}));
{\nwrbrace}

\nwidentuses{\\{{\nwixident{bool}}{bool}}\\{{\nwixident{cycle}}{cycle}}\\{{\nwixident{ex}}{ex}}\\{{\nwixident{matrix}}{matrix}}\\{{\nwixident{op}}{op}}\\{{\nwixident{sl2{\_}similarity}}{sl2:unsimilarity}}}\nwindexuse{\nwixident{bool}}{bool}{NW3gGP3e-1Oxbp0-q}\nwindexuse{\nwixident{cycle}}{cycle}{NW3gGP3e-1Oxbp0-q}\nwindexuse{\nwixident{ex}}{ex}{NW3gGP3e-1Oxbp0-q}\nwindexuse{\nwixident{matrix}}{matrix}{NW3gGP3e-1Oxbp0-q}\nwindexuse{\nwixident{op}}{op}{NW3gGP3e-1Oxbp0-q}\nwindexuse{\nwixident{sl2{\_}similarity}}{sl2:unsimilarity}{NW3gGP3e-1Oxbp0-q}\nwendcode{}\nwbegindocs{580}{\Tt{}\Rm{}{\bf{}cycle}\nwendquote} acts on other {\Tt{}\Rm{}{\bf{}cycle}\nwendquote} by the similarity: \(C: C_1 \mapsto
CC_1C\), see~\cite[~\eqref{E-eq:cycle-conjugation}]{Kisil05a}. If the
metric {\Tt{}\Rm{}{\it{}e}\nwendquote} for similarity is not given, then we use the metric of
\(C_1\) for this.
\nwenddocs{}\nwbegincode{581}\sublabel{NW3gGP3e-1Oxbp0-r}\nwmargintag{{\nwtagstyle{}\subpageref{NW3gGP3e-1Oxbp0-r}}}\moddef{cycle.cpp~{\nwtagstyle{}\subpageref{NW3gGP3e-1Oxbp0-1}}}\plusendmoddef\Rm{}\nwstartdeflinemarkup\nwprevnextdefs{NW3gGP3e-1Oxbp0-q}{NW3gGP3e-1Oxbp0-s}\nwenddeflinemarkup
{\bf{}cycle} {\bf{}cycle}::{\it{}cycle\_similarity}({\bf{}const} {\bf{}cycle} & {\it{}C}, {\bf{}const} {\bf{}ex} & {\it{}e}, {\bf{}const} {\bf{}ex} & {\it{}sign}, {\bf{}const} {\bf{}ex} & {\it{}sign1}, {\bf{}const} {\bf{}ex} & {\it{}sign\_inv}) {\bf{}const}
{\nwlbrace}
//  ex sign\_inv=is\_a\begin{math}<\end{math}matrix\begin{math}>\end{math}(sign)?pow(sign,-1):sign;
 {\bf{}return} {\bf{}cycle}({\it{}ex\_to}\begin{math}<\end{math}{\bf{}matrix}\begin{math}>\end{math}({\it{}canonicalize\_clifford}({\it{}C}.{\it{}mul}({\it{}mul}({\it{}C}, {\it{}e}, {\it{}sign}, {\it{}sign1}.{\it{}is\_zero}()?{\it{}sign}:{\it{}sign1}), {\it{}e}.{\it{}is\_zero}()?{\it{}unit}:{\it{}e}, {\it{}sign1}.{\it{}is\_zero}()?{\it{}sign}:{\it{}sign1}))), {\it{}unit}, {\it{}e}, {\it{}sign\_inv});
{\nwrbrace}

\nwidentuses{\\{{\nwixident{cycle}}{cycle}}\\{{\nwixident{cycle{\_}similarity}}{cycle:unsimilarity}}\\{{\nwixident{ex}}{ex}}\\{{\nwixident{is{\_}zero}}{is:unzero}}\\{{\nwixident{matrix}}{matrix}}\\{{\nwixident{mul}}{mul}}}\nwindexuse{\nwixident{cycle}}{cycle}{NW3gGP3e-1Oxbp0-r}\nwindexuse{\nwixident{cycle{\_}similarity}}{cycle:unsimilarity}{NW3gGP3e-1Oxbp0-r}\nwindexuse{\nwixident{ex}}{ex}{NW3gGP3e-1Oxbp0-r}\nwindexuse{\nwixident{is{\_}zero}}{is:unzero}{NW3gGP3e-1Oxbp0-r}\nwindexuse{\nwixident{matrix}}{matrix}{NW3gGP3e-1Oxbp0-r}\nwindexuse{\nwixident{mul}}{mul}{NW3gGP3e-1Oxbp0-r}\nwendcode{}\nwbegindocs{582}\nwdocspar
\nwenddocs{}\nwbegincode{583}\sublabel{NW3gGP3e-1Oxbp0-s}\nwmargintag{{\nwtagstyle{}\subpageref{NW3gGP3e-1Oxbp0-s}}}\moddef{cycle.cpp~{\nwtagstyle{}\subpageref{NW3gGP3e-1Oxbp0-1}}}\plusendmoddef\Rm{}\nwstartdeflinemarkup\nwprevnextdefs{NW3gGP3e-1Oxbp0-r}{NW3gGP3e-1Oxbp0-t}\nwenddeflinemarkup
{\bf{}ex} {\bf{}cycle}::{\it{}is\_f\_orthogonal}({\bf{}const} {\bf{}cycle} & {\it{}C}, {\bf{}const} {\bf{}ex} & {\it{}e}, {\bf{}const} {\bf{}ex} & {\it{}sign}, {\bf{}const} {\bf{}ex} & {\it{}sign1}, {\bf{}const} {\bf{}ex} & {\it{}sign\_inv}) {\bf{}const}
{\nwlbrace}
    {\bf{}ex} {\it{}ec}={\it{}e}.{\it{}is\_zero}()?{\it{}unit}:{\it{}e};
    {\bf{}ex} {\it{}signc}={\it{}sign1}.{\it{}is\_zero}()?{\it{}sign}:{\it{}sign1};

    {\bf{}return} ({\bf{}cycle}({\it{}ex\_to}\begin{math}<\end{math}{\bf{}matrix}\begin{math}>\end{math}({\it{}canonicalize\_clifford}({\it{}mul}({\it{}C}.{\it{}mul}(\begin{math}\ast\end{math}{\it{}this}, {\it{}ec}, {\it{}sign}, {\it{}signc}), {\it{}ec}, {\it{}signc}))), {\it{}ec}, {\it{}ec}, {\it{}sign\_inv}).{\it{}get\_l}({\it{}get\_dim}()-1).{\it{}normal}() \begin{math}\equiv\end{math} 0);
//  return (C.cycle\_similarity(*this, e, sign, sign1).get\_l(get\_dim()-1).normal() == 0);
{\nwrbrace}

\nwidentuses{\\{{\nwixident{cycle}}{cycle}}\\{{\nwixident{cycle{\_}similarity}}{cycle:unsimilarity}}\\{{\nwixident{ex}}{ex}}\\{{\nwixident{get{\_}dim}}{get:undim}}\\{{\nwixident{get{\_}l}}{get:unl}}\\{{\nwixident{is{\_}f{\_}orthogonal}}{is:unf:unorthogonal}}\\{{\nwixident{is{\_}zero}}{is:unzero}}\\{{\nwixident{matrix}}{matrix}}\\{{\nwixident{mul}}{mul}}\\{{\nwixident{normal}}{normal}}}\nwindexuse{\nwixident{cycle}}{cycle}{NW3gGP3e-1Oxbp0-s}\nwindexuse{\nwixident{cycle{\_}similarity}}{cycle:unsimilarity}{NW3gGP3e-1Oxbp0-s}\nwindexuse{\nwixident{ex}}{ex}{NW3gGP3e-1Oxbp0-s}\nwindexuse{\nwixident{get{\_}dim}}{get:undim}{NW3gGP3e-1Oxbp0-s}\nwindexuse{\nwixident{get{\_}l}}{get:unl}{NW3gGP3e-1Oxbp0-s}\nwindexuse{\nwixident{is{\_}f{\_}orthogonal}}{is:unf:unorthogonal}{NW3gGP3e-1Oxbp0-s}\nwindexuse{\nwixident{is{\_}zero}}{is:unzero}{NW3gGP3e-1Oxbp0-s}\nwindexuse{\nwixident{matrix}}{matrix}{NW3gGP3e-1Oxbp0-s}\nwindexuse{\nwixident{mul}}{mul}{NW3gGP3e-1Oxbp0-s}\nwindexuse{\nwixident{normal}}{normal}{NW3gGP3e-1Oxbp0-s}\nwendcode{}\nwbegindocs{584}\nwdocspar
\subsection[Implementation of the cycle2D class]{Implementation of the {\Tt{}\Rm{}{\bf{}cycle2D}\nwendquote} class}
\label{sec:impl-cycle2D-class}

The derived class {\Tt{}\Rm{}{\bf{}cycle2D}\nwendquote} for two dimensional cycles. Here
constructors, archiving, and comparison come first.
\nwenddocs{}\nwbegincode{585}\sublabel{NW3gGP3e-1Oxbp0-t}\nwmargintag{{\nwtagstyle{}\subpageref{NW3gGP3e-1Oxbp0-t}}}\moddef{cycle.cpp~{\nwtagstyle{}\subpageref{NW3gGP3e-1Oxbp0-1}}}\plusendmoddef\Rm{}\nwstartdeflinemarkup\nwprevnextdefs{NW3gGP3e-1Oxbp0-s}{NW3gGP3e-1Oxbp0-u}\nwenddeflinemarkup
{\bf{}cycle2D}::{\bf{}cycle2D}() : {\it{}inherited}()
{\nwlbrace}
{\bf{}\char35{}if}{\tt{} GINAC\_VERSION\_ATLEAST(1,5)}
{\bf{}\char35{}else}{\tt{}}
 {\it{}tinfo\_key} = &{\bf{}cycle2D}::{\it{}tinfo\_static};
{\bf{}\char35{}endif}{\tt{}}
{\nwrbrace}

\nwidentuses{\\{{\nwixident{cycle2D}}{cycle2D}}\\{{\nwixident{GINAC{\_}VERSION{\_}ATLEAST}}{GINAC:unVERSION:unATLEAST}}}\nwindexuse{\nwixident{cycle2D}}{cycle2D}{NW3gGP3e-1Oxbp0-t}\nwindexuse{\nwixident{GINAC{\_}VERSION{\_}ATLEAST}}{GINAC:unVERSION:unATLEAST}{NW3gGP3e-1Oxbp0-t}\nwendcode{}\nwbegindocs{586}\nwdocspar
\nwenddocs{}\nwbegincode{587}\sublabel{NW3gGP3e-1Oxbp0-u}\nwmargintag{{\nwtagstyle{}\subpageref{NW3gGP3e-1Oxbp0-u}}}\moddef{cycle.cpp~{\nwtagstyle{}\subpageref{NW3gGP3e-1Oxbp0-1}}}\plusendmoddef\Rm{}\nwstartdeflinemarkup\nwprevnextdefs{NW3gGP3e-1Oxbp0-t}{NW3gGP3e-1Oxbp0-v}\nwenddeflinemarkup
{\bf{}cycle2D}::{\bf{}cycle2D}({\bf{}const} {\bf{}ex} & {\it{}k1}, {\bf{}const} {\bf{}ex} & {\it{}l1}, {\bf{}const} {\bf{}ex} & {\it{}m1}, {\bf{}const} {\bf{}ex} & {\it{}metr})
 : {\it{}inherited}({\it{}k1}, {\it{}l1}, {\it{}m1}, {\it{}metr})
{\nwlbrace}
 {\bf{}if} ({\it{}get\_dim}() \begin{math}\neq\end{math} 2)
  {\bf{}throw}({\it{}std}::{\it{}invalid\_argument}({\tt{}"cycle2D::cycle2D(): class cycle2D is defined in two dimensions"}));
{\bf{}\char35{}if}{\tt{} GINAC\_VERSION\_ATLEAST(1,5)}
{\bf{}\char35{}else}{\tt{}}
 {\it{}tinfo\_key} = &{\bf{}cycle2D}::{\it{}tinfo\_static};
{\bf{}\char35{}endif}{\tt{}}
{\nwrbrace}

\nwidentuses{\\{{\nwixident{cycle2D}}{cycle2D}}\\{{\nwixident{ex}}{ex}}\\{{\nwixident{get{\_}dim}}{get:undim}}\\{{\nwixident{GINAC{\_}VERSION{\_}ATLEAST}}{GINAC:unVERSION:unATLEAST}}\\{{\nwixident{metr}}{metr}}}\nwindexuse{\nwixident{cycle2D}}{cycle2D}{NW3gGP3e-1Oxbp0-u}\nwindexuse{\nwixident{ex}}{ex}{NW3gGP3e-1Oxbp0-u}\nwindexuse{\nwixident{get{\_}dim}}{get:undim}{NW3gGP3e-1Oxbp0-u}\nwindexuse{\nwixident{GINAC{\_}VERSION{\_}ATLEAST}}{GINAC:unVERSION:unATLEAST}{NW3gGP3e-1Oxbp0-u}\nwindexuse{\nwixident{metr}}{metr}{NW3gGP3e-1Oxbp0-u}\nwendcode{}\nwbegindocs{588}\nwdocspar
\nwenddocs{}\nwbegincode{589}\sublabel{NW3gGP3e-1Oxbp0-v}\nwmargintag{{\nwtagstyle{}\subpageref{NW3gGP3e-1Oxbp0-v}}}\moddef{cycle.cpp~{\nwtagstyle{}\subpageref{NW3gGP3e-1Oxbp0-1}}}\plusendmoddef\Rm{}\nwstartdeflinemarkup\nwprevnextdefs{NW3gGP3e-1Oxbp0-u}{NW3gGP3e-1Oxbp0-w}\nwenddeflinemarkup
{\bf{}cycle2D}::{\bf{}cycle2D}({\bf{}const} {\bf{}lst} & {\it{}l}, {\bf{}const} {\bf{}ex} & {\it{}r\_squared}, {\bf{}const} {\bf{}ex} & {\it{}metr}, {\bf{}const} {\bf{}ex} & {\it{}e}, {\bf{}const} {\bf{}ex} & {\it{}sign})
 : {\it{}inherited}({\it{}l}, {\it{}r\_squared}, {\it{}metr}, {\it{}e}, {\it{}sign})
{\nwlbrace}
 {\bf{}if} ({\it{}get\_dim}() \begin{math}\neq\end{math} 2)
  {\bf{}throw}({\it{}std}::{\it{}invalid\_argument}({\tt{}"cycle2D::cycle2D(): class cycle2D is defined in two dimensions"}));
{\bf{}\char35{}if}{\tt{} GINAC\_VERSION\_ATLEAST(1,5)}
{\bf{}\char35{}else}{\tt{}}
 {\it{}tinfo\_key} = &{\bf{}cycle2D}::{\it{}tinfo\_static};
{\bf{}\char35{}endif}{\tt{}}
{\nwrbrace}

\nwidentuses{\\{{\nwixident{cycle2D}}{cycle2D}}\\{{\nwixident{ex}}{ex}}\\{{\nwixident{get{\_}dim}}{get:undim}}\\{{\nwixident{GINAC{\_}VERSION{\_}ATLEAST}}{GINAC:unVERSION:unATLEAST}}\\{{\nwixident{l}}{l}}\\{{\nwixident{lst}}{lst}}\\{{\nwixident{metr}}{metr}}}\nwindexuse{\nwixident{cycle2D}}{cycle2D}{NW3gGP3e-1Oxbp0-v}\nwindexuse{\nwixident{ex}}{ex}{NW3gGP3e-1Oxbp0-v}\nwindexuse{\nwixident{get{\_}dim}}{get:undim}{NW3gGP3e-1Oxbp0-v}\nwindexuse{\nwixident{GINAC{\_}VERSION{\_}ATLEAST}}{GINAC:unVERSION:unATLEAST}{NW3gGP3e-1Oxbp0-v}\nwindexuse{\nwixident{l}}{l}{NW3gGP3e-1Oxbp0-v}\nwindexuse{\nwixident{lst}}{lst}{NW3gGP3e-1Oxbp0-v}\nwindexuse{\nwixident{metr}}{metr}{NW3gGP3e-1Oxbp0-v}\nwendcode{}\nwbegindocs{590}\nwdocspar
\nwenddocs{}\nwbegincode{591}\sublabel{NW3gGP3e-1Oxbp0-w}\nwmargintag{{\nwtagstyle{}\subpageref{NW3gGP3e-1Oxbp0-w}}}\moddef{cycle.cpp~{\nwtagstyle{}\subpageref{NW3gGP3e-1Oxbp0-1}}}\plusendmoddef\Rm{}\nwstartdeflinemarkup\nwprevnextdefs{NW3gGP3e-1Oxbp0-v}{NW3gGP3e-1Oxbp0-x}\nwenddeflinemarkup
{\bf{}cycle2D}::{\bf{}cycle2D}({\bf{}const} {\bf{}matrix} & {\it{}M}, {\bf{}const} {\bf{}ex} & {\it{}metr}, {\bf{}const} {\bf{}ex} & {\it{}e}, {\bf{}const} {\bf{}ex} & {\it{}sign})
 : {\it{}inherited}({\it{}M}, {\it{}metr}, {\it{}e}, {\it{}sign})
{\nwlbrace}
 {\bf{}if} ({\it{}get\_dim}() \begin{math}\neq\end{math} 2)
  {\bf{}throw}({\it{}std}::{\it{}invalid\_argument}({\tt{}"cycle2D::cycle2D(): class cycle2D is defined in two dimensions"}));
{\bf{}\char35{}if}{\tt{} GINAC\_VERSION\_ATLEAST(1,5)}
{\bf{}\char35{}else}{\tt{}}
 {\it{}tinfo\_key} = &{\bf{}cycle2D}::{\it{}tinfo\_static};
{\bf{}\char35{}endif}{\tt{}}
{\nwrbrace}

\nwidentuses{\\{{\nwixident{cycle2D}}{cycle2D}}\\{{\nwixident{ex}}{ex}}\\{{\nwixident{get{\_}dim}}{get:undim}}\\{{\nwixident{GINAC{\_}VERSION{\_}ATLEAST}}{GINAC:unVERSION:unATLEAST}}\\{{\nwixident{matrix}}{matrix}}\\{{\nwixident{metr}}{metr}}}\nwindexuse{\nwixident{cycle2D}}{cycle2D}{NW3gGP3e-1Oxbp0-w}\nwindexuse{\nwixident{ex}}{ex}{NW3gGP3e-1Oxbp0-w}\nwindexuse{\nwixident{get{\_}dim}}{get:undim}{NW3gGP3e-1Oxbp0-w}\nwindexuse{\nwixident{GINAC{\_}VERSION{\_}ATLEAST}}{GINAC:unVERSION:unATLEAST}{NW3gGP3e-1Oxbp0-w}\nwindexuse{\nwixident{matrix}}{matrix}{NW3gGP3e-1Oxbp0-w}\nwindexuse{\nwixident{metr}}{metr}{NW3gGP3e-1Oxbp0-w}\nwendcode{}\nwbegindocs{592}\nwdocspar
\nwenddocs{}\nwbegincode{593}\sublabel{NW3gGP3e-1Oxbp0-x}\nwmargintag{{\nwtagstyle{}\subpageref{NW3gGP3e-1Oxbp0-x}}}\moddef{cycle.cpp~{\nwtagstyle{}\subpageref{NW3gGP3e-1Oxbp0-1}}}\plusendmoddef\Rm{}\nwstartdeflinemarkup\nwprevnextdefs{NW3gGP3e-1Oxbp0-w}{NW3gGP3e-1Oxbp0-y}\nwenddeflinemarkup
{\bf{}cycle2D}::{\bf{}cycle2D}({\bf{}const} {\bf{}cycle} & {\it{}C}, {\bf{}const} {\bf{}ex} & {\it{}metr})
{\nwlbrace}
 (\begin{math}\ast\end{math}{\it{}this}) = {\bf{}cycle2D}({\it{}C}.{\it{}get\_k}(), {\it{}C}.{\it{}get\_l}(), {\it{}C}.{\it{}get\_m}(), ({\it{}metr}.{\it{}is\_zero}()? {\it{}C}.{\it{}get\_unit}(): {\it{}metr}));
{\nwrbrace}

\nwidentuses{\\{{\nwixident{cycle}}{cycle}}\\{{\nwixident{cycle2D}}{cycle2D}}\\{{\nwixident{ex}}{ex}}\\{{\nwixident{get{\_}k}}{get:unk}}\\{{\nwixident{get{\_}l}}{get:unl}}\\{{\nwixident{get{\_}m}}{get:unm}}\\{{\nwixident{get{\_}unit}}{get:ununit}}\\{{\nwixident{is{\_}zero}}{is:unzero}}\\{{\nwixident{metr}}{metr}}}\nwindexuse{\nwixident{cycle}}{cycle}{NW3gGP3e-1Oxbp0-x}\nwindexuse{\nwixident{cycle2D}}{cycle2D}{NW3gGP3e-1Oxbp0-x}\nwindexuse{\nwixident{ex}}{ex}{NW3gGP3e-1Oxbp0-x}\nwindexuse{\nwixident{get{\_}k}}{get:unk}{NW3gGP3e-1Oxbp0-x}\nwindexuse{\nwixident{get{\_}l}}{get:unl}{NW3gGP3e-1Oxbp0-x}\nwindexuse{\nwixident{get{\_}m}}{get:unm}{NW3gGP3e-1Oxbp0-x}\nwindexuse{\nwixident{get{\_}unit}}{get:ununit}{NW3gGP3e-1Oxbp0-x}\nwindexuse{\nwixident{is{\_}zero}}{is:unzero}{NW3gGP3e-1Oxbp0-x}\nwindexuse{\nwixident{metr}}{metr}{NW3gGP3e-1Oxbp0-x}\nwendcode{}\nwbegindocs{594}\nwdocspar
\nwenddocs{}\nwbegincode{595}\sublabel{NW3gGP3e-1Oxbp0-y}\nwmargintag{{\nwtagstyle{}\subpageref{NW3gGP3e-1Oxbp0-y}}}\moddef{cycle.cpp~{\nwtagstyle{}\subpageref{NW3gGP3e-1Oxbp0-1}}}\plusendmoddef\Rm{}\nwstartdeflinemarkup\nwprevnextdefs{NW3gGP3e-1Oxbp0-x}{NW3gGP3e-1Oxbp0-z}\nwenddeflinemarkup
{\bf{}void} {\bf{}cycle2D}::{\it{}archive}({\it{}archive\_node} &{\it{}n}) {\bf{}const}\nwindexdefn{\nwixident{cycle2D}}{cycle2D}{NW3gGP3e-1Oxbp0-y}
{\nwlbrace}
    {\it{}inherited}::{\it{}archive}({\it{}n});
{\nwrbrace}

//cycle2D::cycle2D(const archive\_node &n, lst &sym\_lst) : inherited(n, sym\_lst) {\nwlbrace}; {\nwrbrace}

{\bf{}void} {\bf{}cycle2D}::{\it{}read\_archive}({\bf{}const} {\it{}archive\_node} &{\it{}n}, {\bf{}lst} &{\it{}sym\_lst})\nwindexdefn{\nwixident{cycle2D}}{cycle2D}{NW3gGP3e-1Oxbp0-y}
{\nwlbrace}
    {\it{}inherited}::{\it{}read\_archive}({\it{}n}, {\it{}sym\_lst});
{\nwrbrace}
{\it{}GINAC\_BIND\_UNARCHIVER}({\bf{}cycle2D});\nwindexdefn{\nwixident{cycle2D}}{cycle2D}{NW3gGP3e-1Oxbp0-y}

{\bf{}int} {\bf{}cycle2D}::{\it{}compare\_same\_type}({\bf{}const} {\bf{}basic} &{\it{}other}) {\bf{}const}\nwindexdefn{\nwixident{cycle2D}}{cycle2D}{NW3gGP3e-1Oxbp0-y}
{\nwlbrace}
       {\it{}GINAC\_ASSERT}({\it{}is\_a}\begin{math}<\end{math}{\bf{}cycle2D}\begin{math}>\end{math}({\it{}other}));
    {\bf{}return} {\it{}inherited}::{\it{}compare\_same\_type}({\it{}other});
{\nwrbrace}

//const char *cycle2D::get\_class\_name() {\nwlbrace} return "cycle2D"; {\nwrbrace}

\nwidentdefs{\\{{\nwixident{cycle2D}}{cycle2D}}}\nwidentuses{\\{{\nwixident{lst}}{lst}}}\nwindexuse{\nwixident{lst}}{lst}{NW3gGP3e-1Oxbp0-y}\nwendcode{}\nwbegindocs{596}Real and imaginary part of the representing vector.
\nwenddocs{}\nwbegincode{597}\sublabel{NW3gGP3e-1Oxbp0-z}\nwmargintag{{\nwtagstyle{}\subpageref{NW3gGP3e-1Oxbp0-z}}}\moddef{cycle.cpp~{\nwtagstyle{}\subpageref{NW3gGP3e-1Oxbp0-1}}}\plusendmoddef\Rm{}\nwstartdeflinemarkup\nwprevnextdefs{NW3gGP3e-1Oxbp0-y}{NW3gGP3e-1Oxbp0-10}\nwenddeflinemarkup
{\bf{}ex} {\bf{}cycle2D}::{\it{}real\_part}() {\bf{}const}
{\nwlbrace}
    {\bf{}return} {\bf{}cycle2D}({\it{}k}.{\it{}real\_part}(),{\bf{}lst}({\it{}get\_l}(0).{\it{}real\_part}(),{\it{}get\_l}(1).{\it{}real\_part}()),{\it{}m}.{\it{}real\_part}(),{\it{}unit});
{\nwrbrace}

{\bf{}ex} {\bf{}cycle2D}::{\it{}imag\_part}() {\bf{}const}
{\nwlbrace}
    {\bf{}return} {\bf{}cycle2D}({\it{}k}.{\it{}imag\_part}(),{\bf{}lst}({\it{}get\_l}(0).{\it{}imag\_part}(),{\it{}get\_l}(1).{\it{}imag\_part}()),{\it{}m}.{\it{}imag\_part}(),{\it{}unit});
{\nwrbrace}

\nwidentuses{\\{{\nwixident{cycle2D}}{cycle2D}}\\{{\nwixident{ex}}{ex}}\\{{\nwixident{get{\_}l}}{get:unl}}\\{{\nwixident{k}}{k}}\\{{\nwixident{lst}}{lst}}\\{{\nwixident{m}}{m}}}\nwindexuse{\nwixident{cycle2D}}{cycle2D}{NW3gGP3e-1Oxbp0-z}\nwindexuse{\nwixident{ex}}{ex}{NW3gGP3e-1Oxbp0-z}\nwindexuse{\nwixident{get{\_}l}}{get:unl}{NW3gGP3e-1Oxbp0-z}\nwindexuse{\nwixident{k}}{k}{NW3gGP3e-1Oxbp0-z}\nwindexuse{\nwixident{lst}}{lst}{NW3gGP3e-1Oxbp0-z}\nwindexuse{\nwixident{m}}{m}{NW3gGP3e-1Oxbp0-z}\nwendcode{}\nwbegindocs{598}\nwdocspar
\subsubsection[The member functions of the derived class cycle2D]{The member functions of the derived class {\Tt{}\Rm{}{\bf{}cycle2D}\nwendquote}}
\label{sec:memb-funct-deriv}

The standard definition of the focus for a parabola is
\begin{displaymath}
  \left(\frac{l}{k}, \frac{m}{2n} - \frac{l^2}{2nk} + \frac{n}{2k}\right).
\end{displaymath}
We calculate focus of a cycle based on its determinant in the
corresponding metric.
\nwenddocs{}\nwbegincode{599}\sublabel{NW3gGP3e-1Oxbp0-10}\nwmargintag{{\nwtagstyle{}\subpageref{NW3gGP3e-1Oxbp0-10}}}\moddef{cycle.cpp~{\nwtagstyle{}\subpageref{NW3gGP3e-1Oxbp0-1}}}\plusendmoddef\Rm{}\nwstartdeflinemarkup\nwprevnextdefs{NW3gGP3e-1Oxbp0-z}{NW3gGP3e-1Oxbp0-11}\nwenddeflinemarkup
{\bf{}ex} {\bf{}cycle2D}::{\it{}focus}({\bf{}const} {\bf{}ex} & {\it{}e}, {\bf{}bool} {\it{}return\_matrix}) {\bf{}const}
{\nwlbrace}
    {\bf{}lst} {\it{}f}={\bf{}lst}(//jump\_fnct(-get\_metric(varidx(0, 2), varidx(0, 2)))*
        {\it{}get\_l}(0)\begin{math}\div\end{math}{\it{}k},
        (-{\it{}det}({\it{}e})\begin{math}\div\end{math}({\bf{}numeric}(2)\begin{math}\ast\end{math}{\it{}get\_l}(1)\begin{math}\ast\end{math}{\it{}k})).{\it{}normal}());
    {\bf{}return} ({\it{}return\_matrix}? ({\bf{}ex}){\bf{}matrix}(2, 1, {\it{}f}) : ({\bf{}ex}){\it{}f});
{\nwrbrace}

\nwidentuses{\\{{\nwixident{bool}}{bool}}\\{{\nwixident{cycle2D}}{cycle2D}}\\{{\nwixident{det}}{det}}\\{{\nwixident{ex}}{ex}}\\{{\nwixident{focus}}{focus}}\\{{\nwixident{get{\_}l}}{get:unl}}\\{{\nwixident{get{\_}metric}}{get:unmetric}}\\{{\nwixident{jump{\_}fnct}}{jump:unfnct}}\\{{\nwixident{k}}{k}}\\{{\nwixident{lst}}{lst}}\\{{\nwixident{matrix}}{matrix}}\\{{\nwixident{normal}}{normal}}\\{{\nwixident{numeric}}{numeric}}\\{{\nwixident{varidx}}{varidx}}}\nwindexuse{\nwixident{bool}}{bool}{NW3gGP3e-1Oxbp0-10}\nwindexuse{\nwixident{cycle2D}}{cycle2D}{NW3gGP3e-1Oxbp0-10}\nwindexuse{\nwixident{det}}{det}{NW3gGP3e-1Oxbp0-10}\nwindexuse{\nwixident{ex}}{ex}{NW3gGP3e-1Oxbp0-10}\nwindexuse{\nwixident{focus}}{focus}{NW3gGP3e-1Oxbp0-10}\nwindexuse{\nwixident{get{\_}l}}{get:unl}{NW3gGP3e-1Oxbp0-10}\nwindexuse{\nwixident{get{\_}metric}}{get:unmetric}{NW3gGP3e-1Oxbp0-10}\nwindexuse{\nwixident{jump{\_}fnct}}{jump:unfnct}{NW3gGP3e-1Oxbp0-10}\nwindexuse{\nwixident{k}}{k}{NW3gGP3e-1Oxbp0-10}\nwindexuse{\nwixident{lst}}{lst}{NW3gGP3e-1Oxbp0-10}\nwindexuse{\nwixident{matrix}}{matrix}{NW3gGP3e-1Oxbp0-10}\nwindexuse{\nwixident{normal}}{normal}{NW3gGP3e-1Oxbp0-10}\nwindexuse{\nwixident{numeric}}{numeric}{NW3gGP3e-1Oxbp0-10}\nwindexuse{\nwixident{varidx}}{varidx}{NW3gGP3e-1Oxbp0-10}\nwendcode{}\nwbegindocs{600}\nwdocspar
\nwenddocs{}\nwbegincode{601}\sublabel{NW3gGP3e-1Oxbp0-11}\nwmargintag{{\nwtagstyle{}\subpageref{NW3gGP3e-1Oxbp0-11}}}\moddef{cycle.cpp~{\nwtagstyle{}\subpageref{NW3gGP3e-1Oxbp0-1}}}\plusendmoddef\Rm{}\nwstartdeflinemarkup\nwprevnextdefs{NW3gGP3e-1Oxbp0-10}{NW3gGP3e-1Oxbp0-12}\nwenddeflinemarkup
{\bf{}lst} {\bf{}cycle2D}::{\it{}roots}({\bf{}const} {\bf{}ex} & {\it{}y}, {\bf{}bool} {\it{}first}) {\bf{}const}
{\nwlbrace}
 {\bf{}ex} {\it{}D} = {\it{}get\_dim}();
 {\bf{}lst} {\it{}k\_sign} = {\bf{}lst}(-{\it{}k}\begin{math}\ast\end{math}{\it{}get\_metric}({\bf{}varidx}(0, {\it{}D}), {\bf{}varidx}(0, {\it{}D})), -{\it{}k}\begin{math}\ast\end{math}{\it{}get\_metric}({\bf{}varidx}(1, {\it{}D}), {\bf{}varidx}(1, {\it{}D})));
 {\bf{}int} {\it{}i0} = ({\it{}first}?0:1), {\it{}i1} = ({\it{}first}?1:0);
 {\bf{}ex} {\it{}c} = {\it{}k\_sign}.{\it{}op}({\it{}i1})\begin{math}\ast\end{math}{\it{}pow}({\it{}y}, 2) - {\bf{}numeric}(2)\begin{math}\ast\end{math}{\it{}get\_l}({\it{}i1})\begin{math}\ast\end{math}{\it{}y}+{\it{}m};
 {\bf{}if} ({\it{}k\_sign}.{\it{}op}({\it{}i0}).{\it{}is\_zero}())
  {\bf{}return} ({\it{}get\_l}({\it{}i0}).{\it{}is\_zero}() ? {\bf{}lst}() : {\bf{}lst}({\it{}c}\begin{math}\div\end{math}{\it{}get\_l}({\it{}i0})\begin{math}\div\end{math}{\bf{}numeric}(2)));
 {\bf{}else} {\nwlbrace}
  {\bf{}ex} {\it{}disc} = {\it{}sqrt}({\it{}pow}({\it{}get\_l}({\it{}i0}), 2) - {\it{}k\_sign}.{\it{}op}({\it{}i0})\begin{math}\ast\end{math}{\it{}c});
  {\bf{}return} {\bf{}lst}(({\it{}get\_l}({\it{}i0})-{\it{}disc})\begin{math}\div\end{math}{\it{}k\_sign}.{\it{}op}({\it{}i0}), ({\it{}get\_l}({\it{}i0})+{\it{}disc})\begin{math}\div\end{math}{\it{}k\_sign}.{\it{}op}({\it{}i0}));
 {\nwrbrace}
{\nwrbrace}

\nwidentuses{\\{{\nwixident{bool}}{bool}}\\{{\nwixident{cycle2D}}{cycle2D}}\\{{\nwixident{ex}}{ex}}\\{{\nwixident{get{\_}dim}}{get:undim}}\\{{\nwixident{get{\_}l}}{get:unl}}\\{{\nwixident{get{\_}metric}}{get:unmetric}}\\{{\nwixident{is{\_}zero}}{is:unzero}}\\{{\nwixident{k}}{k}}\\{{\nwixident{lst}}{lst}}\\{{\nwixident{m}}{m}}\\{{\nwixident{numeric}}{numeric}}\\{{\nwixident{op}}{op}}\\{{\nwixident{roots}}{roots}}\\{{\nwixident{varidx}}{varidx}}}\nwindexuse{\nwixident{bool}}{bool}{NW3gGP3e-1Oxbp0-11}\nwindexuse{\nwixident{cycle2D}}{cycle2D}{NW3gGP3e-1Oxbp0-11}\nwindexuse{\nwixident{ex}}{ex}{NW3gGP3e-1Oxbp0-11}\nwindexuse{\nwixident{get{\_}dim}}{get:undim}{NW3gGP3e-1Oxbp0-11}\nwindexuse{\nwixident{get{\_}l}}{get:unl}{NW3gGP3e-1Oxbp0-11}\nwindexuse{\nwixident{get{\_}metric}}{get:unmetric}{NW3gGP3e-1Oxbp0-11}\nwindexuse{\nwixident{is{\_}zero}}{is:unzero}{NW3gGP3e-1Oxbp0-11}\nwindexuse{\nwixident{k}}{k}{NW3gGP3e-1Oxbp0-11}\nwindexuse{\nwixident{lst}}{lst}{NW3gGP3e-1Oxbp0-11}\nwindexuse{\nwixident{m}}{m}{NW3gGP3e-1Oxbp0-11}\nwindexuse{\nwixident{numeric}}{numeric}{NW3gGP3e-1Oxbp0-11}\nwindexuse{\nwixident{op}}{op}{NW3gGP3e-1Oxbp0-11}\nwindexuse{\nwixident{roots}}{roots}{NW3gGP3e-1Oxbp0-11}\nwindexuse{\nwixident{varidx}}{varidx}{NW3gGP3e-1Oxbp0-11}\nwendcode{}\nwbegindocs{602}\nwdocspar
\nwenddocs{}\nwbegincode{603}\sublabel{NW3gGP3e-1Oxbp0-12}\nwmargintag{{\nwtagstyle{}\subpageref{NW3gGP3e-1Oxbp0-12}}}\moddef{cycle.cpp~{\nwtagstyle{}\subpageref{NW3gGP3e-1Oxbp0-1}}}\plusendmoddef\Rm{}\nwstartdeflinemarkup\nwprevnextdefs{NW3gGP3e-1Oxbp0-11}{NW3gGP3e-1Oxbp0-13}\nwenddeflinemarkup
{\bf{}lst} {\bf{}cycle2D}::{\it{}line\_intersect}({\bf{}const} {\bf{}ex} & {\it{}a}, {\bf{}const} {\bf{}ex} & {\it{}b}) {\bf{}const}
{\nwlbrace}
    {\bf{}ex} {\it{}D} = {\it{}get\_dim}();
    {\bf{}ex} {\it{}pm} = -{\it{}k}\begin{math}\ast\end{math}{\it{}get\_metric}({\bf{}varidx}(1, {\it{}D}), {\bf{}varidx}(1, {\it{}D}));
    {\bf{}return} {\bf{}cycle2D}({\it{}k}\begin{math}\ast\end{math}({\bf{}numeric}(1)+{\it{}pm}\begin{math}\ast\end{math}{\it{}pow}({\it{}a},2)).{\it{}normal}(),
                   {\bf{}lst}(({\it{}get\_l}(0)+{\it{}get\_l}(1)\begin{math}\ast\end{math}{\it{}a}-{\it{}pm}\begin{math}\ast\end{math}{\it{}a}\begin{math}\ast\end{math}{\it{}b}).{\it{}normal}(), 0), 
                   ({\it{}m}-{\bf{}numeric}(2)\begin{math}\ast\end{math}{\it{}get\_l}(1)\begin{math}\ast\end{math}{\it{}b}+{\it{}pm}\begin{math}\ast\end{math}{\it{}pow}({\it{}b},2)).{\it{}normal}()).{\it{}roots}();
 {\nwrbrace}

\nwidentuses{\\{{\nwixident{cycle2D}}{cycle2D}}\\{{\nwixident{ex}}{ex}}\\{{\nwixident{get{\_}dim}}{get:undim}}\\{{\nwixident{get{\_}l}}{get:unl}}\\{{\nwixident{get{\_}metric}}{get:unmetric}}\\{{\nwixident{k}}{k}}\\{{\nwixident{line{\_}intersect}}{line:unintersect}}\\{{\nwixident{lst}}{lst}}\\{{\nwixident{m}}{m}}\\{{\nwixident{normal}}{normal}}\\{{\nwixident{numeric}}{numeric}}\\{{\nwixident{roots}}{roots}}\\{{\nwixident{varidx}}{varidx}}}\nwindexuse{\nwixident{cycle2D}}{cycle2D}{NW3gGP3e-1Oxbp0-12}\nwindexuse{\nwixident{ex}}{ex}{NW3gGP3e-1Oxbp0-12}\nwindexuse{\nwixident{get{\_}dim}}{get:undim}{NW3gGP3e-1Oxbp0-12}\nwindexuse{\nwixident{get{\_}l}}{get:unl}{NW3gGP3e-1Oxbp0-12}\nwindexuse{\nwixident{get{\_}metric}}{get:unmetric}{NW3gGP3e-1Oxbp0-12}\nwindexuse{\nwixident{k}}{k}{NW3gGP3e-1Oxbp0-12}\nwindexuse{\nwixident{line{\_}intersect}}{line:unintersect}{NW3gGP3e-1Oxbp0-12}\nwindexuse{\nwixident{lst}}{lst}{NW3gGP3e-1Oxbp0-12}\nwindexuse{\nwixident{m}}{m}{NW3gGP3e-1Oxbp0-12}\nwindexuse{\nwixident{normal}}{normal}{NW3gGP3e-1Oxbp0-12}\nwindexuse{\nwixident{numeric}}{numeric}{NW3gGP3e-1Oxbp0-12}\nwindexuse{\nwixident{roots}}{roots}{NW3gGP3e-1Oxbp0-12}\nwindexuse{\nwixident{varidx}}{varidx}{NW3gGP3e-1Oxbp0-12}\nwendcode{}\nwbegindocs{604}\nwdocspar
\subsubsection[Drawing cycle2D]{Drawing {\Tt{}\Rm{}{\bf{}cycle2D}\nwendquote}}
\label{sec:drawing-cycle2d}

Some auxilliary functions used for drawing
\nwenddocs{}\nwbegincode{605}\sublabel{NW3gGP3e-1Oxbp0-13}\nwmargintag{{\nwtagstyle{}\subpageref{NW3gGP3e-1Oxbp0-13}}}\moddef{cycle.cpp~{\nwtagstyle{}\subpageref{NW3gGP3e-1Oxbp0-1}}}\plusendmoddef\Rm{}\nwstartdeflinemarkup\nwprevnextdefs{NW3gGP3e-1Oxbp0-12}{NW3gGP3e-1Oxbp0-14}\nwenddeflinemarkup
{\bf{}inline} {\bf{}ex} {\it{}max}({\bf{}const} {\bf{}ex} &{\it{}a}, {\bf{}const} {\bf{}ex} &{\it{}b}) {\nwlbrace}{\bf{}return} {\it{}ex\_to}\begin{math}<\end{math}{\bf{}numeric}\begin{math}>\end{math}(({\it{}a}-{\it{}b}).{\it{}evalf}()).{\it{}is\_positive}()?{\it{}a}:{\it{}b};{\nwrbrace}
{\bf{}inline} {\bf{}ex} {\it{}min}({\bf{}const} {\bf{}ex} &{\it{}a}, {\bf{}const} {\bf{}ex} &{\it{}b}) {\nwlbrace}{\bf{}return} {\it{}ex\_to}\begin{math}<\end{math}{\bf{}numeric}\begin{math}>\end{math}(({\it{}a}-{\it{}b}).{\it{}evalf}()).{\it{}is\_positive}()?{\it{}b}:{\it{}a};{\nwrbrace}

\nwidentuses{\\{{\nwixident{ex}}{ex}}\\{{\nwixident{numeric}}{numeric}}}\nwindexuse{\nwixident{ex}}{ex}{NW3gGP3e-1Oxbp0-13}\nwindexuse{\nwixident{numeric}}{numeric}{NW3gGP3e-1Oxbp0-13}\nwendcode{}\nwbegindocs{606}The most complicated member function in the class {\Tt{}\Rm{}{\bf{}cycle2D}\nwendquote}
\nwenddocs{}\nwbegincode{607}\sublabel{NW3gGP3e-1Oxbp0-14}\nwmargintag{{\nwtagstyle{}\subpageref{NW3gGP3e-1Oxbp0-14}}}\moddef{cycle.cpp~{\nwtagstyle{}\subpageref{NW3gGP3e-1Oxbp0-1}}}\plusendmoddef\Rm{}\nwstartdeflinemarkup\nwprevnextdefs{NW3gGP3e-1Oxbp0-13}{NW3gGP3e-1Oxbp0-15}\nwenddeflinemarkup
{\bf{}\char35{}define}{\tt{} DRAW\_ARC(X, S)   u = X; \begin{math}\backslash\end{math}}\nwindexdefn{\nwixident{DRAW{\_}ARC}}{DRAW:unARC}{NW3gGP3e-1Oxbp0-14}
    {\it{}v} = {\it{}ex\_to}\begin{math}<\end{math}{\bf{}numeric}\begin{math}>\end{math}({\it{}Cf}.{\it{}roots}({\it{}X}, \begin{math}\neg\end{math}{\it{}not\_swapped}).{\it{}op}({\it{}zero\_or\_one}).{\it{}evalf}()).{\it{}to\_double}(); \begin{math}\backslash\end{math}
    {\it{}du} = {\it{}dir}\begin{math}\ast\end{math}(-{\it{}k\_d}\begin{math}\ast\end{math}{\it{}signv}\begin{math}\ast\end{math}{\it{}v}+{\it{}lv});    \begin{math}\backslash\end{math}
    {\it{}dv} = {\it{}dir}\begin{math}\ast\end{math}({\it{}k\_d}\begin{math}\ast\end{math}{\it{}signu}\begin{math}\ast\end{math}{\it{}u}-{\it{}lu});        \begin{math}\backslash\end{math}
    {\bf{}if} ({\it{}not\_swapped})            \begin{math}\backslash\end{math}
     {\it{}ost} \begin{math}\ll\end{math} {\it{}S} \begin{math}\ll\end{math}  {\it{}u} \begin{math}\ll\end{math} {\tt{}","} \begin{math}\ll\end{math} {\it{}v} \begin{math}\ll\end{math} {\tt{}"){\char123}"} \begin{math}\ll\end{math} {\it{}du} \begin{math}\ll\end{math} {\tt{}","} \begin{math}\ll\end{math} {\it{}dv} \begin{math}\ll\end{math} {\tt{}"{\char125}"}; \begin{math}\backslash\end{math}
    {\bf{}else}                \begin{math}\backslash\end{math}
     {\it{}ost} \begin{math}\ll\end{math} {\it{}S} \begin{math}\ll\end{math}  {\it{}v} \begin{math}\ll\end{math} {\tt{}","} \begin{math}\ll\end{math} {\it{}u} \begin{math}\ll\end{math} {\tt{}"){\char123}"} \begin{math}\ll\end{math} ({\it{}sign} \begin{math}\equiv\end{math} 0? {\it{}dv} : -{\it{}dv}) \begin{math}\ll\end{math} {\tt{}","} \begin{math}\ll\end{math} ({\it{}sign} \begin{math}\equiv\end{math} 0? {\it{}du} : -{\it{}du}) \begin{math}\ll\end{math} {\tt{}"{\char125}"};

\nwidentdefs{\\{{\nwixident{DRAW{\_}ARC}}{DRAW:unARC}}}\nwidentuses{\\{{\nwixident{du}}{du}}\\{{\nwixident{dv}}{dv}}\\{{\nwixident{k{\_}d}}{k:und}}\\{{\nwixident{numeric}}{numeric}}\\{{\nwixident{op}}{op}}\\{{\nwixident{roots}}{roots}}\\{{\nwixident{u}}{u}}\\{{\nwixident{v}}{v}}\\{{\nwixident{zero{\_}or{\_}one}}{zero:unor:unone}}}\nwindexuse{\nwixident{du}}{du}{NW3gGP3e-1Oxbp0-14}\nwindexuse{\nwixident{dv}}{dv}{NW3gGP3e-1Oxbp0-14}\nwindexuse{\nwixident{k{\_}d}}{k:und}{NW3gGP3e-1Oxbp0-14}\nwindexuse{\nwixident{numeric}}{numeric}{NW3gGP3e-1Oxbp0-14}\nwindexuse{\nwixident{op}}{op}{NW3gGP3e-1Oxbp0-14}\nwindexuse{\nwixident{roots}}{roots}{NW3gGP3e-1Oxbp0-14}\nwindexuse{\nwixident{u}}{u}{NW3gGP3e-1Oxbp0-14}\nwindexuse{\nwixident{v}}{v}{NW3gGP3e-1Oxbp0-14}\nwindexuse{\nwixident{zero{\_}or{\_}one}}{zero:unor:unone}{NW3gGP3e-1Oxbp0-14}\nwendcode{}\nwbegindocs{608}an auxillary function to find small numbers
\nwenddocs{}\nwbegincode{609}\sublabel{NW3gGP3e-1Oxbp0-15}\nwmargintag{{\nwtagstyle{}\subpageref{NW3gGP3e-1Oxbp0-15}}}\moddef{cycle.cpp~{\nwtagstyle{}\subpageref{NW3gGP3e-1Oxbp0-1}}}\plusendmoddef\Rm{}\nwstartdeflinemarkup\nwprevnextdefs{NW3gGP3e-1Oxbp0-14}{NW3gGP3e-1Oxbp0-16}\nwenddeflinemarkup
{\bf{}bool} {\it{}is\_almost\_zero}({\bf{}const} {\bf{}ex} & {\it{}x})
{\nwlbrace}
    {\bf{}if} ({\it{}is\_a}\begin{math}<\end{math}{\bf{}numeric}\begin{math}>\end{math}({\it{}x}))
        {\bf{}return} ({\it{}abs}({\it{}ex\_to}\begin{math}<\end{math}{\bf{}numeric}\begin{math}>\end{math}({\it{}x}).{\it{}to\_double}()) \begin{math}<\end{math} 0.0000000001);
    {\bf{}else}
        {\bf{}return} {\it{}x}.{\it{}is\_zero}();
{\nwrbrace}

\nwidentuses{\\{{\nwixident{bool}}{bool}}\\{{\nwixident{ex}}{ex}}\\{{\nwixident{is{\_}zero}}{is:unzero}}\\{{\nwixident{numeric}}{numeric}}}\nwindexuse{\nwixident{bool}}{bool}{NW3gGP3e-1Oxbp0-15}\nwindexuse{\nwixident{ex}}{ex}{NW3gGP3e-1Oxbp0-15}\nwindexuse{\nwixident{is{\_}zero}}{is:unzero}{NW3gGP3e-1Oxbp0-15}\nwindexuse{\nwixident{numeric}}{numeric}{NW3gGP3e-1Oxbp0-15}\nwendcode{}\nwbegindocs{610}an auxillary function to find almost numbers
\nwenddocs{}\nwbegincode{611}\sublabel{NW3gGP3e-1Oxbp0-16}\nwmargintag{{\nwtagstyle{}\subpageref{NW3gGP3e-1Oxbp0-16}}}\moddef{cycle.cpp~{\nwtagstyle{}\subpageref{NW3gGP3e-1Oxbp0-1}}}\plusendmoddef\Rm{}\nwstartdeflinemarkup\nwprevnextdefs{NW3gGP3e-1Oxbp0-15}{NW3gGP3e-1Oxbp0-17}\nwenddeflinemarkup
{\bf{}bool} {\it{}is\_almost\_negative}({\bf{}const} {\bf{}ex} & {\it{}x})
{\nwlbrace}
    {\bf{}if} ({\it{}is\_a}\begin{math}<\end{math}{\bf{}numeric}\begin{math}>\end{math}({\it{}x}))
        {\bf{}return} ({\it{}ex\_to}\begin{math}<\end{math}{\bf{}numeric}\begin{math}>\end{math}({\it{}x}.{\it{}evalf}()).{\it{}to\_double}() \begin{math}<\end{math} 0.0000000001);
    {\bf{}else}
        {\bf{}return} {\it{}x}.{\it{}is\_zero}();
{\nwrbrace}

\nwidentuses{\\{{\nwixident{bool}}{bool}}\\{{\nwixident{ex}}{ex}}\\{{\nwixident{is{\_}zero}}{is:unzero}}\\{{\nwixident{numeric}}{numeric}}}\nwindexuse{\nwixident{bool}}{bool}{NW3gGP3e-1Oxbp0-16}\nwindexuse{\nwixident{ex}}{ex}{NW3gGP3e-1Oxbp0-16}\nwindexuse{\nwixident{is{\_}zero}}{is:unzero}{NW3gGP3e-1Oxbp0-16}\nwindexuse{\nwixident{numeric}}{numeric}{NW3gGP3e-1Oxbp0-16}\nwendcode{}\nwbegindocs{612}The main drawing routine for {\Tt{}\Rm{}{\bf{}cycle2D}\nwendquote}.
\nwenddocs{}\nwbegincode{613}\sublabel{NW3gGP3e-1Oxbp0-17}\nwmargintag{{\nwtagstyle{}\subpageref{NW3gGP3e-1Oxbp0-17}}}\moddef{cycle.cpp~{\nwtagstyle{}\subpageref{NW3gGP3e-1Oxbp0-1}}}\plusendmoddef\Rm{}\nwstartdeflinemarkup\nwprevnextdefs{NW3gGP3e-1Oxbp0-16}{NW3gGP3e-1Oxbp0-18}\nwenddeflinemarkup
{\bf{}void} {\bf{}cycle2D}::{\it{}metapost\_draw}({\it{}ostream} & {\it{}ost}, {\bf{}const} {\bf{}ex} & {\it{}xmin}, {\bf{}const} {\bf{}ex} & {\it{}xmax}, {\bf{}const} {\bf{}ex} & {\it{}ymin}, {\bf{}const} {\bf{}ex} & {\it{}ymax},\nwindexdefn{\nwixident{cycle2D}}{cycle2D}{NW3gGP3e-1Oxbp0-17}
 {\bf{}const} {\bf{}lst} & {\it{}color}, {\bf{}const} {\it{}string} {\it{}more\_options}, {\bf{}bool} {\it{}with\_header},
 {\bf{}int} {\it{}points\_per\_arc}, {\bf{}bool} {\it{}asymptote}, {\bf{}const} {\it{}string} {\it{}picture}, {\bf{}bool} {\it{}only\_path},
 {\bf{}bool} {\it{}is\_continuation}, {\bf{}const} {\it{}string} {\it{}imaginary\_options}) {\bf{}const}
{\nwlbrace}
 {\bf{}ex} {\it{}D} = {\it{}get\_dim}();
 {\it{}ostringstream} {\it{}draw\_start}, {\it{}draw\_options};
 {\it{}string} {\it{}already\_drawn} =({\it{}is\_continuation}? {\tt{}"^^("} : {\tt{}"("} ); // Was any arc already drawn?
{\it{}draw\_start} \begin{math}\ll\end{math} {\tt{}"draw"} \begin{math}\ll\end{math} ({\it{}asymptote} ? {\tt{}"("} : {\tt{}" "}) \begin{math}\ll\end{math} {\it{}picture} \begin{math}\ll\end{math} ({\it{}picture}.{\it{}size}()\begin{math}\equiv\end{math}0? {\tt{}""} : {\tt{}","}) \begin{math}\ll\end{math} {\tt{}"("};
 {\it{}ios\_base}::{\it{}fmtflags} {\it{}keep\_flags} = {\it{}ost}.{\it{}flags}(); // Keep stream's flags to be restored on the exit
 {\it{}draw\_options}.{\it{}flags}({\it{}keep\_flags}); // Synchronise flags between the streams
 {\it{}draw\_options}.{\it{}precision}({\it{}ost}.{\it{}precision}()); // Synchronise flags between the streams

\nwidentdefs{\\{{\nwixident{cycle2D}}{cycle2D}}}\nwidentuses{\\{{\nwixident{bool}}{bool}}\\{{\nwixident{ex}}{ex}}\\{{\nwixident{get{\_}dim}}{get:undim}}\\{{\nwixident{lst}}{lst}}\\{{\nwixident{metapost{\_}draw}}{metapost:undraw}}\\{{\nwixident{string}}{string}}}\nwindexuse{\nwixident{bool}}{bool}{NW3gGP3e-1Oxbp0-17}\nwindexuse{\nwixident{ex}}{ex}{NW3gGP3e-1Oxbp0-17}\nwindexuse{\nwixident{get{\_}dim}}{get:undim}{NW3gGP3e-1Oxbp0-17}\nwindexuse{\nwixident{lst}}{lst}{NW3gGP3e-1Oxbp0-17}\nwindexuse{\nwixident{metapost{\_}draw}}{metapost:undraw}{NW3gGP3e-1Oxbp0-17}\nwindexuse{\nwixident{string}}{string}{NW3gGP3e-1Oxbp0-17}\nwendcode{}\nwbegindocs{614}Each drawing command is concluded by options containing color,
etc. They are formatted differently for \Asymptote\ and \MetaPost.
\nwenddocs{}\nwbegincode{615}\sublabel{NW3gGP3e-1Oxbp0-18}\nwmargintag{{\nwtagstyle{}\subpageref{NW3gGP3e-1Oxbp0-18}}}\moddef{cycle.cpp~{\nwtagstyle{}\subpageref{NW3gGP3e-1Oxbp0-1}}}\plusendmoddef\Rm{}\nwstartdeflinemarkup\nwprevnextdefs{NW3gGP3e-1Oxbp0-17}{NW3gGP3e-1Oxbp0-19}\nwenddeflinemarkup
 {\it{}ost} \begin{math}\ll\end{math} {\it{}fixed};
 {\it{}draw\_options} \begin{math}\ll\end{math} {\it{}fixed};
 {\bf{}if} ({\it{}color}.{\it{}nops}() \begin{math}\equiv\end{math} 3) {\nwlbrace}
  {\bf{}if} ({\it{}asymptote})
   {\it{}draw\_options}  \begin{math}\ll\end{math} {\tt{}",rgb("}
       \begin{math}\ll\end{math} {\it{}ex\_to}\begin{math}<\end{math}{\bf{}numeric}\begin{math}>\end{math}({\it{}color}.{\it{}op}(0)).{\it{}to\_double}() \begin{math}\ll\end{math} {\tt{}","}
       \begin{math}\ll\end{math} {\it{}ex\_to}\begin{math}<\end{math}{\bf{}numeric}\begin{math}>\end{math}({\it{}color}.{\it{}op}(1)).{\it{}to\_double}() \begin{math}\ll\end{math}{\tt{}","}
       \begin{math}\ll\end{math} {\it{}ex\_to}\begin{math}<\end{math}{\bf{}numeric}\begin{math}>\end{math}({\it{}color}.{\it{}op}(2)).{\it{}to\_double}() \begin{math}\ll\end{math} {\tt{}")"};
  {\bf{}else}
   {\it{}draw\_options}  \begin{math}\ll\end{math} {\it{}showpos} \begin{math}\ll\end{math} {\tt{}" withcolor "}
       \begin{math}\ll\end{math} {\it{}ex\_to}\begin{math}<\end{math}{\bf{}numeric}\begin{math}>\end{math}({\it{}color}.{\it{}op}(0)).{\it{}to\_double}() \begin{math}\ll\end{math} {\tt{}"*red"}
       \begin{math}\ll\end{math} {\it{}ex\_to}\begin{math}<\end{math}{\bf{}numeric}\begin{math}>\end{math}({\it{}color}.{\it{}op}(1)).{\it{}to\_double}() \begin{math}\ll\end{math}{\tt{}"*green"}
       \begin{math}\ll\end{math} {\it{}ex\_to}\begin{math}<\end{math}{\bf{}numeric}\begin{math}>\end{math}({\it{}color}.{\it{}op}(2)).{\it{}to\_double}() \begin{math}\ll\end{math} {\tt{}"*blue "};
 {\nwrbrace}
 {\bf{}if} ({\it{}more\_options} \begin{math}\neq\end{math} {\tt{}""}) {\nwlbrace}
     {\bf{}if} ({\it{}color}.{\it{}nops}() \begin{math}\equiv\end{math} 3)
         {\it{}draw\_options} \begin{math}\ll\end{math} {\tt{}"+"};
     {\bf{}else}
         {\it{}draw\_options} \begin{math}\ll\end{math} {\tt{}","};
     {\it{}draw\_options} \begin{math}\ll\end{math} {\it{}more\_options};
 {\nwrbrace}
 {\it{}draw\_options} \begin{math}\ll\end{math} ({\it{}asymptote} ? {\tt{}");"} : {\tt{}";"}) \begin{math}\ll\end{math} {\it{}endl};

\nwidentuses{\\{{\nwixident{nops}}{nops}}\\{{\nwixident{numeric}}{numeric}}\\{{\nwixident{op}}{op}}}\nwindexuse{\nwixident{nops}}{nops}{NW3gGP3e-1Oxbp0-18}\nwindexuse{\nwixident{numeric}}{numeric}{NW3gGP3e-1Oxbp0-18}\nwindexuse{\nwixident{op}}{op}{NW3gGP3e-1Oxbp0-18}\nwendcode{}\nwbegindocs{616}A drawing command can be also preceded by a human-readable comment
describing the cycle to be drawn.
\nwenddocs{}\nwbegincode{617}\sublabel{NW3gGP3e-1Oxbp0-19}\nwmargintag{{\nwtagstyle{}\subpageref{NW3gGP3e-1Oxbp0-19}}}\moddef{cycle.cpp~{\nwtagstyle{}\subpageref{NW3gGP3e-1Oxbp0-1}}}\plusendmoddef\Rm{}\nwstartdeflinemarkup\nwprevnextdefs{NW3gGP3e-1Oxbp0-18}{NW3gGP3e-1Oxbp0-1A}\nwenddeflinemarkup
{\bf{}if} ({\it{}with\_header}) {\nwlbrace}
  {\it{}ost} \begin{math}\ll\end{math} ({\it{}asymptote} ? {\tt{}"// Asymptote"} : {\tt{}"
   \begin{math}\ll\end{math} {\it{}xmax} \begin{math}\ll\end{math} {\tt{}"]x["} \begin{math}\ll\end{math} {\it{}ymin} \begin{math}\ll\end{math} {\tt{}","}
      \begin{math}\ll\end{math} {\it{}ymax} \begin{math}\ll\end{math} {\tt{}"] for "};

 {\it{}ostringstream} {\it{}equat};
 {\it{}equat} \begin{math}\ll\end{math} ({\bf{}ex}){\it{}passing}({\bf{}lst}({\bf{}symbol}({\tt{}"u"}), {\bf{}symbol}({\tt{}"v"})));
 {\bf{}if} ({\it{}equat}.{\it{}str}().{\it{}length}()\begin{math}<\end{math} 256)
     {\it{}ost} \begin{math}\ll\end{math} {\it{}equat}.{\it{}str}();
 {\bf{}else}
     {\it{}ost} \begin{math}\ll\end{math} {\tt{}" [approx.] "} \begin{math}\ll\end{math} ({\bf{}ex}){\it{}evalf}().{\it{}passing}({\bf{}lst}({\bf{}symbol}({\tt{}"u"}), {\bf{}symbol}({\tt{}"v"})));
 {\nwrbrace}

 {\bf{}if} ({\it{}k}.{\it{}is\_zero}() \begin{math}\wedge\end{math} {\it{}l}.{\it{}subs}({\it{}l}.{\it{}op}(1) \begin{math}\equiv\end{math} 0).{\it{}is\_zero}() \begin{math}\wedge\end{math} {\it{}l}.{\it{}subs}({\it{}l}.{\it{}op}(1) \begin{math}\equiv\end{math} 1).{\it{}is\_zero}() \begin{math}\wedge\end{math} \begin{math}\backslash\end{math}
{\it{}m}.{\it{}is\_zero}()) {\nwlbrace}
  {\it{}ost} \begin{math}\ll\end{math} {\tt{}" zero cycle, (whole plane) "} \begin{math}\ll\end{math} {\it{}endl};
  {\it{}ost}.{\it{}flags}({\it{}keep\_flags});
  {\bf{}return};
 {\nwrbrace}

\nwidentuses{\\{{\nwixident{cycle}}{cycle}}\\{{\nwixident{ex}}{ex}}\\{{\nwixident{is{\_}zero}}{is:unzero}}\\{{\nwixident{k}}{k}}\\{{\nwixident{l}}{l}}\\{{\nwixident{lst}}{lst}}\\{{\nwixident{m}}{m}}\\{{\nwixident{op}}{op}}\\{{\nwixident{passing}}{passing}}\\{{\nwixident{subs}}{subs}}\\{{\nwixident{u}}{u}}\\{{\nwixident{v}}{v}}}\nwindexuse{\nwixident{cycle}}{cycle}{NW3gGP3e-1Oxbp0-19}\nwindexuse{\nwixident{ex}}{ex}{NW3gGP3e-1Oxbp0-19}\nwindexuse{\nwixident{is{\_}zero}}{is:unzero}{NW3gGP3e-1Oxbp0-19}\nwindexuse{\nwixident{k}}{k}{NW3gGP3e-1Oxbp0-19}\nwindexuse{\nwixident{l}}{l}{NW3gGP3e-1Oxbp0-19}\nwindexuse{\nwixident{lst}}{lst}{NW3gGP3e-1Oxbp0-19}\nwindexuse{\nwixident{m}}{m}{NW3gGP3e-1Oxbp0-19}\nwindexuse{\nwixident{op}}{op}{NW3gGP3e-1Oxbp0-19}\nwindexuse{\nwixident{passing}}{passing}{NW3gGP3e-1Oxbp0-19}\nwindexuse{\nwixident{subs}}{subs}{NW3gGP3e-1Oxbp0-19}\nwindexuse{\nwixident{u}}{u}{NW3gGP3e-1Oxbp0-19}\nwindexuse{\nwixident{v}}{v}{NW3gGP3e-1Oxbp0-19}\nwendcode{}\nwbegindocs{618}There are several parameters which control the output. Their values
depend from either we draw {\Tt{}\Rm{}{\bf{}cycle}\nwendquote} in the original coordinates or
swap the {\Tt{}\Rm{}{\it{}u}\nwendquote} and {\Tt{}\Rm{}{\it{}v}\nwendquote}
\nwenddocs{}\nwbegincode{619}\sublabel{NW3gGP3e-1Oxbp0-1A}\nwmargintag{{\nwtagstyle{}\subpageref{NW3gGP3e-1Oxbp0-1A}}}\moddef{cycle.cpp~{\nwtagstyle{}\subpageref{NW3gGP3e-1Oxbp0-1}}}\plusendmoddef\Rm{}\nwstartdeflinemarkup\nwprevnextdefs{NW3gGP3e-1Oxbp0-19}{NW3gGP3e-1Oxbp0-1B}\nwenddeflinemarkup
    {\bf{}cycle2D} {\it{}Cf}={\it{}ex\_to}\begin{math}<\end{math}{\bf{}cycle2D}\begin{math}>\end{math}({\it{}evalf}().{\it{}normalize}());
    {\bf{}double}  {\it{}xc} = {\it{}ex\_to}\begin{math}<\end{math}{\bf{}numeric}\begin{math}>\end{math}({\it{}Cf}.{\it{}center}().{\it{}op}(0)).{\it{}to\_double}(),
        {\it{}yc} = {\it{}ex\_to}\begin{math}<\end{math}{\bf{}numeric}\begin{math}>\end{math}({\it{}Cf}.{\it{}center}().{\it{}op}(1)).{\it{}to\_double}(); // the center of cycle
    {\bf{}double} {\it{}sign0} = {\it{}ex\_to}\begin{math}<\end{math}{\bf{}numeric}\begin{math}>\end{math}(-{\it{}get\_metric}({\bf{}varidx}(0, {\it{}D}), {\bf{}varidx}(0, {\it{}D})).{\it{}evalf}()).{\it{}to\_double}(),
    {\it{}sign1} = {\it{}ex\_to}\begin{math}<\end{math}{\bf{}numeric}\begin{math}>\end{math}(-{\it{}get\_metric}({\bf{}varidx}(1, {\it{}D}), {\bf{}varidx}(1, {\it{}D})).{\it{}evalf}()).{\it{}to\_double}(),
    {\it{}sign} = {\it{}sign0} \begin{math}\ast\end{math} {\it{}sign1};
    {\bf{}double} {\it{}determinant} = {\it{}ex\_to}\begin{math}<\end{math}{\bf{}numeric}\begin{math}>\end{math}({\it{}Cf}.{\it{}radius\_sq}()).{\it{}to\_double}(),
        {\it{}r}={\it{}sqrt}({\it{}abs}({\it{}determinant}));
    {\bf{}double} {\it{}epsilon}=0.0000000001;
    {\bf{}bool} {\it{}not\_swapped} = ({\it{}sign}\begin{math}>\end{math}0 \begin{math}\vee\end{math} {\it{}sign1}\begin{math}\equiv\end{math}0 \begin{math}\vee\end{math} (({\it{}sign} \begin{math}<\end{math}0) \begin{math}\wedge\end{math} ({\it{}determinant} \begin{math}<\end{math} {\it{}epsilon})));
    {\bf{}double} {\it{}signu} = ({\it{}not\_swapped}?{\it{}sign0}:{\it{}sign1}), {\it{}signv} = ({\it{}not\_swapped}?{\it{}sign1}:{\it{}sign0});
    {\bf{}int} {\it{}iu} = ({\it{}not\_swapped}?0:1), {\it{}iv} = ({\it{}not\_swapped}?1:0);
    {\bf{}double} {\it{}umin} =  {\it{}ex\_to}\begin{math}<\end{math}{\bf{}numeric}\begin{math}>\end{math}(({\it{}not\_swapped} ? {\it{}xmin} : {\it{}ymin}).{\it{}evalf}()).{\it{}to\_double}(),
        {\it{}umax} =  {\it{}ex\_to}\begin{math}<\end{math}{\bf{}numeric}\begin{math}>\end{math}(({\it{}not\_swapped} ? {\it{}xmax} : {\it{}ymax}).{\it{}evalf}()).{\it{}to\_double}(),
        {\it{}vmin} =  {\it{}ex\_to}\begin{math}<\end{math}{\bf{}numeric}\begin{math}>\end{math}(({\it{}not\_swapped} ? {\it{}ymin}:  {\it{}xmin}).{\it{}evalf}()).{\it{}to\_double}(),
        {\it{}vmax} =  {\it{}ex\_to}\begin{math}<\end{math}{\bf{}numeric}\begin{math}>\end{math}(({\it{}not\_swapped} ? {\it{}ymax} : {\it{}xmax}).{\it{}evalf}()).{\it{}to\_double}(),
        {\it{}uc} =  ({\it{}not\_swapped} ? {\it{}xc}:  {\it{}yc}), {\it{}vc} =  ({\it{}not\_swapped} ? {\it{}yc} : {\it{}xc});
    {\bf{}lst} {\it{}b\_roots} = {\it{}ex\_to}\begin{math}<\end{math}{\bf{}lst}\begin{math}>\end{math}({\it{}Cf}.{\it{}roots}({\it{}vmin}, {\it{}not\_swapped}).{\it{}evalf}()),
        {\it{}t\_roots} = {\it{}ex\_to}\begin{math}<\end{math}{\bf{}lst}\begin{math}>\end{math}({\it{}Cf}.{\it{}roots}({\it{}vmax}, {\it{}not\_swapped}).{\it{}evalf}());

\nwidentuses{\\{{\nwixident{bool}}{bool}}\\{{\nwixident{center}}{center}}\\{{\nwixident{cycle}}{cycle}}\\{{\nwixident{cycle2D}}{cycle2D}}\\{{\nwixident{get{\_}metric}}{get:unmetric}}\\{{\nwixident{lst}}{lst}}\\{{\nwixident{normalize}}{normalize}}\\{{\nwixident{numeric}}{numeric}}\\{{\nwixident{op}}{op}}\\{{\nwixident{radius{\_}sq}}{radius:unsq}}\\{{\nwixident{roots}}{roots}}\\{{\nwixident{varidx}}{varidx}}}\nwindexuse{\nwixident{bool}}{bool}{NW3gGP3e-1Oxbp0-1A}\nwindexuse{\nwixident{center}}{center}{NW3gGP3e-1Oxbp0-1A}\nwindexuse{\nwixident{cycle}}{cycle}{NW3gGP3e-1Oxbp0-1A}\nwindexuse{\nwixident{cycle2D}}{cycle2D}{NW3gGP3e-1Oxbp0-1A}\nwindexuse{\nwixident{get{\_}metric}}{get:unmetric}{NW3gGP3e-1Oxbp0-1A}\nwindexuse{\nwixident{lst}}{lst}{NW3gGP3e-1Oxbp0-1A}\nwindexuse{\nwixident{normalize}}{normalize}{NW3gGP3e-1Oxbp0-1A}\nwindexuse{\nwixident{numeric}}{numeric}{NW3gGP3e-1Oxbp0-1A}\nwindexuse{\nwixident{op}}{op}{NW3gGP3e-1Oxbp0-1A}\nwindexuse{\nwixident{radius{\_}sq}}{radius:unsq}{NW3gGP3e-1Oxbp0-1A}\nwindexuse{\nwixident{roots}}{roots}{NW3gGP3e-1Oxbp0-1A}\nwindexuse{\nwixident{varidx}}{varidx}{NW3gGP3e-1Oxbp0-1A}\nwendcode{}\nwbegindocs{620}Here is the outline of the rest of the method. It effectively splits
into several cases depending from the space metric and degeneracy of
{\Tt{}\Rm{}{\bf{}cycle2D}\nwendquote}.
\nwenddocs{}\nwbegincode{621}\sublabel{NW3gGP3e-1Oxbp0-1B}\nwmargintag{{\nwtagstyle{}\subpageref{NW3gGP3e-1Oxbp0-1B}}}\moddef{cycle.cpp~{\nwtagstyle{}\subpageref{NW3gGP3e-1Oxbp0-1}}}\plusendmoddef\Rm{}\nwstartdeflinemarkup\nwprevnextdefs{NW3gGP3e-1Oxbp0-1A}{NW3gGP3e-1Oxbp0-1C}\nwenddeflinemarkup
    \LA{}Imaginary coefficients~{\nwtagstyle{}\subpageref{NW3gGP3e-5M2Zk-1}}\RA{}
    \LA{}Draw a straight line~{\nwtagstyle{}\subpageref{NW3gGP3e-1iDH4d-1}}\RA{}
    \LA{}Find intersection points with the boundary~{\nwtagstyle{}\subpageref{NW3gGP3e-2elNFX-1}}\RA{}
    {\bf{}if} ({\it{}sign} \begin{math}>\end{math} 0) {\nwlbrace} // elliptic metric
        \LA{}Draw a circle~{\nwtagstyle{}\subpageref{NW3gGP3e-4WQ5O5-1}}\RA{}
            {\nwrbrace} {\bf{}else} {\nwlbrace} // parabolic or hyperbolic  metric
        \LA{}Draw a parabola or hyperbola~{\nwtagstyle{}\subpageref{NW3gGP3e-4H4ZFN-1}}\RA{}
            {\nwrbrace}
{\it{}ost} \begin{math}\ll\end{math} {\it{}endl};
{\it{}ost}.{\it{}flags}({\it{}keep\_flags});
{\bf{}return};
{\nwrbrace}

\nwendcode{}\nwbegindocs{622}If line is detected we identify its visible portion.
\nwenddocs{}\nwbegincode{623}\sublabel{NW3gGP3e-1iDH4d-1}\nwmargintag{{\nwtagstyle{}\subpageref{NW3gGP3e-1iDH4d-1}}}\moddef{Draw a straight line~{\nwtagstyle{}\subpageref{NW3gGP3e-1iDH4d-1}}}\endmoddef\Rm{}\nwstartdeflinemarkup\nwusesondefline{\\{NW3gGP3e-1Oxbp0-1B}}\nwprevnextdefs{\relax}{NW3gGP3e-1iDH4d-2}\nwenddeflinemarkup
{\bf{}if} ({\it{}b\_roots}.{\it{}nops}() \begin{math}\neq\end{math} 2) {\nwlbrace} // a linear object
    {\bf{}if} ({\it{}Cf}.{\it{}get\_k}().{\it{}is\_zero}() \begin{math}\wedge\end{math} {\it{}Cf}.{\it{}get\_l}(0).{\it{}is\_zero}() \begin{math}\wedge\end{math} {\it{}Cf}.{\it{}get\_l}(1).{\it{}is\_zero}()) {\nwlbrace}
        {\bf{}if} ({\it{}with\_header})
            {\it{}ost} \begin{math}\ll\end{math} {\tt{}" the zero-radius cycle at infinity"} \begin{math}\ll\end{math} {\it{}endl};
        {\bf{}return};
    {\nwrbrace}
    {\bf{}if} ({\it{}with\_header})
        {\it{}ost} \begin{math}\ll\end{math} {\tt{}" (straight line)"} \begin{math}\ll\end{math} {\it{}endl};
    {\bf{}double} {\it{}u1}, {\it{}u2}, {\it{}v1}, {\it{}v2};
    {\bf{}if} ({\it{}b\_roots}.{\it{}nops}() \begin{math}\equiv\end{math} 1){\nwlbrace} // a "non-horisontal" line
        {\it{}u1} = {\it{}std}::{\it{}max}({\it{}std}::{\it{}min}({\it{}ex\_to}\begin{math}<\end{math}{\bf{}numeric}\begin{math}>\end{math}({\it{}b\_roots}.{\it{}op}(0)).{\it{}to\_double}(), {\it{}umax}), {\it{}umin});
        {\it{}u2} = {\it{}std}::{\it{}min}({\it{}std}::{\it{}max}({\it{}ex\_to}\begin{math}<\end{math}{\bf{}numeric}\begin{math}>\end{math}({\it{}t\_roots}.{\it{}op}(0)).{\it{}to\_double}(), {\it{}umin}), {\it{}umax});
    {\nwrbrace} {\bf{}else} {\nwlbrace} // a "horisontal" line
        {\it{}u1} = {\it{}umin};
        {\it{}u2} = {\it{}umax};
    {\nwrbrace}

\nwalsodefined{\\{NW3gGP3e-1iDH4d-2}\\{NW3gGP3e-1iDH4d-3}\\{NW3gGP3e-1iDH4d-4}}\nwused{\\{NW3gGP3e-1Oxbp0-1B}}\nwidentuses{\\{{\nwixident{cycle}}{cycle}}\\{{\nwixident{get{\_}k}}{get:unk}}\\{{\nwixident{get{\_}l}}{get:unl}}\\{{\nwixident{is{\_}zero}}{is:unzero}}\\{{\nwixident{nops}}{nops}}\\{{\nwixident{numeric}}{numeric}}\\{{\nwixident{op}}{op}}}\nwindexuse{\nwixident{cycle}}{cycle}{NW3gGP3e-1iDH4d-1}\nwindexuse{\nwixident{get{\_}k}}{get:unk}{NW3gGP3e-1iDH4d-1}\nwindexuse{\nwixident{get{\_}l}}{get:unl}{NW3gGP3e-1iDH4d-1}\nwindexuse{\nwixident{is{\_}zero}}{is:unzero}{NW3gGP3e-1iDH4d-1}\nwindexuse{\nwixident{nops}}{nops}{NW3gGP3e-1iDH4d-1}\nwindexuse{\nwixident{numeric}}{numeric}{NW3gGP3e-1iDH4d-1}\nwindexuse{\nwixident{op}}{op}{NW3gGP3e-1iDH4d-1}\nwendcode{}\nwbegindocs{624}Vertical lines case.
\nwenddocs{}\nwbegincode{625}\sublabel{NW3gGP3e-1iDH4d-2}\nwmargintag{{\nwtagstyle{}\subpageref{NW3gGP3e-1iDH4d-2}}}\moddef{Draw a straight line~{\nwtagstyle{}\subpageref{NW3gGP3e-1iDH4d-1}}}\plusendmoddef\Rm{}\nwstartdeflinemarkup\nwusesondefline{\\{NW3gGP3e-1Oxbp0-1B}}\nwprevnextdefs{NW3gGP3e-1iDH4d-1}{NW3gGP3e-1iDH4d-3}\nwenddeflinemarkup
{\bf{}if} ({\it{}Cf}.{\it{}get\_l}({\it{}iv}).{\it{}is\_zero}()) {\nwlbrace} // a vertical line
    {\bf{}if} ({\it{}ex\_to}\begin{math}<\end{math}{\bf{}numeric}\begin{math}>\end{math}({\it{}b\_roots}.{\it{}op}(0)- {\it{}umin}).{\it{}to\_double}() \begin{math}>\end{math} -{\it{}epsilon}
        \begin{math}\wedge\end{math} {\it{}ex\_to}\begin{math}<\end{math}{\bf{}numeric}\begin{math}>\end{math}({\it{}umax}-{\it{}b\_roots}.{\it{}op}(0)).{\it{}to\_double}() \begin{math}>\end{math} -{\it{}epsilon} ) {\nwlbrace}
        {\it{}v1} =  {\it{}vmin};
        {\it{}v2} =  {\it{}vmax};
    {\nwrbrace} {\bf{}else} {\nwlbrace} // out of scope
        {\it{}ost}.{\it{}flags}({\it{}keep\_flags});
        {\bf{}return};
    {\nwrbrace}

\nwused{\\{NW3gGP3e-1Oxbp0-1B}}\nwidentuses{\\{{\nwixident{get{\_}l}}{get:unl}}\\{{\nwixident{is{\_}zero}}{is:unzero}}\\{{\nwixident{numeric}}{numeric}}\\{{\nwixident{op}}{op}}}\nwindexuse{\nwixident{get{\_}l}}{get:unl}{NW3gGP3e-1iDH4d-2}\nwindexuse{\nwixident{is{\_}zero}}{is:unzero}{NW3gGP3e-1iDH4d-2}\nwindexuse{\nwixident{numeric}}{numeric}{NW3gGP3e-1iDH4d-2}\nwindexuse{\nwixident{op}}{op}{NW3gGP3e-1iDH4d-2}\nwendcode{}\nwbegindocs{626}Look for the visible portion of generic line.
\nwenddocs{}\nwbegincode{627}\sublabel{NW3gGP3e-1iDH4d-3}\nwmargintag{{\nwtagstyle{}\subpageref{NW3gGP3e-1iDH4d-3}}}\moddef{Draw a straight line~{\nwtagstyle{}\subpageref{NW3gGP3e-1iDH4d-1}}}\plusendmoddef\Rm{}\nwstartdeflinemarkup\nwusesondefline{\\{NW3gGP3e-1Oxbp0-1B}}\nwprevnextdefs{NW3gGP3e-1iDH4d-2}{NW3gGP3e-1iDH4d-4}\nwenddeflinemarkup
{\nwrbrace} {\bf{}else} {\nwlbrace}
    {\it{}v1} = {\it{}ex\_to}\begin{math}<\end{math}{\bf{}numeric}\begin{math}>\end{math}({\it{}Cf}.{\it{}roots}({\it{}u1}, \begin{math}\neg\end{math}{\it{}not\_swapped}).{\it{}op}(0)).{\it{}to\_double}();
    {\it{}v2} = {\it{}ex\_to}\begin{math}<\end{math}{\bf{}numeric}\begin{math}>\end{math}({\it{}Cf}.{\it{}roots}({\it{}u2}, \begin{math}\neg\end{math}{\it{}not\_swapped}).{\it{}op}(0)).{\it{}to\_double}();
    {\bf{}if} (({\it{}std}::{\it{}max}({\it{}v1}, {\it{}v2})-{\it{}vmax} \begin{math}>\end{math} {\it{}epsilon}) \begin{math}\vee\end{math} ({\it{}std}::{\it{}min}({\it{}v1}, {\it{}v2}) -{\it{}vmin} \begin{math}<\end{math} -{\it{}epsilon} )) {\nwlbrace}
        {\it{}ost}.{\it{}flags}({\it{}keep\_flags});
        {\bf{}return}; //out of scope
    {\nwrbrace}
{\nwrbrace}

\nwused{\\{NW3gGP3e-1Oxbp0-1B}}\nwidentuses{\\{{\nwixident{numeric}}{numeric}}\\{{\nwixident{op}}{op}}\\{{\nwixident{roots}}{roots}}}\nwindexuse{\nwixident{numeric}}{numeric}{NW3gGP3e-1iDH4d-3}\nwindexuse{\nwixident{op}}{op}{NW3gGP3e-1iDH4d-3}\nwindexuse{\nwixident{roots}}{roots}{NW3gGP3e-1iDH4d-3}\nwendcode{}\nwbegindocs{628}Actual drawing of the line.
\nwenddocs{}\nwbegincode{629}\sublabel{NW3gGP3e-1iDH4d-4}\nwmargintag{{\nwtagstyle{}\subpageref{NW3gGP3e-1iDH4d-4}}}\moddef{Draw a straight line~{\nwtagstyle{}\subpageref{NW3gGP3e-1iDH4d-1}}}\plusendmoddef\Rm{}\nwstartdeflinemarkup\nwusesondefline{\\{NW3gGP3e-1Oxbp0-1B}}\nwprevnextdefs{NW3gGP3e-1iDH4d-3}{\relax}\nwenddeflinemarkup
    {\it{}ost} \begin{math}\ll\end{math}  ({\it{}only\_path} ? {\it{}already\_drawn} : {\it{}draw\_start}.{\it{}str}())
        \begin{math}\ll\end{math} ({\it{}not\_swapped}? {\it{}u1}: {\it{}v1}) \begin{math}\ll\end{math} {\tt{}","} \begin{math}\ll\end{math} ({\it{}not\_swapped} ? {\it{}v1}: {\it{}u1})
        \begin{math}\ll\end{math} {\tt{}")--("} \begin{math}\ll\end{math} ({\it{}not\_swapped} ? {\it{}u2}: {\it{}v2}) \begin{math}\ll\end{math} {\tt{}","} \begin{math}\ll\end{math} ({\it{}not\_swapped} ? {\it{}v2}: {\it{}u2}) \begin{math}\ll\end{math} {\tt{}")"}
        \begin{math}\ll\end{math} ({\it{}only\_path} ? {\tt{}""} : {\it{}draw\_options}.{\it{}str}());
{\it{}already\_drawn}={\tt{}"^^("};
{\bf{}if} ({\it{}with\_header})
    {\it{}ost} \begin{math}\ll\end{math} {\it{}endl};
{\it{}ost}.{\it{}flags}({\it{}keep\_flags});
{\bf{}return};
{\nwrbrace}

\nwused{\\{NW3gGP3e-1Oxbp0-1B}}\nwendcode{}\nwbegindocs{630}Make initially this intervals (left[i], right[i]) irrelevant for
drawing by default, if necessary, it will be redefined letter on.
\nwenddocs{}\nwbegincode{631}\sublabel{NW3gGP3e-2elNFX-1}\nwmargintag{{\nwtagstyle{}\subpageref{NW3gGP3e-2elNFX-1}}}\moddef{Find intersection points with the boundary~{\nwtagstyle{}\subpageref{NW3gGP3e-2elNFX-1}}}\endmoddef\Rm{}\nwstartdeflinemarkup\nwusesondefline{\\{NW3gGP3e-1Oxbp0-1B}}\nwenddeflinemarkup
{\bf{}double} {\it{}left}[2] = {\nwlbrace}{\it{}std}::{\it{}max}({\it{}std}::{\it{}min}({\it{}uc}, {\it{}umax}), {\it{}umin}),\nwindexdefn{\nwixident{left}}{left}{NW3gGP3e-2elNFX-1}
                   {\it{}std}::{\it{}max}({\it{}std}::{\it{}min}({\it{}uc}, {\it{}umax}), {\it{}umin}){\nwrbrace},
    {\it{}right}[2] = {\nwlbrace}{\it{}std}::{\it{}max}({\it{}std}::{\it{}min}({\it{}uc}, {\it{}umax}), {\it{}umin}),
                {\it{}std}::{\it{}max}({\it{}std}::{\it{}min}({\it{}uc}, {\it{}umax}), {\it{}umin}){\nwrbrace};

    {\bf{}if} ({\it{}ex\_to}\begin{math}<\end{math}{\bf{}numeric}\begin{math}>\end{math}({\it{}b\_roots}.{\it{}op}(0).{\it{}evalf}()).{\it{}is\_real}()) {\nwlbrace}
        {\bf{}if} ({\it{}ex\_to}\begin{math}<\end{math}{\bf{}numeric}\begin{math}>\end{math}(({\it{}b\_roots}.{\it{}op}(0)-{\it{}b\_roots}.{\it{}op}(1)).{\it{}evalf}()).{\it{}is\_positive}())
            {\it{}b\_roots} = {\bf{}lst}({\it{}b\_roots}.{\it{}op}(1), {\it{}b\_roots}.{\it{}op}(0)); // rearrange to have minimum value first
        {\it{}left}[0] = {\it{}std}::{\it{}min}({\it{}std}::{\it{}max}({\it{}ex\_to}\begin{math}<\end{math}{\bf{}numeric}\begin{math}>\end{math}({\it{}b\_roots}.{\it{}op}(0)).{\it{}to\_double}(), {\it{}umin}), {\it{}umax});
        {\it{}right}[0] = {\it{}std}::{\it{}max}({\it{}std}::{\it{}min}({\it{}ex\_to}\begin{math}<\end{math}{\bf{}numeric}\begin{math}>\end{math}({\it{}b\_roots}.{\it{}op}(1)).{\it{}to\_double}(), {\it{}umax}), {\it{}umin});
    {\nwrbrace}
    {\bf{}if} ({\it{}ex\_to}\begin{math}<\end{math}{\bf{}numeric}\begin{math}>\end{math}({\it{}t\_roots}.{\it{}op}(0).{\it{}evalf}()).{\it{}is\_real}()) {\nwlbrace}
        {\bf{}if} ({\it{}ex\_to}\begin{math}<\end{math}{\bf{}numeric}\begin{math}>\end{math}(({\it{}t\_roots}.{\it{}op}(0)-{\it{}t\_roots}.{\it{}op}(1)).{\it{}evalf}()).{\it{}is\_positive}())
            {\it{}t\_roots} = {\bf{}lst}({\it{}t\_roots}.{\it{}op}(1), {\it{}t\_roots}.{\it{}op}(0)); // rearrange to have minimum value first
        {\it{}left}[1] = {\it{}std}::{\it{}min}({\it{}std}::{\it{}max}({\it{}ex\_to}\begin{math}<\end{math}{\bf{}numeric}\begin{math}>\end{math}({\it{}t\_roots}.{\it{}op}(0)).{\it{}to\_double}(), {\it{}umin}), {\it{}umax});
        {\it{}right}[1] = {\it{}std}::{\it{}max}({\it{}std}::{\it{}min}({\it{}ex\_to}\begin{math}<\end{math}{\bf{}numeric}\begin{math}>\end{math}({\it{}t\_roots}.{\it{}op}(1)).{\it{}to\_double}(), {\it{}umax}), {\it{}umin});
    {\nwrbrace}

\nwused{\\{NW3gGP3e-1Oxbp0-1B}}\nwidentdefs{\\{{\nwixident{left}}{left}}}\nwidentuses{\\{{\nwixident{lst}}{lst}}\\{{\nwixident{numeric}}{numeric}}\\{{\nwixident{op}}{op}}}\nwindexuse{\nwixident{lst}}{lst}{NW3gGP3e-2elNFX-1}\nwindexuse{\nwixident{numeric}}{numeric}{NW3gGP3e-2elNFX-1}\nwindexuse{\nwixident{op}}{op}{NW3gGP3e-2elNFX-1}\nwendcode{}\nwbegindocs{632}If a {\Tt{}\Rm{}{\bf{}cycle2D}\nwendquote} has complex coefficients it still may intersect the
real plain in a couple of points. To find them we first solve the
linear equation.
\nwenddocs{}\nwbegincode{633}\sublabel{NW3gGP3e-5M2Zk-1}\nwmargintag{{\nwtagstyle{}\subpageref{NW3gGP3e-5M2Zk-1}}}\moddef{Imaginary coefficients~{\nwtagstyle{}\subpageref{NW3gGP3e-5M2Zk-1}}}\endmoddef\Rm{}\nwstartdeflinemarkup\nwusesondefline{\\{NW3gGP3e-1Oxbp0-1B}}\nwprevnextdefs{\relax}{NW3gGP3e-5M2Zk-2}\nwenddeflinemarkup
    {\bf{}if} (\begin{math}\neg\end{math} ({\it{}Cf}.{\it{}get\_k}().{\it{}imag\_part}().{\it{}is\_zero}() \begin{math}\wedge\end{math} {\it{}Cf}.{\it{}get\_l}(0).{\it{}imag\_part}().{\it{}is\_zero}()
           \begin{math}\wedge\end{math} {\it{}Cf}.{\it{}get\_l}(1).{\it{}imag\_part}().{\it{}is\_zero}() \begin{math}\wedge\end{math} {\it{}Cf}.{\it{}get\_m}().{\it{}imag\_part}().{\it{}is\_zero}())) {\nwlbrace}
        {\bf{}if} ({\it{}imaginary\_options} \begin{math}\equiv\end{math} {\tt{}"invisible"})
            {\bf{}return};
        {\bf{}realsymbol} {\it{}x1}({\tt{}"x1"}), {\it{}y1}({\tt{}"y1"});
        {\bf{}cycle2D} {\it{}CI}={\it{}ex\_to}\begin{math}<\end{math}{\bf{}cycle2D}\begin{math}>\end{math}({\it{}Cf}.{\it{}imag\_part}());
        {\bf{}lst} {\it{}sol}={\it{}ex\_to}\begin{math}<\end{math}{\bf{}lst}\begin{math}>\end{math}({\it{}lsolve}({\bf{}lst}({\it{}CI}.{\it{}val}({\bf{}lst}({\it{}x1},{\it{}y1}))\begin{math}\equiv\end{math}0), {\bf{}lst}({\it{}x1},{\it{}y1})));

\nwalsodefined{\\{NW3gGP3e-5M2Zk-2}\\{NW3gGP3e-5M2Zk-3}\\{NW3gGP3e-5M2Zk-4}\\{NW3gGP3e-5M2Zk-5}\\{NW3gGP3e-5M2Zk-6}}\nwused{\\{NW3gGP3e-1Oxbp0-1B}}\nwidentuses{\\{{\nwixident{cycle2D}}{cycle2D}}\\{{\nwixident{get{\_}k}}{get:unk}}\\{{\nwixident{get{\_}l}}{get:unl}}\\{{\nwixident{get{\_}m}}{get:unm}}\\{{\nwixident{is{\_}zero}}{is:unzero}}\\{{\nwixident{lst}}{lst}}\\{{\nwixident{realsymbol}}{realsymbol}}\\{{\nwixident{val}}{val}}}\nwindexuse{\nwixident{cycle2D}}{cycle2D}{NW3gGP3e-5M2Zk-1}\nwindexuse{\nwixident{get{\_}k}}{get:unk}{NW3gGP3e-5M2Zk-1}\nwindexuse{\nwixident{get{\_}l}}{get:unl}{NW3gGP3e-5M2Zk-1}\nwindexuse{\nwixident{get{\_}m}}{get:unm}{NW3gGP3e-5M2Zk-1}\nwindexuse{\nwixident{is{\_}zero}}{is:unzero}{NW3gGP3e-5M2Zk-1}\nwindexuse{\nwixident{lst}}{lst}{NW3gGP3e-5M2Zk-1}\nwindexuse{\nwixident{realsymbol}}{realsymbol}{NW3gGP3e-5M2Zk-1}\nwindexuse{\nwixident{val}}{val}{NW3gGP3e-5M2Zk-1}\nwendcode{}\nwbegindocs{634} Then we use the linear substitution to solve the quadratic equation.
\nwenddocs{}\nwbegincode{635}\sublabel{NW3gGP3e-5M2Zk-2}\nwmargintag{{\nwtagstyle{}\subpageref{NW3gGP3e-5M2Zk-2}}}\moddef{Imaginary coefficients~{\nwtagstyle{}\subpageref{NW3gGP3e-5M2Zk-1}}}\plusendmoddef\Rm{}\nwstartdeflinemarkup\nwusesondefline{\\{NW3gGP3e-1Oxbp0-1B}}\nwprevnextdefs{NW3gGP3e-5M2Zk-1}{NW3gGP3e-5M2Zk-3}\nwenddeflinemarkup
        {\it{}CI}={\it{}ex\_to}\begin{math}<\end{math}{\bf{}cycle2D}\begin{math}>\end{math}({\it{}Cf}.{\it{}normalize}().{\it{}real\_part}());
        {\bf{}ex} {\it{}eq}=({\it{}CI}.{\it{}val}({\bf{}lst}({\it{}x1},{\it{}y1})).{\it{}subs}({\it{}sol})).{\it{}normal}();
        {\bf{}ex} {\it{}t}=({\it{}eq}.{\it{}has}({\it{}x1})?{\it{}x1}:{\it{}y1}),  {\it{}s}=({\it{}eq}.{\it{}has}({\it{}x1})?{\it{}y1}:{\it{}x1});
        {\bf{}double} {\it{}A}, {\it{}B}, {\it{}C}, {\it{}D};
        {\it{}A}={\it{}ex\_to}\begin{math}<\end{math}{\bf{}numeric}\begin{math}>\end{math}({\it{}eq}.{\it{}coeff}({\it{}ex\_to}\begin{math}<\end{math}{\bf{}symbol}\begin{math}>\end{math}({\it{}t}),2)).{\it{}to\_double}();
        {\it{}B}={\it{}ex\_to}\begin{math}<\end{math}{\bf{}numeric}\begin{math}>\end{math}({\it{}eq}.{\it{}coeff}({\it{}ex\_to}\begin{math}<\end{math}{\bf{}symbol}\begin{math}>\end{math}({\it{}t}),1)).{\it{}to\_double}();
        {\it{}C}={\it{}ex\_to}\begin{math}<\end{math}{\bf{}numeric}\begin{math}>\end{math}({\it{}eq}.{\it{}coeff}({\it{}ex\_to}\begin{math}<\end{math}{\bf{}symbol}\begin{math}>\end{math}({\it{}t}),0)).{\it{}to\_double}();
        {\it{}D}={\it{}B}\begin{math}\ast\end{math}{\it{}B}-4\begin{math}\ast\end{math}{\it{}A}\begin{math}\ast\end{math}{\it{}C};

\nwused{\\{NW3gGP3e-1Oxbp0-1B}}\nwidentuses{\\{{\nwixident{cycle2D}}{cycle2D}}\\{{\nwixident{ex}}{ex}}\\{{\nwixident{lst}}{lst}}\\{{\nwixident{normal}}{normal}}\\{{\nwixident{normalize}}{normalize}}\\{{\nwixident{numeric}}{numeric}}\\{{\nwixident{subs}}{subs}}\\{{\nwixident{val}}{val}}}\nwindexuse{\nwixident{cycle2D}}{cycle2D}{NW3gGP3e-5M2Zk-2}\nwindexuse{\nwixident{ex}}{ex}{NW3gGP3e-5M2Zk-2}\nwindexuse{\nwixident{lst}}{lst}{NW3gGP3e-5M2Zk-2}\nwindexuse{\nwixident{normal}}{normal}{NW3gGP3e-5M2Zk-2}\nwindexuse{\nwixident{normalize}}{normalize}{NW3gGP3e-5M2Zk-2}\nwindexuse{\nwixident{numeric}}{numeric}{NW3gGP3e-5M2Zk-2}\nwindexuse{\nwixident{subs}}{subs}{NW3gGP3e-5M2Zk-2}\nwindexuse{\nwixident{val}}{val}{NW3gGP3e-5M2Zk-2}\nwendcode{}\nwbegindocs{636}If the quadratic equation has real roots we draw respective points.
\nwenddocs{}\nwbegincode{637}\sublabel{NW3gGP3e-5M2Zk-3}\nwmargintag{{\nwtagstyle{}\subpageref{NW3gGP3e-5M2Zk-3}}}\moddef{Imaginary coefficients~{\nwtagstyle{}\subpageref{NW3gGP3e-5M2Zk-1}}}\plusendmoddef\Rm{}\nwstartdeflinemarkup\nwusesondefline{\\{NW3gGP3e-1Oxbp0-1B}}\nwprevnextdefs{NW3gGP3e-5M2Zk-2}{NW3gGP3e-5M2Zk-4}\nwenddeflinemarkup
        {\bf{}if} ({\it{}abs}({\it{}A})\begin{math}<\end{math}{\it{}epsilon} \begin{math}\vee\end{math} {\it{}D}\begin{math}\geq\end{math}0){\nwlbrace}
            {\bf{}if} ({\it{}with\_header})
                {\it{}ost} \begin{math}\ll\end{math} {\it{}endl} \begin{math}\ll\end{math} {\tt{}"// imaginary coefficients, the intersection with the real plane is dots only"};

\nwused{\\{NW3gGP3e-1Oxbp0-1B}}\nwendcode{}\nwbegindocs{638}Two roots are follow.
\nwenddocs{}\nwbegincode{639}\sublabel{NW3gGP3e-5M2Zk-4}\nwmargintag{{\nwtagstyle{}\subpageref{NW3gGP3e-5M2Zk-4}}}\moddef{Imaginary coefficients~{\nwtagstyle{}\subpageref{NW3gGP3e-5M2Zk-1}}}\plusendmoddef\Rm{}\nwstartdeflinemarkup\nwusesondefline{\\{NW3gGP3e-1Oxbp0-1B}}\nwprevnextdefs{NW3gGP3e-5M2Zk-3}{NW3gGP3e-5M2Zk-5}\nwenddeflinemarkup
            {\bf{}for}({\bf{}int} {\it{}i}=-1; {\it{}i}\begin{math}<\end{math}2; {\it{}i}+=2) {\nwlbrace}
                {\bf{}double} {\it{}t1};
                {\bf{}if} ({\it{}abs}({\it{}A})\begin{math}<\end{math}{\it{}epsilon}) {\nwlbrace}
                    {\it{}i}=1; // No need for second pass
                    {\bf{}if} ({\it{}abs}({\it{}B})\begin{math}<\end{math}{\it{}epsilon})
                        {\bf{}return}; // trivial identity
                    {\bf{}else}
                        {\it{}t1}=-{\it{}C}\begin{math}\div\end{math}{\it{}B};
                {\nwrbrace} {\bf{}else}
                    {\it{}t1}= (-{\it{}B}+{\it{}i}\begin{math}\ast\end{math}{\it{}sqrt}({\it{}D}))\begin{math}\div\end{math}2.0\begin{math}\div\end{math}{\it{}A};
                {\it{}exmap} {\it{}em};
                {\it{}em}.{\it{}insert}({\it{}std}::{\it{}make\_pair}({\it{}t}, {\it{}t1}));
                {\bf{}ex} {\it{}s1}={\it{}s}.{\it{}subs}({\it{}sol}.{\it{}subs}({\it{}em}));
                {\it{}uc}={\it{}ex\_to}\begin{math}<\end{math}{\bf{}numeric}\begin{math}>\end{math}({\it{}eq}.{\it{}has}({\it{}x1})? {\it{}t1} : {\it{}s1}).{\it{}to\_double}();
                {\it{}vc}={\it{}ex\_to}\begin{math}<\end{math}{\bf{}numeric}\begin{math}>\end{math}({\it{}eq}.{\it{}has}({\it{}x1})? {\it{}s1} : {\it{}t1}).{\it{}to\_double}();

\nwused{\\{NW3gGP3e-1Oxbp0-1B}}\nwidentuses{\\{{\nwixident{ex}}{ex}}\\{{\nwixident{numeric}}{numeric}}\\{{\nwixident{subs}}{subs}}}\nwindexuse{\nwixident{ex}}{ex}{NW3gGP3e-5M2Zk-4}\nwindexuse{\nwixident{numeric}}{numeric}{NW3gGP3e-5M2Zk-4}\nwindexuse{\nwixident{subs}}{subs}{NW3gGP3e-5M2Zk-4}\nwendcode{}\nwbegindocs{640}After the double check, we reset the drawing style to the hard-coded
style for imaginary objects.
\nwenddocs{}\nwbegincode{641}\sublabel{NW3gGP3e-5M2Zk-5}\nwmargintag{{\nwtagstyle{}\subpageref{NW3gGP3e-5M2Zk-5}}}\moddef{Imaginary coefficients~{\nwtagstyle{}\subpageref{NW3gGP3e-5M2Zk-1}}}\plusendmoddef\Rm{}\nwstartdeflinemarkup\nwusesondefline{\\{NW3gGP3e-1Oxbp0-1B}}\nwprevnextdefs{NW3gGP3e-5M2Zk-4}{NW3gGP3e-5M2Zk-6}\nwenddeflinemarkup
                {\bf{}if} ({\it{}abs}({\it{}ex\_to}\begin{math}<\end{math}{\bf{}numeric}\begin{math}>\end{math}({\it{}Cf}.{\it{}val}({\bf{}lst}({\it{}uc},{\it{}vc})).{\it{}evalf}()).{\it{}to\_double}()) \begin{math}<\end{math} {\it{}epsilon}) {\nwlbrace}
                    {\bf{}if} ({\it{}asymptote})
                        {\it{}draw\_options}.{\it{}str}({\tt{}","}+{\it{}imaginary\_options}+{\tt{}");"});
                    {\bf{}else}
                        {\it{}draw\_options}.{\it{}str}({\tt{}" "}+{\it{}imaginary\_options}+{\tt{}";"});
                    {\it{}ost} \begin{math}\ll\end{math} {\it{}endl};
                    {\nwlbrace}\LA{}place a dot~{\nwtagstyle{}\subpageref{NW3gGP3e-FZdRR-1}}\RA{}{\nwrbrace}
                {\nwrbrace} {\bf{}else} {\nwlbrace}
                    {\it{}std}::{\it{}cerr} \begin{math}\ll\end{math} {\tt{}"Calculation of dots in imaginary cycle is inaccurate"} \begin{math}\ll\end{math} {\it{}std}::{\it{}endl};
                {\nwrbrace}
            {\nwrbrace}

\nwused{\\{NW3gGP3e-1Oxbp0-1B}}\nwidentuses{\\{{\nwixident{cycle}}{cycle}}\\{{\nwixident{lst}}{lst}}\\{{\nwixident{numeric}}{numeric}}\\{{\nwixident{val}}{val}}}\nwindexuse{\nwixident{cycle}}{cycle}{NW3gGP3e-5M2Zk-5}\nwindexuse{\nwixident{lst}}{lst}{NW3gGP3e-5M2Zk-5}\nwindexuse{\nwixident{numeric}}{numeric}{NW3gGP3e-5M2Zk-5}\nwindexuse{\nwixident{val}}{val}{NW3gGP3e-5M2Zk-5}\nwendcode{}\nwbegindocs{642}If the quadratic equation does not have real roots we draw respective points.
\nwenddocs{}\nwbegincode{643}\sublabel{NW3gGP3e-5M2Zk-6}\nwmargintag{{\nwtagstyle{}\subpageref{NW3gGP3e-5M2Zk-6}}}\moddef{Imaginary coefficients~{\nwtagstyle{}\subpageref{NW3gGP3e-5M2Zk-1}}}\plusendmoddef\Rm{}\nwstartdeflinemarkup\nwusesondefline{\\{NW3gGP3e-1Oxbp0-1B}}\nwprevnextdefs{NW3gGP3e-5M2Zk-5}{\relax}\nwenddeflinemarkup
    {\nwrbrace} {\bf{}else}
        {\bf{}if} ({\it{}with\_header})
            {\it{}ost} \begin{math}\ll\end{math} {\it{}endl} \begin{math}\ll\end{math} {\tt{}"// imaginary coefficients, no intersection with the real plane"} \begin{math}\ll\end{math} {\it{}endl};
    {\it{}ost} \begin{math}\ll\end{math} {\it{}endl};
    {\bf{}return};
{\nwrbrace}

\nwused{\\{NW3gGP3e-1Oxbp0-1B}}\nwendcode{}\nwbegindocs{644}We start from the most involved case of a circle with a positive
radius. To this end we calculate coordinates {\Tt{}\Rm{}{\it{}u}[2][4]\nwendquote} and
{\Tt{}\Rm{}{\it{}v}[2][4]\nwendquote} of endpoints for up to four arcs making the circle.  The
\(x\)-components of intersection points with vertical boundaries are
rearranged appropriately.
\nwenddocs{}\nwbegincode{645}\sublabel{NW3gGP3e-4WQ5O5-1}\nwmargintag{{\nwtagstyle{}\subpageref{NW3gGP3e-4WQ5O5-1}}}\moddef{Draw a circle~{\nwtagstyle{}\subpageref{NW3gGP3e-4WQ5O5-1}}}\endmoddef\Rm{}\nwstartdeflinemarkup\nwusesondefline{\\{NW3gGP3e-1Oxbp0-1B}}\nwprevnextdefs{\relax}{NW3gGP3e-4WQ5O5-2}\nwenddeflinemarkup
{\bf{}if} ({\it{}determinant} \begin{math}>\end{math} {\it{}epsilon}) {\nwlbrace}
 {\bf{}double} {\it{}u}[2][4], {\it{}v}[2][4];
 {\bf{}if} ({\it{}with\_header})
  {\it{}ost} \begin{math}\ll\end{math} {\tt{}" /circle of radius "} \begin{math}\ll\end{math} {\it{}r}  \begin{math}\ll\end{math} {\it{}endl};
 {\bf{}if} ({\it{}uc}+{\it{}r} \begin{math}<\end{math} {\it{}umin} \begin{math}\vee\end{math} {\it{}uc}-{\it{}r} \begin{math}>\end{math} {\it{}umax} \begin{math}\vee\end{math} {\it{}vc}+{\it{}r}\begin{math}<\end{math} {\it{}vmin} \begin{math}\vee\end{math} {\it{}vc}-{\it{}r} \begin{math}>\end{math} {\it{}vmax} \begin{math}\vee\end{math}
     {\it{}pow}({\it{}std}::{\it{}max}({\it{}umax}-{\it{}uc},{\it{}uc}-{\it{}umin}),2.0)+{\it{}pow}({\it{}std}::{\it{}max}({\it{}vmax}-{\it{}vc},{\it{}vc}-{\it{}vmin}),2.0)\begin{math}<\end{math}{\it{}determinant}) {\nwlbrace}
      {\bf{}if} ({\it{}with\_header})
          {\it{}ost} \begin{math}\ll\end{math} {\tt{}"   // out of the window "} \begin{math}\ll\end{math} {\it{}endl};
 {\nwrbrace} {\bf{}else} {\nwlbrace}

\nwalsodefined{\\{NW3gGP3e-4WQ5O5-2}\\{NW3gGP3e-4WQ5O5-3}\\{NW3gGP3e-4WQ5O5-4}\\{NW3gGP3e-4WQ5O5-5}\\{NW3gGP3e-4WQ5O5-6}\\{NW3gGP3e-4WQ5O5-7}\\{NW3gGP3e-4WQ5O5-8}}\nwused{\\{NW3gGP3e-1Oxbp0-1B}}\nwidentuses{\\{{\nwixident{u}}{u}}\\{{\nwixident{v}}{v}}}\nwindexuse{\nwixident{u}}{u}{NW3gGP3e-4WQ5O5-1}\nwindexuse{\nwixident{v}}{v}{NW3gGP3e-4WQ5O5-1}\nwendcode{}\nwbegindocs{646}Depending from the y-position of the centre we draw different
arcs. The first case is the centre is above the horizontal strip.
\nwenddocs{}\nwbegincode{647}\sublabel{NW3gGP3e-4WQ5O5-2}\nwmargintag{{\nwtagstyle{}\subpageref{NW3gGP3e-4WQ5O5-2}}}\moddef{Draw a circle~{\nwtagstyle{}\subpageref{NW3gGP3e-4WQ5O5-1}}}\plusendmoddef\Rm{}\nwstartdeflinemarkup\nwusesondefline{\\{NW3gGP3e-1Oxbp0-1B}}\nwprevnextdefs{NW3gGP3e-4WQ5O5-1}{NW3gGP3e-4WQ5O5-3}\nwenddeflinemarkup
    {\bf{}if} ( {\it{}vc}-{\it{}vmax} \begin{math}>\end{math} {\it{}epsilon}) {\nwlbrace}
        {\it{}u}[0][2] = {\it{}left}[1]; {\it{}u}[0][3] = {\it{}right}[1];
        {\it{}u}[1][2] = {\it{}left}[0]; {\it{}u}[1][3] = {\it{}right}[0];
        {\it{}u}[0][0] = {\it{}u}[1][0] = {\it{}uc};
        {\it{}u}[0][1] = {\it{}u}[1][1] = {\it{}uc};

\nwused{\\{NW3gGP3e-1Oxbp0-1B}}\nwidentuses{\\{{\nwixident{left}}{left}}\\{{\nwixident{u}}{u}}}\nwindexuse{\nwixident{left}}{left}{NW3gGP3e-4WQ5O5-2}\nwindexuse{\nwixident{u}}{u}{NW3gGP3e-4WQ5O5-2}\nwendcode{}\nwbegindocs{648}The case when the centre is in the  the horizontal strip.
\nwenddocs{}\nwbegincode{649}\sublabel{NW3gGP3e-4WQ5O5-3}\nwmargintag{{\nwtagstyle{}\subpageref{NW3gGP3e-4WQ5O5-3}}}\moddef{Draw a circle~{\nwtagstyle{}\subpageref{NW3gGP3e-4WQ5O5-1}}}\plusendmoddef\Rm{}\nwstartdeflinemarkup\nwusesondefline{\\{NW3gGP3e-1Oxbp0-1B}}\nwprevnextdefs{NW3gGP3e-4WQ5O5-2}{NW3gGP3e-4WQ5O5-4}\nwenddeflinemarkup
    {\nwrbrace} {\bf{}else} {\bf{}if} ({\it{}vc}-{\it{}vmin} \begin{math}>\end{math} {\it{}epsilon}) {\nwlbrace}
        {\it{}u}[0][0] = {\it{}left}[1]; {\it{}u}[0][1] = {\it{}right}[1];
        {\it{}u}[0][2] = {\it{}right}[0]; {\it{}u}[0][3] = {\it{}left}[0];

        {\bf{}if} ({\it{}uc}-{\it{}r}-{\it{}umin} \begin{math}>\end{math} {\it{}epsilon})
            {\it{}u}[1][0] = {\it{}u}[1][3] = {\it{}uc}-{\it{}r};
        {\bf{}else}
            {\it{}u}[1][0] = {\it{}u}[1][3] = {\it{}umin};

        {\bf{}if} ({\it{}umax}-{\it{}uc}-{\it{}r} \begin{math}>\end{math} {\it{}epsilon})
            {\it{}u}[1][1] = {\it{}u}[1][2] = {\it{}uc}+{\it{}r};
        {\bf{}else}
            {\it{}u}[1][1] = {\it{}u}[1][2] = {\it{}umax};

\nwused{\\{NW3gGP3e-1Oxbp0-1B}}\nwidentuses{\\{{\nwixident{left}}{left}}\\{{\nwixident{u}}{u}}}\nwindexuse{\nwixident{left}}{left}{NW3gGP3e-4WQ5O5-3}\nwindexuse{\nwixident{u}}{u}{NW3gGP3e-4WQ5O5-3}\nwendcode{}\nwbegindocs{650}Finally, the centre is below the horizontal strip.
\nwenddocs{}\nwbegincode{651}\sublabel{NW3gGP3e-4WQ5O5-4}\nwmargintag{{\nwtagstyle{}\subpageref{NW3gGP3e-4WQ5O5-4}}}\moddef{Draw a circle~{\nwtagstyle{}\subpageref{NW3gGP3e-4WQ5O5-1}}}\plusendmoddef\Rm{}\nwstartdeflinemarkup\nwusesondefline{\\{NW3gGP3e-1Oxbp0-1B}}\nwprevnextdefs{NW3gGP3e-4WQ5O5-3}{NW3gGP3e-4WQ5O5-5}\nwenddeflinemarkup
 {\nwrbrace} {\bf{}else} {\nwlbrace}
  {\it{}u}[0][0] = {\it{}left}[1]; {\it{}u}[0][1] = {\it{}right}[1];
  {\it{}u}[1][0] = {\it{}left}[0]; {\it{}u}[1][1] = {\it{}right}[0];
  {\it{}u}[0][2] = {\it{}u}[1][2] = {\it{}uc};
  {\it{}u}[0][3] = {\it{}u}[1][3] = {\it{}uc};
 {\nwrbrace}

\nwused{\\{NW3gGP3e-1Oxbp0-1B}}\nwidentuses{\\{{\nwixident{left}}{left}}\\{{\nwixident{u}}{u}}}\nwindexuse{\nwixident{left}}{left}{NW3gGP3e-4WQ5O5-4}\nwindexuse{\nwixident{u}}{u}{NW3gGP3e-4WQ5O5-4}\nwendcode{}\nwbegindocs{652}We calculate now the \(y\)-components of the endpoints corresponding
to \(x\)-components found before.
\nwenddocs{}\nwbegincode{653}\sublabel{NW3gGP3e-4WQ5O5-5}\nwmargintag{{\nwtagstyle{}\subpageref{NW3gGP3e-4WQ5O5-5}}}\moddef{Draw a circle~{\nwtagstyle{}\subpageref{NW3gGP3e-4WQ5O5-1}}}\plusendmoddef\Rm{}\nwstartdeflinemarkup\nwusesondefline{\\{NW3gGP3e-1Oxbp0-1B}}\nwprevnextdefs{NW3gGP3e-4WQ5O5-4}{NW3gGP3e-4WQ5O5-6}\nwenddeflinemarkup
 {\bf{}lst} {\it{}y\_roots};
 {\bf{}for} ({\bf{}int} {\it{}j}=0; {\it{}j}\begin{math}<\end{math}2; {\it{}j}\protect\PP)
  {\bf{}for} ({\bf{}int} {\it{}i}=0; {\it{}i}\begin{math}<\end{math}4; {\it{}i}\protect\PP)
   {\bf{}if} ({\it{}abs}({\it{}u}[{\it{}j}][{\it{}i}]-{\it{}uc}) \begin{math}<\end{math} {\it{}epsilon}) // Touch the horizontal boundary?
    {\it{}v}[{\it{}j}][{\it{}i}] = ({\it{}i}\begin{math}\equiv\end{math}0 \begin{math}\vee\end{math} {\it{}i} \begin{math}\equiv\end{math}1? {\it{}vc}+{\it{}r} : {\it{}vc}-{\it{}r});
   {\bf{}else} {\bf{}if} ({\it{}abs}({\it{}u}[{\it{}j}][{\it{}i}]-{\it{}uc}-{\it{}r})\begin{math}<\end{math}{\it{}epsilon} \begin{math}\vee\end{math} {\it{}abs}({\it{}u}[{\it{}j}][{\it{}i}]-{\it{}uc}+{\it{}r})\begin{math}<\end{math}{\it{}epsilon}) // Touch the vertical boundary?
    {\it{}v}[{\it{}j}][{\it{}i}] = {\it{}vc};
   {\bf{}else} {\nwlbrace}
    {\it{}y\_roots} = {\it{}Cf}.{\it{}roots}({\it{}u}[{\it{}j}][{\it{}i}], {\bf{}false});
    {\bf{}if} ({\it{}ex\_to}\begin{math}<\end{math}{\bf{}numeric}\begin{math}>\end{math}({\it{}y\_roots}.{\it{}op}(0)).{\it{}is\_real}()) {\nwlbrace} // does circle intersect the boundary?
     {\bf{}if} ({\it{}i}\begin{math}<\end{math}2)
         {\it{}v}[{\it{}j}][{\it{}i}] = {\it{}std}::{\it{}min}({\it{}ex\_to}\begin{math}<\end{math}{\bf{}numeric}\begin{math}>\end{math}({\it{}std}::{\it{}max}({\it{}y\_roots}.{\it{}op}(0), {\it{}y\_roots}.{\it{}op}(1))).{\it{}to\_double}(), {\it{}vmax});
     {\bf{}else}
         {\it{}v}[{\it{}j}][{\it{}i}] = {\it{}std}::{\it{}max}({\it{}ex\_to}\begin{math}<\end{math}{\bf{}numeric}\begin{math}>\end{math}({\it{}std}::{\it{}min}({\it{}y\_roots}.{\it{}op}(0), {\it{}y\_roots}.{\it{}op}(1))).{\it{}to\_double}(), {\it{}vmin});
    {\nwrbrace} {\bf{}else}
     {\it{}v}[{\it{}j}][{\it{}i}] = {\it{}vc};
   {\nwrbrace}

\nwused{\\{NW3gGP3e-1Oxbp0-1B}}\nwidentuses{\\{{\nwixident{lst}}{lst}}\\{{\nwixident{numeric}}{numeric}}\\{{\nwixident{op}}{op}}\\{{\nwixident{roots}}{roots}}\\{{\nwixident{u}}{u}}\\{{\nwixident{v}}{v}}}\nwindexuse{\nwixident{lst}}{lst}{NW3gGP3e-4WQ5O5-5}\nwindexuse{\nwixident{numeric}}{numeric}{NW3gGP3e-4WQ5O5-5}\nwindexuse{\nwixident{op}}{op}{NW3gGP3e-4WQ5O5-5}\nwindexuse{\nwixident{roots}}{roots}{NW3gGP3e-4WQ5O5-5}\nwindexuse{\nwixident{u}}{u}{NW3gGP3e-4WQ5O5-5}\nwindexuse{\nwixident{v}}{v}{NW3gGP3e-4WQ5O5-5}\nwendcode{}\nwbegindocs{654}Now we drawing up to four arcs which make the visible part of the
circle. Each arc is defined through its two endpoints and tangent
vector in them.
\nwenddocs{}\nwbegincode{655}\sublabel{NW3gGP3e-4WQ5O5-6}\nwmargintag{{\nwtagstyle{}\subpageref{NW3gGP3e-4WQ5O5-6}}}\moddef{Draw a circle~{\nwtagstyle{}\subpageref{NW3gGP3e-4WQ5O5-1}}}\plusendmoddef\Rm{}\nwstartdeflinemarkup\nwusesondefline{\\{NW3gGP3e-1Oxbp0-1B}}\nwprevnextdefs{NW3gGP3e-4WQ5O5-5}{NW3gGP3e-4WQ5O5-7}\nwenddeflinemarkup
 {\bf{}for} ({\bf{}int} {\it{}i}=0; {\it{}i}\begin{math}<\end{math}4; {\it{}i}\protect\PP) {\nwlbrace}// actual drawing of four arcs
  {\bf{}int} {\it{}s} = ({\it{}i}\begin{math}\equiv\end{math}0 \begin{math}\vee\end{math} {\it{}i} \begin{math}\equiv\end{math}2? -1:1);
  {\bf{}if} (({\it{}u}[0][{\it{}i}] \begin{math}\neq\end{math} {\it{}u}[1][{\it{}i}]) \begin{math}\vee\end{math} ({\it{}v}[0][{\it{}i}] \begin{math}\neq\end{math} {\it{}v}[1][{\it{}i}])) {\nwlbrace}// do not draw empty arc
   {\it{}ost} \begin{math}\ll\end{math} {\tt{}"  "} \begin{math}\ll\end{math} ({\it{}only\_path} ? {\it{}already\_drawn} : {\it{}draw\_start}.{\it{}str}()) \begin{math}\ll\end{math} {\it{}u}[0][{\it{}i}] \begin{math}\ll\end{math}{\tt{}","}
       \begin{math}\ll\end{math} {\it{}v}[0][{\it{}i}] \begin{math}\ll\end{math} {\tt{}"){\char123}"} \begin{math}\ll\end{math} {\it{}s}\begin{math}\ast\end{math}({\it{}v}[0][{\it{}i}]-{\it{}vc}) \begin{math}\ll\end{math} {\tt{}","} \begin{math}\ll\end{math} {\it{}s}\begin{math}\ast\end{math}({\it{}uc}-{\it{}u}[0][{\it{}i}])
       \begin{math}\ll\end{math} ({\it{}asymptote} ? {\tt{}"{\char125}::{\char123}"} : {\tt{}"{\char125}...{\char123}"})
       \begin{math}\ll\end{math} {\it{}s}\begin{math}\ast\end{math}({\it{}v}[1][{\it{}i}]-{\it{}vc}) \begin{math}\ll\end{math} {\tt{}","} \begin{math}\ll\end{math} {\it{}s}\begin{math}\ast\end{math}({\it{}uc}-{\it{}u}[1][{\it{}i}]) \begin{math}\ll\end{math} {\tt{}"{\char125}("} \begin{math}\ll\end{math} {\it{}u}[1][{\it{}i}] \begin{math}\ll\end{math}{\tt{}","} \begin{math}\ll\end{math} {\it{}v}[1][{\it{}i}] \begin{math}\ll\end{math} {\tt{}")"}
       \begin{math}\ll\end{math} ({\it{}only\_path} ? {\tt{}""} : {\it{}draw\_options}.{\it{}str}());
   {\it{}already\_drawn}={\tt{}"^^("};
  {\nwrbrace}
 {\nwrbrace}
 {\nwrbrace}

\nwused{\\{NW3gGP3e-1Oxbp0-1B}}\nwidentuses{\\{{\nwixident{u}}{u}}\\{{\nwixident{v}}{v}}}\nwindexuse{\nwixident{u}}{u}{NW3gGP3e-4WQ5O5-6}\nwindexuse{\nwixident{v}}{v}{NW3gGP3e-4WQ5O5-6}\nwendcode{}\nwbegindocs{656}Finally, for zero-radius circles we draw a point and do not draw
anything for circles with an imaginary radius.
\nwenddocs{}\nwbegincode{657}\sublabel{NW3gGP3e-4WQ5O5-7}\nwmargintag{{\nwtagstyle{}\subpageref{NW3gGP3e-4WQ5O5-7}}}\moddef{Draw a circle~{\nwtagstyle{}\subpageref{NW3gGP3e-4WQ5O5-1}}}\plusendmoddef\Rm{}\nwstartdeflinemarkup\nwusesondefline{\\{NW3gGP3e-1Oxbp0-1B}}\nwprevnextdefs{NW3gGP3e-4WQ5O5-6}{NW3gGP3e-4WQ5O5-8}\nwenddeflinemarkup
{\nwrbrace} {\bf{}else} {\bf{}if} ({\it{}is\_almost\_zero}({\it{}determinant})) {\nwlbrace}
    {\bf{}if} ({\it{}with\_header})
        {\it{}ost} \begin{math}\ll\end{math} {\tt{}" /circle of zero-radius"} \begin{math}\ll\end{math} {\it{}endl};
    \LA{}place a dot~{\nwtagstyle{}\subpageref{NW3gGP3e-FZdRR-1}}\RA{}

\nwused{\\{NW3gGP3e-1Oxbp0-1B}}\nwendcode{}\nwbegindocs{658}This code places a dot at the point {\Tt{}\Rm{}({\it{}U},{\it{}V})\nwendquote}.
\nwenddocs{}\nwbegincode{659}\sublabel{NW3gGP3e-FZdRR-1}\nwmargintag{{\nwtagstyle{}\subpageref{NW3gGP3e-FZdRR-1}}}\moddef{place a dot~{\nwtagstyle{}\subpageref{NW3gGP3e-FZdRR-1}}}\endmoddef\Rm{}\nwstartdeflinemarkup\nwusesondefline{\\{NW3gGP3e-5M2Zk-5}\\{NW3gGP3e-4WQ5O5-7}\\{NW3gGP3e-2yW1fi-2}}\nwprevnextdefs{\relax}{NW3gGP3e-FZdRR-2}\nwenddeflinemarkup
    {\bf{}double} {\it{}U}={\it{}ex\_to}\begin{math}<\end{math}{\bf{}numeric}\begin{math}>\end{math}({\it{}uc}).{\it{}to\_double}();
    {\bf{}double} {\it{}V}={\it{}ex\_to}\begin{math}<\end{math}{\bf{}numeric}\begin{math}>\end{math}({\it{}vc}).{\it{}to\_double}();
    {\bf{}if} (({\it{}umin} \begin{math}\leq\end{math}{\it{}U}) \begin{math}\wedge\end{math} ({\it{}umax}\begin{math}\geq\end{math}{\it{}U}) \begin{math}\wedge\end{math} ({\it{}vmin}\begin{math}\leq\end{math}{\it{}V}) \begin{math}\wedge\end{math} ({\it{}vmax}\begin{math}\geq\end{math}{\it{}V})) {\nwlbrace}
        {\it{}ost} \begin{math}\ll\end{math} ({\it{}asymptote} ? ({\it{}only\_path} ? {\it{}already\_drawn} : {\tt{}"dot("}) : {\tt{}"draw "} )
            \begin{math}\ll\end{math} {\it{}picture} \begin{math}\ll\end{math} ({\it{}picture}.{\it{}size}()\begin{math}\equiv\end{math}0? {\tt{}""} : {\tt{}","})
            \begin{math}\ll\end{math} ({\it{}only\_path} ? {\tt{}""} : {\tt{}"("})
            \begin{math}\ll\end{math} {\it{}uc} \begin{math}\ll\end{math} {\tt{}","} \begin{math}\ll\end{math} {\it{}vc} \begin{math}\ll\end{math} {\tt{}")"} \begin{math}\ll\end{math} ({\it{}only\_path} ? {\tt{}""} : {\it{}draw\_options}.{\it{}str}());
    {\it{}already\_drawn}={\tt{}"^^("};

\nwalsodefined{\\{NW3gGP3e-FZdRR-2}}\nwused{\\{NW3gGP3e-5M2Zk-5}\\{NW3gGP3e-4WQ5O5-7}\\{NW3gGP3e-2yW1fi-2}}\nwidentuses{\\{{\nwixident{numeric}}{numeric}}}\nwindexuse{\nwixident{numeric}}{numeric}{NW3gGP3e-FZdRR-1}\nwendcode{}\nwbegindocs{660}\nwdocspar
\nwenddocs{}\nwbegincode{661}\sublabel{NW3gGP3e-FZdRR-2}\nwmargintag{{\nwtagstyle{}\subpageref{NW3gGP3e-FZdRR-2}}}\moddef{place a dot~{\nwtagstyle{}\subpageref{NW3gGP3e-FZdRR-1}}}\plusendmoddef\Rm{}\nwstartdeflinemarkup\nwusesondefline{\\{NW3gGP3e-5M2Zk-5}\\{NW3gGP3e-4WQ5O5-7}\\{NW3gGP3e-2yW1fi-2}}\nwprevnextdefs{NW3gGP3e-FZdRR-1}{\relax}\nwenddeflinemarkup
    {\nwrbrace} {\bf{}else}
        {\bf{}if} ({\it{}with\_header})
            {\it{}ost} \begin{math}\ll\end{math} {\tt{}"// the vertex is out of range"} \begin{math}\ll\end{math} {\it{}endl};

\nwused{\\{NW3gGP3e-5M2Zk-5}\\{NW3gGP3e-4WQ5O5-7}\\{NW3gGP3e-2yW1fi-2}}\nwendcode{}\nwbegindocs{662}\nwdocspar
\nwenddocs{}\nwbegincode{663}\sublabel{NW3gGP3e-4WQ5O5-8}\nwmargintag{{\nwtagstyle{}\subpageref{NW3gGP3e-4WQ5O5-8}}}\moddef{Draw a circle~{\nwtagstyle{}\subpageref{NW3gGP3e-4WQ5O5-1}}}\plusendmoddef\Rm{}\nwstartdeflinemarkup\nwusesondefline{\\{NW3gGP3e-1Oxbp0-1B}}\nwprevnextdefs{NW3gGP3e-4WQ5O5-7}{\relax}\nwenddeflinemarkup
{\nwrbrace} {\bf{}else}
    {\bf{}if} ({\it{}with\_header})
        {\it{}ost} \begin{math}\ll\end{math} {\tt{}" /circle of imaginary radius--not drawing"} \begin{math}\ll\end{math} {\it{}endl};

\nwused{\\{NW3gGP3e-1Oxbp0-1B}}\nwendcode{}\nwbegindocs{664}First we look if the parabola or hyperbola are degenerates into two
lines, then treat two types of cycles separately.
\nwenddocs{}\nwbegincode{665}\sublabel{NW3gGP3e-4H4ZFN-1}\nwmargintag{{\nwtagstyle{}\subpageref{NW3gGP3e-4H4ZFN-1}}}\moddef{Draw a parabola or hyperbola~{\nwtagstyle{}\subpageref{NW3gGP3e-4H4ZFN-1}}}\endmoddef\Rm{}\nwstartdeflinemarkup\nwusesondefline{\\{NW3gGP3e-1Oxbp0-1B}}\nwenddeflinemarkup
{\bf{}double} {\it{}u}, {\it{}v}, {\it{}du}, {\it{}dv}, {\it{}k\_d} = {\it{}ex\_to}\begin{math}<\end{math}{\bf{}numeric}\begin{math}>\end{math}({\it{}Cf}.{\it{}get\_k}()).{\it{}to\_double}(),\nwindexdefn{\nwixident{u}}{u}{NW3gGP3e-4H4ZFN-1}\nwindexdefn{\nwixident{v}}{v}{NW3gGP3e-4H4ZFN-1}\nwindexdefn{\nwixident{du}}{du}{NW3gGP3e-4H4ZFN-1}\nwindexdefn{\nwixident{dv}}{dv}{NW3gGP3e-4H4ZFN-1}\nwindexdefn{\nwixident{k{\_}d}}{k:und}{NW3gGP3e-4H4ZFN-1}
                 {\it{}lu} = {\it{}ex\_to}\begin{math}<\end{math}{\bf{}numeric}\begin{math}>\end{math}({\it{}Cf}.{\it{}get\_l}({\it{}iu})).{\it{}to\_double}(),
                 {\it{}lv} = {\it{}ex\_to}\begin{math}<\end{math}{\bf{}numeric}\begin{math}>\end{math}({\it{}Cf}.{\it{}get\_l}({\it{}iv})).{\it{}to\_double}();

{\bf{}bool} {\it{}change\_branch} = ({\it{}sign} \begin{math}\neq\end{math} 0); // either to do a swap of branches
{\bf{}int} {\it{}zero\_or\_one} = ({\it{}sign} \begin{math}\equiv\end{math} 0 \begin{math}\vee\end{math} {\it{}k\_d}\begin{math}\ast\end{math}{\it{}signv} \begin{math}>\end{math} 0 ? 0 : 1); // for parabola and positive k take first\nwindexdefn{\nwixident{zero{\_}or{\_}one}}{zero:unor:unone}{NW3gGP3e-4H4ZFN-1}

{\bf{}if} ({\it{}sign} \begin{math}\equiv\end{math} 0) {\nwlbrace}
 \LA{}Treating a parabola~{\nwtagstyle{}\subpageref{NW3gGP3e-24b8UV-1}}\RA{}
{\nwrbrace} {\bf{}else} {\nwlbrace}
 \LA{}Treating a hyperbola~{\nwtagstyle{}\subpageref{NW3gGP3e-2yW1fi-1}}\RA{}
{\nwrbrace}

\nwused{\\{NW3gGP3e-1Oxbp0-1B}}\nwidentdefs{\\{{\nwixident{du}}{du}}\\{{\nwixident{dv}}{dv}}\\{{\nwixident{k{\_}d}}{k:und}}\\{{\nwixident{u}}{u}}\\{{\nwixident{v}}{v}}\\{{\nwixident{zero{\_}or{\_}one}}{zero:unor:unone}}}\nwidentuses{\\{{\nwixident{bool}}{bool}}\\{{\nwixident{get{\_}k}}{get:unk}}\\{{\nwixident{get{\_}l}}{get:unl}}\\{{\nwixident{k}}{k}}\\{{\nwixident{numeric}}{numeric}}}\nwindexuse{\nwixident{bool}}{bool}{NW3gGP3e-4H4ZFN-1}\nwindexuse{\nwixident{get{\_}k}}{get:unk}{NW3gGP3e-4H4ZFN-1}\nwindexuse{\nwixident{get{\_}l}}{get:unl}{NW3gGP3e-4H4ZFN-1}\nwindexuse{\nwixident{k}}{k}{NW3gGP3e-4H4ZFN-1}\nwindexuse{\nwixident{numeric}}{numeric}{NW3gGP3e-4H4ZFN-1}\nwendcode{}\nwbegindocs{666}For parabolas degenerated into two parallel lines we draw them by
the recursive call of this function\\
{\Tt{}\Rm{}{\bf{}cycle2D}::{\it{}metapost\_draw}()\nwendquote}.
\nwenddocs{}\nwbegincode{667}\sublabel{NW3gGP3e-24b8UV-1}\nwmargintag{{\nwtagstyle{}\subpageref{NW3gGP3e-24b8UV-1}}}\moddef{Treating a parabola~{\nwtagstyle{}\subpageref{NW3gGP3e-24b8UV-1}}}\endmoddef\Rm{}\nwstartdeflinemarkup\nwusesondefline{\\{NW3gGP3e-4H4ZFN-1}}\nwprevnextdefs{\relax}{NW3gGP3e-24b8UV-2}\nwenddeflinemarkup
{\bf{}if} ({\it{}sign0} \begin{math}\equiv\end{math} 0 \begin{math}\wedge\end{math} {\it{}Cf}.{\it{}get\_l}(0).{\it{}is\_zero}()) {\nwlbrace}
    {\bf{}if} ({\it{}with\_header})
        {\it{}ost} \begin{math}\ll\end{math} {\tt{}" /parabola degenerated into two horizontal lines"} \begin{math}\ll\end{math} {\it{}endl};
    {\bf{}cycle2D}(0, {\bf{}lst}(0, 1), 2\begin{math}\ast\end{math}{\it{}b\_roots}.{\it{}op}(0), {\it{}unit}).{\it{}metapost\_draw}({\it{}ost}, {\it{}xmin}, {\it{}xmax}, {\it{}ymin}, {\it{}ymax}, {\it{}color}, {\it{}more\_options},
                                                                       {\bf{}false}, 0, {\it{}asymptote}, {\it{}picture}, {\it{}only\_path}, {\it{}is\_continuation});
    {\bf{}cycle2D}(0, {\bf{}lst}(0, 1), 2\begin{math}\ast\end{math}{\it{}b\_roots}.{\it{}op}(1), {\it{}unit}).{\it{}metapost\_draw}({\it{}ost}, {\it{}xmin}, {\it{}xmax}, {\it{}ymin}, {\it{}ymax}, {\it{}color}, {\it{}more\_options},
                                                                       {\bf{}false}, 0, {\it{}asymptote}, {\it{}picture}, {\it{}only\_path}, {\bf{}true});
    {\bf{}if} ({\it{}with\_header})
        {\it{}ost} \begin{math}\ll\end{math} {\it{}endl};
    {\it{}ost}.{\it{}flags}({\it{}keep\_flags});
    {\bf{}return};

\nwalsodefined{\\{NW3gGP3e-24b8UV-2}\\{NW3gGP3e-24b8UV-3}\\{NW3gGP3e-24b8UV-4}\\{NW3gGP3e-24b8UV-5}\\{NW3gGP3e-24b8UV-6}}\nwused{\\{NW3gGP3e-4H4ZFN-1}}\nwidentuses{\\{{\nwixident{cycle2D}}{cycle2D}}\\{{\nwixident{get{\_}l}}{get:unl}}\\{{\nwixident{is{\_}zero}}{is:unzero}}\\{{\nwixident{lst}}{lst}}\\{{\nwixident{metapost{\_}draw}}{metapost:undraw}}\\{{\nwixident{op}}{op}}}\nwindexuse{\nwixident{cycle2D}}{cycle2D}{NW3gGP3e-24b8UV-1}\nwindexuse{\nwixident{get{\_}l}}{get:unl}{NW3gGP3e-24b8UV-1}\nwindexuse{\nwixident{is{\_}zero}}{is:unzero}{NW3gGP3e-24b8UV-1}\nwindexuse{\nwixident{lst}}{lst}{NW3gGP3e-24b8UV-1}\nwindexuse{\nwixident{metapost{\_}draw}}{metapost:undraw}{NW3gGP3e-24b8UV-1}\nwindexuse{\nwixident{op}}{op}{NW3gGP3e-24b8UV-1}\nwendcode{}\nwbegindocs{668}  Two vertical lines are drawn here
\nwenddocs{}\nwbegincode{669}\sublabel{NW3gGP3e-24b8UV-2}\nwmargintag{{\nwtagstyle{}\subpageref{NW3gGP3e-24b8UV-2}}}\moddef{Treating a parabola~{\nwtagstyle{}\subpageref{NW3gGP3e-24b8UV-1}}}\plusendmoddef\Rm{}\nwstartdeflinemarkup\nwusesondefline{\\{NW3gGP3e-4H4ZFN-1}}\nwprevnextdefs{NW3gGP3e-24b8UV-1}{NW3gGP3e-24b8UV-3}\nwenddeflinemarkup
{\nwrbrace} {\bf{}else} {\bf{}if} ({\it{}sign1} \begin{math}\equiv\end{math} 0 \begin{math}\wedge\end{math} {\it{}Cf}.{\it{}get\_l}(1).{\it{}is\_zero}()) {\nwlbrace}
    {\bf{}if} ({\it{}with\_header})
        {\it{}ost} \begin{math}\ll\end{math} {\tt{}" /parabola degenerated into two vertical lines"} \begin{math}\ll\end{math} {\it{}endl};
    {\bf{}cycle2D}(0, {\bf{}lst}(1, 0), 2\begin{math}\ast\end{math}{\it{}b\_roots}.{\it{}op}(0), {\it{}unit}).{\it{}metapost\_draw}({\it{}ost}, {\it{}xmin}, {\it{}xmax}, {\it{}ymin}, {\it{}ymax}, {\it{}color}, {\it{}more\_options},
                                                                       {\bf{}false}, 0, {\it{}asymptote}, {\it{}picture}, {\it{}only\_path}, {\it{}is\_continuation});
    {\bf{}cycle2D}(0, {\bf{}lst}(1, 0), 2\begin{math}\ast\end{math}{\it{}b\_roots}.{\it{}op}(1), {\it{}unit}).{\it{}metapost\_draw}({\it{}ost}, {\it{}xmin}, {\it{}xmax}, {\it{}ymin}, {\it{}ymax}, {\it{}color}, {\it{}more\_options},
                                                                       {\bf{}false}, 0, {\it{}asymptote}, {\it{}picture}, {\it{}only\_path}, {\bf{}true});
    {\bf{}if} ({\it{}with\_header})
        {\it{}ost} \begin{math}\ll\end{math} {\it{}endl};
    {\it{}ost}.{\it{}flags}({\it{}keep\_flags});
    {\bf{}return};
{\nwrbrace}

\nwused{\\{NW3gGP3e-4H4ZFN-1}}\nwidentuses{\\{{\nwixident{cycle2D}}{cycle2D}}\\{{\nwixident{get{\_}l}}{get:unl}}\\{{\nwixident{is{\_}zero}}{is:unzero}}\\{{\nwixident{lst}}{lst}}\\{{\nwixident{metapost{\_}draw}}{metapost:undraw}}\\{{\nwixident{op}}{op}}}\nwindexuse{\nwixident{cycle2D}}{cycle2D}{NW3gGP3e-24b8UV-2}\nwindexuse{\nwixident{get{\_}l}}{get:unl}{NW3gGP3e-24b8UV-2}\nwindexuse{\nwixident{is{\_}zero}}{is:unzero}{NW3gGP3e-24b8UV-2}\nwindexuse{\nwixident{lst}}{lst}{NW3gGP3e-24b8UV-2}\nwindexuse{\nwixident{metapost{\_}draw}}{metapost:undraw}{NW3gGP3e-24b8UV-2}\nwindexuse{\nwixident{op}}{op}{NW3gGP3e-24b8UV-2}\nwendcode{}\nwbegindocs{670}If a proper parabola is detected we rearrange intervals
appropriately in order to draw pieces properly.
\nwenddocs{}\nwbegincode{671}\sublabel{NW3gGP3e-24b8UV-3}\nwmargintag{{\nwtagstyle{}\subpageref{NW3gGP3e-24b8UV-3}}}\moddef{Treating a parabola~{\nwtagstyle{}\subpageref{NW3gGP3e-24b8UV-1}}}\plusendmoddef\Rm{}\nwstartdeflinemarkup\nwusesondefline{\\{NW3gGP3e-4H4ZFN-1}}\nwprevnextdefs{NW3gGP3e-24b8UV-2}{NW3gGP3e-24b8UV-4}\nwenddeflinemarkup
{\bf{}if} ({\it{}with\_header})
    {\it{}ost} \begin{math}\ll\end{math} {\tt{}" /parabola"} \begin{math}\ll\end{math} {\it{}endl};
 {\bf{}if} ({\it{}right}[0]-{\it{}left}[0] \begin{math}>\end{math} {\it{}epsilon} \begin{math}\wedge\end{math} {\it{}right}[1]-{\it{}left}[1] \begin{math}>\end{math} {\it{}epsilon}) {\nwlbrace}
    {\bf{}if} ({\it{}k\_d}\begin{math}\ast\end{math}({\it{}signu}\begin{math}\ast\end{math}{\it{}lv}+{\it{}signv}\begin{math}\ast\end{math}{\it{}lu}) \begin{math}>\end{math} 0) {\nwlbrace} //rearrange intervals
        {\bf{}double} {\it{}e} = {\it{}left}[1]; {\it{}left}[1] = {\it{}right}[0]; {\it{}right}[0] = {\it{}left}[0]; {\it{}left}[0] ={\it{}e};
    {\nwrbrace} {\bf{}else} {\nwlbrace}
        {\bf{}double} {\it{}e} = {\it{}left}[1]; {\it{}left}[1] = {\it{}right}[1]; {\it{}right}[1] = {\it{}right}[0]; {\it{}right}[0] ={\it{}e};
    {\nwrbrace}
 {\nwrbrace}

\nwused{\\{NW3gGP3e-4H4ZFN-1}}\nwidentuses{\\{{\nwixident{k{\_}d}}{k:und}}\\{{\nwixident{left}}{left}}}\nwindexuse{\nwixident{k{\_}d}}{k:und}{NW3gGP3e-24b8UV-3}\nwindexuse{\nwixident{left}}{left}{NW3gGP3e-24b8UV-3}\nwendcode{}\nwbegindocs{672}Parabolas can be exactly represented by a cubic B\'ezier arc if the
second and third control points correspondingly are:
\begin{eqnarray*}
  &&\left(\frac{2}{3} x_0+\frac{1}{3} x_1, \frac{1}{n}\left(\frac{1}{6} x_0^2 k + \frac{1}{3} x_0
  x_1 k- \frac{2}{3} x_0 l- \frac{1}{3} l x_1+\frac{1}{2}m \right)\right),\\
  &&\left(\frac{1}{3} x_0+\frac{2}{3} x_1, \frac{1}{n}\left(\frac{1}{3} x_0 k x_1-\frac{1}{3} x_0
  l-\frac{2}{3} l x_1+\frac{1}{6} k x_1^2+\frac{1}{2}m \right)\right).
\end{eqnarray*}
\nwenddocs{}\nwbegincode{673}\sublabel{NW3gGP3e-24b8UV-4}\nwmargintag{{\nwtagstyle{}\subpageref{NW3gGP3e-24b8UV-4}}}\moddef{Treating a parabola~{\nwtagstyle{}\subpageref{NW3gGP3e-24b8UV-1}}}\plusendmoddef\Rm{}\nwstartdeflinemarkup\nwusesondefline{\\{NW3gGP3e-4H4ZFN-1}}\nwprevnextdefs{NW3gGP3e-24b8UV-3}{NW3gGP3e-24b8UV-5}\nwenddeflinemarkup
{\bf{}for} ({\bf{}int} {\it{}i} =0; {\it{}i} \begin{math}<\end{math} 2; {\it{}i}\protect\PP) {\nwlbrace}
    {\bf{}if} ({\it{}right}[{\it{}i}]-{\it{}left}[{\it{}i}] \begin{math}>\end{math} {\it{}epsilon}) {\nwlbrace} // a proper branch of a parabola
        {\bf{}double} {\it{}cp}[8];
        {\bf{}if} ({\it{}not\_swapped}) {\nwlbrace}
            {\it{}cp}[0] = {\it{}left}[{\it{}i}];
            {\it{}cp}[1] = {\it{}ex\_to}\begin{math}<\end{math}{\bf{}numeric}\begin{math}>\end{math}({\it{}Cf}.{\it{}val}({\bf{}lst}({\it{}cp}[0],0))\begin{math}\div\end{math}2.0\begin{math}\div\end{math}{\it{}Cf}.{\it{}get\_l}(1)).{\it{}to\_double}();
            {\it{}cp}[6] = {\it{}right}[{\it{}i}];
            {\it{}cp}[7] = {\it{}ex\_to}\begin{math}<\end{math}{\bf{}numeric}\begin{math}>\end{math}({\it{}Cf}.{\it{}val}({\bf{}lst}({\it{}cp}[6],0))\begin{math}\div\end{math}2.0\begin{math}\div\end{math}{\it{}Cf}.{\it{}get\_l}(1)).{\it{}to\_double}();
            {\it{}cp}[2] = 2.0\begin{math}\div\end{math}3.0\begin{math}\ast\end{math}{\it{}cp}[0]+1.0\begin{math}\div\end{math}3.0\begin{math}\ast\end{math}{\it{}cp}[6];
            {\it{}cp}[3] = {\it{}ex\_to}\begin{math}<\end{math}{\bf{}numeric}\begin{math}>\end{math}(({\bf{}numeric}(1,6)\begin{math}\ast\end{math}{\it{}cp}[0]\begin{math}\ast\end{math}{\it{}cp}[0]\begin{math}\ast\end{math}{\it{}Cf}.{\it{}get\_k}() + 1.0\begin{math}\div\end{math}3.0\begin{math}\ast\end{math}{\it{}cp}[0]\begin{math}\ast\end{math}{\it{}cp}[6]\begin{math}\ast\end{math}{\it{}Cf}.{\it{}get\_k}()
                                    - 2.0\begin{math}\div\end{math}3.0\begin{math}\ast\end{math}{\it{}cp}[0]\begin{math}\ast\end{math}{\it{}Cf}.{\it{}get\_l}(0)- 1.0\begin{math}\div\end{math}3.0\begin{math}\ast\end{math}{\it{}Cf}.{\it{}get\_l}(0)\begin{math}\ast\end{math}{\it{}cp}[6]+{\it{}Cf}.{\it{}get\_m}()\begin{math}\div\end{math}2.0)\begin{math}\div\end{math}{\it{}Cf}.{\it{}get\_l}(1)).{\it{}to\_double}();
            {\it{}cp}[4] = 1.0\begin{math}\div\end{math}3.0\begin{math}\ast\end{math}{\it{}cp}[0]+2.0\begin{math}\div\end{math}3.0\begin{math}\ast\end{math}{\it{}cp}[6];
            {\it{}cp}[5] = {\it{}ex\_to}\begin{math}<\end{math}{\bf{}numeric}\begin{math}>\end{math}((1.0\begin{math}\div\end{math}3.0\begin{math}\ast\end{math}{\it{}cp}[0]\begin{math}\ast\end{math}{\it{}Cf}.{\it{}get\_k}()\begin{math}\ast\end{math}{\it{}cp}[6]-1.0\begin{math}\div\end{math}3.0\begin{math}\ast\end{math}{\it{}cp}[0]\begin{math}\ast\end{math}{\it{}Cf}.{\it{}get\_l}(0)
                                    -2.0\begin{math}\div\end{math}3.0\begin{math}\ast\end{math}{\it{}Cf}.{\it{}get\_l}(0)\begin{math}\ast\end{math}{\it{}cp}[6]+{\bf{}numeric}(1,6)\begin{math}\ast\end{math}{\it{}Cf}.{\it{}get\_k}()\begin{math}\ast\end{math}{\it{}cp}[6]\begin{math}\ast\end{math}{\it{}cp}[6]+{\it{}Cf}.{\it{}get\_m}()\begin{math}\div\end{math}2.0)\begin{math}\div\end{math}{\it{}Cf}.{\it{}get\_l}(1)).{\it{}to\_double}();

\nwused{\\{NW3gGP3e-4H4ZFN-1}}\nwidentuses{\\{{\nwixident{get{\_}k}}{get:unk}}\\{{\nwixident{get{\_}l}}{get:unl}}\\{{\nwixident{get{\_}m}}{get:unm}}\\{{\nwixident{left}}{left}}\\{{\nwixident{lst}}{lst}}\\{{\nwixident{numeric}}{numeric}}\\{{\nwixident{val}}{val}}}\nwindexuse{\nwixident{get{\_}k}}{get:unk}{NW3gGP3e-24b8UV-4}\nwindexuse{\nwixident{get{\_}l}}{get:unl}{NW3gGP3e-24b8UV-4}\nwindexuse{\nwixident{get{\_}m}}{get:unm}{NW3gGP3e-24b8UV-4}\nwindexuse{\nwixident{left}}{left}{NW3gGP3e-24b8UV-4}\nwindexuse{\nwixident{lst}}{lst}{NW3gGP3e-24b8UV-4}\nwindexuse{\nwixident{numeric}}{numeric}{NW3gGP3e-24b8UV-4}\nwindexuse{\nwixident{val}}{val}{NW3gGP3e-24b8UV-4}\nwendcode{}\nwbegindocs{674}The similar formulae for swapped drawing.
\nwenddocs{}\nwbegincode{675}\sublabel{NW3gGP3e-24b8UV-5}\nwmargintag{{\nwtagstyle{}\subpageref{NW3gGP3e-24b8UV-5}}}\moddef{Treating a parabola~{\nwtagstyle{}\subpageref{NW3gGP3e-24b8UV-1}}}\plusendmoddef\Rm{}\nwstartdeflinemarkup\nwusesondefline{\\{NW3gGP3e-4H4ZFN-1}}\nwprevnextdefs{NW3gGP3e-24b8UV-4}{NW3gGP3e-24b8UV-6}\nwenddeflinemarkup
  {\nwrbrace} {\bf{}else} {\nwlbrace}
            {\it{}cp}[1] = {\it{}left}[{\it{}i}];
            {\it{}cp}[0] = {\it{}ex\_to}\begin{math}<\end{math}{\bf{}numeric}\begin{math}>\end{math}({\it{}Cf}.{\it{}val}({\bf{}lst}(0,{\it{}cp}[1]))\begin{math}\div\end{math}2.0\begin{math}\div\end{math}{\it{}Cf}.{\it{}get\_l}(0)).{\it{}to\_double}();
            {\it{}cp}[7] = {\it{}right}[{\it{}i}];
            {\it{}cp}[6] = {\it{}ex\_to}\begin{math}<\end{math}{\bf{}numeric}\begin{math}>\end{math}({\it{}Cf}.{\it{}val}({\bf{}lst}(0,{\it{}cp}[7]))\begin{math}\div\end{math}2.0\begin{math}\div\end{math}{\it{}Cf}.{\it{}get\_l}(0)).{\it{}to\_double}();
            {\it{}cp}[3] = 2.0\begin{math}\div\end{math}3.0\begin{math}\ast\end{math}{\it{}cp}[1]+1.0\begin{math}\div\end{math}3.0\begin{math}\ast\end{math}{\it{}cp}[7];
            {\it{}cp}[2] = {\it{}ex\_to}\begin{math}<\end{math}{\bf{}numeric}\begin{math}>\end{math}(({\bf{}numeric}(1,6)\begin{math}\ast\end{math}{\it{}cp}[1]\begin{math}\ast\end{math}{\it{}cp}[1]\begin{math}\ast\end{math}{\it{}Cf}.{\it{}get\_k}() + 1.0\begin{math}\div\end{math}3.0\begin{math}\ast\end{math}{\it{}cp}[1]\begin{math}\ast\end{math}{\it{}cp}[7]\begin{math}\ast\end{math}{\it{}Cf}.{\it{}get\_k}()
                                    - 2.0\begin{math}\div\end{math}3.0\begin{math}\ast\end{math}{\it{}cp}[1]\begin{math}\ast\end{math}{\it{}Cf}.{\it{}get\_l}(1)- 1.0\begin{math}\div\end{math}3.0\begin{math}\ast\end{math}{\it{}Cf}.{\it{}get\_l}(1)\begin{math}\ast\end{math}{\it{}cp}[7]+{\it{}Cf}.{\it{}get\_m}()\begin{math}\div\end{math}2.0)\begin{math}\div\end{math}{\it{}Cf}.{\it{}get\_l}(0)).{\it{}to\_double}();
            {\it{}cp}[5] = 1.0\begin{math}\div\end{math}3.0\begin{math}\ast\end{math}{\it{}cp}[1]+2.0\begin{math}\div\end{math}3.0\begin{math}\ast\end{math}{\it{}cp}[7];
            {\it{}cp}[4] = {\it{}ex\_to}\begin{math}<\end{math}{\bf{}numeric}\begin{math}>\end{math}((1.0\begin{math}\div\end{math}3.0\begin{math}\ast\end{math}{\it{}cp}[1]\begin{math}\ast\end{math}{\it{}Cf}.{\it{}get\_k}()\begin{math}\ast\end{math}{\it{}cp}[7]-1.0\begin{math}\div\end{math}3.0\begin{math}\ast\end{math}{\it{}cp}[1]\begin{math}\ast\end{math}{\it{}Cf}.{\it{}get\_l}(1)
                                    -2.0\begin{math}\div\end{math}3.0\begin{math}\ast\end{math}{\it{}Cf}.{\it{}get\_l}(1)\begin{math}\ast\end{math}{\it{}cp}[7]+{\bf{}numeric}(1,6)\begin{math}\ast\end{math}{\it{}Cf}.{\it{}get\_k}()\begin{math}\ast\end{math}{\it{}cp}[7]\begin{math}\ast\end{math}{\it{}cp}[7]+{\it{}Cf}.{\it{}get\_m}()\begin{math}\div\end{math}2.0)\begin{math}\div\end{math}{\it{}Cf}.{\it{}get\_l}(0)).{\it{}to\_double}();
  {\nwrbrace}

\nwused{\\{NW3gGP3e-4H4ZFN-1}}\nwidentuses{\\{{\nwixident{get{\_}k}}{get:unk}}\\{{\nwixident{get{\_}l}}{get:unl}}\\{{\nwixident{get{\_}m}}{get:unm}}\\{{\nwixident{left}}{left}}\\{{\nwixident{lst}}{lst}}\\{{\nwixident{numeric}}{numeric}}\\{{\nwixident{val}}{val}}}\nwindexuse{\nwixident{get{\_}k}}{get:unk}{NW3gGP3e-24b8UV-5}\nwindexuse{\nwixident{get{\_}l}}{get:unl}{NW3gGP3e-24b8UV-5}\nwindexuse{\nwixident{get{\_}m}}{get:unm}{NW3gGP3e-24b8UV-5}\nwindexuse{\nwixident{left}}{left}{NW3gGP3e-24b8UV-5}\nwindexuse{\nwixident{lst}}{lst}{NW3gGP3e-24b8UV-5}\nwindexuse{\nwixident{numeric}}{numeric}{NW3gGP3e-24b8UV-5}\nwindexuse{\nwixident{val}}{val}{NW3gGP3e-24b8UV-5}\nwendcode{}\nwbegindocs{676}The actual drawing of the parabola arcs.
\nwenddocs{}\nwbegincode{677}\sublabel{NW3gGP3e-24b8UV-6}\nwmargintag{{\nwtagstyle{}\subpageref{NW3gGP3e-24b8UV-6}}}\moddef{Treating a parabola~{\nwtagstyle{}\subpageref{NW3gGP3e-24b8UV-1}}}\plusendmoddef\Rm{}\nwstartdeflinemarkup\nwusesondefline{\\{NW3gGP3e-4H4ZFN-1}}\nwprevnextdefs{NW3gGP3e-24b8UV-5}{\relax}\nwenddeflinemarkup
    {\it{}ost} \begin{math}\ll\end{math}  ({\it{}only\_path} ? {\it{}already\_drawn} : {\it{}draw\_start}.{\it{}str}()) \begin{math}\ll\end{math} {\it{}cp}[0] \begin{math}\ll\end{math} {\tt{}","} \begin{math}\ll\end{math} {\it{}cp}[1] \begin{math}\ll\end{math} {\tt{}") .. controls ("};
{\bf{}if} ({\it{}asymptote})
    {\it{}ost} \begin{math}\ll\end{math}  {\it{}cp}[2] \begin{math}\ll\end{math} {\tt{}","} \begin{math}\ll\end{math} {\it{}cp}[3] \begin{math}\ll\end{math} {\tt{}") and ("} \begin{math}\ll\end{math}  {\it{}cp}[4] \begin{math}\ll\end{math} {\tt{}","} \begin{math}\ll\end{math} {\it{}cp}[5] \begin{math}\ll\end{math} {\tt{}") .. ("};
{\bf{}else}
    {\it{}ost} \begin{math}\ll\end{math}  {\tt{}"("} \begin{math}\ll\end{math} {\it{}cp}[2] \begin{math}\ll\end{math} {\tt{}","} \begin{math}\ll\end{math} {\it{}cp}[3] \begin{math}\ll\end{math} {\tt{}")) and (("} \begin{math}\ll\end{math}  {\it{}cp}[4] \begin{math}\ll\end{math} {\tt{}","} \begin{math}\ll\end{math} {\it{}cp}[5] \begin{math}\ll\end{math} {\tt{}")) .. ("};
{\it{}ost} \begin{math}\ll\end{math}  {\it{}cp}[6] \begin{math}\ll\end{math} {\tt{}","} \begin{math}\ll\end{math} {\it{}cp}[7] \begin{math}\ll\end{math} {\tt{}")"} \begin{math}\ll\end{math} ({\it{}only\_path} ? {\tt{}""} : {\it{}draw\_options}.{\it{}str}());
{\it{}already\_drawn}={\tt{}"^^("};
    {\nwrbrace}
{\nwrbrace}

\nwused{\\{NW3gGP3e-4H4ZFN-1}}\nwendcode{}\nwbegindocs{678}If a hyperbola degenerates into a light cone we draw it as two
separate lines.
\nwenddocs{}\nwbegincode{679}\sublabel{NW3gGP3e-2yW1fi-1}\nwmargintag{{\nwtagstyle{}\subpageref{NW3gGP3e-2yW1fi-1}}}\moddef{Treating a hyperbola~{\nwtagstyle{}\subpageref{NW3gGP3e-2yW1fi-1}}}\endmoddef\Rm{}\nwstartdeflinemarkup\nwusesondefline{\\{NW3gGP3e-4H4ZFN-1}}\nwprevnextdefs{\relax}{NW3gGP3e-2yW1fi-2}\nwenddeflinemarkup
{\bf{}if} ({\it{}abs}({\it{}determinant})\begin{math}<\end{math}{\it{}epsilon}) {\nwlbrace}
    {\bf{}if} ({\it{}with\_header})
        {\it{}ost} \begin{math}\ll\end{math} {\tt{}" / a light cone at ("} \begin{math}\ll\end{math} {\it{}xc} \begin{math}\ll\end{math} {\tt{}","} \begin{math}\ll\end{math} {\it{}yc} \begin{math}\ll\end{math}{\tt{}")"} \begin{math}\ll\end{math} {\it{}endl};
    {\bf{}cycle2D}(0, {\bf{}lst}(1, 1), 2\begin{math}\ast\end{math}({\it{}uc}+{\it{}vc}), {\it{}unit}).{\it{}metapost\_draw}({\it{}ost}, {\it{}xmin}, {\it{}xmax}, {\it{}ymin}, {\it{}ymax}, {\it{}color}, {\it{}more\_options},
                                                                 {\bf{}false}, 0, {\it{}asymptote}, {\it{}picture}, {\it{}only\_path}, {\it{}is\_continuation});
    {\bf{}cycle2D}(0, {\bf{}lst}(1, -1), 2\begin{math}\ast\end{math}({\it{}uc}-{\it{}vc}), {\it{}unit}).{\it{}metapost\_draw}({\it{}ost}, {\it{}xmin}, {\it{}xmax}, {\it{}ymin}, {\it{}ymax}, {\it{}color}, {\it{}more\_options},
                                                                  {\bf{}false}, 0, {\it{}asymptote}, {\it{}picture}, {\it{}only\_path}, {\bf{}true});

\nwalsodefined{\\{NW3gGP3e-2yW1fi-2}\\{NW3gGP3e-2yW1fi-3}\\{NW3gGP3e-2yW1fi-4}\\{NW3gGP3e-2yW1fi-5}}\nwused{\\{NW3gGP3e-4H4ZFN-1}}\nwidentuses{\\{{\nwixident{cycle2D}}{cycle2D}}\\{{\nwixident{lst}}{lst}}\\{{\nwixident{metapost{\_}draw}}{metapost:undraw}}}\nwindexuse{\nwixident{cycle2D}}{cycle2D}{NW3gGP3e-2yW1fi-1}\nwindexuse{\nwixident{lst}}{lst}{NW3gGP3e-2yW1fi-1}\nwindexuse{\nwixident{metapost{\_}draw}}{metapost:undraw}{NW3gGP3e-2yW1fi-1}\nwendcode{}\nwbegindocs{680}We also put a dot to single out the light cone vertex.
\nwenddocs{}\nwbegincode{681}\sublabel{NW3gGP3e-2yW1fi-2}\nwmargintag{{\nwtagstyle{}\subpageref{NW3gGP3e-2yW1fi-2}}}\moddef{Treating a hyperbola~{\nwtagstyle{}\subpageref{NW3gGP3e-2yW1fi-1}}}\plusendmoddef\Rm{}\nwstartdeflinemarkup\nwusesondefline{\\{NW3gGP3e-4H4ZFN-1}}\nwprevnextdefs{NW3gGP3e-2yW1fi-1}{NW3gGP3e-2yW1fi-3}\nwenddeflinemarkup
    {\bf{}if} (\begin{math}\neg\end{math} {\it{}only\_path}) {\nwlbrace}
        \LA{}place a dot~{\nwtagstyle{}\subpageref{NW3gGP3e-FZdRR-1}}\RA{}
        {\bf{}if} ({\it{}with\_header})
            {\it{}ost} \begin{math}\ll\end{math} {\it{}endl};
    {\nwrbrace}
    {\it{}ost}.{\it{}flags}({\it{}keep\_flags});
    {\bf{}return};

\nwused{\\{NW3gGP3e-4H4ZFN-1}}\nwendcode{}\nwbegindocs{682}Otherwise we rearrange the interwals for hyperbola branches.
\nwenddocs{}\nwbegincode{683}\sublabel{NW3gGP3e-2yW1fi-3}\nwmargintag{{\nwtagstyle{}\subpageref{NW3gGP3e-2yW1fi-3}}}\moddef{Treating a hyperbola~{\nwtagstyle{}\subpageref{NW3gGP3e-2yW1fi-1}}}\plusendmoddef\Rm{}\nwstartdeflinemarkup\nwusesondefline{\\{NW3gGP3e-4H4ZFN-1}}\nwprevnextdefs{NW3gGP3e-2yW1fi-2}{NW3gGP3e-2yW1fi-4}\nwenddeflinemarkup
{\nwrbrace} {\bf{}else} {\nwlbrace}
    {\bf{}if} ({\it{}with\_header})
        {\it{}ost} \begin{math}\ll\end{math} {\tt{}" /hyperbola"} \begin{math}\ll\end{math} {\it{}endl};
    {\bf{}if} ({\it{}vmin}-{\it{}vc} \begin{math}>\end{math} {\it{}epsilon}) {\nwlbrace}
        {\bf{}double} {\it{}e} = {\it{}left}[1]; {\it{}left}[1] = {\it{}right}[0]; {\it{}right}[0] = {\it{}left}[0]; {\it{}left}[0] ={\it{}e};
        {\it{}change\_branch} = {\bf{}false};
        {\it{}zero\_or\_one} = ({\it{}k\_d}\begin{math}\ast\end{math}{\it{}signv} \begin{math}>\end{math} 0 ? 1 : 0);
    {\nwrbrace}
    {\bf{}if} ({\it{}vc}-{\it{}vmax} \begin{math}>\end{math} {\it{}epsilon}) {\nwlbrace}
        {\bf{}double} {\it{}e} = {\it{}left}[1]; {\it{}left}[1] = {\it{}right}[1]; {\it{}right}[1] = {\it{}right}[0]; {\it{}right}[0] ={\it{}e};
        {\it{}change\_branch} = {\bf{}false};
        {\it{}zero\_or\_one} = ({\it{}k\_d}\begin{math}\ast\end{math}{\it{}signv} \begin{math}>\end{math} 0 ? 0 : 1);
    {\nwrbrace}
{\nwrbrace}

\nwused{\\{NW3gGP3e-4H4ZFN-1}}\nwidentuses{\\{{\nwixident{k{\_}d}}{k:und}}\\{{\nwixident{left}}{left}}\\{{\nwixident{zero{\_}or{\_}one}}{zero:unor:unone}}}\nwindexuse{\nwixident{k{\_}d}}{k:und}{NW3gGP3e-2yW1fi-3}\nwindexuse{\nwixident{left}}{left}{NW3gGP3e-2yW1fi-3}\nwindexuse{\nwixident{zero{\_}or{\_}one}}{zero:unor:unone}{NW3gGP3e-2yW1fi-3}\nwendcode{}\nwbegindocs{684}Two arcs of the hyperbola are drown now
\nwenddocs{}\nwbegincode{685}\sublabel{NW3gGP3e-2yW1fi-4}\nwmargintag{{\nwtagstyle{}\subpageref{NW3gGP3e-2yW1fi-4}}}\moddef{Treating a hyperbola~{\nwtagstyle{}\subpageref{NW3gGP3e-2yW1fi-1}}}\plusendmoddef\Rm{}\nwstartdeflinemarkup\nwusesondefline{\\{NW3gGP3e-4H4ZFN-1}}\nwprevnextdefs{NW3gGP3e-2yW1fi-3}{NW3gGP3e-2yW1fi-5}\nwenddeflinemarkup
{\bf{}int} {\it{}points} = ({\it{}points\_per\_arc} \begin{math}\equiv\end{math} 0? 7 : {\it{}points\_per\_arc});\nwindexdefn{\nwixident{points}}{points}{NW3gGP3e-2yW1fi-4}
{\bf{}for} ({\bf{}int} {\it{}i} =0; {\it{}i} \begin{math}<\end{math} 2; {\it{}i}\protect\PP) {\nwlbrace}
    {\bf{}double} {\it{}dir} = {\it{}ex\_to}\begin{math}<\end{math}{\bf{}numeric}\begin{math}>\end{math}({\it{}csgn}({\it{}signv}\begin{math}\ast\end{math}(2\begin{math}\ast\end{math}{\it{}zero\_or\_one}-1))).{\it{}to\_double}(); //direction of the tangent vectors
    // double dir = ((sign == 0? lv : signv*(2*zero\_or\_one-1))\begin{math}<\end{math}0?-1:1); direction of the tangent vectors (second alternative)
    {\bf{}if} ({\it{}right}[{\it{}i}]-{\it{}left}[{\it{}i}] \begin{math}>\end{math} {\it{}epsilon} ) {\nwlbrace} // a proper branch of the hyperbola

\nwused{\\{NW3gGP3e-4H4ZFN-1}}\nwidentdefs{\\{{\nwixident{points}}{points}}}\nwidentuses{\\{{\nwixident{left}}{left}}\\{{\nwixident{numeric}}{numeric}}\\{{\nwixident{zero{\_}or{\_}one}}{zero:unor:unone}}}\nwindexuse{\nwixident{left}}{left}{NW3gGP3e-2yW1fi-4}\nwindexuse{\nwixident{numeric}}{numeric}{NW3gGP3e-2yW1fi-4}\nwindexuse{\nwixident{zero{\_}or{\_}one}}{zero:unor:unone}{NW3gGP3e-2yW1fi-4}\nwendcode{}\nwbegindocs{686}Points for the spline are placed equally spaced in the hyperbolic
angle parameter.
\nwenddocs{}\nwbegincode{687}\sublabel{NW3gGP3e-2yW1fi-5}\nwmargintag{{\nwtagstyle{}\subpageref{NW3gGP3e-2yW1fi-5}}}\moddef{Treating a hyperbola~{\nwtagstyle{}\subpageref{NW3gGP3e-2yW1fi-1}}}\plusendmoddef\Rm{}\nwstartdeflinemarkup\nwusesondefline{\\{NW3gGP3e-4H4ZFN-1}}\nwprevnextdefs{NW3gGP3e-2yW1fi-4}{\relax}\nwenddeflinemarkup
        {\bf{}double} {\it{}f\_left}={\it{}asinh}(({\it{}left}[{\it{}i}]-{\it{}uc})\begin{math}\div\end{math}{\it{}r}), {\it{}f\_right}={\it{}asinh}(({\it{}right}[{\it{}i}]-{\it{}uc})\begin{math}\div\end{math}{\it{}r});
        {\it{}DRAW\_ARC}({\it{}sinh}({\it{}f\_left})\begin{math}\ast\end{math}{\it{}r}+{\it{}uc}, ({\it{}only\_path} ? {\it{}already\_drawn} : {\it{}draw\_start}.{\it{}str}()));
        {\bf{}for} ({\bf{}int} {\it{}j}=1; {\it{}j}\begin{math}<\end{math}{\it{}points}; {\it{}j}\protect\PP) {\nwlbrace}
            {\it{}DRAW\_ARC}({\it{}sinh}({\it{}f\_left}\begin{math}\ast\end{math}(1.0-{\it{}j}\begin{math}\div\end{math}({\it{}points}-1.0))+{\it{}f\_right}\begin{math}\ast\end{math}{\it{}j}\begin{math}\div\end{math}({\it{}points}-1.0))\begin{math}\ast\end{math}{\it{}r}+{\it{}uc},
                     ({\it{}asymptote} ? {\tt{}"::("} : {\tt{}"...("}) );
        {\nwrbrace}
        {\it{}ost} \begin{math}\ll\end{math} ({\it{}only\_path} ? {\tt{}""} : {\it{}draw\_options}.{\it{}str}());
        {\it{}already\_drawn}={\tt{}"^^("};
    {\nwrbrace}
    {\bf{}if} ({\it{}change\_branch})
        {\it{}zero\_or\_one} = 1 - {\it{}zero\_or\_one}; // make a swap for the next branch of hyperbola
 {\nwrbrace}

\nwused{\\{NW3gGP3e-4H4ZFN-1}}\nwidentuses{\\{{\nwixident{DRAW{\_}ARC}}{DRAW:unARC}}\\{{\nwixident{left}}{left}}\\{{\nwixident{points}}{points}}\\{{\nwixident{zero{\_}or{\_}one}}{zero:unor:unone}}}\nwindexuse{\nwixident{DRAW{\_}ARC}}{DRAW:unARC}{NW3gGP3e-2yW1fi-5}\nwindexuse{\nwixident{left}}{left}{NW3gGP3e-2yW1fi-5}\nwindexuse{\nwixident{points}}{points}{NW3gGP3e-2yW1fi-5}\nwindexuse{\nwixident{zero{\_}or{\_}one}}{zero:unor:unone}{NW3gGP3e-2yW1fi-5}\nwendcode{}\nwbegindocs{688}\nwdocspar
\subsection{Auxiliary functions implementation}
The auxillary functions defined as well.

\nwenddocs{}\nwbegindocs{689}\nwdocspar

\subsubsection{Heaviside function}
\label{sec:heaviside-function}

We define Heaviside function: \(\chi(x)=1\) for \(x\geq0\) and \(\chi(x)=0\) for \(x<0\).
\nwenddocs{}\nwbegincode{690}\sublabel{NW3gGP3e-1Oxbp0-1C}\nwmargintag{{\nwtagstyle{}\subpageref{NW3gGP3e-1Oxbp0-1C}}}\moddef{cycle.cpp~{\nwtagstyle{}\subpageref{NW3gGP3e-1Oxbp0-1}}}\plusendmoddef\Rm{}\nwstartdeflinemarkup\nwprevnextdefs{NW3gGP3e-1Oxbp0-1B}{NW3gGP3e-1Oxbp0-1D}\nwenddeflinemarkup
//////////
// Jump function
//////////

{\bf{}static} {\bf{}ex} {\it{}jump\_fnct\_evalf}({\bf{}const} {\bf{}ex} & {\it{}arg})\nwindexdefn{\nwixident{ex}}{ex}{NW3gGP3e-1Oxbp0-1C}
{\nwlbrace}
 {\bf{}if} ({\it{}is\_exactly\_a}\begin{math}<\end{math}{\bf{}numeric}\begin{math}>\end{math}({\it{}arg})) {\nwlbrace}
  {\bf{}if} (({\it{}ex\_to}\begin{math}<\end{math}{\bf{}numeric}\begin{math}>\end{math}({\it{}arg}).{\it{}is\_real}() \begin{math}\wedge\end{math} {\it{}ex\_to}\begin{math}<\end{math}{\bf{}numeric}\begin{math}>\end{math}({\it{}arg}).{\it{}is\_positive}()) \begin{math}\vee\end{math} {\it{}ex\_to}\begin{math}<\end{math}{\bf{}numeric}\begin{math}>\end{math}({\it{}arg}).{\it{}is\_zero}())
   {\bf{}return} {\bf{}numeric}(1);
  {\bf{}else}
   {\bf{}return} {\bf{}numeric}(-1);
 {\nwrbrace}

 {\bf{}return} {\it{}jump\_fnct}({\it{}arg}).{\it{}hold}();
{\nwrbrace}

\nwidentdefs{\\{{\nwixident{ex}}{ex}}}\nwidentuses{\\{{\nwixident{is{\_}zero}}{is:unzero}}\\{{\nwixident{jump{\_}fnct}}{jump:unfnct}}\\{{\nwixident{numeric}}{numeric}}}\nwindexuse{\nwixident{is{\_}zero}}{is:unzero}{NW3gGP3e-1Oxbp0-1C}\nwindexuse{\nwixident{jump{\_}fnct}}{jump:unfnct}{NW3gGP3e-1Oxbp0-1C}\nwindexuse{\nwixident{numeric}}{numeric}{NW3gGP3e-1Oxbp0-1C}\nwendcode{}\nwbegindocs{691}\nwdocspar
\nwenddocs{}\nwbegincode{692}\sublabel{NW3gGP3e-1Oxbp0-1D}\nwmargintag{{\nwtagstyle{}\subpageref{NW3gGP3e-1Oxbp0-1D}}}\moddef{cycle.cpp~{\nwtagstyle{}\subpageref{NW3gGP3e-1Oxbp0-1}}}\plusendmoddef\Rm{}\nwstartdeflinemarkup\nwprevnextdefs{NW3gGP3e-1Oxbp0-1C}{NW3gGP3e-1Oxbp0-1E}\nwenddeflinemarkup
{\bf{}static} {\bf{}ex} {\it{}jump\_fnct\_eval}({\bf{}const} {\bf{}ex} & {\it{}arg})\nwindexdefn{\nwixident{ex}}{ex}{NW3gGP3e-1Oxbp0-1D}
{\nwlbrace}
 {\bf{}if} ({\it{}is\_exactly\_a}\begin{math}<\end{math}{\bf{}numeric}\begin{math}>\end{math}({\it{}arg})) {\nwlbrace}
  {\bf{}if} (({\it{}ex\_to}\begin{math}<\end{math}{\bf{}numeric}\begin{math}>\end{math}({\it{}arg}).{\it{}is\_real}() \begin{math}\wedge\end{math} {\it{}ex\_to}\begin{math}<\end{math}{\bf{}numeric}\begin{math}>\end{math}({\it{}arg}).{\it{}is\_positive}()) \begin{math}\vee\end{math} {\it{}ex\_to}\begin{math}<\end{math}{\bf{}numeric}\begin{math}>\end{math}({\it{}arg}).{\it{}is\_zero}())
   {\bf{}return} {\bf{}numeric}(1);
  {\bf{}else}
   {\bf{}return} {\bf{}numeric}(-1);
 {\nwrbrace} {\bf{}else} {\bf{}if} ({\it{}is\_exactly\_a}\begin{math}<\end{math}{\it{}mul}\begin{math}>\end{math}({\it{}arg}) \begin{math}\wedge\end{math}
      {\it{}is\_exactly\_a}\begin{math}<\end{math}{\bf{}numeric}\begin{math}>\end{math}({\it{}arg}.{\it{}op}({\it{}arg}.{\it{}nops}()-1))) {\nwlbrace}
  {\bf{}numeric} {\it{}oc} = {\it{}ex\_to}\begin{math}<\end{math}{\bf{}numeric}\begin{math}>\end{math}({\it{}arg}.{\it{}op}({\it{}arg}.{\it{}nops}()-1));
  {\bf{}if} ({\it{}oc}.{\it{}is\_real}()) {\nwlbrace}
   {\bf{}if} ({\it{}oc} \begin{math}>\end{math} 0)
    // jump\_fnct(42*x) -\begin{math}>\end{math} jump\_fnct(x)
    {\bf{}return} {\it{}jump\_fnct}({\it{}arg}\begin{math}\div\end{math}{\it{}oc}).{\it{}hold}();
   {\bf{}else}
    // jump\_fnct(-42*x) -\begin{math}>\end{math} jump\_fnct(-x)
    {\bf{}return} {\it{}jump\_fnct}(-{\it{}arg}\begin{math}\div\end{math}{\it{}oc}).{\it{}hold}();
  {\nwrbrace}
 {\nwrbrace}
 {\bf{}return} {\it{}jump\_fnct}({\it{}arg}).{\it{}hold}();
{\nwrbrace}

\nwidentdefs{\\{{\nwixident{ex}}{ex}}}\nwidentuses{\\{{\nwixident{is{\_}zero}}{is:unzero}}\\{{\nwixident{jump{\_}fnct}}{jump:unfnct}}\\{{\nwixident{mul}}{mul}}\\{{\nwixident{nops}}{nops}}\\{{\nwixident{numeric}}{numeric}}\\{{\nwixident{op}}{op}}}\nwindexuse{\nwixident{is{\_}zero}}{is:unzero}{NW3gGP3e-1Oxbp0-1D}\nwindexuse{\nwixident{jump{\_}fnct}}{jump:unfnct}{NW3gGP3e-1Oxbp0-1D}\nwindexuse{\nwixident{mul}}{mul}{NW3gGP3e-1Oxbp0-1D}\nwindexuse{\nwixident{nops}}{nops}{NW3gGP3e-1Oxbp0-1D}\nwindexuse{\nwixident{numeric}}{numeric}{NW3gGP3e-1Oxbp0-1D}\nwindexuse{\nwixident{op}}{op}{NW3gGP3e-1Oxbp0-1D}\nwendcode{}\nwbegindocs{693}\nwdocspar
\nwenddocs{}\nwbegincode{694}\sublabel{NW3gGP3e-1Oxbp0-1E}\nwmargintag{{\nwtagstyle{}\subpageref{NW3gGP3e-1Oxbp0-1E}}}\moddef{cycle.cpp~{\nwtagstyle{}\subpageref{NW3gGP3e-1Oxbp0-1}}}\plusendmoddef\Rm{}\nwstartdeflinemarkup\nwprevnextdefs{NW3gGP3e-1Oxbp0-1D}{NW3gGP3e-1Oxbp0-1F}\nwenddeflinemarkup
{\bf{}static} {\bf{}ex} {\it{}jump\_fnct\_conjugate}({\bf{}const} {\bf{}ex} & {\it{}arg})\nwindexdefn{\nwixident{ex}}{ex}{NW3gGP3e-1Oxbp0-1E}
{\nwlbrace}
 {\bf{}return} {\it{}jump\_fnct}({\it{}arg});
{\nwrbrace}

\nwidentdefs{\\{{\nwixident{ex}}{ex}}}\nwidentuses{\\{{\nwixident{jump{\_}fnct}}{jump:unfnct}}}\nwindexuse{\nwixident{jump{\_}fnct}}{jump:unfnct}{NW3gGP3e-1Oxbp0-1E}\nwendcode{}\nwbegindocs{695}\nwdocspar
\nwenddocs{}\nwbegincode{696}\sublabel{NW3gGP3e-1Oxbp0-1F}\nwmargintag{{\nwtagstyle{}\subpageref{NW3gGP3e-1Oxbp0-1F}}}\moddef{cycle.cpp~{\nwtagstyle{}\subpageref{NW3gGP3e-1Oxbp0-1}}}\plusendmoddef\Rm{}\nwstartdeflinemarkup\nwprevnextdefs{NW3gGP3e-1Oxbp0-1E}{NW3gGP3e-1Oxbp0-1G}\nwenddeflinemarkup
{\bf{}static} {\bf{}ex} {\it{}jump\_fnct\_power}({\bf{}const} {\bf{}ex} & {\it{}arg}, {\bf{}const} {\bf{}ex} & {\it{}exp})\nwindexdefn{\nwixident{ex}}{ex}{NW3gGP3e-1Oxbp0-1F}
{\nwlbrace}
 {\bf{}if} ({\it{}is\_a}\begin{math}<\end{math}{\bf{}numeric}\begin{math}>\end{math}({\it{}exp}) \begin{math}\wedge\end{math} {\it{}ex\_to}\begin{math}<\end{math}{\bf{}numeric}\begin{math}>\end{math}({\it{}exp}).{\it{}is\_integer}()) {\nwlbrace}
  {\bf{}if} ({\it{}ex\_to}\begin{math}<\end{math}{\bf{}numeric}\begin{math}>\end{math}({\it{}exp}).{\it{}is\_even}())
   {\bf{}return} {\bf{}numeric}(1);
  {\bf{}else}
   {\bf{}return} {\it{}jump\_fnct}({\it{}arg});
 {\nwrbrace}
 {\bf{}if} ({\it{}is\_a}\begin{math}<\end{math}{\bf{}numeric}\begin{math}>\end{math}({\it{}exp}) \begin{math}\wedge\end{math} {\it{}ex\_to}\begin{math}<\end{math}{\bf{}numeric}\begin{math}>\end{math}(-{\it{}exp}).{\it{}is\_positive}())
  {\bf{}return} {\it{}ex\_to}\begin{math}<\end{math}{\bf{}basic}\begin{math}>\end{math}({\it{}pow}({\it{}jump\_fnct}({\it{}arg}), -{\it{}exp})).{\it{}hold}();
 {\bf{}return} {\it{}ex\_to}\begin{math}<\end{math}{\bf{}basic}\begin{math}>\end{math}({\it{}pow}({\it{}jump\_fnct}({\it{}arg}), {\it{}exp})).{\it{}hold}();
{\nwrbrace}

\nwidentdefs{\\{{\nwixident{ex}}{ex}}}\nwidentuses{\\{{\nwixident{jump{\_}fnct}}{jump:unfnct}}\\{{\nwixident{numeric}}{numeric}}}\nwindexuse{\nwixident{jump{\_}fnct}}{jump:unfnct}{NW3gGP3e-1Oxbp0-1F}\nwindexuse{\nwixident{numeric}}{numeric}{NW3gGP3e-1Oxbp0-1F}\nwendcode{}\nwbegindocs{697}\nwdocspar
\nwenddocs{}\nwbegincode{698}\sublabel{NW3gGP3e-1Oxbp0-1G}\nwmargintag{{\nwtagstyle{}\subpageref{NW3gGP3e-1Oxbp0-1G}}}\moddef{cycle.cpp~{\nwtagstyle{}\subpageref{NW3gGP3e-1Oxbp0-1}}}\plusendmoddef\Rm{}\nwstartdeflinemarkup\nwprevnextdefs{NW3gGP3e-1Oxbp0-1F}{NW3gGP3e-1Oxbp0-1H}\nwenddeflinemarkup
{\bf{}static} {\bf{}void} {\it{}jump\_fnct\_print\_dflt\_text}({\bf{}const} {\bf{}ex} & {\it{}x}, {\bf{}const} {\it{}print\_context} & {\it{}c})\nwindexdefn{\nwixident{jump{\_}fnct{\_}print{\_}dflt{\_}text}}{jump:unfnct:unprint:undflt:untext}{NW3gGP3e-1Oxbp0-1G}
{\nwlbrace}
 {\it{}c}.{\it{}s} \begin{math}\ll\end{math} {\tt{}"H("}; {\it{}x}.{\it{}print}({\it{}c}); {\it{}c}.{\it{}s} \begin{math}\ll\end{math} {\tt{}")"};
{\nwrbrace}

{\bf{}static} {\bf{}void} {\it{}jump\_fnct\_print\_latex}({\bf{}const} {\bf{}ex} & {\it{}x}, {\bf{}const} {\it{}print\_context} & {\it{}c})\nwindexdefn{\nwixident{jump{\_}fnct{\_}print{\_}latex}}{jump:unfnct:unprint:unlatex}{NW3gGP3e-1Oxbp0-1G}
{\nwlbrace}
 {\it{}c}.{\it{}s} \begin{math}\ll\end{math} {\tt{}"{\char92}{\char92}chi("}; {\it{}x}.{\it{}print}({\it{}c}); {\it{}c}.{\it{}s} \begin{math}\ll\end{math} {\tt{}")"};
{\nwrbrace}

\nwidentdefs{\\{{\nwixident{jump{\_}fnct{\_}print{\_}dflt{\_}text}}{jump:unfnct:unprint:undflt:untext}}\\{{\nwixident{jump{\_}fnct{\_}print{\_}latex}}{jump:unfnct:unprint:unlatex}}}\nwidentuses{\\{{\nwixident{ex}}{ex}}}\nwindexuse{\nwixident{ex}}{ex}{NW3gGP3e-1Oxbp0-1G}\nwendcode{}\nwbegindocs{699}All above methods are used to register the function now.
\nwenddocs{}\nwbegincode{700}\sublabel{NW3gGP3e-1Oxbp0-1H}\nwmargintag{{\nwtagstyle{}\subpageref{NW3gGP3e-1Oxbp0-1H}}}\moddef{cycle.cpp~{\nwtagstyle{}\subpageref{NW3gGP3e-1Oxbp0-1}}}\plusendmoddef\Rm{}\nwstartdeflinemarkup\nwprevnextdefs{NW3gGP3e-1Oxbp0-1G}{NW3gGP3e-1Oxbp0-1I}\nwenddeflinemarkup
{\it{}REGISTER\_FUNCTION}({\it{}jump\_fnct}, {\it{}eval\_func}({\it{}jump\_fnct\_eval}).
      {\it{}evalf\_func}({\it{}jump\_fnct\_evalf}).
      {\it{}latex\_name}({\tt{}"{\char92}{\char92}chi"}).
      //text\_name("H").
      {\it{}print\_func}\begin{math}<\end{math}{\it{}print\_dflt}\begin{math}>\end{math}({\it{}jump\_fnct\_print\_dflt\_text}).
      {\it{}print\_func}\begin{math}<\end{math}{\it{}print\_latex}\begin{math}>\end{math}({\it{}jump\_fnct\_print\_latex}).
      //derivative\_func(2*delta).
      {\it{}power\_func}({\it{}jump\_fnct\_power}).
      {\it{}conjugate\_func}({\it{}jump\_fnct\_conjugate}));

\nwidentuses{\\{{\nwixident{jump{\_}fnct}}{jump:unfnct}}\\{{\nwixident{jump{\_}fnct{\_}print{\_}dflt{\_}text}}{jump:unfnct:unprint:undflt:untext}}\\{{\nwixident{jump{\_}fnct{\_}print{\_}latex}}{jump:unfnct:unprint:unlatex}}}\nwindexuse{\nwixident{jump{\_}fnct}}{jump:unfnct}{NW3gGP3e-1Oxbp0-1H}\nwindexuse{\nwixident{jump{\_}fnct{\_}print{\_}dflt{\_}text}}{jump:unfnct:unprint:undflt:untext}{NW3gGP3e-1Oxbp0-1H}\nwindexuse{\nwixident{jump{\_}fnct{\_}print{\_}latex}}{jump:unfnct:unprint:unlatex}{NW3gGP3e-1Oxbp0-1H}\nwendcode{}\nwbegindocs{701}This function prints if its parameter is zero in a prominent way.
\nwenddocs{}\nwbegincode{702}\sublabel{NW3gGP3e-1Oxbp0-1I}\nwmargintag{{\nwtagstyle{}\subpageref{NW3gGP3e-1Oxbp0-1I}}}\moddef{cycle.cpp~{\nwtagstyle{}\subpageref{NW3gGP3e-1Oxbp0-1}}}\plusendmoddef\Rm{}\nwstartdeflinemarkup\nwprevnextdefs{NW3gGP3e-1Oxbp0-1H}{NW3gGP3e-1Oxbp0-1J}\nwenddeflinemarkup
{\bf{}const} {\it{}string} {\it{}equality}({\bf{}const} {\bf{}ex} & {\it{}E})\nwindexdefn{\nwixident{string}}{string}{NW3gGP3e-1Oxbp0-1I}
{\nwlbrace}
 {\bf{}if} ({\it{}normal}({\it{}E}).{\it{}is\_zero}())
  {\bf{}return} {\tt{}"-equal-"};
 {\bf{}else}
  {\bf{}return} {\tt{}"DIFFERENT!!!"};
{\nwrbrace}

\nwidentdefs{\\{{\nwixident{string}}{string}}}\nwidentuses{\\{{\nwixident{ex}}{ex}}\\{{\nwixident{is{\_}zero}}{is:unzero}}\\{{\nwixident{normal}}{normal}}}\nwindexuse{\nwixident{ex}}{ex}{NW3gGP3e-1Oxbp0-1I}\nwindexuse{\nwixident{is{\_}zero}}{is:unzero}{NW3gGP3e-1Oxbp0-1I}\nwindexuse{\nwixident{normal}}{normal}{NW3gGP3e-1Oxbp0-1I}\nwendcode{}\nwbegindocs{703}This function decodes metric sign into human-readable form.
\nwenddocs{}\nwbegincode{704}\sublabel{NW3gGP3e-1Oxbp0-1J}\nwmargintag{{\nwtagstyle{}\subpageref{NW3gGP3e-1Oxbp0-1J}}}\moddef{cycle.cpp~{\nwtagstyle{}\subpageref{NW3gGP3e-1Oxbp0-1}}}\plusendmoddef\Rm{}\nwstartdeflinemarkup\nwprevnextdefs{NW3gGP3e-1Oxbp0-1I}{NW3gGP3e-1Oxbp0-1K}\nwenddeflinemarkup
{\bf{}const} {\it{}string} {\it{}eph\_case}({\bf{}const} {\bf{}numeric} & {\it{}sign})\nwindexdefn{\nwixident{string}}{string}{NW3gGP3e-1Oxbp0-1J}
{\nwlbrace}
 {\bf{}if} ({\bf{}numeric}({\it{}sign}-(-1)).{\it{}is\_zero}())
  {\bf{}return} {\tt{}"Elliptic case (sign = -1)"};
 {\bf{}if} ({\bf{}numeric}({\it{}sign}).{\it{}is\_zero}())
  {\bf{}return} {\tt{}"Parabolic case (sign = 0)"};
 {\bf{}if} ({\bf{}numeric}({\it{}sign}-1).{\it{}is\_zero}())
  {\bf{}return} {\tt{}"Hyperbolic case (sign = 1)"};
 {\bf{}return} {\tt{}"Unknown case!!!!"};
{\nwrbrace}

\nwidentdefs{\\{{\nwixident{string}}{string}}}\nwidentuses{\\{{\nwixident{is{\_}zero}}{is:unzero}}\\{{\nwixident{numeric}}{numeric}}}\nwindexuse{\nwixident{is{\_}zero}}{is:unzero}{NW3gGP3e-1Oxbp0-1J}\nwindexuse{\nwixident{numeric}}{numeric}{NW3gGP3e-1Oxbp0-1J}\nwendcode{}\nwbegindocs{705}Elements of \(\SL\) are transformed into appropriate ``cliffordian''
matrix.
\nwenddocs{}\nwbegincode{706}\sublabel{NW3gGP3e-1Oxbp0-1K}\nwmargintag{{\nwtagstyle{}\subpageref{NW3gGP3e-1Oxbp0-1K}}}\moddef{cycle.cpp~{\nwtagstyle{}\subpageref{NW3gGP3e-1Oxbp0-1}}}\plusendmoddef\Rm{}\nwstartdeflinemarkup\nwprevnextdefs{NW3gGP3e-1Oxbp0-1J}{NW3gGP3e-1Oxbp0-1L}\nwenddeflinemarkup
{\bf{}matrix} {\it{}sl2\_clifford}({\bf{}const} {\bf{}ex} & {\it{}a}, {\bf{}const} {\bf{}ex} & {\it{}b}, {\bf{}const} {\bf{}ex} & {\it{}c}, {\bf{}const} {\bf{}ex} & {\it{}d}, {\bf{}const} {\bf{}ex} & {\it{}e}, {\bf{}bool} {\it{}not\_inverse})
{\nwlbrace}
 {\bf{}if} ({\it{}is\_a}\begin{math}<\end{math}{\bf{}clifford}\begin{math}>\end{math}({\it{}e})) {\nwlbrace}
  {\bf{}ex} {\it{}e0} = {\it{}e}.{\it{}subs}({\it{}e}.{\it{}op}(1) \begin{math}\equiv\end{math} 0);
  {\bf{}ex} {\it{}one} = {\it{}dirac\_ONE}({\it{}ex\_to}\begin{math}<\end{math}{\bf{}clifford}\begin{math}>\end{math}({\it{}e}).{\it{}get\_representation\_label}());
  {\bf{}if} ({\it{}not\_inverse})
   {\bf{}return} {\bf{}matrix}(2, 2,
        {\bf{}lst}({\it{}a} \begin{math}\ast\end{math} {\it{}one}, {\it{}b} \begin{math}\ast\end{math} {\it{}e0},
         {\it{}c} \begin{math}\ast\end{math} {\it{}pow}({\it{}e0}, 3), {\it{}d} \begin{math}\ast\end{math} {\it{}one}));
  {\bf{}else}
   {\bf{}return} {\bf{}matrix}(2, 2,
        {\bf{}lst}({\it{}d} \begin{math}\ast\end{math} {\it{}one}, -{\it{}b} \begin{math}\ast\end{math} {\it{}e0},
         -{\it{}c} \begin{math}\ast\end{math} {\it{}pow}({\it{}e0}, 3), {\it{}a} \begin{math}\ast\end{math} {\it{}one}));
 {\nwrbrace} {\bf{}else}
  {\bf{}throw}({\it{}std}::{\it{}invalid\_argument}({\tt{}"sl2\_clifford(): expect a clifford numeber as a parameter"}));
{\nwrbrace}

\nwidentuses{\\{{\nwixident{bool}}{bool}}\\{{\nwixident{ex}}{ex}}\\{{\nwixident{lst}}{lst}}\\{{\nwixident{matrix}}{matrix}}\\{{\nwixident{op}}{op}}\\{{\nwixident{subs}}{subs}}}\nwindexuse{\nwixident{bool}}{bool}{NW3gGP3e-1Oxbp0-1K}\nwindexuse{\nwixident{ex}}{ex}{NW3gGP3e-1Oxbp0-1K}\nwindexuse{\nwixident{lst}}{lst}{NW3gGP3e-1Oxbp0-1K}\nwindexuse{\nwixident{matrix}}{matrix}{NW3gGP3e-1Oxbp0-1K}\nwindexuse{\nwixident{op}}{op}{NW3gGP3e-1Oxbp0-1K}\nwindexuse{\nwixident{subs}}{subs}{NW3gGP3e-1Oxbp0-1K}\nwendcode{}\nwbegindocs{707}We are trying find a scalar part of the given expression.
\nwenddocs{}\nwbegincode{708}\sublabel{NW3gGP3e-1Oxbp0-1L}\nwmargintag{{\nwtagstyle{}\subpageref{NW3gGP3e-1Oxbp0-1L}}}\moddef{cycle.cpp~{\nwtagstyle{}\subpageref{NW3gGP3e-1Oxbp0-1}}}\plusendmoddef\Rm{}\nwstartdeflinemarkup\nwprevnextdefs{NW3gGP3e-1Oxbp0-1K}{NW3gGP3e-1Oxbp0-1M}\nwenddeflinemarkup
{\bf{}ex} {\it{}scalar\_part}({\bf{}const} {\bf{}ex} & {\it{}e}) {\nwlbrace}
    {\bf{}ex} {\it{}given}={\it{}canonicalize\_clifford}({\it{}e}.{\it{}expand}()), 
        {\it{}out}=0, {\it{}term};
    {\bf{}if} ({\it{}is\_a}\begin{math}<\end{math}{\it{}add}\begin{math}>\end{math}({\it{}given})){\nwlbrace}
        {\bf{}for} ({\it{}size\_t} {\it{}i}=0; {\it{}i}\begin{math}<\end{math}{\it{}given}.{\it{}nops}(); {\it{}i}\protect\PP) {\nwlbrace} 
            {\bf{}try} {\nwlbrace}
                {\it{}term}={\it{}remove\_dirac\_ONE}({\it{}given}.{\it{}op}({\it{}i}));
            {\nwrbrace} {\bf{}catch} ({\it{}exception} &{\it{}p}) {\nwlbrace}
                {\it{}term}=0;
            {\nwrbrace}
            {\it{}out}+={\it{}term};
        {\nwrbrace}
        {\bf{}return} {\it{}out}.{\it{}normal}();
    {\nwrbrace} {\bf{}else}{\nwlbrace}
        {\bf{}try} {\nwlbrace}
            {\bf{}return} {\it{}remove\_dirac\_ONE}({\it{}given});
        {\nwrbrace} {\bf{}catch} ({\it{}exception} &{\it{}p}) {\nwlbrace}
            {\bf{}return} 0;
        {\nwrbrace}
    {\nwrbrace}
{\nwrbrace}

\nwidentuses{\\{{\nwixident{add}}{add}}\\{{\nwixident{catch}}{catch}}\\{{\nwixident{ex}}{ex}}\\{{\nwixident{expand}}{expand}}\\{{\nwixident{nops}}{nops}}\\{{\nwixident{normal}}{normal}}\\{{\nwixident{op}}{op}}}\nwindexuse{\nwixident{add}}{add}{NW3gGP3e-1Oxbp0-1L}\nwindexuse{\nwixident{catch}}{catch}{NW3gGP3e-1Oxbp0-1L}\nwindexuse{\nwixident{ex}}{ex}{NW3gGP3e-1Oxbp0-1L}\nwindexuse{\nwixident{expand}}{expand}{NW3gGP3e-1Oxbp0-1L}\nwindexuse{\nwixident{nops}}{nops}{NW3gGP3e-1Oxbp0-1L}\nwindexuse{\nwixident{normal}}{normal}{NW3gGP3e-1Oxbp0-1L}\nwindexuse{\nwixident{op}}{op}{NW3gGP3e-1Oxbp0-1L}\nwendcode{}\nwbegindocs{709}\nwdocspar
\nwenddocs{}\nwbegincode{710}\sublabel{NW3gGP3e-1Oxbp0-1M}\nwmargintag{{\nwtagstyle{}\subpageref{NW3gGP3e-1Oxbp0-1M}}}\moddef{cycle.cpp~{\nwtagstyle{}\subpageref{NW3gGP3e-1Oxbp0-1}}}\plusendmoddef\Rm{}\nwstartdeflinemarkup\nwprevnextdefs{NW3gGP3e-1Oxbp0-1L}{\relax}\nwenddeflinemarkup
{\bf{}matrix} {\it{}sl2\_clifford}({\bf{}const} {\bf{}ex} & {\it{}M}, {\bf{}const} {\bf{}ex} & {\it{}e}, {\bf{}bool} {\it{}not\_inverse})
{\nwlbrace}
 {\bf{}if} ({\it{}is\_a}\begin{math}<\end{math}{\bf{}matrix}\begin{math}>\end{math}({\it{}M}) \begin{math}\vee\end{math} {\it{}M}.{\it{}info}({\it{}info\_flags}::{\it{}list}))
  {\bf{}return} {\it{}sl2\_clifford}({\it{}M}.{\it{}op}(0), {\it{}M}.{\it{}op}(1), {\it{}M}.{\it{}op}(2), {\it{}M}.{\it{}op}(3), {\it{}e}, {\it{}not\_inverse});
 {\bf{}else}
  {\bf{}throw}({\it{}std}::{\it{}invalid\_argument}({\tt{}"sl2\_clifford(): expect a list or matrix as the first parameter"}));
{\nwrbrace}
{\nwrbrace} // namespace MoebInv

\nwidentuses{\\{{\nwixident{bool}}{bool}}\\{{\nwixident{ex}}{ex}}\\{{\nwixident{matrix}}{matrix}}\\{{\nwixident{MoebInv}}{MoebInv}}\\{{\nwixident{op}}{op}}}\nwindexuse{\nwixident{bool}}{bool}{NW3gGP3e-1Oxbp0-1M}\nwindexuse{\nwixident{ex}}{ex}{NW3gGP3e-1Oxbp0-1M}\nwindexuse{\nwixident{matrix}}{matrix}{NW3gGP3e-1Oxbp0-1M}\nwindexuse{\nwixident{MoebInv}}{MoebInv}{NW3gGP3e-1Oxbp0-1M}\nwindexuse{\nwixident{op}}{op}{NW3gGP3e-1Oxbp0-1M}\nwendcode{}\nwbegindocs{711}\nwdocspar
\section{License}
\label{sec:license}
This programme is distributed under GNU GPLv3~\cite{GNUGPL}.
\nwenddocs{}\nwbegincode{712}\sublabel{NW3gGP3e-ZXuKx-1}\nwmargintag{{\nwtagstyle{}\subpageref{NW3gGP3e-ZXuKx-1}}}\moddef{license~{\nwtagstyle{}\subpageref{NW3gGP3e-ZXuKx-1}}}\endmoddef\Rm{}\nwstartdeflinemarkup\nwusesondefline{\\{NW3gGP3e-1p0Y9w-1}\\{NW3gGP3e-4Ef0r4-1}\\{NW3gGP3e-1Oxbp0-1}}\nwenddeflinemarkup
// The library to operate cycles in non-Euclidean geometry
//
//  Copyright (C) 2004-2015 Vladimir V. Kisil
//
//  This program is free software: you can redistribute it and/or modify
//  it under the terms of the GNU General Public License as published by
//  the Free Software Foundation, either version 3 of the License, or
//  (at your option) any later version.
//  
//  This program is distributed in the hope that it will be useful,
//  but WITHOUT ANY WARRANTY; without even the implied warranty of
//  MERCHANTABILITY or FITNESS FOR A PARTICULAR PURPOSE.  See the
//  GNU General Public License for more details.
//  
//  You should have received a copy of the GNU General Public License
//  along with this program.  If not, see \begin{math}<\end{math}http://www.gnu.org/licenses/\begin{math}>\end{math}.

\nwused{\\{NW3gGP3e-1p0Y9w-1}\\{NW3gGP3e-4Ef0r4-1}\\{NW3gGP3e-1Oxbp0-1}}\nwendcode{}\nwbegindocs{713}\nwdocspar
\section{Index of Identifiers}
\label{sec:index-identifiers}
\small
\nowebindex

\nwenddocs{}\nwbegindocs{714}\nwdocspar
\bibliographystyle{plain}
\bibliography{arare,aclifford,abbrevmr,akisil,ageometry,analyse,aphysics}

\nwenddocs{}

\nwixlogsorted{c}{{*}{NW3gGP3e-1p0Y9w-1}{\nwixd{NW3gGP3e-1p0Y9w-1}\nwixd{NW3gGP3e-1p0Y9w-2}\nwixd{NW3gGP3e-1p0Y9w-3}\nwixd{NW3gGP3e-1p0Y9w-4}\nwixd{NW3gGP3e-1p0Y9w-5}}}%
\nwixlogsorted{c}{{accessing the data of a cycle}{NW3gGP3e-rnSJR-1}{\nwixd{NW3gGP3e-rnSJR-1}\nwixd{NW3gGP3e-rnSJR-2}\nwixd{NW3gGP3e-rnSJR-3}\nwixd{NW3gGP3e-rnSJR-4}\nwixu{NW3gGP3e-2wwyff-1}}}%
\nwixlogsorted{c}{{Auxiliary functions headers}{NW3gGP3e-ugGZb-1}{\nwixu{NW3gGP3e-4Ef0r4-4}\nwixd{NW3gGP3e-ugGZb-1}}}%
\nwixlogsorted{c}{{Cayley transform pictures}{NW3gGP3e-q6Aul-1}{\nwixu{NW3gGP3e-3gefqu-1}\nwixd{NW3gGP3e-q6Aul-1}\nwixd{NW3gGP3e-q6Aul-2}\nwixd{NW3gGP3e-q6Aul-3}\nwixd{NW3gGP3e-q6Aul-4}\nwixd{NW3gGP3e-q6Aul-5}}}%
\nwixlogsorted{c}{{Centres and foci of parabolas}{NW3gGP3e-3gnOun-1}{\nwixu{NW3gGP3e-3gefqu-1}\nwixd{NW3gGP3e-3gnOun-1}}}%
\nwixlogsorted{c}{{Check conformal property}{NW3gGP3e-2eoTsH-1}{\nwixu{NW3gGP3e-1Z2pUX-1}\nwixd{NW3gGP3e-2eoTsH-1}\nwixd{NW3gGP3e-2eoTsH-2}\nwixd{NW3gGP3e-2eoTsH-3}\nwixd{NW3gGP3e-2eoTsH-4}\nwixd{NW3gGP3e-2eoTsH-5}\nwixd{NW3gGP3e-2eoTsH-6}\nwixd{NW3gGP3e-2eoTsH-7}\nwixd{NW3gGP3e-2eoTsH-8}}}%
\nwixlogsorted{c}{{Check independence}{NW3gGP3e-3N1PZV-1}{\nwixu{NW3gGP3e-2eoTsH-7}\nwixd{NW3gGP3e-3N1PZV-1}}}%
\nwixlogsorted{c}{{Check Moebius transformations of zero cycles}{NW3gGP3e-19ZdJC-1}{\nwixu{NW3gGP3e-1zkOI-1}\nwixd{NW3gGP3e-19ZdJC-1}\nwixd{NW3gGP3e-19ZdJC-2}\nwixd{NW3gGP3e-19ZdJC-3}\nwixd{NW3gGP3e-19ZdJC-4}\nwixd{NW3gGP3e-19ZdJC-5}}}%
\nwixlogsorted{c}{{Check transformations of zero cycles by conjugation}{NW3gGP3e-3O7JaU-1}{\nwixu{NW3gGP3e-1zkOI-1}\nwixd{NW3gGP3e-3O7JaU-1}\nwixd{NW3gGP3e-3O7JaU-2}}}%
\nwixlogsorted{c}{{constructors of the class cycle2D}{NW3gGP3e-3e3TWq-1}{\nwixd{NW3gGP3e-3e3TWq-1}\nwixd{NW3gGP3e-3e3TWq-2}\nwixd{NW3gGP3e-3e3TWq-3}\nwixd{NW3gGP3e-3e3TWq-4}\nwixu{NW3gGP3e-2ARAe1-1}}}%
\nwixlogsorted{c}{{Create a Clifford unit }{NW3gGP3e-4RDj8q-1}{\nwixu{NW3gGP3e-1Oxbp0-A}\nwixd{NW3gGP3e-4RDj8q-1}}}%
\nwixlogsorted{c}{{cycle class}{NW3gGP3e-2wwyff-1}{\nwixu{NW3gGP3e-4Ef0r4-4}\nwixd{NW3gGP3e-2wwyff-1}}}%
\nwixlogsorted{c}{{cycle class constructors}{NW3gGP3e-2EuhSt-1}{\nwixd{NW3gGP3e-2EuhSt-1}\nwixd{NW3gGP3e-2EuhSt-2}\nwixd{NW3gGP3e-2EuhSt-3}\nwixd{NW3gGP3e-2EuhSt-4}\nwixu{NW3gGP3e-2wwyff-1}}}%
\nwixlogsorted{c}{{cycle.cpp}{NW3gGP3e-1Oxbp0-1}{\nwixd{NW3gGP3e-1Oxbp0-1}\nwixd{NW3gGP3e-1Oxbp0-2}\nwixd{NW3gGP3e-1Oxbp0-3}\nwixd{NW3gGP3e-1Oxbp0-4}\nwixd{NW3gGP3e-1Oxbp0-5}\nwixd{NW3gGP3e-1Oxbp0-6}\nwixd{NW3gGP3e-1Oxbp0-7}\nwixd{NW3gGP3e-1Oxbp0-8}\nwixd{NW3gGP3e-1Oxbp0-9}\nwixd{NW3gGP3e-1Oxbp0-A}\nwixd{NW3gGP3e-1Oxbp0-B}\nwixd{NW3gGP3e-1Oxbp0-C}\nwixd{NW3gGP3e-1Oxbp0-D}\nwixd{NW3gGP3e-1Oxbp0-E}\nwixd{NW3gGP3e-1Oxbp0-F}\nwixd{NW3gGP3e-1Oxbp0-G}\nwixd{NW3gGP3e-1Oxbp0-H}\nwixd{NW3gGP3e-1Oxbp0-I}\nwixd{NW3gGP3e-1Oxbp0-J}\nwixd{NW3gGP3e-1Oxbp0-K}\nwixd{NW3gGP3e-1Oxbp0-L}\nwixd{NW3gGP3e-1Oxbp0-M}\nwixd{NW3gGP3e-1Oxbp0-N}\nwixd{NW3gGP3e-1Oxbp0-O}\nwixd{NW3gGP3e-1Oxbp0-P}\nwixd{NW3gGP3e-1Oxbp0-Q}\nwixd{NW3gGP3e-1Oxbp0-R}\nwixd{NW3gGP3e-1Oxbp0-S}\nwixd{NW3gGP3e-1Oxbp0-T}\nwixd{NW3gGP3e-1Oxbp0-U}\nwixd{NW3gGP3e-1Oxbp0-V}\nwixd{NW3gGP3e-1Oxbp0-W}\nwixd{NW3gGP3e-1Oxbp0-X}\nwixd{NW3gGP3e-1Oxbp0-Y}\nwixd{NW3gGP3e-1Oxbp0-Z}\nwixd{NW3gGP3e-1Oxbp0-a}\nwixd{NW3gGP3e-1Oxbp0-b}\nwixd{NW3gGP3e-1Oxbp0-c}\nwixd{NW3gGP3e-1Oxbp0-d}\nwixd{NW3gGP3e-1Oxbp0-e}\nwixd{NW3gGP3e-1Oxbp0-f}\nwixd{NW3gGP3e-1Oxbp0-g}\nwixd{NW3gGP3e-1Oxbp0-h}\nwixd{NW3gGP3e-1Oxbp0-i}\nwixd{NW3gGP3e-1Oxbp0-j}\nwixd{NW3gGP3e-1Oxbp0-k}\nwixd{NW3gGP3e-1Oxbp0-l}\nwixd{NW3gGP3e-1Oxbp0-m}\nwixd{NW3gGP3e-1Oxbp0-n}\nwixd{NW3gGP3e-1Oxbp0-o}\nwixd{NW3gGP3e-1Oxbp0-p}\nwixd{NW3gGP3e-1Oxbp0-q}\nwixd{NW3gGP3e-1Oxbp0-r}\nwixd{NW3gGP3e-1Oxbp0-s}\nwixd{NW3gGP3e-1Oxbp0-t}\nwixd{NW3gGP3e-1Oxbp0-u}\nwixd{NW3gGP3e-1Oxbp0-v}\nwixd{NW3gGP3e-1Oxbp0-w}\nwixd{NW3gGP3e-1Oxbp0-x}\nwixd{NW3gGP3e-1Oxbp0-y}\nwixd{NW3gGP3e-1Oxbp0-z}\nwixd{NW3gGP3e-1Oxbp0-10}\nwixd{NW3gGP3e-1Oxbp0-11}\nwixd{NW3gGP3e-1Oxbp0-12}\nwixd{NW3gGP3e-1Oxbp0-13}\nwixd{NW3gGP3e-1Oxbp0-14}\nwixd{NW3gGP3e-1Oxbp0-15}\nwixd{NW3gGP3e-1Oxbp0-16}\nwixd{NW3gGP3e-1Oxbp0-17}\nwixd{NW3gGP3e-1Oxbp0-18}\nwixd{NW3gGP3e-1Oxbp0-19}\nwixd{NW3gGP3e-1Oxbp0-1A}\nwixd{NW3gGP3e-1Oxbp0-1B}\nwixd{NW3gGP3e-1Oxbp0-1C}\nwixd{NW3gGP3e-1Oxbp0-1D}\nwixd{NW3gGP3e-1Oxbp0-1E}\nwixd{NW3gGP3e-1Oxbp0-1F}\nwixd{NW3gGP3e-1Oxbp0-1G}\nwixd{NW3gGP3e-1Oxbp0-1H}\nwixd{NW3gGP3e-1Oxbp0-1I}\nwixd{NW3gGP3e-1Oxbp0-1J}\nwixd{NW3gGP3e-1Oxbp0-1K}\nwixd{NW3gGP3e-1Oxbp0-1L}\nwixd{NW3gGP3e-1Oxbp0-1M}}}%
\nwixlogsorted{c}{{cycle.h}{NW3gGP3e-4Ef0r4-1}{\nwixd{NW3gGP3e-4Ef0r4-1}\nwixd{NW3gGP3e-4Ef0r4-2}\nwixd{NW3gGP3e-4Ef0r4-3}\nwixd{NW3gGP3e-4Ef0r4-4}}}%
\nwixlogsorted{c}{{cycle2D class}{NW3gGP3e-2ARAe1-1}{\nwixu{NW3gGP3e-4Ef0r4-4}\nwixd{NW3gGP3e-2ARAe1-1}}}%
\nwixlogsorted{c}{{Declaration of variables}{NW3gGP3e-3hvAAH-1}{\nwixu{NW3gGP3e-1p0Y9w-3}\nwixd{NW3gGP3e-3hvAAH-1}\nwixd{NW3gGP3e-3hvAAH-2}\nwixd{NW3gGP3e-3hvAAH-3}\nwixd{NW3gGP3e-3hvAAH-4}\nwixd{NW3gGP3e-3hvAAH-5}\nwixd{NW3gGP3e-3hvAAH-6}\nwixd{NW3gGP3e-3hvAAH-7}\nwixd{NW3gGP3e-3hvAAH-8}\nwixd{NW3gGP3e-3hvAAH-9}\nwixd{NW3gGP3e-3hvAAH-A}\nwixd{NW3gGP3e-3hvAAH-B}\nwixd{NW3gGP3e-3hvAAH-C}}}%
\nwixlogsorted{c}{{Distance as an extremum}{NW3gGP3e-1g9SsP-1}{\nwixu{NW3gGP3e-3gefqu-1}\nwixd{NW3gGP3e-1g9SsP-1}}}%
\nwixlogsorted{c}{{Distances from cycles}{NW3gGP3e-O1KCX-1}{\nwixu{NW3gGP3e-1zkOI-4}\nwixd{NW3gGP3e-O1KCX-1}\nwixd{NW3gGP3e-O1KCX-2}\nwixd{NW3gGP3e-O1KCX-3}\nwixd{NW3gGP3e-O1KCX-4}\nwixd{NW3gGP3e-O1KCX-5}\nwixd{NW3gGP3e-O1KCX-6}\nwixd{NW3gGP3e-O1KCX-7}\nwixd{NW3gGP3e-O1KCX-8}}}%
\nwixlogsorted{c}{{Draw a circle}{NW3gGP3e-4WQ5O5-1}{\nwixu{NW3gGP3e-1Oxbp0-1B}\nwixd{NW3gGP3e-4WQ5O5-1}\nwixd{NW3gGP3e-4WQ5O5-2}\nwixd{NW3gGP3e-4WQ5O5-3}\nwixd{NW3gGP3e-4WQ5O5-4}\nwixd{NW3gGP3e-4WQ5O5-5}\nwixd{NW3gGP3e-4WQ5O5-6}\nwixd{NW3gGP3e-4WQ5O5-7}\nwixd{NW3gGP3e-4WQ5O5-8}}}%
\nwixlogsorted{c}{{Draw a parabola or hyperbola}{NW3gGP3e-4H4ZFN-1}{\nwixu{NW3gGP3e-1Oxbp0-1B}\nwixd{NW3gGP3e-4H4ZFN-1}}}%
\nwixlogsorted{c}{{Draw a straight line}{NW3gGP3e-1iDH4d-1}{\nwixu{NW3gGP3e-1Oxbp0-1B}\nwixd{NW3gGP3e-1iDH4d-1}\nwixd{NW3gGP3e-1iDH4d-2}\nwixd{NW3gGP3e-1iDH4d-3}\nwixd{NW3gGP3e-1iDH4d-4}}}%
\nwixlogsorted{c}{{Draw Asymptote pictures}{NW3gGP3e-Xmoi0-1}{\nwixu{NW3gGP3e-1p0Y9w-5}\nwixd{NW3gGP3e-Xmoi0-1}\nwixd{NW3gGP3e-Xmoi0-2}\nwixd{NW3gGP3e-Xmoi0-3}}}%
\nwixlogsorted{c}{{Draw axes}{NW3gGP3e-ph2PF-1}{\nwixu{NW3gGP3e-3iFOPx-2}\nwixd{NW3gGP3e-ph2PF-1}\nwixu{NW3gGP3e-4CF7DS-1}\nwixu{NW3gGP3e-3gnOun-1}\nwixu{NW3gGP3e-QUZd0-1}\nwixu{NW3gGP3e-Y6OSa-1}\nwixu{NW3gGP3e-1g9SsP-1}\nwixu{NW3gGP3e-FS4xg-1}\nwixu{NW3gGP3e-q6Aul-5}\nwixu{NW3gGP3e-3xVrpR-1}\nwixu{NW3gGP3e-3xVrpR-3}}}%
\nwixlogsorted{c}{{Drawing first orthogonality}{NW3gGP3e-1VpO8V-1}{\nwixu{NW3gGP3e-Xmoi0-2}\nwixd{NW3gGP3e-1VpO8V-1}}}%
\nwixlogsorted{c}{{Drawing focal orthogonality}{NW3gGP3e-384dQC-1}{\nwixu{NW3gGP3e-Xmoi0-2}\nwixd{NW3gGP3e-384dQC-1}}}%
\nwixlogsorted{c}{{Drawing orthogonal cycles}{NW3gGP3e-3iFOPx-1}{\nwixu{NW3gGP3e-1VpO8V-1}\nwixd{NW3gGP3e-3iFOPx-1}\nwixd{NW3gGP3e-3iFOPx-2}\nwixu{NW3gGP3e-384dQC-1}}}%
\nwixlogsorted{c}{{duplicated linear operation on cycle2D}{NW3gGP3e-1Ihica-1}{\nwixu{NW3gGP3e-2ARAe1-1}\nwixd{NW3gGP3e-1Ihica-1}}}%
\nwixlogsorted{c}{{duplicated methods for class cycle2D}{NW3gGP3e-xs0zU-1}{\nwixu{NW3gGP3e-2ARAe1-1}\nwixd{NW3gGP3e-xs0zU-1}\nwixd{NW3gGP3e-xs0zU-2}}}%
\nwixlogsorted{c}{{Evaluate the fraction}{NW3gGP3e-5o8iC-1}{\nwixu{NW3gGP3e-2eoTsH-1}\nwixd{NW3gGP3e-5o8iC-1}}}%
\nwixlogsorted{c}{{Extra pictures from Asymptote}{NW3gGP3e-3gefqu-1}{\nwixu{NW3gGP3e-Xmoi0-3}\nwixd{NW3gGP3e-3gefqu-1}}}%
\nwixlogsorted{c}{{f-inversion in cycle}{NW3gGP3e-22oRI1-1}{\nwixu{NW3gGP3e-1zkOI-3}\nwixd{NW3gGP3e-22oRI1-1}\nwixd{NW3gGP3e-22oRI1-2}\nwixd{NW3gGP3e-22oRI1-3}\nwixd{NW3gGP3e-22oRI1-4}}}%
\nwixlogsorted{c}{{f-orthogonal line}{NW3gGP3e-3jX7bo-1}{\nwixu{NW3gGP3e-1zkOI-3}\nwixd{NW3gGP3e-3jX7bo-1}\nwixd{NW3gGP3e-3jX7bo-2}}}%
\nwixlogsorted{c}{{Find intersection points with the boundary}{NW3gGP3e-2elNFX-1}{\nwixu{NW3gGP3e-1Oxbp0-1B}\nwixd{NW3gGP3e-2elNFX-1}}}%
\nwixlogsorted{c}{{Find the limit}{NW3gGP3e-2wjYgT-1}{\nwixu{NW3gGP3e-2eoTsH-7}\nwixd{NW3gGP3e-2wjYgT-1}}}%
\nwixlogsorted{c}{{Focal length checks}{NW3gGP3e-2gYzHL-1}{\nwixu{NW3gGP3e-1Z2pUX-1}\nwixd{NW3gGP3e-2gYzHL-1}\nwixd{NW3gGP3e-2gYzHL-2}\nwixd{NW3gGP3e-2gYzHL-3}\nwixd{NW3gGP3e-2gYzHL-4}}}%
\nwixlogsorted{c}{{Focal orthogonality conditions}{NW3gGP3e-2oH953-1}{\nwixu{NW3gGP3e-1zkOI-3}\nwixd{NW3gGP3e-2oH953-1}\nwixd{NW3gGP3e-2oH953-2}\nwixd{NW3gGP3e-2oH953-3}\nwixd{NW3gGP3e-2oH953-4}\nwixd{NW3gGP3e-2oH953-5}\nwixd{NW3gGP3e-2oH953-6}}}%
\nwixlogsorted{c}{{Hyperbolic inversion of a ball}{NW3gGP3e-2DcUMV-1}{\nwixu{NW3gGP3e-3gefqu-1}\nwixd{NW3gGP3e-2DcUMV-1}\nwixd{NW3gGP3e-2DcUMV-2}\nwixd{NW3gGP3e-2DcUMV-3}\nwixd{NW3gGP3e-2DcUMV-4}\nwixd{NW3gGP3e-2DcUMV-5}\nwixd{NW3gGP3e-2DcUMV-6}\nwixd{NW3gGP3e-2DcUMV-7}}}%
\nwixlogsorted{c}{{Imaginary coefficients}{NW3gGP3e-5M2Zk-1}{\nwixu{NW3gGP3e-1Oxbp0-1B}\nwixd{NW3gGP3e-5M2Zk-1}\nwixd{NW3gGP3e-5M2Zk-2}\nwixd{NW3gGP3e-5M2Zk-3}\nwixd{NW3gGP3e-5M2Zk-4}\nwixd{NW3gGP3e-5M2Zk-5}\nwixd{NW3gGP3e-5M2Zk-6}}}%
\nwixlogsorted{c}{{Infinitesimal cycle}{NW3gGP3e-2Xuors-1}{\nwixu{NW3gGP3e-1zkOI-4}\nwixd{NW3gGP3e-2Xuors-1}\nwixd{NW3gGP3e-2Xuors-2}}}%
\nwixlogsorted{c}{{Infinitesimal cycle calculations}{NW3gGP3e-4WB4RW-1}{\nwixu{NW3gGP3e-1Z2pUX-1}\nwixd{NW3gGP3e-4WB4RW-1}\nwixd{NW3gGP3e-4WB4RW-2}\nwixd{NW3gGP3e-4WB4RW-3}\nwixd{NW3gGP3e-4WB4RW-4}\nwixd{NW3gGP3e-4WB4RW-5}\nwixd{NW3gGP3e-4WB4RW-6}\nwixd{NW3gGP3e-4WB4RW-7}\nwixd{NW3gGP3e-4WB4RW-8}\nwixd{NW3gGP3e-4WB4RW-9}\nwixd{NW3gGP3e-4WB4RW-A}}}%
\nwixlogsorted{c}{{Infinitesimal cycles draw}{NW3gGP3e-FS4xg-1}{\nwixu{NW3gGP3e-3gefqu-1}\nwixd{NW3gGP3e-FS4xg-1}}}%
\nwixlogsorted{c}{{Inversion in cycle}{NW3gGP3e-47k9wZ-1}{\nwixu{NW3gGP3e-1zkOI-2}\nwixd{NW3gGP3e-47k9wZ-1}\nwixd{NW3gGP3e-47k9wZ-2}\nwixd{NW3gGP3e-47k9wZ-3}\nwixd{NW3gGP3e-47k9wZ-4}}}%
\nwixlogsorted{c}{{K-orbit invariance}{NW3gGP3e-3pbeiz-1}{\nwixu{NW3gGP3e-1zkOI-1}\nwixd{NW3gGP3e-3pbeiz-1}\nwixd{NW3gGP3e-3pbeiz-2}}}%
\nwixlogsorted{c}{{Lengths from centre}{NW3gGP3e-20e6xZ-1}{\nwixu{NW3gGP3e-1zkOI-4}\nwixd{NW3gGP3e-20e6xZ-1}\nwixd{NW3gGP3e-20e6xZ-2}}}%
\nwixlogsorted{c}{{Lengths from focus}{NW3gGP3e-Nadcx-1}{\nwixu{NW3gGP3e-1zkOI-4}\nwixd{NW3gGP3e-Nadcx-1}\nwixd{NW3gGP3e-Nadcx-2}\nwixd{NW3gGP3e-Nadcx-3}}}%
\nwixlogsorted{c}{{license}{NW3gGP3e-ZXuKx-1}{\nwixu{NW3gGP3e-1p0Y9w-1}\nwixu{NW3gGP3e-4Ef0r4-1}\nwixu{NW3gGP3e-1Oxbp0-1}\nwixd{NW3gGP3e-ZXuKx-1}}}%
\nwixlogsorted{c}{{Linear operation as cycle methods}{NW3gGP3e-3Bzv5Q-1}{\nwixd{NW3gGP3e-3Bzv5Q-1}\nwixu{NW3gGP3e-2wwyff-1}}}%
\nwixlogsorted{c}{{Linear operation on cycles}{NW3gGP3e-3LRIX-1}{\nwixd{NW3gGP3e-3LRIX-1}\nwixd{NW3gGP3e-3LRIX-2}\nwixu{NW3gGP3e-2wwyff-1}}}%
\nwixlogsorted{c}{{List of symbolic calculations}{NW3gGP3e-1zkOI-1}{\nwixd{NW3gGP3e-1zkOI-1}\nwixd{NW3gGP3e-1zkOI-2}\nwixd{NW3gGP3e-1zkOI-3}\nwixd{NW3gGP3e-1zkOI-4}\nwixd{NW3gGP3e-1zkOI-5}\nwixu{NW3gGP3e-1p0Y9w-4}}}%
\nwixlogsorted{c}{{methods specific for class cycle2D}{NW3gGP3e-447MWQ-1}{\nwixd{NW3gGP3e-447MWQ-1}\nwixd{NW3gGP3e-447MWQ-2}\nwixd{NW3gGP3e-447MWQ-3}\nwixd{NW3gGP3e-447MWQ-4}\nwixd{NW3gGP3e-447MWQ-5}\nwixd{NW3gGP3e-447MWQ-6}\nwixd{NW3gGP3e-447MWQ-7}\nwixu{NW3gGP3e-2ARAe1-1}}}%
\nwixlogsorted{c}{{Moebius transformation of cycles}{NW3gGP3e-2em4FU-1}{\nwixd{NW3gGP3e-2em4FU-1}\nwixd{NW3gGP3e-2em4FU-2}\nwixu{NW3gGP3e-1zkOI-1}}}%
\nwixlogsorted{c}{{Moebius transforms of W}{NW3gGP3e-XJsVP-1}{\nwixd{NW3gGP3e-XJsVP-1}\nwixu{NW3gGP3e-3hvAAH-9}}}%
\nwixlogsorted{c}{{One point and f-orthogonality}{NW3gGP3e-P6T5o-1}{\nwixu{NW3gGP3e-1zkOI-3}\nwixd{NW3gGP3e-P6T5o-1}}}%
\nwixlogsorted{c}{{One point and orthogonality}{NW3gGP3e-mJ5pz-1}{\nwixu{NW3gGP3e-1zkOI-2}\nwixd{NW3gGP3e-mJ5pz-1}\nwixd{NW3gGP3e-mJ5pz-2}\nwixd{NW3gGP3e-mJ5pz-3}\nwixd{NW3gGP3e-mJ5pz-4}}}%
\nwixlogsorted{c}{{Orthogonal line}{NW3gGP3e-3uJ41M-1}{\nwixu{NW3gGP3e-1zkOI-2}\nwixd{NW3gGP3e-3uJ41M-1}\nwixd{NW3gGP3e-3uJ41M-2}\nwixd{NW3gGP3e-3uJ41M-3}}}%
\nwixlogsorted{c}{{Orthogonality conditions}{NW3gGP3e-3aTNrh-1}{\nwixu{NW3gGP3e-1zkOI-2}\nwixd{NW3gGP3e-3aTNrh-1}\nwixd{NW3gGP3e-3aTNrh-2}\nwixd{NW3gGP3e-3aTNrh-3}\nwixd{NW3gGP3e-3aTNrh-4}}}%
\nwixlogsorted{c}{{Parabolic Cayley transform of cycles}{NW3gGP3e-3vL8tu-1}{\nwixu{NW3gGP3e-1Z2pUX-1}\nwixd{NW3gGP3e-3vL8tu-1}}}%
\nwixlogsorted{c}{{Parabolic diameters}{NW3gGP3e-Y6OSa-1}{\nwixu{NW3gGP3e-3gefqu-1}\nwixd{NW3gGP3e-Y6OSa-1}}}%
\nwixlogsorted{c}{{place a dot}{NW3gGP3e-FZdRR-1}{\nwixu{NW3gGP3e-5M2Zk-5}\nwixu{NW3gGP3e-4WQ5O5-7}\nwixd{NW3gGP3e-FZdRR-1}\nwixd{NW3gGP3e-FZdRR-2}\nwixu{NW3gGP3e-2yW1fi-2}}}%
\nwixlogsorted{c}{{Print perpendicular}{NW3gGP3e-4PKbLl-1}{\nwixu{NW3gGP3e-1Z2pUX-1}\nwixd{NW3gGP3e-4PKbLl-1}}}%
\nwixlogsorted{c}{{Put labels on 22-23}{NW3gGP3e-3wMlZE-1}{\nwixu{NW3gGP3e-Y6OSa-1}\nwixd{NW3gGP3e-3wMlZE-1}\nwixu{NW3gGP3e-1g9SsP-1}}}%
\nwixlogsorted{c}{{Put units}{NW3gGP3e-1r99tr-1}{\nwixu{NW3gGP3e-3iFOPx-2}\nwixd{NW3gGP3e-1r99tr-1}\nwixu{NW3gGP3e-q6Aul-5}}}%
\nwixlogsorted{c}{{Reflection in cycle}{NW3gGP3e-2W1nIA-1}{\nwixu{NW3gGP3e-1zkOI-2}\nwixd{NW3gGP3e-2W1nIA-1}\nwixd{NW3gGP3e-2W1nIA-2}}}%
\nwixlogsorted{c}{{service functions for class cycle}{NW3gGP3e-2AZq7I-1}{\nwixu{NW3gGP3e-2wwyff-1}\nwixd{NW3gGP3e-2AZq7I-1}\nwixd{NW3gGP3e-2AZq7I-2}\nwixd{NW3gGP3e-2AZq7I-3}}}%
\nwixlogsorted{c}{{Set tinfo to dimension}{NW3gGP3e-MkLdY-1}{\nwixu{NW3gGP3e-1Oxbp0-7}\nwixd{NW3gGP3e-MkLdY-1}}}%
\nwixlogsorted{c}{{specific methods of the class cycle}{NW3gGP3e-mq5nH-1}{\nwixd{NW3gGP3e-mq5nH-1}\nwixd{NW3gGP3e-mq5nH-2}\nwixd{NW3gGP3e-mq5nH-3}\nwixd{NW3gGP3e-mq5nH-4}\nwixd{NW3gGP3e-mq5nH-5}\nwixd{NW3gGP3e-mq5nH-6}\nwixd{NW3gGP3e-mq5nH-7}\nwixd{NW3gGP3e-mq5nH-8}\nwixd{NW3gGP3e-mq5nH-9}\nwixd{NW3gGP3e-mq5nH-A}\nwixd{NW3gGP3e-mq5nH-B}\nwixd{NW3gGP3e-mq5nH-C}\nwixd{NW3gGP3e-mq5nH-D}\nwixd{NW3gGP3e-mq5nH-E}\nwixd{NW3gGP3e-mq5nH-F}\nwixd{NW3gGP3e-mq5nH-G}\nwixd{NW3gGP3e-mq5nH-H}\nwixd{NW3gGP3e-mq5nH-I}\nwixd{NW3gGP3e-mq5nH-J}\nwixd{NW3gGP3e-mq5nH-K}\nwixu{NW3gGP3e-2wwyff-1}}}%
\nwixlogsorted{c}{{Subroutines definitions}{NW3gGP3e-1Z2pUX-1}{\nwixu{NW3gGP3e-1p0Y9w-3}\nwixd{NW3gGP3e-1Z2pUX-1}}}%
\nwixlogsorted{c}{{Three images of the same cycle}{NW3gGP3e-4CF7DS-1}{\nwixu{NW3gGP3e-3gefqu-1}\nwixd{NW3gGP3e-4CF7DS-1}}}%
\nwixlogsorted{c}{{Three inversions}{NW3gGP3e-3xVrpR-1}{\nwixu{NW3gGP3e-3gefqu-1}\nwixd{NW3gGP3e-3xVrpR-1}\nwixd{NW3gGP3e-3xVrpR-2}\nwixd{NW3gGP3e-3xVrpR-3}}}%
\nwixlogsorted{c}{{Treating a hyperbola}{NW3gGP3e-2yW1fi-1}{\nwixu{NW3gGP3e-4H4ZFN-1}\nwixd{NW3gGP3e-2yW1fi-1}\nwixd{NW3gGP3e-2yW1fi-2}\nwixd{NW3gGP3e-2yW1fi-3}\nwixd{NW3gGP3e-2yW1fi-4}\nwixd{NW3gGP3e-2yW1fi-5}}}%
\nwixlogsorted{c}{{Treating a parabola}{NW3gGP3e-24b8UV-1}{\nwixu{NW3gGP3e-4H4ZFN-1}\nwixd{NW3gGP3e-24b8UV-1}\nwixd{NW3gGP3e-24b8UV-2}\nwixd{NW3gGP3e-24b8UV-3}\nwixd{NW3gGP3e-24b8UV-4}\nwixd{NW3gGP3e-24b8UV-5}\nwixd{NW3gGP3e-24b8UV-6}}}%
\nwixlogsorted{c}{{Two points and orthogonality}{NW3gGP3e-3nCzcM-1}{\nwixu{NW3gGP3e-1zkOI-2}\nwixd{NW3gGP3e-3nCzcM-1}\nwixd{NW3gGP3e-3nCzcM-2}}}%
\nwixlogsorted{c}{{Yaglom inversion}{NW3gGP3e-4Lsx4s-1}{\nwixu{NW3gGP3e-1zkOI-2}\nwixd{NW3gGP3e-4Lsx4s-1}}}%
\nwixlogsorted{c}{{Zero-radius cycle implementations}{NW3gGP3e-QUZd0-1}{\nwixu{NW3gGP3e-3gefqu-1}\nwixd{NW3gGP3e-QUZd0-1}}}%
\nwixlogsorted{i}{{\nwixident{add}}{add}}%
\nwixlogsorted{i}{{\nwixident{asy{\_}draw}}{asy:undraw}}%
\nwixlogsorted{i}{{\nwixident{asy{\_}path}}{asy:unpath}}%
\nwixlogsorted{i}{{\nwixident{bool}}{bool}}%
\nwixlogsorted{i}{{\nwixident{catch}}{catch}}%
\nwixlogsorted{i}{{\nwixident{center}}{center}}%
\nwixlogsorted{i}{{\nwixident{check{\_}conformality}}{check:unconformality}}%
\nwixlogsorted{i}{{\nwixident{cycle}}{cycle}}%
\nwixlogsorted{i}{{\nwixident{cycle2D}}{cycle2D}}%
\nwixlogsorted{i}{{\nwixident{cycle{\_}product}}{cycle:unproduct}}%
\nwixlogsorted{i}{{\nwixident{cycle{\_}similarity}}{cycle:unsimilarity}}%
\nwixlogsorted{i}{{\nwixident{debug}}{debug}}%
\nwixlogsorted{i}{{\nwixident{det}}{det}}%
\nwixlogsorted{i}{{\nwixident{div}}{div}}%
\nwixlogsorted{i}{{\nwixident{DRAW{\_}ARC}}{DRAW:unARC}}%
\nwixlogsorted{i}{{\nwixident{du}}{du}}%
\nwixlogsorted{i}{{\nwixident{dv}}{dv}}%
\nwixlogsorted{i}{{\nwixident{ex}}{ex}}%
\nwixlogsorted{i}{{\nwixident{exmul}}{exmul}}%
\nwixlogsorted{i}{{\nwixident{expand}}{expand}}%
\nwixlogsorted{i}{{\nwixident{focal{\_}length}}{focal:unlength}}%
\nwixlogsorted{i}{{\nwixident{focal{\_}length{\_}check}}{focal:unlength:uncheck}}%
\nwixlogsorted{i}{{\nwixident{focus}}{focus}}%
\nwixlogsorted{i}{{\nwixident{frames}}{frames}}%
\nwixlogsorted{i}{{\nwixident{get{\_}dim}}{get:undim}}%
\nwixlogsorted{i}{{\nwixident{get{\_}k}}{get:unk}}%
\nwixlogsorted{i}{{\nwixident{get{\_}l}}{get:unl}}%
\nwixlogsorted{i}{{\nwixident{get{\_}m}}{get:unm}}%
\nwixlogsorted{i}{{\nwixident{get{\_}metric}}{get:unmetric}}%
\nwixlogsorted{i}{{\nwixident{get{\_}unit}}{get:ununit}}%
\nwixlogsorted{i}{{\nwixident{GINAC{\_}VERSION{\_}ATLEAST}}{GINAC:unVERSION:unATLEAST}}%
\nwixlogsorted{i}{{\nwixident{hdet}}{hdet}}%
\nwixlogsorted{i}{{\nwixident{hyp{\_}matr}}{hyp:unmatr}}%
\nwixlogsorted{i}{{\nwixident{infinitesimal{\_}calculations}}{infinitesimal:uncalculations}}%
\nwixlogsorted{i}{{\nwixident{is{\_}equal}}{is:unequal}}%
\nwixlogsorted{i}{{\nwixident{is{\_}f{\_}orthogonal}}{is:unf:unorthogonal}}%
\nwixlogsorted{i}{{\nwixident{is{\_}linear}}{is:unlinear}}%
\nwixlogsorted{i}{{\nwixident{is{\_}normalized}}{is:unnormalized}}%
\nwixlogsorted{i}{{\nwixident{is{\_}orthogonal}}{is:unorthogonal}}%
\nwixlogsorted{i}{{\nwixident{is{\_}zero}}{is:unzero}}%
\nwixlogsorted{i}{{\nwixident{jump{\_}fnct}}{jump:unfnct}}%
\nwixlogsorted{i}{{\nwixident{jump{\_}fnct{\_}print{\_}dflt{\_}text}}{jump:unfnct:unprint:undflt:untext}}%
\nwixlogsorted{i}{{\nwixident{jump{\_}fnct{\_}print{\_}latex}}{jump:unfnct:unprint:unlatex}}%
\nwixlogsorted{i}{{\nwixident{k}}{k}}%
\nwixlogsorted{i}{{\nwixident{k{\_}d}}{k:und}}%
\nwixlogsorted{i}{{\nwixident{l}}{l}}%
\nwixlogsorted{i}{{\nwixident{left}}{left}}%
\nwixlogsorted{i}{{\nwixident{let{\_}op}}{let:unop}}%
\nwixlogsorted{i}{{\nwixident{line{\_}intersect}}{line:unintersect}}%
\nwixlogsorted{i}{{\nwixident{lst}}{lst}}%
\nwixlogsorted{i}{{\nwixident{m}}{m}}%
\nwixlogsorted{i}{{\nwixident{main}}{main}}%
\nwixlogsorted{i}{{\nwixident{math{\_}string}}{math:unstring}}%
\nwixlogsorted{i}{{\nwixident{matrix}}{matrix}}%
\nwixlogsorted{i}{{\nwixident{matrix{\_}similarity}}{matrix:unsimilarity}}%
\nwixlogsorted{i}{{\nwixident{metapost{\_}draw}}{metapost:undraw}}%
\nwixlogsorted{i}{{\nwixident{metr}}{metr}}%
\nwixlogsorted{i}{{\nwixident{MoebInv}}{MoebInv}}%
\nwixlogsorted{i}{{\nwixident{MOEBINV{\_}MAJOR{\_}VERSION}}{MOEBINV:unMAJOR:unVERSION}}%
\nwixlogsorted{i}{{\nwixident{MOEBINV{\_}MINOR{\_}VERSION}}{MOEBINV:unMINOR:unVERSION}}%
\nwixlogsorted{i}{{\nwixident{moebius{\_}map}}{moebius:unmap}}%
\nwixlogsorted{i}{{\nwixident{mul}}{mul}}%
\nwixlogsorted{i}{{\nwixident{nops}}{nops}}%
\nwixlogsorted{i}{{\nwixident{normal}}{normal}}%
\nwixlogsorted{i}{{\nwixident{normalize}}{normalize}}%
\nwixlogsorted{i}{{\nwixident{normalize{\_}det}}{normalize:undet}}%
\nwixlogsorted{i}{{\nwixident{normalize{\_}norm}}{normalize:unnorm}}%
\nwixlogsorted{i}{{\nwixident{numeric}}{numeric}}%
\nwixlogsorted{i}{{\nwixident{op}}{op}}%
\nwixlogsorted{i}{{\nwixident{operator*}}{operator*}}%
\nwixlogsorted{i}{{\nwixident{operator+}}{operator+}}%
\nwixlogsorted{i}{{\nwixident{operator-}}{operator-}}%
\nwixlogsorted{i}{{\nwixident{operator/}}{operator/}}%
\nwixlogsorted{i}{{\nwixident{par{\_}matr}}{par:unmatr}}%
\nwixlogsorted{i}{{\nwixident{passing}}{passing}}%
\nwixlogsorted{i}{{\nwixident{points}}{points}}%
\nwixlogsorted{i}{{\nwixident{PRINT{\_}CYCLE}}{PRINT:unCYCLE}}%
\nwixlogsorted{i}{{\nwixident{print{\_}perpendicular}}{print:unperpendicular}}%
\nwixlogsorted{i}{{\nwixident{r1}}{r1}}%
\nwixlogsorted{i}{{\nwixident{radius{\_}sq}}{radius:unsq}}%
\nwixlogsorted{i}{{\nwixident{realsymbol}}{realsymbol}}%
\nwixlogsorted{i}{{\nwixident{roots}}{roots}}%
\nwixlogsorted{i}{{\nwixident{si}}{si}}%
\nwixlogsorted{i}{{\nwixident{si1}}{si1}}%
\nwixlogsorted{i}{{\nwixident{sl2{\_}similarity}}{sl2:unsimilarity}}%
\nwixlogsorted{i}{{\nwixident{string}}{string}}%
\nwixlogsorted{i}{{\nwixident{sub}}{sub}}%
\nwixlogsorted{i}{{\nwixident{subject{\_}to}}{subject:unto}}%
\nwixlogsorted{i}{{\nwixident{subs}}{subs}}%
\nwixlogsorted{i}{{\nwixident{to{\_}matrix}}{to:unmatrix}}%
\nwixlogsorted{i}{{\nwixident{u}}{u}}%
\nwixlogsorted{i}{{\nwixident{v}}{v}}%
\nwixlogsorted{i}{{\nwixident{val}}{val}}%
\nwixlogsorted{i}{{\nwixident{varidx}}{varidx}}%
\nwixlogsorted{i}{{\nwixident{wspaces}}{wspaces}}%
\nwixlogsorted{i}{{\nwixident{zero{\_}or{\_}one}}{zero:unor:unone}}%
\nwbegindocs{715}\nwdocspar

\end{document}